\pdfoutput=1
\documentclass[12pt]{article}
\usepackage[margin=1in]{geometry}
\usepackage{setspace}
\usepackage{amsmath,amssymb}
\usepackage{booktabs,threeparttable,array,adjustbox}
\usepackage[round,authoryear]{natbib}
\usepackage[unicode,hidelinks]{hyperref}
\usepackage{caption}
\usepackage{pgfplots}
\pgfplotsset{compat=1.18}
\usepackage[section]{placeins}
\newcommand{\EERnum}[1]{\ifcsname EER.#1\endcsname\csname EER.#1\endcsname\else\PackageError{EER numbers}{Missing result #1}{Rerun the complete Stata display exporter.}\fi}
\expandafter\def\csname EER.model.restricted.linear.k1.hr\endcsname{3.26}
\expandafter\def\csname EER.model.restricted.linear.k1.lo\endcsname{2.67}
\expandafter\def\csname EER.model.restricted.linear.k1.hi\endcsname{3.97}
\expandafter\def\csname EER.model.restricted.linear.k1.ci\endcsname{[2.67, 3.97]}
\expandafter\def\csname EER.model.restricted.linear.k1.increase_pct\endcsname{226}
\expandafter\def\csname EER.model.restricted.linear.k1.women\endcsname{2,749}
\expandafter\def\csname EER.model.restricted.linear.k1.months\endcsname{57,991}
\expandafter\def\csname EER.model.restricted.linear.k1.events\endcsname{501}
\expandafter\def\csname EER.model.restricted.linear.k1.conceptions\endcsname{443}
\expandafter\def\csname EER.model.restricted.linear.k1.exposed\endcsname{5,700}
\expandafter\def\csname EER.model.restricted.linear.k2.hr\endcsname{3.08}
\expandafter\def\csname EER.model.restricted.linear.k2.lo\endcsname{2.36}
\expandafter\def\csname EER.model.restricted.linear.k2.hi\endcsname{4.04}
\expandafter\def\csname EER.model.restricted.linear.k2.ci\endcsname{[2.36, 4.04]}
\expandafter\def\csname EER.model.restricted.linear.k2.increase_pct\endcsname{208}
\expandafter\def\csname EER.model.restricted.linear.k2.women\endcsname{2,749}
\expandafter\def\csname EER.model.restricted.linear.k2.months\endcsname{57,991}
\expandafter\def\csname EER.model.restricted.linear.k2.events\endcsname{501}
\expandafter\def\csname EER.model.restricted.linear.k2.conceptions\endcsname{443}
\expandafter\def\csname EER.model.restricted.linear.k2.exposed\endcsname{5,700}
\expandafter\def\csname EER.model.restricted.flexible.k1.hr\endcsname{3.25}
\expandafter\def\csname EER.model.restricted.flexible.k1.lo\endcsname{2.66}
\expandafter\def\csname EER.model.restricted.flexible.k1.hi\endcsname{3.97}
\expandafter\def\csname EER.model.restricted.flexible.k1.ci\endcsname{[2.66, 3.97]}
\expandafter\def\csname EER.model.restricted.flexible.k1.increase_pct\endcsname{225}
\expandafter\def\csname EER.model.restricted.flexible.k1.women\endcsname{2,749}
\expandafter\def\csname EER.model.restricted.flexible.k1.months\endcsname{57,991}
\expandafter\def\csname EER.model.restricted.flexible.k1.events\endcsname{501}
\expandafter\def\csname EER.model.restricted.flexible.k1.conceptions\endcsname{443}
\expandafter\def\csname EER.model.restricted.flexible.k1.exposed\endcsname{5,700}
\expandafter\def\csname EER.model.restricted.flexible.k2.hr\endcsname{3.10}
\expandafter\def\csname EER.model.restricted.flexible.k2.lo\endcsname{2.37}
\expandafter\def\csname EER.model.restricted.flexible.k2.hi\endcsname{4.07}
\expandafter\def\csname EER.model.restricted.flexible.k2.ci\endcsname{[2.37, 4.07]}
\expandafter\def\csname EER.model.restricted.flexible.k2.increase_pct\endcsname{210}
\expandafter\def\csname EER.model.restricted.flexible.k2.women\endcsname{2,749}
\expandafter\def\csname EER.model.restricted.flexible.k2.months\endcsname{57,991}
\expandafter\def\csname EER.model.restricted.flexible.k2.events\endcsname{501}
\expandafter\def\csname EER.model.restricted.flexible.k2.conceptions\endcsname{443}
\expandafter\def\csname EER.model.restricted.flexible.k2.exposed\endcsname{5,700}
\expandafter\def\csname EER.model.full.linear.k1.hr\endcsname{3.35}
\expandafter\def\csname EER.model.full.linear.k1.lo\endcsname{2.75}
\expandafter\def\csname EER.model.full.linear.k1.hi\endcsname{4.08}
\expandafter\def\csname EER.model.full.linear.k1.ci\endcsname{[2.75, 4.08]}
\expandafter\def\csname EER.model.full.linear.k1.increase_pct\endcsname{235}
\expandafter\def\csname EER.model.full.linear.k1.women\endcsname{2,951}
\expandafter\def\csname EER.model.full.linear.k1.months\endcsname{61,649}
\expandafter\def\csname EER.model.full.linear.k1.events\endcsname{594}
\expandafter\def\csname EER.model.full.linear.k1.conceptions\endcsname{443}
\expandafter\def\csname EER.model.full.linear.k1.exposed\endcsname{8,593}
\expandafter\def\csname EER.model.full.linear.k2.hr\endcsname{3.23}
\expandafter\def\csname EER.model.full.linear.k2.lo\endcsname{2.48}
\expandafter\def\csname EER.model.full.linear.k2.hi\endcsname{4.21}
\expandafter\def\csname EER.model.full.linear.k2.ci\endcsname{[2.48, 4.21]}
\expandafter\def\csname EER.model.full.linear.k2.increase_pct\endcsname{223}
\expandafter\def\csname EER.model.full.linear.k2.women\endcsname{2,951}
\expandafter\def\csname EER.model.full.linear.k2.months\endcsname{61,649}
\expandafter\def\csname EER.model.full.linear.k2.events\endcsname{594}
\expandafter\def\csname EER.model.full.linear.k2.conceptions\endcsname{443}
\expandafter\def\csname EER.model.full.linear.k2.exposed\endcsname{8,593}
\expandafter\def\csname EER.model.full.flexible.k1.hr\endcsname{3.34}
\expandafter\def\csname EER.model.full.flexible.k1.lo\endcsname{2.74}
\expandafter\def\csname EER.model.full.flexible.k1.hi\endcsname{4.07}
\expandafter\def\csname EER.model.full.flexible.k1.ci\endcsname{[2.74, 4.07]}
\expandafter\def\csname EER.model.full.flexible.k1.increase_pct\endcsname{234}
\expandafter\def\csname EER.model.full.flexible.k1.women\endcsname{2,951}
\expandafter\def\csname EER.model.full.flexible.k1.months\endcsname{61,649}
\expandafter\def\csname EER.model.full.flexible.k1.events\endcsname{594}
\expandafter\def\csname EER.model.full.flexible.k1.conceptions\endcsname{443}
\expandafter\def\csname EER.model.full.flexible.k1.exposed\endcsname{8,593}
\expandafter\def\csname EER.model.full.flexible.k2.hr\endcsname{3.23}
\expandafter\def\csname EER.model.full.flexible.k2.lo\endcsname{2.47}
\expandafter\def\csname EER.model.full.flexible.k2.hi\endcsname{4.21}
\expandafter\def\csname EER.model.full.flexible.k2.ci\endcsname{[2.47, 4.21]}
\expandafter\def\csname EER.model.full.flexible.k2.increase_pct\endcsname{223}
\expandafter\def\csname EER.model.full.flexible.k2.women\endcsname{2,951}
\expandafter\def\csname EER.model.full.flexible.k2.months\endcsname{61,649}
\expandafter\def\csname EER.model.full.flexible.k2.events\endcsname{594}
\expandafter\def\csname EER.model.full.flexible.k2.conceptions\endcsname{443}
\expandafter\def\csname EER.model.full.flexible.k2.exposed\endcsname{8,593}
\expandafter\def\csname EER.model.debut.linear.k1.hr\endcsname{3.21}
\expandafter\def\csname EER.model.debut.linear.k1.lo\endcsname{2.58}
\expandafter\def\csname EER.model.debut.linear.k1.hi\endcsname{3.99}
\expandafter\def\csname EER.model.debut.linear.k1.ci\endcsname{[2.58, 3.99]}
\expandafter\def\csname EER.model.debut.linear.k1.increase_pct\endcsname{221}
\expandafter\def\csname EER.model.debut.linear.k1.women\endcsname{2,673}
\expandafter\def\csname EER.model.debut.linear.k1.months\endcsname{56,471}
\expandafter\def\csname EER.model.debut.linear.k1.events\endcsname{467}
\expandafter\def\csname EER.model.debut.linear.k1.conceptions\endcsname{368}
\expandafter\def\csname EER.model.debut.linear.k1.exposed\endcsname{4,327}
\expandafter\def\csname EER.model.debut.linear.k2.hr\endcsname{9.01}
\expandafter\def\csname EER.model.debut.linear.k2.lo\endcsname{4.40}
\expandafter\def\csname EER.model.debut.linear.k2.hi\endcsname{18.43}
\expandafter\def\csname EER.model.debut.linear.k2.ci\endcsname{[4.40, 18.43]}
\expandafter\def\csname EER.model.debut.linear.k2.increase_pct\endcsname{801}
\expandafter\def\csname EER.model.debut.linear.k2.women\endcsname{2,673}
\expandafter\def\csname EER.model.debut.linear.k2.months\endcsname{56,471}
\expandafter\def\csname EER.model.debut.linear.k2.events\endcsname{467}
\expandafter\def\csname EER.model.debut.linear.k2.conceptions\endcsname{368}
\expandafter\def\csname EER.model.debut.linear.k2.exposed\endcsname{4,327}
\expandafter\def\csname EER.model.debut.flexible.k1.hr\endcsname{3.22}
\expandafter\def\csname EER.model.debut.flexible.k1.lo\endcsname{2.58}
\expandafter\def\csname EER.model.debut.flexible.k1.hi\endcsname{4.01}
\expandafter\def\csname EER.model.debut.flexible.k1.ci\endcsname{[2.58, 4.01]}
\expandafter\def\csname EER.model.debut.flexible.k1.increase_pct\endcsname{222}
\expandafter\def\csname EER.model.debut.flexible.k1.women\endcsname{2,673}
\expandafter\def\csname EER.model.debut.flexible.k1.months\endcsname{56,471}
\expandafter\def\csname EER.model.debut.flexible.k1.events\endcsname{467}
\expandafter\def\csname EER.model.debut.flexible.k1.conceptions\endcsname{368}
\expandafter\def\csname EER.model.debut.flexible.k1.exposed\endcsname{4,327}
\expandafter\def\csname EER.model.debut.flexible.k2.hr\endcsname{8.42}
\expandafter\def\csname EER.model.debut.flexible.k2.lo\endcsname{4.31}
\expandafter\def\csname EER.model.debut.flexible.k2.hi\endcsname{16.46}
\expandafter\def\csname EER.model.debut.flexible.k2.ci\endcsname{[4.31, 16.46]}
\expandafter\def\csname EER.model.debut.flexible.k2.increase_pct\endcsname{742}
\expandafter\def\csname EER.model.debut.flexible.k2.women\endcsname{2,673}
\expandafter\def\csname EER.model.debut.flexible.k2.months\endcsname{56,471}
\expandafter\def\csname EER.model.debut.flexible.k2.events\endcsname{467}
\expandafter\def\csname EER.model.debut.flexible.k2.conceptions\endcsname{368}
\expandafter\def\csname EER.model.debut.flexible.k2.exposed\endcsname{4,327}
\expandafter\def\csname EER.alternative.school3.full.hr\endcsname{5.52}
\expandafter\def\csname EER.alternative.school3.full.lo\endcsname{4.71}
\expandafter\def\csname EER.alternative.school3.full.hi\endcsname{6.46}
\expandafter\def\csname EER.alternative.school3.full.ci\endcsname{[4.71, 6.46]}
\expandafter\def\csname EER.alternative.school3.full.women\endcsname{4,205}
\expandafter\def\csname EER.alternative.school3.full.months\endcsname{120,582}
\expandafter\def\csname EER.alternative.school3.full.events\endcsname{910}
\expandafter\def\csname EER.alternative.school3.full.exposed\endcsname{5,939}
\expandafter\def\csname EER.alternative.school3.postdebut.hr\endcsname{3.85}
\expandafter\def\csname EER.alternative.school3.postdebut.lo\endcsname{3.29}
\expandafter\def\csname EER.alternative.school3.postdebut.hi\endcsname{4.51}
\expandafter\def\csname EER.alternative.school3.postdebut.ci\endcsname{[3.29, 4.51]}
\expandafter\def\csname EER.alternative.school3.postdebut.women\endcsname{2,828}
\expandafter\def\csname EER.alternative.school3.postdebut.months\endcsname{48,953}
\expandafter\def\csname EER.alternative.school3.postdebut.events\endcsname{711}
\expandafter\def\csname EER.alternative.school3.postdebut.exposed\endcsname{5,939}
\expandafter\def\csname EER.alternative.school3.unmodified_debut.hr\endcsname{3.83}
\expandafter\def\csname EER.alternative.school3.unmodified_debut.lo\endcsname{3.21}
\expandafter\def\csname EER.alternative.school3.unmodified_debut.hi\endcsname{4.58}
\expandafter\def\csname EER.alternative.school3.unmodified_debut.ci\endcsname{[3.21, 4.58]}
\expandafter\def\csname EER.alternative.school3.unmodified_debut.women\endcsname{2,708}
\expandafter\def\csname EER.alternative.school3.unmodified_debut.months\endcsname{47,300}
\expandafter\def\csname EER.alternative.school3.unmodified_debut.events\endcsname{643}
\expandafter\def\csname EER.alternative.school3.unmodified_debut.exposed\endcsname{4,411}
\expandafter\def\csname EER.alternative.school3.preonset_enrollment.hr\endcsname{3.79}
\expandafter\def\csname EER.alternative.school3.preonset_enrollment.lo\endcsname{3.16}
\expandafter\def\csname EER.alternative.school3.preonset_enrollment.hi\endcsname{4.56}
\expandafter\def\csname EER.alternative.school3.preonset_enrollment.ci\endcsname{[3.16, 4.56]}
\expandafter\def\csname EER.alternative.school3.preonset_enrollment.women\endcsname{2,672}
\expandafter\def\csname EER.alternative.school3.preonset_enrollment.months\endcsname{46,811}
\expandafter\def\csname EER.alternative.school3.preonset_enrollment.events\endcsname{628}
\expandafter\def\csname EER.alternative.school3.preonset_enrollment.exposed\endcsname{3,927}
\expandafter\def\csname EER.alternative.calendar4.full.hr\endcsname{5.37}
\expandafter\def\csname EER.alternative.calendar4.full.lo\endcsname{4.59}
\expandafter\def\csname EER.alternative.calendar4.full.hi\endcsname{6.28}
\expandafter\def\csname EER.alternative.calendar4.full.ci\endcsname{[4.59, 6.28]}
\expandafter\def\csname EER.alternative.calendar4.full.women\endcsname{4,205}
\expandafter\def\csname EER.alternative.calendar4.full.months\endcsname{119,261}
\expandafter\def\csname EER.alternative.calendar4.full.events\endcsname{950}
\expandafter\def\csname EER.alternative.calendar4.full.exposed\endcsname{5,804}
\expandafter\def\csname EER.alternative.calendar4.postdebut.hr\endcsname{3.72}
\expandafter\def\csname EER.alternative.calendar4.postdebut.lo\endcsname{3.18}
\expandafter\def\csname EER.alternative.calendar4.postdebut.hi\endcsname{4.35}
\expandafter\def\csname EER.alternative.calendar4.postdebut.ci\endcsname{[3.18, 4.35]}
\expandafter\def\csname EER.alternative.calendar4.postdebut.women\endcsname{2,809}
\expandafter\def\csname EER.alternative.calendar4.postdebut.months\endcsname{48,120}
\expandafter\def\csname EER.alternative.calendar4.postdebut.events\endcsname{726}
\expandafter\def\csname EER.alternative.calendar4.postdebut.exposed\endcsname{5,804}
\expandafter\def\csname EER.alternative.calendar4.unmodified_debut.hr\endcsname{3.73}
\expandafter\def\csname EER.alternative.calendar4.unmodified_debut.lo\endcsname{3.13}
\expandafter\def\csname EER.alternative.calendar4.unmodified_debut.hi\endcsname{4.46}
\expandafter\def\csname EER.alternative.calendar4.unmodified_debut.ci\endcsname{[3.13, 4.46]}
\expandafter\def\csname EER.alternative.calendar4.unmodified_debut.women\endcsname{2,689}
\expandafter\def\csname EER.alternative.calendar4.unmodified_debut.months\endcsname{46,482}
\expandafter\def\csname EER.alternative.calendar4.unmodified_debut.events\endcsname{659}
\expandafter\def\csname EER.alternative.calendar4.unmodified_debut.exposed\endcsname{4,291}
\expandafter\def\csname EER.alternative.calendar4.preonset_enrollment.hr\endcsname{3.61}
\expandafter\def\csname EER.alternative.calendar4.preonset_enrollment.lo\endcsname{3.01}
\expandafter\def\csname EER.alternative.calendar4.preonset_enrollment.hi\endcsname{4.35}
\expandafter\def\csname EER.alternative.calendar4.preonset_enrollment.ci\endcsname{[3.01, 4.35]}
\expandafter\def\csname EER.alternative.calendar4.preonset_enrollment.women\endcsname{2,655}
\expandafter\def\csname EER.alternative.calendar4.preonset_enrollment.months\endcsname{46,059}
\expandafter\def\csname EER.alternative.calendar4.preonset_enrollment.events\endcsname{641}
\expandafter\def\csname EER.alternative.calendar4.preonset_enrollment.exposed\endcsname{3,830}
\expandafter\def\csname EER.birth.calendar4.1.phase1.hr\endcsname{4.35}
\expandafter\def\csname EER.birth.calendar4.1.phase1.lo\endcsname{3.19}
\expandafter\def\csname EER.birth.calendar4.1.phase1.hi\endcsname{5.93}
\expandafter\def\csname EER.birth.calendar4.1.phase1.ci\endcsname{[3.19, 5.93]}
\expandafter\def\csname EER.birth.calendar4.1.phase1.p\endcsname{}
\expandafter\def\csname EER.birth.calendar4.1.phase1.women\endcsname{2,777}
\expandafter\def\csname EER.birth.calendar4.1.phase1.months\endcsname{47,466}
\expandafter\def\csname EER.birth.calendar4.1.phase1.events\endcsname{709}
\expandafter\def\csname EER.birth.calendar4.1.phase1.exposed\endcsname{5,496}
\expandafter\def\csname EER.birth.calendar4.1.phase2.hr\endcsname{2.61}
\expandafter\def\csname EER.birth.calendar4.1.phase2.lo\endcsname{1.67}
\expandafter\def\csname EER.birth.calendar4.1.phase2.hi\endcsname{4.06}
\expandafter\def\csname EER.birth.calendar4.1.phase2.ci\endcsname{[1.67, 4.06]}
\expandafter\def\csname EER.birth.calendar4.1.phase2.p\endcsname{}
\expandafter\def\csname EER.birth.calendar4.1.phase2.women\endcsname{2,777}
\expandafter\def\csname EER.birth.calendar4.1.phase2.months\endcsname{47,466}
\expandafter\def\csname EER.birth.calendar4.1.phase2.events\endcsname{709}
\expandafter\def\csname EER.birth.calendar4.1.phase2.exposed\endcsname{5,496}
\expandafter\def\csname EER.birth.calendar4.1.phase3.hr\endcsname{4.61}
\expandafter\def\csname EER.birth.calendar4.1.phase3.lo\endcsname{3.04}
\expandafter\def\csname EER.birth.calendar4.1.phase3.hi\endcsname{6.98}
\expandafter\def\csname EER.birth.calendar4.1.phase3.ci\endcsname{[3.04, 6.98]}
\expandafter\def\csname EER.birth.calendar4.1.phase3.p\endcsname{}
\expandafter\def\csname EER.birth.calendar4.1.phase3.women\endcsname{2,777}
\expandafter\def\csname EER.birth.calendar4.1.phase3.months\endcsname{47,466}
\expandafter\def\csname EER.birth.calendar4.1.phase3.events\endcsname{709}
\expandafter\def\csname EER.birth.calendar4.1.phase3.exposed\endcsname{5,496}
\expandafter\def\csname EER.birth.calendar4.1.phase4.hr\endcsname{4.93}
\expandafter\def\csname EER.birth.calendar4.1.phase4.lo\endcsname{3.42}
\expandafter\def\csname EER.birth.calendar4.1.phase4.hi\endcsname{7.12}
\expandafter\def\csname EER.birth.calendar4.1.phase4.ci\endcsname{[3.42, 7.12]}
\expandafter\def\csname EER.birth.calendar4.1.phase4.p\endcsname{}
\expandafter\def\csname EER.birth.calendar4.1.phase4.women\endcsname{2,777}
\expandafter\def\csname EER.birth.calendar4.1.phase4.months\endcsname{47,466}
\expandafter\def\csname EER.birth.calendar4.1.phase4.events\endcsname{709}
\expandafter\def\csname EER.birth.calendar4.1.phase4.exposed\endcsname{5,496}
\expandafter\def\csname EER.birth.calendar4.1.phase5.hr\endcsname{2.53}
\expandafter\def\csname EER.birth.calendar4.1.phase5.lo\endcsname{1.51}
\expandafter\def\csname EER.birth.calendar4.1.phase5.hi\endcsname{4.23}
\expandafter\def\csname EER.birth.calendar4.1.phase5.ci\endcsname{[1.51, 4.23]}
\expandafter\def\csname EER.birth.calendar4.1.phase5.p\endcsname{}
\expandafter\def\csname EER.birth.calendar4.1.phase5.women\endcsname{2,777}
\expandafter\def\csname EER.birth.calendar4.1.phase5.months\endcsname{47,466}
\expandafter\def\csname EER.birth.calendar4.1.phase5.events\endcsname{709}
\expandafter\def\csname EER.birth.calendar4.1.phase5.exposed\endcsname{5,496}
\expandafter\def\csname EER.birth.calendar4.1.phase6.hr\endcsname{3.49}
\expandafter\def\csname EER.birth.calendar4.1.phase6.lo\endcsname{2.81}
\expandafter\def\csname EER.birth.calendar4.1.phase6.hi\endcsname{4.33}
\expandafter\def\csname EER.birth.calendar4.1.phase6.ci\endcsname{[2.81, 4.33]}
\expandafter\def\csname EER.birth.calendar4.1.phase6.p\endcsname{}
\expandafter\def\csname EER.birth.calendar4.1.phase6.women\endcsname{2,777}
\expandafter\def\csname EER.birth.calendar4.1.phase6.months\endcsname{47,466}
\expandafter\def\csname EER.birth.calendar4.1.phase6.events\endcsname{709}
\expandafter\def\csname EER.birth.calendar4.1.phase6.exposed\endcsname{5,496}
\expandafter\def\csname EER.birth.calendar4.1.prenatal.hr\endcsname{3.81}
\expandafter\def\csname EER.birth.calendar4.1.prenatal.lo\endcsname{3.03}
\expandafter\def\csname EER.birth.calendar4.1.prenatal.hi\endcsname{4.80}
\expandafter\def\csname EER.birth.calendar4.1.prenatal.ci\endcsname{[3.03, 4.80]}
\expandafter\def\csname EER.birth.calendar4.1.prenatal.p\endcsname{0.654}
\expandafter\def\csname EER.birth.calendar4.1.prenatal.women\endcsname{2,777}
\expandafter\def\csname EER.birth.calendar4.1.prenatal.months\endcsname{47,466}
\expandafter\def\csname EER.birth.calendar4.1.prenatal.events\endcsname{709}
\expandafter\def\csname EER.birth.calendar4.1.prenatal.exposed\endcsname{5,496}
\expandafter\def\csname EER.birth.calendar4.1.birth_or_after.hr\endcsname{3.58}
\expandafter\def\csname EER.birth.calendar4.1.birth_or_after.lo\endcsname{2.96}
\expandafter\def\csname EER.birth.calendar4.1.birth_or_after.hi\endcsname{4.33}
\expandafter\def\csname EER.birth.calendar4.1.birth_or_after.ci\endcsname{[2.96, 4.33]}
\expandafter\def\csname EER.birth.calendar4.1.birth_or_after.p\endcsname{0.654}
\expandafter\def\csname EER.birth.calendar4.1.birth_or_after.women\endcsname{2,777}
\expandafter\def\csname EER.birth.calendar4.1.birth_or_after.months\endcsname{47,466}
\expandafter\def\csname EER.birth.calendar4.1.birth_or_after.events\endcsname{709}
\expandafter\def\csname EER.birth.calendar4.1.birth_or_after.exposed\endcsname{5,496}
\expandafter\def\csname EER.birth.calendar4.0.phase1.hr\endcsname{4.51}
\expandafter\def\csname EER.birth.calendar4.0.phase1.lo\endcsname{3.28}
\expandafter\def\csname EER.birth.calendar4.0.phase1.hi\endcsname{6.22}
\expandafter\def\csname EER.birth.calendar4.0.phase1.ci\endcsname{[3.28, 6.22]}
\expandafter\def\csname EER.birth.calendar4.0.phase1.p\endcsname{}
\expandafter\def\csname EER.birth.calendar4.0.phase1.women\endcsname{2,703}
\expandafter\def\csname EER.birth.calendar4.0.phase1.months\endcsname{45,954}
\expandafter\def\csname EER.birth.calendar4.0.phase1.events\endcsname{677}
\expandafter\def\csname EER.birth.calendar4.0.phase1.exposed\endcsname{4,992}
\expandafter\def\csname EER.birth.calendar4.0.phase2.hr\endcsname{2.03}
\expandafter\def\csname EER.birth.calendar4.0.phase2.lo\endcsname{1.19}
\expandafter\def\csname EER.birth.calendar4.0.phase2.hi\endcsname{3.47}
\expandafter\def\csname EER.birth.calendar4.0.phase2.ci\endcsname{[1.19, 3.47]}
\expandafter\def\csname EER.birth.calendar4.0.phase2.p\endcsname{}
\expandafter\def\csname EER.birth.calendar4.0.phase2.women\endcsname{2,703}
\expandafter\def\csname EER.birth.calendar4.0.phase2.months\endcsname{45,954}
\expandafter\def\csname EER.birth.calendar4.0.phase2.events\endcsname{677}
\expandafter\def\csname EER.birth.calendar4.0.phase2.exposed\endcsname{4,992}
\expandafter\def\csname EER.birth.calendar4.0.phase3.hr\endcsname{3.93}
\expandafter\def\csname EER.birth.calendar4.0.phase3.lo\endcsname{2.43}
\expandafter\def\csname EER.birth.calendar4.0.phase3.hi\endcsname{6.35}
\expandafter\def\csname EER.birth.calendar4.0.phase3.ci\endcsname{[2.43, 6.35]}
\expandafter\def\csname EER.birth.calendar4.0.phase3.p\endcsname{}
\expandafter\def\csname EER.birth.calendar4.0.phase3.women\endcsname{2,703}
\expandafter\def\csname EER.birth.calendar4.0.phase3.months\endcsname{45,954}
\expandafter\def\csname EER.birth.calendar4.0.phase3.events\endcsname{677}
\expandafter\def\csname EER.birth.calendar4.0.phase3.exposed\endcsname{4,992}
\expandafter\def\csname EER.birth.calendar4.0.phase4.hr\endcsname{4.81}
\expandafter\def\csname EER.birth.calendar4.0.phase4.lo\endcsname{3.25}
\expandafter\def\csname EER.birth.calendar4.0.phase4.hi\endcsname{7.11}
\expandafter\def\csname EER.birth.calendar4.0.phase4.ci\endcsname{[3.25, 7.11]}
\expandafter\def\csname EER.birth.calendar4.0.phase4.p\endcsname{}
\expandafter\def\csname EER.birth.calendar4.0.phase4.women\endcsname{2,703}
\expandafter\def\csname EER.birth.calendar4.0.phase4.months\endcsname{45,954}
\expandafter\def\csname EER.birth.calendar4.0.phase4.events\endcsname{677}
\expandafter\def\csname EER.birth.calendar4.0.phase4.exposed\endcsname{4,992}
\expandafter\def\csname EER.birth.calendar4.0.phase5.hr\endcsname{2.57}
\expandafter\def\csname EER.birth.calendar4.0.phase5.lo\endcsname{1.50}
\expandafter\def\csname EER.birth.calendar4.0.phase5.hi\endcsname{4.38}
\expandafter\def\csname EER.birth.calendar4.0.phase5.ci\endcsname{[1.50, 4.38]}
\expandafter\def\csname EER.birth.calendar4.0.phase5.p\endcsname{}
\expandafter\def\csname EER.birth.calendar4.0.phase5.women\endcsname{2,703}
\expandafter\def\csname EER.birth.calendar4.0.phase5.months\endcsname{45,954}
\expandafter\def\csname EER.birth.calendar4.0.phase5.events\endcsname{677}
\expandafter\def\csname EER.birth.calendar4.0.phase5.exposed\endcsname{4,992}
\expandafter\def\csname EER.birth.calendar4.0.phase6.hr\endcsname{3.44}
\expandafter\def\csname EER.birth.calendar4.0.phase6.lo\endcsname{2.76}
\expandafter\def\csname EER.birth.calendar4.0.phase6.hi\endcsname{4.28}
\expandafter\def\csname EER.birth.calendar4.0.phase6.ci\endcsname{[2.76, 4.28]}
\expandafter\def\csname EER.birth.calendar4.0.phase6.p\endcsname{}
\expandafter\def\csname EER.birth.calendar4.0.phase6.women\endcsname{2,703}
\expandafter\def\csname EER.birth.calendar4.0.phase6.months\endcsname{45,954}
\expandafter\def\csname EER.birth.calendar4.0.phase6.events\endcsname{677}
\expandafter\def\csname EER.birth.calendar4.0.phase6.exposed\endcsname{4,992}
\expandafter\def\csname EER.birth.calendar4.0.prenatal.hr\endcsname{3.54}
\expandafter\def\csname EER.birth.calendar4.0.prenatal.lo\endcsname{2.75}
\expandafter\def\csname EER.birth.calendar4.0.prenatal.hi\endcsname{4.56}
\expandafter\def\csname EER.birth.calendar4.0.prenatal.ci\endcsname{[2.75, 4.56]}
\expandafter\def\csname EER.birth.calendar4.0.prenatal.p\endcsname{0.982}
\expandafter\def\csname EER.birth.calendar4.0.prenatal.women\endcsname{2,703}
\expandafter\def\csname EER.birth.calendar4.0.prenatal.months\endcsname{45,954}
\expandafter\def\csname EER.birth.calendar4.0.prenatal.events\endcsname{677}
\expandafter\def\csname EER.birth.calendar4.0.prenatal.exposed\endcsname{4,992}
\expandafter\def\csname EER.birth.calendar4.0.birth_or_after.hr\endcsname{3.53}
\expandafter\def\csname EER.birth.calendar4.0.birth_or_after.lo\endcsname{2.91}
\expandafter\def\csname EER.birth.calendar4.0.birth_or_after.hi\endcsname{4.29}
\expandafter\def\csname EER.birth.calendar4.0.birth_or_after.ci\endcsname{[2.91, 4.29]}
\expandafter\def\csname EER.birth.calendar4.0.birth_or_after.p\endcsname{0.982}
\expandafter\def\csname EER.birth.calendar4.0.birth_or_after.women\endcsname{2,703}
\expandafter\def\csname EER.birth.calendar4.0.birth_or_after.months\endcsname{45,954}
\expandafter\def\csname EER.birth.calendar4.0.birth_or_after.events\endcsname{677}
\expandafter\def\csname EER.birth.calendar4.0.birth_or_after.exposed\endcsname{4,992}
\expandafter\def\csname EER.birth.calendar4.3.phase1.hr\endcsname{4.27}
\expandafter\def\csname EER.birth.calendar4.3.phase1.lo\endcsname{3.13}
\expandafter\def\csname EER.birth.calendar4.3.phase1.hi\endcsname{5.82}
\expandafter\def\csname EER.birth.calendar4.3.phase1.ci\endcsname{[3.13, 5.82]}
\expandafter\def\csname EER.birth.calendar4.3.phase1.p\endcsname{}
\expandafter\def\csname EER.birth.calendar4.3.phase1.women\endcsname{2,790}
\expandafter\def\csname EER.birth.calendar4.3.phase1.months\endcsname{47,731}
\expandafter\def\csname EER.birth.calendar4.3.phase1.events\endcsname{717}
\expandafter\def\csname EER.birth.calendar4.3.phase1.exposed\endcsname{5,617}
\expandafter\def\csname EER.birth.calendar4.3.phase2.hr\endcsname{2.79}
\expandafter\def\csname EER.birth.calendar4.3.phase2.lo\endcsname{1.83}
\expandafter\def\csname EER.birth.calendar4.3.phase2.hi\endcsname{4.26}
\expandafter\def\csname EER.birth.calendar4.3.phase2.ci\endcsname{[1.83, 4.26]}
\expandafter\def\csname EER.birth.calendar4.3.phase2.p\endcsname{}
\expandafter\def\csname EER.birth.calendar4.3.phase2.women\endcsname{2,790}
\expandafter\def\csname EER.birth.calendar4.3.phase2.months\endcsname{47,731}
\expandafter\def\csname EER.birth.calendar4.3.phase2.events\endcsname{717}
\expandafter\def\csname EER.birth.calendar4.3.phase2.exposed\endcsname{5,617}
\expandafter\def\csname EER.birth.calendar4.3.phase3.hr\endcsname{4.49}
\expandafter\def\csname EER.birth.calendar4.3.phase3.lo\endcsname{2.99}
\expandafter\def\csname EER.birth.calendar4.3.phase3.hi\endcsname{6.75}
\expandafter\def\csname EER.birth.calendar4.3.phase3.ci\endcsname{[2.99, 6.75]}
\expandafter\def\csname EER.birth.calendar4.3.phase3.p\endcsname{}
\expandafter\def\csname EER.birth.calendar4.3.phase3.women\endcsname{2,790}
\expandafter\def\csname EER.birth.calendar4.3.phase3.months\endcsname{47,731}
\expandafter\def\csname EER.birth.calendar4.3.phase3.events\endcsname{717}
\expandafter\def\csname EER.birth.calendar4.3.phase3.exposed\endcsname{5,617}
\expandafter\def\csname EER.birth.calendar4.3.phase4.hr\endcsname{5.01}
\expandafter\def\csname EER.birth.calendar4.3.phase4.lo\endcsname{3.49}
\expandafter\def\csname EER.birth.calendar4.3.phase4.hi\endcsname{7.19}
\expandafter\def\csname EER.birth.calendar4.3.phase4.ci\endcsname{[3.49, 7.19]}
\expandafter\def\csname EER.birth.calendar4.3.phase4.p\endcsname{}
\expandafter\def\csname EER.birth.calendar4.3.phase4.women\endcsname{2,790}
\expandafter\def\csname EER.birth.calendar4.3.phase4.months\endcsname{47,731}
\expandafter\def\csname EER.birth.calendar4.3.phase4.events\endcsname{717}
\expandafter\def\csname EER.birth.calendar4.3.phase4.exposed\endcsname{5,617}
\expandafter\def\csname EER.birth.calendar4.3.phase5.hr\endcsname{2.82}
\expandafter\def\csname EER.birth.calendar4.3.phase5.lo\endcsname{1.74}
\expandafter\def\csname EER.birth.calendar4.3.phase5.hi\endcsname{4.57}
\expandafter\def\csname EER.birth.calendar4.3.phase5.ci\endcsname{[1.74, 4.57]}
\expandafter\def\csname EER.birth.calendar4.3.phase5.p\endcsname{}
\expandafter\def\csname EER.birth.calendar4.3.phase5.women\endcsname{2,790}
\expandafter\def\csname EER.birth.calendar4.3.phase5.months\endcsname{47,731}
\expandafter\def\csname EER.birth.calendar4.3.phase5.events\endcsname{717}
\expandafter\def\csname EER.birth.calendar4.3.phase5.exposed\endcsname{5,617}
\expandafter\def\csname EER.birth.calendar4.3.phase6.hr\endcsname{3.51}
\expandafter\def\csname EER.birth.calendar4.3.phase6.lo\endcsname{2.83}
\expandafter\def\csname EER.birth.calendar4.3.phase6.hi\endcsname{4.35}
\expandafter\def\csname EER.birth.calendar4.3.phase6.ci\endcsname{[2.83, 4.35]}
\expandafter\def\csname EER.birth.calendar4.3.phase6.p\endcsname{}
\expandafter\def\csname EER.birth.calendar4.3.phase6.women\endcsname{2,790}
\expandafter\def\csname EER.birth.calendar4.3.phase6.months\endcsname{47,731}
\expandafter\def\csname EER.birth.calendar4.3.phase6.events\endcsname{717}
\expandafter\def\csname EER.birth.calendar4.3.phase6.exposed\endcsname{5,617}
\expandafter\def\csname EER.birth.calendar4.3.prenatal.hr\endcsname{3.82}
\expandafter\def\csname EER.birth.calendar4.3.prenatal.lo\endcsname{3.04}
\expandafter\def\csname EER.birth.calendar4.3.prenatal.hi\endcsname{4.79}
\expandafter\def\csname EER.birth.calendar4.3.prenatal.ci\endcsname{[3.04, 4.79]}
\expandafter\def\csname EER.birth.calendar4.3.prenatal.p\endcsname{0.746}
\expandafter\def\csname EER.birth.calendar4.3.prenatal.women\endcsname{2,790}
\expandafter\def\csname EER.birth.calendar4.3.prenatal.months\endcsname{47,731}
\expandafter\def\csname EER.birth.calendar4.3.prenatal.events\endcsname{717}
\expandafter\def\csname EER.birth.calendar4.3.prenatal.exposed\endcsname{5,617}
\expandafter\def\csname EER.birth.calendar4.3.birth_or_after.hr\endcsname{3.65}
\expandafter\def\csname EER.birth.calendar4.3.birth_or_after.lo\endcsname{3.03}
\expandafter\def\csname EER.birth.calendar4.3.birth_or_after.hi\endcsname{4.41}
\expandafter\def\csname EER.birth.calendar4.3.birth_or_after.ci\endcsname{[3.03, 4.41]}
\expandafter\def\csname EER.birth.calendar4.3.birth_or_after.p\endcsname{0.746}
\expandafter\def\csname EER.birth.calendar4.3.birth_or_after.women\endcsname{2,790}
\expandafter\def\csname EER.birth.calendar4.3.birth_or_after.months\endcsname{47,731}
\expandafter\def\csname EER.birth.calendar4.3.birth_or_after.events\endcsname{717}
\expandafter\def\csname EER.birth.calendar4.3.birth_or_after.exposed\endcsname{5,617}
\expandafter\def\csname EER.birth.school3.1.phase1.hr\endcsname{4.59}
\expandafter\def\csname EER.birth.school3.1.phase1.lo\endcsname{3.37}
\expandafter\def\csname EER.birth.school3.1.phase1.hi\endcsname{6.24}
\expandafter\def\csname EER.birth.school3.1.phase1.ci\endcsname{[3.37, 6.24]}
\expandafter\def\csname EER.birth.school3.1.phase1.p\endcsname{}
\expandafter\def\csname EER.birth.school3.1.phase1.women\endcsname{2,796}
\expandafter\def\csname EER.birth.school3.1.phase1.months\endcsname{48,299}
\expandafter\def\csname EER.birth.school3.1.phase1.events\endcsname{694}
\expandafter\def\csname EER.birth.school3.1.phase1.exposed\endcsname{5,631}
\expandafter\def\csname EER.birth.school3.1.phase2.hr\endcsname{2.67}
\expandafter\def\csname EER.birth.school3.1.phase2.lo\endcsname{1.71}
\expandafter\def\csname EER.birth.school3.1.phase2.hi\endcsname{4.15}
\expandafter\def\csname EER.birth.school3.1.phase2.ci\endcsname{[1.71, 4.15]}
\expandafter\def\csname EER.birth.school3.1.phase2.p\endcsname{}
\expandafter\def\csname EER.birth.school3.1.phase2.women\endcsname{2,796}
\expandafter\def\csname EER.birth.school3.1.phase2.months\endcsname{48,299}
\expandafter\def\csname EER.birth.school3.1.phase2.events\endcsname{694}
\expandafter\def\csname EER.birth.school3.1.phase2.exposed\endcsname{5,631}
\expandafter\def\csname EER.birth.school3.1.phase3.hr\endcsname{5.03}
\expandafter\def\csname EER.birth.school3.1.phase3.lo\endcsname{3.35}
\expandafter\def\csname EER.birth.school3.1.phase3.hi\endcsname{7.55}
\expandafter\def\csname EER.birth.school3.1.phase3.ci\endcsname{[3.35, 7.55]}
\expandafter\def\csname EER.birth.school3.1.phase3.p\endcsname{}
\expandafter\def\csname EER.birth.school3.1.phase3.women\endcsname{2,796}
\expandafter\def\csname EER.birth.school3.1.phase3.months\endcsname{48,299}
\expandafter\def\csname EER.birth.school3.1.phase3.events\endcsname{694}
\expandafter\def\csname EER.birth.school3.1.phase3.exposed\endcsname{5,631}
\expandafter\def\csname EER.birth.school3.1.phase4.hr\endcsname{4.82}
\expandafter\def\csname EER.birth.school3.1.phase4.lo\endcsname{3.31}
\expandafter\def\csname EER.birth.school3.1.phase4.hi\endcsname{7.02}
\expandafter\def\csname EER.birth.school3.1.phase4.ci\endcsname{[3.31, 7.02]}
\expandafter\def\csname EER.birth.school3.1.phase4.p\endcsname{}
\expandafter\def\csname EER.birth.school3.1.phase4.women\endcsname{2,796}
\expandafter\def\csname EER.birth.school3.1.phase4.months\endcsname{48,299}
\expandafter\def\csname EER.birth.school3.1.phase4.events\endcsname{694}
\expandafter\def\csname EER.birth.school3.1.phase4.exposed\endcsname{5,631}
\expandafter\def\csname EER.birth.school3.1.phase5.hr\endcsname{2.77}
\expandafter\def\csname EER.birth.school3.1.phase5.lo\endcsname{1.69}
\expandafter\def\csname EER.birth.school3.1.phase5.hi\endcsname{4.56}
\expandafter\def\csname EER.birth.school3.1.phase5.ci\endcsname{[1.69, 4.56]}
\expandafter\def\csname EER.birth.school3.1.phase5.p\endcsname{}
\expandafter\def\csname EER.birth.school3.1.phase5.women\endcsname{2,796}
\expandafter\def\csname EER.birth.school3.1.phase5.months\endcsname{48,299}
\expandafter\def\csname EER.birth.school3.1.phase5.events\endcsname{694}
\expandafter\def\csname EER.birth.school3.1.phase5.exposed\endcsname{5,631}
\expandafter\def\csname EER.birth.school3.1.phase6.hr\endcsname{3.58}
\expandafter\def\csname EER.birth.school3.1.phase6.lo\endcsname{2.88}
\expandafter\def\csname EER.birth.school3.1.phase6.hi\endcsname{4.43}
\expandafter\def\csname EER.birth.school3.1.phase6.ci\endcsname{[2.88, 4.43]}
\expandafter\def\csname EER.birth.school3.1.phase6.p\endcsname{}
\expandafter\def\csname EER.birth.school3.1.phase6.women\endcsname{2,796}
\expandafter\def\csname EER.birth.school3.1.phase6.months\endcsname{48,299}
\expandafter\def\csname EER.birth.school3.1.phase6.events\endcsname{694}
\expandafter\def\csname EER.birth.school3.1.phase6.exposed\endcsname{5,631}
\expandafter\def\csname EER.birth.school3.1.prenatal.hr\endcsname{4.02}
\expandafter\def\csname EER.birth.school3.1.prenatal.lo\endcsname{3.20}
\expandafter\def\csname EER.birth.school3.1.prenatal.hi\endcsname{5.06}
\expandafter\def\csname EER.birth.school3.1.prenatal.ci\endcsname{[3.20, 5.06]}
\expandafter\def\csname EER.birth.school3.1.prenatal.p\endcsname{0.486}
\expandafter\def\csname EER.birth.school3.1.prenatal.women\endcsname{2,796}
\expandafter\def\csname EER.birth.school3.1.prenatal.months\endcsname{48,299}
\expandafter\def\csname EER.birth.school3.1.prenatal.events\endcsname{694}
\expandafter\def\csname EER.birth.school3.1.prenatal.exposed\endcsname{5,631}
\expandafter\def\csname EER.birth.school3.1.birth_or_after.hr\endcsname{3.66}
\expandafter\def\csname EER.birth.school3.1.birth_or_after.lo\endcsname{3.03}
\expandafter\def\csname EER.birth.school3.1.birth_or_after.hi\endcsname{4.42}
\expandafter\def\csname EER.birth.school3.1.birth_or_after.ci\endcsname{[3.03, 4.42]}
\expandafter\def\csname EER.birth.school3.1.birth_or_after.p\endcsname{0.486}
\expandafter\def\csname EER.birth.school3.1.birth_or_after.women\endcsname{2,796}
\expandafter\def\csname EER.birth.school3.1.birth_or_after.months\endcsname{48,299}
\expandafter\def\csname EER.birth.school3.1.birth_or_after.events\endcsname{694}
\expandafter\def\csname EER.birth.school3.1.birth_or_after.exposed\endcsname{5,631}
\expandafter\def\csname EER.birth.school3.0.phase1.hr\endcsname{4.80}
\expandafter\def\csname EER.birth.school3.0.phase1.lo\endcsname{3.50}
\expandafter\def\csname EER.birth.school3.0.phase1.hi\endcsname{6.59}
\expandafter\def\csname EER.birth.school3.0.phase1.ci\endcsname{[3.50, 6.59]}
\expandafter\def\csname EER.birth.school3.0.phase1.p\endcsname{}
\expandafter\def\csname EER.birth.school3.0.phase1.women\endcsname{2,718}
\expandafter\def\csname EER.birth.school3.0.phase1.months\endcsname{46,670}
\expandafter\def\csname EER.birth.school3.0.phase1.events\endcsname{659}
\expandafter\def\csname EER.birth.school3.0.phase1.exposed\endcsname{5,122}
\expandafter\def\csname EER.birth.school3.0.phase2.hr\endcsname{2.07}
\expandafter\def\csname EER.birth.school3.0.phase2.lo\endcsname{1.21}
\expandafter\def\csname EER.birth.school3.0.phase2.hi\endcsname{3.54}
\expandafter\def\csname EER.birth.school3.0.phase2.ci\endcsname{[1.21, 3.54]}
\expandafter\def\csname EER.birth.school3.0.phase2.p\endcsname{}
\expandafter\def\csname EER.birth.school3.0.phase2.women\endcsname{2,718}
\expandafter\def\csname EER.birth.school3.0.phase2.months\endcsname{46,670}
\expandafter\def\csname EER.birth.school3.0.phase2.events\endcsname{659}
\expandafter\def\csname EER.birth.school3.0.phase2.exposed\endcsname{5,122}
\expandafter\def\csname EER.birth.school3.0.phase3.hr\endcsname{4.12}
\expandafter\def\csname EER.birth.school3.0.phase3.lo\endcsname{2.55}
\expandafter\def\csname EER.birth.school3.0.phase3.hi\endcsname{6.66}
\expandafter\def\csname EER.birth.school3.0.phase3.ci\endcsname{[2.55, 6.66]}
\expandafter\def\csname EER.birth.school3.0.phase3.p\endcsname{}
\expandafter\def\csname EER.birth.school3.0.phase3.women\endcsname{2,718}
\expandafter\def\csname EER.birth.school3.0.phase3.months\endcsname{46,670}
\expandafter\def\csname EER.birth.school3.0.phase3.events\endcsname{659}
\expandafter\def\csname EER.birth.school3.0.phase3.exposed\endcsname{5,122}
\expandafter\def\csname EER.birth.school3.0.phase4.hr\endcsname{4.47}
\expandafter\def\csname EER.birth.school3.0.phase4.lo\endcsname{2.96}
\expandafter\def\csname EER.birth.school3.0.phase4.hi\endcsname{6.74}
\expandafter\def\csname EER.birth.school3.0.phase4.ci\endcsname{[2.96, 6.74]}
\expandafter\def\csname EER.birth.school3.0.phase4.p\endcsname{}
\expandafter\def\csname EER.birth.school3.0.phase4.women\endcsname{2,718}
\expandafter\def\csname EER.birth.school3.0.phase4.months\endcsname{46,670}
\expandafter\def\csname EER.birth.school3.0.phase4.events\endcsname{659}
\expandafter\def\csname EER.birth.school3.0.phase4.exposed\endcsname{5,122}
\expandafter\def\csname EER.birth.school3.0.phase5.hr\endcsname{2.79}
\expandafter\def\csname EER.birth.school3.0.phase5.lo\endcsname{1.67}
\expandafter\def\csname EER.birth.school3.0.phase5.hi\endcsname{4.68}
\expandafter\def\csname EER.birth.school3.0.phase5.ci\endcsname{[1.67, 4.68]}
\expandafter\def\csname EER.birth.school3.0.phase5.p\endcsname{}
\expandafter\def\csname EER.birth.school3.0.phase5.women\endcsname{2,718}
\expandafter\def\csname EER.birth.school3.0.phase5.months\endcsname{46,670}
\expandafter\def\csname EER.birth.school3.0.phase5.events\endcsname{659}
\expandafter\def\csname EER.birth.school3.0.phase5.exposed\endcsname{5,122}
\expandafter\def\csname EER.birth.school3.0.phase6.hr\endcsname{3.46}
\expandafter\def\csname EER.birth.school3.0.phase6.lo\endcsname{2.78}
\expandafter\def\csname EER.birth.school3.0.phase6.hi\endcsname{4.30}
\expandafter\def\csname EER.birth.school3.0.phase6.ci\endcsname{[2.78, 4.30]}
\expandafter\def\csname EER.birth.school3.0.phase6.p\endcsname{}
\expandafter\def\csname EER.birth.school3.0.phase6.women\endcsname{2,718}
\expandafter\def\csname EER.birth.school3.0.phase6.months\endcsname{46,670}
\expandafter\def\csname EER.birth.school3.0.phase6.events\endcsname{659}
\expandafter\def\csname EER.birth.school3.0.phase6.exposed\endcsname{5,122}
\expandafter\def\csname EER.birth.school3.0.prenatal.hr\endcsname{3.71}
\expandafter\def\csname EER.birth.school3.0.prenatal.lo\endcsname{2.88}
\expandafter\def\csname EER.birth.school3.0.prenatal.hi\endcsname{4.79}
\expandafter\def\csname EER.birth.school3.0.prenatal.ci\endcsname{[2.88, 4.79]}
\expandafter\def\csname EER.birth.school3.0.prenatal.p\endcsname{0.715}
\expandafter\def\csname EER.birth.school3.0.prenatal.women\endcsname{2,718}
\expandafter\def\csname EER.birth.school3.0.prenatal.months\endcsname{46,670}
\expandafter\def\csname EER.birth.school3.0.prenatal.events\endcsname{659}
\expandafter\def\csname EER.birth.school3.0.prenatal.exposed\endcsname{5,122}
\expandafter\def\csname EER.birth.school3.0.birth_or_after.hr\endcsname{3.52}
\expandafter\def\csname EER.birth.school3.0.birth_or_after.lo\endcsname{2.90}
\expandafter\def\csname EER.birth.school3.0.birth_or_after.hi\endcsname{4.27}
\expandafter\def\csname EER.birth.school3.0.birth_or_after.ci\endcsname{[2.90, 4.27]}
\expandafter\def\csname EER.birth.school3.0.birth_or_after.p\endcsname{0.715}
\expandafter\def\csname EER.birth.school3.0.birth_or_after.women\endcsname{2,718}
\expandafter\def\csname EER.birth.school3.0.birth_or_after.months\endcsname{46,670}
\expandafter\def\csname EER.birth.school3.0.birth_or_after.events\endcsname{659}
\expandafter\def\csname EER.birth.school3.0.birth_or_after.exposed\endcsname{5,122}
\expandafter\def\csname EER.birth.school3.3.phase1.hr\endcsname{4.50}
\expandafter\def\csname EER.birth.school3.3.phase1.lo\endcsname{3.31}
\expandafter\def\csname EER.birth.school3.3.phase1.hi\endcsname{6.12}
\expandafter\def\csname EER.birth.school3.3.phase1.ci\endcsname{[3.31, 6.12]}
\expandafter\def\csname EER.birth.school3.3.phase1.p\endcsname{}
\expandafter\def\csname EER.birth.school3.3.phase1.women\endcsname{2,809}
\expandafter\def\csname EER.birth.school3.3.phase1.months\endcsname{48,564}
\expandafter\def\csname EER.birth.school3.3.phase1.events\endcsname{702}
\expandafter\def\csname EER.birth.school3.3.phase1.exposed\endcsname{5,752}
\expandafter\def\csname EER.birth.school3.3.phase2.hr\endcsname{2.86}
\expandafter\def\csname EER.birth.school3.3.phase2.lo\endcsname{1.87}
\expandafter\def\csname EER.birth.school3.3.phase2.hi\endcsname{4.37}
\expandafter\def\csname EER.birth.school3.3.phase2.ci\endcsname{[1.87, 4.37]}
\expandafter\def\csname EER.birth.school3.3.phase2.p\endcsname{}
\expandafter\def\csname EER.birth.school3.3.phase2.women\endcsname{2,809}
\expandafter\def\csname EER.birth.school3.3.phase2.months\endcsname{48,564}
\expandafter\def\csname EER.birth.school3.3.phase2.events\endcsname{702}
\expandafter\def\csname EER.birth.school3.3.phase2.exposed\endcsname{5,752}
\expandafter\def\csname EER.birth.school3.3.phase3.hr\endcsname{4.89}
\expandafter\def\csname EER.birth.school3.3.phase3.lo\endcsname{3.28}
\expandafter\def\csname EER.birth.school3.3.phase3.hi\endcsname{7.29}
\expandafter\def\csname EER.birth.school3.3.phase3.ci\endcsname{[3.28, 7.29]}
\expandafter\def\csname EER.birth.school3.3.phase3.p\endcsname{}
\expandafter\def\csname EER.birth.school3.3.phase3.women\endcsname{2,809}
\expandafter\def\csname EER.birth.school3.3.phase3.months\endcsname{48,564}
\expandafter\def\csname EER.birth.school3.3.phase3.events\endcsname{702}
\expandafter\def\csname EER.birth.school3.3.phase3.exposed\endcsname{5,752}
\expandafter\def\csname EER.birth.school3.3.phase4.hr\endcsname{4.91}
\expandafter\def\csname EER.birth.school3.3.phase4.lo\endcsname{3.39}
\expandafter\def\csname EER.birth.school3.3.phase4.hi\endcsname{7.10}
\expandafter\def\csname EER.birth.school3.3.phase4.ci\endcsname{[3.39, 7.10]}
\expandafter\def\csname EER.birth.school3.3.phase4.p\endcsname{}
\expandafter\def\csname EER.birth.school3.3.phase4.women\endcsname{2,809}
\expandafter\def\csname EER.birth.school3.3.phase4.months\endcsname{48,564}
\expandafter\def\csname EER.birth.school3.3.phase4.events\endcsname{702}
\expandafter\def\csname EER.birth.school3.3.phase4.exposed\endcsname{5,752}
\expandafter\def\csname EER.birth.school3.3.phase5.hr\endcsname{3.07}
\expandafter\def\csname EER.birth.school3.3.phase5.lo\endcsname{1.92}
\expandafter\def\csname EER.birth.school3.3.phase5.hi\endcsname{4.91}
\expandafter\def\csname EER.birth.school3.3.phase5.ci\endcsname{[1.92, 4.91]}
\expandafter\def\csname EER.birth.school3.3.phase5.p\endcsname{}
\expandafter\def\csname EER.birth.school3.3.phase5.women\endcsname{2,809}
\expandafter\def\csname EER.birth.school3.3.phase5.months\endcsname{48,564}
\expandafter\def\csname EER.birth.school3.3.phase5.events\endcsname{702}
\expandafter\def\csname EER.birth.school3.3.phase5.exposed\endcsname{5,752}
\expandafter\def\csname EER.birth.school3.3.phase6.hr\endcsname{3.60}
\expandafter\def\csname EER.birth.school3.3.phase6.lo\endcsname{2.91}
\expandafter\def\csname EER.birth.school3.3.phase6.hi\endcsname{4.46}
\expandafter\def\csname EER.birth.school3.3.phase6.ci\endcsname{[2.91, 4.46]}
\expandafter\def\csname EER.birth.school3.3.phase6.p\endcsname{}
\expandafter\def\csname EER.birth.school3.3.phase6.women\endcsname{2,809}
\expandafter\def\csname EER.birth.school3.3.phase6.months\endcsname{48,564}
\expandafter\def\csname EER.birth.school3.3.phase6.events\endcsname{702}
\expandafter\def\csname EER.birth.school3.3.phase6.exposed\endcsname{5,752}
\expandafter\def\csname EER.birth.school3.3.prenatal.hr\endcsname{4.03}
\expandafter\def\csname EER.birth.school3.3.prenatal.lo\endcsname{3.21}
\expandafter\def\csname EER.birth.school3.3.prenatal.hi\endcsname{5.05}
\expandafter\def\csname EER.birth.school3.3.prenatal.ci\endcsname{[3.21, 5.05]}
\expandafter\def\csname EER.birth.school3.3.prenatal.p\endcsname{0.571}
\expandafter\def\csname EER.birth.school3.3.prenatal.women\endcsname{2,809}
\expandafter\def\csname EER.birth.school3.3.prenatal.months\endcsname{48,564}
\expandafter\def\csname EER.birth.school3.3.prenatal.events\endcsname{702}
\expandafter\def\csname EER.birth.school3.3.prenatal.exposed\endcsname{5,752}
\expandafter\def\csname EER.birth.school3.3.birth_or_after.hr\endcsname{3.73}
\expandafter\def\csname EER.birth.school3.3.birth_or_after.lo\endcsname{3.10}
\expandafter\def\csname EER.birth.school3.3.birth_or_after.hi\endcsname{4.50}
\expandafter\def\csname EER.birth.school3.3.birth_or_after.ci\endcsname{[3.10, 4.50]}
\expandafter\def\csname EER.birth.school3.3.birth_or_after.p\endcsname{0.571}
\expandafter\def\csname EER.birth.school3.3.birth_or_after.women\endcsname{2,809}
\expandafter\def\csname EER.birth.school3.3.birth_or_after.months\endcsname{48,564}
\expandafter\def\csname EER.birth.school3.3.birth_or_after.events\endcsname{702}
\expandafter\def\csname EER.birth.school3.3.birth_or_after.exposed\endcsname{5,752}
\expandafter\def\csname EER.birth.recorded.1.phase1.hr\endcsname{4.31}
\expandafter\def\csname EER.birth.recorded.1.phase1.lo\endcsname{3.00}
\expandafter\def\csname EER.birth.recorded.1.phase1.hi\endcsname{6.19}
\expandafter\def\csname EER.birth.recorded.1.phase1.ci\endcsname{[3.00, 6.19]}
\expandafter\def\csname EER.birth.recorded.1.phase1.p\endcsname{}
\expandafter\def\csname EER.birth.recorded.1.phase1.women\endcsname{2,914}
\expandafter\def\csname EER.birth.recorded.1.phase1.months\endcsname{60,719}
\expandafter\def\csname EER.birth.recorded.1.phase1.events\endcsname{577}
\expandafter\def\csname EER.birth.recorded.1.phase1.exposed\endcsname{8,133}
\expandafter\def\csname EER.birth.recorded.1.phase2.hr\endcsname{4.23}
\expandafter\def\csname EER.birth.recorded.1.phase2.lo\endcsname{2.89}
\expandafter\def\csname EER.birth.recorded.1.phase2.hi\endcsname{6.20}
\expandafter\def\csname EER.birth.recorded.1.phase2.ci\endcsname{[2.89, 6.20]}
\expandafter\def\csname EER.birth.recorded.1.phase2.p\endcsname{}
\expandafter\def\csname EER.birth.recorded.1.phase2.women\endcsname{2,914}
\expandafter\def\csname EER.birth.recorded.1.phase2.months\endcsname{60,719}
\expandafter\def\csname EER.birth.recorded.1.phase2.events\endcsname{577}
\expandafter\def\csname EER.birth.recorded.1.phase2.exposed\endcsname{8,133}
\expandafter\def\csname EER.birth.recorded.1.phase3.hr\endcsname{5.07}
\expandafter\def\csname EER.birth.recorded.1.phase3.lo\endcsname{3.31}
\expandafter\def\csname EER.birth.recorded.1.phase3.hi\endcsname{7.76}
\expandafter\def\csname EER.birth.recorded.1.phase3.ci\endcsname{[3.31, 7.76]}
\expandafter\def\csname EER.birth.recorded.1.phase3.p\endcsname{}
\expandafter\def\csname EER.birth.recorded.1.phase3.women\endcsname{2,914}
\expandafter\def\csname EER.birth.recorded.1.phase3.months\endcsname{60,719}
\expandafter\def\csname EER.birth.recorded.1.phase3.events\endcsname{577}
\expandafter\def\csname EER.birth.recorded.1.phase3.exposed\endcsname{8,133}
\expandafter\def\csname EER.birth.recorded.1.phase4.hr\endcsname{3.67}
\expandafter\def\csname EER.birth.recorded.1.phase4.lo\endcsname{2.32}
\expandafter\def\csname EER.birth.recorded.1.phase4.hi\endcsname{5.81}
\expandafter\def\csname EER.birth.recorded.1.phase4.ci\endcsname{[2.32, 5.81]}
\expandafter\def\csname EER.birth.recorded.1.phase4.p\endcsname{}
\expandafter\def\csname EER.birth.recorded.1.phase4.women\endcsname{2,914}
\expandafter\def\csname EER.birth.recorded.1.phase4.months\endcsname{60,719}
\expandafter\def\csname EER.birth.recorded.1.phase4.events\endcsname{577}
\expandafter\def\csname EER.birth.recorded.1.phase4.exposed\endcsname{8,133}
\expandafter\def\csname EER.birth.recorded.1.phase5.hr\endcsname{2.94}
\expandafter\def\csname EER.birth.recorded.1.phase5.lo\endcsname{1.77}
\expandafter\def\csname EER.birth.recorded.1.phase5.hi\endcsname{4.89}
\expandafter\def\csname EER.birth.recorded.1.phase5.ci\endcsname{[1.77, 4.89]}
\expandafter\def\csname EER.birth.recorded.1.phase5.p\endcsname{}
\expandafter\def\csname EER.birth.recorded.1.phase5.women\endcsname{2,914}
\expandafter\def\csname EER.birth.recorded.1.phase5.months\endcsname{60,719}
\expandafter\def\csname EER.birth.recorded.1.phase5.events\endcsname{577}
\expandafter\def\csname EER.birth.recorded.1.phase5.exposed\endcsname{8,133}
\expandafter\def\csname EER.birth.recorded.1.phase6.hr\endcsname{3.27}
\expandafter\def\csname EER.birth.recorded.1.phase6.lo\endcsname{2.61}
\expandafter\def\csname EER.birth.recorded.1.phase6.hi\endcsname{4.10}
\expandafter\def\csname EER.birth.recorded.1.phase6.ci\endcsname{[2.61, 4.10]}
\expandafter\def\csname EER.birth.recorded.1.phase6.p\endcsname{}
\expandafter\def\csname EER.birth.recorded.1.phase6.women\endcsname{2,914}
\expandafter\def\csname EER.birth.recorded.1.phase6.months\endcsname{60,719}
\expandafter\def\csname EER.birth.recorded.1.phase6.events\endcsname{577}
\expandafter\def\csname EER.birth.recorded.1.phase6.exposed\endcsname{8,133}
\expandafter\def\csname EER.birth.recorded.1.prenatal.hr\endcsname{4.46}
\expandafter\def\csname EER.birth.recorded.1.prenatal.lo\endcsname{3.51}
\expandafter\def\csname EER.birth.recorded.1.prenatal.hi\endcsname{5.69}
\expandafter\def\csname EER.birth.recorded.1.prenatal.ci\endcsname{[3.51, 5.69]}
\expandafter\def\csname EER.birth.recorded.1.prenatal.p\endcsname{0.027}
\expandafter\def\csname EER.birth.recorded.1.prenatal.women\endcsname{2,914}
\expandafter\def\csname EER.birth.recorded.1.prenatal.months\endcsname{60,719}
\expandafter\def\csname EER.birth.recorded.1.prenatal.events\endcsname{577}
\expandafter\def\csname EER.birth.recorded.1.prenatal.exposed\endcsname{8,133}
\expandafter\def\csname EER.birth.recorded.1.birth_or_after.hr\endcsname{3.28}
\expandafter\def\csname EER.birth.recorded.1.birth_or_after.lo\endcsname{2.67}
\expandafter\def\csname EER.birth.recorded.1.birth_or_after.hi\endcsname{4.03}
\expandafter\def\csname EER.birth.recorded.1.birth_or_after.ci\endcsname{[2.67, 4.03]}
\expandafter\def\csname EER.birth.recorded.1.birth_or_after.p\endcsname{0.027}
\expandafter\def\csname EER.birth.recorded.1.birth_or_after.women\endcsname{2,914}
\expandafter\def\csname EER.birth.recorded.1.birth_or_after.months\endcsname{60,719}
\expandafter\def\csname EER.birth.recorded.1.birth_or_after.events\endcsname{577}
\expandafter\def\csname EER.birth.recorded.1.birth_or_after.exposed\endcsname{8,133}
\expandafter\def\csname EER.birth.recorded.0.phase1.hr\endcsname{4.32}
\expandafter\def\csname EER.birth.recorded.0.phase1.lo\endcsname{2.95}
\expandafter\def\csname EER.birth.recorded.0.phase1.hi\endcsname{6.34}
\expandafter\def\csname EER.birth.recorded.0.phase1.ci\endcsname{[2.95, 6.34]}
\expandafter\def\csname EER.birth.recorded.0.phase1.p\endcsname{}
\expandafter\def\csname EER.birth.recorded.0.phase1.women\endcsname{2,839}
\expandafter\def\csname EER.birth.recorded.0.phase1.months\endcsname{58,746}
\expandafter\def\csname EER.birth.recorded.0.phase1.events\endcsname{549}
\expandafter\def\csname EER.birth.recorded.0.phase1.exposed\endcsname{7,363}
\expandafter\def\csname EER.birth.recorded.0.phase2.hr\endcsname{4.20}
\expandafter\def\csname EER.birth.recorded.0.phase2.lo\endcsname{2.79}
\expandafter\def\csname EER.birth.recorded.0.phase2.hi\endcsname{6.31}
\expandafter\def\csname EER.birth.recorded.0.phase2.ci\endcsname{[2.79, 6.31]}
\expandafter\def\csname EER.birth.recorded.0.phase2.p\endcsname{}
\expandafter\def\csname EER.birth.recorded.0.phase2.women\endcsname{2,839}
\expandafter\def\csname EER.birth.recorded.0.phase2.months\endcsname{58,746}
\expandafter\def\csname EER.birth.recorded.0.phase2.events\endcsname{549}
\expandafter\def\csname EER.birth.recorded.0.phase2.exposed\endcsname{7,363}
\expandafter\def\csname EER.birth.recorded.0.phase3.hr\endcsname{4.28}
\expandafter\def\csname EER.birth.recorded.0.phase3.lo\endcsname{2.59}
\expandafter\def\csname EER.birth.recorded.0.phase3.hi\endcsname{7.08}
\expandafter\def\csname EER.birth.recorded.0.phase3.ci\endcsname{[2.59, 7.08]}
\expandafter\def\csname EER.birth.recorded.0.phase3.p\endcsname{}
\expandafter\def\csname EER.birth.recorded.0.phase3.women\endcsname{2,839}
\expandafter\def\csname EER.birth.recorded.0.phase3.months\endcsname{58,746}
\expandafter\def\csname EER.birth.recorded.0.phase3.events\endcsname{549}
\expandafter\def\csname EER.birth.recorded.0.phase3.exposed\endcsname{7,363}
\expandafter\def\csname EER.birth.recorded.0.phase4.hr\endcsname{3.96}
\expandafter\def\csname EER.birth.recorded.0.phase4.lo\endcsname{2.47}
\expandafter\def\csname EER.birth.recorded.0.phase4.hi\endcsname{6.33}
\expandafter\def\csname EER.birth.recorded.0.phase4.ci\endcsname{[2.47, 6.33]}
\expandafter\def\csname EER.birth.recorded.0.phase4.p\endcsname{}
\expandafter\def\csname EER.birth.recorded.0.phase4.women\endcsname{2,839}
\expandafter\def\csname EER.birth.recorded.0.phase4.months\endcsname{58,746}
\expandafter\def\csname EER.birth.recorded.0.phase4.events\endcsname{549}
\expandafter\def\csname EER.birth.recorded.0.phase4.exposed\endcsname{7,363}
\expandafter\def\csname EER.birth.recorded.0.phase5.hr\endcsname{2.86}
\expandafter\def\csname EER.birth.recorded.0.phase5.lo\endcsname{1.66}
\expandafter\def\csname EER.birth.recorded.0.phase5.hi\endcsname{4.92}
\expandafter\def\csname EER.birth.recorded.0.phase5.ci\endcsname{[1.66, 4.92]}
\expandafter\def\csname EER.birth.recorded.0.phase5.p\endcsname{}
\expandafter\def\csname EER.birth.recorded.0.phase5.women\endcsname{2,839}
\expandafter\def\csname EER.birth.recorded.0.phase5.months\endcsname{58,746}
\expandafter\def\csname EER.birth.recorded.0.phase5.events\endcsname{549}
\expandafter\def\csname EER.birth.recorded.0.phase5.exposed\endcsname{7,363}
\expandafter\def\csname EER.birth.recorded.0.phase6.hr\endcsname{3.12}
\expandafter\def\csname EER.birth.recorded.0.phase6.lo\endcsname{2.48}
\expandafter\def\csname EER.birth.recorded.0.phase6.hi\endcsname{3.94}
\expandafter\def\csname EER.birth.recorded.0.phase6.ci\endcsname{[2.48, 3.94]}
\expandafter\def\csname EER.birth.recorded.0.phase6.p\endcsname{}
\expandafter\def\csname EER.birth.recorded.0.phase6.women\endcsname{2,839}
\expandafter\def\csname EER.birth.recorded.0.phase6.months\endcsname{58,746}
\expandafter\def\csname EER.birth.recorded.0.phase6.events\endcsname{549}
\expandafter\def\csname EER.birth.recorded.0.phase6.exposed\endcsname{7,363}
\expandafter\def\csname EER.birth.recorded.0.prenatal.hr\endcsname{4.27}
\expandafter\def\csname EER.birth.recorded.0.prenatal.lo\endcsname{3.29}
\expandafter\def\csname EER.birth.recorded.0.prenatal.hi\endcsname{5.55}
\expandafter\def\csname EER.birth.recorded.0.prenatal.ci\endcsname{[3.29, 5.55]}
\expandafter\def\csname EER.birth.recorded.0.prenatal.p\endcsname{0.054}
\expandafter\def\csname EER.birth.recorded.0.prenatal.women\endcsname{2,839}
\expandafter\def\csname EER.birth.recorded.0.prenatal.months\endcsname{58,746}
\expandafter\def\csname EER.birth.recorded.0.prenatal.events\endcsname{549}
\expandafter\def\csname EER.birth.recorded.0.prenatal.exposed\endcsname{7,363}
\expandafter\def\csname EER.birth.recorded.0.birth_or_after.hr\endcsname{3.19}
\expandafter\def\csname EER.birth.recorded.0.birth_or_after.lo\endcsname{2.58}
\expandafter\def\csname EER.birth.recorded.0.birth_or_after.hi\endcsname{3.94}
\expandafter\def\csname EER.birth.recorded.0.birth_or_after.ci\endcsname{[2.58, 3.94]}
\expandafter\def\csname EER.birth.recorded.0.birth_or_after.p\endcsname{0.054}
\expandafter\def\csname EER.birth.recorded.0.birth_or_after.women\endcsname{2,839}
\expandafter\def\csname EER.birth.recorded.0.birth_or_after.months\endcsname{58,746}
\expandafter\def\csname EER.birth.recorded.0.birth_or_after.events\endcsname{549}
\expandafter\def\csname EER.birth.recorded.0.birth_or_after.exposed\endcsname{7,363}
\expandafter\def\csname EER.birth.recorded.3.phase1.hr\endcsname{4.20}
\expandafter\def\csname EER.birth.recorded.3.phase1.lo\endcsname{2.93}
\expandafter\def\csname EER.birth.recorded.3.phase1.hi\endcsname{6.03}
\expandafter\def\csname EER.birth.recorded.3.phase1.ci\endcsname{[2.93, 6.03]}
\expandafter\def\csname EER.birth.recorded.3.phase1.p\endcsname{}
\expandafter\def\csname EER.birth.recorded.3.phase1.women\endcsname{2,927}
\expandafter\def\csname EER.birth.recorded.3.phase1.months\endcsname{61,035}
\expandafter\def\csname EER.birth.recorded.3.phase1.events\endcsname{583}
\expandafter\def\csname EER.birth.recorded.3.phase1.exposed\endcsname{8,279}
\expandafter\def\csname EER.birth.recorded.3.phase2.hr\endcsname{4.32}
\expandafter\def\csname EER.birth.recorded.3.phase2.lo\endcsname{2.97}
\expandafter\def\csname EER.birth.recorded.3.phase2.hi\endcsname{6.29}
\expandafter\def\csname EER.birth.recorded.3.phase2.ci\endcsname{[2.97, 6.29]}
\expandafter\def\csname EER.birth.recorded.3.phase2.p\endcsname{}
\expandafter\def\csname EER.birth.recorded.3.phase2.women\endcsname{2,927}
\expandafter\def\csname EER.birth.recorded.3.phase2.months\endcsname{61,035}
\expandafter\def\csname EER.birth.recorded.3.phase2.events\endcsname{583}
\expandafter\def\csname EER.birth.recorded.3.phase2.exposed\endcsname{8,279}
\expandafter\def\csname EER.birth.recorded.3.phase3.hr\endcsname{5.09}
\expandafter\def\csname EER.birth.recorded.3.phase3.lo\endcsname{3.35}
\expandafter\def\csname EER.birth.recorded.3.phase3.hi\endcsname{7.73}
\expandafter\def\csname EER.birth.recorded.3.phase3.ci\endcsname{[3.35, 7.73]}
\expandafter\def\csname EER.birth.recorded.3.phase3.p\endcsname{}
\expandafter\def\csname EER.birth.recorded.3.phase3.women\endcsname{2,927}
\expandafter\def\csname EER.birth.recorded.3.phase3.months\endcsname{61,035}
\expandafter\def\csname EER.birth.recorded.3.phase3.events\endcsname{583}
\expandafter\def\csname EER.birth.recorded.3.phase3.exposed\endcsname{8,279}
\expandafter\def\csname EER.birth.recorded.3.phase4.hr\endcsname{3.82}
\expandafter\def\csname EER.birth.recorded.3.phase4.lo\endcsname{2.44}
\expandafter\def\csname EER.birth.recorded.3.phase4.hi\endcsname{5.98}
\expandafter\def\csname EER.birth.recorded.3.phase4.ci\endcsname{[2.44, 5.98]}
\expandafter\def\csname EER.birth.recorded.3.phase4.p\endcsname{}
\expandafter\def\csname EER.birth.recorded.3.phase4.women\endcsname{2,927}
\expandafter\def\csname EER.birth.recorded.3.phase4.months\endcsname{61,035}
\expandafter\def\csname EER.birth.recorded.3.phase4.events\endcsname{583}
\expandafter\def\csname EER.birth.recorded.3.phase4.exposed\endcsname{8,279}
\expandafter\def\csname EER.birth.recorded.3.phase5.hr\endcsname{3.08}
\expandafter\def\csname EER.birth.recorded.3.phase5.lo\endcsname{1.88}
\expandafter\def\csname EER.birth.recorded.3.phase5.hi\endcsname{5.04}
\expandafter\def\csname EER.birth.recorded.3.phase5.ci\endcsname{[1.88, 5.04]}
\expandafter\def\csname EER.birth.recorded.3.phase5.p\endcsname{}
\expandafter\def\csname EER.birth.recorded.3.phase5.women\endcsname{2,927}
\expandafter\def\csname EER.birth.recorded.3.phase5.months\endcsname{61,035}
\expandafter\def\csname EER.birth.recorded.3.phase5.events\endcsname{583}
\expandafter\def\csname EER.birth.recorded.3.phase5.exposed\endcsname{8,279}
\expandafter\def\csname EER.birth.recorded.3.phase6.hr\endcsname{3.31}
\expandafter\def\csname EER.birth.recorded.3.phase6.lo\endcsname{2.64}
\expandafter\def\csname EER.birth.recorded.3.phase6.hi\endcsname{4.13}
\expandafter\def\csname EER.birth.recorded.3.phase6.ci\endcsname{[2.64, 4.13]}
\expandafter\def\csname EER.birth.recorded.3.phase6.p\endcsname{}
\expandafter\def\csname EER.birth.recorded.3.phase6.women\endcsname{2,927}
\expandafter\def\csname EER.birth.recorded.3.phase6.months\endcsname{61,035}
\expandafter\def\csname EER.birth.recorded.3.phase6.events\endcsname{583}
\expandafter\def\csname EER.birth.recorded.3.phase6.exposed\endcsname{8,279}
\expandafter\def\csname EER.birth.recorded.3.prenatal.hr\endcsname{4.46}
\expandafter\def\csname EER.birth.recorded.3.prenatal.lo\endcsname{3.51}
\expandafter\def\csname EER.birth.recorded.3.prenatal.hi\endcsname{5.67}
\expandafter\def\csname EER.birth.recorded.3.prenatal.ci\endcsname{[3.51, 5.67]}
\expandafter\def\csname EER.birth.recorded.3.prenatal.p\endcsname{0.036}
\expandafter\def\csname EER.birth.recorded.3.prenatal.women\endcsname{2,927}
\expandafter\def\csname EER.birth.recorded.3.prenatal.months\endcsname{61,035}
\expandafter\def\csname EER.birth.recorded.3.prenatal.events\endcsname{583}
\expandafter\def\csname EER.birth.recorded.3.prenatal.exposed\endcsname{8,279}
\expandafter\def\csname EER.birth.recorded.3.birth_or_after.hr\endcsname{3.34}
\expandafter\def\csname EER.birth.recorded.3.birth_or_after.lo\endcsname{2.73}
\expandafter\def\csname EER.birth.recorded.3.birth_or_after.hi\endcsname{4.10}
\expandafter\def\csname EER.birth.recorded.3.birth_or_after.ci\endcsname{[2.73, 4.10]}
\expandafter\def\csname EER.birth.recorded.3.birth_or_after.p\endcsname{0.036}
\expandafter\def\csname EER.birth.recorded.3.birth_or_after.women\endcsname{2,927}
\expandafter\def\csname EER.birth.recorded.3.birth_or_after.months\endcsname{61,035}
\expandafter\def\csname EER.birth.recorded.3.birth_or_after.events\endcsname{583}
\expandafter\def\csname EER.birth.recorded.3.birth_or_after.exposed\endcsname{8,279}
\expandafter\def\csname EER.birth.calendar4.0.later.min\endcsname{2.03}
\expandafter\def\csname EER.birth.calendar4.0.later.max\endcsname{4.81}
\expandafter\def\csname EER.birth.calendar4.1.later.min\endcsname{2.53}
\expandafter\def\csname EER.birth.calendar4.1.later.max\endcsname{4.93}
\expandafter\def\csname EER.birth.calendar4.3.later.min\endcsname{2.79}
\expandafter\def\csname EER.birth.calendar4.3.later.max\endcsname{5.01}
\expandafter\def\csname EER.birth.recorded.0.later.min\endcsname{2.86}
\expandafter\def\csname EER.birth.recorded.0.later.max\endcsname{4.28}
\expandafter\def\csname EER.birth.recorded.1.later.min\endcsname{2.94}
\expandafter\def\csname EER.birth.recorded.1.later.max\endcsname{5.07}
\expandafter\def\csname EER.birth.recorded.3.later.min\endcsname{3.08}
\expandafter\def\csname EER.birth.recorded.3.later.max\endcsname{5.09}
\expandafter\def\csname EER.birth.school3.0.later.min\endcsname{2.07}
\expandafter\def\csname EER.birth.school3.0.later.max\endcsname{4.47}
\expandafter\def\csname EER.birth.school3.1.later.min\endcsname{2.67}
\expandafter\def\csname EER.birth.school3.1.later.max\endcsname{5.03}
\expandafter\def\csname EER.birth.school3.3.later.min\endcsname{2.86}
\expandafter\def\csname EER.birth.school3.3.later.max\endcsname{4.91}
\expandafter\def\csname EER.followup.all_origins.recorded.12.0.postevent_enrollment.pct\endcsname{2.9}
\expandafter\def\csname EER.followup.all_origins.recorded.12.0.postevent_enrollment.total\endcsname{434}
\expandafter\def\csname EER.followup.all_origins.recorded.12.0.postevent_enrollment.observed\endcsname{419}
\expandafter\def\csname EER.followup.all_origins.recorded.12.0.postevent_enrollment.yes\endcsname{12}
\expandafter\def\csname EER.followup.all_origins.recorded.12.0.postevent_enrollment.missing\endcsname{15}
\expandafter\def\csname EER.followup.all_origins.recorded.12.0.postevent_enrollment.fraction\endcsname{12/419}
\expandafter\def\csname EER.followup.all_origins.recorded.12.0.observed_resumption.pct\endcsname{1.2}
\expandafter\def\csname EER.followup.all_origins.recorded.12.0.observed_resumption.total\endcsname{434}
\expandafter\def\csname EER.followup.all_origins.recorded.12.0.observed_resumption.observed\endcsname{419}
\expandafter\def\csname EER.followup.all_origins.recorded.12.0.observed_resumption.yes\endcsname{5}
\expandafter\def\csname EER.followup.all_origins.recorded.12.0.observed_resumption.missing\endcsname{15}
\expandafter\def\csname EER.followup.all_origins.recorded.12.0.observed_resumption.fraction\endcsname{5/419}
\expandafter\def\csname EER.followup.all_origins.recorded.12.0.any_credential.pct\endcsname{14.5}
\expandafter\def\csname EER.followup.all_origins.recorded.12.0.any_credential.total\endcsname{434}
\expandafter\def\csname EER.followup.all_origins.recorded.12.0.any_credential.observed\endcsname{428}
\expandafter\def\csname EER.followup.all_origins.recorded.12.0.any_credential.yes\endcsname{62}
\expandafter\def\csname EER.followup.all_origins.recorded.12.0.any_credential.missing\endcsname{6}
\expandafter\def\csname EER.followup.all_origins.recorded.12.0.any_credential.fraction\endcsname{62/428}
\expandafter\def\csname EER.followup.all_origins.recorded.12.0.diploma.pct\endcsname{0.0}
\expandafter\def\csname EER.followup.all_origins.recorded.12.0.diploma.total\endcsname{434}
\expandafter\def\csname EER.followup.all_origins.recorded.12.0.diploma.observed\endcsname{428}
\expandafter\def\csname EER.followup.all_origins.recorded.12.0.diploma.yes\endcsname{0}
\expandafter\def\csname EER.followup.all_origins.recorded.12.0.diploma.missing\endcsname{6}
\expandafter\def\csname EER.followup.all_origins.recorded.12.0.diploma.fraction\endcsname{0/428}
\expandafter\def\csname EER.followup.all_origins.recorded.12.0.ged.pct\endcsname{14.5}
\expandafter\def\csname EER.followup.all_origins.recorded.12.0.ged.total\endcsname{434}
\expandafter\def\csname EER.followup.all_origins.recorded.12.0.ged.observed\endcsname{428}
\expandafter\def\csname EER.followup.all_origins.recorded.12.0.ged.yes\endcsname{62}
\expandafter\def\csname EER.followup.all_origins.recorded.12.0.ged.missing\endcsname{6}
\expandafter\def\csname EER.followup.all_origins.recorded.12.0.ged.fraction\endcsname{62/428}
\expandafter\def\csname EER.followup.all_origins.recorded.12.0.missing_interview.pct\endcsname{1.4}
\expandafter\def\csname EER.followup.all_origins.recorded.12.0.missing_interview.total\endcsname{434}
\expandafter\def\csname EER.followup.all_origins.recorded.12.0.missing_interview.observed\endcsname{434}
\expandafter\def\csname EER.followup.all_origins.recorded.12.0.missing_interview.yes\endcsname{6}
\expandafter\def\csname EER.followup.all_origins.recorded.12.0.missing_interview.missing\endcsname{0}
\expandafter\def\csname EER.followup.all_origins.recorded.12.0.missing_interview.fraction\endcsname{6/434}
\expandafter\def\csname EER.followup.all_origins.recorded.12.0.missing_array.pct\endcsname{1.4}
\expandafter\def\csname EER.followup.all_origins.recorded.12.0.missing_array.total\endcsname{434}
\expandafter\def\csname EER.followup.all_origins.recorded.12.0.missing_array.observed\endcsname{434}
\expandafter\def\csname EER.followup.all_origins.recorded.12.0.missing_array.yes\endcsname{6}
\expandafter\def\csname EER.followup.all_origins.recorded.12.0.missing_array.missing\endcsname{0}
\expandafter\def\csname EER.followup.all_origins.recorded.12.0.missing_array.fraction\endcsname{6/434}
\expandafter\def\csname EER.followup.all_origins.recorded.12.0.missing_status.pct\endcsname{2.1}
\expandafter\def\csname EER.followup.all_origins.recorded.12.0.missing_status.total\endcsname{434}
\expandafter\def\csname EER.followup.all_origins.recorded.12.0.missing_status.observed\endcsname{434}
\expandafter\def\csname EER.followup.all_origins.recorded.12.0.missing_status.yes\endcsname{9}
\expandafter\def\csname EER.followup.all_origins.recorded.12.0.missing_status.missing\endcsname{0}
\expandafter\def\csname EER.followup.all_origins.recorded.12.0.missing_status.fraction\endcsname{9/434}
\expandafter\def\csname EER.followup.all_origins.recorded.12.1.postevent_enrollment.pct\endcsname{3.3}
\expandafter\def\csname EER.followup.all_origins.recorded.12.1.postevent_enrollment.total\endcsname{246}
\expandafter\def\csname EER.followup.all_origins.recorded.12.1.postevent_enrollment.observed\endcsname{242}
\expandafter\def\csname EER.followup.all_origins.recorded.12.1.postevent_enrollment.yes\endcsname{8}
\expandafter\def\csname EER.followup.all_origins.recorded.12.1.postevent_enrollment.missing\endcsname{4}
\expandafter\def\csname EER.followup.all_origins.recorded.12.1.postevent_enrollment.fraction\endcsname{8/242}
\expandafter\def\csname EER.followup.all_origins.recorded.12.1.observed_resumption.pct\endcsname{1.7}
\expandafter\def\csname EER.followup.all_origins.recorded.12.1.observed_resumption.total\endcsname{246}
\expandafter\def\csname EER.followup.all_origins.recorded.12.1.observed_resumption.observed\endcsname{242}
\expandafter\def\csname EER.followup.all_origins.recorded.12.1.observed_resumption.yes\endcsname{4}
\expandafter\def\csname EER.followup.all_origins.recorded.12.1.observed_resumption.missing\endcsname{4}
\expandafter\def\csname EER.followup.all_origins.recorded.12.1.observed_resumption.fraction\endcsname{4/242}
\expandafter\def\csname EER.followup.all_origins.recorded.12.1.any_credential.pct\endcsname{10.2}
\expandafter\def\csname EER.followup.all_origins.recorded.12.1.any_credential.total\endcsname{246}
\expandafter\def\csname EER.followup.all_origins.recorded.12.1.any_credential.observed\endcsname{246}
\expandafter\def\csname EER.followup.all_origins.recorded.12.1.any_credential.yes\endcsname{25}
\expandafter\def\csname EER.followup.all_origins.recorded.12.1.any_credential.missing\endcsname{0}
\expandafter\def\csname EER.followup.all_origins.recorded.12.1.any_credential.fraction\endcsname{25/246}
\expandafter\def\csname EER.followup.all_origins.recorded.12.1.diploma.pct\endcsname{0.0}
\expandafter\def\csname EER.followup.all_origins.recorded.12.1.diploma.total\endcsname{246}
\expandafter\def\csname EER.followup.all_origins.recorded.12.1.diploma.observed\endcsname{246}
\expandafter\def\csname EER.followup.all_origins.recorded.12.1.diploma.yes\endcsname{0}
\expandafter\def\csname EER.followup.all_origins.recorded.12.1.diploma.missing\endcsname{0}
\expandafter\def\csname EER.followup.all_origins.recorded.12.1.diploma.fraction\endcsname{0/246}
\expandafter\def\csname EER.followup.all_origins.recorded.12.1.ged.pct\endcsname{10.2}
\expandafter\def\csname EER.followup.all_origins.recorded.12.1.ged.total\endcsname{246}
\expandafter\def\csname EER.followup.all_origins.recorded.12.1.ged.observed\endcsname{246}
\expandafter\def\csname EER.followup.all_origins.recorded.12.1.ged.yes\endcsname{25}
\expandafter\def\csname EER.followup.all_origins.recorded.12.1.ged.missing\endcsname{0}
\expandafter\def\csname EER.followup.all_origins.recorded.12.1.ged.fraction\endcsname{25/246}
\expandafter\def\csname EER.followup.all_origins.recorded.12.1.missing_interview.pct\endcsname{0.0}
\expandafter\def\csname EER.followup.all_origins.recorded.12.1.missing_interview.total\endcsname{246}
\expandafter\def\csname EER.followup.all_origins.recorded.12.1.missing_interview.observed\endcsname{246}
\expandafter\def\csname EER.followup.all_origins.recorded.12.1.missing_interview.yes\endcsname{0}
\expandafter\def\csname EER.followup.all_origins.recorded.12.1.missing_interview.missing\endcsname{0}
\expandafter\def\csname EER.followup.all_origins.recorded.12.1.missing_interview.fraction\endcsname{0/246}
\expandafter\def\csname EER.followup.all_origins.recorded.12.1.missing_array.pct\endcsname{0.0}
\expandafter\def\csname EER.followup.all_origins.recorded.12.1.missing_array.total\endcsname{246}
\expandafter\def\csname EER.followup.all_origins.recorded.12.1.missing_array.observed\endcsname{246}
\expandafter\def\csname EER.followup.all_origins.recorded.12.1.missing_array.yes\endcsname{0}
\expandafter\def\csname EER.followup.all_origins.recorded.12.1.missing_array.missing\endcsname{0}
\expandafter\def\csname EER.followup.all_origins.recorded.12.1.missing_array.fraction\endcsname{0/246}
\expandafter\def\csname EER.followup.all_origins.recorded.12.1.missing_status.pct\endcsname{1.6}
\expandafter\def\csname EER.followup.all_origins.recorded.12.1.missing_status.total\endcsname{246}
\expandafter\def\csname EER.followup.all_origins.recorded.12.1.missing_status.observed\endcsname{246}
\expandafter\def\csname EER.followup.all_origins.recorded.12.1.missing_status.yes\endcsname{4}
\expandafter\def\csname EER.followup.all_origins.recorded.12.1.missing_status.missing\endcsname{0}
\expandafter\def\csname EER.followup.all_origins.recorded.12.1.missing_status.fraction\endcsname{4/246}
\expandafter\def\csname EER.followup.all_origins.recorded.24.0.postevent_enrollment.pct\endcsname{2.6}
\expandafter\def\csname EER.followup.all_origins.recorded.24.0.postevent_enrollment.total\endcsname{434}
\expandafter\def\csname EER.followup.all_origins.recorded.24.0.postevent_enrollment.observed\endcsname{416}
\expandafter\def\csname EER.followup.all_origins.recorded.24.0.postevent_enrollment.yes\endcsname{11}
\expandafter\def\csname EER.followup.all_origins.recorded.24.0.postevent_enrollment.missing\endcsname{18}
\expandafter\def\csname EER.followup.all_origins.recorded.24.0.postevent_enrollment.fraction\endcsname{11/416}
\expandafter\def\csname EER.followup.all_origins.recorded.24.0.observed_resumption.pct\endcsname{1.2}
\expandafter\def\csname EER.followup.all_origins.recorded.24.0.observed_resumption.total\endcsname{434}
\expandafter\def\csname EER.followup.all_origins.recorded.24.0.observed_resumption.observed\endcsname{416}
\expandafter\def\csname EER.followup.all_origins.recorded.24.0.observed_resumption.yes\endcsname{5}
\expandafter\def\csname EER.followup.all_origins.recorded.24.0.observed_resumption.missing\endcsname{18}
\expandafter\def\csname EER.followup.all_origins.recorded.24.0.observed_resumption.fraction\endcsname{5/416}
\expandafter\def\csname EER.followup.all_origins.recorded.24.0.any_credential.pct\endcsname{24.1}
\expandafter\def\csname EER.followup.all_origins.recorded.24.0.any_credential.total\endcsname{434}
\expandafter\def\csname EER.followup.all_origins.recorded.24.0.any_credential.observed\endcsname{427}
\expandafter\def\csname EER.followup.all_origins.recorded.24.0.any_credential.yes\endcsname{103}
\expandafter\def\csname EER.followup.all_origins.recorded.24.0.any_credential.missing\endcsname{7}
\expandafter\def\csname EER.followup.all_origins.recorded.24.0.any_credential.fraction\endcsname{103/427}
\expandafter\def\csname EER.followup.all_origins.recorded.24.0.diploma.pct\endcsname{0.0}
\expandafter\def\csname EER.followup.all_origins.recorded.24.0.diploma.total\endcsname{434}
\expandafter\def\csname EER.followup.all_origins.recorded.24.0.diploma.observed\endcsname{427}
\expandafter\def\csname EER.followup.all_origins.recorded.24.0.diploma.yes\endcsname{0}
\expandafter\def\csname EER.followup.all_origins.recorded.24.0.diploma.missing\endcsname{7}
\expandafter\def\csname EER.followup.all_origins.recorded.24.0.diploma.fraction\endcsname{0/427}
\expandafter\def\csname EER.followup.all_origins.recorded.24.0.ged.pct\endcsname{24.1}
\expandafter\def\csname EER.followup.all_origins.recorded.24.0.ged.total\endcsname{434}
\expandafter\def\csname EER.followup.all_origins.recorded.24.0.ged.observed\endcsname{427}
\expandafter\def\csname EER.followup.all_origins.recorded.24.0.ged.yes\endcsname{103}
\expandafter\def\csname EER.followup.all_origins.recorded.24.0.ged.missing\endcsname{7}
\expandafter\def\csname EER.followup.all_origins.recorded.24.0.ged.fraction\endcsname{103/427}
\expandafter\def\csname EER.followup.all_origins.recorded.24.0.missing_interview.pct\endcsname{1.6}
\expandafter\def\csname EER.followup.all_origins.recorded.24.0.missing_interview.total\endcsname{434}
\expandafter\def\csname EER.followup.all_origins.recorded.24.0.missing_interview.observed\endcsname{434}
\expandafter\def\csname EER.followup.all_origins.recorded.24.0.missing_interview.yes\endcsname{7}
\expandafter\def\csname EER.followup.all_origins.recorded.24.0.missing_interview.missing\endcsname{0}
\expandafter\def\csname EER.followup.all_origins.recorded.24.0.missing_interview.fraction\endcsname{7/434}
\expandafter\def\csname EER.followup.all_origins.recorded.24.0.missing_array.pct\endcsname{1.6}
\expandafter\def\csname EER.followup.all_origins.recorded.24.0.missing_array.total\endcsname{434}
\expandafter\def\csname EER.followup.all_origins.recorded.24.0.missing_array.observed\endcsname{434}
\expandafter\def\csname EER.followup.all_origins.recorded.24.0.missing_array.yes\endcsname{7}
\expandafter\def\csname EER.followup.all_origins.recorded.24.0.missing_array.missing\endcsname{0}
\expandafter\def\csname EER.followup.all_origins.recorded.24.0.missing_array.fraction\endcsname{7/434}
\expandafter\def\csname EER.followup.all_origins.recorded.24.0.missing_status.pct\endcsname{2.5}
\expandafter\def\csname EER.followup.all_origins.recorded.24.0.missing_status.total\endcsname{434}
\expandafter\def\csname EER.followup.all_origins.recorded.24.0.missing_status.observed\endcsname{434}
\expandafter\def\csname EER.followup.all_origins.recorded.24.0.missing_status.yes\endcsname{11}
\expandafter\def\csname EER.followup.all_origins.recorded.24.0.missing_status.missing\endcsname{0}
\expandafter\def\csname EER.followup.all_origins.recorded.24.0.missing_status.fraction\endcsname{11/434}
\expandafter\def\csname EER.followup.all_origins.recorded.24.1.postevent_enrollment.pct\endcsname{3.7}
\expandafter\def\csname EER.followup.all_origins.recorded.24.1.postevent_enrollment.total\endcsname{246}
\expandafter\def\csname EER.followup.all_origins.recorded.24.1.postevent_enrollment.observed\endcsname{241}
\expandafter\def\csname EER.followup.all_origins.recorded.24.1.postevent_enrollment.yes\endcsname{9}
\expandafter\def\csname EER.followup.all_origins.recorded.24.1.postevent_enrollment.missing\endcsname{5}
\expandafter\def\csname EER.followup.all_origins.recorded.24.1.postevent_enrollment.fraction\endcsname{9/241}
\expandafter\def\csname EER.followup.all_origins.recorded.24.1.observed_resumption.pct\endcsname{2.5}
\expandafter\def\csname EER.followup.all_origins.recorded.24.1.observed_resumption.total\endcsname{246}
\expandafter\def\csname EER.followup.all_origins.recorded.24.1.observed_resumption.observed\endcsname{241}
\expandafter\def\csname EER.followup.all_origins.recorded.24.1.observed_resumption.yes\endcsname{6}
\expandafter\def\csname EER.followup.all_origins.recorded.24.1.observed_resumption.missing\endcsname{5}
\expandafter\def\csname EER.followup.all_origins.recorded.24.1.observed_resumption.fraction\endcsname{6/241}
\expandafter\def\csname EER.followup.all_origins.recorded.24.1.any_credential.pct\endcsname{19.1}
\expandafter\def\csname EER.followup.all_origins.recorded.24.1.any_credential.total\endcsname{246}
\expandafter\def\csname EER.followup.all_origins.recorded.24.1.any_credential.observed\endcsname{246}
\expandafter\def\csname EER.followup.all_origins.recorded.24.1.any_credential.yes\endcsname{47}
\expandafter\def\csname EER.followup.all_origins.recorded.24.1.any_credential.missing\endcsname{0}
\expandafter\def\csname EER.followup.all_origins.recorded.24.1.any_credential.fraction\endcsname{47/246}
\expandafter\def\csname EER.followup.all_origins.recorded.24.1.diploma.pct\endcsname{0.4}
\expandafter\def\csname EER.followup.all_origins.recorded.24.1.diploma.total\endcsname{246}
\expandafter\def\csname EER.followup.all_origins.recorded.24.1.diploma.observed\endcsname{246}
\expandafter\def\csname EER.followup.all_origins.recorded.24.1.diploma.yes\endcsname{1}
\expandafter\def\csname EER.followup.all_origins.recorded.24.1.diploma.missing\endcsname{0}
\expandafter\def\csname EER.followup.all_origins.recorded.24.1.diploma.fraction\endcsname{1/246}
\expandafter\def\csname EER.followup.all_origins.recorded.24.1.ged.pct\endcsname{18.7}
\expandafter\def\csname EER.followup.all_origins.recorded.24.1.ged.total\endcsname{246}
\expandafter\def\csname EER.followup.all_origins.recorded.24.1.ged.observed\endcsname{246}
\expandafter\def\csname EER.followup.all_origins.recorded.24.1.ged.yes\endcsname{46}
\expandafter\def\csname EER.followup.all_origins.recorded.24.1.ged.missing\endcsname{0}
\expandafter\def\csname EER.followup.all_origins.recorded.24.1.ged.fraction\endcsname{46/246}
\expandafter\def\csname EER.followup.all_origins.recorded.24.1.missing_interview.pct\endcsname{0.0}
\expandafter\def\csname EER.followup.all_origins.recorded.24.1.missing_interview.total\endcsname{246}
\expandafter\def\csname EER.followup.all_origins.recorded.24.1.missing_interview.observed\endcsname{246}
\expandafter\def\csname EER.followup.all_origins.recorded.24.1.missing_interview.yes\endcsname{0}
\expandafter\def\csname EER.followup.all_origins.recorded.24.1.missing_interview.missing\endcsname{0}
\expandafter\def\csname EER.followup.all_origins.recorded.24.1.missing_interview.fraction\endcsname{0/246}
\expandafter\def\csname EER.followup.all_origins.recorded.24.1.missing_array.pct\endcsname{0.0}
\expandafter\def\csname EER.followup.all_origins.recorded.24.1.missing_array.total\endcsname{246}
\expandafter\def\csname EER.followup.all_origins.recorded.24.1.missing_array.observed\endcsname{246}
\expandafter\def\csname EER.followup.all_origins.recorded.24.1.missing_array.yes\endcsname{0}
\expandafter\def\csname EER.followup.all_origins.recorded.24.1.missing_array.missing\endcsname{0}
\expandafter\def\csname EER.followup.all_origins.recorded.24.1.missing_array.fraction\endcsname{0/246}
\expandafter\def\csname EER.followup.all_origins.recorded.24.1.missing_status.pct\endcsname{2.0}
\expandafter\def\csname EER.followup.all_origins.recorded.24.1.missing_status.total\endcsname{246}
\expandafter\def\csname EER.followup.all_origins.recorded.24.1.missing_status.observed\endcsname{246}
\expandafter\def\csname EER.followup.all_origins.recorded.24.1.missing_status.yes\endcsname{5}
\expandafter\def\csname EER.followup.all_origins.recorded.24.1.missing_status.missing\endcsname{0}
\expandafter\def\csname EER.followup.all_origins.recorded.24.1.missing_status.fraction\endcsname{5/246}
\expandafter\def\csname EER.followup.all_origins.school3.12.0.postevent_enrollment.pct\endcsname{32.4}
\expandafter\def\csname EER.followup.all_origins.school3.12.0.postevent_enrollment.total\endcsname{649}
\expandafter\def\csname EER.followup.all_origins.school3.12.0.postevent_enrollment.observed\endcsname{646}
\expandafter\def\csname EER.followup.all_origins.school3.12.0.postevent_enrollment.yes\endcsname{209}
\expandafter\def\csname EER.followup.all_origins.school3.12.0.postevent_enrollment.missing\endcsname{3}
\expandafter\def\csname EER.followup.all_origins.school3.12.0.postevent_enrollment.fraction\endcsname{209/646}
\expandafter\def\csname EER.followup.all_origins.school3.12.0.observed_resumption.pct\endcsname{32.4}
\expandafter\def\csname EER.followup.all_origins.school3.12.0.observed_resumption.total\endcsname{649}
\expandafter\def\csname EER.followup.all_origins.school3.12.0.observed_resumption.observed\endcsname{646}
\expandafter\def\csname EER.followup.all_origins.school3.12.0.observed_resumption.yes\endcsname{209}
\expandafter\def\csname EER.followup.all_origins.school3.12.0.observed_resumption.missing\endcsname{3}
\expandafter\def\csname EER.followup.all_origins.school3.12.0.observed_resumption.fraction\endcsname{209/646}
\expandafter\def\csname EER.followup.all_origins.school3.12.0.any_credential.pct\endcsname{21.3}
\expandafter\def\csname EER.followup.all_origins.school3.12.0.any_credential.total\endcsname{649}
\expandafter\def\csname EER.followup.all_origins.school3.12.0.any_credential.observed\endcsname{649}
\expandafter\def\csname EER.followup.all_origins.school3.12.0.any_credential.yes\endcsname{138}
\expandafter\def\csname EER.followup.all_origins.school3.12.0.any_credential.missing\endcsname{0}
\expandafter\def\csname EER.followup.all_origins.school3.12.0.any_credential.fraction\endcsname{138/649}
\expandafter\def\csname EER.followup.all_origins.school3.12.0.diploma.pct\endcsname{14.8}
\expandafter\def\csname EER.followup.all_origins.school3.12.0.diploma.total\endcsname{649}
\expandafter\def\csname EER.followup.all_origins.school3.12.0.diploma.observed\endcsname{649}
\expandafter\def\csname EER.followup.all_origins.school3.12.0.diploma.yes\endcsname{96}
\expandafter\def\csname EER.followup.all_origins.school3.12.0.diploma.missing\endcsname{0}
\expandafter\def\csname EER.followup.all_origins.school3.12.0.diploma.fraction\endcsname{96/649}
\expandafter\def\csname EER.followup.all_origins.school3.12.0.ged.pct\endcsname{6.5}
\expandafter\def\csname EER.followup.all_origins.school3.12.0.ged.total\endcsname{649}
\expandafter\def\csname EER.followup.all_origins.school3.12.0.ged.observed\endcsname{649}
\expandafter\def\csname EER.followup.all_origins.school3.12.0.ged.yes\endcsname{42}
\expandafter\def\csname EER.followup.all_origins.school3.12.0.ged.missing\endcsname{0}
\expandafter\def\csname EER.followup.all_origins.school3.12.0.ged.fraction\endcsname{42/649}
\expandafter\def\csname EER.followup.all_origins.school3.12.0.missing_interview.pct\endcsname{0.0}
\expandafter\def\csname EER.followup.all_origins.school3.12.0.missing_interview.total\endcsname{649}
\expandafter\def\csname EER.followup.all_origins.school3.12.0.missing_interview.observed\endcsname{649}
\expandafter\def\csname EER.followup.all_origins.school3.12.0.missing_interview.yes\endcsname{0}
\expandafter\def\csname EER.followup.all_origins.school3.12.0.missing_interview.missing\endcsname{0}
\expandafter\def\csname EER.followup.all_origins.school3.12.0.missing_interview.fraction\endcsname{0/649}
\expandafter\def\csname EER.followup.all_origins.school3.12.0.missing_array.pct\endcsname{0.0}
\expandafter\def\csname EER.followup.all_origins.school3.12.0.missing_array.total\endcsname{649}
\expandafter\def\csname EER.followup.all_origins.school3.12.0.missing_array.observed\endcsname{649}
\expandafter\def\csname EER.followup.all_origins.school3.12.0.missing_array.yes\endcsname{0}
\expandafter\def\csname EER.followup.all_origins.school3.12.0.missing_array.missing\endcsname{0}
\expandafter\def\csname EER.followup.all_origins.school3.12.0.missing_array.fraction\endcsname{0/649}
\expandafter\def\csname EER.followup.all_origins.school3.12.0.missing_status.pct\endcsname{0.5}
\expandafter\def\csname EER.followup.all_origins.school3.12.0.missing_status.total\endcsname{649}
\expandafter\def\csname EER.followup.all_origins.school3.12.0.missing_status.observed\endcsname{649}
\expandafter\def\csname EER.followup.all_origins.school3.12.0.missing_status.yes\endcsname{3}
\expandafter\def\csname EER.followup.all_origins.school3.12.0.missing_status.missing\endcsname{0}
\expandafter\def\csname EER.followup.all_origins.school3.12.0.missing_status.fraction\endcsname{3/649}
\expandafter\def\csname EER.followup.all_origins.school3.12.1.postevent_enrollment.pct\endcsname{20.4}
\expandafter\def\csname EER.followup.all_origins.school3.12.1.postevent_enrollment.total\endcsname{262}
\expandafter\def\csname EER.followup.all_origins.school3.12.1.postevent_enrollment.observed\endcsname{260}
\expandafter\def\csname EER.followup.all_origins.school3.12.1.postevent_enrollment.yes\endcsname{53}
\expandafter\def\csname EER.followup.all_origins.school3.12.1.postevent_enrollment.missing\endcsname{2}
\expandafter\def\csname EER.followup.all_origins.school3.12.1.postevent_enrollment.fraction\endcsname{53/260}
\expandafter\def\csname EER.followup.all_origins.school3.12.1.observed_resumption.pct\endcsname{20.4}
\expandafter\def\csname EER.followup.all_origins.school3.12.1.observed_resumption.total\endcsname{262}
\expandafter\def\csname EER.followup.all_origins.school3.12.1.observed_resumption.observed\endcsname{260}
\expandafter\def\csname EER.followup.all_origins.school3.12.1.observed_resumption.yes\endcsname{53}
\expandafter\def\csname EER.followup.all_origins.school3.12.1.observed_resumption.missing\endcsname{2}
\expandafter\def\csname EER.followup.all_origins.school3.12.1.observed_resumption.fraction\endcsname{53/260}
\expandafter\def\csname EER.followup.all_origins.school3.12.1.any_credential.pct\endcsname{15.6}
\expandafter\def\csname EER.followup.all_origins.school3.12.1.any_credential.total\endcsname{262}
\expandafter\def\csname EER.followup.all_origins.school3.12.1.any_credential.observed\endcsname{262}
\expandafter\def\csname EER.followup.all_origins.school3.12.1.any_credential.yes\endcsname{41}
\expandafter\def\csname EER.followup.all_origins.school3.12.1.any_credential.missing\endcsname{0}
\expandafter\def\csname EER.followup.all_origins.school3.12.1.any_credential.fraction\endcsname{41/262}
\expandafter\def\csname EER.followup.all_origins.school3.12.1.diploma.pct\endcsname{7.6}
\expandafter\def\csname EER.followup.all_origins.school3.12.1.diploma.total\endcsname{262}
\expandafter\def\csname EER.followup.all_origins.school3.12.1.diploma.observed\endcsname{262}
\expandafter\def\csname EER.followup.all_origins.school3.12.1.diploma.yes\endcsname{20}
\expandafter\def\csname EER.followup.all_origins.school3.12.1.diploma.missing\endcsname{0}
\expandafter\def\csname EER.followup.all_origins.school3.12.1.diploma.fraction\endcsname{20/262}
\expandafter\def\csname EER.followup.all_origins.school3.12.1.ged.pct\endcsname{8.0}
\expandafter\def\csname EER.followup.all_origins.school3.12.1.ged.total\endcsname{262}
\expandafter\def\csname EER.followup.all_origins.school3.12.1.ged.observed\endcsname{262}
\expandafter\def\csname EER.followup.all_origins.school3.12.1.ged.yes\endcsname{21}
\expandafter\def\csname EER.followup.all_origins.school3.12.1.ged.missing\endcsname{0}
\expandafter\def\csname EER.followup.all_origins.school3.12.1.ged.fraction\endcsname{21/262}
\expandafter\def\csname EER.followup.all_origins.school3.12.1.missing_interview.pct\endcsname{0.0}
\expandafter\def\csname EER.followup.all_origins.school3.12.1.missing_interview.total\endcsname{262}
\expandafter\def\csname EER.followup.all_origins.school3.12.1.missing_interview.observed\endcsname{262}
\expandafter\def\csname EER.followup.all_origins.school3.12.1.missing_interview.yes\endcsname{0}
\expandafter\def\csname EER.followup.all_origins.school3.12.1.missing_interview.missing\endcsname{0}
\expandafter\def\csname EER.followup.all_origins.school3.12.1.missing_interview.fraction\endcsname{0/262}
\expandafter\def\csname EER.followup.all_origins.school3.12.1.missing_array.pct\endcsname{0.0}
\expandafter\def\csname EER.followup.all_origins.school3.12.1.missing_array.total\endcsname{262}
\expandafter\def\csname EER.followup.all_origins.school3.12.1.missing_array.observed\endcsname{262}
\expandafter\def\csname EER.followup.all_origins.school3.12.1.missing_array.yes\endcsname{0}
\expandafter\def\csname EER.followup.all_origins.school3.12.1.missing_array.missing\endcsname{0}
\expandafter\def\csname EER.followup.all_origins.school3.12.1.missing_array.fraction\endcsname{0/262}
\expandafter\def\csname EER.followup.all_origins.school3.12.1.missing_status.pct\endcsname{0.8}
\expandafter\def\csname EER.followup.all_origins.school3.12.1.missing_status.total\endcsname{262}
\expandafter\def\csname EER.followup.all_origins.school3.12.1.missing_status.observed\endcsname{262}
\expandafter\def\csname EER.followup.all_origins.school3.12.1.missing_status.yes\endcsname{2}
\expandafter\def\csname EER.followup.all_origins.school3.12.1.missing_status.missing\endcsname{0}
\expandafter\def\csname EER.followup.all_origins.school3.12.1.missing_status.fraction\endcsname{2/262}
\expandafter\def\csname EER.followup.all_origins.school3.24.0.postevent_enrollment.pct\endcsname{34.1}
\expandafter\def\csname EER.followup.all_origins.school3.24.0.postevent_enrollment.total\endcsname{649}
\expandafter\def\csname EER.followup.all_origins.school3.24.0.postevent_enrollment.observed\endcsname{640}
\expandafter\def\csname EER.followup.all_origins.school3.24.0.postevent_enrollment.yes\endcsname{218}
\expandafter\def\csname EER.followup.all_origins.school3.24.0.postevent_enrollment.missing\endcsname{9}
\expandafter\def\csname EER.followup.all_origins.school3.24.0.postevent_enrollment.fraction\endcsname{218/640}
\expandafter\def\csname EER.followup.all_origins.school3.24.0.observed_resumption.pct\endcsname{34.1}
\expandafter\def\csname EER.followup.all_origins.school3.24.0.observed_resumption.total\endcsname{649}
\expandafter\def\csname EER.followup.all_origins.school3.24.0.observed_resumption.observed\endcsname{640}
\expandafter\def\csname EER.followup.all_origins.school3.24.0.observed_resumption.yes\endcsname{218}
\expandafter\def\csname EER.followup.all_origins.school3.24.0.observed_resumption.missing\endcsname{9}
\expandafter\def\csname EER.followup.all_origins.school3.24.0.observed_resumption.fraction\endcsname{218/640}
\expandafter\def\csname EER.followup.all_origins.school3.24.0.any_credential.pct\endcsname{35.2}
\expandafter\def\csname EER.followup.all_origins.school3.24.0.any_credential.total\endcsname{649}
\expandafter\def\csname EER.followup.all_origins.school3.24.0.any_credential.observed\endcsname{645}
\expandafter\def\csname EER.followup.all_origins.school3.24.0.any_credential.yes\endcsname{227}
\expandafter\def\csname EER.followup.all_origins.school3.24.0.any_credential.missing\endcsname{4}
\expandafter\def\csname EER.followup.all_origins.school3.24.0.any_credential.fraction\endcsname{227/645}
\expandafter\def\csname EER.followup.all_origins.school3.24.0.diploma.pct\endcsname{22.2}
\expandafter\def\csname EER.followup.all_origins.school3.24.0.diploma.total\endcsname{649}
\expandafter\def\csname EER.followup.all_origins.school3.24.0.diploma.observed\endcsname{645}
\expandafter\def\csname EER.followup.all_origins.school3.24.0.diploma.yes\endcsname{143}
\expandafter\def\csname EER.followup.all_origins.school3.24.0.diploma.missing\endcsname{4}
\expandafter\def\csname EER.followup.all_origins.school3.24.0.diploma.fraction\endcsname{143/645}
\expandafter\def\csname EER.followup.all_origins.school3.24.0.ged.pct\endcsname{13.0}
\expandafter\def\csname EER.followup.all_origins.school3.24.0.ged.total\endcsname{649}
\expandafter\def\csname EER.followup.all_origins.school3.24.0.ged.observed\endcsname{645}
\expandafter\def\csname EER.followup.all_origins.school3.24.0.ged.yes\endcsname{84}
\expandafter\def\csname EER.followup.all_origins.school3.24.0.ged.missing\endcsname{4}
\expandafter\def\csname EER.followup.all_origins.school3.24.0.ged.fraction\endcsname{84/645}
\expandafter\def\csname EER.followup.all_origins.school3.24.0.missing_interview.pct\endcsname{0.6}
\expandafter\def\csname EER.followup.all_origins.school3.24.0.missing_interview.total\endcsname{649}
\expandafter\def\csname EER.followup.all_origins.school3.24.0.missing_interview.observed\endcsname{649}
\expandafter\def\csname EER.followup.all_origins.school3.24.0.missing_interview.yes\endcsname{4}
\expandafter\def\csname EER.followup.all_origins.school3.24.0.missing_interview.missing\endcsname{0}
\expandafter\def\csname EER.followup.all_origins.school3.24.0.missing_interview.fraction\endcsname{4/649}
\expandafter\def\csname EER.followup.all_origins.school3.24.0.missing_array.pct\endcsname{0.6}
\expandafter\def\csname EER.followup.all_origins.school3.24.0.missing_array.total\endcsname{649}
\expandafter\def\csname EER.followup.all_origins.school3.24.0.missing_array.observed\endcsname{649}
\expandafter\def\csname EER.followup.all_origins.school3.24.0.missing_array.yes\endcsname{4}
\expandafter\def\csname EER.followup.all_origins.school3.24.0.missing_array.missing\endcsname{0}
\expandafter\def\csname EER.followup.all_origins.school3.24.0.missing_array.fraction\endcsname{4/649}
\expandafter\def\csname EER.followup.all_origins.school3.24.0.missing_status.pct\endcsname{0.8}
\expandafter\def\csname EER.followup.all_origins.school3.24.0.missing_status.total\endcsname{649}
\expandafter\def\csname EER.followup.all_origins.school3.24.0.missing_status.observed\endcsname{649}
\expandafter\def\csname EER.followup.all_origins.school3.24.0.missing_status.yes\endcsname{5}
\expandafter\def\csname EER.followup.all_origins.school3.24.0.missing_status.missing\endcsname{0}
\expandafter\def\csname EER.followup.all_origins.school3.24.0.missing_status.fraction\endcsname{5/649}
\expandafter\def\csname EER.followup.all_origins.school3.24.1.postevent_enrollment.pct\endcsname{20.3}
\expandafter\def\csname EER.followup.all_origins.school3.24.1.postevent_enrollment.total\endcsname{262}
\expandafter\def\csname EER.followup.all_origins.school3.24.1.postevent_enrollment.observed\endcsname{256}
\expandafter\def\csname EER.followup.all_origins.school3.24.1.postevent_enrollment.yes\endcsname{52}
\expandafter\def\csname EER.followup.all_origins.school3.24.1.postevent_enrollment.missing\endcsname{6}
\expandafter\def\csname EER.followup.all_origins.school3.24.1.postevent_enrollment.fraction\endcsname{52/256}
\expandafter\def\csname EER.followup.all_origins.school3.24.1.observed_resumption.pct\endcsname{20.3}
\expandafter\def\csname EER.followup.all_origins.school3.24.1.observed_resumption.total\endcsname{262}
\expandafter\def\csname EER.followup.all_origins.school3.24.1.observed_resumption.observed\endcsname{256}
\expandafter\def\csname EER.followup.all_origins.school3.24.1.observed_resumption.yes\endcsname{52}
\expandafter\def\csname EER.followup.all_origins.school3.24.1.observed_resumption.missing\endcsname{6}
\expandafter\def\csname EER.followup.all_origins.school3.24.1.observed_resumption.fraction\endcsname{52/256}
\expandafter\def\csname EER.followup.all_origins.school3.24.1.any_credential.pct\endcsname{26.1}
\expandafter\def\csname EER.followup.all_origins.school3.24.1.any_credential.total\endcsname{262}
\expandafter\def\csname EER.followup.all_origins.school3.24.1.any_credential.observed\endcsname{261}
\expandafter\def\csname EER.followup.all_origins.school3.24.1.any_credential.yes\endcsname{68}
\expandafter\def\csname EER.followup.all_origins.school3.24.1.any_credential.missing\endcsname{1}
\expandafter\def\csname EER.followup.all_origins.school3.24.1.any_credential.fraction\endcsname{68/261}
\expandafter\def\csname EER.followup.all_origins.school3.24.1.diploma.pct\endcsname{11.5}
\expandafter\def\csname EER.followup.all_origins.school3.24.1.diploma.total\endcsname{262}
\expandafter\def\csname EER.followup.all_origins.school3.24.1.diploma.observed\endcsname{261}
\expandafter\def\csname EER.followup.all_origins.school3.24.1.diploma.yes\endcsname{30}
\expandafter\def\csname EER.followup.all_origins.school3.24.1.diploma.missing\endcsname{1}
\expandafter\def\csname EER.followup.all_origins.school3.24.1.diploma.fraction\endcsname{30/261}
\expandafter\def\csname EER.followup.all_origins.school3.24.1.ged.pct\endcsname{14.6}
\expandafter\def\csname EER.followup.all_origins.school3.24.1.ged.total\endcsname{262}
\expandafter\def\csname EER.followup.all_origins.school3.24.1.ged.observed\endcsname{261}
\expandafter\def\csname EER.followup.all_origins.school3.24.1.ged.yes\endcsname{38}
\expandafter\def\csname EER.followup.all_origins.school3.24.1.ged.missing\endcsname{1}
\expandafter\def\csname EER.followup.all_origins.school3.24.1.ged.fraction\endcsname{38/261}
\expandafter\def\csname EER.followup.all_origins.school3.24.1.missing_interview.pct\endcsname{0.4}
\expandafter\def\csname EER.followup.all_origins.school3.24.1.missing_interview.total\endcsname{262}
\expandafter\def\csname EER.followup.all_origins.school3.24.1.missing_interview.observed\endcsname{262}
\expandafter\def\csname EER.followup.all_origins.school3.24.1.missing_interview.yes\endcsname{1}
\expandafter\def\csname EER.followup.all_origins.school3.24.1.missing_interview.missing\endcsname{0}
\expandafter\def\csname EER.followup.all_origins.school3.24.1.missing_interview.fraction\endcsname{1/262}
\expandafter\def\csname EER.followup.all_origins.school3.24.1.missing_array.pct\endcsname{0.4}
\expandafter\def\csname EER.followup.all_origins.school3.24.1.missing_array.total\endcsname{262}
\expandafter\def\csname EER.followup.all_origins.school3.24.1.missing_array.observed\endcsname{262}
\expandafter\def\csname EER.followup.all_origins.school3.24.1.missing_array.yes\endcsname{1}
\expandafter\def\csname EER.followup.all_origins.school3.24.1.missing_array.missing\endcsname{0}
\expandafter\def\csname EER.followup.all_origins.school3.24.1.missing_array.fraction\endcsname{1/262}
\expandafter\def\csname EER.followup.all_origins.school3.24.1.missing_status.pct\endcsname{1.9}
\expandafter\def\csname EER.followup.all_origins.school3.24.1.missing_status.total\endcsname{262}
\expandafter\def\csname EER.followup.all_origins.school3.24.1.missing_status.observed\endcsname{262}
\expandafter\def\csname EER.followup.all_origins.school3.24.1.missing_status.yes\endcsname{5}
\expandafter\def\csname EER.followup.all_origins.school3.24.1.missing_status.missing\endcsname{0}
\expandafter\def\csname EER.followup.all_origins.school3.24.1.missing_status.fraction\endcsname{5/262}
\expandafter\def\csname EER.followup.all_origins.calendar4.12.0.postevent_enrollment.pct\endcsname{38.4}
\expandafter\def\csname EER.followup.all_origins.calendar4.12.0.postevent_enrollment.total\endcsname{692}
\expandafter\def\csname EER.followup.all_origins.calendar4.12.0.postevent_enrollment.observed\endcsname{690}
\expandafter\def\csname EER.followup.all_origins.calendar4.12.0.postevent_enrollment.yes\endcsname{265}
\expandafter\def\csname EER.followup.all_origins.calendar4.12.0.postevent_enrollment.missing\endcsname{2}
\expandafter\def\csname EER.followup.all_origins.calendar4.12.0.postevent_enrollment.fraction\endcsname{265/690}
\expandafter\def\csname EER.followup.all_origins.calendar4.12.0.observed_resumption.pct\endcsname{38.4}
\expandafter\def\csname EER.followup.all_origins.calendar4.12.0.observed_resumption.total\endcsname{692}
\expandafter\def\csname EER.followup.all_origins.calendar4.12.0.observed_resumption.observed\endcsname{690}
\expandafter\def\csname EER.followup.all_origins.calendar4.12.0.observed_resumption.yes\endcsname{265}
\expandafter\def\csname EER.followup.all_origins.calendar4.12.0.observed_resumption.missing\endcsname{2}
\expandafter\def\csname EER.followup.all_origins.calendar4.12.0.observed_resumption.fraction\endcsname{265/690}
\expandafter\def\csname EER.followup.all_origins.calendar4.12.0.any_credential.pct\endcsname{20.2}
\expandafter\def\csname EER.followup.all_origins.calendar4.12.0.any_credential.total\endcsname{692}
\expandafter\def\csname EER.followup.all_origins.calendar4.12.0.any_credential.observed\endcsname{692}
\expandafter\def\csname EER.followup.all_origins.calendar4.12.0.any_credential.yes\endcsname{140}
\expandafter\def\csname EER.followup.all_origins.calendar4.12.0.any_credential.missing\endcsname{0}
\expandafter\def\csname EER.followup.all_origins.calendar4.12.0.any_credential.fraction\endcsname{140/692}
\expandafter\def\csname EER.followup.all_origins.calendar4.12.0.diploma.pct\endcsname{13.7}
\expandafter\def\csname EER.followup.all_origins.calendar4.12.0.diploma.total\endcsname{692}
\expandafter\def\csname EER.followup.all_origins.calendar4.12.0.diploma.observed\endcsname{692}
\expandafter\def\csname EER.followup.all_origins.calendar4.12.0.diploma.yes\endcsname{95}
\expandafter\def\csname EER.followup.all_origins.calendar4.12.0.diploma.missing\endcsname{0}
\expandafter\def\csname EER.followup.all_origins.calendar4.12.0.diploma.fraction\endcsname{95/692}
\expandafter\def\csname EER.followup.all_origins.calendar4.12.0.ged.pct\endcsname{6.5}
\expandafter\def\csname EER.followup.all_origins.calendar4.12.0.ged.total\endcsname{692}
\expandafter\def\csname EER.followup.all_origins.calendar4.12.0.ged.observed\endcsname{692}
\expandafter\def\csname EER.followup.all_origins.calendar4.12.0.ged.yes\endcsname{45}
\expandafter\def\csname EER.followup.all_origins.calendar4.12.0.ged.missing\endcsname{0}
\expandafter\def\csname EER.followup.all_origins.calendar4.12.0.ged.fraction\endcsname{45/692}
\expandafter\def\csname EER.followup.all_origins.calendar4.12.0.missing_interview.pct\endcsname{0.0}
\expandafter\def\csname EER.followup.all_origins.calendar4.12.0.missing_interview.total\endcsname{692}
\expandafter\def\csname EER.followup.all_origins.calendar4.12.0.missing_interview.observed\endcsname{692}
\expandafter\def\csname EER.followup.all_origins.calendar4.12.0.missing_interview.yes\endcsname{0}
\expandafter\def\csname EER.followup.all_origins.calendar4.12.0.missing_interview.missing\endcsname{0}
\expandafter\def\csname EER.followup.all_origins.calendar4.12.0.missing_interview.fraction\endcsname{0/692}
\expandafter\def\csname EER.followup.all_origins.calendar4.12.0.missing_array.pct\endcsname{0.0}
\expandafter\def\csname EER.followup.all_origins.calendar4.12.0.missing_array.total\endcsname{692}
\expandafter\def\csname EER.followup.all_origins.calendar4.12.0.missing_array.observed\endcsname{692}
\expandafter\def\csname EER.followup.all_origins.calendar4.12.0.missing_array.yes\endcsname{0}
\expandafter\def\csname EER.followup.all_origins.calendar4.12.0.missing_array.missing\endcsname{0}
\expandafter\def\csname EER.followup.all_origins.calendar4.12.0.missing_array.fraction\endcsname{0/692}
\expandafter\def\csname EER.followup.all_origins.calendar4.12.0.missing_status.pct\endcsname{0.3}
\expandafter\def\csname EER.followup.all_origins.calendar4.12.0.missing_status.total\endcsname{692}
\expandafter\def\csname EER.followup.all_origins.calendar4.12.0.missing_status.observed\endcsname{692}
\expandafter\def\csname EER.followup.all_origins.calendar4.12.0.missing_status.yes\endcsname{2}
\expandafter\def\csname EER.followup.all_origins.calendar4.12.0.missing_status.missing\endcsname{0}
\expandafter\def\csname EER.followup.all_origins.calendar4.12.0.missing_status.fraction\endcsname{2/692}
\expandafter\def\csname EER.followup.all_origins.calendar4.12.1.postevent_enrollment.pct\endcsname{21.9}
\expandafter\def\csname EER.followup.all_origins.calendar4.12.1.postevent_enrollment.total\endcsname{259}
\expandafter\def\csname EER.followup.all_origins.calendar4.12.1.postevent_enrollment.observed\endcsname{256}
\expandafter\def\csname EER.followup.all_origins.calendar4.12.1.postevent_enrollment.yes\endcsname{56}
\expandafter\def\csname EER.followup.all_origins.calendar4.12.1.postevent_enrollment.missing\endcsname{3}
\expandafter\def\csname EER.followup.all_origins.calendar4.12.1.postevent_enrollment.fraction\endcsname{56/256}
\expandafter\def\csname EER.followup.all_origins.calendar4.12.1.observed_resumption.pct\endcsname{21.9}
\expandafter\def\csname EER.followup.all_origins.calendar4.12.1.observed_resumption.total\endcsname{259}
\expandafter\def\csname EER.followup.all_origins.calendar4.12.1.observed_resumption.observed\endcsname{256}
\expandafter\def\csname EER.followup.all_origins.calendar4.12.1.observed_resumption.yes\endcsname{56}
\expandafter\def\csname EER.followup.all_origins.calendar4.12.1.observed_resumption.missing\endcsname{3}
\expandafter\def\csname EER.followup.all_origins.calendar4.12.1.observed_resumption.fraction\endcsname{56/256}
\expandafter\def\csname EER.followup.all_origins.calendar4.12.1.any_credential.pct\endcsname{14.3}
\expandafter\def\csname EER.followup.all_origins.calendar4.12.1.any_credential.total\endcsname{259}
\expandafter\def\csname EER.followup.all_origins.calendar4.12.1.any_credential.observed\endcsname{259}
\expandafter\def\csname EER.followup.all_origins.calendar4.12.1.any_credential.yes\endcsname{37}
\expandafter\def\csname EER.followup.all_origins.calendar4.12.1.any_credential.missing\endcsname{0}
\expandafter\def\csname EER.followup.all_origins.calendar4.12.1.any_credential.fraction\endcsname{37/259}
\expandafter\def\csname EER.followup.all_origins.calendar4.12.1.diploma.pct\endcsname{6.9}
\expandafter\def\csname EER.followup.all_origins.calendar4.12.1.diploma.total\endcsname{259}
\expandafter\def\csname EER.followup.all_origins.calendar4.12.1.diploma.observed\endcsname{259}
\expandafter\def\csname EER.followup.all_origins.calendar4.12.1.diploma.yes\endcsname{18}
\expandafter\def\csname EER.followup.all_origins.calendar4.12.1.diploma.missing\endcsname{0}
\expandafter\def\csname EER.followup.all_origins.calendar4.12.1.diploma.fraction\endcsname{18/259}
\expandafter\def\csname EER.followup.all_origins.calendar4.12.1.ged.pct\endcsname{7.3}
\expandafter\def\csname EER.followup.all_origins.calendar4.12.1.ged.total\endcsname{259}
\expandafter\def\csname EER.followup.all_origins.calendar4.12.1.ged.observed\endcsname{259}
\expandafter\def\csname EER.followup.all_origins.calendar4.12.1.ged.yes\endcsname{19}
\expandafter\def\csname EER.followup.all_origins.calendar4.12.1.ged.missing\endcsname{0}
\expandafter\def\csname EER.followup.all_origins.calendar4.12.1.ged.fraction\endcsname{19/259}
\expandafter\def\csname EER.followup.all_origins.calendar4.12.1.missing_interview.pct\endcsname{0.0}
\expandafter\def\csname EER.followup.all_origins.calendar4.12.1.missing_interview.total\endcsname{259}
\expandafter\def\csname EER.followup.all_origins.calendar4.12.1.missing_interview.observed\endcsname{259}
\expandafter\def\csname EER.followup.all_origins.calendar4.12.1.missing_interview.yes\endcsname{0}
\expandafter\def\csname EER.followup.all_origins.calendar4.12.1.missing_interview.missing\endcsname{0}
\expandafter\def\csname EER.followup.all_origins.calendar4.12.1.missing_interview.fraction\endcsname{0/259}
\expandafter\def\csname EER.followup.all_origins.calendar4.12.1.missing_array.pct\endcsname{0.0}
\expandafter\def\csname EER.followup.all_origins.calendar4.12.1.missing_array.total\endcsname{259}
\expandafter\def\csname EER.followup.all_origins.calendar4.12.1.missing_array.observed\endcsname{259}
\expandafter\def\csname EER.followup.all_origins.calendar4.12.1.missing_array.yes\endcsname{0}
\expandafter\def\csname EER.followup.all_origins.calendar4.12.1.missing_array.missing\endcsname{0}
\expandafter\def\csname EER.followup.all_origins.calendar4.12.1.missing_array.fraction\endcsname{0/259}
\expandafter\def\csname EER.followup.all_origins.calendar4.12.1.missing_status.pct\endcsname{1.2}
\expandafter\def\csname EER.followup.all_origins.calendar4.12.1.missing_status.total\endcsname{259}
\expandafter\def\csname EER.followup.all_origins.calendar4.12.1.missing_status.observed\endcsname{259}
\expandafter\def\csname EER.followup.all_origins.calendar4.12.1.missing_status.yes\endcsname{3}
\expandafter\def\csname EER.followup.all_origins.calendar4.12.1.missing_status.missing\endcsname{0}
\expandafter\def\csname EER.followup.all_origins.calendar4.12.1.missing_status.fraction\endcsname{3/259}
\expandafter\def\csname EER.followup.all_origins.calendar4.24.0.postevent_enrollment.pct\endcsname{39.5}
\expandafter\def\csname EER.followup.all_origins.calendar4.24.0.postevent_enrollment.total\endcsname{692}
\expandafter\def\csname EER.followup.all_origins.calendar4.24.0.postevent_enrollment.observed\endcsname{681}
\expandafter\def\csname EER.followup.all_origins.calendar4.24.0.postevent_enrollment.yes\endcsname{269}
\expandafter\def\csname EER.followup.all_origins.calendar4.24.0.postevent_enrollment.missing\endcsname{11}
\expandafter\def\csname EER.followup.all_origins.calendar4.24.0.postevent_enrollment.fraction\endcsname{269/681}
\expandafter\def\csname EER.followup.all_origins.calendar4.24.0.observed_resumption.pct\endcsname{39.5}
\expandafter\def\csname EER.followup.all_origins.calendar4.24.0.observed_resumption.total\endcsname{692}
\expandafter\def\csname EER.followup.all_origins.calendar4.24.0.observed_resumption.observed\endcsname{681}
\expandafter\def\csname EER.followup.all_origins.calendar4.24.0.observed_resumption.yes\endcsname{269}
\expandafter\def\csname EER.followup.all_origins.calendar4.24.0.observed_resumption.missing\endcsname{11}
\expandafter\def\csname EER.followup.all_origins.calendar4.24.0.observed_resumption.fraction\endcsname{269/681}
\expandafter\def\csname EER.followup.all_origins.calendar4.24.0.any_credential.pct\endcsname{34.3}
\expandafter\def\csname EER.followup.all_origins.calendar4.24.0.any_credential.total\endcsname{692}
\expandafter\def\csname EER.followup.all_origins.calendar4.24.0.any_credential.observed\endcsname{688}
\expandafter\def\csname EER.followup.all_origins.calendar4.24.0.any_credential.yes\endcsname{236}
\expandafter\def\csname EER.followup.all_origins.calendar4.24.0.any_credential.missing\endcsname{4}
\expandafter\def\csname EER.followup.all_origins.calendar4.24.0.any_credential.fraction\endcsname{236/688}
\expandafter\def\csname EER.followup.all_origins.calendar4.24.0.diploma.pct\endcsname{21.9}
\expandafter\def\csname EER.followup.all_origins.calendar4.24.0.diploma.total\endcsname{692}
\expandafter\def\csname EER.followup.all_origins.calendar4.24.0.diploma.observed\endcsname{688}
\expandafter\def\csname EER.followup.all_origins.calendar4.24.0.diploma.yes\endcsname{151}
\expandafter\def\csname EER.followup.all_origins.calendar4.24.0.diploma.missing\endcsname{4}
\expandafter\def\csname EER.followup.all_origins.calendar4.24.0.diploma.fraction\endcsname{151/688}
\expandafter\def\csname EER.followup.all_origins.calendar4.24.0.ged.pct\endcsname{12.4}
\expandafter\def\csname EER.followup.all_origins.calendar4.24.0.ged.total\endcsname{692}
\expandafter\def\csname EER.followup.all_origins.calendar4.24.0.ged.observed\endcsname{688}
\expandafter\def\csname EER.followup.all_origins.calendar4.24.0.ged.yes\endcsname{85}
\expandafter\def\csname EER.followup.all_origins.calendar4.24.0.ged.missing\endcsname{4}
\expandafter\def\csname EER.followup.all_origins.calendar4.24.0.ged.fraction\endcsname{85/688}
\expandafter\def\csname EER.followup.all_origins.calendar4.24.0.missing_interview.pct\endcsname{0.6}
\expandafter\def\csname EER.followup.all_origins.calendar4.24.0.missing_interview.total\endcsname{692}
\expandafter\def\csname EER.followup.all_origins.calendar4.24.0.missing_interview.observed\endcsname{692}
\expandafter\def\csname EER.followup.all_origins.calendar4.24.0.missing_interview.yes\endcsname{4}
\expandafter\def\csname EER.followup.all_origins.calendar4.24.0.missing_interview.missing\endcsname{0}
\expandafter\def\csname EER.followup.all_origins.calendar4.24.0.missing_interview.fraction\endcsname{4/692}
\expandafter\def\csname EER.followup.all_origins.calendar4.24.0.missing_array.pct\endcsname{0.6}
\expandafter\def\csname EER.followup.all_origins.calendar4.24.0.missing_array.total\endcsname{692}
\expandafter\def\csname EER.followup.all_origins.calendar4.24.0.missing_array.observed\endcsname{692}
\expandafter\def\csname EER.followup.all_origins.calendar4.24.0.missing_array.yes\endcsname{4}
\expandafter\def\csname EER.followup.all_origins.calendar4.24.0.missing_array.missing\endcsname{0}
\expandafter\def\csname EER.followup.all_origins.calendar4.24.0.missing_array.fraction\endcsname{4/692}
\expandafter\def\csname EER.followup.all_origins.calendar4.24.0.missing_status.pct\endcsname{1.0}
\expandafter\def\csname EER.followup.all_origins.calendar4.24.0.missing_status.total\endcsname{692}
\expandafter\def\csname EER.followup.all_origins.calendar4.24.0.missing_status.observed\endcsname{692}
\expandafter\def\csname EER.followup.all_origins.calendar4.24.0.missing_status.yes\endcsname{7}
\expandafter\def\csname EER.followup.all_origins.calendar4.24.0.missing_status.missing\endcsname{0}
\expandafter\def\csname EER.followup.all_origins.calendar4.24.0.missing_status.fraction\endcsname{7/692}
\expandafter\def\csname EER.followup.all_origins.calendar4.24.1.postevent_enrollment.pct\endcsname{22.0}
\expandafter\def\csname EER.followup.all_origins.calendar4.24.1.postevent_enrollment.total\endcsname{259}
\expandafter\def\csname EER.followup.all_origins.calendar4.24.1.postevent_enrollment.observed\endcsname{254}
\expandafter\def\csname EER.followup.all_origins.calendar4.24.1.postevent_enrollment.yes\endcsname{56}
\expandafter\def\csname EER.followup.all_origins.calendar4.24.1.postevent_enrollment.missing\endcsname{5}
\expandafter\def\csname EER.followup.all_origins.calendar4.24.1.postevent_enrollment.fraction\endcsname{56/254}
\expandafter\def\csname EER.followup.all_origins.calendar4.24.1.observed_resumption.pct\endcsname{22.0}
\expandafter\def\csname EER.followup.all_origins.calendar4.24.1.observed_resumption.total\endcsname{259}
\expandafter\def\csname EER.followup.all_origins.calendar4.24.1.observed_resumption.observed\endcsname{254}
\expandafter\def\csname EER.followup.all_origins.calendar4.24.1.observed_resumption.yes\endcsname{56}
\expandafter\def\csname EER.followup.all_origins.calendar4.24.1.observed_resumption.missing\endcsname{5}
\expandafter\def\csname EER.followup.all_origins.calendar4.24.1.observed_resumption.fraction\endcsname{56/254}
\expandafter\def\csname EER.followup.all_origins.calendar4.24.1.any_credential.pct\endcsname{24.8}
\expandafter\def\csname EER.followup.all_origins.calendar4.24.1.any_credential.total\endcsname{259}
\expandafter\def\csname EER.followup.all_origins.calendar4.24.1.any_credential.observed\endcsname{258}
\expandafter\def\csname EER.followup.all_origins.calendar4.24.1.any_credential.yes\endcsname{64}
\expandafter\def\csname EER.followup.all_origins.calendar4.24.1.any_credential.missing\endcsname{1}
\expandafter\def\csname EER.followup.all_origins.calendar4.24.1.any_credential.fraction\endcsname{64/258}
\expandafter\def\csname EER.followup.all_origins.calendar4.24.1.diploma.pct\endcsname{11.2}
\expandafter\def\csname EER.followup.all_origins.calendar4.24.1.diploma.total\endcsname{259}
\expandafter\def\csname EER.followup.all_origins.calendar4.24.1.diploma.observed\endcsname{258}
\expandafter\def\csname EER.followup.all_origins.calendar4.24.1.diploma.yes\endcsname{29}
\expandafter\def\csname EER.followup.all_origins.calendar4.24.1.diploma.missing\endcsname{1}
\expandafter\def\csname EER.followup.all_origins.calendar4.24.1.diploma.fraction\endcsname{29/258}
\expandafter\def\csname EER.followup.all_origins.calendar4.24.1.ged.pct\endcsname{13.6}
\expandafter\def\csname EER.followup.all_origins.calendar4.24.1.ged.total\endcsname{259}
\expandafter\def\csname EER.followup.all_origins.calendar4.24.1.ged.observed\endcsname{258}
\expandafter\def\csname EER.followup.all_origins.calendar4.24.1.ged.yes\endcsname{35}
\expandafter\def\csname EER.followup.all_origins.calendar4.24.1.ged.missing\endcsname{1}
\expandafter\def\csname EER.followup.all_origins.calendar4.24.1.ged.fraction\endcsname{35/258}
\expandafter\def\csname EER.followup.all_origins.calendar4.24.1.missing_interview.pct\endcsname{0.4}
\expandafter\def\csname EER.followup.all_origins.calendar4.24.1.missing_interview.total\endcsname{259}
\expandafter\def\csname EER.followup.all_origins.calendar4.24.1.missing_interview.observed\endcsname{259}
\expandafter\def\csname EER.followup.all_origins.calendar4.24.1.missing_interview.yes\endcsname{1}
\expandafter\def\csname EER.followup.all_origins.calendar4.24.1.missing_interview.missing\endcsname{0}
\expandafter\def\csname EER.followup.all_origins.calendar4.24.1.missing_interview.fraction\endcsname{1/259}
\expandafter\def\csname EER.followup.all_origins.calendar4.24.1.missing_array.pct\endcsname{0.4}
\expandafter\def\csname EER.followup.all_origins.calendar4.24.1.missing_array.total\endcsname{259}
\expandafter\def\csname EER.followup.all_origins.calendar4.24.1.missing_array.observed\endcsname{259}
\expandafter\def\csname EER.followup.all_origins.calendar4.24.1.missing_array.yes\endcsname{1}
\expandafter\def\csname EER.followup.all_origins.calendar4.24.1.missing_array.missing\endcsname{0}
\expandafter\def\csname EER.followup.all_origins.calendar4.24.1.missing_array.fraction\endcsname{1/259}
\expandafter\def\csname EER.followup.all_origins.calendar4.24.1.missing_status.pct\endcsname{1.5}
\expandafter\def\csname EER.followup.all_origins.calendar4.24.1.missing_status.total\endcsname{259}
\expandafter\def\csname EER.followup.all_origins.calendar4.24.1.missing_status.observed\endcsname{259}
\expandafter\def\csname EER.followup.all_origins.calendar4.24.1.missing_status.yes\endcsname{4}
\expandafter\def\csname EER.followup.all_origins.calendar4.24.1.missing_status.missing\endcsname{0}
\expandafter\def\csname EER.followup.all_origins.calendar4.24.1.missing_status.fraction\endcsname{4/259}
\expandafter\def\csname EER.followup.original_sample.recorded.12.0.postevent_enrollment.pct\endcsname{2.9}
\expandafter\def\csname EER.followup.original_sample.recorded.12.0.postevent_enrollment.total\endcsname{434}
\expandafter\def\csname EER.followup.original_sample.recorded.12.0.postevent_enrollment.observed\endcsname{419}
\expandafter\def\csname EER.followup.original_sample.recorded.12.0.postevent_enrollment.yes\endcsname{12}
\expandafter\def\csname EER.followup.original_sample.recorded.12.0.postevent_enrollment.missing\endcsname{15}
\expandafter\def\csname EER.followup.original_sample.recorded.12.0.postevent_enrollment.fraction\endcsname{12/419}
\expandafter\def\csname EER.followup.original_sample.recorded.12.0.observed_resumption.pct\endcsname{1.2}
\expandafter\def\csname EER.followup.original_sample.recorded.12.0.observed_resumption.total\endcsname{434}
\expandafter\def\csname EER.followup.original_sample.recorded.12.0.observed_resumption.observed\endcsname{419}
\expandafter\def\csname EER.followup.original_sample.recorded.12.0.observed_resumption.yes\endcsname{5}
\expandafter\def\csname EER.followup.original_sample.recorded.12.0.observed_resumption.missing\endcsname{15}
\expandafter\def\csname EER.followup.original_sample.recorded.12.0.observed_resumption.fraction\endcsname{5/419}
\expandafter\def\csname EER.followup.original_sample.recorded.12.0.any_credential.pct\endcsname{14.5}
\expandafter\def\csname EER.followup.original_sample.recorded.12.0.any_credential.total\endcsname{434}
\expandafter\def\csname EER.followup.original_sample.recorded.12.0.any_credential.observed\endcsname{428}
\expandafter\def\csname EER.followup.original_sample.recorded.12.0.any_credential.yes\endcsname{62}
\expandafter\def\csname EER.followup.original_sample.recorded.12.0.any_credential.missing\endcsname{6}
\expandafter\def\csname EER.followup.original_sample.recorded.12.0.any_credential.fraction\endcsname{62/428}
\expandafter\def\csname EER.followup.original_sample.recorded.12.0.diploma.pct\endcsname{0.0}
\expandafter\def\csname EER.followup.original_sample.recorded.12.0.diploma.total\endcsname{434}
\expandafter\def\csname EER.followup.original_sample.recorded.12.0.diploma.observed\endcsname{428}
\expandafter\def\csname EER.followup.original_sample.recorded.12.0.diploma.yes\endcsname{0}
\expandafter\def\csname EER.followup.original_sample.recorded.12.0.diploma.missing\endcsname{6}
\expandafter\def\csname EER.followup.original_sample.recorded.12.0.diploma.fraction\endcsname{0/428}
\expandafter\def\csname EER.followup.original_sample.recorded.12.0.ged.pct\endcsname{14.5}
\expandafter\def\csname EER.followup.original_sample.recorded.12.0.ged.total\endcsname{434}
\expandafter\def\csname EER.followup.original_sample.recorded.12.0.ged.observed\endcsname{428}
\expandafter\def\csname EER.followup.original_sample.recorded.12.0.ged.yes\endcsname{62}
\expandafter\def\csname EER.followup.original_sample.recorded.12.0.ged.missing\endcsname{6}
\expandafter\def\csname EER.followup.original_sample.recorded.12.0.ged.fraction\endcsname{62/428}
\expandafter\def\csname EER.followup.original_sample.recorded.12.0.missing_interview.pct\endcsname{1.4}
\expandafter\def\csname EER.followup.original_sample.recorded.12.0.missing_interview.total\endcsname{434}
\expandafter\def\csname EER.followup.original_sample.recorded.12.0.missing_interview.observed\endcsname{434}
\expandafter\def\csname EER.followup.original_sample.recorded.12.0.missing_interview.yes\endcsname{6}
\expandafter\def\csname EER.followup.original_sample.recorded.12.0.missing_interview.missing\endcsname{0}
\expandafter\def\csname EER.followup.original_sample.recorded.12.0.missing_interview.fraction\endcsname{6/434}
\expandafter\def\csname EER.followup.original_sample.recorded.12.0.missing_array.pct\endcsname{1.4}
\expandafter\def\csname EER.followup.original_sample.recorded.12.0.missing_array.total\endcsname{434}
\expandafter\def\csname EER.followup.original_sample.recorded.12.0.missing_array.observed\endcsname{434}
\expandafter\def\csname EER.followup.original_sample.recorded.12.0.missing_array.yes\endcsname{6}
\expandafter\def\csname EER.followup.original_sample.recorded.12.0.missing_array.missing\endcsname{0}
\expandafter\def\csname EER.followup.original_sample.recorded.12.0.missing_array.fraction\endcsname{6/434}
\expandafter\def\csname EER.followup.original_sample.recorded.12.0.missing_status.pct\endcsname{2.1}
\expandafter\def\csname EER.followup.original_sample.recorded.12.0.missing_status.total\endcsname{434}
\expandafter\def\csname EER.followup.original_sample.recorded.12.0.missing_status.observed\endcsname{434}
\expandafter\def\csname EER.followup.original_sample.recorded.12.0.missing_status.yes\endcsname{9}
\expandafter\def\csname EER.followup.original_sample.recorded.12.0.missing_status.missing\endcsname{0}
\expandafter\def\csname EER.followup.original_sample.recorded.12.0.missing_status.fraction\endcsname{9/434}
\expandafter\def\csname EER.followup.original_sample.recorded.12.1.postevent_enrollment.pct\endcsname{3.3}
\expandafter\def\csname EER.followup.original_sample.recorded.12.1.postevent_enrollment.total\endcsname{246}
\expandafter\def\csname EER.followup.original_sample.recorded.12.1.postevent_enrollment.observed\endcsname{242}
\expandafter\def\csname EER.followup.original_sample.recorded.12.1.postevent_enrollment.yes\endcsname{8}
\expandafter\def\csname EER.followup.original_sample.recorded.12.1.postevent_enrollment.missing\endcsname{4}
\expandafter\def\csname EER.followup.original_sample.recorded.12.1.postevent_enrollment.fraction\endcsname{8/242}
\expandafter\def\csname EER.followup.original_sample.recorded.12.1.observed_resumption.pct\endcsname{1.7}
\expandafter\def\csname EER.followup.original_sample.recorded.12.1.observed_resumption.total\endcsname{246}
\expandafter\def\csname EER.followup.original_sample.recorded.12.1.observed_resumption.observed\endcsname{242}
\expandafter\def\csname EER.followup.original_sample.recorded.12.1.observed_resumption.yes\endcsname{4}
\expandafter\def\csname EER.followup.original_sample.recorded.12.1.observed_resumption.missing\endcsname{4}
\expandafter\def\csname EER.followup.original_sample.recorded.12.1.observed_resumption.fraction\endcsname{4/242}
\expandafter\def\csname EER.followup.original_sample.recorded.12.1.any_credential.pct\endcsname{10.2}
\expandafter\def\csname EER.followup.original_sample.recorded.12.1.any_credential.total\endcsname{246}
\expandafter\def\csname EER.followup.original_sample.recorded.12.1.any_credential.observed\endcsname{246}
\expandafter\def\csname EER.followup.original_sample.recorded.12.1.any_credential.yes\endcsname{25}
\expandafter\def\csname EER.followup.original_sample.recorded.12.1.any_credential.missing\endcsname{0}
\expandafter\def\csname EER.followup.original_sample.recorded.12.1.any_credential.fraction\endcsname{25/246}
\expandafter\def\csname EER.followup.original_sample.recorded.12.1.diploma.pct\endcsname{0.0}
\expandafter\def\csname EER.followup.original_sample.recorded.12.1.diploma.total\endcsname{246}
\expandafter\def\csname EER.followup.original_sample.recorded.12.1.diploma.observed\endcsname{246}
\expandafter\def\csname EER.followup.original_sample.recorded.12.1.diploma.yes\endcsname{0}
\expandafter\def\csname EER.followup.original_sample.recorded.12.1.diploma.missing\endcsname{0}
\expandafter\def\csname EER.followup.original_sample.recorded.12.1.diploma.fraction\endcsname{0/246}
\expandafter\def\csname EER.followup.original_sample.recorded.12.1.ged.pct\endcsname{10.2}
\expandafter\def\csname EER.followup.original_sample.recorded.12.1.ged.total\endcsname{246}
\expandafter\def\csname EER.followup.original_sample.recorded.12.1.ged.observed\endcsname{246}
\expandafter\def\csname EER.followup.original_sample.recorded.12.1.ged.yes\endcsname{25}
\expandafter\def\csname EER.followup.original_sample.recorded.12.1.ged.missing\endcsname{0}
\expandafter\def\csname EER.followup.original_sample.recorded.12.1.ged.fraction\endcsname{25/246}
\expandafter\def\csname EER.followup.original_sample.recorded.12.1.missing_interview.pct\endcsname{0.0}
\expandafter\def\csname EER.followup.original_sample.recorded.12.1.missing_interview.total\endcsname{246}
\expandafter\def\csname EER.followup.original_sample.recorded.12.1.missing_interview.observed\endcsname{246}
\expandafter\def\csname EER.followup.original_sample.recorded.12.1.missing_interview.yes\endcsname{0}
\expandafter\def\csname EER.followup.original_sample.recorded.12.1.missing_interview.missing\endcsname{0}
\expandafter\def\csname EER.followup.original_sample.recorded.12.1.missing_interview.fraction\endcsname{0/246}
\expandafter\def\csname EER.followup.original_sample.recorded.12.1.missing_array.pct\endcsname{0.0}
\expandafter\def\csname EER.followup.original_sample.recorded.12.1.missing_array.total\endcsname{246}
\expandafter\def\csname EER.followup.original_sample.recorded.12.1.missing_array.observed\endcsname{246}
\expandafter\def\csname EER.followup.original_sample.recorded.12.1.missing_array.yes\endcsname{0}
\expandafter\def\csname EER.followup.original_sample.recorded.12.1.missing_array.missing\endcsname{0}
\expandafter\def\csname EER.followup.original_sample.recorded.12.1.missing_array.fraction\endcsname{0/246}
\expandafter\def\csname EER.followup.original_sample.recorded.12.1.missing_status.pct\endcsname{1.6}
\expandafter\def\csname EER.followup.original_sample.recorded.12.1.missing_status.total\endcsname{246}
\expandafter\def\csname EER.followup.original_sample.recorded.12.1.missing_status.observed\endcsname{246}
\expandafter\def\csname EER.followup.original_sample.recorded.12.1.missing_status.yes\endcsname{4}
\expandafter\def\csname EER.followup.original_sample.recorded.12.1.missing_status.missing\endcsname{0}
\expandafter\def\csname EER.followup.original_sample.recorded.12.1.missing_status.fraction\endcsname{4/246}
\expandafter\def\csname EER.followup.original_sample.recorded.24.0.postevent_enrollment.pct\endcsname{2.6}
\expandafter\def\csname EER.followup.original_sample.recorded.24.0.postevent_enrollment.total\endcsname{434}
\expandafter\def\csname EER.followup.original_sample.recorded.24.0.postevent_enrollment.observed\endcsname{416}
\expandafter\def\csname EER.followup.original_sample.recorded.24.0.postevent_enrollment.yes\endcsname{11}
\expandafter\def\csname EER.followup.original_sample.recorded.24.0.postevent_enrollment.missing\endcsname{18}
\expandafter\def\csname EER.followup.original_sample.recorded.24.0.postevent_enrollment.fraction\endcsname{11/416}
\expandafter\def\csname EER.followup.original_sample.recorded.24.0.observed_resumption.pct\endcsname{1.2}
\expandafter\def\csname EER.followup.original_sample.recorded.24.0.observed_resumption.total\endcsname{434}
\expandafter\def\csname EER.followup.original_sample.recorded.24.0.observed_resumption.observed\endcsname{416}
\expandafter\def\csname EER.followup.original_sample.recorded.24.0.observed_resumption.yes\endcsname{5}
\expandafter\def\csname EER.followup.original_sample.recorded.24.0.observed_resumption.missing\endcsname{18}
\expandafter\def\csname EER.followup.original_sample.recorded.24.0.observed_resumption.fraction\endcsname{5/416}
\expandafter\def\csname EER.followup.original_sample.recorded.24.0.any_credential.pct\endcsname{24.1}
\expandafter\def\csname EER.followup.original_sample.recorded.24.0.any_credential.total\endcsname{434}
\expandafter\def\csname EER.followup.original_sample.recorded.24.0.any_credential.observed\endcsname{427}
\expandafter\def\csname EER.followup.original_sample.recorded.24.0.any_credential.yes\endcsname{103}
\expandafter\def\csname EER.followup.original_sample.recorded.24.0.any_credential.missing\endcsname{7}
\expandafter\def\csname EER.followup.original_sample.recorded.24.0.any_credential.fraction\endcsname{103/427}
\expandafter\def\csname EER.followup.original_sample.recorded.24.0.diploma.pct\endcsname{0.0}
\expandafter\def\csname EER.followup.original_sample.recorded.24.0.diploma.total\endcsname{434}
\expandafter\def\csname EER.followup.original_sample.recorded.24.0.diploma.observed\endcsname{427}
\expandafter\def\csname EER.followup.original_sample.recorded.24.0.diploma.yes\endcsname{0}
\expandafter\def\csname EER.followup.original_sample.recorded.24.0.diploma.missing\endcsname{7}
\expandafter\def\csname EER.followup.original_sample.recorded.24.0.diploma.fraction\endcsname{0/427}
\expandafter\def\csname EER.followup.original_sample.recorded.24.0.ged.pct\endcsname{24.1}
\expandafter\def\csname EER.followup.original_sample.recorded.24.0.ged.total\endcsname{434}
\expandafter\def\csname EER.followup.original_sample.recorded.24.0.ged.observed\endcsname{427}
\expandafter\def\csname EER.followup.original_sample.recorded.24.0.ged.yes\endcsname{103}
\expandafter\def\csname EER.followup.original_sample.recorded.24.0.ged.missing\endcsname{7}
\expandafter\def\csname EER.followup.original_sample.recorded.24.0.ged.fraction\endcsname{103/427}
\expandafter\def\csname EER.followup.original_sample.recorded.24.0.missing_interview.pct\endcsname{1.6}
\expandafter\def\csname EER.followup.original_sample.recorded.24.0.missing_interview.total\endcsname{434}
\expandafter\def\csname EER.followup.original_sample.recorded.24.0.missing_interview.observed\endcsname{434}
\expandafter\def\csname EER.followup.original_sample.recorded.24.0.missing_interview.yes\endcsname{7}
\expandafter\def\csname EER.followup.original_sample.recorded.24.0.missing_interview.missing\endcsname{0}
\expandafter\def\csname EER.followup.original_sample.recorded.24.0.missing_interview.fraction\endcsname{7/434}
\expandafter\def\csname EER.followup.original_sample.recorded.24.0.missing_array.pct\endcsname{1.6}
\expandafter\def\csname EER.followup.original_sample.recorded.24.0.missing_array.total\endcsname{434}
\expandafter\def\csname EER.followup.original_sample.recorded.24.0.missing_array.observed\endcsname{434}
\expandafter\def\csname EER.followup.original_sample.recorded.24.0.missing_array.yes\endcsname{7}
\expandafter\def\csname EER.followup.original_sample.recorded.24.0.missing_array.missing\endcsname{0}
\expandafter\def\csname EER.followup.original_sample.recorded.24.0.missing_array.fraction\endcsname{7/434}
\expandafter\def\csname EER.followup.original_sample.recorded.24.0.missing_status.pct\endcsname{2.5}
\expandafter\def\csname EER.followup.original_sample.recorded.24.0.missing_status.total\endcsname{434}
\expandafter\def\csname EER.followup.original_sample.recorded.24.0.missing_status.observed\endcsname{434}
\expandafter\def\csname EER.followup.original_sample.recorded.24.0.missing_status.yes\endcsname{11}
\expandafter\def\csname EER.followup.original_sample.recorded.24.0.missing_status.missing\endcsname{0}
\expandafter\def\csname EER.followup.original_sample.recorded.24.0.missing_status.fraction\endcsname{11/434}
\expandafter\def\csname EER.followup.original_sample.recorded.24.1.postevent_enrollment.pct\endcsname{3.7}
\expandafter\def\csname EER.followup.original_sample.recorded.24.1.postevent_enrollment.total\endcsname{246}
\expandafter\def\csname EER.followup.original_sample.recorded.24.1.postevent_enrollment.observed\endcsname{241}
\expandafter\def\csname EER.followup.original_sample.recorded.24.1.postevent_enrollment.yes\endcsname{9}
\expandafter\def\csname EER.followup.original_sample.recorded.24.1.postevent_enrollment.missing\endcsname{5}
\expandafter\def\csname EER.followup.original_sample.recorded.24.1.postevent_enrollment.fraction\endcsname{9/241}
\expandafter\def\csname EER.followup.original_sample.recorded.24.1.observed_resumption.pct\endcsname{2.5}
\expandafter\def\csname EER.followup.original_sample.recorded.24.1.observed_resumption.total\endcsname{246}
\expandafter\def\csname EER.followup.original_sample.recorded.24.1.observed_resumption.observed\endcsname{241}
\expandafter\def\csname EER.followup.original_sample.recorded.24.1.observed_resumption.yes\endcsname{6}
\expandafter\def\csname EER.followup.original_sample.recorded.24.1.observed_resumption.missing\endcsname{5}
\expandafter\def\csname EER.followup.original_sample.recorded.24.1.observed_resumption.fraction\endcsname{6/241}
\expandafter\def\csname EER.followup.original_sample.recorded.24.1.any_credential.pct\endcsname{19.1}
\expandafter\def\csname EER.followup.original_sample.recorded.24.1.any_credential.total\endcsname{246}
\expandafter\def\csname EER.followup.original_sample.recorded.24.1.any_credential.observed\endcsname{246}
\expandafter\def\csname EER.followup.original_sample.recorded.24.1.any_credential.yes\endcsname{47}
\expandafter\def\csname EER.followup.original_sample.recorded.24.1.any_credential.missing\endcsname{0}
\expandafter\def\csname EER.followup.original_sample.recorded.24.1.any_credential.fraction\endcsname{47/246}
\expandafter\def\csname EER.followup.original_sample.recorded.24.1.diploma.pct\endcsname{0.4}
\expandafter\def\csname EER.followup.original_sample.recorded.24.1.diploma.total\endcsname{246}
\expandafter\def\csname EER.followup.original_sample.recorded.24.1.diploma.observed\endcsname{246}
\expandafter\def\csname EER.followup.original_sample.recorded.24.1.diploma.yes\endcsname{1}
\expandafter\def\csname EER.followup.original_sample.recorded.24.1.diploma.missing\endcsname{0}
\expandafter\def\csname EER.followup.original_sample.recorded.24.1.diploma.fraction\endcsname{1/246}
\expandafter\def\csname EER.followup.original_sample.recorded.24.1.ged.pct\endcsname{18.7}
\expandafter\def\csname EER.followup.original_sample.recorded.24.1.ged.total\endcsname{246}
\expandafter\def\csname EER.followup.original_sample.recorded.24.1.ged.observed\endcsname{246}
\expandafter\def\csname EER.followup.original_sample.recorded.24.1.ged.yes\endcsname{46}
\expandafter\def\csname EER.followup.original_sample.recorded.24.1.ged.missing\endcsname{0}
\expandafter\def\csname EER.followup.original_sample.recorded.24.1.ged.fraction\endcsname{46/246}
\expandafter\def\csname EER.followup.original_sample.recorded.24.1.missing_interview.pct\endcsname{0.0}
\expandafter\def\csname EER.followup.original_sample.recorded.24.1.missing_interview.total\endcsname{246}
\expandafter\def\csname EER.followup.original_sample.recorded.24.1.missing_interview.observed\endcsname{246}
\expandafter\def\csname EER.followup.original_sample.recorded.24.1.missing_interview.yes\endcsname{0}
\expandafter\def\csname EER.followup.original_sample.recorded.24.1.missing_interview.missing\endcsname{0}
\expandafter\def\csname EER.followup.original_sample.recorded.24.1.missing_interview.fraction\endcsname{0/246}
\expandafter\def\csname EER.followup.original_sample.recorded.24.1.missing_array.pct\endcsname{0.0}
\expandafter\def\csname EER.followup.original_sample.recorded.24.1.missing_array.total\endcsname{246}
\expandafter\def\csname EER.followup.original_sample.recorded.24.1.missing_array.observed\endcsname{246}
\expandafter\def\csname EER.followup.original_sample.recorded.24.1.missing_array.yes\endcsname{0}
\expandafter\def\csname EER.followup.original_sample.recorded.24.1.missing_array.missing\endcsname{0}
\expandafter\def\csname EER.followup.original_sample.recorded.24.1.missing_array.fraction\endcsname{0/246}
\expandafter\def\csname EER.followup.original_sample.recorded.24.1.missing_status.pct\endcsname{2.0}
\expandafter\def\csname EER.followup.original_sample.recorded.24.1.missing_status.total\endcsname{246}
\expandafter\def\csname EER.followup.original_sample.recorded.24.1.missing_status.observed\endcsname{246}
\expandafter\def\csname EER.followup.original_sample.recorded.24.1.missing_status.yes\endcsname{5}
\expandafter\def\csname EER.followup.original_sample.recorded.24.1.missing_status.missing\endcsname{0}
\expandafter\def\csname EER.followup.original_sample.recorded.24.1.missing_status.fraction\endcsname{5/246}
\expandafter\def\csname EER.followup.original_sample.school3.12.0.postevent_enrollment.pct\endcsname{32.4}
\expandafter\def\csname EER.followup.original_sample.school3.12.0.postevent_enrollment.total\endcsname{649}
\expandafter\def\csname EER.followup.original_sample.school3.12.0.postevent_enrollment.observed\endcsname{646}
\expandafter\def\csname EER.followup.original_sample.school3.12.0.postevent_enrollment.yes\endcsname{209}
\expandafter\def\csname EER.followup.original_sample.school3.12.0.postevent_enrollment.missing\endcsname{3}
\expandafter\def\csname EER.followup.original_sample.school3.12.0.postevent_enrollment.fraction\endcsname{209/646}
\expandafter\def\csname EER.followup.original_sample.school3.12.0.observed_resumption.pct\endcsname{32.4}
\expandafter\def\csname EER.followup.original_sample.school3.12.0.observed_resumption.total\endcsname{649}
\expandafter\def\csname EER.followup.original_sample.school3.12.0.observed_resumption.observed\endcsname{646}
\expandafter\def\csname EER.followup.original_sample.school3.12.0.observed_resumption.yes\endcsname{209}
\expandafter\def\csname EER.followup.original_sample.school3.12.0.observed_resumption.missing\endcsname{3}
\expandafter\def\csname EER.followup.original_sample.school3.12.0.observed_resumption.fraction\endcsname{209/646}
\expandafter\def\csname EER.followup.original_sample.school3.12.0.any_credential.pct\endcsname{21.3}
\expandafter\def\csname EER.followup.original_sample.school3.12.0.any_credential.total\endcsname{649}
\expandafter\def\csname EER.followup.original_sample.school3.12.0.any_credential.observed\endcsname{649}
\expandafter\def\csname EER.followup.original_sample.school3.12.0.any_credential.yes\endcsname{138}
\expandafter\def\csname EER.followup.original_sample.school3.12.0.any_credential.missing\endcsname{0}
\expandafter\def\csname EER.followup.original_sample.school3.12.0.any_credential.fraction\endcsname{138/649}
\expandafter\def\csname EER.followup.original_sample.school3.12.0.diploma.pct\endcsname{14.8}
\expandafter\def\csname EER.followup.original_sample.school3.12.0.diploma.total\endcsname{649}
\expandafter\def\csname EER.followup.original_sample.school3.12.0.diploma.observed\endcsname{649}
\expandafter\def\csname EER.followup.original_sample.school3.12.0.diploma.yes\endcsname{96}
\expandafter\def\csname EER.followup.original_sample.school3.12.0.diploma.missing\endcsname{0}
\expandafter\def\csname EER.followup.original_sample.school3.12.0.diploma.fraction\endcsname{96/649}
\expandafter\def\csname EER.followup.original_sample.school3.12.0.ged.pct\endcsname{6.5}
\expandafter\def\csname EER.followup.original_sample.school3.12.0.ged.total\endcsname{649}
\expandafter\def\csname EER.followup.original_sample.school3.12.0.ged.observed\endcsname{649}
\expandafter\def\csname EER.followup.original_sample.school3.12.0.ged.yes\endcsname{42}
\expandafter\def\csname EER.followup.original_sample.school3.12.0.ged.missing\endcsname{0}
\expandafter\def\csname EER.followup.original_sample.school3.12.0.ged.fraction\endcsname{42/649}
\expandafter\def\csname EER.followup.original_sample.school3.12.0.missing_interview.pct\endcsname{0.0}
\expandafter\def\csname EER.followup.original_sample.school3.12.0.missing_interview.total\endcsname{649}
\expandafter\def\csname EER.followup.original_sample.school3.12.0.missing_interview.observed\endcsname{649}
\expandafter\def\csname EER.followup.original_sample.school3.12.0.missing_interview.yes\endcsname{0}
\expandafter\def\csname EER.followup.original_sample.school3.12.0.missing_interview.missing\endcsname{0}
\expandafter\def\csname EER.followup.original_sample.school3.12.0.missing_interview.fraction\endcsname{0/649}
\expandafter\def\csname EER.followup.original_sample.school3.12.0.missing_array.pct\endcsname{0.0}
\expandafter\def\csname EER.followup.original_sample.school3.12.0.missing_array.total\endcsname{649}
\expandafter\def\csname EER.followup.original_sample.school3.12.0.missing_array.observed\endcsname{649}
\expandafter\def\csname EER.followup.original_sample.school3.12.0.missing_array.yes\endcsname{0}
\expandafter\def\csname EER.followup.original_sample.school3.12.0.missing_array.missing\endcsname{0}
\expandafter\def\csname EER.followup.original_sample.school3.12.0.missing_array.fraction\endcsname{0/649}
\expandafter\def\csname EER.followup.original_sample.school3.12.0.missing_status.pct\endcsname{0.5}
\expandafter\def\csname EER.followup.original_sample.school3.12.0.missing_status.total\endcsname{649}
\expandafter\def\csname EER.followup.original_sample.school3.12.0.missing_status.observed\endcsname{649}
\expandafter\def\csname EER.followup.original_sample.school3.12.0.missing_status.yes\endcsname{3}
\expandafter\def\csname EER.followup.original_sample.school3.12.0.missing_status.missing\endcsname{0}
\expandafter\def\csname EER.followup.original_sample.school3.12.0.missing_status.fraction\endcsname{3/649}
\expandafter\def\csname EER.followup.original_sample.school3.12.1.postevent_enrollment.pct\endcsname{20.5}
\expandafter\def\csname EER.followup.original_sample.school3.12.1.postevent_enrollment.total\endcsname{261}
\expandafter\def\csname EER.followup.original_sample.school3.12.1.postevent_enrollment.observed\endcsname{259}
\expandafter\def\csname EER.followup.original_sample.school3.12.1.postevent_enrollment.yes\endcsname{53}
\expandafter\def\csname EER.followup.original_sample.school3.12.1.postevent_enrollment.missing\endcsname{2}
\expandafter\def\csname EER.followup.original_sample.school3.12.1.postevent_enrollment.fraction\endcsname{53/259}
\expandafter\def\csname EER.followup.original_sample.school3.12.1.observed_resumption.pct\endcsname{20.5}
\expandafter\def\csname EER.followup.original_sample.school3.12.1.observed_resumption.total\endcsname{261}
\expandafter\def\csname EER.followup.original_sample.school3.12.1.observed_resumption.observed\endcsname{259}
\expandafter\def\csname EER.followup.original_sample.school3.12.1.observed_resumption.yes\endcsname{53}
\expandafter\def\csname EER.followup.original_sample.school3.12.1.observed_resumption.missing\endcsname{2}
\expandafter\def\csname EER.followup.original_sample.school3.12.1.observed_resumption.fraction\endcsname{53/259}
\expandafter\def\csname EER.followup.original_sample.school3.12.1.any_credential.pct\endcsname{15.7}
\expandafter\def\csname EER.followup.original_sample.school3.12.1.any_credential.total\endcsname{261}
\expandafter\def\csname EER.followup.original_sample.school3.12.1.any_credential.observed\endcsname{261}
\expandafter\def\csname EER.followup.original_sample.school3.12.1.any_credential.yes\endcsname{41}
\expandafter\def\csname EER.followup.original_sample.school3.12.1.any_credential.missing\endcsname{0}
\expandafter\def\csname EER.followup.original_sample.school3.12.1.any_credential.fraction\endcsname{41/261}
\expandafter\def\csname EER.followup.original_sample.school3.12.1.diploma.pct\endcsname{7.7}
\expandafter\def\csname EER.followup.original_sample.school3.12.1.diploma.total\endcsname{261}
\expandafter\def\csname EER.followup.original_sample.school3.12.1.diploma.observed\endcsname{261}
\expandafter\def\csname EER.followup.original_sample.school3.12.1.diploma.yes\endcsname{20}
\expandafter\def\csname EER.followup.original_sample.school3.12.1.diploma.missing\endcsname{0}
\expandafter\def\csname EER.followup.original_sample.school3.12.1.diploma.fraction\endcsname{20/261}
\expandafter\def\csname EER.followup.original_sample.school3.12.1.ged.pct\endcsname{8.0}
\expandafter\def\csname EER.followup.original_sample.school3.12.1.ged.total\endcsname{261}
\expandafter\def\csname EER.followup.original_sample.school3.12.1.ged.observed\endcsname{261}
\expandafter\def\csname EER.followup.original_sample.school3.12.1.ged.yes\endcsname{21}
\expandafter\def\csname EER.followup.original_sample.school3.12.1.ged.missing\endcsname{0}
\expandafter\def\csname EER.followup.original_sample.school3.12.1.ged.fraction\endcsname{21/261}
\expandafter\def\csname EER.followup.original_sample.school3.12.1.missing_interview.pct\endcsname{0.0}
\expandafter\def\csname EER.followup.original_sample.school3.12.1.missing_interview.total\endcsname{261}
\expandafter\def\csname EER.followup.original_sample.school3.12.1.missing_interview.observed\endcsname{261}
\expandafter\def\csname EER.followup.original_sample.school3.12.1.missing_interview.yes\endcsname{0}
\expandafter\def\csname EER.followup.original_sample.school3.12.1.missing_interview.missing\endcsname{0}
\expandafter\def\csname EER.followup.original_sample.school3.12.1.missing_interview.fraction\endcsname{0/261}
\expandafter\def\csname EER.followup.original_sample.school3.12.1.missing_array.pct\endcsname{0.0}
\expandafter\def\csname EER.followup.original_sample.school3.12.1.missing_array.total\endcsname{261}
\expandafter\def\csname EER.followup.original_sample.school3.12.1.missing_array.observed\endcsname{261}
\expandafter\def\csname EER.followup.original_sample.school3.12.1.missing_array.yes\endcsname{0}
\expandafter\def\csname EER.followup.original_sample.school3.12.1.missing_array.missing\endcsname{0}
\expandafter\def\csname EER.followup.original_sample.school3.12.1.missing_array.fraction\endcsname{0/261}
\expandafter\def\csname EER.followup.original_sample.school3.12.1.missing_status.pct\endcsname{0.8}
\expandafter\def\csname EER.followup.original_sample.school3.12.1.missing_status.total\endcsname{261}
\expandafter\def\csname EER.followup.original_sample.school3.12.1.missing_status.observed\endcsname{261}
\expandafter\def\csname EER.followup.original_sample.school3.12.1.missing_status.yes\endcsname{2}
\expandafter\def\csname EER.followup.original_sample.school3.12.1.missing_status.missing\endcsname{0}
\expandafter\def\csname EER.followup.original_sample.school3.12.1.missing_status.fraction\endcsname{2/261}
\expandafter\def\csname EER.followup.original_sample.school3.24.0.postevent_enrollment.pct\endcsname{34.1}
\expandafter\def\csname EER.followup.original_sample.school3.24.0.postevent_enrollment.total\endcsname{649}
\expandafter\def\csname EER.followup.original_sample.school3.24.0.postevent_enrollment.observed\endcsname{640}
\expandafter\def\csname EER.followup.original_sample.school3.24.0.postevent_enrollment.yes\endcsname{218}
\expandafter\def\csname EER.followup.original_sample.school3.24.0.postevent_enrollment.missing\endcsname{9}
\expandafter\def\csname EER.followup.original_sample.school3.24.0.postevent_enrollment.fraction\endcsname{218/640}
\expandafter\def\csname EER.followup.original_sample.school3.24.0.observed_resumption.pct\endcsname{34.1}
\expandafter\def\csname EER.followup.original_sample.school3.24.0.observed_resumption.total\endcsname{649}
\expandafter\def\csname EER.followup.original_sample.school3.24.0.observed_resumption.observed\endcsname{640}
\expandafter\def\csname EER.followup.original_sample.school3.24.0.observed_resumption.yes\endcsname{218}
\expandafter\def\csname EER.followup.original_sample.school3.24.0.observed_resumption.missing\endcsname{9}
\expandafter\def\csname EER.followup.original_sample.school3.24.0.observed_resumption.fraction\endcsname{218/640}
\expandafter\def\csname EER.followup.original_sample.school3.24.0.any_credential.pct\endcsname{35.2}
\expandafter\def\csname EER.followup.original_sample.school3.24.0.any_credential.total\endcsname{649}
\expandafter\def\csname EER.followup.original_sample.school3.24.0.any_credential.observed\endcsname{645}
\expandafter\def\csname EER.followup.original_sample.school3.24.0.any_credential.yes\endcsname{227}
\expandafter\def\csname EER.followup.original_sample.school3.24.0.any_credential.missing\endcsname{4}
\expandafter\def\csname EER.followup.original_sample.school3.24.0.any_credential.fraction\endcsname{227/645}
\expandafter\def\csname EER.followup.original_sample.school3.24.0.diploma.pct\endcsname{22.2}
\expandafter\def\csname EER.followup.original_sample.school3.24.0.diploma.total\endcsname{649}
\expandafter\def\csname EER.followup.original_sample.school3.24.0.diploma.observed\endcsname{645}
\expandafter\def\csname EER.followup.original_sample.school3.24.0.diploma.yes\endcsname{143}
\expandafter\def\csname EER.followup.original_sample.school3.24.0.diploma.missing\endcsname{4}
\expandafter\def\csname EER.followup.original_sample.school3.24.0.diploma.fraction\endcsname{143/645}
\expandafter\def\csname EER.followup.original_sample.school3.24.0.ged.pct\endcsname{13.0}
\expandafter\def\csname EER.followup.original_sample.school3.24.0.ged.total\endcsname{649}
\expandafter\def\csname EER.followup.original_sample.school3.24.0.ged.observed\endcsname{645}
\expandafter\def\csname EER.followup.original_sample.school3.24.0.ged.yes\endcsname{84}
\expandafter\def\csname EER.followup.original_sample.school3.24.0.ged.missing\endcsname{4}
\expandafter\def\csname EER.followup.original_sample.school3.24.0.ged.fraction\endcsname{84/645}
\expandafter\def\csname EER.followup.original_sample.school3.24.0.missing_interview.pct\endcsname{0.6}
\expandafter\def\csname EER.followup.original_sample.school3.24.0.missing_interview.total\endcsname{649}
\expandafter\def\csname EER.followup.original_sample.school3.24.0.missing_interview.observed\endcsname{649}
\expandafter\def\csname EER.followup.original_sample.school3.24.0.missing_interview.yes\endcsname{4}
\expandafter\def\csname EER.followup.original_sample.school3.24.0.missing_interview.missing\endcsname{0}
\expandafter\def\csname EER.followup.original_sample.school3.24.0.missing_interview.fraction\endcsname{4/649}
\expandafter\def\csname EER.followup.original_sample.school3.24.0.missing_array.pct\endcsname{0.6}
\expandafter\def\csname EER.followup.original_sample.school3.24.0.missing_array.total\endcsname{649}
\expandafter\def\csname EER.followup.original_sample.school3.24.0.missing_array.observed\endcsname{649}
\expandafter\def\csname EER.followup.original_sample.school3.24.0.missing_array.yes\endcsname{4}
\expandafter\def\csname EER.followup.original_sample.school3.24.0.missing_array.missing\endcsname{0}
\expandafter\def\csname EER.followup.original_sample.school3.24.0.missing_array.fraction\endcsname{4/649}
\expandafter\def\csname EER.followup.original_sample.school3.24.0.missing_status.pct\endcsname{0.8}
\expandafter\def\csname EER.followup.original_sample.school3.24.0.missing_status.total\endcsname{649}
\expandafter\def\csname EER.followup.original_sample.school3.24.0.missing_status.observed\endcsname{649}
\expandafter\def\csname EER.followup.original_sample.school3.24.0.missing_status.yes\endcsname{5}
\expandafter\def\csname EER.followup.original_sample.school3.24.0.missing_status.missing\endcsname{0}
\expandafter\def\csname EER.followup.original_sample.school3.24.0.missing_status.fraction\endcsname{5/649}
\expandafter\def\csname EER.followup.original_sample.school3.24.1.postevent_enrollment.pct\endcsname{20.4}
\expandafter\def\csname EER.followup.original_sample.school3.24.1.postevent_enrollment.total\endcsname{261}
\expandafter\def\csname EER.followup.original_sample.school3.24.1.postevent_enrollment.observed\endcsname{255}
\expandafter\def\csname EER.followup.original_sample.school3.24.1.postevent_enrollment.yes\endcsname{52}
\expandafter\def\csname EER.followup.original_sample.school3.24.1.postevent_enrollment.missing\endcsname{6}
\expandafter\def\csname EER.followup.original_sample.school3.24.1.postevent_enrollment.fraction\endcsname{52/255}
\expandafter\def\csname EER.followup.original_sample.school3.24.1.observed_resumption.pct\endcsname{20.4}
\expandafter\def\csname EER.followup.original_sample.school3.24.1.observed_resumption.total\endcsname{261}
\expandafter\def\csname EER.followup.original_sample.school3.24.1.observed_resumption.observed\endcsname{255}
\expandafter\def\csname EER.followup.original_sample.school3.24.1.observed_resumption.yes\endcsname{52}
\expandafter\def\csname EER.followup.original_sample.school3.24.1.observed_resumption.missing\endcsname{6}
\expandafter\def\csname EER.followup.original_sample.school3.24.1.observed_resumption.fraction\endcsname{52/255}
\expandafter\def\csname EER.followup.original_sample.school3.24.1.any_credential.pct\endcsname{26.2}
\expandafter\def\csname EER.followup.original_sample.school3.24.1.any_credential.total\endcsname{261}
\expandafter\def\csname EER.followup.original_sample.school3.24.1.any_credential.observed\endcsname{260}
\expandafter\def\csname EER.followup.original_sample.school3.24.1.any_credential.yes\endcsname{68}
\expandafter\def\csname EER.followup.original_sample.school3.24.1.any_credential.missing\endcsname{1}
\expandafter\def\csname EER.followup.original_sample.school3.24.1.any_credential.fraction\endcsname{68/260}
\expandafter\def\csname EER.followup.original_sample.school3.24.1.diploma.pct\endcsname{11.5}
\expandafter\def\csname EER.followup.original_sample.school3.24.1.diploma.total\endcsname{261}
\expandafter\def\csname EER.followup.original_sample.school3.24.1.diploma.observed\endcsname{260}
\expandafter\def\csname EER.followup.original_sample.school3.24.1.diploma.yes\endcsname{30}
\expandafter\def\csname EER.followup.original_sample.school3.24.1.diploma.missing\endcsname{1}
\expandafter\def\csname EER.followup.original_sample.school3.24.1.diploma.fraction\endcsname{30/260}
\expandafter\def\csname EER.followup.original_sample.school3.24.1.ged.pct\endcsname{14.6}
\expandafter\def\csname EER.followup.original_sample.school3.24.1.ged.total\endcsname{261}
\expandafter\def\csname EER.followup.original_sample.school3.24.1.ged.observed\endcsname{260}
\expandafter\def\csname EER.followup.original_sample.school3.24.1.ged.yes\endcsname{38}
\expandafter\def\csname EER.followup.original_sample.school3.24.1.ged.missing\endcsname{1}
\expandafter\def\csname EER.followup.original_sample.school3.24.1.ged.fraction\endcsname{38/260}
\expandafter\def\csname EER.followup.original_sample.school3.24.1.missing_interview.pct\endcsname{0.4}
\expandafter\def\csname EER.followup.original_sample.school3.24.1.missing_interview.total\endcsname{261}
\expandafter\def\csname EER.followup.original_sample.school3.24.1.missing_interview.observed\endcsname{261}
\expandafter\def\csname EER.followup.original_sample.school3.24.1.missing_interview.yes\endcsname{1}
\expandafter\def\csname EER.followup.original_sample.school3.24.1.missing_interview.missing\endcsname{0}
\expandafter\def\csname EER.followup.original_sample.school3.24.1.missing_interview.fraction\endcsname{1/261}
\expandafter\def\csname EER.followup.original_sample.school3.24.1.missing_array.pct\endcsname{0.4}
\expandafter\def\csname EER.followup.original_sample.school3.24.1.missing_array.total\endcsname{261}
\expandafter\def\csname EER.followup.original_sample.school3.24.1.missing_array.observed\endcsname{261}
\expandafter\def\csname EER.followup.original_sample.school3.24.1.missing_array.yes\endcsname{1}
\expandafter\def\csname EER.followup.original_sample.school3.24.1.missing_array.missing\endcsname{0}
\expandafter\def\csname EER.followup.original_sample.school3.24.1.missing_array.fraction\endcsname{1/261}
\expandafter\def\csname EER.followup.original_sample.school3.24.1.missing_status.pct\endcsname{1.9}
\expandafter\def\csname EER.followup.original_sample.school3.24.1.missing_status.total\endcsname{261}
\expandafter\def\csname EER.followup.original_sample.school3.24.1.missing_status.observed\endcsname{261}
\expandafter\def\csname EER.followup.original_sample.school3.24.1.missing_status.yes\endcsname{5}
\expandafter\def\csname EER.followup.original_sample.school3.24.1.missing_status.missing\endcsname{0}
\expandafter\def\csname EER.followup.original_sample.school3.24.1.missing_status.fraction\endcsname{5/261}
\expandafter\def\csname EER.followup.original_sample.calendar4.12.0.postevent_enrollment.pct\endcsname{38.4}
\expandafter\def\csname EER.followup.original_sample.calendar4.12.0.postevent_enrollment.total\endcsname{692}
\expandafter\def\csname EER.followup.original_sample.calendar4.12.0.postevent_enrollment.observed\endcsname{690}
\expandafter\def\csname EER.followup.original_sample.calendar4.12.0.postevent_enrollment.yes\endcsname{265}
\expandafter\def\csname EER.followup.original_sample.calendar4.12.0.postevent_enrollment.missing\endcsname{2}
\expandafter\def\csname EER.followup.original_sample.calendar4.12.0.postevent_enrollment.fraction\endcsname{265/690}
\expandafter\def\csname EER.followup.original_sample.calendar4.12.0.observed_resumption.pct\endcsname{38.4}
\expandafter\def\csname EER.followup.original_sample.calendar4.12.0.observed_resumption.total\endcsname{692}
\expandafter\def\csname EER.followup.original_sample.calendar4.12.0.observed_resumption.observed\endcsname{690}
\expandafter\def\csname EER.followup.original_sample.calendar4.12.0.observed_resumption.yes\endcsname{265}
\expandafter\def\csname EER.followup.original_sample.calendar4.12.0.observed_resumption.missing\endcsname{2}
\expandafter\def\csname EER.followup.original_sample.calendar4.12.0.observed_resumption.fraction\endcsname{265/690}
\expandafter\def\csname EER.followup.original_sample.calendar4.12.0.any_credential.pct\endcsname{20.2}
\expandafter\def\csname EER.followup.original_sample.calendar4.12.0.any_credential.total\endcsname{692}
\expandafter\def\csname EER.followup.original_sample.calendar4.12.0.any_credential.observed\endcsname{692}
\expandafter\def\csname EER.followup.original_sample.calendar4.12.0.any_credential.yes\endcsname{140}
\expandafter\def\csname EER.followup.original_sample.calendar4.12.0.any_credential.missing\endcsname{0}
\expandafter\def\csname EER.followup.original_sample.calendar4.12.0.any_credential.fraction\endcsname{140/692}
\expandafter\def\csname EER.followup.original_sample.calendar4.12.0.diploma.pct\endcsname{13.7}
\expandafter\def\csname EER.followup.original_sample.calendar4.12.0.diploma.total\endcsname{692}
\expandafter\def\csname EER.followup.original_sample.calendar4.12.0.diploma.observed\endcsname{692}
\expandafter\def\csname EER.followup.original_sample.calendar4.12.0.diploma.yes\endcsname{95}
\expandafter\def\csname EER.followup.original_sample.calendar4.12.0.diploma.missing\endcsname{0}
\expandafter\def\csname EER.followup.original_sample.calendar4.12.0.diploma.fraction\endcsname{95/692}
\expandafter\def\csname EER.followup.original_sample.calendar4.12.0.ged.pct\endcsname{6.5}
\expandafter\def\csname EER.followup.original_sample.calendar4.12.0.ged.total\endcsname{692}
\expandafter\def\csname EER.followup.original_sample.calendar4.12.0.ged.observed\endcsname{692}
\expandafter\def\csname EER.followup.original_sample.calendar4.12.0.ged.yes\endcsname{45}
\expandafter\def\csname EER.followup.original_sample.calendar4.12.0.ged.missing\endcsname{0}
\expandafter\def\csname EER.followup.original_sample.calendar4.12.0.ged.fraction\endcsname{45/692}
\expandafter\def\csname EER.followup.original_sample.calendar4.12.0.missing_interview.pct\endcsname{0.0}
\expandafter\def\csname EER.followup.original_sample.calendar4.12.0.missing_interview.total\endcsname{692}
\expandafter\def\csname EER.followup.original_sample.calendar4.12.0.missing_interview.observed\endcsname{692}
\expandafter\def\csname EER.followup.original_sample.calendar4.12.0.missing_interview.yes\endcsname{0}
\expandafter\def\csname EER.followup.original_sample.calendar4.12.0.missing_interview.missing\endcsname{0}
\expandafter\def\csname EER.followup.original_sample.calendar4.12.0.missing_interview.fraction\endcsname{0/692}
\expandafter\def\csname EER.followup.original_sample.calendar4.12.0.missing_array.pct\endcsname{0.0}
\expandafter\def\csname EER.followup.original_sample.calendar4.12.0.missing_array.total\endcsname{692}
\expandafter\def\csname EER.followup.original_sample.calendar4.12.0.missing_array.observed\endcsname{692}
\expandafter\def\csname EER.followup.original_sample.calendar4.12.0.missing_array.yes\endcsname{0}
\expandafter\def\csname EER.followup.original_sample.calendar4.12.0.missing_array.missing\endcsname{0}
\expandafter\def\csname EER.followup.original_sample.calendar4.12.0.missing_array.fraction\endcsname{0/692}
\expandafter\def\csname EER.followup.original_sample.calendar4.12.0.missing_status.pct\endcsname{0.3}
\expandafter\def\csname EER.followup.original_sample.calendar4.12.0.missing_status.total\endcsname{692}
\expandafter\def\csname EER.followup.original_sample.calendar4.12.0.missing_status.observed\endcsname{692}
\expandafter\def\csname EER.followup.original_sample.calendar4.12.0.missing_status.yes\endcsname{2}
\expandafter\def\csname EER.followup.original_sample.calendar4.12.0.missing_status.missing\endcsname{0}
\expandafter\def\csname EER.followup.original_sample.calendar4.12.0.missing_status.fraction\endcsname{2/692}
\expandafter\def\csname EER.followup.original_sample.calendar4.12.1.postevent_enrollment.pct\endcsname{22.0}
\expandafter\def\csname EER.followup.original_sample.calendar4.12.1.postevent_enrollment.total\endcsname{258}
\expandafter\def\csname EER.followup.original_sample.calendar4.12.1.postevent_enrollment.observed\endcsname{255}
\expandafter\def\csname EER.followup.original_sample.calendar4.12.1.postevent_enrollment.yes\endcsname{56}
\expandafter\def\csname EER.followup.original_sample.calendar4.12.1.postevent_enrollment.missing\endcsname{3}
\expandafter\def\csname EER.followup.original_sample.calendar4.12.1.postevent_enrollment.fraction\endcsname{56/255}
\expandafter\def\csname EER.followup.original_sample.calendar4.12.1.observed_resumption.pct\endcsname{22.0}
\expandafter\def\csname EER.followup.original_sample.calendar4.12.1.observed_resumption.total\endcsname{258}
\expandafter\def\csname EER.followup.original_sample.calendar4.12.1.observed_resumption.observed\endcsname{255}
\expandafter\def\csname EER.followup.original_sample.calendar4.12.1.observed_resumption.yes\endcsname{56}
\expandafter\def\csname EER.followup.original_sample.calendar4.12.1.observed_resumption.missing\endcsname{3}
\expandafter\def\csname EER.followup.original_sample.calendar4.12.1.observed_resumption.fraction\endcsname{56/255}
\expandafter\def\csname EER.followup.original_sample.calendar4.12.1.any_credential.pct\endcsname{14.3}
\expandafter\def\csname EER.followup.original_sample.calendar4.12.1.any_credential.total\endcsname{258}
\expandafter\def\csname EER.followup.original_sample.calendar4.12.1.any_credential.observed\endcsname{258}
\expandafter\def\csname EER.followup.original_sample.calendar4.12.1.any_credential.yes\endcsname{37}
\expandafter\def\csname EER.followup.original_sample.calendar4.12.1.any_credential.missing\endcsname{0}
\expandafter\def\csname EER.followup.original_sample.calendar4.12.1.any_credential.fraction\endcsname{37/258}
\expandafter\def\csname EER.followup.original_sample.calendar4.12.1.diploma.pct\endcsname{7.0}
\expandafter\def\csname EER.followup.original_sample.calendar4.12.1.diploma.total\endcsname{258}
\expandafter\def\csname EER.followup.original_sample.calendar4.12.1.diploma.observed\endcsname{258}
\expandafter\def\csname EER.followup.original_sample.calendar4.12.1.diploma.yes\endcsname{18}
\expandafter\def\csname EER.followup.original_sample.calendar4.12.1.diploma.missing\endcsname{0}
\expandafter\def\csname EER.followup.original_sample.calendar4.12.1.diploma.fraction\endcsname{18/258}
\expandafter\def\csname EER.followup.original_sample.calendar4.12.1.ged.pct\endcsname{7.4}
\expandafter\def\csname EER.followup.original_sample.calendar4.12.1.ged.total\endcsname{258}
\expandafter\def\csname EER.followup.original_sample.calendar4.12.1.ged.observed\endcsname{258}
\expandafter\def\csname EER.followup.original_sample.calendar4.12.1.ged.yes\endcsname{19}
\expandafter\def\csname EER.followup.original_sample.calendar4.12.1.ged.missing\endcsname{0}
\expandafter\def\csname EER.followup.original_sample.calendar4.12.1.ged.fraction\endcsname{19/258}
\expandafter\def\csname EER.followup.original_sample.calendar4.12.1.missing_interview.pct\endcsname{0.0}
\expandafter\def\csname EER.followup.original_sample.calendar4.12.1.missing_interview.total\endcsname{258}
\expandafter\def\csname EER.followup.original_sample.calendar4.12.1.missing_interview.observed\endcsname{258}
\expandafter\def\csname EER.followup.original_sample.calendar4.12.1.missing_interview.yes\endcsname{0}
\expandafter\def\csname EER.followup.original_sample.calendar4.12.1.missing_interview.missing\endcsname{0}
\expandafter\def\csname EER.followup.original_sample.calendar4.12.1.missing_interview.fraction\endcsname{0/258}
\expandafter\def\csname EER.followup.original_sample.calendar4.12.1.missing_array.pct\endcsname{0.0}
\expandafter\def\csname EER.followup.original_sample.calendar4.12.1.missing_array.total\endcsname{258}
\expandafter\def\csname EER.followup.original_sample.calendar4.12.1.missing_array.observed\endcsname{258}
\expandafter\def\csname EER.followup.original_sample.calendar4.12.1.missing_array.yes\endcsname{0}
\expandafter\def\csname EER.followup.original_sample.calendar4.12.1.missing_array.missing\endcsname{0}
\expandafter\def\csname EER.followup.original_sample.calendar4.12.1.missing_array.fraction\endcsname{0/258}
\expandafter\def\csname EER.followup.original_sample.calendar4.12.1.missing_status.pct\endcsname{1.2}
\expandafter\def\csname EER.followup.original_sample.calendar4.12.1.missing_status.total\endcsname{258}
\expandafter\def\csname EER.followup.original_sample.calendar4.12.1.missing_status.observed\endcsname{258}
\expandafter\def\csname EER.followup.original_sample.calendar4.12.1.missing_status.yes\endcsname{3}
\expandafter\def\csname EER.followup.original_sample.calendar4.12.1.missing_status.missing\endcsname{0}
\expandafter\def\csname EER.followup.original_sample.calendar4.12.1.missing_status.fraction\endcsname{3/258}
\expandafter\def\csname EER.followup.original_sample.calendar4.24.0.postevent_enrollment.pct\endcsname{39.5}
\expandafter\def\csname EER.followup.original_sample.calendar4.24.0.postevent_enrollment.total\endcsname{692}
\expandafter\def\csname EER.followup.original_sample.calendar4.24.0.postevent_enrollment.observed\endcsname{681}
\expandafter\def\csname EER.followup.original_sample.calendar4.24.0.postevent_enrollment.yes\endcsname{269}
\expandafter\def\csname EER.followup.original_sample.calendar4.24.0.postevent_enrollment.missing\endcsname{11}
\expandafter\def\csname EER.followup.original_sample.calendar4.24.0.postevent_enrollment.fraction\endcsname{269/681}
\expandafter\def\csname EER.followup.original_sample.calendar4.24.0.observed_resumption.pct\endcsname{39.5}
\expandafter\def\csname EER.followup.original_sample.calendar4.24.0.observed_resumption.total\endcsname{692}
\expandafter\def\csname EER.followup.original_sample.calendar4.24.0.observed_resumption.observed\endcsname{681}
\expandafter\def\csname EER.followup.original_sample.calendar4.24.0.observed_resumption.yes\endcsname{269}
\expandafter\def\csname EER.followup.original_sample.calendar4.24.0.observed_resumption.missing\endcsname{11}
\expandafter\def\csname EER.followup.original_sample.calendar4.24.0.observed_resumption.fraction\endcsname{269/681}
\expandafter\def\csname EER.followup.original_sample.calendar4.24.0.any_credential.pct\endcsname{34.3}
\expandafter\def\csname EER.followup.original_sample.calendar4.24.0.any_credential.total\endcsname{692}
\expandafter\def\csname EER.followup.original_sample.calendar4.24.0.any_credential.observed\endcsname{688}
\expandafter\def\csname EER.followup.original_sample.calendar4.24.0.any_credential.yes\endcsname{236}
\expandafter\def\csname EER.followup.original_sample.calendar4.24.0.any_credential.missing\endcsname{4}
\expandafter\def\csname EER.followup.original_sample.calendar4.24.0.any_credential.fraction\endcsname{236/688}
\expandafter\def\csname EER.followup.original_sample.calendar4.24.0.diploma.pct\endcsname{21.9}
\expandafter\def\csname EER.followup.original_sample.calendar4.24.0.diploma.total\endcsname{692}
\expandafter\def\csname EER.followup.original_sample.calendar4.24.0.diploma.observed\endcsname{688}
\expandafter\def\csname EER.followup.original_sample.calendar4.24.0.diploma.yes\endcsname{151}
\expandafter\def\csname EER.followup.original_sample.calendar4.24.0.diploma.missing\endcsname{4}
\expandafter\def\csname EER.followup.original_sample.calendar4.24.0.diploma.fraction\endcsname{151/688}
\expandafter\def\csname EER.followup.original_sample.calendar4.24.0.ged.pct\endcsname{12.4}
\expandafter\def\csname EER.followup.original_sample.calendar4.24.0.ged.total\endcsname{692}
\expandafter\def\csname EER.followup.original_sample.calendar4.24.0.ged.observed\endcsname{688}
\expandafter\def\csname EER.followup.original_sample.calendar4.24.0.ged.yes\endcsname{85}
\expandafter\def\csname EER.followup.original_sample.calendar4.24.0.ged.missing\endcsname{4}
\expandafter\def\csname EER.followup.original_sample.calendar4.24.0.ged.fraction\endcsname{85/688}
\expandafter\def\csname EER.followup.original_sample.calendar4.24.0.missing_interview.pct\endcsname{0.6}
\expandafter\def\csname EER.followup.original_sample.calendar4.24.0.missing_interview.total\endcsname{692}
\expandafter\def\csname EER.followup.original_sample.calendar4.24.0.missing_interview.observed\endcsname{692}
\expandafter\def\csname EER.followup.original_sample.calendar4.24.0.missing_interview.yes\endcsname{4}
\expandafter\def\csname EER.followup.original_sample.calendar4.24.0.missing_interview.missing\endcsname{0}
\expandafter\def\csname EER.followup.original_sample.calendar4.24.0.missing_interview.fraction\endcsname{4/692}
\expandafter\def\csname EER.followup.original_sample.calendar4.24.0.missing_array.pct\endcsname{0.6}
\expandafter\def\csname EER.followup.original_sample.calendar4.24.0.missing_array.total\endcsname{692}
\expandafter\def\csname EER.followup.original_sample.calendar4.24.0.missing_array.observed\endcsname{692}
\expandafter\def\csname EER.followup.original_sample.calendar4.24.0.missing_array.yes\endcsname{4}
\expandafter\def\csname EER.followup.original_sample.calendar4.24.0.missing_array.missing\endcsname{0}
\expandafter\def\csname EER.followup.original_sample.calendar4.24.0.missing_array.fraction\endcsname{4/692}
\expandafter\def\csname EER.followup.original_sample.calendar4.24.0.missing_status.pct\endcsname{1.0}
\expandafter\def\csname EER.followup.original_sample.calendar4.24.0.missing_status.total\endcsname{692}
\expandafter\def\csname EER.followup.original_sample.calendar4.24.0.missing_status.observed\endcsname{692}
\expandafter\def\csname EER.followup.original_sample.calendar4.24.0.missing_status.yes\endcsname{7}
\expandafter\def\csname EER.followup.original_sample.calendar4.24.0.missing_status.missing\endcsname{0}
\expandafter\def\csname EER.followup.original_sample.calendar4.24.0.missing_status.fraction\endcsname{7/692}
\expandafter\def\csname EER.followup.original_sample.calendar4.24.1.postevent_enrollment.pct\endcsname{22.1}
\expandafter\def\csname EER.followup.original_sample.calendar4.24.1.postevent_enrollment.total\endcsname{258}
\expandafter\def\csname EER.followup.original_sample.calendar4.24.1.postevent_enrollment.observed\endcsname{253}
\expandafter\def\csname EER.followup.original_sample.calendar4.24.1.postevent_enrollment.yes\endcsname{56}
\expandafter\def\csname EER.followup.original_sample.calendar4.24.1.postevent_enrollment.missing\endcsname{5}
\expandafter\def\csname EER.followup.original_sample.calendar4.24.1.postevent_enrollment.fraction\endcsname{56/253}
\expandafter\def\csname EER.followup.original_sample.calendar4.24.1.observed_resumption.pct\endcsname{22.1}
\expandafter\def\csname EER.followup.original_sample.calendar4.24.1.observed_resumption.total\endcsname{258}
\expandafter\def\csname EER.followup.original_sample.calendar4.24.1.observed_resumption.observed\endcsname{253}
\expandafter\def\csname EER.followup.original_sample.calendar4.24.1.observed_resumption.yes\endcsname{56}
\expandafter\def\csname EER.followup.original_sample.calendar4.24.1.observed_resumption.missing\endcsname{5}
\expandafter\def\csname EER.followup.original_sample.calendar4.24.1.observed_resumption.fraction\endcsname{56/253}
\expandafter\def\csname EER.followup.original_sample.calendar4.24.1.any_credential.pct\endcsname{24.9}
\expandafter\def\csname EER.followup.original_sample.calendar4.24.1.any_credential.total\endcsname{258}
\expandafter\def\csname EER.followup.original_sample.calendar4.24.1.any_credential.observed\endcsname{257}
\expandafter\def\csname EER.followup.original_sample.calendar4.24.1.any_credential.yes\endcsname{64}
\expandafter\def\csname EER.followup.original_sample.calendar4.24.1.any_credential.missing\endcsname{1}
\expandafter\def\csname EER.followup.original_sample.calendar4.24.1.any_credential.fraction\endcsname{64/257}
\expandafter\def\csname EER.followup.original_sample.calendar4.24.1.diploma.pct\endcsname{11.3}
\expandafter\def\csname EER.followup.original_sample.calendar4.24.1.diploma.total\endcsname{258}
\expandafter\def\csname EER.followup.original_sample.calendar4.24.1.diploma.observed\endcsname{257}
\expandafter\def\csname EER.followup.original_sample.calendar4.24.1.diploma.yes\endcsname{29}
\expandafter\def\csname EER.followup.original_sample.calendar4.24.1.diploma.missing\endcsname{1}
\expandafter\def\csname EER.followup.original_sample.calendar4.24.1.diploma.fraction\endcsname{29/257}
\expandafter\def\csname EER.followup.original_sample.calendar4.24.1.ged.pct\endcsname{13.6}
\expandafter\def\csname EER.followup.original_sample.calendar4.24.1.ged.total\endcsname{258}
\expandafter\def\csname EER.followup.original_sample.calendar4.24.1.ged.observed\endcsname{257}
\expandafter\def\csname EER.followup.original_sample.calendar4.24.1.ged.yes\endcsname{35}
\expandafter\def\csname EER.followup.original_sample.calendar4.24.1.ged.missing\endcsname{1}
\expandafter\def\csname EER.followup.original_sample.calendar4.24.1.ged.fraction\endcsname{35/257}
\expandafter\def\csname EER.followup.original_sample.calendar4.24.1.missing_interview.pct\endcsname{0.4}
\expandafter\def\csname EER.followup.original_sample.calendar4.24.1.missing_interview.total\endcsname{258}
\expandafter\def\csname EER.followup.original_sample.calendar4.24.1.missing_interview.observed\endcsname{258}
\expandafter\def\csname EER.followup.original_sample.calendar4.24.1.missing_interview.yes\endcsname{1}
\expandafter\def\csname EER.followup.original_sample.calendar4.24.1.missing_interview.missing\endcsname{0}
\expandafter\def\csname EER.followup.original_sample.calendar4.24.1.missing_interview.fraction\endcsname{1/258}
\expandafter\def\csname EER.followup.original_sample.calendar4.24.1.missing_array.pct\endcsname{0.4}
\expandafter\def\csname EER.followup.original_sample.calendar4.24.1.missing_array.total\endcsname{258}
\expandafter\def\csname EER.followup.original_sample.calendar4.24.1.missing_array.observed\endcsname{258}
\expandafter\def\csname EER.followup.original_sample.calendar4.24.1.missing_array.yes\endcsname{1}
\expandafter\def\csname EER.followup.original_sample.calendar4.24.1.missing_array.missing\endcsname{0}
\expandafter\def\csname EER.followup.original_sample.calendar4.24.1.missing_array.fraction\endcsname{1/258}
\expandafter\def\csname EER.followup.original_sample.calendar4.24.1.missing_status.pct\endcsname{1.6}
\expandafter\def\csname EER.followup.original_sample.calendar4.24.1.missing_status.total\endcsname{258}
\expandafter\def\csname EER.followup.original_sample.calendar4.24.1.missing_status.observed\endcsname{258}
\expandafter\def\csname EER.followup.original_sample.calendar4.24.1.missing_status.yes\endcsname{4}
\expandafter\def\csname EER.followup.original_sample.calendar4.24.1.missing_status.missing\endcsname{0}
\expandafter\def\csname EER.followup.original_sample.calendar4.24.1.missing_status.fraction\endcsname{4/258}
\expandafter\def\csname EER.followup.original_postdebut.recorded.12.0.postevent_enrollment.pct\endcsname{2.3}
\expandafter\def\csname EER.followup.original_postdebut.recorded.12.0.postevent_enrollment.total\endcsname{348}
\expandafter\def\csname EER.followup.original_postdebut.recorded.12.0.postevent_enrollment.observed\endcsname{342}
\expandafter\def\csname EER.followup.original_postdebut.recorded.12.0.postevent_enrollment.yes\endcsname{8}
\expandafter\def\csname EER.followup.original_postdebut.recorded.12.0.postevent_enrollment.missing\endcsname{6}
\expandafter\def\csname EER.followup.original_postdebut.recorded.12.0.postevent_enrollment.fraction\endcsname{8/342}
\expandafter\def\csname EER.followup.original_postdebut.recorded.12.0.observed_resumption.pct\endcsname{0.9}
\expandafter\def\csname EER.followup.original_postdebut.recorded.12.0.observed_resumption.total\endcsname{348}
\expandafter\def\csname EER.followup.original_postdebut.recorded.12.0.observed_resumption.observed\endcsname{342}
\expandafter\def\csname EER.followup.original_postdebut.recorded.12.0.observed_resumption.yes\endcsname{3}
\expandafter\def\csname EER.followup.original_postdebut.recorded.12.0.observed_resumption.missing\endcsname{6}
\expandafter\def\csname EER.followup.original_postdebut.recorded.12.0.observed_resumption.fraction\endcsname{3/342}
\expandafter\def\csname EER.followup.original_postdebut.recorded.12.0.any_credential.pct\endcsname{16.4}
\expandafter\def\csname EER.followup.original_postdebut.recorded.12.0.any_credential.total\endcsname{348}
\expandafter\def\csname EER.followup.original_postdebut.recorded.12.0.any_credential.observed\endcsname{347}
\expandafter\def\csname EER.followup.original_postdebut.recorded.12.0.any_credential.yes\endcsname{57}
\expandafter\def\csname EER.followup.original_postdebut.recorded.12.0.any_credential.missing\endcsname{1}
\expandafter\def\csname EER.followup.original_postdebut.recorded.12.0.any_credential.fraction\endcsname{57/347}
\expandafter\def\csname EER.followup.original_postdebut.recorded.12.0.diploma.pct\endcsname{0.0}
\expandafter\def\csname EER.followup.original_postdebut.recorded.12.0.diploma.total\endcsname{348}
\expandafter\def\csname EER.followup.original_postdebut.recorded.12.0.diploma.observed\endcsname{347}
\expandafter\def\csname EER.followup.original_postdebut.recorded.12.0.diploma.yes\endcsname{0}
\expandafter\def\csname EER.followup.original_postdebut.recorded.12.0.diploma.missing\endcsname{1}
\expandafter\def\csname EER.followup.original_postdebut.recorded.12.0.diploma.fraction\endcsname{0/347}
\expandafter\def\csname EER.followup.original_postdebut.recorded.12.0.ged.pct\endcsname{16.4}
\expandafter\def\csname EER.followup.original_postdebut.recorded.12.0.ged.total\endcsname{348}
\expandafter\def\csname EER.followup.original_postdebut.recorded.12.0.ged.observed\endcsname{347}
\expandafter\def\csname EER.followup.original_postdebut.recorded.12.0.ged.yes\endcsname{57}
\expandafter\def\csname EER.followup.original_postdebut.recorded.12.0.ged.missing\endcsname{1}
\expandafter\def\csname EER.followup.original_postdebut.recorded.12.0.ged.fraction\endcsname{57/347}
\expandafter\def\csname EER.followup.original_postdebut.recorded.12.0.missing_interview.pct\endcsname{0.3}
\expandafter\def\csname EER.followup.original_postdebut.recorded.12.0.missing_interview.total\endcsname{348}
\expandafter\def\csname EER.followup.original_postdebut.recorded.12.0.missing_interview.observed\endcsname{348}
\expandafter\def\csname EER.followup.original_postdebut.recorded.12.0.missing_interview.yes\endcsname{1}
\expandafter\def\csname EER.followup.original_postdebut.recorded.12.0.missing_interview.missing\endcsname{0}
\expandafter\def\csname EER.followup.original_postdebut.recorded.12.0.missing_interview.fraction\endcsname{1/348}
\expandafter\def\csname EER.followup.original_postdebut.recorded.12.0.missing_array.pct\endcsname{0.3}
\expandafter\def\csname EER.followup.original_postdebut.recorded.12.0.missing_array.total\endcsname{348}
\expandafter\def\csname EER.followup.original_postdebut.recorded.12.0.missing_array.observed\endcsname{348}
\expandafter\def\csname EER.followup.original_postdebut.recorded.12.0.missing_array.yes\endcsname{1}
\expandafter\def\csname EER.followup.original_postdebut.recorded.12.0.missing_array.missing\endcsname{0}
\expandafter\def\csname EER.followup.original_postdebut.recorded.12.0.missing_array.fraction\endcsname{1/348}
\expandafter\def\csname EER.followup.original_postdebut.recorded.12.0.missing_status.pct\endcsname{1.4}
\expandafter\def\csname EER.followup.original_postdebut.recorded.12.0.missing_status.total\endcsname{348}
\expandafter\def\csname EER.followup.original_postdebut.recorded.12.0.missing_status.observed\endcsname{348}
\expandafter\def\csname EER.followup.original_postdebut.recorded.12.0.missing_status.yes\endcsname{5}
\expandafter\def\csname EER.followup.original_postdebut.recorded.12.0.missing_status.missing\endcsname{0}
\expandafter\def\csname EER.followup.original_postdebut.recorded.12.0.missing_status.fraction\endcsname{5/348}
\expandafter\def\csname EER.followup.original_postdebut.recorded.12.1.postevent_enrollment.pct\endcsname{3.3}
\expandafter\def\csname EER.followup.original_postdebut.recorded.12.1.postevent_enrollment.total\endcsname{246}
\expandafter\def\csname EER.followup.original_postdebut.recorded.12.1.postevent_enrollment.observed\endcsname{242}
\expandafter\def\csname EER.followup.original_postdebut.recorded.12.1.postevent_enrollment.yes\endcsname{8}
\expandafter\def\csname EER.followup.original_postdebut.recorded.12.1.postevent_enrollment.missing\endcsname{4}
\expandafter\def\csname EER.followup.original_postdebut.recorded.12.1.postevent_enrollment.fraction\endcsname{8/242}
\expandafter\def\csname EER.followup.original_postdebut.recorded.12.1.observed_resumption.pct\endcsname{1.7}
\expandafter\def\csname EER.followup.original_postdebut.recorded.12.1.observed_resumption.total\endcsname{246}
\expandafter\def\csname EER.followup.original_postdebut.recorded.12.1.observed_resumption.observed\endcsname{242}
\expandafter\def\csname EER.followup.original_postdebut.recorded.12.1.observed_resumption.yes\endcsname{4}
\expandafter\def\csname EER.followup.original_postdebut.recorded.12.1.observed_resumption.missing\endcsname{4}
\expandafter\def\csname EER.followup.original_postdebut.recorded.12.1.observed_resumption.fraction\endcsname{4/242}
\expandafter\def\csname EER.followup.original_postdebut.recorded.12.1.any_credential.pct\endcsname{10.2}
\expandafter\def\csname EER.followup.original_postdebut.recorded.12.1.any_credential.total\endcsname{246}
\expandafter\def\csname EER.followup.original_postdebut.recorded.12.1.any_credential.observed\endcsname{246}
\expandafter\def\csname EER.followup.original_postdebut.recorded.12.1.any_credential.yes\endcsname{25}
\expandafter\def\csname EER.followup.original_postdebut.recorded.12.1.any_credential.missing\endcsname{0}
\expandafter\def\csname EER.followup.original_postdebut.recorded.12.1.any_credential.fraction\endcsname{25/246}
\expandafter\def\csname EER.followup.original_postdebut.recorded.12.1.diploma.pct\endcsname{0.0}
\expandafter\def\csname EER.followup.original_postdebut.recorded.12.1.diploma.total\endcsname{246}
\expandafter\def\csname EER.followup.original_postdebut.recorded.12.1.diploma.observed\endcsname{246}
\expandafter\def\csname EER.followup.original_postdebut.recorded.12.1.diploma.yes\endcsname{0}
\expandafter\def\csname EER.followup.original_postdebut.recorded.12.1.diploma.missing\endcsname{0}
\expandafter\def\csname EER.followup.original_postdebut.recorded.12.1.diploma.fraction\endcsname{0/246}
\expandafter\def\csname EER.followup.original_postdebut.recorded.12.1.ged.pct\endcsname{10.2}
\expandafter\def\csname EER.followup.original_postdebut.recorded.12.1.ged.total\endcsname{246}
\expandafter\def\csname EER.followup.original_postdebut.recorded.12.1.ged.observed\endcsname{246}
\expandafter\def\csname EER.followup.original_postdebut.recorded.12.1.ged.yes\endcsname{25}
\expandafter\def\csname EER.followup.original_postdebut.recorded.12.1.ged.missing\endcsname{0}
\expandafter\def\csname EER.followup.original_postdebut.recorded.12.1.ged.fraction\endcsname{25/246}
\expandafter\def\csname EER.followup.original_postdebut.recorded.12.1.missing_interview.pct\endcsname{0.0}
\expandafter\def\csname EER.followup.original_postdebut.recorded.12.1.missing_interview.total\endcsname{246}
\expandafter\def\csname EER.followup.original_postdebut.recorded.12.1.missing_interview.observed\endcsname{246}
\expandafter\def\csname EER.followup.original_postdebut.recorded.12.1.missing_interview.yes\endcsname{0}
\expandafter\def\csname EER.followup.original_postdebut.recorded.12.1.missing_interview.missing\endcsname{0}
\expandafter\def\csname EER.followup.original_postdebut.recorded.12.1.missing_interview.fraction\endcsname{0/246}
\expandafter\def\csname EER.followup.original_postdebut.recorded.12.1.missing_array.pct\endcsname{0.0}
\expandafter\def\csname EER.followup.original_postdebut.recorded.12.1.missing_array.total\endcsname{246}
\expandafter\def\csname EER.followup.original_postdebut.recorded.12.1.missing_array.observed\endcsname{246}
\expandafter\def\csname EER.followup.original_postdebut.recorded.12.1.missing_array.yes\endcsname{0}
\expandafter\def\csname EER.followup.original_postdebut.recorded.12.1.missing_array.missing\endcsname{0}
\expandafter\def\csname EER.followup.original_postdebut.recorded.12.1.missing_array.fraction\endcsname{0/246}
\expandafter\def\csname EER.followup.original_postdebut.recorded.12.1.missing_status.pct\endcsname{1.6}
\expandafter\def\csname EER.followup.original_postdebut.recorded.12.1.missing_status.total\endcsname{246}
\expandafter\def\csname EER.followup.original_postdebut.recorded.12.1.missing_status.observed\endcsname{246}
\expandafter\def\csname EER.followup.original_postdebut.recorded.12.1.missing_status.yes\endcsname{4}
\expandafter\def\csname EER.followup.original_postdebut.recorded.12.1.missing_status.missing\endcsname{0}
\expandafter\def\csname EER.followup.original_postdebut.recorded.12.1.missing_status.fraction\endcsname{4/246}
\expandafter\def\csname EER.followup.original_postdebut.recorded.24.0.postevent_enrollment.pct\endcsname{2.1}
\expandafter\def\csname EER.followup.original_postdebut.recorded.24.0.postevent_enrollment.total\endcsname{348}
\expandafter\def\csname EER.followup.original_postdebut.recorded.24.0.postevent_enrollment.observed\endcsname{339}
\expandafter\def\csname EER.followup.original_postdebut.recorded.24.0.postevent_enrollment.yes\endcsname{7}
\expandafter\def\csname EER.followup.original_postdebut.recorded.24.0.postevent_enrollment.missing\endcsname{9}
\expandafter\def\csname EER.followup.original_postdebut.recorded.24.0.postevent_enrollment.fraction\endcsname{7/339}
\expandafter\def\csname EER.followup.original_postdebut.recorded.24.0.observed_resumption.pct\endcsname{0.9}
\expandafter\def\csname EER.followup.original_postdebut.recorded.24.0.observed_resumption.total\endcsname{348}
\expandafter\def\csname EER.followup.original_postdebut.recorded.24.0.observed_resumption.observed\endcsname{339}
\expandafter\def\csname EER.followup.original_postdebut.recorded.24.0.observed_resumption.yes\endcsname{3}
\expandafter\def\csname EER.followup.original_postdebut.recorded.24.0.observed_resumption.missing\endcsname{9}
\expandafter\def\csname EER.followup.original_postdebut.recorded.24.0.observed_resumption.fraction\endcsname{3/339}
\expandafter\def\csname EER.followup.original_postdebut.recorded.24.0.any_credential.pct\endcsname{26.6}
\expandafter\def\csname EER.followup.original_postdebut.recorded.24.0.any_credential.total\endcsname{348}
\expandafter\def\csname EER.followup.original_postdebut.recorded.24.0.any_credential.observed\endcsname{346}
\expandafter\def\csname EER.followup.original_postdebut.recorded.24.0.any_credential.yes\endcsname{92}
\expandafter\def\csname EER.followup.original_postdebut.recorded.24.0.any_credential.missing\endcsname{2}
\expandafter\def\csname EER.followup.original_postdebut.recorded.24.0.any_credential.fraction\endcsname{92/346}
\expandafter\def\csname EER.followup.original_postdebut.recorded.24.0.diploma.pct\endcsname{0.0}
\expandafter\def\csname EER.followup.original_postdebut.recorded.24.0.diploma.total\endcsname{348}
\expandafter\def\csname EER.followup.original_postdebut.recorded.24.0.diploma.observed\endcsname{346}
\expandafter\def\csname EER.followup.original_postdebut.recorded.24.0.diploma.yes\endcsname{0}
\expandafter\def\csname EER.followup.original_postdebut.recorded.24.0.diploma.missing\endcsname{2}
\expandafter\def\csname EER.followup.original_postdebut.recorded.24.0.diploma.fraction\endcsname{0/346}
\expandafter\def\csname EER.followup.original_postdebut.recorded.24.0.ged.pct\endcsname{26.6}
\expandafter\def\csname EER.followup.original_postdebut.recorded.24.0.ged.total\endcsname{348}
\expandafter\def\csname EER.followup.original_postdebut.recorded.24.0.ged.observed\endcsname{346}
\expandafter\def\csname EER.followup.original_postdebut.recorded.24.0.ged.yes\endcsname{92}
\expandafter\def\csname EER.followup.original_postdebut.recorded.24.0.ged.missing\endcsname{2}
\expandafter\def\csname EER.followup.original_postdebut.recorded.24.0.ged.fraction\endcsname{92/346}
\expandafter\def\csname EER.followup.original_postdebut.recorded.24.0.missing_interview.pct\endcsname{0.6}
\expandafter\def\csname EER.followup.original_postdebut.recorded.24.0.missing_interview.total\endcsname{348}
\expandafter\def\csname EER.followup.original_postdebut.recorded.24.0.missing_interview.observed\endcsname{348}
\expandafter\def\csname EER.followup.original_postdebut.recorded.24.0.missing_interview.yes\endcsname{2}
\expandafter\def\csname EER.followup.original_postdebut.recorded.24.0.missing_interview.missing\endcsname{0}
\expandafter\def\csname EER.followup.original_postdebut.recorded.24.0.missing_interview.fraction\endcsname{2/348}
\expandafter\def\csname EER.followup.original_postdebut.recorded.24.0.missing_array.pct\endcsname{0.6}
\expandafter\def\csname EER.followup.original_postdebut.recorded.24.0.missing_array.total\endcsname{348}
\expandafter\def\csname EER.followup.original_postdebut.recorded.24.0.missing_array.observed\endcsname{348}
\expandafter\def\csname EER.followup.original_postdebut.recorded.24.0.missing_array.yes\endcsname{2}
\expandafter\def\csname EER.followup.original_postdebut.recorded.24.0.missing_array.missing\endcsname{0}
\expandafter\def\csname EER.followup.original_postdebut.recorded.24.0.missing_array.fraction\endcsname{2/348}
\expandafter\def\csname EER.followup.original_postdebut.recorded.24.0.missing_status.pct\endcsname{2.0}
\expandafter\def\csname EER.followup.original_postdebut.recorded.24.0.missing_status.total\endcsname{348}
\expandafter\def\csname EER.followup.original_postdebut.recorded.24.0.missing_status.observed\endcsname{348}
\expandafter\def\csname EER.followup.original_postdebut.recorded.24.0.missing_status.yes\endcsname{7}
\expandafter\def\csname EER.followup.original_postdebut.recorded.24.0.missing_status.missing\endcsname{0}
\expandafter\def\csname EER.followup.original_postdebut.recorded.24.0.missing_status.fraction\endcsname{7/348}
\expandafter\def\csname EER.followup.original_postdebut.recorded.24.1.postevent_enrollment.pct\endcsname{3.7}
\expandafter\def\csname EER.followup.original_postdebut.recorded.24.1.postevent_enrollment.total\endcsname{246}
\expandafter\def\csname EER.followup.original_postdebut.recorded.24.1.postevent_enrollment.observed\endcsname{241}
\expandafter\def\csname EER.followup.original_postdebut.recorded.24.1.postevent_enrollment.yes\endcsname{9}
\expandafter\def\csname EER.followup.original_postdebut.recorded.24.1.postevent_enrollment.missing\endcsname{5}
\expandafter\def\csname EER.followup.original_postdebut.recorded.24.1.postevent_enrollment.fraction\endcsname{9/241}
\expandafter\def\csname EER.followup.original_postdebut.recorded.24.1.observed_resumption.pct\endcsname{2.5}
\expandafter\def\csname EER.followup.original_postdebut.recorded.24.1.observed_resumption.total\endcsname{246}
\expandafter\def\csname EER.followup.original_postdebut.recorded.24.1.observed_resumption.observed\endcsname{241}
\expandafter\def\csname EER.followup.original_postdebut.recorded.24.1.observed_resumption.yes\endcsname{6}
\expandafter\def\csname EER.followup.original_postdebut.recorded.24.1.observed_resumption.missing\endcsname{5}
\expandafter\def\csname EER.followup.original_postdebut.recorded.24.1.observed_resumption.fraction\endcsname{6/241}
\expandafter\def\csname EER.followup.original_postdebut.recorded.24.1.any_credential.pct\endcsname{19.1}
\expandafter\def\csname EER.followup.original_postdebut.recorded.24.1.any_credential.total\endcsname{246}
\expandafter\def\csname EER.followup.original_postdebut.recorded.24.1.any_credential.observed\endcsname{246}
\expandafter\def\csname EER.followup.original_postdebut.recorded.24.1.any_credential.yes\endcsname{47}
\expandafter\def\csname EER.followup.original_postdebut.recorded.24.1.any_credential.missing\endcsname{0}
\expandafter\def\csname EER.followup.original_postdebut.recorded.24.1.any_credential.fraction\endcsname{47/246}
\expandafter\def\csname EER.followup.original_postdebut.recorded.24.1.diploma.pct\endcsname{0.4}
\expandafter\def\csname EER.followup.original_postdebut.recorded.24.1.diploma.total\endcsname{246}
\expandafter\def\csname EER.followup.original_postdebut.recorded.24.1.diploma.observed\endcsname{246}
\expandafter\def\csname EER.followup.original_postdebut.recorded.24.1.diploma.yes\endcsname{1}
\expandafter\def\csname EER.followup.original_postdebut.recorded.24.1.diploma.missing\endcsname{0}
\expandafter\def\csname EER.followup.original_postdebut.recorded.24.1.diploma.fraction\endcsname{1/246}
\expandafter\def\csname EER.followup.original_postdebut.recorded.24.1.ged.pct\endcsname{18.7}
\expandafter\def\csname EER.followup.original_postdebut.recorded.24.1.ged.total\endcsname{246}
\expandafter\def\csname EER.followup.original_postdebut.recorded.24.1.ged.observed\endcsname{246}
\expandafter\def\csname EER.followup.original_postdebut.recorded.24.1.ged.yes\endcsname{46}
\expandafter\def\csname EER.followup.original_postdebut.recorded.24.1.ged.missing\endcsname{0}
\expandafter\def\csname EER.followup.original_postdebut.recorded.24.1.ged.fraction\endcsname{46/246}
\expandafter\def\csname EER.followup.original_postdebut.recorded.24.1.missing_interview.pct\endcsname{0.0}
\expandafter\def\csname EER.followup.original_postdebut.recorded.24.1.missing_interview.total\endcsname{246}
\expandafter\def\csname EER.followup.original_postdebut.recorded.24.1.missing_interview.observed\endcsname{246}
\expandafter\def\csname EER.followup.original_postdebut.recorded.24.1.missing_interview.yes\endcsname{0}
\expandafter\def\csname EER.followup.original_postdebut.recorded.24.1.missing_interview.missing\endcsname{0}
\expandafter\def\csname EER.followup.original_postdebut.recorded.24.1.missing_interview.fraction\endcsname{0/246}
\expandafter\def\csname EER.followup.original_postdebut.recorded.24.1.missing_array.pct\endcsname{0.0}
\expandafter\def\csname EER.followup.original_postdebut.recorded.24.1.missing_array.total\endcsname{246}
\expandafter\def\csname EER.followup.original_postdebut.recorded.24.1.missing_array.observed\endcsname{246}
\expandafter\def\csname EER.followup.original_postdebut.recorded.24.1.missing_array.yes\endcsname{0}
\expandafter\def\csname EER.followup.original_postdebut.recorded.24.1.missing_array.missing\endcsname{0}
\expandafter\def\csname EER.followup.original_postdebut.recorded.24.1.missing_array.fraction\endcsname{0/246}
\expandafter\def\csname EER.followup.original_postdebut.recorded.24.1.missing_status.pct\endcsname{2.0}
\expandafter\def\csname EER.followup.original_postdebut.recorded.24.1.missing_status.total\endcsname{246}
\expandafter\def\csname EER.followup.original_postdebut.recorded.24.1.missing_status.observed\endcsname{246}
\expandafter\def\csname EER.followup.original_postdebut.recorded.24.1.missing_status.yes\endcsname{5}
\expandafter\def\csname EER.followup.original_postdebut.recorded.24.1.missing_status.missing\endcsname{0}
\expandafter\def\csname EER.followup.original_postdebut.recorded.24.1.missing_status.fraction\endcsname{5/246}
\expandafter\def\csname EER.followup.original_postdebut.school3.12.0.postevent_enrollment.pct\endcsname{27.0}
\expandafter\def\csname EER.followup.original_postdebut.school3.12.0.postevent_enrollment.total\endcsname{450}
\expandafter\def\csname EER.followup.original_postdebut.school3.12.0.postevent_enrollment.observed\endcsname{448}
\expandafter\def\csname EER.followup.original_postdebut.school3.12.0.postevent_enrollment.yes\endcsname{121}
\expandafter\def\csname EER.followup.original_postdebut.school3.12.0.postevent_enrollment.missing\endcsname{2}
\expandafter\def\csname EER.followup.original_postdebut.school3.12.0.postevent_enrollment.fraction\endcsname{121/448}
\expandafter\def\csname EER.followup.original_postdebut.school3.12.0.observed_resumption.pct\endcsname{27.0}
\expandafter\def\csname EER.followup.original_postdebut.school3.12.0.observed_resumption.total\endcsname{450}
\expandafter\def\csname EER.followup.original_postdebut.school3.12.0.observed_resumption.observed\endcsname{448}
\expandafter\def\csname EER.followup.original_postdebut.school3.12.0.observed_resumption.yes\endcsname{121}
\expandafter\def\csname EER.followup.original_postdebut.school3.12.0.observed_resumption.missing\endcsname{2}
\expandafter\def\csname EER.followup.original_postdebut.school3.12.0.observed_resumption.fraction\endcsname{121/448}
\expandafter\def\csname EER.followup.original_postdebut.school3.12.0.any_credential.pct\endcsname{22.0}
\expandafter\def\csname EER.followup.original_postdebut.school3.12.0.any_credential.total\endcsname{450}
\expandafter\def\csname EER.followup.original_postdebut.school3.12.0.any_credential.observed\endcsname{450}
\expandafter\def\csname EER.followup.original_postdebut.school3.12.0.any_credential.yes\endcsname{99}
\expandafter\def\csname EER.followup.original_postdebut.school3.12.0.any_credential.missing\endcsname{0}
\expandafter\def\csname EER.followup.original_postdebut.school3.12.0.any_credential.fraction\endcsname{99/450}
\expandafter\def\csname EER.followup.original_postdebut.school3.12.0.diploma.pct\endcsname{14.0}
\expandafter\def\csname EER.followup.original_postdebut.school3.12.0.diploma.total\endcsname{450}
\expandafter\def\csname EER.followup.original_postdebut.school3.12.0.diploma.observed\endcsname{450}
\expandafter\def\csname EER.followup.original_postdebut.school3.12.0.diploma.yes\endcsname{63}
\expandafter\def\csname EER.followup.original_postdebut.school3.12.0.diploma.missing\endcsname{0}
\expandafter\def\csname EER.followup.original_postdebut.school3.12.0.diploma.fraction\endcsname{63/450}
\expandafter\def\csname EER.followup.original_postdebut.school3.12.0.ged.pct\endcsname{8.0}
\expandafter\def\csname EER.followup.original_postdebut.school3.12.0.ged.total\endcsname{450}
\expandafter\def\csname EER.followup.original_postdebut.school3.12.0.ged.observed\endcsname{450}
\expandafter\def\csname EER.followup.original_postdebut.school3.12.0.ged.yes\endcsname{36}
\expandafter\def\csname EER.followup.original_postdebut.school3.12.0.ged.missing\endcsname{0}
\expandafter\def\csname EER.followup.original_postdebut.school3.12.0.ged.fraction\endcsname{36/450}
\expandafter\def\csname EER.followup.original_postdebut.school3.12.0.missing_interview.pct\endcsname{0.0}
\expandafter\def\csname EER.followup.original_postdebut.school3.12.0.missing_interview.total\endcsname{450}
\expandafter\def\csname EER.followup.original_postdebut.school3.12.0.missing_interview.observed\endcsname{450}
\expandafter\def\csname EER.followup.original_postdebut.school3.12.0.missing_interview.yes\endcsname{0}
\expandafter\def\csname EER.followup.original_postdebut.school3.12.0.missing_interview.missing\endcsname{0}
\expandafter\def\csname EER.followup.original_postdebut.school3.12.0.missing_interview.fraction\endcsname{0/450}
\expandafter\def\csname EER.followup.original_postdebut.school3.12.0.missing_array.pct\endcsname{0.0}
\expandafter\def\csname EER.followup.original_postdebut.school3.12.0.missing_array.total\endcsname{450}
\expandafter\def\csname EER.followup.original_postdebut.school3.12.0.missing_array.observed\endcsname{450}
\expandafter\def\csname EER.followup.original_postdebut.school3.12.0.missing_array.yes\endcsname{0}
\expandafter\def\csname EER.followup.original_postdebut.school3.12.0.missing_array.missing\endcsname{0}
\expandafter\def\csname EER.followup.original_postdebut.school3.12.0.missing_array.fraction\endcsname{0/450}
\expandafter\def\csname EER.followup.original_postdebut.school3.12.0.missing_status.pct\endcsname{0.4}
\expandafter\def\csname EER.followup.original_postdebut.school3.12.0.missing_status.total\endcsname{450}
\expandafter\def\csname EER.followup.original_postdebut.school3.12.0.missing_status.observed\endcsname{450}
\expandafter\def\csname EER.followup.original_postdebut.school3.12.0.missing_status.yes\endcsname{2}
\expandafter\def\csname EER.followup.original_postdebut.school3.12.0.missing_status.missing\endcsname{0}
\expandafter\def\csname EER.followup.original_postdebut.school3.12.0.missing_status.fraction\endcsname{2/450}
\expandafter\def\csname EER.followup.original_postdebut.school3.12.1.postevent_enrollment.pct\endcsname{20.5}
\expandafter\def\csname EER.followup.original_postdebut.school3.12.1.postevent_enrollment.total\endcsname{261}
\expandafter\def\csname EER.followup.original_postdebut.school3.12.1.postevent_enrollment.observed\endcsname{259}
\expandafter\def\csname EER.followup.original_postdebut.school3.12.1.postevent_enrollment.yes\endcsname{53}
\expandafter\def\csname EER.followup.original_postdebut.school3.12.1.postevent_enrollment.missing\endcsname{2}
\expandafter\def\csname EER.followup.original_postdebut.school3.12.1.postevent_enrollment.fraction\endcsname{53/259}
\expandafter\def\csname EER.followup.original_postdebut.school3.12.1.observed_resumption.pct\endcsname{20.5}
\expandafter\def\csname EER.followup.original_postdebut.school3.12.1.observed_resumption.total\endcsname{261}
\expandafter\def\csname EER.followup.original_postdebut.school3.12.1.observed_resumption.observed\endcsname{259}
\expandafter\def\csname EER.followup.original_postdebut.school3.12.1.observed_resumption.yes\endcsname{53}
\expandafter\def\csname EER.followup.original_postdebut.school3.12.1.observed_resumption.missing\endcsname{2}
\expandafter\def\csname EER.followup.original_postdebut.school3.12.1.observed_resumption.fraction\endcsname{53/259}
\expandafter\def\csname EER.followup.original_postdebut.school3.12.1.any_credential.pct\endcsname{15.7}
\expandafter\def\csname EER.followup.original_postdebut.school3.12.1.any_credential.total\endcsname{261}
\expandafter\def\csname EER.followup.original_postdebut.school3.12.1.any_credential.observed\endcsname{261}
\expandafter\def\csname EER.followup.original_postdebut.school3.12.1.any_credential.yes\endcsname{41}
\expandafter\def\csname EER.followup.original_postdebut.school3.12.1.any_credential.missing\endcsname{0}
\expandafter\def\csname EER.followup.original_postdebut.school3.12.1.any_credential.fraction\endcsname{41/261}
\expandafter\def\csname EER.followup.original_postdebut.school3.12.1.diploma.pct\endcsname{7.7}
\expandafter\def\csname EER.followup.original_postdebut.school3.12.1.diploma.total\endcsname{261}
\expandafter\def\csname EER.followup.original_postdebut.school3.12.1.diploma.observed\endcsname{261}
\expandafter\def\csname EER.followup.original_postdebut.school3.12.1.diploma.yes\endcsname{20}
\expandafter\def\csname EER.followup.original_postdebut.school3.12.1.diploma.missing\endcsname{0}
\expandafter\def\csname EER.followup.original_postdebut.school3.12.1.diploma.fraction\endcsname{20/261}
\expandafter\def\csname EER.followup.original_postdebut.school3.12.1.ged.pct\endcsname{8.0}
\expandafter\def\csname EER.followup.original_postdebut.school3.12.1.ged.total\endcsname{261}
\expandafter\def\csname EER.followup.original_postdebut.school3.12.1.ged.observed\endcsname{261}
\expandafter\def\csname EER.followup.original_postdebut.school3.12.1.ged.yes\endcsname{21}
\expandafter\def\csname EER.followup.original_postdebut.school3.12.1.ged.missing\endcsname{0}
\expandafter\def\csname EER.followup.original_postdebut.school3.12.1.ged.fraction\endcsname{21/261}
\expandafter\def\csname EER.followup.original_postdebut.school3.12.1.missing_interview.pct\endcsname{0.0}
\expandafter\def\csname EER.followup.original_postdebut.school3.12.1.missing_interview.total\endcsname{261}
\expandafter\def\csname EER.followup.original_postdebut.school3.12.1.missing_interview.observed\endcsname{261}
\expandafter\def\csname EER.followup.original_postdebut.school3.12.1.missing_interview.yes\endcsname{0}
\expandafter\def\csname EER.followup.original_postdebut.school3.12.1.missing_interview.missing\endcsname{0}
\expandafter\def\csname EER.followup.original_postdebut.school3.12.1.missing_interview.fraction\endcsname{0/261}
\expandafter\def\csname EER.followup.original_postdebut.school3.12.1.missing_array.pct\endcsname{0.0}
\expandafter\def\csname EER.followup.original_postdebut.school3.12.1.missing_array.total\endcsname{261}
\expandafter\def\csname EER.followup.original_postdebut.school3.12.1.missing_array.observed\endcsname{261}
\expandafter\def\csname EER.followup.original_postdebut.school3.12.1.missing_array.yes\endcsname{0}
\expandafter\def\csname EER.followup.original_postdebut.school3.12.1.missing_array.missing\endcsname{0}
\expandafter\def\csname EER.followup.original_postdebut.school3.12.1.missing_array.fraction\endcsname{0/261}
\expandafter\def\csname EER.followup.original_postdebut.school3.12.1.missing_status.pct\endcsname{0.8}
\expandafter\def\csname EER.followup.original_postdebut.school3.12.1.missing_status.total\endcsname{261}
\expandafter\def\csname EER.followup.original_postdebut.school3.12.1.missing_status.observed\endcsname{261}
\expandafter\def\csname EER.followup.original_postdebut.school3.12.1.missing_status.yes\endcsname{2}
\expandafter\def\csname EER.followup.original_postdebut.school3.12.1.missing_status.missing\endcsname{0}
\expandafter\def\csname EER.followup.original_postdebut.school3.12.1.missing_status.fraction\endcsname{2/261}
\expandafter\def\csname EER.followup.original_postdebut.school3.24.0.postevent_enrollment.pct\endcsname{27.6}
\expandafter\def\csname EER.followup.original_postdebut.school3.24.0.postevent_enrollment.total\endcsname{450}
\expandafter\def\csname EER.followup.original_postdebut.school3.24.0.postevent_enrollment.observed\endcsname{445}
\expandafter\def\csname EER.followup.original_postdebut.school3.24.0.postevent_enrollment.yes\endcsname{123}
\expandafter\def\csname EER.followup.original_postdebut.school3.24.0.postevent_enrollment.missing\endcsname{5}
\expandafter\def\csname EER.followup.original_postdebut.school3.24.0.postevent_enrollment.fraction\endcsname{123/445}
\expandafter\def\csname EER.followup.original_postdebut.school3.24.0.observed_resumption.pct\endcsname{27.6}
\expandafter\def\csname EER.followup.original_postdebut.school3.24.0.observed_resumption.total\endcsname{450}
\expandafter\def\csname EER.followup.original_postdebut.school3.24.0.observed_resumption.observed\endcsname{445}
\expandafter\def\csname EER.followup.original_postdebut.school3.24.0.observed_resumption.yes\endcsname{123}
\expandafter\def\csname EER.followup.original_postdebut.school3.24.0.observed_resumption.missing\endcsname{5}
\expandafter\def\csname EER.followup.original_postdebut.school3.24.0.observed_resumption.fraction\endcsname{123/445}
\expandafter\def\csname EER.followup.original_postdebut.school3.24.0.any_credential.pct\endcsname{35.0}
\expandafter\def\csname EER.followup.original_postdebut.school3.24.0.any_credential.total\endcsname{450}
\expandafter\def\csname EER.followup.original_postdebut.school3.24.0.any_credential.observed\endcsname{448}
\expandafter\def\csname EER.followup.original_postdebut.school3.24.0.any_credential.yes\endcsname{157}
\expandafter\def\csname EER.followup.original_postdebut.school3.24.0.any_credential.missing\endcsname{2}
\expandafter\def\csname EER.followup.original_postdebut.school3.24.0.any_credential.fraction\endcsname{157/448}
\expandafter\def\csname EER.followup.original_postdebut.school3.24.0.diploma.pct\endcsname{19.6}
\expandafter\def\csname EER.followup.original_postdebut.school3.24.0.diploma.total\endcsname{450}
\expandafter\def\csname EER.followup.original_postdebut.school3.24.0.diploma.observed\endcsname{448}
\expandafter\def\csname EER.followup.original_postdebut.school3.24.0.diploma.yes\endcsname{88}
\expandafter\def\csname EER.followup.original_postdebut.school3.24.0.diploma.missing\endcsname{2}
\expandafter\def\csname EER.followup.original_postdebut.school3.24.0.diploma.fraction\endcsname{88/448}
\expandafter\def\csname EER.followup.original_postdebut.school3.24.0.ged.pct\endcsname{15.4}
\expandafter\def\csname EER.followup.original_postdebut.school3.24.0.ged.total\endcsname{450}
\expandafter\def\csname EER.followup.original_postdebut.school3.24.0.ged.observed\endcsname{448}
\expandafter\def\csname EER.followup.original_postdebut.school3.24.0.ged.yes\endcsname{69}
\expandafter\def\csname EER.followup.original_postdebut.school3.24.0.ged.missing\endcsname{2}
\expandafter\def\csname EER.followup.original_postdebut.school3.24.0.ged.fraction\endcsname{69/448}
\expandafter\def\csname EER.followup.original_postdebut.school3.24.0.missing_interview.pct\endcsname{0.4}
\expandafter\def\csname EER.followup.original_postdebut.school3.24.0.missing_interview.total\endcsname{450}
\expandafter\def\csname EER.followup.original_postdebut.school3.24.0.missing_interview.observed\endcsname{450}
\expandafter\def\csname EER.followup.original_postdebut.school3.24.0.missing_interview.yes\endcsname{2}
\expandafter\def\csname EER.followup.original_postdebut.school3.24.0.missing_interview.missing\endcsname{0}
\expandafter\def\csname EER.followup.original_postdebut.school3.24.0.missing_interview.fraction\endcsname{2/450}
\expandafter\def\csname EER.followup.original_postdebut.school3.24.0.missing_array.pct\endcsname{0.4}
\expandafter\def\csname EER.followup.original_postdebut.school3.24.0.missing_array.total\endcsname{450}
\expandafter\def\csname EER.followup.original_postdebut.school3.24.0.missing_array.observed\endcsname{450}
\expandafter\def\csname EER.followup.original_postdebut.school3.24.0.missing_array.yes\endcsname{2}
\expandafter\def\csname EER.followup.original_postdebut.school3.24.0.missing_array.missing\endcsname{0}
\expandafter\def\csname EER.followup.original_postdebut.school3.24.0.missing_array.fraction\endcsname{2/450}
\expandafter\def\csname EER.followup.original_postdebut.school3.24.0.missing_status.pct\endcsname{0.7}
\expandafter\def\csname EER.followup.original_postdebut.school3.24.0.missing_status.total\endcsname{450}
\expandafter\def\csname EER.followup.original_postdebut.school3.24.0.missing_status.observed\endcsname{450}
\expandafter\def\csname EER.followup.original_postdebut.school3.24.0.missing_status.yes\endcsname{3}
\expandafter\def\csname EER.followup.original_postdebut.school3.24.0.missing_status.missing\endcsname{0}
\expandafter\def\csname EER.followup.original_postdebut.school3.24.0.missing_status.fraction\endcsname{3/450}
\expandafter\def\csname EER.followup.original_postdebut.school3.24.1.postevent_enrollment.pct\endcsname{20.4}
\expandafter\def\csname EER.followup.original_postdebut.school3.24.1.postevent_enrollment.total\endcsname{261}
\expandafter\def\csname EER.followup.original_postdebut.school3.24.1.postevent_enrollment.observed\endcsname{255}
\expandafter\def\csname EER.followup.original_postdebut.school3.24.1.postevent_enrollment.yes\endcsname{52}
\expandafter\def\csname EER.followup.original_postdebut.school3.24.1.postevent_enrollment.missing\endcsname{6}
\expandafter\def\csname EER.followup.original_postdebut.school3.24.1.postevent_enrollment.fraction\endcsname{52/255}
\expandafter\def\csname EER.followup.original_postdebut.school3.24.1.observed_resumption.pct\endcsname{20.4}
\expandafter\def\csname EER.followup.original_postdebut.school3.24.1.observed_resumption.total\endcsname{261}
\expandafter\def\csname EER.followup.original_postdebut.school3.24.1.observed_resumption.observed\endcsname{255}
\expandafter\def\csname EER.followup.original_postdebut.school3.24.1.observed_resumption.yes\endcsname{52}
\expandafter\def\csname EER.followup.original_postdebut.school3.24.1.observed_resumption.missing\endcsname{6}
\expandafter\def\csname EER.followup.original_postdebut.school3.24.1.observed_resumption.fraction\endcsname{52/255}
\expandafter\def\csname EER.followup.original_postdebut.school3.24.1.any_credential.pct\endcsname{26.2}
\expandafter\def\csname EER.followup.original_postdebut.school3.24.1.any_credential.total\endcsname{261}
\expandafter\def\csname EER.followup.original_postdebut.school3.24.1.any_credential.observed\endcsname{260}
\expandafter\def\csname EER.followup.original_postdebut.school3.24.1.any_credential.yes\endcsname{68}
\expandafter\def\csname EER.followup.original_postdebut.school3.24.1.any_credential.missing\endcsname{1}
\expandafter\def\csname EER.followup.original_postdebut.school3.24.1.any_credential.fraction\endcsname{68/260}
\expandafter\def\csname EER.followup.original_postdebut.school3.24.1.diploma.pct\endcsname{11.5}
\expandafter\def\csname EER.followup.original_postdebut.school3.24.1.diploma.total\endcsname{261}
\expandafter\def\csname EER.followup.original_postdebut.school3.24.1.diploma.observed\endcsname{260}
\expandafter\def\csname EER.followup.original_postdebut.school3.24.1.diploma.yes\endcsname{30}
\expandafter\def\csname EER.followup.original_postdebut.school3.24.1.diploma.missing\endcsname{1}
\expandafter\def\csname EER.followup.original_postdebut.school3.24.1.diploma.fraction\endcsname{30/260}
\expandafter\def\csname EER.followup.original_postdebut.school3.24.1.ged.pct\endcsname{14.6}
\expandafter\def\csname EER.followup.original_postdebut.school3.24.1.ged.total\endcsname{261}
\expandafter\def\csname EER.followup.original_postdebut.school3.24.1.ged.observed\endcsname{260}
\expandafter\def\csname EER.followup.original_postdebut.school3.24.1.ged.yes\endcsname{38}
\expandafter\def\csname EER.followup.original_postdebut.school3.24.1.ged.missing\endcsname{1}
\expandafter\def\csname EER.followup.original_postdebut.school3.24.1.ged.fraction\endcsname{38/260}
\expandafter\def\csname EER.followup.original_postdebut.school3.24.1.missing_interview.pct\endcsname{0.4}
\expandafter\def\csname EER.followup.original_postdebut.school3.24.1.missing_interview.total\endcsname{261}
\expandafter\def\csname EER.followup.original_postdebut.school3.24.1.missing_interview.observed\endcsname{261}
\expandafter\def\csname EER.followup.original_postdebut.school3.24.1.missing_interview.yes\endcsname{1}
\expandafter\def\csname EER.followup.original_postdebut.school3.24.1.missing_interview.missing\endcsname{0}
\expandafter\def\csname EER.followup.original_postdebut.school3.24.1.missing_interview.fraction\endcsname{1/261}
\expandafter\def\csname EER.followup.original_postdebut.school3.24.1.missing_array.pct\endcsname{0.4}
\expandafter\def\csname EER.followup.original_postdebut.school3.24.1.missing_array.total\endcsname{261}
\expandafter\def\csname EER.followup.original_postdebut.school3.24.1.missing_array.observed\endcsname{261}
\expandafter\def\csname EER.followup.original_postdebut.school3.24.1.missing_array.yes\endcsname{1}
\expandafter\def\csname EER.followup.original_postdebut.school3.24.1.missing_array.missing\endcsname{0}
\expandafter\def\csname EER.followup.original_postdebut.school3.24.1.missing_array.fraction\endcsname{1/261}
\expandafter\def\csname EER.followup.original_postdebut.school3.24.1.missing_status.pct\endcsname{1.9}
\expandafter\def\csname EER.followup.original_postdebut.school3.24.1.missing_status.total\endcsname{261}
\expandafter\def\csname EER.followup.original_postdebut.school3.24.1.missing_status.observed\endcsname{261}
\expandafter\def\csname EER.followup.original_postdebut.school3.24.1.missing_status.yes\endcsname{5}
\expandafter\def\csname EER.followup.original_postdebut.school3.24.1.missing_status.missing\endcsname{0}
\expandafter\def\csname EER.followup.original_postdebut.school3.24.1.missing_status.fraction\endcsname{5/261}
\expandafter\def\csname EER.followup.original_postdebut.calendar4.12.0.postevent_enrollment.pct\endcsname{31.9}
\expandafter\def\csname EER.followup.original_postdebut.calendar4.12.0.postevent_enrollment.total\endcsname{468}
\expandafter\def\csname EER.followup.original_postdebut.calendar4.12.0.postevent_enrollment.observed\endcsname{467}
\expandafter\def\csname EER.followup.original_postdebut.calendar4.12.0.postevent_enrollment.yes\endcsname{149}
\expandafter\def\csname EER.followup.original_postdebut.calendar4.12.0.postevent_enrollment.missing\endcsname{1}
\expandafter\def\csname EER.followup.original_postdebut.calendar4.12.0.postevent_enrollment.fraction\endcsname{149/467}
\expandafter\def\csname EER.followup.original_postdebut.calendar4.12.0.observed_resumption.pct\endcsname{31.9}
\expandafter\def\csname EER.followup.original_postdebut.calendar4.12.0.observed_resumption.total\endcsname{468}
\expandafter\def\csname EER.followup.original_postdebut.calendar4.12.0.observed_resumption.observed\endcsname{467}
\expandafter\def\csname EER.followup.original_postdebut.calendar4.12.0.observed_resumption.yes\endcsname{149}
\expandafter\def\csname EER.followup.original_postdebut.calendar4.12.0.observed_resumption.missing\endcsname{1}
\expandafter\def\csname EER.followup.original_postdebut.calendar4.12.0.observed_resumption.fraction\endcsname{149/467}
\expandafter\def\csname EER.followup.original_postdebut.calendar4.12.0.any_credential.pct\endcsname{22.0}
\expandafter\def\csname EER.followup.original_postdebut.calendar4.12.0.any_credential.total\endcsname{468}
\expandafter\def\csname EER.followup.original_postdebut.calendar4.12.0.any_credential.observed\endcsname{468}
\expandafter\def\csname EER.followup.original_postdebut.calendar4.12.0.any_credential.yes\endcsname{103}
\expandafter\def\csname EER.followup.original_postdebut.calendar4.12.0.any_credential.missing\endcsname{0}
\expandafter\def\csname EER.followup.original_postdebut.calendar4.12.0.any_credential.fraction\endcsname{103/468}
\expandafter\def\csname EER.followup.original_postdebut.calendar4.12.0.diploma.pct\endcsname{13.5}
\expandafter\def\csname EER.followup.original_postdebut.calendar4.12.0.diploma.total\endcsname{468}
\expandafter\def\csname EER.followup.original_postdebut.calendar4.12.0.diploma.observed\endcsname{468}
\expandafter\def\csname EER.followup.original_postdebut.calendar4.12.0.diploma.yes\endcsname{63}
\expandafter\def\csname EER.followup.original_postdebut.calendar4.12.0.diploma.missing\endcsname{0}
\expandafter\def\csname EER.followup.original_postdebut.calendar4.12.0.diploma.fraction\endcsname{63/468}
\expandafter\def\csname EER.followup.original_postdebut.calendar4.12.0.ged.pct\endcsname{8.5}
\expandafter\def\csname EER.followup.original_postdebut.calendar4.12.0.ged.total\endcsname{468}
\expandafter\def\csname EER.followup.original_postdebut.calendar4.12.0.ged.observed\endcsname{468}
\expandafter\def\csname EER.followup.original_postdebut.calendar4.12.0.ged.yes\endcsname{40}
\expandafter\def\csname EER.followup.original_postdebut.calendar4.12.0.ged.missing\endcsname{0}
\expandafter\def\csname EER.followup.original_postdebut.calendar4.12.0.ged.fraction\endcsname{40/468}
\expandafter\def\csname EER.followup.original_postdebut.calendar4.12.0.missing_interview.pct\endcsname{0.0}
\expandafter\def\csname EER.followup.original_postdebut.calendar4.12.0.missing_interview.total\endcsname{468}
\expandafter\def\csname EER.followup.original_postdebut.calendar4.12.0.missing_interview.observed\endcsname{468}
\expandafter\def\csname EER.followup.original_postdebut.calendar4.12.0.missing_interview.yes\endcsname{0}
\expandafter\def\csname EER.followup.original_postdebut.calendar4.12.0.missing_interview.missing\endcsname{0}
\expandafter\def\csname EER.followup.original_postdebut.calendar4.12.0.missing_interview.fraction\endcsname{0/468}
\expandafter\def\csname EER.followup.original_postdebut.calendar4.12.0.missing_array.pct\endcsname{0.0}
\expandafter\def\csname EER.followup.original_postdebut.calendar4.12.0.missing_array.total\endcsname{468}
\expandafter\def\csname EER.followup.original_postdebut.calendar4.12.0.missing_array.observed\endcsname{468}
\expandafter\def\csname EER.followup.original_postdebut.calendar4.12.0.missing_array.yes\endcsname{0}
\expandafter\def\csname EER.followup.original_postdebut.calendar4.12.0.missing_array.missing\endcsname{0}
\expandafter\def\csname EER.followup.original_postdebut.calendar4.12.0.missing_array.fraction\endcsname{0/468}
\expandafter\def\csname EER.followup.original_postdebut.calendar4.12.0.missing_status.pct\endcsname{0.2}
\expandafter\def\csname EER.followup.original_postdebut.calendar4.12.0.missing_status.total\endcsname{468}
\expandafter\def\csname EER.followup.original_postdebut.calendar4.12.0.missing_status.observed\endcsname{468}
\expandafter\def\csname EER.followup.original_postdebut.calendar4.12.0.missing_status.yes\endcsname{1}
\expandafter\def\csname EER.followup.original_postdebut.calendar4.12.0.missing_status.missing\endcsname{0}
\expandafter\def\csname EER.followup.original_postdebut.calendar4.12.0.missing_status.fraction\endcsname{1/468}
\expandafter\def\csname EER.followup.original_postdebut.calendar4.12.1.postevent_enrollment.pct\endcsname{22.0}
\expandafter\def\csname EER.followup.original_postdebut.calendar4.12.1.postevent_enrollment.total\endcsname{258}
\expandafter\def\csname EER.followup.original_postdebut.calendar4.12.1.postevent_enrollment.observed\endcsname{255}
\expandafter\def\csname EER.followup.original_postdebut.calendar4.12.1.postevent_enrollment.yes\endcsname{56}
\expandafter\def\csname EER.followup.original_postdebut.calendar4.12.1.postevent_enrollment.missing\endcsname{3}
\expandafter\def\csname EER.followup.original_postdebut.calendar4.12.1.postevent_enrollment.fraction\endcsname{56/255}
\expandafter\def\csname EER.followup.original_postdebut.calendar4.12.1.observed_resumption.pct\endcsname{22.0}
\expandafter\def\csname EER.followup.original_postdebut.calendar4.12.1.observed_resumption.total\endcsname{258}
\expandafter\def\csname EER.followup.original_postdebut.calendar4.12.1.observed_resumption.observed\endcsname{255}
\expandafter\def\csname EER.followup.original_postdebut.calendar4.12.1.observed_resumption.yes\endcsname{56}
\expandafter\def\csname EER.followup.original_postdebut.calendar4.12.1.observed_resumption.missing\endcsname{3}
\expandafter\def\csname EER.followup.original_postdebut.calendar4.12.1.observed_resumption.fraction\endcsname{56/255}
\expandafter\def\csname EER.followup.original_postdebut.calendar4.12.1.any_credential.pct\endcsname{14.3}
\expandafter\def\csname EER.followup.original_postdebut.calendar4.12.1.any_credential.total\endcsname{258}
\expandafter\def\csname EER.followup.original_postdebut.calendar4.12.1.any_credential.observed\endcsname{258}
\expandafter\def\csname EER.followup.original_postdebut.calendar4.12.1.any_credential.yes\endcsname{37}
\expandafter\def\csname EER.followup.original_postdebut.calendar4.12.1.any_credential.missing\endcsname{0}
\expandafter\def\csname EER.followup.original_postdebut.calendar4.12.1.any_credential.fraction\endcsname{37/258}
\expandafter\def\csname EER.followup.original_postdebut.calendar4.12.1.diploma.pct\endcsname{7.0}
\expandafter\def\csname EER.followup.original_postdebut.calendar4.12.1.diploma.total\endcsname{258}
\expandafter\def\csname EER.followup.original_postdebut.calendar4.12.1.diploma.observed\endcsname{258}
\expandafter\def\csname EER.followup.original_postdebut.calendar4.12.1.diploma.yes\endcsname{18}
\expandafter\def\csname EER.followup.original_postdebut.calendar4.12.1.diploma.missing\endcsname{0}
\expandafter\def\csname EER.followup.original_postdebut.calendar4.12.1.diploma.fraction\endcsname{18/258}
\expandafter\def\csname EER.followup.original_postdebut.calendar4.12.1.ged.pct\endcsname{7.4}
\expandafter\def\csname EER.followup.original_postdebut.calendar4.12.1.ged.total\endcsname{258}
\expandafter\def\csname EER.followup.original_postdebut.calendar4.12.1.ged.observed\endcsname{258}
\expandafter\def\csname EER.followup.original_postdebut.calendar4.12.1.ged.yes\endcsname{19}
\expandafter\def\csname EER.followup.original_postdebut.calendar4.12.1.ged.missing\endcsname{0}
\expandafter\def\csname EER.followup.original_postdebut.calendar4.12.1.ged.fraction\endcsname{19/258}
\expandafter\def\csname EER.followup.original_postdebut.calendar4.12.1.missing_interview.pct\endcsname{0.0}
\expandafter\def\csname EER.followup.original_postdebut.calendar4.12.1.missing_interview.total\endcsname{258}
\expandafter\def\csname EER.followup.original_postdebut.calendar4.12.1.missing_interview.observed\endcsname{258}
\expandafter\def\csname EER.followup.original_postdebut.calendar4.12.1.missing_interview.yes\endcsname{0}
\expandafter\def\csname EER.followup.original_postdebut.calendar4.12.1.missing_interview.missing\endcsname{0}
\expandafter\def\csname EER.followup.original_postdebut.calendar4.12.1.missing_interview.fraction\endcsname{0/258}
\expandafter\def\csname EER.followup.original_postdebut.calendar4.12.1.missing_array.pct\endcsname{0.0}
\expandafter\def\csname EER.followup.original_postdebut.calendar4.12.1.missing_array.total\endcsname{258}
\expandafter\def\csname EER.followup.original_postdebut.calendar4.12.1.missing_array.observed\endcsname{258}
\expandafter\def\csname EER.followup.original_postdebut.calendar4.12.1.missing_array.yes\endcsname{0}
\expandafter\def\csname EER.followup.original_postdebut.calendar4.12.1.missing_array.missing\endcsname{0}
\expandafter\def\csname EER.followup.original_postdebut.calendar4.12.1.missing_array.fraction\endcsname{0/258}
\expandafter\def\csname EER.followup.original_postdebut.calendar4.12.1.missing_status.pct\endcsname{1.2}
\expandafter\def\csname EER.followup.original_postdebut.calendar4.12.1.missing_status.total\endcsname{258}
\expandafter\def\csname EER.followup.original_postdebut.calendar4.12.1.missing_status.observed\endcsname{258}
\expandafter\def\csname EER.followup.original_postdebut.calendar4.12.1.missing_status.yes\endcsname{3}
\expandafter\def\csname EER.followup.original_postdebut.calendar4.12.1.missing_status.missing\endcsname{0}
\expandafter\def\csname EER.followup.original_postdebut.calendar4.12.1.missing_status.fraction\endcsname{3/258}
\expandafter\def\csname EER.followup.original_postdebut.calendar4.24.0.postevent_enrollment.pct\endcsname{31.9}
\expandafter\def\csname EER.followup.original_postdebut.calendar4.24.0.postevent_enrollment.total\endcsname{468}
\expandafter\def\csname EER.followup.original_postdebut.calendar4.24.0.postevent_enrollment.observed\endcsname{461}
\expandafter\def\csname EER.followup.original_postdebut.calendar4.24.0.postevent_enrollment.yes\endcsname{147}
\expandafter\def\csname EER.followup.original_postdebut.calendar4.24.0.postevent_enrollment.missing\endcsname{7}
\expandafter\def\csname EER.followup.original_postdebut.calendar4.24.0.postevent_enrollment.fraction\endcsname{147/461}
\expandafter\def\csname EER.followup.original_postdebut.calendar4.24.0.observed_resumption.pct\endcsname{31.9}
\expandafter\def\csname EER.followup.original_postdebut.calendar4.24.0.observed_resumption.total\endcsname{468}
\expandafter\def\csname EER.followup.original_postdebut.calendar4.24.0.observed_resumption.observed\endcsname{461}
\expandafter\def\csname EER.followup.original_postdebut.calendar4.24.0.observed_resumption.yes\endcsname{147}
\expandafter\def\csname EER.followup.original_postdebut.calendar4.24.0.observed_resumption.missing\endcsname{7}
\expandafter\def\csname EER.followup.original_postdebut.calendar4.24.0.observed_resumption.fraction\endcsname{147/461}
\expandafter\def\csname EER.followup.original_postdebut.calendar4.24.0.any_credential.pct\endcsname{35.6}
\expandafter\def\csname EER.followup.original_postdebut.calendar4.24.0.any_credential.total\endcsname{468}
\expandafter\def\csname EER.followup.original_postdebut.calendar4.24.0.any_credential.observed\endcsname{466}
\expandafter\def\csname EER.followup.original_postdebut.calendar4.24.0.any_credential.yes\endcsname{166}
\expandafter\def\csname EER.followup.original_postdebut.calendar4.24.0.any_credential.missing\endcsname{2}
\expandafter\def\csname EER.followup.original_postdebut.calendar4.24.0.any_credential.fraction\endcsname{166/466}
\expandafter\def\csname EER.followup.original_postdebut.calendar4.24.0.diploma.pct\endcsname{20.4}
\expandafter\def\csname EER.followup.original_postdebut.calendar4.24.0.diploma.total\endcsname{468}
\expandafter\def\csname EER.followup.original_postdebut.calendar4.24.0.diploma.observed\endcsname{466}
\expandafter\def\csname EER.followup.original_postdebut.calendar4.24.0.diploma.yes\endcsname{95}
\expandafter\def\csname EER.followup.original_postdebut.calendar4.24.0.diploma.missing\endcsname{2}
\expandafter\def\csname EER.followup.original_postdebut.calendar4.24.0.diploma.fraction\endcsname{95/466}
\expandafter\def\csname EER.followup.original_postdebut.calendar4.24.0.ged.pct\endcsname{15.2}
\expandafter\def\csname EER.followup.original_postdebut.calendar4.24.0.ged.total\endcsname{468}
\expandafter\def\csname EER.followup.original_postdebut.calendar4.24.0.ged.observed\endcsname{466}
\expandafter\def\csname EER.followup.original_postdebut.calendar4.24.0.ged.yes\endcsname{71}
\expandafter\def\csname EER.followup.original_postdebut.calendar4.24.0.ged.missing\endcsname{2}
\expandafter\def\csname EER.followup.original_postdebut.calendar4.24.0.ged.fraction\endcsname{71/466}
\expandafter\def\csname EER.followup.original_postdebut.calendar4.24.0.missing_interview.pct\endcsname{0.4}
\expandafter\def\csname EER.followup.original_postdebut.calendar4.24.0.missing_interview.total\endcsname{468}
\expandafter\def\csname EER.followup.original_postdebut.calendar4.24.0.missing_interview.observed\endcsname{468}
\expandafter\def\csname EER.followup.original_postdebut.calendar4.24.0.missing_interview.yes\endcsname{2}
\expandafter\def\csname EER.followup.original_postdebut.calendar4.24.0.missing_interview.missing\endcsname{0}
\expandafter\def\csname EER.followup.original_postdebut.calendar4.24.0.missing_interview.fraction\endcsname{2/468}
\expandafter\def\csname EER.followup.original_postdebut.calendar4.24.0.missing_array.pct\endcsname{0.4}
\expandafter\def\csname EER.followup.original_postdebut.calendar4.24.0.missing_array.total\endcsname{468}
\expandafter\def\csname EER.followup.original_postdebut.calendar4.24.0.missing_array.observed\endcsname{468}
\expandafter\def\csname EER.followup.original_postdebut.calendar4.24.0.missing_array.yes\endcsname{2}
\expandafter\def\csname EER.followup.original_postdebut.calendar4.24.0.missing_array.missing\endcsname{0}
\expandafter\def\csname EER.followup.original_postdebut.calendar4.24.0.missing_array.fraction\endcsname{2/468}
\expandafter\def\csname EER.followup.original_postdebut.calendar4.24.0.missing_status.pct\endcsname{1.1}
\expandafter\def\csname EER.followup.original_postdebut.calendar4.24.0.missing_status.total\endcsname{468}
\expandafter\def\csname EER.followup.original_postdebut.calendar4.24.0.missing_status.observed\endcsname{468}
\expandafter\def\csname EER.followup.original_postdebut.calendar4.24.0.missing_status.yes\endcsname{5}
\expandafter\def\csname EER.followup.original_postdebut.calendar4.24.0.missing_status.missing\endcsname{0}
\expandafter\def\csname EER.followup.original_postdebut.calendar4.24.0.missing_status.fraction\endcsname{5/468}
\expandafter\def\csname EER.followup.original_postdebut.calendar4.24.1.postevent_enrollment.pct\endcsname{22.1}
\expandafter\def\csname EER.followup.original_postdebut.calendar4.24.1.postevent_enrollment.total\endcsname{258}
\expandafter\def\csname EER.followup.original_postdebut.calendar4.24.1.postevent_enrollment.observed\endcsname{253}
\expandafter\def\csname EER.followup.original_postdebut.calendar4.24.1.postevent_enrollment.yes\endcsname{56}
\expandafter\def\csname EER.followup.original_postdebut.calendar4.24.1.postevent_enrollment.missing\endcsname{5}
\expandafter\def\csname EER.followup.original_postdebut.calendar4.24.1.postevent_enrollment.fraction\endcsname{56/253}
\expandafter\def\csname EER.followup.original_postdebut.calendar4.24.1.observed_resumption.pct\endcsname{22.1}
\expandafter\def\csname EER.followup.original_postdebut.calendar4.24.1.observed_resumption.total\endcsname{258}
\expandafter\def\csname EER.followup.original_postdebut.calendar4.24.1.observed_resumption.observed\endcsname{253}
\expandafter\def\csname EER.followup.original_postdebut.calendar4.24.1.observed_resumption.yes\endcsname{56}
\expandafter\def\csname EER.followup.original_postdebut.calendar4.24.1.observed_resumption.missing\endcsname{5}
\expandafter\def\csname EER.followup.original_postdebut.calendar4.24.1.observed_resumption.fraction\endcsname{56/253}
\expandafter\def\csname EER.followup.original_postdebut.calendar4.24.1.any_credential.pct\endcsname{24.9}
\expandafter\def\csname EER.followup.original_postdebut.calendar4.24.1.any_credential.total\endcsname{258}
\expandafter\def\csname EER.followup.original_postdebut.calendar4.24.1.any_credential.observed\endcsname{257}
\expandafter\def\csname EER.followup.original_postdebut.calendar4.24.1.any_credential.yes\endcsname{64}
\expandafter\def\csname EER.followup.original_postdebut.calendar4.24.1.any_credential.missing\endcsname{1}
\expandafter\def\csname EER.followup.original_postdebut.calendar4.24.1.any_credential.fraction\endcsname{64/257}
\expandafter\def\csname EER.followup.original_postdebut.calendar4.24.1.diploma.pct\endcsname{11.3}
\expandafter\def\csname EER.followup.original_postdebut.calendar4.24.1.diploma.total\endcsname{258}
\expandafter\def\csname EER.followup.original_postdebut.calendar4.24.1.diploma.observed\endcsname{257}
\expandafter\def\csname EER.followup.original_postdebut.calendar4.24.1.diploma.yes\endcsname{29}
\expandafter\def\csname EER.followup.original_postdebut.calendar4.24.1.diploma.missing\endcsname{1}
\expandafter\def\csname EER.followup.original_postdebut.calendar4.24.1.diploma.fraction\endcsname{29/257}
\expandafter\def\csname EER.followup.original_postdebut.calendar4.24.1.ged.pct\endcsname{13.6}
\expandafter\def\csname EER.followup.original_postdebut.calendar4.24.1.ged.total\endcsname{258}
\expandafter\def\csname EER.followup.original_postdebut.calendar4.24.1.ged.observed\endcsname{257}
\expandafter\def\csname EER.followup.original_postdebut.calendar4.24.1.ged.yes\endcsname{35}
\expandafter\def\csname EER.followup.original_postdebut.calendar4.24.1.ged.missing\endcsname{1}
\expandafter\def\csname EER.followup.original_postdebut.calendar4.24.1.ged.fraction\endcsname{35/257}
\expandafter\def\csname EER.followup.original_postdebut.calendar4.24.1.missing_interview.pct\endcsname{0.4}
\expandafter\def\csname EER.followup.original_postdebut.calendar4.24.1.missing_interview.total\endcsname{258}
\expandafter\def\csname EER.followup.original_postdebut.calendar4.24.1.missing_interview.observed\endcsname{258}
\expandafter\def\csname EER.followup.original_postdebut.calendar4.24.1.missing_interview.yes\endcsname{1}
\expandafter\def\csname EER.followup.original_postdebut.calendar4.24.1.missing_interview.missing\endcsname{0}
\expandafter\def\csname EER.followup.original_postdebut.calendar4.24.1.missing_interview.fraction\endcsname{1/258}
\expandafter\def\csname EER.followup.original_postdebut.calendar4.24.1.missing_array.pct\endcsname{0.4}
\expandafter\def\csname EER.followup.original_postdebut.calendar4.24.1.missing_array.total\endcsname{258}
\expandafter\def\csname EER.followup.original_postdebut.calendar4.24.1.missing_array.observed\endcsname{258}
\expandafter\def\csname EER.followup.original_postdebut.calendar4.24.1.missing_array.yes\endcsname{1}
\expandafter\def\csname EER.followup.original_postdebut.calendar4.24.1.missing_array.missing\endcsname{0}
\expandafter\def\csname EER.followup.original_postdebut.calendar4.24.1.missing_array.fraction\endcsname{1/258}
\expandafter\def\csname EER.followup.original_postdebut.calendar4.24.1.missing_status.pct\endcsname{1.6}
\expandafter\def\csname EER.followup.original_postdebut.calendar4.24.1.missing_status.total\endcsname{258}
\expandafter\def\csname EER.followup.original_postdebut.calendar4.24.1.missing_status.observed\endcsname{258}
\expandafter\def\csname EER.followup.original_postdebut.calendar4.24.1.missing_status.yes\endcsname{4}
\expandafter\def\csname EER.followup.original_postdebut.calendar4.24.1.missing_status.missing\endcsname{0}
\expandafter\def\csname EER.followup.original_postdebut.calendar4.24.1.missing_status.fraction\endcsname{4/258}
\expandafter\def\csname EER.followup.all_origins.calendar4.12.all.any_credential.missing_all\endcsname{0}
\expandafter\def\csname EER.followup.all_origins.calendar4.12.all.diploma.missing_all\endcsname{0}
\expandafter\def\csname EER.followup.all_origins.calendar4.12.all.ged.missing_all\endcsname{0}
\expandafter\def\csname EER.followup.all_origins.calendar4.12.all.missing_array.missing_all\endcsname{0}
\expandafter\def\csname EER.followup.all_origins.calendar4.12.all.missing_interview.missing_all\endcsname{0}
\expandafter\def\csname EER.followup.all_origins.calendar4.12.all.missing_status.missing_all\endcsname{0}
\expandafter\def\csname EER.followup.all_origins.calendar4.12.all.observed_resumption.missing_all\endcsname{5}
\expandafter\def\csname EER.followup.all_origins.calendar4.12.all.postevent_enrollment.missing_all\endcsname{5}
\expandafter\def\csname EER.followup.all_origins.calendar4.24.all.any_credential.missing_all\endcsname{5}
\expandafter\def\csname EER.followup.all_origins.calendar4.24.all.diploma.missing_all\endcsname{5}
\expandafter\def\csname EER.followup.all_origins.calendar4.24.all.ged.missing_all\endcsname{5}
\expandafter\def\csname EER.followup.all_origins.calendar4.24.all.missing_array.missing_all\endcsname{0}
\expandafter\def\csname EER.followup.all_origins.calendar4.24.all.missing_interview.missing_all\endcsname{0}
\expandafter\def\csname EER.followup.all_origins.calendar4.24.all.missing_status.missing_all\endcsname{0}
\expandafter\def\csname EER.followup.all_origins.calendar4.24.all.observed_resumption.missing_all\endcsname{16}
\expandafter\def\csname EER.followup.all_origins.calendar4.24.all.postevent_enrollment.missing_all\endcsname{16}
\expandafter\def\csname EER.followup.all_origins.recorded.12.all.any_credential.missing_all\endcsname{6}
\expandafter\def\csname EER.followup.all_origins.recorded.12.all.diploma.missing_all\endcsname{6}
\expandafter\def\csname EER.followup.all_origins.recorded.12.all.ged.missing_all\endcsname{6}
\expandafter\def\csname EER.followup.all_origins.recorded.12.all.missing_array.missing_all\endcsname{0}
\expandafter\def\csname EER.followup.all_origins.recorded.12.all.missing_interview.missing_all\endcsname{0}
\expandafter\def\csname EER.followup.all_origins.recorded.12.all.missing_status.missing_all\endcsname{0}
\expandafter\def\csname EER.followup.all_origins.recorded.12.all.observed_resumption.missing_all\endcsname{19}
\expandafter\def\csname EER.followup.all_origins.recorded.12.all.postevent_enrollment.missing_all\endcsname{19}
\expandafter\def\csname EER.followup.all_origins.recorded.24.all.any_credential.missing_all\endcsname{7}
\expandafter\def\csname EER.followup.all_origins.recorded.24.all.diploma.missing_all\endcsname{7}
\expandafter\def\csname EER.followup.all_origins.recorded.24.all.ged.missing_all\endcsname{7}
\expandafter\def\csname EER.followup.all_origins.recorded.24.all.missing_array.missing_all\endcsname{0}
\expandafter\def\csname EER.followup.all_origins.recorded.24.all.missing_interview.missing_all\endcsname{0}
\expandafter\def\csname EER.followup.all_origins.recorded.24.all.missing_status.missing_all\endcsname{0}
\expandafter\def\csname EER.followup.all_origins.recorded.24.all.observed_resumption.missing_all\endcsname{23}
\expandafter\def\csname EER.followup.all_origins.recorded.24.all.postevent_enrollment.missing_all\endcsname{23}
\expandafter\def\csname EER.followup.all_origins.school3.12.all.any_credential.missing_all\endcsname{0}
\expandafter\def\csname EER.followup.all_origins.school3.12.all.diploma.missing_all\endcsname{0}
\expandafter\def\csname EER.followup.all_origins.school3.12.all.ged.missing_all\endcsname{0}
\expandafter\def\csname EER.followup.all_origins.school3.12.all.missing_array.missing_all\endcsname{0}
\expandafter\def\csname EER.followup.all_origins.school3.12.all.missing_interview.missing_all\endcsname{0}
\expandafter\def\csname EER.followup.all_origins.school3.12.all.missing_status.missing_all\endcsname{0}
\expandafter\def\csname EER.followup.all_origins.school3.12.all.observed_resumption.missing_all\endcsname{5}
\expandafter\def\csname EER.followup.all_origins.school3.12.all.postevent_enrollment.missing_all\endcsname{5}
\expandafter\def\csname EER.followup.all_origins.school3.24.all.any_credential.missing_all\endcsname{5}
\expandafter\def\csname EER.followup.all_origins.school3.24.all.diploma.missing_all\endcsname{5}
\expandafter\def\csname EER.followup.all_origins.school3.24.all.ged.missing_all\endcsname{5}
\expandafter\def\csname EER.followup.all_origins.school3.24.all.missing_array.missing_all\endcsname{0}
\expandafter\def\csname EER.followup.all_origins.school3.24.all.missing_interview.missing_all\endcsname{0}
\expandafter\def\csname EER.followup.all_origins.school3.24.all.missing_status.missing_all\endcsname{0}
\expandafter\def\csname EER.followup.all_origins.school3.24.all.observed_resumption.missing_all\endcsname{15}
\expandafter\def\csname EER.followup.all_origins.school3.24.all.postevent_enrollment.missing_all\endcsname{15}
\expandafter\def\csname EER.followup.original_postdebut.calendar4.12.all.any_credential.missing_all\endcsname{0}
\expandafter\def\csname EER.followup.original_postdebut.calendar4.12.all.diploma.missing_all\endcsname{0}
\expandafter\def\csname EER.followup.original_postdebut.calendar4.12.all.ged.missing_all\endcsname{0}
\expandafter\def\csname EER.followup.original_postdebut.calendar4.12.all.missing_array.missing_all\endcsname{0}
\expandafter\def\csname EER.followup.original_postdebut.calendar4.12.all.missing_interview.missing_all\endcsname{0}
\expandafter\def\csname EER.followup.original_postdebut.calendar4.12.all.missing_status.missing_all\endcsname{0}
\expandafter\def\csname EER.followup.original_postdebut.calendar4.12.all.observed_resumption.missing_all\endcsname{4}
\expandafter\def\csname EER.followup.original_postdebut.calendar4.12.all.postevent_enrollment.missing_all\endcsname{4}
\expandafter\def\csname EER.followup.original_postdebut.calendar4.24.all.any_credential.missing_all\endcsname{3}
\expandafter\def\csname EER.followup.original_postdebut.calendar4.24.all.diploma.missing_all\endcsname{3}
\expandafter\def\csname EER.followup.original_postdebut.calendar4.24.all.ged.missing_all\endcsname{3}
\expandafter\def\csname EER.followup.original_postdebut.calendar4.24.all.missing_array.missing_all\endcsname{0}
\expandafter\def\csname EER.followup.original_postdebut.calendar4.24.all.missing_interview.missing_all\endcsname{0}
\expandafter\def\csname EER.followup.original_postdebut.calendar4.24.all.missing_status.missing_all\endcsname{0}
\expandafter\def\csname EER.followup.original_postdebut.calendar4.24.all.observed_resumption.missing_all\endcsname{12}
\expandafter\def\csname EER.followup.original_postdebut.calendar4.24.all.postevent_enrollment.missing_all\endcsname{12}
\expandafter\def\csname EER.followup.original_postdebut.recorded.12.all.any_credential.missing_all\endcsname{1}
\expandafter\def\csname EER.followup.original_postdebut.recorded.12.all.diploma.missing_all\endcsname{1}
\expandafter\def\csname EER.followup.original_postdebut.recorded.12.all.ged.missing_all\endcsname{1}
\expandafter\def\csname EER.followup.original_postdebut.recorded.12.all.missing_array.missing_all\endcsname{0}
\expandafter\def\csname EER.followup.original_postdebut.recorded.12.all.missing_interview.missing_all\endcsname{0}
\expandafter\def\csname EER.followup.original_postdebut.recorded.12.all.missing_status.missing_all\endcsname{0}
\expandafter\def\csname EER.followup.original_postdebut.recorded.12.all.observed_resumption.missing_all\endcsname{10}
\expandafter\def\csname EER.followup.original_postdebut.recorded.12.all.postevent_enrollment.missing_all\endcsname{10}
\expandafter\def\csname EER.followup.original_postdebut.recorded.24.all.any_credential.missing_all\endcsname{2}
\expandafter\def\csname EER.followup.original_postdebut.recorded.24.all.diploma.missing_all\endcsname{2}
\expandafter\def\csname EER.followup.original_postdebut.recorded.24.all.ged.missing_all\endcsname{2}
\expandafter\def\csname EER.followup.original_postdebut.recorded.24.all.missing_array.missing_all\endcsname{0}
\expandafter\def\csname EER.followup.original_postdebut.recorded.24.all.missing_interview.missing_all\endcsname{0}
\expandafter\def\csname EER.followup.original_postdebut.recorded.24.all.missing_status.missing_all\endcsname{0}
\expandafter\def\csname EER.followup.original_postdebut.recorded.24.all.observed_resumption.missing_all\endcsname{14}
\expandafter\def\csname EER.followup.original_postdebut.recorded.24.all.postevent_enrollment.missing_all\endcsname{14}
\expandafter\def\csname EER.followup.original_postdebut.school3.12.all.any_credential.missing_all\endcsname{0}
\expandafter\def\csname EER.followup.original_postdebut.school3.12.all.diploma.missing_all\endcsname{0}
\expandafter\def\csname EER.followup.original_postdebut.school3.12.all.ged.missing_all\endcsname{0}
\expandafter\def\csname EER.followup.original_postdebut.school3.12.all.missing_array.missing_all\endcsname{0}
\expandafter\def\csname EER.followup.original_postdebut.school3.12.all.missing_interview.missing_all\endcsname{0}
\expandafter\def\csname EER.followup.original_postdebut.school3.12.all.missing_status.missing_all\endcsname{0}
\expandafter\def\csname EER.followup.original_postdebut.school3.12.all.observed_resumption.missing_all\endcsname{4}
\expandafter\def\csname EER.followup.original_postdebut.school3.12.all.postevent_enrollment.missing_all\endcsname{4}
\expandafter\def\csname EER.followup.original_postdebut.school3.24.all.any_credential.missing_all\endcsname{3}
\expandafter\def\csname EER.followup.original_postdebut.school3.24.all.diploma.missing_all\endcsname{3}
\expandafter\def\csname EER.followup.original_postdebut.school3.24.all.ged.missing_all\endcsname{3}
\expandafter\def\csname EER.followup.original_postdebut.school3.24.all.missing_array.missing_all\endcsname{0}
\expandafter\def\csname EER.followup.original_postdebut.school3.24.all.missing_interview.missing_all\endcsname{0}
\expandafter\def\csname EER.followup.original_postdebut.school3.24.all.missing_status.missing_all\endcsname{0}
\expandafter\def\csname EER.followup.original_postdebut.school3.24.all.observed_resumption.missing_all\endcsname{11}
\expandafter\def\csname EER.followup.original_postdebut.school3.24.all.postevent_enrollment.missing_all\endcsname{11}
\expandafter\def\csname EER.followup.original_sample.calendar4.12.all.any_credential.missing_all\endcsname{0}
\expandafter\def\csname EER.followup.original_sample.calendar4.12.all.diploma.missing_all\endcsname{0}
\expandafter\def\csname EER.followup.original_sample.calendar4.12.all.ged.missing_all\endcsname{0}
\expandafter\def\csname EER.followup.original_sample.calendar4.12.all.missing_array.missing_all\endcsname{0}
\expandafter\def\csname EER.followup.original_sample.calendar4.12.all.missing_interview.missing_all\endcsname{0}
\expandafter\def\csname EER.followup.original_sample.calendar4.12.all.missing_status.missing_all\endcsname{0}
\expandafter\def\csname EER.followup.original_sample.calendar4.12.all.observed_resumption.missing_all\endcsname{5}
\expandafter\def\csname EER.followup.original_sample.calendar4.12.all.postevent_enrollment.missing_all\endcsname{5}
\expandafter\def\csname EER.followup.original_sample.calendar4.24.all.any_credential.missing_all\endcsname{5}
\expandafter\def\csname EER.followup.original_sample.calendar4.24.all.diploma.missing_all\endcsname{5}
\expandafter\def\csname EER.followup.original_sample.calendar4.24.all.ged.missing_all\endcsname{5}
\expandafter\def\csname EER.followup.original_sample.calendar4.24.all.missing_array.missing_all\endcsname{0}
\expandafter\def\csname EER.followup.original_sample.calendar4.24.all.missing_interview.missing_all\endcsname{0}
\expandafter\def\csname EER.followup.original_sample.calendar4.24.all.missing_status.missing_all\endcsname{0}
\expandafter\def\csname EER.followup.original_sample.calendar4.24.all.observed_resumption.missing_all\endcsname{16}
\expandafter\def\csname EER.followup.original_sample.calendar4.24.all.postevent_enrollment.missing_all\endcsname{16}
\expandafter\def\csname EER.followup.original_sample.recorded.12.all.any_credential.missing_all\endcsname{6}
\expandafter\def\csname EER.followup.original_sample.recorded.12.all.diploma.missing_all\endcsname{6}
\expandafter\def\csname EER.followup.original_sample.recorded.12.all.ged.missing_all\endcsname{6}
\expandafter\def\csname EER.followup.original_sample.recorded.12.all.missing_array.missing_all\endcsname{0}
\expandafter\def\csname EER.followup.original_sample.recorded.12.all.missing_interview.missing_all\endcsname{0}
\expandafter\def\csname EER.followup.original_sample.recorded.12.all.missing_status.missing_all\endcsname{0}
\expandafter\def\csname EER.followup.original_sample.recorded.12.all.observed_resumption.missing_all\endcsname{19}
\expandafter\def\csname EER.followup.original_sample.recorded.12.all.postevent_enrollment.missing_all\endcsname{19}
\expandafter\def\csname EER.followup.original_sample.recorded.24.all.any_credential.missing_all\endcsname{7}
\expandafter\def\csname EER.followup.original_sample.recorded.24.all.diploma.missing_all\endcsname{7}
\expandafter\def\csname EER.followup.original_sample.recorded.24.all.ged.missing_all\endcsname{7}
\expandafter\def\csname EER.followup.original_sample.recorded.24.all.missing_array.missing_all\endcsname{0}
\expandafter\def\csname EER.followup.original_sample.recorded.24.all.missing_interview.missing_all\endcsname{0}
\expandafter\def\csname EER.followup.original_sample.recorded.24.all.missing_status.missing_all\endcsname{0}
\expandafter\def\csname EER.followup.original_sample.recorded.24.all.observed_resumption.missing_all\endcsname{23}
\expandafter\def\csname EER.followup.original_sample.recorded.24.all.postevent_enrollment.missing_all\endcsname{23}
\expandafter\def\csname EER.followup.original_sample.school3.12.all.any_credential.missing_all\endcsname{0}
\expandafter\def\csname EER.followup.original_sample.school3.12.all.diploma.missing_all\endcsname{0}
\expandafter\def\csname EER.followup.original_sample.school3.12.all.ged.missing_all\endcsname{0}
\expandafter\def\csname EER.followup.original_sample.school3.12.all.missing_array.missing_all\endcsname{0}
\expandafter\def\csname EER.followup.original_sample.school3.12.all.missing_interview.missing_all\endcsname{0}
\expandafter\def\csname EER.followup.original_sample.school3.12.all.missing_status.missing_all\endcsname{0}
\expandafter\def\csname EER.followup.original_sample.school3.12.all.observed_resumption.missing_all\endcsname{5}
\expandafter\def\csname EER.followup.original_sample.school3.12.all.postevent_enrollment.missing_all\endcsname{5}
\expandafter\def\csname EER.followup.original_sample.school3.24.all.any_credential.missing_all\endcsname{5}
\expandafter\def\csname EER.followup.original_sample.school3.24.all.diploma.missing_all\endcsname{5}
\expandafter\def\csname EER.followup.original_sample.school3.24.all.ged.missing_all\endcsname{5}
\expandafter\def\csname EER.followup.original_sample.school3.24.all.missing_array.missing_all\endcsname{0}
\expandafter\def\csname EER.followup.original_sample.school3.24.all.missing_interview.missing_all\endcsname{0}
\expandafter\def\csname EER.followup.original_sample.school3.24.all.missing_status.missing_all\endcsname{0}
\expandafter\def\csname EER.followup.original_sample.school3.24.all.observed_resumption.missing_all\endcsname{15}
\expandafter\def\csname EER.followup.original_sample.school3.24.all.postevent_enrollment.missing_all\endcsname{15}
\expandafter\def\csname EER.timing.original_retained_months.all.value\endcsname{147,956}
\expandafter\def\csname EER.timing.original_retained_months.all.denom\endcsname{}
\expandafter\def\csname EER.timing.original_retained_months.all.pct\endcsname{}
\expandafter\def\csname EER.timing.original_retained_months.post_debut.value\endcsname{61,649}
\expandafter\def\csname EER.timing.original_retained_months.post_debut.denom\endcsname{}
\expandafter\def\csname EER.timing.original_retained_months.post_debut.pct\endcsname{}
\expandafter\def\csname EER.timing.original_events.all.value\endcsname{680}
\expandafter\def\csname EER.timing.original_events.all.denom\endcsname{}
\expandafter\def\csname EER.timing.original_events.all.pct\endcsname{}
\expandafter\def\csname EER.timing.original_events.post_debut.value\endcsname{594}
\expandafter\def\csname EER.timing.original_events.post_debut.denom\endcsname{}
\expandafter\def\csname EER.timing.original_events.post_debut.pct\endcsname{}
\expandafter\def\csname EER.timing.persons.all.value\endcsname{4,235}
\expandafter\def\csname EER.timing.persons.all.denom\endcsname{}
\expandafter\def\csname EER.timing.persons.all.pct\endcsname{}
\expandafter\def\csname EER.timing.legacy_gap4.all.value\endcsname{160}
\expandafter\def\csname EER.timing.legacy_gap4.all.denom\endcsname{4,235}
\expandafter\def\csname EER.timing.legacy_gap4.all.pct\endcsname{3.8}
\expandafter\def\csname EER.timing.legacy_unbounded.all.value\endcsname{37}
\expandafter\def\csname EER.timing.legacy_unbounded.all.denom\endcsname{4,235}
\expandafter\def\csname EER.timing.legacy_unbounded.all.pct\endcsname{0.9}
\expandafter\def\csname EER.timing.legacy_inside.all.value\endcsname{20}
\expandafter\def\csname EER.timing.legacy_inside.all.denom\endcsname{4,235}
\expandafter\def\csname EER.timing.legacy_inside.all.pct\endcsname{0.5}
\expandafter\def\csname EER.timing.legacy_after.all.value\endcsname{20}
\expandafter\def\csname EER.timing.legacy_after.all.denom\endcsname{4,235}
\expandafter\def\csname EER.timing.legacy_after.all.pct\endcsname{0.5}
\expandafter\def\csname EER.timing.legacy_after_only.all.value\endcsname{17}
\expandafter\def\csname EER.timing.legacy_after_only.all.denom\endcsname{4,235}
\expandafter\def\csname EER.timing.legacy_after_only.all.pct\endcsname{0.4}
\expandafter\def\csname EER.timing.S_i_adjusted.all.value\endcsname{148}
\expandafter\def\csname EER.timing.S_i_adjusted.all.denom\endcsname{4,235}
\expandafter\def\csname EER.timing.S_i_adjusted.all.pct\endcsname{3.5}
\expandafter\def\csname EER.timing.S_i_imputed.all.value\endcsname{90}
\expandafter\def\csname EER.timing.S_i_imputed.all.denom\endcsname{4,235}
\expandafter\def\csname EER.timing.S_i_imputed.all.pct\endcsname{2.1}
\expandafter\def\csname EER.timing.debut_unmodified.all.value\endcsname{3,663}
\expandafter\def\csname EER.timing.debut_unmodified.all.denom\endcsname{4,235}
\expandafter\def\csname EER.timing.debut_unmodified.all.pct\endcsname{86.5}
\expandafter\def\csname EER.timing.continuous_prior3.all.value\endcsname{528}
\expandafter\def\csname EER.timing.continuous_prior3.all.denom\endcsname{4,235}
\expandafter\def\csname EER.timing.continuous_prior3.all.pct\endcsname{12.5}
\expandafter\def\csname EER.timing.preg_in_prior_school3.all.value\endcsname{18}
\expandafter\def\csname EER.timing.preg_in_prior_school3.all.denom\endcsname{4,235}
\expandafter\def\csname EER.timing.preg_in_prior_school3.all.pct\endcsname{0.4}
\expandafter\def\csname EER.timing.preg_after_prior_school3.all.value\endcsname{13}
\expandafter\def\csname EER.timing.preg_after_prior_school3.all.denom\endcsname{4,235}
\expandafter\def\csname EER.timing.preg_after_prior_school3.all.pct\endcsname{0.3}
\expandafter\def\csname EER.timing.preg_in_prior_calendar4.all.value\endcsname{18}
\expandafter\def\csname EER.timing.preg_in_prior_calendar4.all.denom\endcsname{4,235}
\expandafter\def\csname EER.timing.preg_in_prior_calendar4.all.pct\endcsname{0.4}
\expandafter\def\csname EER.timing.preg_after_prior_calendar4.all.value\endcsname{16}
\expandafter\def\csname EER.timing.preg_after_prior_calendar4.all.denom\endcsname{4,235}
\expandafter\def\csname EER.timing.preg_after_prior_calendar4.all.pct\endcsname{0.4}
\expandafter\def\csname EER.timing.has_school3.all.value\endcsname{910}
\expandafter\def\csname EER.timing.has_school3.all.denom\endcsname{4,235}
\expandafter\def\csname EER.timing.has_school3.all.pct\endcsname{21.5}
\expandafter\def\csname EER.timing.school3_before_recorded.all.value\endcsname{112}
\expandafter\def\csname EER.timing.school3_before_recorded.all.denom\endcsname{4,235}
\expandafter\def\csname EER.timing.school3_before_recorded.all.pct\endcsname{2.6}
\expandafter\def\csname EER.timing.has_calendar4.all.value\endcsname{950}
\expandafter\def\csname EER.timing.has_calendar4.all.denom\endcsname{4,235}
\expandafter\def\csname EER.timing.has_calendar4.all.pct\endcsname{22.4}
\expandafter\def\csname EER.timing.calendar4_before_recorded.all.value\endcsname{127}
\expandafter\def\csname EER.timing.calendar4_before_recorded.all.denom\endcsname{4,235}
\expandafter\def\csname EER.timing.calendar4_before_recorded.all.pct\endcsname{3.0}
\expandafter\def\csname EER.timing.persons.post_debut.value\endcsname{2,951}
\expandafter\def\csname EER.timing.persons.post_debut.denom\endcsname{}
\expandafter\def\csname EER.timing.persons.post_debut.pct\endcsname{}
\expandafter\def\csname EER.timing.legacy_gap4.post_debut.value\endcsname{151}
\expandafter\def\csname EER.timing.legacy_gap4.post_debut.denom\endcsname{2,951}
\expandafter\def\csname EER.timing.legacy_gap4.post_debut.pct\endcsname{5.1}
\expandafter\def\csname EER.timing.legacy_unbounded.post_debut.value\endcsname{37}
\expandafter\def\csname EER.timing.legacy_unbounded.post_debut.denom\endcsname{2,951}
\expandafter\def\csname EER.timing.legacy_unbounded.post_debut.pct\endcsname{1.3}
\expandafter\def\csname EER.timing.legacy_inside.post_debut.value\endcsname{20}
\expandafter\def\csname EER.timing.legacy_inside.post_debut.denom\endcsname{2,951}
\expandafter\def\csname EER.timing.legacy_inside.post_debut.pct\endcsname{0.7}
\expandafter\def\csname EER.timing.legacy_after.post_debut.value\endcsname{20}
\expandafter\def\csname EER.timing.legacy_after.post_debut.denom\endcsname{2,951}
\expandafter\def\csname EER.timing.legacy_after.post_debut.pct\endcsname{0.7}
\expandafter\def\csname EER.timing.legacy_after_only.post_debut.value\endcsname{17}
\expandafter\def\csname EER.timing.legacy_after_only.post_debut.denom\endcsname{2,951}
\expandafter\def\csname EER.timing.legacy_after_only.post_debut.pct\endcsname{0.6}
\expandafter\def\csname EER.timing.S_i_adjusted.post_debut.value\endcsname{118}
\expandafter\def\csname EER.timing.S_i_adjusted.post_debut.denom\endcsname{2,951}
\expandafter\def\csname EER.timing.S_i_adjusted.post_debut.pct\endcsname{4.0}
\expandafter\def\csname EER.timing.S_i_imputed.post_debut.value\endcsname{11}
\expandafter\def\csname EER.timing.S_i_imputed.post_debut.denom\endcsname{2,951}
\expandafter\def\csname EER.timing.S_i_imputed.post_debut.pct\endcsname{0.4}
\expandafter\def\csname EER.timing.debut_unmodified.post_debut.value\endcsname{2,822}
\expandafter\def\csname EER.timing.debut_unmodified.post_debut.denom\endcsname{2,951}
\expandafter\def\csname EER.timing.debut_unmodified.post_debut.pct\endcsname{95.6}
\expandafter\def\csname EER.timing.continuous_prior3.post_debut.value\endcsname{519}
\expandafter\def\csname EER.timing.continuous_prior3.post_debut.denom\endcsname{2,951}
\expandafter\def\csname EER.timing.continuous_prior3.post_debut.pct\endcsname{17.6}
\expandafter\def\csname EER.timing.preg_in_prior_school3.post_debut.value\endcsname{18}
\expandafter\def\csname EER.timing.preg_in_prior_school3.post_debut.denom\endcsname{2,951}
\expandafter\def\csname EER.timing.preg_in_prior_school3.post_debut.pct\endcsname{0.6}
\expandafter\def\csname EER.timing.preg_after_prior_school3.post_debut.value\endcsname{13}
\expandafter\def\csname EER.timing.preg_after_prior_school3.post_debut.denom\endcsname{2,951}
\expandafter\def\csname EER.timing.preg_after_prior_school3.post_debut.pct\endcsname{0.4}
\expandafter\def\csname EER.timing.preg_in_prior_calendar4.post_debut.value\endcsname{18}
\expandafter\def\csname EER.timing.preg_in_prior_calendar4.post_debut.denom\endcsname{2,951}
\expandafter\def\csname EER.timing.preg_in_prior_calendar4.post_debut.pct\endcsname{0.6}
\expandafter\def\csname EER.timing.preg_after_prior_calendar4.post_debut.value\endcsname{16}
\expandafter\def\csname EER.timing.preg_after_prior_calendar4.post_debut.denom\endcsname{2,951}
\expandafter\def\csname EER.timing.preg_after_prior_calendar4.post_debut.pct\endcsname{0.5}
\expandafter\def\csname EER.timing.has_school3.post_debut.value\endcsname{764}
\expandafter\def\csname EER.timing.has_school3.post_debut.denom\endcsname{2,951}
\expandafter\def\csname EER.timing.has_school3.post_debut.pct\endcsname{25.9}
\expandafter\def\csname EER.timing.school3_before_recorded.post_debut.value\endcsname{103}
\expandafter\def\csname EER.timing.school3_before_recorded.post_debut.denom\endcsname{2,951}
\expandafter\def\csname EER.timing.school3_before_recorded.post_debut.pct\endcsname{3.5}
\expandafter\def\csname EER.timing.has_calendar4.post_debut.value\endcsname{798}
\expandafter\def\csname EER.timing.has_calendar4.post_debut.denom\endcsname{2,951}
\expandafter\def\csname EER.timing.has_calendar4.post_debut.pct\endcsname{27.0}
\expandafter\def\csname EER.timing.calendar4_before_recorded.post_debut.value\endcsname{118}
\expandafter\def\csname EER.timing.calendar4_before_recorded.post_debut.denom\endcsname{2,951}
\expandafter\def\csname EER.timing.calendar4_before_recorded.post_debut.pct\endcsname{4.0}
\expandafter\def\csname EER.timing.persons.original_events.value\endcsname{680}
\expandafter\def\csname EER.timing.persons.original_events.denom\endcsname{}
\expandafter\def\csname EER.timing.persons.original_events.pct\endcsname{}
\expandafter\def\csname EER.timing.legacy_gap4.original_events.value\endcsname{153}
\expandafter\def\csname EER.timing.legacy_gap4.original_events.denom\endcsname{680}
\expandafter\def\csname EER.timing.legacy_gap4.original_events.pct\endcsname{22.5}
\expandafter\def\csname EER.timing.legacy_unbounded.original_events.value\endcsname{33}
\expandafter\def\csname EER.timing.legacy_unbounded.original_events.denom\endcsname{680}
\expandafter\def\csname EER.timing.legacy_unbounded.original_events.pct\endcsname{4.9}
\expandafter\def\csname EER.timing.legacy_inside.original_events.value\endcsname{16}
\expandafter\def\csname EER.timing.legacy_inside.original_events.denom\endcsname{680}
\expandafter\def\csname EER.timing.legacy_inside.original_events.pct\endcsname{2.4}
\expandafter\def\csname EER.timing.legacy_after.original_events.value\endcsname{20}
\expandafter\def\csname EER.timing.legacy_after.original_events.denom\endcsname{680}
\expandafter\def\csname EER.timing.legacy_after.original_events.pct\endcsname{2.9}
\expandafter\def\csname EER.timing.legacy_after_only.original_events.value\endcsname{17}
\expandafter\def\csname EER.timing.legacy_after_only.original_events.denom\endcsname{680}
\expandafter\def\csname EER.timing.legacy_after_only.original_events.pct\endcsname{2.5}
\expandafter\def\csname EER.timing.S_i_adjusted.original_events.value\endcsname{67}
\expandafter\def\csname EER.timing.S_i_adjusted.original_events.denom\endcsname{680}
\expandafter\def\csname EER.timing.S_i_adjusted.original_events.pct\endcsname{9.9}
\expandafter\def\csname EER.timing.S_i_imputed.original_events.value\endcsname{14}
\expandafter\def\csname EER.timing.S_i_imputed.original_events.denom\endcsname{680}
\expandafter\def\csname EER.timing.S_i_imputed.original_events.pct\endcsname{2.1}
\expandafter\def\csname EER.timing.debut_unmodified.original_events.value\endcsname{587}
\expandafter\def\csname EER.timing.debut_unmodified.original_events.denom\endcsname{680}
\expandafter\def\csname EER.timing.debut_unmodified.original_events.pct\endcsname{86.3}
\expandafter\def\csname EER.timing.continuous_prior3.original_events.value\endcsname{189}
\expandafter\def\csname EER.timing.continuous_prior3.original_events.denom\endcsname{680}
\expandafter\def\csname EER.timing.continuous_prior3.original_events.pct\endcsname{27.8}
\expandafter\def\csname EER.timing.preg_in_prior_school3.original_events.value\endcsname{14}
\expandafter\def\csname EER.timing.preg_in_prior_school3.original_events.denom\endcsname{680}
\expandafter\def\csname EER.timing.preg_in_prior_school3.original_events.pct\endcsname{2.1}
\expandafter\def\csname EER.timing.preg_after_prior_school3.original_events.value\endcsname{13}
\expandafter\def\csname EER.timing.preg_after_prior_school3.original_events.denom\endcsname{680}
\expandafter\def\csname EER.timing.preg_after_prior_school3.original_events.pct\endcsname{1.9}
\expandafter\def\csname EER.timing.preg_in_prior_calendar4.original_events.value\endcsname{14}
\expandafter\def\csname EER.timing.preg_in_prior_calendar4.original_events.denom\endcsname{680}
\expandafter\def\csname EER.timing.preg_in_prior_calendar4.original_events.pct\endcsname{2.1}
\expandafter\def\csname EER.timing.preg_after_prior_calendar4.original_events.value\endcsname{16}
\expandafter\def\csname EER.timing.preg_after_prior_calendar4.original_events.denom\endcsname{680}
\expandafter\def\csname EER.timing.preg_after_prior_calendar4.original_events.pct\endcsname{2.4}
\expandafter\def\csname EER.timing.has_school3.original_events.value\endcsname{573}
\expandafter\def\csname EER.timing.has_school3.original_events.denom\endcsname{680}
\expandafter\def\csname EER.timing.has_school3.original_events.pct\endcsname{84.3}
\expandafter\def\csname EER.timing.school3_before_recorded.original_events.value\endcsname{105}
\expandafter\def\csname EER.timing.school3_before_recorded.original_events.denom\endcsname{680}
\expandafter\def\csname EER.timing.school3_before_recorded.original_events.pct\endcsname{15.4}
\expandafter\def\csname EER.timing.has_calendar4.original_events.value\endcsname{580}
\expandafter\def\csname EER.timing.has_calendar4.original_events.denom\endcsname{680}
\expandafter\def\csname EER.timing.has_calendar4.original_events.pct\endcsname{85.3}
\expandafter\def\csname EER.timing.calendar4_before_recorded.original_events.value\endcsname{120}
\expandafter\def\csname EER.timing.calendar4_before_recorded.original_events.denom\endcsname{680}
\expandafter\def\csname EER.timing.calendar4_before_recorded.original_events.pct\endcsname{17.6}
\expandafter\def\csname EER.timing.persons.exposed_events.value\endcsname{246}
\expandafter\def\csname EER.timing.persons.exposed_events.denom\endcsname{}
\expandafter\def\csname EER.timing.persons.exposed_events.pct\endcsname{}
\expandafter\def\csname EER.timing.legacy_gap4.exposed_events.value\endcsname{65}
\expandafter\def\csname EER.timing.legacy_gap4.exposed_events.denom\endcsname{246}
\expandafter\def\csname EER.timing.legacy_gap4.exposed_events.pct\endcsname{26.4}
\expandafter\def\csname EER.timing.legacy_unbounded.exposed_events.value\endcsname{33}
\expandafter\def\csname EER.timing.legacy_unbounded.exposed_events.denom\endcsname{246}
\expandafter\def\csname EER.timing.legacy_unbounded.exposed_events.pct\endcsname{13.4}
\expandafter\def\csname EER.timing.legacy_inside.exposed_events.value\endcsname{16}
\expandafter\def\csname EER.timing.legacy_inside.exposed_events.denom\endcsname{246}
\expandafter\def\csname EER.timing.legacy_inside.exposed_events.pct\endcsname{6.5}
\expandafter\def\csname EER.timing.legacy_after.exposed_events.value\endcsname{20}
\expandafter\def\csname EER.timing.legacy_after.exposed_events.denom\endcsname{246}
\expandafter\def\csname EER.timing.legacy_after.exposed_events.pct\endcsname{8.1}
\expandafter\def\csname EER.timing.legacy_after_only.exposed_events.value\endcsname{17}
\expandafter\def\csname EER.timing.legacy_after_only.exposed_events.denom\endcsname{246}
\expandafter\def\csname EER.timing.legacy_after_only.exposed_events.pct\endcsname{6.9}
\expandafter\def\csname EER.timing.S_i_adjusted.exposed_events.value\endcsname{56}
\expandafter\def\csname EER.timing.S_i_adjusted.exposed_events.denom\endcsname{246}
\expandafter\def\csname EER.timing.S_i_adjusted.exposed_events.pct\endcsname{22.8}
\expandafter\def\csname EER.timing.S_i_imputed.exposed_events.value\endcsname{5}
\expandafter\def\csname EER.timing.S_i_imputed.exposed_events.denom\endcsname{246}
\expandafter\def\csname EER.timing.S_i_imputed.exposed_events.pct\endcsname{2.0}
\expandafter\def\csname EER.timing.debut_unmodified.exposed_events.value\endcsname{185}
\expandafter\def\csname EER.timing.debut_unmodified.exposed_events.denom\endcsname{246}
\expandafter\def\csname EER.timing.debut_unmodified.exposed_events.pct\endcsname{75.2}
\expandafter\def\csname EER.timing.continuous_prior3.exposed_events.value\endcsname{156}
\expandafter\def\csname EER.timing.continuous_prior3.exposed_events.denom\endcsname{246}
\expandafter\def\csname EER.timing.continuous_prior3.exposed_events.pct\endcsname{63.4}
\expandafter\def\csname EER.timing.preg_in_prior_school3.exposed_events.value\endcsname{14}
\expandafter\def\csname EER.timing.preg_in_prior_school3.exposed_events.denom\endcsname{246}
\expandafter\def\csname EER.timing.preg_in_prior_school3.exposed_events.pct\endcsname{5.7}
\expandafter\def\csname EER.timing.preg_after_prior_school3.exposed_events.value\endcsname{13}
\expandafter\def\csname EER.timing.preg_after_prior_school3.exposed_events.denom\endcsname{246}
\expandafter\def\csname EER.timing.preg_after_prior_school3.exposed_events.pct\endcsname{5.3}
\expandafter\def\csname EER.timing.preg_in_prior_calendar4.exposed_events.value\endcsname{14}
\expandafter\def\csname EER.timing.preg_in_prior_calendar4.exposed_events.denom\endcsname{246}
\expandafter\def\csname EER.timing.preg_in_prior_calendar4.exposed_events.pct\endcsname{5.7}
\expandafter\def\csname EER.timing.preg_after_prior_calendar4.exposed_events.value\endcsname{16}
\expandafter\def\csname EER.timing.preg_after_prior_calendar4.exposed_events.denom\endcsname{246}
\expandafter\def\csname EER.timing.preg_after_prior_calendar4.exposed_events.pct\endcsname{6.5}
\expandafter\def\csname EER.timing.has_school3.exposed_events.value\endcsname{207}
\expandafter\def\csname EER.timing.has_school3.exposed_events.denom\endcsname{246}
\expandafter\def\csname EER.timing.has_school3.exposed_events.pct\endcsname{84.1}
\expandafter\def\csname EER.timing.school3_before_recorded.exposed_events.value\endcsname{47}
\expandafter\def\csname EER.timing.school3_before_recorded.exposed_events.denom\endcsname{246}
\expandafter\def\csname EER.timing.school3_before_recorded.exposed_events.pct\endcsname{19.1}
\expandafter\def\csname EER.timing.has_calendar4.exposed_events.value\endcsname{208}
\expandafter\def\csname EER.timing.has_calendar4.exposed_events.denom\endcsname{246}
\expandafter\def\csname EER.timing.has_calendar4.exposed_events.pct\endcsname{84.6}
\expandafter\def\csname EER.timing.calendar4_before_recorded.exposed_events.value\endcsname{53}
\expandafter\def\csname EER.timing.calendar4_before_recorded.exposed_events.denom\endcsname{246}
\expandafter\def\csname EER.timing.calendar4_before_recorded.exposed_events.pct\endcsname{21.5}
\expandafter\def\csname EER.timing.alternative_persons.school3.value\endcsname{4,207}
\expandafter\def\csname EER.timing.alternative_persons.school3.denom\endcsname{}
\expandafter\def\csname EER.timing.alternative_persons.school3.pct\endcsname{}
\expandafter\def\csname EER.timing.alternative_calendar_months.school3.value\endcsname{160,715}
\expandafter\def\csname EER.timing.alternative_calendar_months.school3.denom\endcsname{}
\expandafter\def\csname EER.timing.alternative_calendar_months.school3.pct\endcsname{}
\expandafter\def\csname EER.timing.alternative_events.school3.value\endcsname{911}
\expandafter\def\csname EER.timing.alternative_events.school3.denom\endcsname{}
\expandafter\def\csname EER.timing.alternative_events.school3.pct\endcsname{}
\expandafter\def\csname EER.timing.alternative_exposed_events.school3.value\endcsname{262}
\expandafter\def\csname EER.timing.alternative_exposed_events.school3.denom\endcsname{}
\expandafter\def\csname EER.timing.alternative_exposed_events.school3.pct\endcsname{}
\expandafter\def\csname EER.timing.alternative_postdebut_months.school3.value\endcsname{64,635}
\expandafter\def\csname EER.timing.alternative_postdebut_months.school3.denom\endcsname{}
\expandafter\def\csname EER.timing.alternative_postdebut_months.school3.pct\endcsname{}
\expandafter\def\csname EER.timing.alternative_persons.calendar4.value\endcsname{4,207}
\expandafter\def\csname EER.timing.alternative_persons.calendar4.denom\endcsname{}
\expandafter\def\csname EER.timing.alternative_persons.calendar4.pct\endcsname{}
\expandafter\def\csname EER.timing.alternative_calendar_months.calendar4.value\endcsname{158,995}
\expandafter\def\csname EER.timing.alternative_calendar_months.calendar4.denom\endcsname{}
\expandafter\def\csname EER.timing.alternative_calendar_months.calendar4.pct\endcsname{}
\expandafter\def\csname EER.timing.alternative_events.calendar4.value\endcsname{951}
\expandafter\def\csname EER.timing.alternative_events.calendar4.denom\endcsname{}
\expandafter\def\csname EER.timing.alternative_events.calendar4.pct\endcsname{}
\expandafter\def\csname EER.timing.alternative_exposed_events.calendar4.value\endcsname{259}
\expandafter\def\csname EER.timing.alternative_exposed_events.calendar4.denom\endcsname{}
\expandafter\def\csname EER.timing.alternative_exposed_events.calendar4.pct\endcsname{}
\expandafter\def\csname EER.timing.alternative_postdebut_months.calendar4.value\endcsname{63,542}
\expandafter\def\csname EER.timing.alternative_postdebut_months.calendar4.denom\endcsname{}
\expandafter\def\csname EER.timing.alternative_postdebut_months.calendar4.pct\endcsname{}
\newcommand{\EERMainHR}{\EERnum{model.full.linear.k2.hr}}
\newcommand{\EERRestrictedHR}{\EERnum{model.restricted.linear.k2.hr}}
\newcommand{\EERInterruptionHR}{\EERnum{alternative.school3.postdebut.hr}}
\newcommand{\EERPrebirthHR}{\EERnum{birth.recorded.1.prenatal.hr}}
\newcommand{\EERPostbirthHR}{\EERnum{birth.recorded.1.birth_or_after.hr}}
\newcommand{\EERPrebirthInterruptionHR}{\EERnum{birth.school3.1.prenatal.hr}}
\newcommand{\EERPostbirthInterruptionHR}{\EERnum{birth.school3.1.birth_or_after.hr}}

\newcommand{\EERInterruptionPriorHR}{\EERnum{alternative.school3.preonset_enrollment.hr}}

\newcommand{\EERReturnExposed}{\EERnum{followup.original_postdebut.school3.24.1.observed_resumption.pct}}
\newcommand{\EERCredentialExposed}{\EERnum{followup.original_postdebut.school3.24.1.any_credential.pct}}
\newcommand{\EERReturnUnexposed}{\EERnum{followup.original_postdebut.school3.24.0.observed_resumption.pct}}
\newcommand{\EERCredentialUnexposed}{\EERnum{followup.original_postdebut.school3.24.0.any_credential.pct}}

\newcommand{\PregHistoryBaselineHR}{3.85}
\newcommand{\PregHistoryHR}{3.31}
\newcommand{\PregHistoryCILower}{2.85}
\newcommand{\PregHistoryCIUpper}{3.85}
\newcommand{\PregHistoryAltMin}{3.30}
\newcommand{\PregHistoryAltMax}{3.41}

\newcommand{\EERdiag}[1]{\csname EERdiag@#1\endcsname}
\expandafter\def\csname EERdiag@debut_flexible.raw_lambda_p\endcsname{-0.900}
\expandafter\def\csname EERdiag@debut_flexible.raw_lambda_d\endcsname{2.474}
\expandafter\def\csname EERdiag@debut_flexible.raw_eta2\endcsname{-0.186}
\expandafter\def\csname EERdiag@debut_flexible.lambda_p\endcsname{0.900}
\expandafter\def\csname EERdiag@debut_flexible.lambda_d\endcsname{-2.474}
\expandafter\def\csname EERdiag@debut_flexible.loading_product\endcsname{-2.227}
\expandafter\def\csname EERdiag@debut_flexible.low_share_pct\endcsname{45.4}
\expandafter\def\csname EERdiag@debut_flexible.high_share_pct\endcsname{54.6}
\expandafter\def\csname EERdiag@debut_flexible.conception_ratio\endcsname{2.46}
\expandafter\def\csname EERdiag@debut_flexible.exit_ratio\endcsname{0.08}
\expandafter\def\csname EERdiag@debut_linear.raw_lambda_p\endcsname{0.980}
\expandafter\def\csname EERdiag@debut_linear.raw_lambda_d\endcsname{-2.468}
\expandafter\def\csname EERdiag@debut_linear.raw_eta2\endcsname{0.257}
\expandafter\def\csname EERdiag@debut_linear.lambda_p\endcsname{0.980}
\expandafter\def\csname EERdiag@debut_linear.lambda_d\endcsname{-2.468}
\expandafter\def\csname EERdiag@debut_linear.loading_product\endcsname{-2.418}
\expandafter\def\csname EERdiag@debut_linear.low_share_pct\endcsname{43.6}
\expandafter\def\csname EERdiag@debut_linear.high_share_pct\endcsname{56.4}
\expandafter\def\csname EERdiag@debut_linear.conception_ratio\endcsname{2.66}
\expandafter\def\csname EERdiag@debut_linear.exit_ratio\endcsname{0.08}
\expandafter\def\csname EERdiag@full_flexible.raw_lambda_p\endcsname{4.175}
\expandafter\def\csname EERdiag@full_flexible.raw_lambda_d\endcsname{0.129}
\expandafter\def\csname EERdiag@full_flexible.raw_eta2\endcsname{-3.337}
\expandafter\def\csname EERdiag@full_flexible.lambda_p\endcsname{4.175}
\expandafter\def\csname EERdiag@full_flexible.lambda_d\endcsname{0.129}
\expandafter\def\csname EERdiag@full_flexible.loading_product\endcsname{0.540}
\expandafter\def\csname EERdiag@full_flexible.low_share_pct\endcsname{96.6}
\expandafter\def\csname EERdiag@full_flexible.high_share_pct\endcsname{3.4}
\expandafter\def\csname EERdiag@full_flexible.conception_ratio\endcsname{65.03}
\expandafter\def\csname EERdiag@full_flexible.exit_ratio\endcsname{1.14}
\expandafter\def\csname EERdiag@full_linear.raw_lambda_p\endcsname{-4.132}
\expandafter\def\csname EERdiag@full_linear.raw_lambda_d\endcsname{-0.158}
\expandafter\def\csname EERdiag@full_linear.raw_eta2\endcsname{3.476}
\expandafter\def\csname EERdiag@full_linear.lambda_p\endcsname{4.132}
\expandafter\def\csname EERdiag@full_linear.lambda_d\endcsname{0.158}
\expandafter\def\csname EERdiag@full_linear.loading_product\endcsname{0.651}
\expandafter\def\csname EERdiag@full_linear.low_share_pct\endcsname{97.0}
\expandafter\def\csname EERdiag@full_linear.high_share_pct\endcsname{3.0}
\expandafter\def\csname EERdiag@full_linear.conception_ratio\endcsname{62.32}
\expandafter\def\csname EERdiag@full_linear.exit_ratio\endcsname{1.17}
\expandafter\def\csname EERdiag@restricted_flexible.raw_lambda_p\endcsname{4.161}
\expandafter\def\csname EERdiag@restricted_flexible.raw_lambda_d\endcsname{0.172}
\expandafter\def\csname EERdiag@restricted_flexible.raw_eta2\endcsname{-3.321}
\expandafter\def\csname EERdiag@restricted_flexible.lambda_p\endcsname{4.161}
\expandafter\def\csname EERdiag@restricted_flexible.lambda_d\endcsname{0.172}
\expandafter\def\csname EERdiag@restricted_flexible.loading_product\endcsname{0.715}
\expandafter\def\csname EERdiag@restricted_flexible.low_share_pct\endcsname{96.5}
\expandafter\def\csname EERdiag@restricted_flexible.high_share_pct\endcsname{3.5}
\expandafter\def\csname EERdiag@restricted_flexible.conception_ratio\endcsname{64.11}
\expandafter\def\csname EERdiag@restricted_flexible.exit_ratio\endcsname{1.19}
\expandafter\def\csname EERdiag@restricted_linear.raw_lambda_p\endcsname{4.123}
\expandafter\def\csname EERdiag@restricted_linear.raw_lambda_d\endcsname{0.230}
\expandafter\def\csname EERdiag@restricted_linear.raw_eta2\endcsname{-3.462}
\expandafter\def\csname EERdiag@restricted_linear.lambda_p\endcsname{4.123}
\expandafter\def\csname EERdiag@restricted_linear.lambda_d\endcsname{0.230}
\expandafter\def\csname EERdiag@restricted_linear.loading_product\endcsname{0.950}
\expandafter\def\csname EERdiag@restricted_linear.low_share_pct\endcsname{97.0}
\expandafter\def\csname EERdiag@restricted_linear.high_share_pct\endcsname{3.0}
\expandafter\def\csname EERdiag@restricted_linear.conception_ratio\endcsname{61.72}
\expandafter\def\csname EERdiag@restricted_linear.exit_ratio\endcsname{1.26}
\expandafter\def\csname EERdiag@debut_flexible.ll1\endcsname{-4765.62}
\expandafter\def\csname EERdiag@debut_flexible.ll2\endcsname{-4762.62}
\expandafter\def\csname EERdiag@debut_flexible.ll_gain\endcsname{3.00}
\expandafter\def\csname EERdiag@debut_flexible.aic1\endcsname{9577.23}
\expandafter\def\csname EERdiag@debut_flexible.aic2\endcsname{9577.24}
\expandafter\def\csname EERdiag@debut_flexible.bic_persons1\endcsname{9712.72}
\expandafter\def\csname EERdiag@debut_flexible.bic_persons2\endcsname{9730.40}
\expandafter\def\csname EERdiag@debut_flexible.aic_difference\endcsname{0.01}
\expandafter\def\csname EERdiag@debut_flexible.bic_difference\endcsname{17.68}
\expandafter\def\csname EERdiag@debut_flexible.kparams1\endcsname{23}
\expandafter\def\csname EERdiag@debut_flexible.kparams2\endcsname{26}
\expandafter\def\csname EERdiag@debut_flexible.women\endcsname{2,673}
\expandafter\def\csname EERdiag@debut_flexible.person_months\endcsname{56,471}
\expandafter\def\csname EERdiag@debut_flexible.dropout_events\endcsname{467}
\expandafter\def\csname EERdiag@debut_flexible.pregnancy_events\endcsname{368}
\expandafter\def\csname EERdiag@debut_flexible.exposed_months\endcsname{4,327}
\expandafter\def\csname EERdiag@debut_flexible.aic_preferred_k\endcsname{1}
\expandafter\def\csname EERdiag@debut_flexible.bic_preferred_k\endcsname{1}
\expandafter\def\csname EERdiag@debut_linear.ll1\endcsname{-4780.31}
\expandafter\def\csname EERdiag@debut_linear.ll2\endcsname{-4777.22}
\expandafter\def\csname EERdiag@debut_linear.ll_gain\endcsname{3.09}
\expandafter\def\csname EERdiag@debut_linear.aic1\endcsname{9574.62}
\expandafter\def\csname EERdiag@debut_linear.aic2\endcsname{9574.44}
\expandafter\def\csname EERdiag@debut_linear.bic_persons1\endcsname{9615.86}
\expandafter\def\csname EERdiag@debut_linear.bic_persons2\endcsname{9633.35}
\expandafter\def\csname EERdiag@debut_linear.aic_difference\endcsname{-0.18}
\expandafter\def\csname EERdiag@debut_linear.bic_difference\endcsname{17.49}
\expandafter\def\csname EERdiag@debut_linear.kparams1\endcsname{7}
\expandafter\def\csname EERdiag@debut_linear.kparams2\endcsname{10}
\expandafter\def\csname EERdiag@debut_linear.women\endcsname{2,673}
\expandafter\def\csname EERdiag@debut_linear.person_months\endcsname{56,471}
\expandafter\def\csname EERdiag@debut_linear.dropout_events\endcsname{467}
\expandafter\def\csname EERdiag@debut_linear.pregnancy_events\endcsname{368}
\expandafter\def\csname EERdiag@debut_linear.exposed_months\endcsname{4,327}
\expandafter\def\csname EERdiag@debut_linear.aic_preferred_k\endcsname{2}
\expandafter\def\csname EERdiag@debut_linear.bic_preferred_k\endcsname{1}
\expandafter\def\csname EERdiag@full_flexible.ll1\endcsname{-5714.83}
\expandafter\def\csname EERdiag@full_flexible.ll2\endcsname{-5698.65}
\expandafter\def\csname EERdiag@full_flexible.ll_gain\endcsname{16.18}
\expandafter\def\csname EERdiag@full_flexible.aic1\endcsname{11477.65}
\expandafter\def\csname EERdiag@full_flexible.aic2\endcsname{11451.30}
\expandafter\def\csname EERdiag@full_flexible.bic_persons1\endcsname{11621.41}
\expandafter\def\csname EERdiag@full_flexible.bic_persons2\endcsname{11613.02}
\expandafter\def\csname EERdiag@full_flexible.aic_difference\endcsname{-26.36}
\expandafter\def\csname EERdiag@full_flexible.bic_difference\endcsname{-8.39}
\expandafter\def\csname EERdiag@full_flexible.kparams1\endcsname{24}
\expandafter\def\csname EERdiag@full_flexible.kparams2\endcsname{27}
\expandafter\def\csname EERdiag@full_flexible.women\endcsname{2,951}
\expandafter\def\csname EERdiag@full_flexible.person_months\endcsname{61,649}
\expandafter\def\csname EERdiag@full_flexible.dropout_events\endcsname{594}
\expandafter\def\csname EERdiag@full_flexible.pregnancy_events\endcsname{443}
\expandafter\def\csname EERdiag@full_flexible.exposed_months\endcsname{8,593}
\expandafter\def\csname EERdiag@full_flexible.aic_preferred_k\endcsname{2}
\expandafter\def\csname EERdiag@full_flexible.bic_preferred_k\endcsname{2}
\expandafter\def\csname EERdiag@full_linear.ll1\endcsname{-5732.51}
\expandafter\def\csname EERdiag@full_linear.ll2\endcsname{-5718.69}
\expandafter\def\csname EERdiag@full_linear.ll_gain\endcsname{13.82}
\expandafter\def\csname EERdiag@full_linear.aic1\endcsname{11481.02}
\expandafter\def\csname EERdiag@full_linear.aic2\endcsname{11459.38}
\expandafter\def\csname EERdiag@full_linear.bic_persons1\endcsname{11528.93}
\expandafter\def\csname EERdiag@full_linear.bic_persons2\endcsname{11525.26}
\expandafter\def\csname EERdiag@full_linear.aic_difference\endcsname{-21.64}
\expandafter\def\csname EERdiag@full_linear.bic_difference\endcsname{-3.67}
\expandafter\def\csname EERdiag@full_linear.kparams1\endcsname{8}
\expandafter\def\csname EERdiag@full_linear.kparams2\endcsname{11}
\expandafter\def\csname EERdiag@full_linear.women\endcsname{2,951}
\expandafter\def\csname EERdiag@full_linear.person_months\endcsname{61,649}
\expandafter\def\csname EERdiag@full_linear.dropout_events\endcsname{594}
\expandafter\def\csname EERdiag@full_linear.pregnancy_events\endcsname{443}
\expandafter\def\csname EERdiag@full_linear.exposed_months\endcsname{8,593}
\expandafter\def\csname EERdiag@full_linear.aic_preferred_k\endcsname{2}
\expandafter\def\csname EERdiag@full_linear.bic_preferred_k\endcsname{2}
\expandafter\def\csname EERdiag@restricted_flexible.ll1\endcsname{-5296.34}
\expandafter\def\csname EERdiag@restricted_flexible.ll2\endcsname{-5280.13}
\expandafter\def\csname EERdiag@restricted_flexible.ll_gain\endcsname{16.21}
\expandafter\def\csname EERdiag@restricted_flexible.aic1\endcsname{10638.68}
\expandafter\def\csname EERdiag@restricted_flexible.aic2\endcsname{10612.25}
\expandafter\def\csname EERdiag@restricted_flexible.bic_persons1\endcsname{10774.82}
\expandafter\def\csname EERdiag@restricted_flexible.bic_persons2\endcsname{10766.15}
\expandafter\def\csname EERdiag@restricted_flexible.aic_difference\endcsname{-26.43}
\expandafter\def\csname EERdiag@restricted_flexible.bic_difference\endcsname{-8.67}
\expandafter\def\csname EERdiag@restricted_flexible.kparams1\endcsname{23}
\expandafter\def\csname EERdiag@restricted_flexible.kparams2\endcsname{26}
\expandafter\def\csname EERdiag@restricted_flexible.women\endcsname{2,749}
\expandafter\def\csname EERdiag@restricted_flexible.person_months\endcsname{57,991}
\expandafter\def\csname EERdiag@restricted_flexible.dropout_events\endcsname{501}
\expandafter\def\csname EERdiag@restricted_flexible.pregnancy_events\endcsname{443}
\expandafter\def\csname EERdiag@restricted_flexible.exposed_months\endcsname{5,700}
\expandafter\def\csname EERdiag@restricted_flexible.aic_preferred_k\endcsname{2}
\expandafter\def\csname EERdiag@restricted_flexible.bic_preferred_k\endcsname{2}
\expandafter\def\csname EERdiag@restricted_linear.ll1\endcsname{-5311.85}
\expandafter\def\csname EERdiag@restricted_linear.ll2\endcsname{-5297.92}
\expandafter\def\csname EERdiag@restricted_linear.ll_gain\endcsname{13.92}
\expandafter\def\csname EERdiag@restricted_linear.aic1\endcsname{10637.69}
\expandafter\def\csname EERdiag@restricted_linear.aic2\endcsname{10615.84}
\expandafter\def\csname EERdiag@restricted_linear.bic_persons1\endcsname{10679.13}
\expandafter\def\csname EERdiag@restricted_linear.bic_persons2\endcsname{10675.03}
\expandafter\def\csname EERdiag@restricted_linear.aic_difference\endcsname{-21.85}
\expandafter\def\csname EERdiag@restricted_linear.bic_difference\endcsname{-4.09}
\expandafter\def\csname EERdiag@restricted_linear.kparams1\endcsname{7}
\expandafter\def\csname EERdiag@restricted_linear.kparams2\endcsname{10}
\expandafter\def\csname EERdiag@restricted_linear.women\endcsname{2,749}
\expandafter\def\csname EERdiag@restricted_linear.person_months\endcsname{57,991}
\expandafter\def\csname EERdiag@restricted_linear.dropout_events\endcsname{501}
\expandafter\def\csname EERdiag@restricted_linear.pregnancy_events\endcsname{443}
\expandafter\def\csname EERdiag@restricted_linear.exposed_months\endcsname{5,700}
\expandafter\def\csname EERdiag@restricted_linear.aic_preferred_k\endcsname{2}
\expandafter\def\csname EERdiag@restricted_linear.bic_preferred_k\endcsname{2}
\expandafter\def\csname EERdiag@exclusion.excl_women\endcsname{76}
\expandafter\def\csname EERdiag@exclusion.excl_person_months\endcsname{1,520}
\expandafter\def\csname EERdiag@exclusion.excl_exposed_months\endcsname{1,373}
\expandafter\def\csname EERdiag@exclusion.excl_dropout_events\endcsname{34}
\expandafter\def\csname EERdiag@exclusion.excl_conception_events\endcsname{75}
\expandafter\def\csname EERdiag@exclusion.excl_adjusted_women\endcsname{68}
\expandafter\def\csname EERdiag@exclusion.excl_imputed_women\endcsname{8}
\expandafter\def\csname EERdiag@exclusion.excluded_exposure_pct\endcsname{24.1}

\newcommand{\EERadj}[1]{\csname EERadj@#1\endcsname}
\expandafter\gdef\csname EERadj@adjusted_results.school3.main_baseline.women\endcsname{2,828}
\expandafter\gdef\csname EERadj@adjusted_results.school3.main_core.women\endcsname{2,828}
\expandafter\gdef\csname EERadj@adjusted_results.school3.history_baseline.women\endcsname{2,571}
\expandafter\gdef\csname EERadj@adjusted_results.school3.history_core.women\endcsname{2,571}
\expandafter\gdef\csname EERadj@adjusted_results.school3.later_all_core.women\endcsname{2,313}
\expandafter\gdef\csname EERadj@adjusted_results.school3.later_cc_baseline.women\endcsname{1,733}
\expandafter\gdef\csname EERadj@adjusted_results.school3.later_cc_core.women\endcsname{1,733}
\expandafter\gdef\csname EERadj@adjusted_results.school3.later_cc_rich.women\endcsname{1,733}
\expandafter\gdef\csname EERadj@adjusted_results.calendar4.main_baseline.women\endcsname{2,809}
\expandafter\gdef\csname EERadj@adjusted_results.calendar4.main_core.women\endcsname{2,809}
\expandafter\gdef\csname EERadj@adjusted_results.calendar4.history_baseline.women\endcsname{2,553}
\expandafter\gdef\csname EERadj@adjusted_results.calendar4.history_core.women\endcsname{2,553}
\expandafter\gdef\csname EERadj@adjusted_results.calendar4.later_all_core.women\endcsname{2,291}
\expandafter\gdef\csname EERadj@adjusted_results.calendar4.later_cc_baseline.women\endcsname{1,715}
\expandafter\gdef\csname EERadj@adjusted_results.calendar4.later_cc_core.women\endcsname{1,715}
\expandafter\gdef\csname EERadj@adjusted_results.calendar4.later_cc_rich.women\endcsname{1,715}
\expandafter\gdef\csname EERadj@adjusted_results.school3.main_baseline.months\endcsname{48,953}
\expandafter\gdef\csname EERadj@adjusted_results.school3.main_core.months\endcsname{48,953}
\expandafter\gdef\csname EERadj@adjusted_results.school3.history_baseline.months\endcsname{45,471}
\expandafter\gdef\csname EERadj@adjusted_results.school3.history_core.months\endcsname{45,471}
\expandafter\gdef\csname EERadj@adjusted_results.school3.later_all_core.months\endcsname{35,595}
\expandafter\gdef\csname EERadj@adjusted_results.school3.later_cc_baseline.months\endcsname{26,777}
\expandafter\gdef\csname EERadj@adjusted_results.school3.later_cc_core.months\endcsname{26,777}
\expandafter\gdef\csname EERadj@adjusted_results.school3.later_cc_rich.months\endcsname{26,777}
\expandafter\gdef\csname EERadj@adjusted_results.calendar4.main_baseline.months\endcsname{48,120}
\expandafter\gdef\csname EERadj@adjusted_results.calendar4.main_core.months\endcsname{48,120}
\expandafter\gdef\csname EERadj@adjusted_results.calendar4.history_baseline.months\endcsname{44,724}
\expandafter\gdef\csname EERadj@adjusted_results.calendar4.history_core.months\endcsname{44,724}
\expandafter\gdef\csname EERadj@adjusted_results.calendar4.later_all_core.months\endcsname{34,828}
\expandafter\gdef\csname EERadj@adjusted_results.calendar4.later_cc_baseline.months\endcsname{26,147}
\expandafter\gdef\csname EERadj@adjusted_results.calendar4.later_cc_core.months\endcsname{26,147}
\expandafter\gdef\csname EERadj@adjusted_results.calendar4.later_cc_rich.months\endcsname{26,147}
\expandafter\gdef\csname EERadj@adjusted_results.school3.main_baseline.events\endcsname{711}
\expandafter\gdef\csname EERadj@adjusted_results.school3.main_core.events\endcsname{711}
\expandafter\gdef\csname EERadj@adjusted_results.school3.history_baseline.events\endcsname{570}
\expandafter\gdef\csname EERadj@adjusted_results.school3.history_core.events\endcsname{570}
\expandafter\gdef\csname EERadj@adjusted_results.school3.later_all_core.events\endcsname{521}
\expandafter\gdef\csname EERadj@adjusted_results.school3.later_cc_baseline.events\endcsname{336}
\expandafter\gdef\csname EERadj@adjusted_results.school3.later_cc_core.events\endcsname{336}
\expandafter\gdef\csname EERadj@adjusted_results.school3.later_cc_rich.events\endcsname{336}
\expandafter\gdef\csname EERadj@adjusted_results.calendar4.main_baseline.events\endcsname{726}
\expandafter\gdef\csname EERadj@adjusted_results.calendar4.main_core.events\endcsname{726}
\expandafter\gdef\csname EERadj@adjusted_results.calendar4.history_baseline.events\endcsname{586}
\expandafter\gdef\csname EERadj@adjusted_results.calendar4.history_core.events\endcsname{586}
\expandafter\gdef\csname EERadj@adjusted_results.calendar4.later_all_core.events\endcsname{530}
\expandafter\gdef\csname EERadj@adjusted_results.calendar4.later_cc_baseline.events\endcsname{346}
\expandafter\gdef\csname EERadj@adjusted_results.calendar4.later_cc_core.events\endcsname{346}
\expandafter\gdef\csname EERadj@adjusted_results.calendar4.later_cc_rich.events\endcsname{346}
\expandafter\gdef\csname EERadj@adjusted_results.school3.main_baseline.exposed_months\endcsname{5,939}
\expandafter\gdef\csname EERadj@adjusted_results.school3.main_core.exposed_months\endcsname{5,939}
\expandafter\gdef\csname EERadj@adjusted_results.school3.history_baseline.exposed_months\endcsname{2,680}
\expandafter\gdef\csname EERadj@adjusted_results.school3.history_core.exposed_months\endcsname{2,680}
\expandafter\gdef\csname EERadj@adjusted_results.school3.later_all_core.exposed_months\endcsname{3,002}
\expandafter\gdef\csname EERadj@adjusted_results.school3.later_cc_baseline.exposed_months\endcsname{2,055}
\expandafter\gdef\csname EERadj@adjusted_results.school3.later_cc_core.exposed_months\endcsname{2,055}
\expandafter\gdef\csname EERadj@adjusted_results.school3.later_cc_rich.exposed_months\endcsname{2,055}
\expandafter\gdef\csname EERadj@adjusted_results.calendar4.main_baseline.exposed_months\endcsname{5,804}
\expandafter\gdef\csname EERadj@adjusted_results.calendar4.main_core.exposed_months\endcsname{5,804}
\expandafter\gdef\csname EERadj@adjusted_results.calendar4.history_baseline.exposed_months\endcsname{2,628}
\expandafter\gdef\csname EERadj@adjusted_results.calendar4.history_core.exposed_months\endcsname{2,628}
\expandafter\gdef\csname EERadj@adjusted_results.calendar4.later_all_core.exposed_months\endcsname{2,918}
\expandafter\gdef\csname EERadj@adjusted_results.calendar4.later_cc_baseline.exposed_months\endcsname{2,010}
\expandafter\gdef\csname EERadj@adjusted_results.calendar4.later_cc_core.exposed_months\endcsname{2,010}
\expandafter\gdef\csname EERadj@adjusted_results.calendar4.later_cc_rich.exposed_months\endcsname{2,010}
\expandafter\gdef\csname EERadj@adjusted_results.school3.main_baseline.exposed_events\endcsname{261}
\expandafter\gdef\csname EERadj@adjusted_results.school3.main_core.exposed_events\endcsname{261}
\expandafter\gdef\csname EERadj@adjusted_results.school3.history_baseline.exposed_events\endcsname{122}
\expandafter\gdef\csname EERadj@adjusted_results.school3.history_core.exposed_events\endcsname{122}
\expandafter\gdef\csname EERadj@adjusted_results.school3.later_all_core.exposed_events\endcsname{136}
\expandafter\gdef\csname EERadj@adjusted_results.school3.later_cc_baseline.exposed_events\endcsname{84}
\expandafter\gdef\csname EERadj@adjusted_results.school3.later_cc_core.exposed_events\endcsname{84}
\expandafter\gdef\csname EERadj@adjusted_results.school3.later_cc_rich.exposed_events\endcsname{84}
\expandafter\gdef\csname EERadj@adjusted_results.calendar4.main_baseline.exposed_events\endcsname{258}
\expandafter\gdef\csname EERadj@adjusted_results.calendar4.main_core.exposed_events\endcsname{258}
\expandafter\gdef\csname EERadj@adjusted_results.calendar4.history_baseline.exposed_events\endcsname{120}
\expandafter\gdef\csname EERadj@adjusted_results.calendar4.history_core.exposed_events\endcsname{120}
\expandafter\gdef\csname EERadj@adjusted_results.calendar4.later_all_core.exposed_events\endcsname{130}
\expandafter\gdef\csname EERadj@adjusted_results.calendar4.later_cc_baseline.exposed_events\endcsname{79}
\expandafter\gdef\csname EERadj@adjusted_results.calendar4.later_cc_core.exposed_events\endcsname{79}
\expandafter\gdef\csname EERadj@adjusted_results.calendar4.later_cc_rich.exposed_events\endcsname{79}
\expandafter\gdef\csname EERadj@adjusted_results.school3.main_baseline.b\endcsname{1.348}
\expandafter\gdef\csname EERadj@adjusted_results.school3.main_core.b\endcsname{1.378}
\expandafter\gdef\csname EERadj@adjusted_results.school3.history_baseline.b\endcsname{1.290}
\expandafter\gdef\csname EERadj@adjusted_results.school3.history_core.b\endcsname{1.331}
\expandafter\gdef\csname EERadj@adjusted_results.school3.later_all_core.b\endcsname{1.259}
\expandafter\gdef\csname EERadj@adjusted_results.school3.later_cc_baseline.b\endcsname{1.262}
\expandafter\gdef\csname EERadj@adjusted_results.school3.later_cc_core.b\endcsname{1.308}
\expandafter\gdef\csname EERadj@adjusted_results.school3.later_cc_rich.b\endcsname{1.046}
\expandafter\gdef\csname EERadj@adjusted_results.calendar4.main_baseline.b\endcsname{1.313}
\expandafter\gdef\csname EERadj@adjusted_results.calendar4.main_core.b\endcsname{1.332}
\expandafter\gdef\csname EERadj@adjusted_results.calendar4.history_baseline.b\endcsname{1.242}
\expandafter\gdef\csname EERadj@adjusted_results.calendar4.history_core.b\endcsname{1.271}
\expandafter\gdef\csname EERadj@adjusted_results.calendar4.later_all_core.b\endcsname{1.196}
\expandafter\gdef\csname EERadj@adjusted_results.calendar4.later_cc_baseline.b\endcsname{1.155}
\expandafter\gdef\csname EERadj@adjusted_results.calendar4.later_cc_core.b\endcsname{1.187}
\expandafter\gdef\csname EERadj@adjusted_results.calendar4.later_cc_rich.b\endcsname{0.903}
\expandafter\gdef\csname EERadj@adjusted_results.school3.main_baseline.se\endcsname{0.081}
\expandafter\gdef\csname EERadj@adjusted_results.school3.main_core.se\endcsname{0.085}
\expandafter\gdef\csname EERadj@adjusted_results.school3.history_baseline.se\endcsname{0.112}
\expandafter\gdef\csname EERadj@adjusted_results.school3.history_core.se\endcsname{0.116}
\expandafter\gdef\csname EERadj@adjusted_results.school3.later_all_core.se\endcsname{0.111}
\expandafter\gdef\csname EERadj@adjusted_results.school3.later_cc_baseline.se\endcsname{0.136}
\expandafter\gdef\csname EERadj@adjusted_results.school3.later_cc_core.se\endcsname{0.145}
\expandafter\gdef\csname EERadj@adjusted_results.school3.later_cc_rich.se\endcsname{0.150}
\expandafter\gdef\csname EERadj@adjusted_results.calendar4.main_baseline.se\endcsname{0.080}
\expandafter\gdef\csname EERadj@adjusted_results.calendar4.main_core.se\endcsname{0.084}
\expandafter\gdef\csname EERadj@adjusted_results.calendar4.history_baseline.se\endcsname{0.113}
\expandafter\gdef\csname EERadj@adjusted_results.calendar4.history_core.se\endcsname{0.115}
\expandafter\gdef\csname EERadj@adjusted_results.calendar4.later_all_core.se\endcsname{0.111}
\expandafter\gdef\csname EERadj@adjusted_results.calendar4.later_cc_baseline.se\endcsname{0.138}
\expandafter\gdef\csname EERadj@adjusted_results.calendar4.later_cc_core.se\endcsname{0.145}
\expandafter\gdef\csname EERadj@adjusted_results.calendar4.later_cc_rich.se\endcsname{0.151}
\expandafter\gdef\csname EERadj@adjusted_results.school3.main_baseline.hr\endcsname{3.85}
\expandafter\gdef\csname EERadj@adjusted_results.school3.main_core.hr\endcsname{3.97}
\expandafter\gdef\csname EERadj@adjusted_results.school3.history_baseline.hr\endcsname{3.63}
\expandafter\gdef\csname EERadj@adjusted_results.school3.history_core.hr\endcsname{3.79}
\expandafter\gdef\csname EERadj@adjusted_results.school3.later_all_core.hr\endcsname{3.52}
\expandafter\gdef\csname EERadj@adjusted_results.school3.later_cc_baseline.hr\endcsname{3.53}
\expandafter\gdef\csname EERadj@adjusted_results.school3.later_cc_core.hr\endcsname{3.70}
\expandafter\gdef\csname EERadj@adjusted_results.school3.later_cc_rich.hr\endcsname{2.85}
\expandafter\gdef\csname EERadj@adjusted_results.calendar4.main_baseline.hr\endcsname{3.72}
\expandafter\gdef\csname EERadj@adjusted_results.calendar4.main_core.hr\endcsname{3.79}
\expandafter\gdef\csname EERadj@adjusted_results.calendar4.history_baseline.hr\endcsname{3.46}
\expandafter\gdef\csname EERadj@adjusted_results.calendar4.history_core.hr\endcsname{3.56}
\expandafter\gdef\csname EERadj@adjusted_results.calendar4.later_all_core.hr\endcsname{3.31}
\expandafter\gdef\csname EERadj@adjusted_results.calendar4.later_cc_baseline.hr\endcsname{3.18}
\expandafter\gdef\csname EERadj@adjusted_results.calendar4.later_cc_core.hr\endcsname{3.28}
\expandafter\gdef\csname EERadj@adjusted_results.calendar4.later_cc_rich.hr\endcsname{2.47}
\expandafter\gdef\csname EERadj@adjusted_results.school3.main_baseline.ci_lo\endcsname{3.29}
\expandafter\gdef\csname EERadj@adjusted_results.school3.main_core.ci_lo\endcsname{3.36}
\expandafter\gdef\csname EERadj@adjusted_results.school3.history_baseline.ci_lo\endcsname{2.91}
\expandafter\gdef\csname EERadj@adjusted_results.school3.history_core.ci_lo\endcsname{3.02}
\expandafter\gdef\csname EERadj@adjusted_results.school3.later_all_core.ci_lo\endcsname{2.83}
\expandafter\gdef\csname EERadj@adjusted_results.school3.later_cc_baseline.ci_lo\endcsname{2.71}
\expandafter\gdef\csname EERadj@adjusted_results.school3.later_cc_core.ci_lo\endcsname{2.79}
\expandafter\gdef\csname EERadj@adjusted_results.school3.later_cc_rich.ci_lo\endcsname{2.12}
\expandafter\gdef\csname EERadj@adjusted_results.calendar4.main_baseline.ci_lo\endcsname{3.18}
\expandafter\gdef\csname EERadj@adjusted_results.calendar4.main_core.ci_lo\endcsname{3.21}
\expandafter\gdef\csname EERadj@adjusted_results.calendar4.history_baseline.ci_lo\endcsname{2.78}
\expandafter\gdef\csname EERadj@adjusted_results.calendar4.history_core.ci_lo\endcsname{2.84}
\expandafter\gdef\csname EERadj@adjusted_results.calendar4.later_all_core.ci_lo\endcsname{2.66}
\expandafter\gdef\csname EERadj@adjusted_results.calendar4.later_cc_baseline.ci_lo\endcsname{2.42}
\expandafter\gdef\csname EERadj@adjusted_results.calendar4.later_cc_core.ci_lo\endcsname{2.46}
\expandafter\gdef\csname EERadj@adjusted_results.calendar4.later_cc_rich.ci_lo\endcsname{1.83}
\expandafter\gdef\csname EERadj@adjusted_results.school3.main_baseline.ci_hi\endcsname{4.51}
\expandafter\gdef\csname EERadj@adjusted_results.school3.main_core.ci_hi\endcsname{4.69}
\expandafter\gdef\csname EERadj@adjusted_results.school3.history_baseline.ci_hi\endcsname{4.53}
\expandafter\gdef\csname EERadj@adjusted_results.school3.history_core.ci_hi\endcsname{4.75}
\expandafter\gdef\csname EERadj@adjusted_results.school3.later_all_core.ci_hi\endcsname{4.38}
\expandafter\gdef\csname EERadj@adjusted_results.school3.later_cc_baseline.ci_hi\endcsname{4.61}
\expandafter\gdef\csname EERadj@adjusted_results.school3.later_cc_core.ci_hi\endcsname{4.91}
\expandafter\gdef\csname EERadj@adjusted_results.school3.later_cc_rich.ci_hi\endcsname{3.82}
\expandafter\gdef\csname EERadj@adjusted_results.calendar4.main_baseline.ci_hi\endcsname{4.35}
\expandafter\gdef\csname EERadj@adjusted_results.calendar4.main_core.ci_hi\endcsname{4.47}
\expandafter\gdef\csname EERadj@adjusted_results.calendar4.history_baseline.ci_hi\endcsname{4.32}
\expandafter\gdef\csname EERadj@adjusted_results.calendar4.history_core.ci_hi\endcsname{4.46}
\expandafter\gdef\csname EERadj@adjusted_results.calendar4.later_all_core.ci_hi\endcsname{4.12}
\expandafter\gdef\csname EERadj@adjusted_results.calendar4.later_cc_baseline.ci_hi\endcsname{4.16}
\expandafter\gdef\csname EERadj@adjusted_results.calendar4.later_cc_core.ci_hi\endcsname{4.36}
\expandafter\gdef\csname EERadj@adjusted_results.calendar4.later_cc_rich.ci_hi\endcsname{3.31}
\expandafter\gdef\csname EERadj@adjusted_results.school3.main_baseline.loglik\endcsname{-3493.26}
\expandafter\gdef\csname EERadj@adjusted_results.school3.main_core.loglik\endcsname{-3471.48}
\expandafter\gdef\csname EERadj@adjusted_results.school3.history_baseline.loglik\endcsname{-2932.95}
\expandafter\gdef\csname EERadj@adjusted_results.school3.history_core.loglik\endcsname{-2909.92}
\expandafter\gdef\csname EERadj@adjusted_results.school3.later_all_core.loglik\endcsname{-2594.40}
\expandafter\gdef\csname EERadj@adjusted_results.school3.later_cc_baseline.loglik\endcsname{-1730.32}
\expandafter\gdef\csname EERadj@adjusted_results.school3.later_cc_core.loglik\endcsname{-1718.76}
\expandafter\gdef\csname EERadj@adjusted_results.school3.later_cc_rich.loglik\endcsname{-1677.38}
\expandafter\gdef\csname EERadj@adjusted_results.calendar4.main_baseline.loglik\endcsname{-3547.91}
\expandafter\gdef\csname EERadj@adjusted_results.calendar4.main_core.loglik\endcsname{-3528.82}
\expandafter\gdef\csname EERadj@adjusted_results.calendar4.history_baseline.loglik\endcsname{-2992.65}
\expandafter\gdef\csname EERadj@adjusted_results.calendar4.history_core.loglik\endcsname{-2974.78}
\expandafter\gdef\csname EERadj@adjusted_results.calendar4.later_all_core.loglik\endcsname{-2631.49}
\expandafter\gdef\csname EERadj@adjusted_results.calendar4.later_cc_baseline.loglik\endcsname{-1773.17}
\expandafter\gdef\csname EERadj@adjusted_results.calendar4.later_cc_core.loglik\endcsname{-1766.31}
\expandafter\gdef\csname EERadj@adjusted_results.calendar4.later_cc_rich.loglik\endcsname{-1714.25}
\expandafter\gdef\csname EERadj@adjusted_results.school3.main_baseline.rank\endcsname{11}
\expandafter\gdef\csname EERadj@adjusted_results.school3.main_core.rank\endcsname{20}
\expandafter\gdef\csname EERadj@adjusted_results.school3.history_baseline.rank\endcsname{11}
\expandafter\gdef\csname EERadj@adjusted_results.school3.history_core.rank\endcsname{20}
\expandafter\gdef\csname EERadj@adjusted_results.school3.later_all_core.rank\endcsname{19}
\expandafter\gdef\csname EERadj@adjusted_results.school3.later_cc_baseline.rank\endcsname{11}
\expandafter\gdef\csname EERadj@adjusted_results.school3.later_cc_core.rank\endcsname{19}
\expandafter\gdef\csname EERadj@adjusted_results.school3.later_cc_rich.rank\endcsname{23}
\expandafter\gdef\csname EERadj@adjusted_results.calendar4.main_baseline.rank\endcsname{11}
\expandafter\gdef\csname EERadj@adjusted_results.calendar4.main_core.rank\endcsname{20}
\expandafter\gdef\csname EERadj@adjusted_results.calendar4.history_baseline.rank\endcsname{11}
\expandafter\gdef\csname EERadj@adjusted_results.calendar4.history_core.rank\endcsname{20}
\expandafter\gdef\csname EERadj@adjusted_results.calendar4.later_all_core.rank\endcsname{19}
\expandafter\gdef\csname EERadj@adjusted_results.calendar4.later_cc_baseline.rank\endcsname{11}
\expandafter\gdef\csname EERadj@adjusted_results.calendar4.later_cc_core.rank\endcsname{19}
\expandafter\gdef\csname EERadj@adjusted_results.calendar4.later_cc_rich.rank\endcsname{23}
\expandafter\gdef\csname EERadj@adjusted_results.school3.main_baseline.converged\endcsname{1}
\expandafter\gdef\csname EERadj@adjusted_results.school3.main_core.converged\endcsname{1}
\expandafter\gdef\csname EERadj@adjusted_results.school3.history_baseline.converged\endcsname{1}
\expandafter\gdef\csname EERadj@adjusted_results.school3.history_core.converged\endcsname{1}
\expandafter\gdef\csname EERadj@adjusted_results.school3.later_all_core.converged\endcsname{1}
\expandafter\gdef\csname EERadj@adjusted_results.school3.later_cc_baseline.converged\endcsname{1}
\expandafter\gdef\csname EERadj@adjusted_results.school3.later_cc_core.converged\endcsname{1}
\expandafter\gdef\csname EERadj@adjusted_results.school3.later_cc_rich.converged\endcsname{1}
\expandafter\gdef\csname EERadj@adjusted_results.calendar4.main_baseline.converged\endcsname{1}
\expandafter\gdef\csname EERadj@adjusted_results.calendar4.main_core.converged\endcsname{1}
\expandafter\gdef\csname EERadj@adjusted_results.calendar4.history_baseline.converged\endcsname{1}
\expandafter\gdef\csname EERadj@adjusted_results.calendar4.history_core.converged\endcsname{1}
\expandafter\gdef\csname EERadj@adjusted_results.calendar4.later_all_core.converged\endcsname{1}
\expandafter\gdef\csname EERadj@adjusted_results.calendar4.later_cc_baseline.converged\endcsname{1}
\expandafter\gdef\csname EERadj@adjusted_results.calendar4.later_cc_core.converged\endcsname{1}
\expandafter\gdef\csname EERadj@adjusted_results.calendar4.later_cc_rich.converged\endcsname{1}
\expandafter\gdef\csname EERadj@adjusted_coeff.school3.main_baseline.dropout_event:treated.b\endcsname{1.348}
\expandafter\gdef\csname EERadj@adjusted_coeff.school3.main_baseline.dropout_event:1b.eer_durbin.b\endcsname{0.000}
\expandafter\gdef\csname EERadj@adjusted_coeff.school3.main_baseline.dropout_event:2.eer_durbin.b\endcsname{-0.526}
\expandafter\gdef\csname EERadj@adjusted_coeff.school3.main_baseline.dropout_event:3.eer_durbin.b\endcsname{0.141}
\expandafter\gdef\csname EERadj@adjusted_coeff.school3.main_baseline.dropout_event:4.eer_durbin.b\endcsname{-0.090}
\expandafter\gdef\csname EERadj@adjusted_coeff.school3.main_baseline.dropout_event:5.eer_durbin.b\endcsname{-0.048}
\expandafter\gdef\csname EERadj@adjusted_coeff.school3.main_baseline.dropout_event:6.eer_durbin.b\endcsname{-0.052}
\expandafter\gdef\csname EERadj@adjusted_coeff.school3.main_baseline.dropout_event:7.eer_durbin.b\endcsname{-0.020}
\expandafter\gdef\csname EERadj@adjusted_coeff.school3.main_baseline.dropout_event:1b.eer_term.b\endcsname{0.000}
\expandafter\gdef\csname EERadj@adjusted_coeff.school3.main_baseline.dropout_event:2.eer_term.b\endcsname{-0.644}
\expandafter\gdef\csname EERadj@adjusted_coeff.school3.main_baseline.dropout_event:eer_age.b\endcsname{0.207}
\expandafter\gdef\csname EERadj@adjusted_coeff.school3.main_baseline.dropout_event:eer_age2.b\endcsname{0.019}
\expandafter\gdef\csname EERadj@adjusted_coeff.school3.main_baseline.dropout_event:_cons.b\endcsname{-4.441}
\expandafter\gdef\csname EERadj@adjusted_coeff.school3.main_core.dropout_event:treated.b\endcsname{1.378}
\expandafter\gdef\csname EERadj@adjusted_coeff.school3.main_core.dropout_event:1b.eer_durbin.b\endcsname{0.000}
\expandafter\gdef\csname EERadj@adjusted_coeff.school3.main_core.dropout_event:2.eer_durbin.b\endcsname{-0.761}
\expandafter\gdef\csname EERadj@adjusted_coeff.school3.main_core.dropout_event:3.eer_durbin.b\endcsname{-0.491}
\expandafter\gdef\csname EERadj@adjusted_coeff.school3.main_core.dropout_event:4.eer_durbin.b\endcsname{-0.941}
\expandafter\gdef\csname EERadj@adjusted_coeff.school3.main_core.dropout_event:5.eer_durbin.b\endcsname{-1.292}
\expandafter\gdef\csname EERadj@adjusted_coeff.school3.main_core.dropout_event:6.eer_durbin.b\endcsname{-1.577}
\expandafter\gdef\csname EERadj@adjusted_coeff.school3.main_core.dropout_event:7.eer_durbin.b\endcsname{-1.787}
\expandafter\gdef\csname EERadj@adjusted_coeff.school3.main_core.dropout_event:1b.eer_term.b\endcsname{0.000}
\expandafter\gdef\csname EERadj@adjusted_coeff.school3.main_core.dropout_event:2.eer_term.b\endcsname{-0.904}
\expandafter\gdef\csname EERadj@adjusted_coeff.school3.main_core.dropout_event:eer_age.b\endcsname{0.206}
\expandafter\gdef\csname EERadj@adjusted_coeff.school3.main_core.dropout_event:eer_age2.b\endcsname{0.009}
\expandafter\gdef\csname EERadj@adjusted_coeff.school3.main_core.dropout_event:1.race_ethnicity.b\endcsname{-0.217}
\expandafter\gdef\csname EERadj@adjusted_coeff.school3.main_core.dropout_event:2.race_ethnicity.b\endcsname{0.157}
\expandafter\gdef\csname EERadj@adjusted_coeff.school3.main_core.dropout_event:3.race_ethnicity.b\endcsname{0.029}
\expandafter\gdef\csname EERadj@adjusted_coeff.school3.main_core.dropout_event:4b.race_ethnicity.b\endcsname{0.000}
\expandafter\gdef\csname EERadj@adjusted_coeff.school3.main_core.dropout_event:1997b.calendar_period.b\endcsname{0.000}
\expandafter\gdef\csname EERadj@adjusted_coeff.school3.main_core.dropout_event:1998.calendar_period.b\endcsname{0.731}
\expandafter\gdef\csname EERadj@adjusted_coeff.school3.main_core.dropout_event:1999.calendar_period.b\endcsname{1.334}
\expandafter\gdef\csname EERadj@adjusted_coeff.school3.main_core.dropout_event:2000.calendar_period.b\endcsname{1.682}
\expandafter\gdef\csname EERadj@adjusted_coeff.school3.main_core.dropout_event:2001.calendar_period.b\endcsname{1.873}
\expandafter\gdef\csname EERadj@adjusted_coeff.school3.main_core.dropout_event:2002.calendar_period.b\endcsname{2.093}
\expandafter\gdef\csname EERadj@adjusted_coeff.school3.main_core.dropout_event:2003.calendar_period.b\endcsname{2.290}
\expandafter\gdef\csname EERadj@adjusted_coeff.school3.main_core.dropout_event:_cons.b\endcsname{-4.505}
\expandafter\gdef\csname EERadj@adjusted_coeff.school3.history_baseline.dropout_event:treated.b\endcsname{1.290}
\expandafter\gdef\csname EERadj@adjusted_coeff.school3.history_baseline.dropout_event:1b.eer_durbin.b\endcsname{0.000}
\expandafter\gdef\csname EERadj@adjusted_coeff.school3.history_baseline.dropout_event:2.eer_durbin.b\endcsname{-0.427}
\expandafter\gdef\csname EERadj@adjusted_coeff.school3.history_baseline.dropout_event:3.eer_durbin.b\endcsname{0.140}
\expandafter\gdef\csname EERadj@adjusted_coeff.school3.history_baseline.dropout_event:4.eer_durbin.b\endcsname{-0.026}
\expandafter\gdef\csname EERadj@adjusted_coeff.school3.history_baseline.dropout_event:5.eer_durbin.b\endcsname{0.042}
\expandafter\gdef\csname EERadj@adjusted_coeff.school3.history_baseline.dropout_event:6.eer_durbin.b\endcsname{0.066}
\expandafter\gdef\csname EERadj@adjusted_coeff.school3.history_baseline.dropout_event:7.eer_durbin.b\endcsname{0.071}
\expandafter\gdef\csname EERadj@adjusted_coeff.school3.history_baseline.dropout_event:1b.eer_term.b\endcsname{0.000}
\expandafter\gdef\csname EERadj@adjusted_coeff.school3.history_baseline.dropout_event:2.eer_term.b\endcsname{-0.590}
\expandafter\gdef\csname EERadj@adjusted_coeff.school3.history_baseline.dropout_event:eer_age.b\endcsname{0.188}
\expandafter\gdef\csname EERadj@adjusted_coeff.school3.history_baseline.dropout_event:eer_age2.b\endcsname{0.019}
\expandafter\gdef\csname EERadj@adjusted_coeff.school3.history_baseline.dropout_event:_cons.b\endcsname{-4.513}
\expandafter\gdef\csname EERadj@adjusted_coeff.school3.history_core.dropout_event:treated.b\endcsname{1.331}
\expandafter\gdef\csname EERadj@adjusted_coeff.school3.history_core.dropout_event:1b.eer_durbin.b\endcsname{0.000}
\expandafter\gdef\csname EERadj@adjusted_coeff.school3.history_core.dropout_event:2.eer_durbin.b\endcsname{-0.934}
\expandafter\gdef\csname EERadj@adjusted_coeff.school3.history_core.dropout_event:3.eer_durbin.b\endcsname{-0.946}
\expandafter\gdef\csname EERadj@adjusted_coeff.school3.history_core.dropout_event:4.eer_durbin.b\endcsname{-1.410}
\expandafter\gdef\csname EERadj@adjusted_coeff.school3.history_core.dropout_event:5.eer_durbin.b\endcsname{-1.905}
\expandafter\gdef\csname EERadj@adjusted_coeff.school3.history_core.dropout_event:6.eer_durbin.b\endcsname{-2.377}
\expandafter\gdef\csname EERadj@adjusted_coeff.school3.history_core.dropout_event:7.eer_durbin.b\endcsname{-2.809}
\expandafter\gdef\csname EERadj@adjusted_coeff.school3.history_core.dropout_event:1b.eer_term.b\endcsname{0.000}
\expandafter\gdef\csname EERadj@adjusted_coeff.school3.history_core.dropout_event:2.eer_term.b\endcsname{-0.981}
\expandafter\gdef\csname EERadj@adjusted_coeff.school3.history_core.dropout_event:eer_age.b\endcsname{0.204}
\expandafter\gdef\csname EERadj@adjusted_coeff.school3.history_core.dropout_event:eer_age2.b\endcsname{0.006}
\expandafter\gdef\csname EERadj@adjusted_coeff.school3.history_core.dropout_event:1.race_ethnicity.b\endcsname{-0.198}
\expandafter\gdef\csname EERadj@adjusted_coeff.school3.history_core.dropout_event:2.race_ethnicity.b\endcsname{0.158}
\expandafter\gdef\csname EERadj@adjusted_coeff.school3.history_core.dropout_event:3.race_ethnicity.b\endcsname{-0.220}
\expandafter\gdef\csname EERadj@adjusted_coeff.school3.history_core.dropout_event:4b.race_ethnicity.b\endcsname{0.000}
\expandafter\gdef\csname EERadj@adjusted_coeff.school3.history_core.dropout_event:1997b.calendar_period.b\endcsname{0.000}
\expandafter\gdef\csname EERadj@adjusted_coeff.school3.history_core.dropout_event:1998.calendar_period.b\endcsname{1.149}
\expandafter\gdef\csname EERadj@adjusted_coeff.school3.history_core.dropout_event:1999.calendar_period.b\endcsname{1.957}
\expandafter\gdef\csname EERadj@adjusted_coeff.school3.history_core.dropout_event:2000.calendar_period.b\endcsname{2.510}
\expandafter\gdef\csname EERadj@adjusted_coeff.school3.history_core.dropout_event:2001.calendar_period.b\endcsname{2.927}
\expandafter\gdef\csname EERadj@adjusted_coeff.school3.history_core.dropout_event:2002.calendar_period.b\endcsname{3.236}
\expandafter\gdef\csname EERadj@adjusted_coeff.school3.history_core.dropout_event:2003.calendar_period.b\endcsname{3.452}
\expandafter\gdef\csname EERadj@adjusted_coeff.school3.history_core.dropout_event:_cons.b\endcsname{-4.562}
\expandafter\gdef\csname EERadj@adjusted_coeff.school3.later_all_core.dropout_event:treated.b\endcsname{1.259}
\expandafter\gdef\csname EERadj@adjusted_coeff.school3.later_all_core.dropout_event:1b.eer_durbin.b\endcsname{0.000}
\expandafter\gdef\csname EERadj@adjusted_coeff.school3.later_all_core.dropout_event:2.eer_durbin.b\endcsname{-0.555}
\expandafter\gdef\csname EERadj@adjusted_coeff.school3.later_all_core.dropout_event:3.eer_durbin.b\endcsname{-0.252}
\expandafter\gdef\csname EERadj@adjusted_coeff.school3.later_all_core.dropout_event:4.eer_durbin.b\endcsname{-0.690}
\expandafter\gdef\csname EERadj@adjusted_coeff.school3.later_all_core.dropout_event:5.eer_durbin.b\endcsname{-0.958}
\expandafter\gdef\csname EERadj@adjusted_coeff.school3.later_all_core.dropout_event:6.eer_durbin.b\endcsname{-1.245}
\expandafter\gdef\csname EERadj@adjusted_coeff.school3.later_all_core.dropout_event:7.eer_durbin.b\endcsname{-1.422}
\expandafter\gdef\csname EERadj@adjusted_coeff.school3.later_all_core.dropout_event:1b.eer_term.b\endcsname{0.000}
\expandafter\gdef\csname EERadj@adjusted_coeff.school3.later_all_core.dropout_event:2.eer_term.b\endcsname{-0.821}
\expandafter\gdef\csname EERadj@adjusted_coeff.school3.later_all_core.dropout_event:eer_age.b\endcsname{0.131}
\expandafter\gdef\csname EERadj@adjusted_coeff.school3.later_all_core.dropout_event:eer_age2.b\endcsname{0.033}
\expandafter\gdef\csname EERadj@adjusted_coeff.school3.later_all_core.dropout_event:1.race_ethnicity.b\endcsname{-0.210}
\expandafter\gdef\csname EERadj@adjusted_coeff.school3.later_all_core.dropout_event:2.race_ethnicity.b\endcsname{0.134}
\expandafter\gdef\csname EERadj@adjusted_coeff.school3.later_all_core.dropout_event:3.race_ethnicity.b\endcsname{-0.145}
\expandafter\gdef\csname EERadj@adjusted_coeff.school3.later_all_core.dropout_event:4b.race_ethnicity.b\endcsname{0.000}
\expandafter\gdef\csname EERadj@adjusted_coeff.school3.later_all_core.dropout_event:1998b.calendar_period.b\endcsname{0.000}
\expandafter\gdef\csname EERadj@adjusted_coeff.school3.later_all_core.dropout_event:1999.calendar_period.b\endcsname{0.523}
\expandafter\gdef\csname EERadj@adjusted_coeff.school3.later_all_core.dropout_event:2000.calendar_period.b\endcsname{0.917}
\expandafter\gdef\csname EERadj@adjusted_coeff.school3.later_all_core.dropout_event:2001.calendar_period.b\endcsname{1.115}
\expandafter\gdef\csname EERadj@adjusted_coeff.school3.later_all_core.dropout_event:2002.calendar_period.b\endcsname{1.284}
\expandafter\gdef\csname EERadj@adjusted_coeff.school3.later_all_core.dropout_event:2003.calendar_period.b\endcsname{1.448}
\expandafter\gdef\csname EERadj@adjusted_coeff.school3.later_all_core.dropout_event:_cons.b\endcsname{-4.032}
\expandafter\gdef\csname EERadj@adjusted_coeff.school3.later_cc_baseline.dropout_event:treated.b\endcsname{1.262}
\expandafter\gdef\csname EERadj@adjusted_coeff.school3.later_cc_baseline.dropout_event:1b.eer_durbin.b\endcsname{0.000}
\expandafter\gdef\csname EERadj@adjusted_coeff.school3.later_cc_baseline.dropout_event:2.eer_durbin.b\endcsname{-1.052}
\expandafter\gdef\csname EERadj@adjusted_coeff.school3.later_cc_baseline.dropout_event:3.eer_durbin.b\endcsname{-1.295}
\expandafter\gdef\csname EERadj@adjusted_coeff.school3.later_cc_baseline.dropout_event:4.eer_durbin.b\endcsname{-1.371}
\expandafter\gdef\csname EERadj@adjusted_coeff.school3.later_cc_baseline.dropout_event:5.eer_durbin.b\endcsname{-1.258}
\expandafter\gdef\csname EERadj@adjusted_coeff.school3.later_cc_baseline.dropout_event:6.eer_durbin.b\endcsname{-1.200}
\expandafter\gdef\csname EERadj@adjusted_coeff.school3.later_cc_baseline.dropout_event:7.eer_durbin.b\endcsname{-1.186}
\expandafter\gdef\csname EERadj@adjusted_coeff.school3.later_cc_baseline.dropout_event:1b.eer_term.b\endcsname{0.000}
\expandafter\gdef\csname EERadj@adjusted_coeff.school3.later_cc_baseline.dropout_event:2.eer_term.b\endcsname{-0.579}
\expandafter\gdef\csname EERadj@adjusted_coeff.school3.later_cc_baseline.dropout_event:eer_age.b\endcsname{-0.014}
\expandafter\gdef\csname EERadj@adjusted_coeff.school3.later_cc_baseline.dropout_event:eer_age2.b\endcsname{0.081}
\expandafter\gdef\csname EERadj@adjusted_coeff.school3.later_cc_baseline.dropout_event:_cons.b\endcsname{-3.257}
\expandafter\gdef\csname EERadj@adjusted_coeff.school3.later_cc_core.dropout_event:treated.b\endcsname{1.308}
\expandafter\gdef\csname EERadj@adjusted_coeff.school3.later_cc_core.dropout_event:1b.eer_durbin.b\endcsname{0.000}
\expandafter\gdef\csname EERadj@adjusted_coeff.school3.later_cc_core.dropout_event:2.eer_durbin.b\endcsname{-0.851}
\expandafter\gdef\csname EERadj@adjusted_coeff.school3.later_cc_core.dropout_event:3.eer_durbin.b\endcsname{-1.042}
\expandafter\gdef\csname EERadj@adjusted_coeff.school3.later_cc_core.dropout_event:4.eer_durbin.b\endcsname{-1.321}
\expandafter\gdef\csname EERadj@adjusted_coeff.school3.later_cc_core.dropout_event:5.eer_durbin.b\endcsname{-1.732}
\expandafter\gdef\csname EERadj@adjusted_coeff.school3.later_cc_core.dropout_event:6.eer_durbin.b\endcsname{-2.219}
\expandafter\gdef\csname EERadj@adjusted_coeff.school3.later_cc_core.dropout_event:7.eer_durbin.b\endcsname{-2.629}
\expandafter\gdef\csname EERadj@adjusted_coeff.school3.later_cc_core.dropout_event:1b.eer_term.b\endcsname{0.000}
\expandafter\gdef\csname EERadj@adjusted_coeff.school3.later_cc_core.dropout_event:2.eer_term.b\endcsname{-0.938}
\expandafter\gdef\csname EERadj@adjusted_coeff.school3.later_cc_core.dropout_event:eer_age.b\endcsname{0.002}
\expandafter\gdef\csname EERadj@adjusted_coeff.school3.later_cc_core.dropout_event:eer_age2.b\endcsname{0.077}
\expandafter\gdef\csname EERadj@adjusted_coeff.school3.later_cc_core.dropout_event:1.race_ethnicity.b\endcsname{-0.339}
\expandafter\gdef\csname EERadj@adjusted_coeff.school3.later_cc_core.dropout_event:2.race_ethnicity.b\endcsname{0.124}
\expandafter\gdef\csname EERadj@adjusted_coeff.school3.later_cc_core.dropout_event:3.race_ethnicity.b\endcsname{-0.625}
\expandafter\gdef\csname EERadj@adjusted_coeff.school3.later_cc_core.dropout_event:4b.race_ethnicity.b\endcsname{0.000}
\expandafter\gdef\csname EERadj@adjusted_coeff.school3.later_cc_core.dropout_event:1998b.calendar_period.b\endcsname{0.000}
\expandafter\gdef\csname EERadj@adjusted_coeff.school3.later_cc_core.dropout_event:1999.calendar_period.b\endcsname{0.737}
\expandafter\gdef\csname EERadj@adjusted_coeff.school3.later_cc_core.dropout_event:2000.calendar_period.b\endcsname{1.243}
\expandafter\gdef\csname EERadj@adjusted_coeff.school3.later_cc_core.dropout_event:2001.calendar_period.b\endcsname{1.830}
\expandafter\gdef\csname EERadj@adjusted_coeff.school3.later_cc_core.dropout_event:2002.calendar_period.b\endcsname{1.941}
\expandafter\gdef\csname EERadj@adjusted_coeff.school3.later_cc_core.dropout_event:2003.calendar_period.b\endcsname{2.086}
\expandafter\gdef\csname EERadj@adjusted_coeff.school3.later_cc_core.dropout_event:_cons.b\endcsname{-3.515}
\expandafter\gdef\csname EERadj@adjusted_coeff.school3.later_cc_rich.dropout_event:treated.b\endcsname{1.046}
\expandafter\gdef\csname EERadj@adjusted_coeff.school3.later_cc_rich.dropout_event:1b.eer_durbin.b\endcsname{0.000}
\expandafter\gdef\csname EERadj@adjusted_coeff.school3.later_cc_rich.dropout_event:2.eer_durbin.b\endcsname{-0.533}
\expandafter\gdef\csname EERadj@adjusted_coeff.school3.later_cc_rich.dropout_event:3.eer_durbin.b\endcsname{-0.685}
\expandafter\gdef\csname EERadj@adjusted_coeff.school3.later_cc_rich.dropout_event:4.eer_durbin.b\endcsname{-0.984}
\expandafter\gdef\csname EERadj@adjusted_coeff.school3.later_cc_rich.dropout_event:5.eer_durbin.b\endcsname{-1.341}
\expandafter\gdef\csname EERadj@adjusted_coeff.school3.later_cc_rich.dropout_event:6.eer_durbin.b\endcsname{-1.747}
\expandafter\gdef\csname EERadj@adjusted_coeff.school3.later_cc_rich.dropout_event:7.eer_durbin.b\endcsname{-2.146}
\expandafter\gdef\csname EERadj@adjusted_coeff.school3.later_cc_rich.dropout_event:1b.eer_term.b\endcsname{0.000}
\expandafter\gdef\csname EERadj@adjusted_coeff.school3.later_cc_rich.dropout_event:2.eer_term.b\endcsname{-0.863}
\expandafter\gdef\csname EERadj@adjusted_coeff.school3.later_cc_rich.dropout_event:eer_age.b\endcsname{0.066}
\expandafter\gdef\csname EERadj@adjusted_coeff.school3.later_cc_rich.dropout_event:eer_age2.b\endcsname{0.037}
\expandafter\gdef\csname EERadj@adjusted_coeff.school3.later_cc_rich.dropout_event:1.race_ethnicity.b\endcsname{-0.799}
\expandafter\gdef\csname EERadj@adjusted_coeff.school3.later_cc_rich.dropout_event:2.race_ethnicity.b\endcsname{-0.235}
\expandafter\gdef\csname EERadj@adjusted_coeff.school3.later_cc_rich.dropout_event:3.race_ethnicity.b\endcsname{-0.695}
\expandafter\gdef\csname EERadj@adjusted_coeff.school3.later_cc_rich.dropout_event:4b.race_ethnicity.b\endcsname{0.000}
\expandafter\gdef\csname EERadj@adjusted_coeff.school3.later_cc_rich.dropout_event:1998b.calendar_period.b\endcsname{0.000}
\expandafter\gdef\csname EERadj@adjusted_coeff.school3.later_cc_rich.dropout_event:1999.calendar_period.b\endcsname{0.700}
\expandafter\gdef\csname EERadj@adjusted_coeff.school3.later_cc_rich.dropout_event:2000.calendar_period.b\endcsname{1.162}
\expandafter\gdef\csname EERadj@adjusted_coeff.school3.later_cc_rich.dropout_event:2001.calendar_period.b\endcsname{1.720}
\expandafter\gdef\csname EERadj@adjusted_coeff.school3.later_cc_rich.dropout_event:2002.calendar_period.b\endcsname{1.817}
\expandafter\gdef\csname EERadj@adjusted_coeff.school3.later_cc_rich.dropout_event:2003.calendar_period.b\endcsname{1.899}
\expandafter\gdef\csname EERadj@adjusted_coeff.school3.later_cc_rich.dropout_event:mother_grade97.b\endcsname{-0.044}
\expandafter\gdef\csname EERadj@adjusted_coeff.school3.later_cc_rich.dropout_event:1b.household_group.b\endcsname{0.000}
\expandafter\gdef\csname EERadj@adjusted_coeff.school3.later_cc_rich.dropout_event:2.household_group.b\endcsname{0.435}
\expandafter\gdef\csname EERadj@adjusted_coeff.school3.later_cc_rich.dropout_event:3.household_group.b\endcsname{0.468}
\expandafter\gdef\csname EERadj@adjusted_coeff.school3.later_cc_rich.dropout_event:asvab10.b\endcsname{-0.167}
\expandafter\gdef\csname EERadj@adjusted_coeff.school3.later_cc_rich.dropout_event:_cons.b\endcsname{-2.756}
\expandafter\gdef\csname EERadj@adjusted_coeff.calendar4.main_baseline.dropout_event:treated.b\endcsname{1.313}
\expandafter\gdef\csname EERadj@adjusted_coeff.calendar4.main_baseline.dropout_event:1b.eer_durbin.b\endcsname{0.000}
\expandafter\gdef\csname EERadj@adjusted_coeff.calendar4.main_baseline.dropout_event:2.eer_durbin.b\endcsname{-0.561}
\expandafter\gdef\csname EERadj@adjusted_coeff.calendar4.main_baseline.dropout_event:3.eer_durbin.b\endcsname{0.227}
\expandafter\gdef\csname EERadj@adjusted_coeff.calendar4.main_baseline.dropout_event:4.eer_durbin.b\endcsname{-0.115}
\expandafter\gdef\csname EERadj@adjusted_coeff.calendar4.main_baseline.dropout_event:5.eer_durbin.b\endcsname{-0.048}
\expandafter\gdef\csname EERadj@adjusted_coeff.calendar4.main_baseline.dropout_event:6.eer_durbin.b\endcsname{-0.002}
\expandafter\gdef\csname EERadj@adjusted_coeff.calendar4.main_baseline.dropout_event:7.eer_durbin.b\endcsname{-0.019}
\expandafter\gdef\csname EERadj@adjusted_coeff.calendar4.main_baseline.dropout_event:1b.eer_term.b\endcsname{0.000}
\expandafter\gdef\csname EERadj@adjusted_coeff.calendar4.main_baseline.dropout_event:2.eer_term.b\endcsname{-0.663}
\expandafter\gdef\csname EERadj@adjusted_coeff.calendar4.main_baseline.dropout_event:eer_age.b\endcsname{0.180}
\expandafter\gdef\csname EERadj@adjusted_coeff.calendar4.main_baseline.dropout_event:eer_age2.b\endcsname{0.024}
\expandafter\gdef\csname EERadj@adjusted_coeff.calendar4.main_baseline.dropout_event:_cons.b\endcsname{-4.366}
\expandafter\gdef\csname EERadj@adjusted_coeff.calendar4.main_core.dropout_event:treated.b\endcsname{1.332}
\expandafter\gdef\csname EERadj@adjusted_coeff.calendar4.main_core.dropout_event:1b.eer_durbin.b\endcsname{0.000}
\expandafter\gdef\csname EERadj@adjusted_coeff.calendar4.main_core.dropout_event:2.eer_durbin.b\endcsname{-0.858}
\expandafter\gdef\csname EERadj@adjusted_coeff.calendar4.main_core.dropout_event:3.eer_durbin.b\endcsname{-0.475}
\expandafter\gdef\csname EERadj@adjusted_coeff.calendar4.main_core.dropout_event:4.eer_durbin.b\endcsname{-0.969}
\expandafter\gdef\csname EERadj@adjusted_coeff.calendar4.main_core.dropout_event:5.eer_durbin.b\endcsname{-1.221}
\expandafter\gdef\csname EERadj@adjusted_coeff.calendar4.main_core.dropout_event:6.eer_durbin.b\endcsname{-1.436}
\expandafter\gdef\csname EERadj@adjusted_coeff.calendar4.main_core.dropout_event:7.eer_durbin.b\endcsname{-1.688}
\expandafter\gdef\csname EERadj@adjusted_coeff.calendar4.main_core.dropout_event:1b.eer_term.b\endcsname{0.000}
\expandafter\gdef\csname EERadj@adjusted_coeff.calendar4.main_core.dropout_event:2.eer_term.b\endcsname{-0.896}
\expandafter\gdef\csname EERadj@adjusted_coeff.calendar4.main_core.dropout_event:eer_age.b\endcsname{0.177}
\expandafter\gdef\csname EERadj@adjusted_coeff.calendar4.main_core.dropout_event:eer_age2.b\endcsname{0.016}
\expandafter\gdef\csname EERadj@adjusted_coeff.calendar4.main_core.dropout_event:1.race_ethnicity.b\endcsname{-0.172}
\expandafter\gdef\csname EERadj@adjusted_coeff.calendar4.main_core.dropout_event:2.race_ethnicity.b\endcsname{0.170}
\expandafter\gdef\csname EERadj@adjusted_coeff.calendar4.main_core.dropout_event:3.race_ethnicity.b\endcsname{0.174}
\expandafter\gdef\csname EERadj@adjusted_coeff.calendar4.main_core.dropout_event:4b.race_ethnicity.b\endcsname{0.000}
\expandafter\gdef\csname EERadj@adjusted_coeff.calendar4.main_core.dropout_event:1997b.calendar_period.b\endcsname{0.000}
\expandafter\gdef\csname EERadj@adjusted_coeff.calendar4.main_core.dropout_event:1998.calendar_period.b\endcsname{0.832}
\expandafter\gdef\csname EERadj@adjusted_coeff.calendar4.main_core.dropout_event:1999.calendar_period.b\endcsname{1.288}
\expandafter\gdef\csname EERadj@adjusted_coeff.calendar4.main_core.dropout_event:2000.calendar_period.b\endcsname{1.582}
\expandafter\gdef\csname EERadj@adjusted_coeff.calendar4.main_core.dropout_event:2001.calendar_period.b\endcsname{1.801}
\expandafter\gdef\csname EERadj@adjusted_coeff.calendar4.main_core.dropout_event:2002.calendar_period.b\endcsname{2.004}
\expandafter\gdef\csname EERadj@adjusted_coeff.calendar4.main_core.dropout_event:2003.calendar_period.b\endcsname{2.098}
\expandafter\gdef\csname EERadj@adjusted_coeff.calendar4.main_core.dropout_event:_cons.b\endcsname{-4.466}
\expandafter\gdef\csname EERadj@adjusted_coeff.calendar4.history_baseline.dropout_event:treated.b\endcsname{1.242}
\expandafter\gdef\csname EERadj@adjusted_coeff.calendar4.history_baseline.dropout_event:1b.eer_durbin.b\endcsname{0.000}
\expandafter\gdef\csname EERadj@adjusted_coeff.calendar4.history_baseline.dropout_event:2.eer_durbin.b\endcsname{-0.458}
\expandafter\gdef\csname EERadj@adjusted_coeff.calendar4.history_baseline.dropout_event:3.eer_durbin.b\endcsname{0.281}
\expandafter\gdef\csname EERadj@adjusted_coeff.calendar4.history_baseline.dropout_event:4.eer_durbin.b\endcsname{-0.054}
\expandafter\gdef\csname EERadj@adjusted_coeff.calendar4.history_baseline.dropout_event:5.eer_durbin.b\endcsname{0.052}
\expandafter\gdef\csname EERadj@adjusted_coeff.calendar4.history_baseline.dropout_event:6.eer_durbin.b\endcsname{0.125}
\expandafter\gdef\csname EERadj@adjusted_coeff.calendar4.history_baseline.dropout_event:7.eer_durbin.b\endcsname{0.099}
\expandafter\gdef\csname EERadj@adjusted_coeff.calendar4.history_baseline.dropout_event:1b.eer_term.b\endcsname{0.000}
\expandafter\gdef\csname EERadj@adjusted_coeff.calendar4.history_baseline.dropout_event:2.eer_term.b\endcsname{-0.622}
\expandafter\gdef\csname EERadj@adjusted_coeff.calendar4.history_baseline.dropout_event:eer_age.b\endcsname{0.156}
\expandafter\gdef\csname EERadj@adjusted_coeff.calendar4.history_baseline.dropout_event:eer_age2.b\endcsname{0.026}
\expandafter\gdef\csname EERadj@adjusted_coeff.calendar4.history_baseline.dropout_event:_cons.b\endcsname{-4.448}
\expandafter\gdef\csname EERadj@adjusted_coeff.calendar4.history_core.dropout_event:treated.b\endcsname{1.271}
\expandafter\gdef\csname EERadj@adjusted_coeff.calendar4.history_core.dropout_event:1b.eer_durbin.b\endcsname{0.000}
\expandafter\gdef\csname EERadj@adjusted_coeff.calendar4.history_core.dropout_event:2.eer_durbin.b\endcsname{-0.995}
\expandafter\gdef\csname EERadj@adjusted_coeff.calendar4.history_core.dropout_event:3.eer_durbin.b\endcsname{-0.800}
\expandafter\gdef\csname EERadj@adjusted_coeff.calendar4.history_core.dropout_event:4.eer_durbin.b\endcsname{-1.350}
\expandafter\gdef\csname EERadj@adjusted_coeff.calendar4.history_core.dropout_event:5.eer_durbin.b\endcsname{-1.687}
\expandafter\gdef\csname EERadj@adjusted_coeff.calendar4.history_core.dropout_event:6.eer_durbin.b\endcsname{-2.028}
\expandafter\gdef\csname EERadj@adjusted_coeff.calendar4.history_core.dropout_event:7.eer_durbin.b\endcsname{-2.457}
\expandafter\gdef\csname EERadj@adjusted_coeff.calendar4.history_core.dropout_event:1b.eer_term.b\endcsname{0.000}
\expandafter\gdef\csname EERadj@adjusted_coeff.calendar4.history_core.dropout_event:2.eer_term.b\endcsname{-0.953}
\expandafter\gdef\csname EERadj@adjusted_coeff.calendar4.history_core.dropout_event:eer_age.b\endcsname{0.168}
\expandafter\gdef\csname EERadj@adjusted_coeff.calendar4.history_core.dropout_event:eer_age2.b\endcsname{0.016}
\expandafter\gdef\csname EERadj@adjusted_coeff.calendar4.history_core.dropout_event:1.race_ethnicity.b\endcsname{-0.145}
\expandafter\gdef\csname EERadj@adjusted_coeff.calendar4.history_core.dropout_event:2.race_ethnicity.b\endcsname{0.142}
\expandafter\gdef\csname EERadj@adjusted_coeff.calendar4.history_core.dropout_event:3.race_ethnicity.b\endcsname{-0.041}
\expandafter\gdef\csname EERadj@adjusted_coeff.calendar4.history_core.dropout_event:4b.race_ethnicity.b\endcsname{0.000}
\expandafter\gdef\csname EERadj@adjusted_coeff.calendar4.history_core.dropout_event:1997b.calendar_period.b\endcsname{0.000}
\expandafter\gdef\csname EERadj@adjusted_coeff.calendar4.history_core.dropout_event:1998.calendar_period.b\endcsname{1.185}
\expandafter\gdef\csname EERadj@adjusted_coeff.calendar4.history_core.dropout_event:1999.calendar_period.b\endcsname{1.801}
\expandafter\gdef\csname EERadj@adjusted_coeff.calendar4.history_core.dropout_event:2000.calendar_period.b\endcsname{2.231}
\expandafter\gdef\csname EERadj@adjusted_coeff.calendar4.history_core.dropout_event:2001.calendar_period.b\endcsname{2.618}
\expandafter\gdef\csname EERadj@adjusted_coeff.calendar4.history_core.dropout_event:2002.calendar_period.b\endcsname{2.933}
\expandafter\gdef\csname EERadj@adjusted_coeff.calendar4.history_core.dropout_event:2003.calendar_period.b\endcsname{3.006}
\expandafter\gdef\csname EERadj@adjusted_coeff.calendar4.history_core.dropout_event:_cons.b\endcsname{-4.533}
\expandafter\gdef\csname EERadj@adjusted_coeff.calendar4.later_all_core.dropout_event:treated.b\endcsname{1.196}
\expandafter\gdef\csname EERadj@adjusted_coeff.calendar4.later_all_core.dropout_event:1b.eer_durbin.b\endcsname{0.000}
\expandafter\gdef\csname EERadj@adjusted_coeff.calendar4.later_all_core.dropout_event:2.eer_durbin.b\endcsname{-0.467}
\expandafter\gdef\csname EERadj@adjusted_coeff.calendar4.later_all_core.dropout_event:3.eer_durbin.b\endcsname{-0.255}
\expandafter\gdef\csname EERadj@adjusted_coeff.calendar4.later_all_core.dropout_event:4.eer_durbin.b\endcsname{-0.800}
\expandafter\gdef\csname EERadj@adjusted_coeff.calendar4.later_all_core.dropout_event:5.eer_durbin.b\endcsname{-0.873}
\expandafter\gdef\csname EERadj@adjusted_coeff.calendar4.later_all_core.dropout_event:6.eer_durbin.b\endcsname{-1.055}
\expandafter\gdef\csname EERadj@adjusted_coeff.calendar4.later_all_core.dropout_event:7.eer_durbin.b\endcsname{-1.255}
\expandafter\gdef\csname EERadj@adjusted_coeff.calendar4.later_all_core.dropout_event:1b.eer_term.b\endcsname{0.000}
\expandafter\gdef\csname EERadj@adjusted_coeff.calendar4.later_all_core.dropout_event:2.eer_term.b\endcsname{-0.764}
\expandafter\gdef\csname EERadj@adjusted_coeff.calendar4.later_all_core.dropout_event:eer_age.b\endcsname{0.096}
\expandafter\gdef\csname EERadj@adjusted_coeff.calendar4.later_all_core.dropout_event:eer_age2.b\endcsname{0.043}
\expandafter\gdef\csname EERadj@adjusted_coeff.calendar4.later_all_core.dropout_event:1.race_ethnicity.b\endcsname{-0.161}
\expandafter\gdef\csname EERadj@adjusted_coeff.calendar4.later_all_core.dropout_event:2.race_ethnicity.b\endcsname{0.126}
\expandafter\gdef\csname EERadj@adjusted_coeff.calendar4.later_all_core.dropout_event:3.race_ethnicity.b\endcsname{0.063}
\expandafter\gdef\csname EERadj@adjusted_coeff.calendar4.later_all_core.dropout_event:4b.race_ethnicity.b\endcsname{0.000}
\expandafter\gdef\csname EERadj@adjusted_coeff.calendar4.later_all_core.dropout_event:1998b.calendar_period.b\endcsname{0.000}
\expandafter\gdef\csname EERadj@adjusted_coeff.calendar4.later_all_core.dropout_event:1999.calendar_period.b\endcsname{0.328}
\expandafter\gdef\csname EERadj@adjusted_coeff.calendar4.later_all_core.dropout_event:2000.calendar_period.b\endcsname{0.623}
\expandafter\gdef\csname EERadj@adjusted_coeff.calendar4.later_all_core.dropout_event:2001.calendar_period.b\endcsname{0.818}
\expandafter\gdef\csname EERadj@adjusted_coeff.calendar4.later_all_core.dropout_event:2002.calendar_period.b\endcsname{0.956}
\expandafter\gdef\csname EERadj@adjusted_coeff.calendar4.later_all_core.dropout_event:2003.calendar_period.b\endcsname{1.002}
\expandafter\gdef\csname EERadj@adjusted_coeff.calendar4.later_all_core.dropout_event:_cons.b\endcsname{-3.847}
\expandafter\gdef\csname EERadj@adjusted_coeff.calendar4.later_cc_baseline.dropout_event:treated.b\endcsname{1.155}
\expandafter\gdef\csname EERadj@adjusted_coeff.calendar4.later_cc_baseline.dropout_event:1b.eer_durbin.b\endcsname{0.000}
\expandafter\gdef\csname EERadj@adjusted_coeff.calendar4.later_cc_baseline.dropout_event:2.eer_durbin.b\endcsname{-0.688}
\expandafter\gdef\csname EERadj@adjusted_coeff.calendar4.later_cc_baseline.dropout_event:3.eer_durbin.b\endcsname{-1.050}
\expandafter\gdef\csname EERadj@adjusted_coeff.calendar4.later_cc_baseline.dropout_event:4.eer_durbin.b\endcsname{-1.437}
\expandafter\gdef\csname EERadj@adjusted_coeff.calendar4.later_cc_baseline.dropout_event:5.eer_durbin.b\endcsname{-1.229}
\expandafter\gdef\csname EERadj@adjusted_coeff.calendar4.later_cc_baseline.dropout_event:6.eer_durbin.b\endcsname{-1.143}
\expandafter\gdef\csname EERadj@adjusted_coeff.calendar4.later_cc_baseline.dropout_event:7.eer_durbin.b\endcsname{-1.134}
\expandafter\gdef\csname EERadj@adjusted_coeff.calendar4.later_cc_baseline.dropout_event:1b.eer_term.b\endcsname{0.000}
\expandafter\gdef\csname EERadj@adjusted_coeff.calendar4.later_cc_baseline.dropout_event:2.eer_term.b\endcsname{-0.541}
\expandafter\gdef\csname EERadj@adjusted_coeff.calendar4.later_cc_baseline.dropout_event:eer_age.b\endcsname{-0.034}
\expandafter\gdef\csname EERadj@adjusted_coeff.calendar4.later_cc_baseline.dropout_event:eer_age2.b\endcsname{0.086}
\expandafter\gdef\csname EERadj@adjusted_coeff.calendar4.later_cc_baseline.dropout_event:_cons.b\endcsname{-3.227}
\expandafter\gdef\csname EERadj@adjusted_coeff.calendar4.later_cc_core.dropout_event:treated.b\endcsname{1.187}
\expandafter\gdef\csname EERadj@adjusted_coeff.calendar4.later_cc_core.dropout_event:1b.eer_durbin.b\endcsname{0.000}
\expandafter\gdef\csname EERadj@adjusted_coeff.calendar4.later_cc_core.dropout_event:2.eer_durbin.b\endcsname{-0.555}
\expandafter\gdef\csname EERadj@adjusted_coeff.calendar4.later_cc_core.dropout_event:3.eer_durbin.b\endcsname{-0.882}
\expandafter\gdef\csname EERadj@adjusted_coeff.calendar4.later_cc_core.dropout_event:4.eer_durbin.b\endcsname{-1.391}
\expandafter\gdef\csname EERadj@adjusted_coeff.calendar4.later_cc_core.dropout_event:5.eer_durbin.b\endcsname{-1.536}
\expandafter\gdef\csname EERadj@adjusted_coeff.calendar4.later_cc_core.dropout_event:6.eer_durbin.b\endcsname{-1.867}
\expandafter\gdef\csname EERadj@adjusted_coeff.calendar4.later_cc_core.dropout_event:7.eer_durbin.b\endcsname{-2.187}
\expandafter\gdef\csname EERadj@adjusted_coeff.calendar4.later_cc_core.dropout_event:1b.eer_term.b\endcsname{0.000}
\expandafter\gdef\csname EERadj@adjusted_coeff.calendar4.later_cc_core.dropout_event:2.eer_term.b\endcsname{-0.796}
\expandafter\gdef\csname EERadj@adjusted_coeff.calendar4.later_cc_core.dropout_event:eer_age.b\endcsname{-0.029}
\expandafter\gdef\csname EERadj@adjusted_coeff.calendar4.later_cc_core.dropout_event:eer_age2.b\endcsname{0.087}
\expandafter\gdef\csname EERadj@adjusted_coeff.calendar4.later_cc_core.dropout_event:1.race_ethnicity.b\endcsname{-0.240}
\expandafter\gdef\csname EERadj@adjusted_coeff.calendar4.later_cc_core.dropout_event:2.race_ethnicity.b\endcsname{0.146}
\expandafter\gdef\csname EERadj@adjusted_coeff.calendar4.later_cc_core.dropout_event:3.race_ethnicity.b\endcsname{-0.170}
\expandafter\gdef\csname EERadj@adjusted_coeff.calendar4.later_cc_core.dropout_event:4b.race_ethnicity.b\endcsname{0.000}
\expandafter\gdef\csname EERadj@adjusted_coeff.calendar4.later_cc_core.dropout_event:1998b.calendar_period.b\endcsname{0.000}
\expandafter\gdef\csname EERadj@adjusted_coeff.calendar4.later_cc_core.dropout_event:1999.calendar_period.b\endcsname{0.480}
\expandafter\gdef\csname EERadj@adjusted_coeff.calendar4.later_cc_core.dropout_event:2000.calendar_period.b\endcsname{0.831}
\expandafter\gdef\csname EERadj@adjusted_coeff.calendar4.later_cc_core.dropout_event:2001.calendar_period.b\endcsname{1.323}
\expandafter\gdef\csname EERadj@adjusted_coeff.calendar4.later_cc_core.dropout_event:2002.calendar_period.b\endcsname{1.411}
\expandafter\gdef\csname EERadj@adjusted_coeff.calendar4.later_cc_core.dropout_event:2003.calendar_period.b\endcsname{1.365}
\expandafter\gdef\csname EERadj@adjusted_coeff.calendar4.later_cc_core.dropout_event:_cons.b\endcsname{-3.402}
\expandafter\gdef\csname EERadj@adjusted_coeff.calendar4.later_cc_rich.dropout_event:treated.b\endcsname{0.903}
\expandafter\gdef\csname EERadj@adjusted_coeff.calendar4.later_cc_rich.dropout_event:1b.eer_durbin.b\endcsname{0.000}
\expandafter\gdef\csname EERadj@adjusted_coeff.calendar4.later_cc_rich.dropout_event:2.eer_durbin.b\endcsname{-0.196}
\expandafter\gdef\csname EERadj@adjusted_coeff.calendar4.later_cc_rich.dropout_event:3.eer_durbin.b\endcsname{-0.472}
\expandafter\gdef\csname EERadj@adjusted_coeff.calendar4.later_cc_rich.dropout_event:4.eer_durbin.b\endcsname{-1.002}
\expandafter\gdef\csname EERadj@adjusted_coeff.calendar4.later_cc_rich.dropout_event:5.eer_durbin.b\endcsname{-1.091}
\expandafter\gdef\csname EERadj@adjusted_coeff.calendar4.later_cc_rich.dropout_event:6.eer_durbin.b\endcsname{-1.326}
\expandafter\gdef\csname EERadj@adjusted_coeff.calendar4.later_cc_rich.dropout_event:7.eer_durbin.b\endcsname{-1.625}
\expandafter\gdef\csname EERadj@adjusted_coeff.calendar4.later_cc_rich.dropout_event:1b.eer_term.b\endcsname{0.000}
\expandafter\gdef\csname EERadj@adjusted_coeff.calendar4.later_cc_rich.dropout_event:2.eer_term.b\endcsname{-0.715}
\expandafter\gdef\csname EERadj@adjusted_coeff.calendar4.later_cc_rich.dropout_event:eer_age.b\endcsname{0.037}
\expandafter\gdef\csname EERadj@adjusted_coeff.calendar4.later_cc_rich.dropout_event:eer_age2.b\endcsname{0.043}
\expandafter\gdef\csname EERadj@adjusted_coeff.calendar4.later_cc_rich.dropout_event:1.race_ethnicity.b\endcsname{-0.778}
\expandafter\gdef\csname EERadj@adjusted_coeff.calendar4.later_cc_rich.dropout_event:2.race_ethnicity.b\endcsname{-0.255}
\expandafter\gdef\csname EERadj@adjusted_coeff.calendar4.later_cc_rich.dropout_event:3.race_ethnicity.b\endcsname{-0.194}
\expandafter\gdef\csname EERadj@adjusted_coeff.calendar4.later_cc_rich.dropout_event:4b.race_ethnicity.b\endcsname{0.000}
\expandafter\gdef\csname EERadj@adjusted_coeff.calendar4.later_cc_rich.dropout_event:1998b.calendar_period.b\endcsname{0.000}
\expandafter\gdef\csname EERadj@adjusted_coeff.calendar4.later_cc_rich.dropout_event:1999.calendar_period.b\endcsname{0.432}
\expandafter\gdef\csname EERadj@adjusted_coeff.calendar4.later_cc_rich.dropout_event:2000.calendar_period.b\endcsname{0.735}
\expandafter\gdef\csname EERadj@adjusted_coeff.calendar4.later_cc_rich.dropout_event:2001.calendar_period.b\endcsname{1.197}
\expandafter\gdef\csname EERadj@adjusted_coeff.calendar4.later_cc_rich.dropout_event:2002.calendar_period.b\endcsname{1.270}
\expandafter\gdef\csname EERadj@adjusted_coeff.calendar4.later_cc_rich.dropout_event:2003.calendar_period.b\endcsname{1.134}
\expandafter\gdef\csname EERadj@adjusted_coeff.calendar4.later_cc_rich.dropout_event:mother_grade97.b\endcsname{-0.039}
\expandafter\gdef\csname EERadj@adjusted_coeff.calendar4.later_cc_rich.dropout_event:1b.household_group.b\endcsname{0.000}
\expandafter\gdef\csname EERadj@adjusted_coeff.calendar4.later_cc_rich.dropout_event:2.household_group.b\endcsname{0.326}
\expandafter\gdef\csname EERadj@adjusted_coeff.calendar4.later_cc_rich.dropout_event:3.household_group.b\endcsname{0.534}
\expandafter\gdef\csname EERadj@adjusted_coeff.calendar4.later_cc_rich.dropout_event:asvab10.b\endcsname{-0.193}
\expandafter\gdef\csname EERadj@adjusted_coeff.calendar4.later_cc_rich.dropout_event:_cons.b\endcsname{-2.640}
\expandafter\gdef\csname EERadj@adjusted_coeff.school3.main_baseline.dropout_event:treated.se\endcsname{0.081}
\expandafter\gdef\csname EERadj@adjusted_coeff.school3.main_baseline.dropout_event:1b.eer_durbin.se\endcsname{0.000}
\expandafter\gdef\csname EERadj@adjusted_coeff.school3.main_baseline.dropout_event:2.eer_durbin.se\endcsname{0.192}
\expandafter\gdef\csname EERadj@adjusted_coeff.school3.main_baseline.dropout_event:3.eer_durbin.se\endcsname{0.172}
\expandafter\gdef\csname EERadj@adjusted_coeff.school3.main_baseline.dropout_event:4.eer_durbin.se\endcsname{0.168}
\expandafter\gdef\csname EERadj@adjusted_coeff.school3.main_baseline.dropout_event:5.eer_durbin.se\endcsname{0.154}
\expandafter\gdef\csname EERadj@adjusted_coeff.school3.main_baseline.dropout_event:6.eer_durbin.se\endcsname{0.164}
\expandafter\gdef\csname EERadj@adjusted_coeff.school3.main_baseline.dropout_event:7.eer_durbin.se\endcsname{0.170}
\expandafter\gdef\csname EERadj@adjusted_coeff.school3.main_baseline.dropout_event:1b.eer_term.se\endcsname{0.000}
\expandafter\gdef\csname EERadj@adjusted_coeff.school3.main_baseline.dropout_event:2.eer_term.se\endcsname{0.078}
\expandafter\gdef\csname EERadj@adjusted_coeff.school3.main_baseline.dropout_event:eer_age.se\endcsname{0.060}
\expandafter\gdef\csname EERadj@adjusted_coeff.school3.main_baseline.dropout_event:eer_age2.se\endcsname{0.019}
\expandafter\gdef\csname EERadj@adjusted_coeff.school3.main_baseline.dropout_event:_cons.se\endcsname{0.131}
\expandafter\gdef\csname EERadj@adjusted_coeff.school3.main_core.dropout_event:treated.se\endcsname{0.085}
\expandafter\gdef\csname EERadj@adjusted_coeff.school3.main_core.dropout_event:1b.eer_durbin.se\endcsname{0.000}
\expandafter\gdef\csname EERadj@adjusted_coeff.school3.main_core.dropout_event:2.eer_durbin.se\endcsname{0.229}
\expandafter\gdef\csname EERadj@adjusted_coeff.school3.main_core.dropout_event:3.eer_durbin.se\endcsname{0.252}
\expandafter\gdef\csname EERadj@adjusted_coeff.school3.main_core.dropout_event:4.eer_durbin.se\endcsname{0.262}
\expandafter\gdef\csname EERadj@adjusted_coeff.school3.main_core.dropout_event:5.eer_durbin.se\endcsname{0.290}
\expandafter\gdef\csname EERadj@adjusted_coeff.school3.main_core.dropout_event:6.eer_durbin.se\endcsname{0.336}
\expandafter\gdef\csname EERadj@adjusted_coeff.school3.main_core.dropout_event:7.eer_durbin.se\endcsname{0.396}
\expandafter\gdef\csname EERadj@adjusted_coeff.school3.main_core.dropout_event:1b.eer_term.se\endcsname{0.000}
\expandafter\gdef\csname EERadj@adjusted_coeff.school3.main_core.dropout_event:2.eer_term.se\endcsname{0.096}
\expandafter\gdef\csname EERadj@adjusted_coeff.school3.main_core.dropout_event:eer_age.se\endcsname{0.062}
\expandafter\gdef\csname EERadj@adjusted_coeff.school3.main_core.dropout_event:eer_age2.se\endcsname{0.020}
\expandafter\gdef\csname EERadj@adjusted_coeff.school3.main_core.dropout_event:1.race_ethnicity.se\endcsname{0.092}
\expandafter\gdef\csname EERadj@adjusted_coeff.school3.main_core.dropout_event:2.race_ethnicity.se\endcsname{0.099}
\expandafter\gdef\csname EERadj@adjusted_coeff.school3.main_core.dropout_event:3.race_ethnicity.se\endcsname{0.293}
\expandafter\gdef\csname EERadj@adjusted_coeff.school3.main_core.dropout_event:4b.race_ethnicity.se\endcsname{0.000}
\expandafter\gdef\csname EERadj@adjusted_coeff.school3.main_core.dropout_event:1997b.calendar_period.se\endcsname{0.000}
\expandafter\gdef\csname EERadj@adjusted_coeff.school3.main_core.dropout_event:1998.calendar_period.se\endcsname{0.236}
\expandafter\gdef\csname EERadj@adjusted_coeff.school3.main_core.dropout_event:1999.calendar_period.se\endcsname{0.286}
\expandafter\gdef\csname EERadj@adjusted_coeff.school3.main_core.dropout_event:2000.calendar_period.se\endcsname{0.331}
\expandafter\gdef\csname EERadj@adjusted_coeff.school3.main_core.dropout_event:2001.calendar_period.se\endcsname{0.388}
\expandafter\gdef\csname EERadj@adjusted_coeff.school3.main_core.dropout_event:2002.calendar_period.se\endcsname{0.435}
\expandafter\gdef\csname EERadj@adjusted_coeff.school3.main_core.dropout_event:2003.calendar_period.se\endcsname{0.496}
\expandafter\gdef\csname EERadj@adjusted_coeff.school3.main_core.dropout_event:_cons.se\endcsname{0.149}
\expandafter\gdef\csname EERadj@adjusted_coeff.school3.history_baseline.dropout_event:treated.se\endcsname{0.112}
\expandafter\gdef\csname EERadj@adjusted_coeff.school3.history_baseline.dropout_event:1b.eer_durbin.se\endcsname{0.000}
\expandafter\gdef\csname EERadj@adjusted_coeff.school3.history_baseline.dropout_event:2.eer_durbin.se\endcsname{0.239}
\expandafter\gdef\csname EERadj@adjusted_coeff.school3.history_baseline.dropout_event:3.eer_durbin.se\endcsname{0.221}
\expandafter\gdef\csname EERadj@adjusted_coeff.school3.history_baseline.dropout_event:4.eer_durbin.se\endcsname{0.214}
\expandafter\gdef\csname EERadj@adjusted_coeff.school3.history_baseline.dropout_event:5.eer_durbin.se\endcsname{0.199}
\expandafter\gdef\csname EERadj@adjusted_coeff.school3.history_baseline.dropout_event:6.eer_durbin.se\endcsname{0.208}
\expandafter\gdef\csname EERadj@adjusted_coeff.school3.history_baseline.dropout_event:7.eer_durbin.se\endcsname{0.216}
\expandafter\gdef\csname EERadj@adjusted_coeff.school3.history_baseline.dropout_event:1b.eer_term.se\endcsname{0.000}
\expandafter\gdef\csname EERadj@adjusted_coeff.school3.history_baseline.dropout_event:2.eer_term.se\endcsname{0.088}
\expandafter\gdef\csname EERadj@adjusted_coeff.school3.history_baseline.dropout_event:eer_age.se\endcsname{0.067}
\expandafter\gdef\csname EERadj@adjusted_coeff.school3.history_baseline.dropout_event:eer_age2.se\endcsname{0.022}
\expandafter\gdef\csname EERadj@adjusted_coeff.school3.history_baseline.dropout_event:_cons.se\endcsname{0.169}
\expandafter\gdef\csname EERadj@adjusted_coeff.school3.history_core.dropout_event:treated.se\endcsname{0.116}
\expandafter\gdef\csname EERadj@adjusted_coeff.school3.history_core.dropout_event:1b.eer_durbin.se\endcsname{0.000}
\expandafter\gdef\csname EERadj@adjusted_coeff.school3.history_core.dropout_event:2.eer_durbin.se\endcsname{0.290}
\expandafter\gdef\csname EERadj@adjusted_coeff.school3.history_core.dropout_event:3.eer_durbin.se\endcsname{0.318}
\expandafter\gdef\csname EERadj@adjusted_coeff.school3.history_core.dropout_event:4.eer_durbin.se\endcsname{0.335}
\expandafter\gdef\csname EERadj@adjusted_coeff.school3.history_core.dropout_event:5.eer_durbin.se\endcsname{0.371}
\expandafter\gdef\csname EERadj@adjusted_coeff.school3.history_core.dropout_event:6.eer_durbin.se\endcsname{0.420}
\expandafter\gdef\csname EERadj@adjusted_coeff.school3.history_core.dropout_event:7.eer_durbin.se\endcsname{0.488}
\expandafter\gdef\csname EERadj@adjusted_coeff.school3.history_core.dropout_event:1b.eer_term.se\endcsname{0.000}
\expandafter\gdef\csname EERadj@adjusted_coeff.school3.history_core.dropout_event:2.eer_term.se\endcsname{0.105}
\expandafter\gdef\csname EERadj@adjusted_coeff.school3.history_core.dropout_event:eer_age.se\endcsname{0.067}
\expandafter\gdef\csname EERadj@adjusted_coeff.school3.history_core.dropout_event:eer_age2.se\endcsname{0.023}
\expandafter\gdef\csname EERadj@adjusted_coeff.school3.history_core.dropout_event:1.race_ethnicity.se\endcsname{0.102}
\expandafter\gdef\csname EERadj@adjusted_coeff.school3.history_core.dropout_event:2.race_ethnicity.se\endcsname{0.110}
\expandafter\gdef\csname EERadj@adjusted_coeff.school3.history_core.dropout_event:3.race_ethnicity.se\endcsname{0.350}
\expandafter\gdef\csname EERadj@adjusted_coeff.school3.history_core.dropout_event:4b.race_ethnicity.se\endcsname{0.000}
\expandafter\gdef\csname EERadj@adjusted_coeff.school3.history_core.dropout_event:1997b.calendar_period.se\endcsname{0.000}
\expandafter\gdef\csname EERadj@adjusted_coeff.school3.history_core.dropout_event:1998.calendar_period.se\endcsname{0.293}
\expandafter\gdef\csname EERadj@adjusted_coeff.school3.history_core.dropout_event:1999.calendar_period.se\endcsname{0.359}
\expandafter\gdef\csname EERadj@adjusted_coeff.school3.history_core.dropout_event:2000.calendar_period.se\endcsname{0.403}
\expandafter\gdef\csname EERadj@adjusted_coeff.school3.history_core.dropout_event:2001.calendar_period.se\endcsname{0.458}
\expandafter\gdef\csname EERadj@adjusted_coeff.school3.history_core.dropout_event:2002.calendar_period.se\endcsname{0.518}
\expandafter\gdef\csname EERadj@adjusted_coeff.school3.history_core.dropout_event:2003.calendar_period.se\endcsname{0.576}
\expandafter\gdef\csname EERadj@adjusted_coeff.school3.history_core.dropout_event:_cons.se\endcsname{0.186}
\expandafter\gdef\csname EERadj@adjusted_coeff.school3.later_all_core.dropout_event:treated.se\endcsname{0.111}
\expandafter\gdef\csname EERadj@adjusted_coeff.school3.later_all_core.dropout_event:1b.eer_durbin.se\endcsname{0.000}
\expandafter\gdef\csname EERadj@adjusted_coeff.school3.later_all_core.dropout_event:2.eer_durbin.se\endcsname{0.711}
\expandafter\gdef\csname EERadj@adjusted_coeff.school3.later_all_core.dropout_event:3.eer_durbin.se\endcsname{0.544}
\expandafter\gdef\csname EERadj@adjusted_coeff.school3.later_all_core.dropout_event:4.eer_durbin.se\endcsname{0.536}
\expandafter\gdef\csname EERadj@adjusted_coeff.school3.later_all_core.dropout_event:5.eer_durbin.se\endcsname{0.519}
\expandafter\gdef\csname EERadj@adjusted_coeff.school3.later_all_core.dropout_event:6.eer_durbin.se\endcsname{0.512}
\expandafter\gdef\csname EERadj@adjusted_coeff.school3.later_all_core.dropout_event:7.eer_durbin.se\endcsname{0.522}
\expandafter\gdef\csname EERadj@adjusted_coeff.school3.later_all_core.dropout_event:1b.eer_term.se\endcsname{0.000}
\expandafter\gdef\csname EERadj@adjusted_coeff.school3.later_all_core.dropout_event:2.eer_term.se\endcsname{0.118}
\expandafter\gdef\csname EERadj@adjusted_coeff.school3.later_all_core.dropout_event:eer_age.se\endcsname{0.084}
\expandafter\gdef\csname EERadj@adjusted_coeff.school3.later_all_core.dropout_event:eer_age2.se\endcsname{0.026}
\expandafter\gdef\csname EERadj@adjusted_coeff.school3.later_all_core.dropout_event:1.race_ethnicity.se\endcsname{0.108}
\expandafter\gdef\csname EERadj@adjusted_coeff.school3.later_all_core.dropout_event:2.race_ethnicity.se\endcsname{0.115}
\expandafter\gdef\csname EERadj@adjusted_coeff.school3.later_all_core.dropout_event:3.race_ethnicity.se\endcsname{0.424}
\expandafter\gdef\csname EERadj@adjusted_coeff.school3.later_all_core.dropout_event:4b.race_ethnicity.se\endcsname{0.000}
\expandafter\gdef\csname EERadj@adjusted_coeff.school3.later_all_core.dropout_event:1998b.calendar_period.se\endcsname{0.000}
\expandafter\gdef\csname EERadj@adjusted_coeff.school3.later_all_core.dropout_event:1999.calendar_period.se\endcsname{0.221}
\expandafter\gdef\csname EERadj@adjusted_coeff.school3.later_all_core.dropout_event:2000.calendar_period.se\endcsname{0.288}
\expandafter\gdef\csname EERadj@adjusted_coeff.school3.later_all_core.dropout_event:2001.calendar_period.se\endcsname{0.361}
\expandafter\gdef\csname EERadj@adjusted_coeff.school3.later_all_core.dropout_event:2002.calendar_period.se\endcsname{0.426}
\expandafter\gdef\csname EERadj@adjusted_coeff.school3.later_all_core.dropout_event:2003.calendar_period.se\endcsname{0.493}
\expandafter\gdef\csname EERadj@adjusted_coeff.school3.later_all_core.dropout_event:_cons.se\endcsname{0.533}
\expandafter\gdef\csname EERadj@adjusted_coeff.school3.later_cc_baseline.dropout_event:treated.se\endcsname{0.136}
\expandafter\gdef\csname EERadj@adjusted_coeff.school3.later_cc_baseline.dropout_event:1b.eer_durbin.se\endcsname{0.000}
\expandafter\gdef\csname EERadj@adjusted_coeff.school3.later_cc_baseline.dropout_event:2.eer_durbin.se\endcsname{0.922}
\expandafter\gdef\csname EERadj@adjusted_coeff.school3.later_cc_baseline.dropout_event:3.eer_durbin.se\endcsname{0.674}
\expandafter\gdef\csname EERadj@adjusted_coeff.school3.later_cc_baseline.dropout_event:4.eer_durbin.se\endcsname{0.673}
\expandafter\gdef\csname EERadj@adjusted_coeff.school3.later_cc_baseline.dropout_event:5.eer_durbin.se\endcsname{0.664}
\expandafter\gdef\csname EERadj@adjusted_coeff.school3.later_cc_baseline.dropout_event:6.eer_durbin.se\endcsname{0.668}
\expandafter\gdef\csname EERadj@adjusted_coeff.school3.later_cc_baseline.dropout_event:7.eer_durbin.se\endcsname{0.674}
\expandafter\gdef\csname EERadj@adjusted_coeff.school3.later_cc_baseline.dropout_event:1b.eer_term.se\endcsname{0.000}
\expandafter\gdef\csname EERadj@adjusted_coeff.school3.later_cc_baseline.dropout_event:2.eer_term.se\endcsname{0.118}
\expandafter\gdef\csname EERadj@adjusted_coeff.school3.later_cc_baseline.dropout_event:eer_age.se\endcsname{0.094}
\expandafter\gdef\csname EERadj@adjusted_coeff.school3.later_cc_baseline.dropout_event:eer_age2.se\endcsname{0.028}
\expandafter\gdef\csname EERadj@adjusted_coeff.school3.later_cc_baseline.dropout_event:_cons.se\endcsname{0.645}
\expandafter\gdef\csname EERadj@adjusted_coeff.school3.later_cc_core.dropout_event:treated.se\endcsname{0.145}
\expandafter\gdef\csname EERadj@adjusted_coeff.school3.later_cc_core.dropout_event:1b.eer_durbin.se\endcsname{0.000}
\expandafter\gdef\csname EERadj@adjusted_coeff.school3.later_cc_core.dropout_event:2.eer_durbin.se\endcsname{0.904}
\expandafter\gdef\csname EERadj@adjusted_coeff.school3.later_cc_core.dropout_event:3.eer_durbin.se\endcsname{0.674}
\expandafter\gdef\csname EERadj@adjusted_coeff.school3.later_cc_core.dropout_event:4.eer_durbin.se\endcsname{0.710}
\expandafter\gdef\csname EERadj@adjusted_coeff.school3.later_cc_core.dropout_event:5.eer_durbin.se\endcsname{0.760}
\expandafter\gdef\csname EERadj@adjusted_coeff.school3.later_cc_core.dropout_event:6.eer_durbin.se\endcsname{0.833}
\expandafter\gdef\csname EERadj@adjusted_coeff.school3.later_cc_core.dropout_event:7.eer_durbin.se\endcsname{0.925}
\expandafter\gdef\csname EERadj@adjusted_coeff.school3.later_cc_core.dropout_event:1b.eer_term.se\endcsname{0.000}
\expandafter\gdef\csname EERadj@adjusted_coeff.school3.later_cc_core.dropout_event:2.eer_term.se\endcsname{0.157}
\expandafter\gdef\csname EERadj@adjusted_coeff.school3.later_cc_core.dropout_event:eer_age.se\endcsname{0.097}
\expandafter\gdef\csname EERadj@adjusted_coeff.school3.later_cc_core.dropout_event:eer_age2.se\endcsname{0.030}
\expandafter\gdef\csname EERadj@adjusted_coeff.school3.later_cc_core.dropout_event:1.race_ethnicity.se\endcsname{0.139}
\expandafter\gdef\csname EERadj@adjusted_coeff.school3.later_cc_core.dropout_event:2.race_ethnicity.se\endcsname{0.145}
\expandafter\gdef\csname EERadj@adjusted_coeff.school3.later_cc_core.dropout_event:3.race_ethnicity.se\endcsname{0.571}
\expandafter\gdef\csname EERadj@adjusted_coeff.school3.later_cc_core.dropout_event:4b.race_ethnicity.se\endcsname{0.000}
\expandafter\gdef\csname EERadj@adjusted_coeff.school3.later_cc_core.dropout_event:1998b.calendar_period.se\endcsname{0.000}
\expandafter\gdef\csname EERadj@adjusted_coeff.school3.later_cc_core.dropout_event:1999.calendar_period.se\endcsname{0.296}
\expandafter\gdef\csname EERadj@adjusted_coeff.school3.later_cc_core.dropout_event:2000.calendar_period.se\endcsname{0.394}
\expandafter\gdef\csname EERadj@adjusted_coeff.school3.later_cc_core.dropout_event:2001.calendar_period.se\endcsname{0.521}
\expandafter\gdef\csname EERadj@adjusted_coeff.school3.later_cc_core.dropout_event:2002.calendar_period.se\endcsname{0.611}
\expandafter\gdef\csname EERadj@adjusted_coeff.school3.later_cc_core.dropout_event:2003.calendar_period.se\endcsname{0.694}
\expandafter\gdef\csname EERadj@adjusted_coeff.school3.later_cc_core.dropout_event:_cons.se\endcsname{0.648}
\expandafter\gdef\csname EERadj@adjusted_coeff.school3.later_cc_rich.dropout_event:treated.se\endcsname{0.150}
\expandafter\gdef\csname EERadj@adjusted_coeff.school3.later_cc_rich.dropout_event:1b.eer_durbin.se\endcsname{0.000}
\expandafter\gdef\csname EERadj@adjusted_coeff.school3.later_cc_rich.dropout_event:2.eer_durbin.se\endcsname{0.853}
\expandafter\gdef\csname EERadj@adjusted_coeff.school3.later_cc_rich.dropout_event:3.eer_durbin.se\endcsname{0.600}
\expandafter\gdef\csname EERadj@adjusted_coeff.school3.later_cc_rich.dropout_event:4.eer_durbin.se\endcsname{0.637}
\expandafter\gdef\csname EERadj@adjusted_coeff.school3.later_cc_rich.dropout_event:5.eer_durbin.se\endcsname{0.698}
\expandafter\gdef\csname EERadj@adjusted_coeff.school3.later_cc_rich.dropout_event:6.eer_durbin.se\endcsname{0.785}
\expandafter\gdef\csname EERadj@adjusted_coeff.school3.later_cc_rich.dropout_event:7.eer_durbin.se\endcsname{0.895}
\expandafter\gdef\csname EERadj@adjusted_coeff.school3.later_cc_rich.dropout_event:1b.eer_term.se\endcsname{0.000}
\expandafter\gdef\csname EERadj@adjusted_coeff.school3.later_cc_rich.dropout_event:2.eer_term.se\endcsname{0.157}
\expandafter\gdef\csname EERadj@adjusted_coeff.school3.later_cc_rich.dropout_event:eer_age.se\endcsname{0.100}
\expandafter\gdef\csname EERadj@adjusted_coeff.school3.later_cc_rich.dropout_event:eer_age2.se\endcsname{0.032}
\expandafter\gdef\csname EERadj@adjusted_coeff.school3.later_cc_rich.dropout_event:1.race_ethnicity.se\endcsname{0.146}
\expandafter\gdef\csname EERadj@adjusted_coeff.school3.later_cc_rich.dropout_event:2.race_ethnicity.se\endcsname{0.159}
\expandafter\gdef\csname EERadj@adjusted_coeff.school3.later_cc_rich.dropout_event:3.race_ethnicity.se\endcsname{0.615}
\expandafter\gdef\csname EERadj@adjusted_coeff.school3.later_cc_rich.dropout_event:4b.race_ethnicity.se\endcsname{0.000}
\expandafter\gdef\csname EERadj@adjusted_coeff.school3.later_cc_rich.dropout_event:1998b.calendar_period.se\endcsname{0.000}
\expandafter\gdef\csname EERadj@adjusted_coeff.school3.later_cc_rich.dropout_event:1999.calendar_period.se\endcsname{0.302}
\expandafter\gdef\csname EERadj@adjusted_coeff.school3.later_cc_rich.dropout_event:2000.calendar_period.se\endcsname{0.402}
\expandafter\gdef\csname EERadj@adjusted_coeff.school3.later_cc_rich.dropout_event:2001.calendar_period.se\endcsname{0.533}
\expandafter\gdef\csname EERadj@adjusted_coeff.school3.later_cc_rich.dropout_event:2002.calendar_period.se\endcsname{0.632}
\expandafter\gdef\csname EERadj@adjusted_coeff.school3.later_cc_rich.dropout_event:2003.calendar_period.se\endcsname{0.715}
\expandafter\gdef\csname EERadj@adjusted_coeff.school3.later_cc_rich.dropout_event:mother_grade97.se\endcsname{0.027}
\expandafter\gdef\csname EERadj@adjusted_coeff.school3.later_cc_rich.dropout_event:1b.household_group.se\endcsname{0.000}
\expandafter\gdef\csname EERadj@adjusted_coeff.school3.later_cc_rich.dropout_event:2.household_group.se\endcsname{0.175}
\expandafter\gdef\csname EERadj@adjusted_coeff.school3.later_cc_rich.dropout_event:3.household_group.se\endcsname{0.134}
\expandafter\gdef\csname EERadj@adjusted_coeff.school3.later_cc_rich.dropout_event:asvab10.se\endcsname{0.026}
\expandafter\gdef\csname EERadj@adjusted_coeff.school3.later_cc_rich.dropout_event:_cons.se\endcsname{0.633}
\expandafter\gdef\csname EERadj@adjusted_coeff.calendar4.main_baseline.dropout_event:treated.se\endcsname{0.080}
\expandafter\gdef\csname EERadj@adjusted_coeff.calendar4.main_baseline.dropout_event:1b.eer_durbin.se\endcsname{0.000}
\expandafter\gdef\csname EERadj@adjusted_coeff.calendar4.main_baseline.dropout_event:2.eer_durbin.se\endcsname{0.191}
\expandafter\gdef\csname EERadj@adjusted_coeff.calendar4.main_baseline.dropout_event:3.eer_durbin.se\endcsname{0.166}
\expandafter\gdef\csname EERadj@adjusted_coeff.calendar4.main_baseline.dropout_event:4.eer_durbin.se\endcsname{0.167}
\expandafter\gdef\csname EERadj@adjusted_coeff.calendar4.main_baseline.dropout_event:5.eer_durbin.se\endcsname{0.152}
\expandafter\gdef\csname EERadj@adjusted_coeff.calendar4.main_baseline.dropout_event:6.eer_durbin.se\endcsname{0.161}
\expandafter\gdef\csname EERadj@adjusted_coeff.calendar4.main_baseline.dropout_event:7.eer_durbin.se\endcsname{0.168}
\expandafter\gdef\csname EERadj@adjusted_coeff.calendar4.main_baseline.dropout_event:1b.eer_term.se\endcsname{0.000}
\expandafter\gdef\csname EERadj@adjusted_coeff.calendar4.main_baseline.dropout_event:2.eer_term.se\endcsname{0.078}
\expandafter\gdef\csname EERadj@adjusted_coeff.calendar4.main_baseline.dropout_event:eer_age.se\endcsname{0.057}
\expandafter\gdef\csname EERadj@adjusted_coeff.calendar4.main_baseline.dropout_event:eer_age2.se\endcsname{0.019}
\expandafter\gdef\csname EERadj@adjusted_coeff.calendar4.main_baseline.dropout_event:_cons.se\endcsname{0.128}
\expandafter\gdef\csname EERadj@adjusted_coeff.calendar4.main_core.dropout_event:treated.se\endcsname{0.084}
\expandafter\gdef\csname EERadj@adjusted_coeff.calendar4.main_core.dropout_event:1b.eer_durbin.se\endcsname{0.000}
\expandafter\gdef\csname EERadj@adjusted_coeff.calendar4.main_core.dropout_event:2.eer_durbin.se\endcsname{0.224}
\expandafter\gdef\csname EERadj@adjusted_coeff.calendar4.main_core.dropout_event:3.eer_durbin.se\endcsname{0.238}
\expandafter\gdef\csname EERadj@adjusted_coeff.calendar4.main_core.dropout_event:4.eer_durbin.se\endcsname{0.250}
\expandafter\gdef\csname EERadj@adjusted_coeff.calendar4.main_core.dropout_event:5.eer_durbin.se\endcsname{0.280}
\expandafter\gdef\csname EERadj@adjusted_coeff.calendar4.main_core.dropout_event:6.eer_durbin.se\endcsname{0.327}
\expandafter\gdef\csname EERadj@adjusted_coeff.calendar4.main_core.dropout_event:7.eer_durbin.se\endcsname{0.386}
\expandafter\gdef\csname EERadj@adjusted_coeff.calendar4.main_core.dropout_event:1b.eer_term.se\endcsname{0.000}
\expandafter\gdef\csname EERadj@adjusted_coeff.calendar4.main_core.dropout_event:2.eer_term.se\endcsname{0.096}
\expandafter\gdef\csname EERadj@adjusted_coeff.calendar4.main_core.dropout_event:eer_age.se\endcsname{0.058}
\expandafter\gdef\csname EERadj@adjusted_coeff.calendar4.main_core.dropout_event:eer_age2.se\endcsname{0.020}
\expandafter\gdef\csname EERadj@adjusted_coeff.calendar4.main_core.dropout_event:1.race_ethnicity.se\endcsname{0.090}
\expandafter\gdef\csname EERadj@adjusted_coeff.calendar4.main_core.dropout_event:2.race_ethnicity.se\endcsname{0.098}
\expandafter\gdef\csname EERadj@adjusted_coeff.calendar4.main_core.dropout_event:3.race_ethnicity.se\endcsname{0.294}
\expandafter\gdef\csname EERadj@adjusted_coeff.calendar4.main_core.dropout_event:4b.race_ethnicity.se\endcsname{0.000}
\expandafter\gdef\csname EERadj@adjusted_coeff.calendar4.main_core.dropout_event:1997b.calendar_period.se\endcsname{0.000}
\expandafter\gdef\csname EERadj@adjusted_coeff.calendar4.main_core.dropout_event:1998.calendar_period.se\endcsname{0.228}
\expandafter\gdef\csname EERadj@adjusted_coeff.calendar4.main_core.dropout_event:1999.calendar_period.se\endcsname{0.280}
\expandafter\gdef\csname EERadj@adjusted_coeff.calendar4.main_core.dropout_event:2000.calendar_period.se\endcsname{0.325}
\expandafter\gdef\csname EERadj@adjusted_coeff.calendar4.main_core.dropout_event:2001.calendar_period.se\endcsname{0.378}
\expandafter\gdef\csname EERadj@adjusted_coeff.calendar4.main_core.dropout_event:2002.calendar_period.se\endcsname{0.430}
\expandafter\gdef\csname EERadj@adjusted_coeff.calendar4.main_core.dropout_event:2003.calendar_period.se\endcsname{0.504}
\expandafter\gdef\csname EERadj@adjusted_coeff.calendar4.main_core.dropout_event:_cons.se\endcsname{0.147}
\expandafter\gdef\csname EERadj@adjusted_coeff.calendar4.history_baseline.dropout_event:treated.se\endcsname{0.113}
\expandafter\gdef\csname EERadj@adjusted_coeff.calendar4.history_baseline.dropout_event:1b.eer_durbin.se\endcsname{0.000}
\expandafter\gdef\csname EERadj@adjusted_coeff.calendar4.history_baseline.dropout_event:2.eer_durbin.se\endcsname{0.236}
\expandafter\gdef\csname EERadj@adjusted_coeff.calendar4.history_baseline.dropout_event:3.eer_durbin.se\endcsname{0.210}
\expandafter\gdef\csname EERadj@adjusted_coeff.calendar4.history_baseline.dropout_event:4.eer_durbin.se\endcsname{0.211}
\expandafter\gdef\csname EERadj@adjusted_coeff.calendar4.history_baseline.dropout_event:5.eer_durbin.se\endcsname{0.195}
\expandafter\gdef\csname EERadj@adjusted_coeff.calendar4.history_baseline.dropout_event:6.eer_durbin.se\endcsname{0.204}
\expandafter\gdef\csname EERadj@adjusted_coeff.calendar4.history_baseline.dropout_event:7.eer_durbin.se\endcsname{0.212}
\expandafter\gdef\csname EERadj@adjusted_coeff.calendar4.history_baseline.dropout_event:1b.eer_term.se\endcsname{0.000}
\expandafter\gdef\csname EERadj@adjusted_coeff.calendar4.history_baseline.dropout_event:2.eer_term.se\endcsname{0.087}
\expandafter\gdef\csname EERadj@adjusted_coeff.calendar4.history_baseline.dropout_event:eer_age.se\endcsname{0.063}
\expandafter\gdef\csname EERadj@adjusted_coeff.calendar4.history_baseline.dropout_event:eer_age2.se\endcsname{0.021}
\expandafter\gdef\csname EERadj@adjusted_coeff.calendar4.history_baseline.dropout_event:_cons.se\endcsname{0.165}
\expandafter\gdef\csname EERadj@adjusted_coeff.calendar4.history_core.dropout_event:treated.se\endcsname{0.115}
\expandafter\gdef\csname EERadj@adjusted_coeff.calendar4.history_core.dropout_event:1b.eer_durbin.se\endcsname{0.000}
\expandafter\gdef\csname EERadj@adjusted_coeff.calendar4.history_core.dropout_event:2.eer_durbin.se\endcsname{0.279}
\expandafter\gdef\csname EERadj@adjusted_coeff.calendar4.history_core.dropout_event:3.eer_durbin.se\endcsname{0.300}
\expandafter\gdef\csname EERadj@adjusted_coeff.calendar4.history_core.dropout_event:4.eer_durbin.se\endcsname{0.321}
\expandafter\gdef\csname EERadj@adjusted_coeff.calendar4.history_core.dropout_event:5.eer_durbin.se\endcsname{0.364}
\expandafter\gdef\csname EERadj@adjusted_coeff.calendar4.history_core.dropout_event:6.eer_durbin.se\endcsname{0.420}
\expandafter\gdef\csname EERadj@adjusted_coeff.calendar4.history_core.dropout_event:7.eer_durbin.se\endcsname{0.489}
\expandafter\gdef\csname EERadj@adjusted_coeff.calendar4.history_core.dropout_event:1b.eer_term.se\endcsname{0.000}
\expandafter\gdef\csname EERadj@adjusted_coeff.calendar4.history_core.dropout_event:2.eer_term.se\endcsname{0.107}
\expandafter\gdef\csname EERadj@adjusted_coeff.calendar4.history_core.dropout_event:eer_age.se\endcsname{0.062}
\expandafter\gdef\csname EERadj@adjusted_coeff.calendar4.history_core.dropout_event:eer_age2.se\endcsname{0.022}
\expandafter\gdef\csname EERadj@adjusted_coeff.calendar4.history_core.dropout_event:1.race_ethnicity.se\endcsname{0.099}
\expandafter\gdef\csname EERadj@adjusted_coeff.calendar4.history_core.dropout_event:2.race_ethnicity.se\endcsname{0.110}
\expandafter\gdef\csname EERadj@adjusted_coeff.calendar4.history_core.dropout_event:3.race_ethnicity.se\endcsname{0.331}
\expandafter\gdef\csname EERadj@adjusted_coeff.calendar4.history_core.dropout_event:4b.race_ethnicity.se\endcsname{0.000}
\expandafter\gdef\csname EERadj@adjusted_coeff.calendar4.history_core.dropout_event:1997b.calendar_period.se\endcsname{0.000}
\expandafter\gdef\csname EERadj@adjusted_coeff.calendar4.history_core.dropout_event:1998.calendar_period.se\endcsname{0.281}
\expandafter\gdef\csname EERadj@adjusted_coeff.calendar4.history_core.dropout_event:1999.calendar_period.se\endcsname{0.351}
\expandafter\gdef\csname EERadj@adjusted_coeff.calendar4.history_core.dropout_event:2000.calendar_period.se\endcsname{0.403}
\expandafter\gdef\csname EERadj@adjusted_coeff.calendar4.history_core.dropout_event:2001.calendar_period.se\endcsname{0.461}
\expandafter\gdef\csname EERadj@adjusted_coeff.calendar4.history_core.dropout_event:2002.calendar_period.se\endcsname{0.523}
\expandafter\gdef\csname EERadj@adjusted_coeff.calendar4.history_core.dropout_event:2003.calendar_period.se\endcsname{0.589}
\expandafter\gdef\csname EERadj@adjusted_coeff.calendar4.history_core.dropout_event:_cons.se\endcsname{0.185}
\expandafter\gdef\csname EERadj@adjusted_coeff.calendar4.later_all_core.dropout_event:treated.se\endcsname{0.111}
\expandafter\gdef\csname EERadj@adjusted_coeff.calendar4.later_all_core.dropout_event:1b.eer_durbin.se\endcsname{0.000}
\expandafter\gdef\csname EERadj@adjusted_coeff.calendar4.later_all_core.dropout_event:2.eer_durbin.se\endcsname{0.640}
\expandafter\gdef\csname EERadj@adjusted_coeff.calendar4.later_all_core.dropout_event:3.eer_durbin.se\endcsname{0.498}
\expandafter\gdef\csname EERadj@adjusted_coeff.calendar4.later_all_core.dropout_event:4.eer_durbin.se\endcsname{0.496}
\expandafter\gdef\csname EERadj@adjusted_coeff.calendar4.later_all_core.dropout_event:5.eer_durbin.se\endcsname{0.477}
\expandafter\gdef\csname EERadj@adjusted_coeff.calendar4.later_all_core.dropout_event:6.eer_durbin.se\endcsname{0.474}
\expandafter\gdef\csname EERadj@adjusted_coeff.calendar4.later_all_core.dropout_event:7.eer_durbin.se\endcsname{0.490}
\expandafter\gdef\csname EERadj@adjusted_coeff.calendar4.later_all_core.dropout_event:1b.eer_term.se\endcsname{0.000}
\expandafter\gdef\csname EERadj@adjusted_coeff.calendar4.later_all_core.dropout_event:2.eer_term.se\endcsname{0.121}
\expandafter\gdef\csname EERadj@adjusted_coeff.calendar4.later_all_core.dropout_event:eer_age.se\endcsname{0.079}
\expandafter\gdef\csname EERadj@adjusted_coeff.calendar4.later_all_core.dropout_event:eer_age2.se\endcsname{0.026}
\expandafter\gdef\csname EERadj@adjusted_coeff.calendar4.later_all_core.dropout_event:1.race_ethnicity.se\endcsname{0.105}
\expandafter\gdef\csname EERadj@adjusted_coeff.calendar4.later_all_core.dropout_event:2.race_ethnicity.se\endcsname{0.115}
\expandafter\gdef\csname EERadj@adjusted_coeff.calendar4.later_all_core.dropout_event:3.race_ethnicity.se\endcsname{0.391}
\expandafter\gdef\csname EERadj@adjusted_coeff.calendar4.later_all_core.dropout_event:4b.race_ethnicity.se\endcsname{0.000}
\expandafter\gdef\csname EERadj@adjusted_coeff.calendar4.later_all_core.dropout_event:1998b.calendar_period.se\endcsname{0.000}
\expandafter\gdef\csname EERadj@adjusted_coeff.calendar4.later_all_core.dropout_event:1999.calendar_period.se\endcsname{0.214}
\expandafter\gdef\csname EERadj@adjusted_coeff.calendar4.later_all_core.dropout_event:2000.calendar_period.se\endcsname{0.288}
\expandafter\gdef\csname EERadj@adjusted_coeff.calendar4.later_all_core.dropout_event:2001.calendar_period.se\endcsname{0.361}
\expandafter\gdef\csname EERadj@adjusted_coeff.calendar4.later_all_core.dropout_event:2002.calendar_period.se\endcsname{0.432}
\expandafter\gdef\csname EERadj@adjusted_coeff.calendar4.later_all_core.dropout_event:2003.calendar_period.se\endcsname{0.511}
\expandafter\gdef\csname EERadj@adjusted_coeff.calendar4.later_all_core.dropout_event:_cons.se\endcsname{0.487}
\expandafter\gdef\csname EERadj@adjusted_coeff.calendar4.later_cc_baseline.dropout_event:treated.se\endcsname{0.138}
\expandafter\gdef\csname EERadj@adjusted_coeff.calendar4.later_cc_baseline.dropout_event:1b.eer_durbin.se\endcsname{0.000}
\expandafter\gdef\csname EERadj@adjusted_coeff.calendar4.later_cc_baseline.dropout_event:2.eer_durbin.se\endcsname{0.820}
\expandafter\gdef\csname EERadj@adjusted_coeff.calendar4.later_cc_baseline.dropout_event:3.eer_durbin.se\endcsname{0.656}
\expandafter\gdef\csname EERadj@adjusted_coeff.calendar4.later_cc_baseline.dropout_event:4.eer_durbin.se\endcsname{0.662}
\expandafter\gdef\csname EERadj@adjusted_coeff.calendar4.later_cc_baseline.dropout_event:5.eer_durbin.se\endcsname{0.653}
\expandafter\gdef\csname EERadj@adjusted_coeff.calendar4.later_cc_baseline.dropout_event:6.eer_durbin.se\endcsname{0.656}
\expandafter\gdef\csname EERadj@adjusted_coeff.calendar4.later_cc_baseline.dropout_event:7.eer_durbin.se\endcsname{0.662}
\expandafter\gdef\csname EERadj@adjusted_coeff.calendar4.later_cc_baseline.dropout_event:1b.eer_term.se\endcsname{0.000}
\expandafter\gdef\csname EERadj@adjusted_coeff.calendar4.later_cc_baseline.dropout_event:2.eer_term.se\endcsname{0.117}
\expandafter\gdef\csname EERadj@adjusted_coeff.calendar4.later_cc_baseline.dropout_event:eer_age.se\endcsname{0.086}
\expandafter\gdef\csname EERadj@adjusted_coeff.calendar4.later_cc_baseline.dropout_event:eer_age2.se\endcsname{0.028}
\expandafter\gdef\csname EERadj@adjusted_coeff.calendar4.later_cc_baseline.dropout_event:_cons.se\endcsname{0.633}
\expandafter\gdef\csname EERadj@adjusted_coeff.calendar4.later_cc_core.dropout_event:treated.se\endcsname{0.145}
\expandafter\gdef\csname EERadj@adjusted_coeff.calendar4.later_cc_core.dropout_event:1b.eer_durbin.se\endcsname{0.000}
\expandafter\gdef\csname EERadj@adjusted_coeff.calendar4.later_cc_core.dropout_event:2.eer_durbin.se\endcsname{0.793}
\expandafter\gdef\csname EERadj@adjusted_coeff.calendar4.later_cc_core.dropout_event:3.eer_durbin.se\endcsname{0.636}
\expandafter\gdef\csname EERadj@adjusted_coeff.calendar4.later_cc_core.dropout_event:4.eer_durbin.se\endcsname{0.676}
\expandafter\gdef\csname EERadj@adjusted_coeff.calendar4.later_cc_core.dropout_event:5.eer_durbin.se\endcsname{0.724}
\expandafter\gdef\csname EERadj@adjusted_coeff.calendar4.later_cc_core.dropout_event:6.eer_durbin.se\endcsname{0.793}
\expandafter\gdef\csname EERadj@adjusted_coeff.calendar4.later_cc_core.dropout_event:7.eer_durbin.se\endcsname{0.881}
\expandafter\gdef\csname EERadj@adjusted_coeff.calendar4.later_cc_core.dropout_event:1b.eer_term.se\endcsname{0.000}
\expandafter\gdef\csname EERadj@adjusted_coeff.calendar4.later_cc_core.dropout_event:2.eer_term.se\endcsname{0.161}
\expandafter\gdef\csname EERadj@adjusted_coeff.calendar4.later_cc_core.dropout_event:eer_age.se\endcsname{0.088}
\expandafter\gdef\csname EERadj@adjusted_coeff.calendar4.later_cc_core.dropout_event:eer_age2.se\endcsname{0.029}
\expandafter\gdef\csname EERadj@adjusted_coeff.calendar4.later_cc_core.dropout_event:1.race_ethnicity.se\endcsname{0.134}
\expandafter\gdef\csname EERadj@adjusted_coeff.calendar4.later_cc_core.dropout_event:2.race_ethnicity.se\endcsname{0.144}
\expandafter\gdef\csname EERadj@adjusted_coeff.calendar4.later_cc_core.dropout_event:3.race_ethnicity.se\endcsname{0.470}
\expandafter\gdef\csname EERadj@adjusted_coeff.calendar4.later_cc_core.dropout_event:4b.race_ethnicity.se\endcsname{0.000}
\expandafter\gdef\csname EERadj@adjusted_coeff.calendar4.later_cc_core.dropout_event:1998b.calendar_period.se\endcsname{0.000}
\expandafter\gdef\csname EERadj@adjusted_coeff.calendar4.later_cc_core.dropout_event:1999.calendar_period.se\endcsname{0.288}
\expandafter\gdef\csname EERadj@adjusted_coeff.calendar4.later_cc_core.dropout_event:2000.calendar_period.se\endcsname{0.395}
\expandafter\gdef\csname EERadj@adjusted_coeff.calendar4.later_cc_core.dropout_event:2001.calendar_period.se\endcsname{0.527}
\expandafter\gdef\csname EERadj@adjusted_coeff.calendar4.later_cc_core.dropout_event:2002.calendar_period.se\endcsname{0.619}
\expandafter\gdef\csname EERadj@adjusted_coeff.calendar4.later_cc_core.dropout_event:2003.calendar_period.se\endcsname{0.716}
\expandafter\gdef\csname EERadj@adjusted_coeff.calendar4.later_cc_core.dropout_event:_cons.se\endcsname{0.613}
\expandafter\gdef\csname EERadj@adjusted_coeff.calendar4.later_cc_rich.dropout_event:treated.se\endcsname{0.151}
\expandafter\gdef\csname EERadj@adjusted_coeff.calendar4.later_cc_rich.dropout_event:1b.eer_durbin.se\endcsname{0.000}
\expandafter\gdef\csname EERadj@adjusted_coeff.calendar4.later_cc_rich.dropout_event:2.eer_durbin.se\endcsname{0.731}
\expandafter\gdef\csname EERadj@adjusted_coeff.calendar4.later_cc_rich.dropout_event:3.eer_durbin.se\endcsname{0.554}
\expandafter\gdef\csname EERadj@adjusted_coeff.calendar4.later_cc_rich.dropout_event:4.eer_durbin.se\endcsname{0.594}
\expandafter\gdef\csname EERadj@adjusted_coeff.calendar4.later_cc_rich.dropout_event:5.eer_durbin.se\endcsname{0.654}
\expandafter\gdef\csname EERadj@adjusted_coeff.calendar4.later_cc_rich.dropout_event:6.eer_durbin.se\endcsname{0.736}
\expandafter\gdef\csname EERadj@adjusted_coeff.calendar4.later_cc_rich.dropout_event:7.eer_durbin.se\endcsname{0.842}
\expandafter\gdef\csname EERadj@adjusted_coeff.calendar4.later_cc_rich.dropout_event:1b.eer_term.se\endcsname{0.000}
\expandafter\gdef\csname EERadj@adjusted_coeff.calendar4.later_cc_rich.dropout_event:2.eer_term.se\endcsname{0.160}
\expandafter\gdef\csname EERadj@adjusted_coeff.calendar4.later_cc_rich.dropout_event:eer_age.se\endcsname{0.091}
\expandafter\gdef\csname EERadj@adjusted_coeff.calendar4.later_cc_rich.dropout_event:eer_age2.se\endcsname{0.031}
\expandafter\gdef\csname EERadj@adjusted_coeff.calendar4.later_cc_rich.dropout_event:1.race_ethnicity.se\endcsname{0.141}
\expandafter\gdef\csname EERadj@adjusted_coeff.calendar4.later_cc_rich.dropout_event:2.race_ethnicity.se\endcsname{0.157}
\expandafter\gdef\csname EERadj@adjusted_coeff.calendar4.later_cc_rich.dropout_event:3.race_ethnicity.se\endcsname{0.447}
\expandafter\gdef\csname EERadj@adjusted_coeff.calendar4.later_cc_rich.dropout_event:4b.race_ethnicity.se\endcsname{0.000}
\expandafter\gdef\csname EERadj@adjusted_coeff.calendar4.later_cc_rich.dropout_event:1998b.calendar_period.se\endcsname{0.000}
\expandafter\gdef\csname EERadj@adjusted_coeff.calendar4.later_cc_rich.dropout_event:1999.calendar_period.se\endcsname{0.293}
\expandafter\gdef\csname EERadj@adjusted_coeff.calendar4.later_cc_rich.dropout_event:2000.calendar_period.se\endcsname{0.399}
\expandafter\gdef\csname EERadj@adjusted_coeff.calendar4.later_cc_rich.dropout_event:2001.calendar_period.se\endcsname{0.532}
\expandafter\gdef\csname EERadj@adjusted_coeff.calendar4.later_cc_rich.dropout_event:2002.calendar_period.se\endcsname{0.631}
\expandafter\gdef\csname EERadj@adjusted_coeff.calendar4.later_cc_rich.dropout_event:2003.calendar_period.se\endcsname{0.725}
\expandafter\gdef\csname EERadj@adjusted_coeff.calendar4.later_cc_rich.dropout_event:mother_grade97.se\endcsname{0.026}
\expandafter\gdef\csname EERadj@adjusted_coeff.calendar4.later_cc_rich.dropout_event:1b.household_group.se\endcsname{0.000}
\expandafter\gdef\csname EERadj@adjusted_coeff.calendar4.later_cc_rich.dropout_event:2.household_group.se\endcsname{0.179}
\expandafter\gdef\csname EERadj@adjusted_coeff.calendar4.later_cc_rich.dropout_event:3.household_group.se\endcsname{0.131}
\expandafter\gdef\csname EERadj@adjusted_coeff.calendar4.later_cc_rich.dropout_event:asvab10.se\endcsname{0.026}
\expandafter\gdef\csname EERadj@adjusted_coeff.calendar4.later_cc_rich.dropout_event:_cons.se\endcsname{0.588}
\expandafter\gdef\csname EERadj@adjusted_coeff.school3.main_baseline.dropout_event:treated.omitted\endcsname{0}
\expandafter\gdef\csname EERadj@adjusted_coeff.school3.main_baseline.dropout_event:1b.eer_durbin.omitted\endcsname{1}
\expandafter\gdef\csname EERadj@adjusted_coeff.school3.main_baseline.dropout_event:2.eer_durbin.omitted\endcsname{0}
\expandafter\gdef\csname EERadj@adjusted_coeff.school3.main_baseline.dropout_event:3.eer_durbin.omitted\endcsname{0}
\expandafter\gdef\csname EERadj@adjusted_coeff.school3.main_baseline.dropout_event:4.eer_durbin.omitted\endcsname{0}
\expandafter\gdef\csname EERadj@adjusted_coeff.school3.main_baseline.dropout_event:5.eer_durbin.omitted\endcsname{0}
\expandafter\gdef\csname EERadj@adjusted_coeff.school3.main_baseline.dropout_event:6.eer_durbin.omitted\endcsname{0}
\expandafter\gdef\csname EERadj@adjusted_coeff.school3.main_baseline.dropout_event:7.eer_durbin.omitted\endcsname{0}
\expandafter\gdef\csname EERadj@adjusted_coeff.school3.main_baseline.dropout_event:1b.eer_term.omitted\endcsname{1}
\expandafter\gdef\csname EERadj@adjusted_coeff.school3.main_baseline.dropout_event:2.eer_term.omitted\endcsname{0}
\expandafter\gdef\csname EERadj@adjusted_coeff.school3.main_baseline.dropout_event:eer_age.omitted\endcsname{0}
\expandafter\gdef\csname EERadj@adjusted_coeff.school3.main_baseline.dropout_event:eer_age2.omitted\endcsname{0}
\expandafter\gdef\csname EERadj@adjusted_coeff.school3.main_baseline.dropout_event:_cons.omitted\endcsname{0}
\expandafter\gdef\csname EERadj@adjusted_coeff.school3.main_core.dropout_event:treated.omitted\endcsname{0}
\expandafter\gdef\csname EERadj@adjusted_coeff.school3.main_core.dropout_event:1b.eer_durbin.omitted\endcsname{1}
\expandafter\gdef\csname EERadj@adjusted_coeff.school3.main_core.dropout_event:2.eer_durbin.omitted\endcsname{0}
\expandafter\gdef\csname EERadj@adjusted_coeff.school3.main_core.dropout_event:3.eer_durbin.omitted\endcsname{0}
\expandafter\gdef\csname EERadj@adjusted_coeff.school3.main_core.dropout_event:4.eer_durbin.omitted\endcsname{0}
\expandafter\gdef\csname EERadj@adjusted_coeff.school3.main_core.dropout_event:5.eer_durbin.omitted\endcsname{0}
\expandafter\gdef\csname EERadj@adjusted_coeff.school3.main_core.dropout_event:6.eer_durbin.omitted\endcsname{0}
\expandafter\gdef\csname EERadj@adjusted_coeff.school3.main_core.dropout_event:7.eer_durbin.omitted\endcsname{0}
\expandafter\gdef\csname EERadj@adjusted_coeff.school3.main_core.dropout_event:1b.eer_term.omitted\endcsname{1}
\expandafter\gdef\csname EERadj@adjusted_coeff.school3.main_core.dropout_event:2.eer_term.omitted\endcsname{0}
\expandafter\gdef\csname EERadj@adjusted_coeff.school3.main_core.dropout_event:eer_age.omitted\endcsname{0}
\expandafter\gdef\csname EERadj@adjusted_coeff.school3.main_core.dropout_event:eer_age2.omitted\endcsname{0}
\expandafter\gdef\csname EERadj@adjusted_coeff.school3.main_core.dropout_event:1.race_ethnicity.omitted\endcsname{0}
\expandafter\gdef\csname EERadj@adjusted_coeff.school3.main_core.dropout_event:2.race_ethnicity.omitted\endcsname{0}
\expandafter\gdef\csname EERadj@adjusted_coeff.school3.main_core.dropout_event:3.race_ethnicity.omitted\endcsname{0}
\expandafter\gdef\csname EERadj@adjusted_coeff.school3.main_core.dropout_event:4b.race_ethnicity.omitted\endcsname{1}
\expandafter\gdef\csname EERadj@adjusted_coeff.school3.main_core.dropout_event:1997b.calendar_period.omitted\endcsname{1}
\expandafter\gdef\csname EERadj@adjusted_coeff.school3.main_core.dropout_event:1998.calendar_period.omitted\endcsname{0}
\expandafter\gdef\csname EERadj@adjusted_coeff.school3.main_core.dropout_event:1999.calendar_period.omitted\endcsname{0}
\expandafter\gdef\csname EERadj@adjusted_coeff.school3.main_core.dropout_event:2000.calendar_period.omitted\endcsname{0}
\expandafter\gdef\csname EERadj@adjusted_coeff.school3.main_core.dropout_event:2001.calendar_period.omitted\endcsname{0}
\expandafter\gdef\csname EERadj@adjusted_coeff.school3.main_core.dropout_event:2002.calendar_period.omitted\endcsname{0}
\expandafter\gdef\csname EERadj@adjusted_coeff.school3.main_core.dropout_event:2003.calendar_period.omitted\endcsname{0}
\expandafter\gdef\csname EERadj@adjusted_coeff.school3.main_core.dropout_event:_cons.omitted\endcsname{0}
\expandafter\gdef\csname EERadj@adjusted_coeff.school3.history_baseline.dropout_event:treated.omitted\endcsname{0}
\expandafter\gdef\csname EERadj@adjusted_coeff.school3.history_baseline.dropout_event:1b.eer_durbin.omitted\endcsname{1}
\expandafter\gdef\csname EERadj@adjusted_coeff.school3.history_baseline.dropout_event:2.eer_durbin.omitted\endcsname{0}
\expandafter\gdef\csname EERadj@adjusted_coeff.school3.history_baseline.dropout_event:3.eer_durbin.omitted\endcsname{0}
\expandafter\gdef\csname EERadj@adjusted_coeff.school3.history_baseline.dropout_event:4.eer_durbin.omitted\endcsname{0}
\expandafter\gdef\csname EERadj@adjusted_coeff.school3.history_baseline.dropout_event:5.eer_durbin.omitted\endcsname{0}
\expandafter\gdef\csname EERadj@adjusted_coeff.school3.history_baseline.dropout_event:6.eer_durbin.omitted\endcsname{0}
\expandafter\gdef\csname EERadj@adjusted_coeff.school3.history_baseline.dropout_event:7.eer_durbin.omitted\endcsname{0}
\expandafter\gdef\csname EERadj@adjusted_coeff.school3.history_baseline.dropout_event:1b.eer_term.omitted\endcsname{1}
\expandafter\gdef\csname EERadj@adjusted_coeff.school3.history_baseline.dropout_event:2.eer_term.omitted\endcsname{0}
\expandafter\gdef\csname EERadj@adjusted_coeff.school3.history_baseline.dropout_event:eer_age.omitted\endcsname{0}
\expandafter\gdef\csname EERadj@adjusted_coeff.school3.history_baseline.dropout_event:eer_age2.omitted\endcsname{0}
\expandafter\gdef\csname EERadj@adjusted_coeff.school3.history_baseline.dropout_event:_cons.omitted\endcsname{0}
\expandafter\gdef\csname EERadj@adjusted_coeff.school3.history_core.dropout_event:treated.omitted\endcsname{0}
\expandafter\gdef\csname EERadj@adjusted_coeff.school3.history_core.dropout_event:1b.eer_durbin.omitted\endcsname{1}
\expandafter\gdef\csname EERadj@adjusted_coeff.school3.history_core.dropout_event:2.eer_durbin.omitted\endcsname{0}
\expandafter\gdef\csname EERadj@adjusted_coeff.school3.history_core.dropout_event:3.eer_durbin.omitted\endcsname{0}
\expandafter\gdef\csname EERadj@adjusted_coeff.school3.history_core.dropout_event:4.eer_durbin.omitted\endcsname{0}
\expandafter\gdef\csname EERadj@adjusted_coeff.school3.history_core.dropout_event:5.eer_durbin.omitted\endcsname{0}
\expandafter\gdef\csname EERadj@adjusted_coeff.school3.history_core.dropout_event:6.eer_durbin.omitted\endcsname{0}
\expandafter\gdef\csname EERadj@adjusted_coeff.school3.history_core.dropout_event:7.eer_durbin.omitted\endcsname{0}
\expandafter\gdef\csname EERadj@adjusted_coeff.school3.history_core.dropout_event:1b.eer_term.omitted\endcsname{1}
\expandafter\gdef\csname EERadj@adjusted_coeff.school3.history_core.dropout_event:2.eer_term.omitted\endcsname{0}
\expandafter\gdef\csname EERadj@adjusted_coeff.school3.history_core.dropout_event:eer_age.omitted\endcsname{0}
\expandafter\gdef\csname EERadj@adjusted_coeff.school3.history_core.dropout_event:eer_age2.omitted\endcsname{0}
\expandafter\gdef\csname EERadj@adjusted_coeff.school3.history_core.dropout_event:1.race_ethnicity.omitted\endcsname{0}
\expandafter\gdef\csname EERadj@adjusted_coeff.school3.history_core.dropout_event:2.race_ethnicity.omitted\endcsname{0}
\expandafter\gdef\csname EERadj@adjusted_coeff.school3.history_core.dropout_event:3.race_ethnicity.omitted\endcsname{0}
\expandafter\gdef\csname EERadj@adjusted_coeff.school3.history_core.dropout_event:4b.race_ethnicity.omitted\endcsname{1}
\expandafter\gdef\csname EERadj@adjusted_coeff.school3.history_core.dropout_event:1997b.calendar_period.omitted\endcsname{1}
\expandafter\gdef\csname EERadj@adjusted_coeff.school3.history_core.dropout_event:1998.calendar_period.omitted\endcsname{0}
\expandafter\gdef\csname EERadj@adjusted_coeff.school3.history_core.dropout_event:1999.calendar_period.omitted\endcsname{0}
\expandafter\gdef\csname EERadj@adjusted_coeff.school3.history_core.dropout_event:2000.calendar_period.omitted\endcsname{0}
\expandafter\gdef\csname EERadj@adjusted_coeff.school3.history_core.dropout_event:2001.calendar_period.omitted\endcsname{0}
\expandafter\gdef\csname EERadj@adjusted_coeff.school3.history_core.dropout_event:2002.calendar_period.omitted\endcsname{0}
\expandafter\gdef\csname EERadj@adjusted_coeff.school3.history_core.dropout_event:2003.calendar_period.omitted\endcsname{0}
\expandafter\gdef\csname EERadj@adjusted_coeff.school3.history_core.dropout_event:_cons.omitted\endcsname{0}
\expandafter\gdef\csname EERadj@adjusted_coeff.school3.later_all_core.dropout_event:treated.omitted\endcsname{0}
\expandafter\gdef\csname EERadj@adjusted_coeff.school3.later_all_core.dropout_event:1b.eer_durbin.omitted\endcsname{1}
\expandafter\gdef\csname EERadj@adjusted_coeff.school3.later_all_core.dropout_event:2.eer_durbin.omitted\endcsname{0}
\expandafter\gdef\csname EERadj@adjusted_coeff.school3.later_all_core.dropout_event:3.eer_durbin.omitted\endcsname{0}
\expandafter\gdef\csname EERadj@adjusted_coeff.school3.later_all_core.dropout_event:4.eer_durbin.omitted\endcsname{0}
\expandafter\gdef\csname EERadj@adjusted_coeff.school3.later_all_core.dropout_event:5.eer_durbin.omitted\endcsname{0}
\expandafter\gdef\csname EERadj@adjusted_coeff.school3.later_all_core.dropout_event:6.eer_durbin.omitted\endcsname{0}
\expandafter\gdef\csname EERadj@adjusted_coeff.school3.later_all_core.dropout_event:7.eer_durbin.omitted\endcsname{0}
\expandafter\gdef\csname EERadj@adjusted_coeff.school3.later_all_core.dropout_event:1b.eer_term.omitted\endcsname{1}
\expandafter\gdef\csname EERadj@adjusted_coeff.school3.later_all_core.dropout_event:2.eer_term.omitted\endcsname{0}
\expandafter\gdef\csname EERadj@adjusted_coeff.school3.later_all_core.dropout_event:eer_age.omitted\endcsname{0}
\expandafter\gdef\csname EERadj@adjusted_coeff.school3.later_all_core.dropout_event:eer_age2.omitted\endcsname{0}
\expandafter\gdef\csname EERadj@adjusted_coeff.school3.later_all_core.dropout_event:1.race_ethnicity.omitted\endcsname{0}
\expandafter\gdef\csname EERadj@adjusted_coeff.school3.later_all_core.dropout_event:2.race_ethnicity.omitted\endcsname{0}
\expandafter\gdef\csname EERadj@adjusted_coeff.school3.later_all_core.dropout_event:3.race_ethnicity.omitted\endcsname{0}
\expandafter\gdef\csname EERadj@adjusted_coeff.school3.later_all_core.dropout_event:4b.race_ethnicity.omitted\endcsname{1}
\expandafter\gdef\csname EERadj@adjusted_coeff.school3.later_all_core.dropout_event:1998b.calendar_period.omitted\endcsname{1}
\expandafter\gdef\csname EERadj@adjusted_coeff.school3.later_all_core.dropout_event:1999.calendar_period.omitted\endcsname{0}
\expandafter\gdef\csname EERadj@adjusted_coeff.school3.later_all_core.dropout_event:2000.calendar_period.omitted\endcsname{0}
\expandafter\gdef\csname EERadj@adjusted_coeff.school3.later_all_core.dropout_event:2001.calendar_period.omitted\endcsname{0}
\expandafter\gdef\csname EERadj@adjusted_coeff.school3.later_all_core.dropout_event:2002.calendar_period.omitted\endcsname{0}
\expandafter\gdef\csname EERadj@adjusted_coeff.school3.later_all_core.dropout_event:2003.calendar_period.omitted\endcsname{0}
\expandafter\gdef\csname EERadj@adjusted_coeff.school3.later_all_core.dropout_event:_cons.omitted\endcsname{0}
\expandafter\gdef\csname EERadj@adjusted_coeff.school3.later_cc_baseline.dropout_event:treated.omitted\endcsname{0}
\expandafter\gdef\csname EERadj@adjusted_coeff.school3.later_cc_baseline.dropout_event:1b.eer_durbin.omitted\endcsname{1}
\expandafter\gdef\csname EERadj@adjusted_coeff.school3.later_cc_baseline.dropout_event:2.eer_durbin.omitted\endcsname{0}
\expandafter\gdef\csname EERadj@adjusted_coeff.school3.later_cc_baseline.dropout_event:3.eer_durbin.omitted\endcsname{0}
\expandafter\gdef\csname EERadj@adjusted_coeff.school3.later_cc_baseline.dropout_event:4.eer_durbin.omitted\endcsname{0}
\expandafter\gdef\csname EERadj@adjusted_coeff.school3.later_cc_baseline.dropout_event:5.eer_durbin.omitted\endcsname{0}
\expandafter\gdef\csname EERadj@adjusted_coeff.school3.later_cc_baseline.dropout_event:6.eer_durbin.omitted\endcsname{0}
\expandafter\gdef\csname EERadj@adjusted_coeff.school3.later_cc_baseline.dropout_event:7.eer_durbin.omitted\endcsname{0}
\expandafter\gdef\csname EERadj@adjusted_coeff.school3.later_cc_baseline.dropout_event:1b.eer_term.omitted\endcsname{1}
\expandafter\gdef\csname EERadj@adjusted_coeff.school3.later_cc_baseline.dropout_event:2.eer_term.omitted\endcsname{0}
\expandafter\gdef\csname EERadj@adjusted_coeff.school3.later_cc_baseline.dropout_event:eer_age.omitted\endcsname{0}
\expandafter\gdef\csname EERadj@adjusted_coeff.school3.later_cc_baseline.dropout_event:eer_age2.omitted\endcsname{0}
\expandafter\gdef\csname EERadj@adjusted_coeff.school3.later_cc_baseline.dropout_event:_cons.omitted\endcsname{0}
\expandafter\gdef\csname EERadj@adjusted_coeff.school3.later_cc_core.dropout_event:treated.omitted\endcsname{0}
\expandafter\gdef\csname EERadj@adjusted_coeff.school3.later_cc_core.dropout_event:1b.eer_durbin.omitted\endcsname{1}
\expandafter\gdef\csname EERadj@adjusted_coeff.school3.later_cc_core.dropout_event:2.eer_durbin.omitted\endcsname{0}
\expandafter\gdef\csname EERadj@adjusted_coeff.school3.later_cc_core.dropout_event:3.eer_durbin.omitted\endcsname{0}
\expandafter\gdef\csname EERadj@adjusted_coeff.school3.later_cc_core.dropout_event:4.eer_durbin.omitted\endcsname{0}
\expandafter\gdef\csname EERadj@adjusted_coeff.school3.later_cc_core.dropout_event:5.eer_durbin.omitted\endcsname{0}
\expandafter\gdef\csname EERadj@adjusted_coeff.school3.later_cc_core.dropout_event:6.eer_durbin.omitted\endcsname{0}
\expandafter\gdef\csname EERadj@adjusted_coeff.school3.later_cc_core.dropout_event:7.eer_durbin.omitted\endcsname{0}
\expandafter\gdef\csname EERadj@adjusted_coeff.school3.later_cc_core.dropout_event:1b.eer_term.omitted\endcsname{1}
\expandafter\gdef\csname EERadj@adjusted_coeff.school3.later_cc_core.dropout_event:2.eer_term.omitted\endcsname{0}
\expandafter\gdef\csname EERadj@adjusted_coeff.school3.later_cc_core.dropout_event:eer_age.omitted\endcsname{0}
\expandafter\gdef\csname EERadj@adjusted_coeff.school3.later_cc_core.dropout_event:eer_age2.omitted\endcsname{0}
\expandafter\gdef\csname EERadj@adjusted_coeff.school3.later_cc_core.dropout_event:1.race_ethnicity.omitted\endcsname{0}
\expandafter\gdef\csname EERadj@adjusted_coeff.school3.later_cc_core.dropout_event:2.race_ethnicity.omitted\endcsname{0}
\expandafter\gdef\csname EERadj@adjusted_coeff.school3.later_cc_core.dropout_event:3.race_ethnicity.omitted\endcsname{0}
\expandafter\gdef\csname EERadj@adjusted_coeff.school3.later_cc_core.dropout_event:4b.race_ethnicity.omitted\endcsname{1}
\expandafter\gdef\csname EERadj@adjusted_coeff.school3.later_cc_core.dropout_event:1998b.calendar_period.omitted\endcsname{1}
\expandafter\gdef\csname EERadj@adjusted_coeff.school3.later_cc_core.dropout_event:1999.calendar_period.omitted\endcsname{0}
\expandafter\gdef\csname EERadj@adjusted_coeff.school3.later_cc_core.dropout_event:2000.calendar_period.omitted\endcsname{0}
\expandafter\gdef\csname EERadj@adjusted_coeff.school3.later_cc_core.dropout_event:2001.calendar_period.omitted\endcsname{0}
\expandafter\gdef\csname EERadj@adjusted_coeff.school3.later_cc_core.dropout_event:2002.calendar_period.omitted\endcsname{0}
\expandafter\gdef\csname EERadj@adjusted_coeff.school3.later_cc_core.dropout_event:2003.calendar_period.omitted\endcsname{0}
\expandafter\gdef\csname EERadj@adjusted_coeff.school3.later_cc_core.dropout_event:_cons.omitted\endcsname{0}
\expandafter\gdef\csname EERadj@adjusted_coeff.school3.later_cc_rich.dropout_event:treated.omitted\endcsname{0}
\expandafter\gdef\csname EERadj@adjusted_coeff.school3.later_cc_rich.dropout_event:1b.eer_durbin.omitted\endcsname{1}
\expandafter\gdef\csname EERadj@adjusted_coeff.school3.later_cc_rich.dropout_event:2.eer_durbin.omitted\endcsname{0}
\expandafter\gdef\csname EERadj@adjusted_coeff.school3.later_cc_rich.dropout_event:3.eer_durbin.omitted\endcsname{0}
\expandafter\gdef\csname EERadj@adjusted_coeff.school3.later_cc_rich.dropout_event:4.eer_durbin.omitted\endcsname{0}
\expandafter\gdef\csname EERadj@adjusted_coeff.school3.later_cc_rich.dropout_event:5.eer_durbin.omitted\endcsname{0}
\expandafter\gdef\csname EERadj@adjusted_coeff.school3.later_cc_rich.dropout_event:6.eer_durbin.omitted\endcsname{0}
\expandafter\gdef\csname EERadj@adjusted_coeff.school3.later_cc_rich.dropout_event:7.eer_durbin.omitted\endcsname{0}
\expandafter\gdef\csname EERadj@adjusted_coeff.school3.later_cc_rich.dropout_event:1b.eer_term.omitted\endcsname{1}
\expandafter\gdef\csname EERadj@adjusted_coeff.school3.later_cc_rich.dropout_event:2.eer_term.omitted\endcsname{0}
\expandafter\gdef\csname EERadj@adjusted_coeff.school3.later_cc_rich.dropout_event:eer_age.omitted\endcsname{0}
\expandafter\gdef\csname EERadj@adjusted_coeff.school3.later_cc_rich.dropout_event:eer_age2.omitted\endcsname{0}
\expandafter\gdef\csname EERadj@adjusted_coeff.school3.later_cc_rich.dropout_event:1.race_ethnicity.omitted\endcsname{0}
\expandafter\gdef\csname EERadj@adjusted_coeff.school3.later_cc_rich.dropout_event:2.race_ethnicity.omitted\endcsname{0}
\expandafter\gdef\csname EERadj@adjusted_coeff.school3.later_cc_rich.dropout_event:3.race_ethnicity.omitted\endcsname{0}
\expandafter\gdef\csname EERadj@adjusted_coeff.school3.later_cc_rich.dropout_event:4b.race_ethnicity.omitted\endcsname{1}
\expandafter\gdef\csname EERadj@adjusted_coeff.school3.later_cc_rich.dropout_event:1998b.calendar_period.omitted\endcsname{1}
\expandafter\gdef\csname EERadj@adjusted_coeff.school3.later_cc_rich.dropout_event:1999.calendar_period.omitted\endcsname{0}
\expandafter\gdef\csname EERadj@adjusted_coeff.school3.later_cc_rich.dropout_event:2000.calendar_period.omitted\endcsname{0}
\expandafter\gdef\csname EERadj@adjusted_coeff.school3.later_cc_rich.dropout_event:2001.calendar_period.omitted\endcsname{0}
\expandafter\gdef\csname EERadj@adjusted_coeff.school3.later_cc_rich.dropout_event:2002.calendar_period.omitted\endcsname{0}
\expandafter\gdef\csname EERadj@adjusted_coeff.school3.later_cc_rich.dropout_event:2003.calendar_period.omitted\endcsname{0}
\expandafter\gdef\csname EERadj@adjusted_coeff.school3.later_cc_rich.dropout_event:mother_grade97.omitted\endcsname{0}
\expandafter\gdef\csname EERadj@adjusted_coeff.school3.later_cc_rich.dropout_event:1b.household_group.omitted\endcsname{1}
\expandafter\gdef\csname EERadj@adjusted_coeff.school3.later_cc_rich.dropout_event:2.household_group.omitted\endcsname{0}
\expandafter\gdef\csname EERadj@adjusted_coeff.school3.later_cc_rich.dropout_event:3.household_group.omitted\endcsname{0}
\expandafter\gdef\csname EERadj@adjusted_coeff.school3.later_cc_rich.dropout_event:asvab10.omitted\endcsname{0}
\expandafter\gdef\csname EERadj@adjusted_coeff.school3.later_cc_rich.dropout_event:_cons.omitted\endcsname{0}
\expandafter\gdef\csname EERadj@adjusted_coeff.calendar4.main_baseline.dropout_event:treated.omitted\endcsname{0}
\expandafter\gdef\csname EERadj@adjusted_coeff.calendar4.main_baseline.dropout_event:1b.eer_durbin.omitted\endcsname{1}
\expandafter\gdef\csname EERadj@adjusted_coeff.calendar4.main_baseline.dropout_event:2.eer_durbin.omitted\endcsname{0}
\expandafter\gdef\csname EERadj@adjusted_coeff.calendar4.main_baseline.dropout_event:3.eer_durbin.omitted\endcsname{0}
\expandafter\gdef\csname EERadj@adjusted_coeff.calendar4.main_baseline.dropout_event:4.eer_durbin.omitted\endcsname{0}
\expandafter\gdef\csname EERadj@adjusted_coeff.calendar4.main_baseline.dropout_event:5.eer_durbin.omitted\endcsname{0}
\expandafter\gdef\csname EERadj@adjusted_coeff.calendar4.main_baseline.dropout_event:6.eer_durbin.omitted\endcsname{0}
\expandafter\gdef\csname EERadj@adjusted_coeff.calendar4.main_baseline.dropout_event:7.eer_durbin.omitted\endcsname{0}
\expandafter\gdef\csname EERadj@adjusted_coeff.calendar4.main_baseline.dropout_event:1b.eer_term.omitted\endcsname{1}
\expandafter\gdef\csname EERadj@adjusted_coeff.calendar4.main_baseline.dropout_event:2.eer_term.omitted\endcsname{0}
\expandafter\gdef\csname EERadj@adjusted_coeff.calendar4.main_baseline.dropout_event:eer_age.omitted\endcsname{0}
\expandafter\gdef\csname EERadj@adjusted_coeff.calendar4.main_baseline.dropout_event:eer_age2.omitted\endcsname{0}
\expandafter\gdef\csname EERadj@adjusted_coeff.calendar4.main_baseline.dropout_event:_cons.omitted\endcsname{0}
\expandafter\gdef\csname EERadj@adjusted_coeff.calendar4.main_core.dropout_event:treated.omitted\endcsname{0}
\expandafter\gdef\csname EERadj@adjusted_coeff.calendar4.main_core.dropout_event:1b.eer_durbin.omitted\endcsname{1}
\expandafter\gdef\csname EERadj@adjusted_coeff.calendar4.main_core.dropout_event:2.eer_durbin.omitted\endcsname{0}
\expandafter\gdef\csname EERadj@adjusted_coeff.calendar4.main_core.dropout_event:3.eer_durbin.omitted\endcsname{0}
\expandafter\gdef\csname EERadj@adjusted_coeff.calendar4.main_core.dropout_event:4.eer_durbin.omitted\endcsname{0}
\expandafter\gdef\csname EERadj@adjusted_coeff.calendar4.main_core.dropout_event:5.eer_durbin.omitted\endcsname{0}
\expandafter\gdef\csname EERadj@adjusted_coeff.calendar4.main_core.dropout_event:6.eer_durbin.omitted\endcsname{0}
\expandafter\gdef\csname EERadj@adjusted_coeff.calendar4.main_core.dropout_event:7.eer_durbin.omitted\endcsname{0}
\expandafter\gdef\csname EERadj@adjusted_coeff.calendar4.main_core.dropout_event:1b.eer_term.omitted\endcsname{1}
\expandafter\gdef\csname EERadj@adjusted_coeff.calendar4.main_core.dropout_event:2.eer_term.omitted\endcsname{0}
\expandafter\gdef\csname EERadj@adjusted_coeff.calendar4.main_core.dropout_event:eer_age.omitted\endcsname{0}
\expandafter\gdef\csname EERadj@adjusted_coeff.calendar4.main_core.dropout_event:eer_age2.omitted\endcsname{0}
\expandafter\gdef\csname EERadj@adjusted_coeff.calendar4.main_core.dropout_event:1.race_ethnicity.omitted\endcsname{0}
\expandafter\gdef\csname EERadj@adjusted_coeff.calendar4.main_core.dropout_event:2.race_ethnicity.omitted\endcsname{0}
\expandafter\gdef\csname EERadj@adjusted_coeff.calendar4.main_core.dropout_event:3.race_ethnicity.omitted\endcsname{0}
\expandafter\gdef\csname EERadj@adjusted_coeff.calendar4.main_core.dropout_event:4b.race_ethnicity.omitted\endcsname{1}
\expandafter\gdef\csname EERadj@adjusted_coeff.calendar4.main_core.dropout_event:1997b.calendar_period.omitted\endcsname{1}
\expandafter\gdef\csname EERadj@adjusted_coeff.calendar4.main_core.dropout_event:1998.calendar_period.omitted\endcsname{0}
\expandafter\gdef\csname EERadj@adjusted_coeff.calendar4.main_core.dropout_event:1999.calendar_period.omitted\endcsname{0}
\expandafter\gdef\csname EERadj@adjusted_coeff.calendar4.main_core.dropout_event:2000.calendar_period.omitted\endcsname{0}
\expandafter\gdef\csname EERadj@adjusted_coeff.calendar4.main_core.dropout_event:2001.calendar_period.omitted\endcsname{0}
\expandafter\gdef\csname EERadj@adjusted_coeff.calendar4.main_core.dropout_event:2002.calendar_period.omitted\endcsname{0}
\expandafter\gdef\csname EERadj@adjusted_coeff.calendar4.main_core.dropout_event:2003.calendar_period.omitted\endcsname{0}
\expandafter\gdef\csname EERadj@adjusted_coeff.calendar4.main_core.dropout_event:_cons.omitted\endcsname{0}
\expandafter\gdef\csname EERadj@adjusted_coeff.calendar4.history_baseline.dropout_event:treated.omitted\endcsname{0}
\expandafter\gdef\csname EERadj@adjusted_coeff.calendar4.history_baseline.dropout_event:1b.eer_durbin.omitted\endcsname{1}
\expandafter\gdef\csname EERadj@adjusted_coeff.calendar4.history_baseline.dropout_event:2.eer_durbin.omitted\endcsname{0}
\expandafter\gdef\csname EERadj@adjusted_coeff.calendar4.history_baseline.dropout_event:3.eer_durbin.omitted\endcsname{0}
\expandafter\gdef\csname EERadj@adjusted_coeff.calendar4.history_baseline.dropout_event:4.eer_durbin.omitted\endcsname{0}
\expandafter\gdef\csname EERadj@adjusted_coeff.calendar4.history_baseline.dropout_event:5.eer_durbin.omitted\endcsname{0}
\expandafter\gdef\csname EERadj@adjusted_coeff.calendar4.history_baseline.dropout_event:6.eer_durbin.omitted\endcsname{0}
\expandafter\gdef\csname EERadj@adjusted_coeff.calendar4.history_baseline.dropout_event:7.eer_durbin.omitted\endcsname{0}
\expandafter\gdef\csname EERadj@adjusted_coeff.calendar4.history_baseline.dropout_event:1b.eer_term.omitted\endcsname{1}
\expandafter\gdef\csname EERadj@adjusted_coeff.calendar4.history_baseline.dropout_event:2.eer_term.omitted\endcsname{0}
\expandafter\gdef\csname EERadj@adjusted_coeff.calendar4.history_baseline.dropout_event:eer_age.omitted\endcsname{0}
\expandafter\gdef\csname EERadj@adjusted_coeff.calendar4.history_baseline.dropout_event:eer_age2.omitted\endcsname{0}
\expandafter\gdef\csname EERadj@adjusted_coeff.calendar4.history_baseline.dropout_event:_cons.omitted\endcsname{0}
\expandafter\gdef\csname EERadj@adjusted_coeff.calendar4.history_core.dropout_event:treated.omitted\endcsname{0}
\expandafter\gdef\csname EERadj@adjusted_coeff.calendar4.history_core.dropout_event:1b.eer_durbin.omitted\endcsname{1}
\expandafter\gdef\csname EERadj@adjusted_coeff.calendar4.history_core.dropout_event:2.eer_durbin.omitted\endcsname{0}
\expandafter\gdef\csname EERadj@adjusted_coeff.calendar4.history_core.dropout_event:3.eer_durbin.omitted\endcsname{0}
\expandafter\gdef\csname EERadj@adjusted_coeff.calendar4.history_core.dropout_event:4.eer_durbin.omitted\endcsname{0}
\expandafter\gdef\csname EERadj@adjusted_coeff.calendar4.history_core.dropout_event:5.eer_durbin.omitted\endcsname{0}
\expandafter\gdef\csname EERadj@adjusted_coeff.calendar4.history_core.dropout_event:6.eer_durbin.omitted\endcsname{0}
\expandafter\gdef\csname EERadj@adjusted_coeff.calendar4.history_core.dropout_event:7.eer_durbin.omitted\endcsname{0}
\expandafter\gdef\csname EERadj@adjusted_coeff.calendar4.history_core.dropout_event:1b.eer_term.omitted\endcsname{1}
\expandafter\gdef\csname EERadj@adjusted_coeff.calendar4.history_core.dropout_event:2.eer_term.omitted\endcsname{0}
\expandafter\gdef\csname EERadj@adjusted_coeff.calendar4.history_core.dropout_event:eer_age.omitted\endcsname{0}
\expandafter\gdef\csname EERadj@adjusted_coeff.calendar4.history_core.dropout_event:eer_age2.omitted\endcsname{0}
\expandafter\gdef\csname EERadj@adjusted_coeff.calendar4.history_core.dropout_event:1.race_ethnicity.omitted\endcsname{0}
\expandafter\gdef\csname EERadj@adjusted_coeff.calendar4.history_core.dropout_event:2.race_ethnicity.omitted\endcsname{0}
\expandafter\gdef\csname EERadj@adjusted_coeff.calendar4.history_core.dropout_event:3.race_ethnicity.omitted\endcsname{0}
\expandafter\gdef\csname EERadj@adjusted_coeff.calendar4.history_core.dropout_event:4b.race_ethnicity.omitted\endcsname{1}
\expandafter\gdef\csname EERadj@adjusted_coeff.calendar4.history_core.dropout_event:1997b.calendar_period.omitted\endcsname{1}
\expandafter\gdef\csname EERadj@adjusted_coeff.calendar4.history_core.dropout_event:1998.calendar_period.omitted\endcsname{0}
\expandafter\gdef\csname EERadj@adjusted_coeff.calendar4.history_core.dropout_event:1999.calendar_period.omitted\endcsname{0}
\expandafter\gdef\csname EERadj@adjusted_coeff.calendar4.history_core.dropout_event:2000.calendar_period.omitted\endcsname{0}
\expandafter\gdef\csname EERadj@adjusted_coeff.calendar4.history_core.dropout_event:2001.calendar_period.omitted\endcsname{0}
\expandafter\gdef\csname EERadj@adjusted_coeff.calendar4.history_core.dropout_event:2002.calendar_period.omitted\endcsname{0}
\expandafter\gdef\csname EERadj@adjusted_coeff.calendar4.history_core.dropout_event:2003.calendar_period.omitted\endcsname{0}
\expandafter\gdef\csname EERadj@adjusted_coeff.calendar4.history_core.dropout_event:_cons.omitted\endcsname{0}
\expandafter\gdef\csname EERadj@adjusted_coeff.calendar4.later_all_core.dropout_event:treated.omitted\endcsname{0}
\expandafter\gdef\csname EERadj@adjusted_coeff.calendar4.later_all_core.dropout_event:1b.eer_durbin.omitted\endcsname{1}
\expandafter\gdef\csname EERadj@adjusted_coeff.calendar4.later_all_core.dropout_event:2.eer_durbin.omitted\endcsname{0}
\expandafter\gdef\csname EERadj@adjusted_coeff.calendar4.later_all_core.dropout_event:3.eer_durbin.omitted\endcsname{0}
\expandafter\gdef\csname EERadj@adjusted_coeff.calendar4.later_all_core.dropout_event:4.eer_durbin.omitted\endcsname{0}
\expandafter\gdef\csname EERadj@adjusted_coeff.calendar4.later_all_core.dropout_event:5.eer_durbin.omitted\endcsname{0}
\expandafter\gdef\csname EERadj@adjusted_coeff.calendar4.later_all_core.dropout_event:6.eer_durbin.omitted\endcsname{0}
\expandafter\gdef\csname EERadj@adjusted_coeff.calendar4.later_all_core.dropout_event:7.eer_durbin.omitted\endcsname{0}
\expandafter\gdef\csname EERadj@adjusted_coeff.calendar4.later_all_core.dropout_event:1b.eer_term.omitted\endcsname{1}
\expandafter\gdef\csname EERadj@adjusted_coeff.calendar4.later_all_core.dropout_event:2.eer_term.omitted\endcsname{0}
\expandafter\gdef\csname EERadj@adjusted_coeff.calendar4.later_all_core.dropout_event:eer_age.omitted\endcsname{0}
\expandafter\gdef\csname EERadj@adjusted_coeff.calendar4.later_all_core.dropout_event:eer_age2.omitted\endcsname{0}
\expandafter\gdef\csname EERadj@adjusted_coeff.calendar4.later_all_core.dropout_event:1.race_ethnicity.omitted\endcsname{0}
\expandafter\gdef\csname EERadj@adjusted_coeff.calendar4.later_all_core.dropout_event:2.race_ethnicity.omitted\endcsname{0}
\expandafter\gdef\csname EERadj@adjusted_coeff.calendar4.later_all_core.dropout_event:3.race_ethnicity.omitted\endcsname{0}
\expandafter\gdef\csname EERadj@adjusted_coeff.calendar4.later_all_core.dropout_event:4b.race_ethnicity.omitted\endcsname{1}
\expandafter\gdef\csname EERadj@adjusted_coeff.calendar4.later_all_core.dropout_event:1998b.calendar_period.omitted\endcsname{1}
\expandafter\gdef\csname EERadj@adjusted_coeff.calendar4.later_all_core.dropout_event:1999.calendar_period.omitted\endcsname{0}
\expandafter\gdef\csname EERadj@adjusted_coeff.calendar4.later_all_core.dropout_event:2000.calendar_period.omitted\endcsname{0}
\expandafter\gdef\csname EERadj@adjusted_coeff.calendar4.later_all_core.dropout_event:2001.calendar_period.omitted\endcsname{0}
\expandafter\gdef\csname EERadj@adjusted_coeff.calendar4.later_all_core.dropout_event:2002.calendar_period.omitted\endcsname{0}
\expandafter\gdef\csname EERadj@adjusted_coeff.calendar4.later_all_core.dropout_event:2003.calendar_period.omitted\endcsname{0}
\expandafter\gdef\csname EERadj@adjusted_coeff.calendar4.later_all_core.dropout_event:_cons.omitted\endcsname{0}
\expandafter\gdef\csname EERadj@adjusted_coeff.calendar4.later_cc_baseline.dropout_event:treated.omitted\endcsname{0}
\expandafter\gdef\csname EERadj@adjusted_coeff.calendar4.later_cc_baseline.dropout_event:1b.eer_durbin.omitted\endcsname{1}
\expandafter\gdef\csname EERadj@adjusted_coeff.calendar4.later_cc_baseline.dropout_event:2.eer_durbin.omitted\endcsname{0}
\expandafter\gdef\csname EERadj@adjusted_coeff.calendar4.later_cc_baseline.dropout_event:3.eer_durbin.omitted\endcsname{0}
\expandafter\gdef\csname EERadj@adjusted_coeff.calendar4.later_cc_baseline.dropout_event:4.eer_durbin.omitted\endcsname{0}
\expandafter\gdef\csname EERadj@adjusted_coeff.calendar4.later_cc_baseline.dropout_event:5.eer_durbin.omitted\endcsname{0}
\expandafter\gdef\csname EERadj@adjusted_coeff.calendar4.later_cc_baseline.dropout_event:6.eer_durbin.omitted\endcsname{0}
\expandafter\gdef\csname EERadj@adjusted_coeff.calendar4.later_cc_baseline.dropout_event:7.eer_durbin.omitted\endcsname{0}
\expandafter\gdef\csname EERadj@adjusted_coeff.calendar4.later_cc_baseline.dropout_event:1b.eer_term.omitted\endcsname{1}
\expandafter\gdef\csname EERadj@adjusted_coeff.calendar4.later_cc_baseline.dropout_event:2.eer_term.omitted\endcsname{0}
\expandafter\gdef\csname EERadj@adjusted_coeff.calendar4.later_cc_baseline.dropout_event:eer_age.omitted\endcsname{0}
\expandafter\gdef\csname EERadj@adjusted_coeff.calendar4.later_cc_baseline.dropout_event:eer_age2.omitted\endcsname{0}
\expandafter\gdef\csname EERadj@adjusted_coeff.calendar4.later_cc_baseline.dropout_event:_cons.omitted\endcsname{0}
\expandafter\gdef\csname EERadj@adjusted_coeff.calendar4.later_cc_core.dropout_event:treated.omitted\endcsname{0}
\expandafter\gdef\csname EERadj@adjusted_coeff.calendar4.later_cc_core.dropout_event:1b.eer_durbin.omitted\endcsname{1}
\expandafter\gdef\csname EERadj@adjusted_coeff.calendar4.later_cc_core.dropout_event:2.eer_durbin.omitted\endcsname{0}
\expandafter\gdef\csname EERadj@adjusted_coeff.calendar4.later_cc_core.dropout_event:3.eer_durbin.omitted\endcsname{0}
\expandafter\gdef\csname EERadj@adjusted_coeff.calendar4.later_cc_core.dropout_event:4.eer_durbin.omitted\endcsname{0}
\expandafter\gdef\csname EERadj@adjusted_coeff.calendar4.later_cc_core.dropout_event:5.eer_durbin.omitted\endcsname{0}
\expandafter\gdef\csname EERadj@adjusted_coeff.calendar4.later_cc_core.dropout_event:6.eer_durbin.omitted\endcsname{0}
\expandafter\gdef\csname EERadj@adjusted_coeff.calendar4.later_cc_core.dropout_event:7.eer_durbin.omitted\endcsname{0}
\expandafter\gdef\csname EERadj@adjusted_coeff.calendar4.later_cc_core.dropout_event:1b.eer_term.omitted\endcsname{1}
\expandafter\gdef\csname EERadj@adjusted_coeff.calendar4.later_cc_core.dropout_event:2.eer_term.omitted\endcsname{0}
\expandafter\gdef\csname EERadj@adjusted_coeff.calendar4.later_cc_core.dropout_event:eer_age.omitted\endcsname{0}
\expandafter\gdef\csname EERadj@adjusted_coeff.calendar4.later_cc_core.dropout_event:eer_age2.omitted\endcsname{0}
\expandafter\gdef\csname EERadj@adjusted_coeff.calendar4.later_cc_core.dropout_event:1.race_ethnicity.omitted\endcsname{0}
\expandafter\gdef\csname EERadj@adjusted_coeff.calendar4.later_cc_core.dropout_event:2.race_ethnicity.omitted\endcsname{0}
\expandafter\gdef\csname EERadj@adjusted_coeff.calendar4.later_cc_core.dropout_event:3.race_ethnicity.omitted\endcsname{0}
\expandafter\gdef\csname EERadj@adjusted_coeff.calendar4.later_cc_core.dropout_event:4b.race_ethnicity.omitted\endcsname{1}
\expandafter\gdef\csname EERadj@adjusted_coeff.calendar4.later_cc_core.dropout_event:1998b.calendar_period.omitted\endcsname{1}
\expandafter\gdef\csname EERadj@adjusted_coeff.calendar4.later_cc_core.dropout_event:1999.calendar_period.omitted\endcsname{0}
\expandafter\gdef\csname EERadj@adjusted_coeff.calendar4.later_cc_core.dropout_event:2000.calendar_period.omitted\endcsname{0}
\expandafter\gdef\csname EERadj@adjusted_coeff.calendar4.later_cc_core.dropout_event:2001.calendar_period.omitted\endcsname{0}
\expandafter\gdef\csname EERadj@adjusted_coeff.calendar4.later_cc_core.dropout_event:2002.calendar_period.omitted\endcsname{0}
\expandafter\gdef\csname EERadj@adjusted_coeff.calendar4.later_cc_core.dropout_event:2003.calendar_period.omitted\endcsname{0}
\expandafter\gdef\csname EERadj@adjusted_coeff.calendar4.later_cc_core.dropout_event:_cons.omitted\endcsname{0}
\expandafter\gdef\csname EERadj@adjusted_coeff.calendar4.later_cc_rich.dropout_event:treated.omitted\endcsname{0}
\expandafter\gdef\csname EERadj@adjusted_coeff.calendar4.later_cc_rich.dropout_event:1b.eer_durbin.omitted\endcsname{1}
\expandafter\gdef\csname EERadj@adjusted_coeff.calendar4.later_cc_rich.dropout_event:2.eer_durbin.omitted\endcsname{0}
\expandafter\gdef\csname EERadj@adjusted_coeff.calendar4.later_cc_rich.dropout_event:3.eer_durbin.omitted\endcsname{0}
\expandafter\gdef\csname EERadj@adjusted_coeff.calendar4.later_cc_rich.dropout_event:4.eer_durbin.omitted\endcsname{0}
\expandafter\gdef\csname EERadj@adjusted_coeff.calendar4.later_cc_rich.dropout_event:5.eer_durbin.omitted\endcsname{0}
\expandafter\gdef\csname EERadj@adjusted_coeff.calendar4.later_cc_rich.dropout_event:6.eer_durbin.omitted\endcsname{0}
\expandafter\gdef\csname EERadj@adjusted_coeff.calendar4.later_cc_rich.dropout_event:7.eer_durbin.omitted\endcsname{0}
\expandafter\gdef\csname EERadj@adjusted_coeff.calendar4.later_cc_rich.dropout_event:1b.eer_term.omitted\endcsname{1}
\expandafter\gdef\csname EERadj@adjusted_coeff.calendar4.later_cc_rich.dropout_event:2.eer_term.omitted\endcsname{0}
\expandafter\gdef\csname EERadj@adjusted_coeff.calendar4.later_cc_rich.dropout_event:eer_age.omitted\endcsname{0}
\expandafter\gdef\csname EERadj@adjusted_coeff.calendar4.later_cc_rich.dropout_event:eer_age2.omitted\endcsname{0}
\expandafter\gdef\csname EERadj@adjusted_coeff.calendar4.later_cc_rich.dropout_event:1.race_ethnicity.omitted\endcsname{0}
\expandafter\gdef\csname EERadj@adjusted_coeff.calendar4.later_cc_rich.dropout_event:2.race_ethnicity.omitted\endcsname{0}
\expandafter\gdef\csname EERadj@adjusted_coeff.calendar4.later_cc_rich.dropout_event:3.race_ethnicity.omitted\endcsname{0}
\expandafter\gdef\csname EERadj@adjusted_coeff.calendar4.later_cc_rich.dropout_event:4b.race_ethnicity.omitted\endcsname{1}
\expandafter\gdef\csname EERadj@adjusted_coeff.calendar4.later_cc_rich.dropout_event:1998b.calendar_period.omitted\endcsname{1}
\expandafter\gdef\csname EERadj@adjusted_coeff.calendar4.later_cc_rich.dropout_event:1999.calendar_period.omitted\endcsname{0}
\expandafter\gdef\csname EERadj@adjusted_coeff.calendar4.later_cc_rich.dropout_event:2000.calendar_period.omitted\endcsname{0}
\expandafter\gdef\csname EERadj@adjusted_coeff.calendar4.later_cc_rich.dropout_event:2001.calendar_period.omitted\endcsname{0}
\expandafter\gdef\csname EERadj@adjusted_coeff.calendar4.later_cc_rich.dropout_event:2002.calendar_period.omitted\endcsname{0}
\expandafter\gdef\csname EERadj@adjusted_coeff.calendar4.later_cc_rich.dropout_event:2003.calendar_period.omitted\endcsname{0}
\expandafter\gdef\csname EERadj@adjusted_coeff.calendar4.later_cc_rich.dropout_event:mother_grade97.omitted\endcsname{0}
\expandafter\gdef\csname EERadj@adjusted_coeff.calendar4.later_cc_rich.dropout_event:1b.household_group.omitted\endcsname{1}
\expandafter\gdef\csname EERadj@adjusted_coeff.calendar4.later_cc_rich.dropout_event:2.household_group.omitted\endcsname{0}
\expandafter\gdef\csname EERadj@adjusted_coeff.calendar4.later_cc_rich.dropout_event:3.household_group.omitted\endcsname{0}
\expandafter\gdef\csname EERadj@adjusted_coeff.calendar4.later_cc_rich.dropout_event:asvab10.omitted\endcsname{0}
\expandafter\gdef\csname EERadj@adjusted_coeff.calendar4.later_cc_rich.dropout_event:_cons.omitted\endcsname{0}
\expandafter\gdef\csname EERadj@adjusted_counts.school3.main_women.value\endcsname{2,828}
\expandafter\gdef\csname EERadj@adjusted_counts.school3.history_supported_women.value\endcsname{2,571}
\expandafter\gdef\csname EERadj@adjusted_counts.school3.later_eligible_women.value\endcsname{2,313}
\expandafter\gdef\csname EERadj@adjusted_counts.school3.later_missing_mother_grade97.value\endcsname{176}
\expandafter\gdef\csname EERadj@adjusted_counts.school3.later_missing_household_group.value\endcsname{9}
\expandafter\gdef\csname EERadj@adjusted_counts.school3.later_missing_asvab10.value\endcsname{456}
\expandafter\gdef\csname EERadj@adjusted_counts.calendar4.main_women.value\endcsname{2,809}
\expandafter\gdef\csname EERadj@adjusted_counts.calendar4.history_supported_women.value\endcsname{2,553}
\expandafter\gdef\csname EERadj@adjusted_counts.calendar4.later_eligible_women.value\endcsname{2,291}
\expandafter\gdef\csname EERadj@adjusted_counts.calendar4.later_missing_mother_grade97.value\endcsname{173}
\expandafter\gdef\csname EERadj@adjusted_counts.calendar4.later_missing_household_group.value\endcsname{9}
\expandafter\gdef\csname EERadj@adjusted_counts.calendar4.later_missing_asvab10.value\endcsname{454}
\expandafter\gdef\csname EERadj@adjusted_rates.school3.0.treated\endcsname{0}
\expandafter\gdef\csname EERadj@adjusted_rates.school3.1.treated\endcsname{1}
\expandafter\gdef\csname EERadj@adjusted_rates.calendar4.0.treated\endcsname{0}
\expandafter\gdef\csname EERadj@adjusted_rates.calendar4.1.treated\endcsname{1}
\expandafter\gdef\csname EERadj@adjusted_rates.school3.0.women\endcsname{2,690}
\expandafter\gdef\csname EERadj@adjusted_rates.school3.1.women\endcsname{562}
\expandafter\gdef\csname EERadj@adjusted_rates.calendar4.0.women\endcsname{2,671}
\expandafter\gdef\csname EERadj@adjusted_rates.calendar4.1.women\endcsname{549}
\expandafter\gdef\csname EERadj@adjusted_rates.school3.0.months\endcsname{43,014}
\expandafter\gdef\csname EERadj@adjusted_rates.school3.1.months\endcsname{5,939}
\expandafter\gdef\csname EERadj@adjusted_rates.calendar4.0.months\endcsname{42,316}
\expandafter\gdef\csname EERadj@adjusted_rates.calendar4.1.months\endcsname{5,804}
\expandafter\gdef\csname EERadj@adjusted_rates.school3.0.events\endcsname{450}
\expandafter\gdef\csname EERadj@adjusted_rates.school3.1.events\endcsname{261}
\expandafter\gdef\csname EERadj@adjusted_rates.calendar4.0.events\endcsname{468}
\expandafter\gdef\csname EERadj@adjusted_rates.calendar4.1.events\endcsname{258}
\expandafter\gdef\csname EERadj@adjusted_rates.school3.0.rate_per_1000\endcsname{10.5}
\expandafter\gdef\csname EERadj@adjusted_rates.school3.1.rate_per_1000\endcsname{43.9}
\expandafter\gdef\csname EERadj@adjusted_rates.calendar4.0.rate_per_1000\endcsname{11.1}
\expandafter\gdef\csname EERadj@adjusted_rates.calendar4.1.rate_per_1000\endcsname{44.5}
\expandafter\gdef\csname EERadj@landmark_results.calendar4.1.group\endcsname{1}
\expandafter\gdef\csname EERadj@landmark_results.calendar4.0.group\endcsname{0}
\expandafter\gdef\csname EERadj@landmark_results.school3.1.group\endcsname{1}
\expandafter\gdef\csname EERadj@landmark_results.school3.0.group\endcsname{0}
\expandafter\gdef\csname EERadj@landmark_results.calendar4.1.index_rows\endcsname{437}
\expandafter\gdef\csname EERadj@landmark_results.calendar4.0.index_rows\endcsname{41,951}
\expandafter\gdef\csname EERadj@landmark_results.school3.1.index_rows\endcsname{451}
\expandafter\gdef\csname EERadj@landmark_results.school3.0.index_rows\endcsname{42,909}
\expandafter\gdef\csname EERadj@landmark_results.calendar4.1.unique_women\endcsname{437}
\expandafter\gdef\csname EERadj@landmark_results.calendar4.0.unique_women\endcsname{2,637}
\expandafter\gdef\csname EERadj@landmark_results.school3.1.unique_women\endcsname{451}
\expandafter\gdef\csname EERadj@landmark_results.school3.0.unique_women\endcsname{2,658}
\expandafter\gdef\csname EERadj@landmark_results.calendar4.1.weighted_index_rows\endcsname{437.000}
\expandafter\gdef\csname EERadj@landmark_results.calendar4.0.weighted_index_rows\endcsname{437.000}
\expandafter\gdef\csname EERadj@landmark_results.school3.1.weighted_index_rows\endcsname{451.000}
\expandafter\gdef\csname EERadj@landmark_results.school3.0.weighted_index_rows\endcsname{451.000}
\expandafter\gdef\csname EERadj@landmark_results.calendar4.1.raw_interruptions\endcsname{98}
\expandafter\gdef\csname EERadj@landmark_results.calendar4.0.raw_interruptions\endcsname{3,000}
\expandafter\gdef\csname EERadj@landmark_results.school3.1.raw_interruptions\endcsname{98}
\expandafter\gdef\csname EERadj@landmark_results.school3.0.raw_interruptions\endcsname{2,941}
\expandafter\gdef\csname EERadj@landmark_results.calendar4.1.raw_credentials\endcsname{85}
\expandafter\gdef\csname EERadj@landmark_results.calendar4.0.raw_credentials\endcsname{10,988}
\expandafter\gdef\csname EERadj@landmark_results.school3.1.raw_credentials\endcsname{89}
\expandafter\gdef\csname EERadj@landmark_results.school3.0.raw_credentials\endcsname{11,218}
\expandafter\gdef\csname EERadj@landmark_results.calendar4.1.raw_early_censors\endcsname{31}
\expandafter\gdef\csname EERadj@landmark_results.calendar4.0.raw_early_censors\endcsname{454}
\expandafter\gdef\csname EERadj@landmark_results.school3.1.raw_early_censors\endcsname{37}
\expandafter\gdef\csname EERadj@landmark_results.school3.0.raw_early_censors\endcsname{484}
\expandafter\gdef\csname EERadj@landmark_results.calendar4.1.raw_full_observed\endcsname{406}
\expandafter\gdef\csname EERadj@landmark_results.calendar4.0.raw_full_observed\endcsname{41,497}
\expandafter\gdef\csname EERadj@landmark_results.school3.1.raw_full_observed\endcsname{414}
\expandafter\gdef\csname EERadj@landmark_results.school3.0.raw_full_observed\endcsname{42,425}
\expandafter\gdef\csname EERadj@landmark_results.calendar4.1.weighted_early_censors\endcsname{31.000}
\expandafter\gdef\csname EERadj@landmark_results.calendar4.0.weighted_early_censors\endcsname{6.793}
\expandafter\gdef\csname EERadj@landmark_results.school3.1.weighted_early_censors\endcsname{37.000}
\expandafter\gdef\csname EERadj@landmark_results.school3.0.weighted_early_censors\endcsname{8.350}
\expandafter\gdef\csname EERadj@landmark_results.calendar4.1.weighted_full_observed\endcsname{406.000}
\expandafter\gdef\csname EERadj@landmark_results.calendar4.0.weighted_full_observed\endcsname{430.207}
\expandafter\gdef\csname EERadj@landmark_results.school3.1.weighted_full_observed\endcsname{414.000}
\expandafter\gdef\csname EERadj@landmark_results.school3.0.weighted_full_observed\endcsname{442.650}
\expandafter\gdef\csname EERadj@landmark_results.calendar4.1.cif_interruption\endcsname{0.2364}
\expandafter\gdef\csname EERadj@landmark_results.calendar4.0.cif_interruption\endcsname{0.0742}
\expandafter\gdef\csname EERadj@landmark_results.school3.1.cif_interruption\endcsname{0.2308}
\expandafter\gdef\csname EERadj@landmark_results.school3.0.cif_interruption\endcsname{0.0720}
\expandafter\gdef\csname EERadj@landmark_results.calendar4.1.cif_interruption_pct\endcsname{23.6}
\expandafter\gdef\csname EERadj@landmark_results.calendar4.0.cif_interruption_pct\endcsname{7.4}
\expandafter\gdef\csname EERadj@landmark_results.school3.1.cif_interruption_pct\endcsname{23.1}
\expandafter\gdef\csname EERadj@landmark_results.school3.0.cif_interruption_pct\endcsname{7.2}
\expandafter\gdef\csname EERadj@landmark_results.calendar4.1.cif_credential\endcsname{0.2043}
\expandafter\gdef\csname EERadj@landmark_results.calendar4.0.cif_credential\endcsname{0.2659}
\expandafter\gdef\csname EERadj@landmark_results.school3.1.cif_credential\endcsname{0.2085}
\expandafter\gdef\csname EERadj@landmark_results.school3.0.cif_credential\endcsname{0.2697}
\expandafter\gdef\csname EERadj@landmark_results.calendar4.1.cif_credential_pct\endcsname{20.4}
\expandafter\gdef\csname EERadj@landmark_results.calendar4.0.cif_credential_pct\endcsname{26.6}
\expandafter\gdef\csname EERadj@landmark_results.school3.1.cif_credential_pct\endcsname{20.9}
\expandafter\gdef\csname EERadj@landmark_results.school3.0.cif_credential_pct\endcsname{27.0}
\expandafter\gdef\csname EERadj@landmark_results.calendar4.1.survival\endcsname{0.5593}
\expandafter\gdef\csname EERadj@landmark_results.calendar4.0.survival\endcsname{0.6599}
\expandafter\gdef\csname EERadj@landmark_results.school3.1.survival\endcsname{0.5606}
\expandafter\gdef\csname EERadj@landmark_results.school3.0.survival\endcsname{0.6583}
\expandafter\gdef\csname EERadj@landmark_results.calendar4.1.survival_pct\endcsname{55.9}
\expandafter\gdef\csname EERadj@landmark_results.calendar4.0.survival_pct\endcsname{66.0}
\expandafter\gdef\csname EERadj@landmark_results.school3.1.survival_pct\endcsname{56.1}
\expandafter\gdef\csname EERadj@landmark_results.school3.0.survival_pct\endcsname{65.8}
\expandafter\gdef\csname EERadj@landmark_results.calendar4.1.raw_interruption_fraction\endcsname{0.2243}
\expandafter\gdef\csname EERadj@landmark_results.calendar4.0.raw_interruption_fraction\endcsname{0.0715}
\expandafter\gdef\csname EERadj@landmark_results.school3.1.raw_interruption_fraction\endcsname{0.2173}
\expandafter\gdef\csname EERadj@landmark_results.school3.0.raw_interruption_fraction\endcsname{0.0685}
\expandafter\gdef\csname EERadj@landmark_results.calendar4.1.raw_interruption_fraction_pct\endcsname{22.4}
\expandafter\gdef\csname EERadj@landmark_results.calendar4.0.raw_interruption_fraction_pct\endcsname{7.2}
\expandafter\gdef\csname EERadj@landmark_results.school3.1.raw_interruption_fraction_pct\endcsname{21.7}
\expandafter\gdef\csname EERadj@landmark_results.school3.0.raw_interruption_fraction_pct\endcsname{6.9}
\expandafter\gdef\csname EERadj@landmark_results.calendar4.1.weighted_observed_fraction\endcsname{0.2243}
\expandafter\gdef\csname EERadj@landmark_results.calendar4.0.weighted_observed_fraction\endcsname{0.0736}
\expandafter\gdef\csname EERadj@landmark_results.school3.1.weighted_observed_fraction\endcsname{0.2173}
\expandafter\gdef\csname EERadj@landmark_results.school3.0.weighted_observed_fraction\endcsname{0.0713}
\expandafter\gdef\csname EERadj@landmark_results.calendar4.1.weighted_observed_fraction_pct\endcsname{22.4}
\expandafter\gdef\csname EERadj@landmark_results.calendar4.0.weighted_observed_fraction_pct\endcsname{7.4}
\expandafter\gdef\csname EERadj@landmark_results.school3.1.weighted_observed_fraction_pct\endcsname{21.7}
\expandafter\gdef\csname EERadj@landmark_results.school3.0.weighted_observed_fraction_pct\endcsname{7.1}
\expandafter\gdef\csname EERadj@landmark_results.calendar4.1.censor_bound_lo\endcsname{0.2243}
\expandafter\gdef\csname EERadj@landmark_results.calendar4.0.censor_bound_lo\endcsname{0.0736}
\expandafter\gdef\csname EERadj@landmark_results.school3.1.censor_bound_lo\endcsname{0.2173}
\expandafter\gdef\csname EERadj@landmark_results.school3.0.censor_bound_lo\endcsname{0.0713}
\expandafter\gdef\csname EERadj@landmark_results.calendar4.1.censor_bound_lo_pct\endcsname{22.4}
\expandafter\gdef\csname EERadj@landmark_results.calendar4.0.censor_bound_lo_pct\endcsname{7.4}
\expandafter\gdef\csname EERadj@landmark_results.school3.1.censor_bound_lo_pct\endcsname{21.7}
\expandafter\gdef\csname EERadj@landmark_results.school3.0.censor_bound_lo_pct\endcsname{7.1}
\expandafter\gdef\csname EERadj@landmark_results.calendar4.1.censor_bound_hi\endcsname{0.2952}
\expandafter\gdef\csname EERadj@landmark_results.calendar4.0.censor_bound_hi\endcsname{0.0891}
\expandafter\gdef\csname EERadj@landmark_results.school3.1.censor_bound_hi\endcsname{0.2993}
\expandafter\gdef\csname EERadj@landmark_results.school3.0.censor_bound_hi\endcsname{0.0898}
\expandafter\gdef\csname EERadj@landmark_results.calendar4.1.censor_bound_hi_pct\endcsname{29.5}
\expandafter\gdef\csname EERadj@landmark_results.calendar4.0.censor_bound_hi_pct\endcsname{8.9}
\expandafter\gdef\csname EERadj@landmark_results.school3.1.censor_bound_hi_pct\endcsname{29.9}
\expandafter\gdef\csname EERadj@landmark_results.school3.0.censor_bound_hi_pct\endcsname{9.0}
\expandafter\gdef\csname EERadj@landmark_trajectories.school3.0.1.group\endcsname{0}
\expandafter\gdef\csname EERadj@landmark_trajectories.school3.0.2.group\endcsname{0}
\expandafter\gdef\csname EERadj@landmark_trajectories.school3.0.3.group\endcsname{0}
\expandafter\gdef\csname EERadj@landmark_trajectories.school3.0.4.group\endcsname{0}
\expandafter\gdef\csname EERadj@landmark_trajectories.school3.0.5.group\endcsname{0}
\expandafter\gdef\csname EERadj@landmark_trajectories.school3.0.6.group\endcsname{0}
\expandafter\gdef\csname EERadj@landmark_trajectories.school3.0.7.group\endcsname{0}
\expandafter\gdef\csname EERadj@landmark_trajectories.school3.0.8.group\endcsname{0}
\expandafter\gdef\csname EERadj@landmark_trajectories.school3.0.9.group\endcsname{0}
\expandafter\gdef\csname EERadj@landmark_trajectories.school3.1.1.group\endcsname{1}
\expandafter\gdef\csname EERadj@landmark_trajectories.school3.1.2.group\endcsname{1}
\expandafter\gdef\csname EERadj@landmark_trajectories.school3.1.3.group\endcsname{1}
\expandafter\gdef\csname EERadj@landmark_trajectories.school3.1.4.group\endcsname{1}
\expandafter\gdef\csname EERadj@landmark_trajectories.school3.1.5.group\endcsname{1}
\expandafter\gdef\csname EERadj@landmark_trajectories.school3.1.6.group\endcsname{1}
\expandafter\gdef\csname EERadj@landmark_trajectories.school3.1.7.group\endcsname{1}
\expandafter\gdef\csname EERadj@landmark_trajectories.school3.1.8.group\endcsname{1}
\expandafter\gdef\csname EERadj@landmark_trajectories.school3.1.9.group\endcsname{1}
\expandafter\gdef\csname EERadj@landmark_trajectories.calendar4.0.1.group\endcsname{0}
\expandafter\gdef\csname EERadj@landmark_trajectories.calendar4.0.2.group\endcsname{0}
\expandafter\gdef\csname EERadj@landmark_trajectories.calendar4.0.3.group\endcsname{0}
\expandafter\gdef\csname EERadj@landmark_trajectories.calendar4.0.4.group\endcsname{0}
\expandafter\gdef\csname EERadj@landmark_trajectories.calendar4.0.5.group\endcsname{0}
\expandafter\gdef\csname EERadj@landmark_trajectories.calendar4.0.6.group\endcsname{0}
\expandafter\gdef\csname EERadj@landmark_trajectories.calendar4.0.7.group\endcsname{0}
\expandafter\gdef\csname EERadj@landmark_trajectories.calendar4.0.8.group\endcsname{0}
\expandafter\gdef\csname EERadj@landmark_trajectories.calendar4.0.9.group\endcsname{0}
\expandafter\gdef\csname EERadj@landmark_trajectories.calendar4.1.1.group\endcsname{1}
\expandafter\gdef\csname EERadj@landmark_trajectories.calendar4.1.2.group\endcsname{1}
\expandafter\gdef\csname EERadj@landmark_trajectories.calendar4.1.3.group\endcsname{1}
\expandafter\gdef\csname EERadj@landmark_trajectories.calendar4.1.4.group\endcsname{1}
\expandafter\gdef\csname EERadj@landmark_trajectories.calendar4.1.5.group\endcsname{1}
\expandafter\gdef\csname EERadj@landmark_trajectories.calendar4.1.6.group\endcsname{1}
\expandafter\gdef\csname EERadj@landmark_trajectories.calendar4.1.7.group\endcsname{1}
\expandafter\gdef\csname EERadj@landmark_trajectories.calendar4.1.8.group\endcsname{1}
\expandafter\gdef\csname EERadj@landmark_trajectories.calendar4.1.9.group\endcsname{1}
\expandafter\gdef\csname EERadj@landmark_trajectories.school3.0.1.month\endcsname{1}
\expandafter\gdef\csname EERadj@landmark_trajectories.school3.0.2.month\endcsname{2}
\expandafter\gdef\csname EERadj@landmark_trajectories.school3.0.3.month\endcsname{3}
\expandafter\gdef\csname EERadj@landmark_trajectories.school3.0.4.month\endcsname{4}
\expandafter\gdef\csname EERadj@landmark_trajectories.school3.0.5.month\endcsname{5}
\expandafter\gdef\csname EERadj@landmark_trajectories.school3.0.6.month\endcsname{6}
\expandafter\gdef\csname EERadj@landmark_trajectories.school3.0.7.month\endcsname{7}
\expandafter\gdef\csname EERadj@landmark_trajectories.school3.0.8.month\endcsname{8}
\expandafter\gdef\csname EERadj@landmark_trajectories.school3.0.9.month\endcsname{9}
\expandafter\gdef\csname EERadj@landmark_trajectories.school3.1.1.month\endcsname{1}
\expandafter\gdef\csname EERadj@landmark_trajectories.school3.1.2.month\endcsname{2}
\expandafter\gdef\csname EERadj@landmark_trajectories.school3.1.3.month\endcsname{3}
\expandafter\gdef\csname EERadj@landmark_trajectories.school3.1.4.month\endcsname{4}
\expandafter\gdef\csname EERadj@landmark_trajectories.school3.1.5.month\endcsname{5}
\expandafter\gdef\csname EERadj@landmark_trajectories.school3.1.6.month\endcsname{6}
\expandafter\gdef\csname EERadj@landmark_trajectories.school3.1.7.month\endcsname{7}
\expandafter\gdef\csname EERadj@landmark_trajectories.school3.1.8.month\endcsname{8}
\expandafter\gdef\csname EERadj@landmark_trajectories.school3.1.9.month\endcsname{9}
\expandafter\gdef\csname EERadj@landmark_trajectories.calendar4.0.1.month\endcsname{1}
\expandafter\gdef\csname EERadj@landmark_trajectories.calendar4.0.2.month\endcsname{2}
\expandafter\gdef\csname EERadj@landmark_trajectories.calendar4.0.3.month\endcsname{3}
\expandafter\gdef\csname EERadj@landmark_trajectories.calendar4.0.4.month\endcsname{4}
\expandafter\gdef\csname EERadj@landmark_trajectories.calendar4.0.5.month\endcsname{5}
\expandafter\gdef\csname EERadj@landmark_trajectories.calendar4.0.6.month\endcsname{6}
\expandafter\gdef\csname EERadj@landmark_trajectories.calendar4.0.7.month\endcsname{7}
\expandafter\gdef\csname EERadj@landmark_trajectories.calendar4.0.8.month\endcsname{8}
\expandafter\gdef\csname EERadj@landmark_trajectories.calendar4.0.9.month\endcsname{9}
\expandafter\gdef\csname EERadj@landmark_trajectories.calendar4.1.1.month\endcsname{1}
\expandafter\gdef\csname EERadj@landmark_trajectories.calendar4.1.2.month\endcsname{2}
\expandafter\gdef\csname EERadj@landmark_trajectories.calendar4.1.3.month\endcsname{3}
\expandafter\gdef\csname EERadj@landmark_trajectories.calendar4.1.4.month\endcsname{4}
\expandafter\gdef\csname EERadj@landmark_trajectories.calendar4.1.5.month\endcsname{5}
\expandafter\gdef\csname EERadj@landmark_trajectories.calendar4.1.6.month\endcsname{6}
\expandafter\gdef\csname EERadj@landmark_trajectories.calendar4.1.7.month\endcsname{7}
\expandafter\gdef\csname EERadj@landmark_trajectories.calendar4.1.8.month\endcsname{8}
\expandafter\gdef\csname EERadj@landmark_trajectories.calendar4.1.9.month\endcsname{9}
\expandafter\gdef\csname EERadj@landmark_trajectories.school3.0.1.at_risk_weight\endcsname{451.000}
\expandafter\gdef\csname EERadj@landmark_trajectories.school3.0.2.at_risk_weight\endcsname{430.688}
\expandafter\gdef\csname EERadj@landmark_trajectories.school3.0.3.at_risk_weight\endcsname{410.410}
\expandafter\gdef\csname EERadj@landmark_trajectories.school3.0.4.at_risk_weight\endcsname{392.432}
\expandafter\gdef\csname EERadj@landmark_trajectories.school3.0.5.at_risk_weight\endcsname{373.018}
\expandafter\gdef\csname EERadj@landmark_trajectories.school3.0.6.at_risk_weight\endcsname{353.746}
\expandafter\gdef\csname EERadj@landmark_trajectories.school3.0.7.at_risk_weight\endcsname{338.870}
\expandafter\gdef\csname EERadj@landmark_trajectories.school3.0.8.at_risk_weight\endcsname{321.984}
\expandafter\gdef\csname EERadj@landmark_trajectories.school3.0.9.at_risk_weight\endcsname{304.609}
\expandafter\gdef\csname EERadj@landmark_trajectories.school3.1.1.at_risk_weight\endcsname{451.000}
\expandafter\gdef\csname EERadj@landmark_trajectories.school3.1.2.at_risk_weight\endcsname{426.000}
\expandafter\gdef\csname EERadj@landmark_trajectories.school3.1.3.at_risk_weight\endcsname{388.000}
\expandafter\gdef\csname EERadj@landmark_trajectories.school3.1.4.at_risk_weight\endcsname{344.000}
\expandafter\gdef\csname EERadj@landmark_trajectories.school3.1.5.at_risk_weight\endcsname{318.000}
\expandafter\gdef\csname EERadj@landmark_trajectories.school3.1.6.at_risk_weight\endcsname{297.000}
\expandafter\gdef\csname EERadj@landmark_trajectories.school3.1.7.at_risk_weight\endcsname{281.000}
\expandafter\gdef\csname EERadj@landmark_trajectories.school3.1.8.at_risk_weight\endcsname{256.000}
\expandafter\gdef\csname EERadj@landmark_trajectories.school3.1.9.at_risk_weight\endcsname{242.000}
\expandafter\gdef\csname EERadj@landmark_trajectories.calendar4.0.1.at_risk_weight\endcsname{437.000}
\expandafter\gdef\csname EERadj@landmark_trajectories.calendar4.0.2.at_risk_weight\endcsname{418.221}
\expandafter\gdef\csname EERadj@landmark_trajectories.calendar4.0.3.at_risk_weight\endcsname{399.232}
\expandafter\gdef\csname EERadj@landmark_trajectories.calendar4.0.4.at_risk_weight\endcsname{381.935}
\expandafter\gdef\csname EERadj@landmark_trajectories.calendar4.0.5.at_risk_weight\endcsname{363.782}
\expandafter\gdef\csname EERadj@landmark_trajectories.calendar4.0.6.at_risk_weight\endcsname{344.447}
\expandafter\gdef\csname EERadj@landmark_trajectories.calendar4.0.7.at_risk_weight\endcsname{330.090}
\expandafter\gdef\csname EERadj@landmark_trajectories.calendar4.0.8.at_risk_weight\endcsname{313.664}
\expandafter\gdef\csname EERadj@landmark_trajectories.calendar4.0.9.at_risk_weight\endcsname{297.219}
\expandafter\gdef\csname EERadj@landmark_trajectories.calendar4.1.1.at_risk_weight\endcsname{437.000}
\expandafter\gdef\csname EERadj@landmark_trajectories.calendar4.1.2.at_risk_weight\endcsname{414.000}
\expandafter\gdef\csname EERadj@landmark_trajectories.calendar4.1.3.at_risk_weight\endcsname{381.000}
\expandafter\gdef\csname EERadj@landmark_trajectories.calendar4.1.4.at_risk_weight\endcsname{339.000}
\expandafter\gdef\csname EERadj@landmark_trajectories.calendar4.1.5.at_risk_weight\endcsname{313.000}
\expandafter\gdef\csname EERadj@landmark_trajectories.calendar4.1.6.at_risk_weight\endcsname{293.000}
\expandafter\gdef\csname EERadj@landmark_trajectories.calendar4.1.7.at_risk_weight\endcsname{277.000}
\expandafter\gdef\csname EERadj@landmark_trajectories.calendar4.1.8.at_risk_weight\endcsname{253.000}
\expandafter\gdef\csname EERadj@landmark_trajectories.calendar4.1.9.at_risk_weight\endcsname{239.000}
\expandafter\gdef\csname EERadj@landmark_trajectories.school3.0.1.interruption_weight\endcsname{4.076}
\expandafter\gdef\csname EERadj@landmark_trajectories.school3.0.2.interruption_weight\endcsname{3.355}
\expandafter\gdef\csname EERadj@landmark_trajectories.school3.0.3.interruption_weight\endcsname{4.494}
\expandafter\gdef\csname EERadj@landmark_trajectories.school3.0.4.interruption_weight\endcsname{2.989}
\expandafter\gdef\csname EERadj@landmark_trajectories.school3.0.5.interruption_weight\endcsname{3.931}
\expandafter\gdef\csname EERadj@landmark_trajectories.school3.0.6.interruption_weight\endcsname{3.342}
\expandafter\gdef\csname EERadj@landmark_trajectories.school3.0.7.interruption_weight\endcsname{3.415}
\expandafter\gdef\csname EERadj@landmark_trajectories.school3.0.8.interruption_weight\endcsname{3.597}
\expandafter\gdef\csname EERadj@landmark_trajectories.school3.0.9.interruption_weight\endcsname{2.950}
\expandafter\gdef\csname EERadj@landmark_trajectories.school3.1.1.interruption_weight\endcsname{11.000}
\expandafter\gdef\csname EERadj@landmark_trajectories.school3.1.2.interruption_weight\endcsname{11.000}
\expandafter\gdef\csname EERadj@landmark_trajectories.school3.1.3.interruption_weight\endcsname{23.000}
\expandafter\gdef\csname EERadj@landmark_trajectories.school3.1.4.interruption_weight\endcsname{4.000}
\expandafter\gdef\csname EERadj@landmark_trajectories.school3.1.5.interruption_weight\endcsname{10.000}
\expandafter\gdef\csname EERadj@landmark_trajectories.school3.1.6.interruption_weight\endcsname{8.000}
\expandafter\gdef\csname EERadj@landmark_trajectories.school3.1.7.interruption_weight\endcsname{14.000}
\expandafter\gdef\csname EERadj@landmark_trajectories.school3.1.8.interruption_weight\endcsname{7.000}
\expandafter\gdef\csname EERadj@landmark_trajectories.school3.1.9.interruption_weight\endcsname{10.000}
\expandafter\gdef\csname EERadj@landmark_trajectories.calendar4.0.1.interruption_weight\endcsname{4.495}
\expandafter\gdef\csname EERadj@landmark_trajectories.calendar4.0.2.interruption_weight\endcsname{3.508}
\expandafter\gdef\csname EERadj@landmark_trajectories.calendar4.0.3.interruption_weight\endcsname{4.028}
\expandafter\gdef\csname EERadj@landmark_trajectories.calendar4.0.4.interruption_weight\endcsname{2.685}
\expandafter\gdef\csname EERadj@landmark_trajectories.calendar4.0.5.interruption_weight\endcsname{3.955}
\expandafter\gdef\csname EERadj@landmark_trajectories.calendar4.0.6.interruption_weight\endcsname{3.498}
\expandafter\gdef\csname EERadj@landmark_trajectories.calendar4.0.7.interruption_weight\endcsname{3.453}
\expandafter\gdef\csname EERadj@landmark_trajectories.calendar4.0.8.interruption_weight\endcsname{3.588}
\expandafter\gdef\csname EERadj@landmark_trajectories.calendar4.0.9.interruption_weight\endcsname{2.941}
\expandafter\gdef\csname EERadj@landmark_trajectories.calendar4.1.1.interruption_weight\endcsname{11.000}
\expandafter\gdef\csname EERadj@landmark_trajectories.calendar4.1.2.interruption_weight\endcsname{10.000}
\expandafter\gdef\csname EERadj@landmark_trajectories.calendar4.1.3.interruption_weight\endcsname{23.000}
\expandafter\gdef\csname EERadj@landmark_trajectories.calendar4.1.4.interruption_weight\endcsname{4.000}
\expandafter\gdef\csname EERadj@landmark_trajectories.calendar4.1.5.interruption_weight\endcsname{9.000}
\expandafter\gdef\csname EERadj@landmark_trajectories.calendar4.1.6.interruption_weight\endcsname{9.000}
\expandafter\gdef\csname EERadj@landmark_trajectories.calendar4.1.7.interruption_weight\endcsname{14.000}
\expandafter\gdef\csname EERadj@landmark_trajectories.calendar4.1.8.interruption_weight\endcsname{7.000}
\expandafter\gdef\csname EERadj@landmark_trajectories.calendar4.1.9.interruption_weight\endcsname{11.000}
\expandafter\gdef\csname EERadj@landmark_trajectories.school3.0.1.credential_weight\endcsname{15.466}
\expandafter\gdef\csname EERadj@landmark_trajectories.school3.0.2.credential_weight\endcsname{15.363}
\expandafter\gdef\csname EERadj@landmark_trajectories.school3.0.3.credential_weight\endcsname{12.725}
\expandafter\gdef\csname EERadj@landmark_trajectories.school3.0.4.credential_weight\endcsname{15.606}
\expandafter\gdef\csname EERadj@landmark_trajectories.school3.0.5.credential_weight\endcsname{14.602}
\expandafter\gdef\csname EERadj@landmark_trajectories.school3.0.6.credential_weight\endcsname{10.308}
\expandafter\gdef\csname EERadj@landmark_trajectories.school3.0.7.credential_weight\endcsname{12.445}
\expandafter\gdef\csname EERadj@landmark_trajectories.school3.0.8.credential_weight\endcsname{12.326}
\expandafter\gdef\csname EERadj@landmark_trajectories.school3.0.9.credential_weight\endcsname{11.635}
\expandafter\gdef\csname EERadj@landmark_trajectories.school3.1.1.credential_weight\endcsname{12.000}
\expandafter\gdef\csname EERadj@landmark_trajectories.school3.1.2.credential_weight\endcsname{17.000}
\expandafter\gdef\csname EERadj@landmark_trajectories.school3.1.3.credential_weight\endcsname{10.000}
\expandafter\gdef\csname EERadj@landmark_trajectories.school3.1.4.credential_weight\endcsname{12.000}
\expandafter\gdef\csname EERadj@landmark_trajectories.school3.1.5.credential_weight\endcsname{10.000}
\expandafter\gdef\csname EERadj@landmark_trajectories.school3.1.6.credential_weight\endcsname{8.000}
\expandafter\gdef\csname EERadj@landmark_trajectories.school3.1.7.credential_weight\endcsname{9.000}
\expandafter\gdef\csname EERadj@landmark_trajectories.school3.1.8.credential_weight\endcsname{6.000}
\expandafter\gdef\csname EERadj@landmark_trajectories.school3.1.9.credential_weight\endcsname{5.000}
\expandafter\gdef\csname EERadj@landmark_trajectories.calendar4.0.1.credential_weight\endcsname{13.679}
\expandafter\gdef\csname EERadj@landmark_trajectories.calendar4.0.2.credential_weight\endcsname{13.902}
\expandafter\gdef\csname EERadj@landmark_trajectories.calendar4.0.3.credential_weight\endcsname{12.523}
\expandafter\gdef\csname EERadj@landmark_trajectories.calendar4.0.4.credential_weight\endcsname{14.830}
\expandafter\gdef\csname EERadj@landmark_trajectories.calendar4.0.5.credential_weight\endcsname{14.676}
\expandafter\gdef\csname EERadj@landmark_trajectories.calendar4.0.6.credential_weight\endcsname{10.215}
\expandafter\gdef\csname EERadj@landmark_trajectories.calendar4.0.7.credential_weight\endcsname{11.833}
\expandafter\gdef\csname EERadj@landmark_trajectories.calendar4.0.8.credential_weight\endcsname{12.120}
\expandafter\gdef\csname EERadj@landmark_trajectories.calendar4.0.9.credential_weight\endcsname{11.380}
\expandafter\gdef\csname EERadj@landmark_trajectories.calendar4.1.1.credential_weight\endcsname{10.000}
\expandafter\gdef\csname EERadj@landmark_trajectories.calendar4.1.2.credential_weight\endcsname{15.000}
\expandafter\gdef\csname EERadj@landmark_trajectories.calendar4.1.3.credential_weight\endcsname{10.000}
\expandafter\gdef\csname EERadj@landmark_trajectories.calendar4.1.4.credential_weight\endcsname{13.000}
\expandafter\gdef\csname EERadj@landmark_trajectories.calendar4.1.5.credential_weight\endcsname{10.000}
\expandafter\gdef\csname EERadj@landmark_trajectories.calendar4.1.6.credential_weight\endcsname{7.000}
\expandafter\gdef\csname EERadj@landmark_trajectories.calendar4.1.7.credential_weight\endcsname{9.000}
\expandafter\gdef\csname EERadj@landmark_trajectories.calendar4.1.8.credential_weight\endcsname{6.000}
\expandafter\gdef\csname EERadj@landmark_trajectories.calendar4.1.9.credential_weight\endcsname{5.000}
\expandafter\gdef\csname EERadj@landmark_trajectories.school3.0.1.censor_weight\endcsname{0.769}
\expandafter\gdef\csname EERadj@landmark_trajectories.school3.0.2.censor_weight\endcsname{1.560}
\expandafter\gdef\csname EERadj@landmark_trajectories.school3.0.3.censor_weight\endcsname{0.760}
\expandafter\gdef\csname EERadj@landmark_trajectories.school3.0.4.censor_weight\endcsname{0.818}
\expandafter\gdef\csname EERadj@landmark_trajectories.school3.0.5.censor_weight\endcsname{0.739}
\expandafter\gdef\csname EERadj@landmark_trajectories.school3.0.6.censor_weight\endcsname{1.226}
\expandafter\gdef\csname EERadj@landmark_trajectories.school3.0.7.censor_weight\endcsname{1.026}
\expandafter\gdef\csname EERadj@landmark_trajectories.school3.0.8.censor_weight\endcsname{1.451}
\expandafter\gdef\csname EERadj@landmark_trajectories.school3.0.9.censor_weight\endcsname{0.000}
\expandafter\gdef\csname EERadj@landmark_trajectories.school3.1.1.censor_weight\endcsname{2.000}
\expandafter\gdef\csname EERadj@landmark_trajectories.school3.1.2.censor_weight\endcsname{10.000}
\expandafter\gdef\csname EERadj@landmark_trajectories.school3.1.3.censor_weight\endcsname{11.000}
\expandafter\gdef\csname EERadj@landmark_trajectories.school3.1.4.censor_weight\endcsname{10.000}
\expandafter\gdef\csname EERadj@landmark_trajectories.school3.1.5.censor_weight\endcsname{1.000}
\expandafter\gdef\csname EERadj@landmark_trajectories.school3.1.6.censor_weight\endcsname{0.000}
\expandafter\gdef\csname EERadj@landmark_trajectories.school3.1.7.censor_weight\endcsname{2.000}
\expandafter\gdef\csname EERadj@landmark_trajectories.school3.1.8.censor_weight\endcsname{1.000}
\expandafter\gdef\csname EERadj@landmark_trajectories.school3.1.9.censor_weight\endcsname{0.000}
\expandafter\gdef\csname EERadj@landmark_trajectories.calendar4.0.1.censor_weight\endcsname{0.605}
\expandafter\gdef\csname EERadj@landmark_trajectories.calendar4.0.2.censor_weight\endcsname{1.579}
\expandafter\gdef\csname EERadj@landmark_trajectories.calendar4.0.3.censor_weight\endcsname{0.746}
\expandafter\gdef\csname EERadj@landmark_trajectories.calendar4.0.4.censor_weight\endcsname{0.638}
\expandafter\gdef\csname EERadj@landmark_trajectories.calendar4.0.5.censor_weight\endcsname{0.704}
\expandafter\gdef\csname EERadj@landmark_trajectories.calendar4.0.6.censor_weight\endcsname{0.644}
\expandafter\gdef\csname EERadj@landmark_trajectories.calendar4.0.7.censor_weight\endcsname{1.140}
\expandafter\gdef\csname EERadj@landmark_trajectories.calendar4.0.8.censor_weight\endcsname{0.737}
\expandafter\gdef\csname EERadj@landmark_trajectories.calendar4.0.9.censor_weight\endcsname{0.000}
\expandafter\gdef\csname EERadj@landmark_trajectories.calendar4.1.1.censor_weight\endcsname{2.000}
\expandafter\gdef\csname EERadj@landmark_trajectories.calendar4.1.2.censor_weight\endcsname{8.000}
\expandafter\gdef\csname EERadj@landmark_trajectories.calendar4.1.3.censor_weight\endcsname{9.000}
\expandafter\gdef\csname EERadj@landmark_trajectories.calendar4.1.4.censor_weight\endcsname{9.000}
\expandafter\gdef\csname EERadj@landmark_trajectories.calendar4.1.5.censor_weight\endcsname{1.000}
\expandafter\gdef\csname EERadj@landmark_trajectories.calendar4.1.6.censor_weight\endcsname{0.000}
\expandafter\gdef\csname EERadj@landmark_trajectories.calendar4.1.7.censor_weight\endcsname{1.000}
\expandafter\gdef\csname EERadj@landmark_trajectories.calendar4.1.8.censor_weight\endcsname{1.000}
\expandafter\gdef\csname EERadj@landmark_trajectories.calendar4.1.9.censor_weight\endcsname{0.000}
\expandafter\gdef\csname EERadj@landmark_trajectories.school3.0.1.survival\endcsname{0.9567}
\expandafter\gdef\csname EERadj@landmark_trajectories.school3.0.2.survival\endcsname{0.9151}
\expandafter\gdef\csname EERadj@landmark_trajectories.school3.0.3.survival\endcsname{0.8767}
\expandafter\gdef\csname EERadj@landmark_trajectories.school3.0.4.survival\endcsname{0.8352}
\expandafter\gdef\csname EERadj@landmark_trajectories.school3.0.5.survival\endcsname{0.7937}
\expandafter\gdef\csname EERadj@landmark_trajectories.school3.0.6.survival\endcsname{0.7630}
\expandafter\gdef\csname EERadj@landmark_trajectories.school3.0.7.survival\endcsname{0.7273}
\expandafter\gdef\csname EERadj@landmark_trajectories.school3.0.8.survival\endcsname{0.6914}
\expandafter\gdef\csname EERadj@landmark_trajectories.school3.0.9.survival\endcsname{0.6583}
\expandafter\gdef\csname EERadj@landmark_trajectories.school3.1.1.survival\endcsname{0.9490}
\expandafter\gdef\csname EERadj@landmark_trajectories.school3.1.2.survival\endcsname{0.8866}
\expandafter\gdef\csname EERadj@landmark_trajectories.school3.1.3.survival\endcsname{0.8112}
\expandafter\gdef\csname EERadj@landmark_trajectories.school3.1.4.survival\endcsname{0.7735}
\expandafter\gdef\csname EERadj@landmark_trajectories.school3.1.5.survival\endcsname{0.7248}
\expandafter\gdef\csname EERadj@landmark_trajectories.school3.1.6.survival\endcsname{0.6858}
\expandafter\gdef\csname EERadj@landmark_trajectories.school3.1.7.survival\endcsname{0.6297}
\expandafter\gdef\csname EERadj@landmark_trajectories.school3.1.8.survival\endcsname{0.5977}
\expandafter\gdef\csname EERadj@landmark_trajectories.school3.1.9.survival\endcsname{0.5606}
\expandafter\gdef\csname EERadj@landmark_trajectories.calendar4.0.1.survival\endcsname{0.9584}
\expandafter\gdef\csname EERadj@landmark_trajectories.calendar4.0.2.survival\endcsname{0.9185}
\expandafter\gdef\csname EERadj@landmark_trajectories.calendar4.0.3.survival\endcsname{0.8804}
\expandafter\gdef\csname EERadj@landmark_trajectories.calendar4.0.4.survival\endcsname{0.8401}
\expandafter\gdef\csname EERadj@landmark_trajectories.calendar4.0.5.survival\endcsname{0.7970}
\expandafter\gdef\csname EERadj@landmark_trajectories.calendar4.0.6.survival\endcsname{0.7653}
\expandafter\gdef\csname EERadj@landmark_trajectories.calendar4.0.7.survival\endcsname{0.7299}
\expandafter\gdef\csname EERadj@landmark_trajectories.calendar4.0.8.survival\endcsname{0.6933}
\expandafter\gdef\csname EERadj@landmark_trajectories.calendar4.0.9.survival\endcsname{0.6599}
\expandafter\gdef\csname EERadj@landmark_trajectories.calendar4.1.1.survival\endcsname{0.9519}
\expandafter\gdef\csname EERadj@landmark_trajectories.calendar4.1.2.survival\endcsname{0.8945}
\expandafter\gdef\csname EERadj@landmark_trajectories.calendar4.1.3.survival\endcsname{0.8170}
\expandafter\gdef\csname EERadj@landmark_trajectories.calendar4.1.4.survival\endcsname{0.7760}
\expandafter\gdef\csname EERadj@landmark_trajectories.calendar4.1.5.survival\endcsname{0.7289}
\expandafter\gdef\csname EERadj@landmark_trajectories.calendar4.1.6.survival\endcsname{0.6891}
\expandafter\gdef\csname EERadj@landmark_trajectories.calendar4.1.7.survival\endcsname{0.6319}
\expandafter\gdef\csname EERadj@landmark_trajectories.calendar4.1.8.survival\endcsname{0.5994}
\expandafter\gdef\csname EERadj@landmark_trajectories.calendar4.1.9.survival\endcsname{0.5593}
\expandafter\gdef\csname EERadj@landmark_trajectories.school3.0.1.survival_pct\endcsname{95.7}
\expandafter\gdef\csname EERadj@landmark_trajectories.school3.0.2.survival_pct\endcsname{91.5}
\expandafter\gdef\csname EERadj@landmark_trajectories.school3.0.3.survival_pct\endcsname{87.7}
\expandafter\gdef\csname EERadj@landmark_trajectories.school3.0.4.survival_pct\endcsname{83.5}
\expandafter\gdef\csname EERadj@landmark_trajectories.school3.0.5.survival_pct\endcsname{79.4}
\expandafter\gdef\csname EERadj@landmark_trajectories.school3.0.6.survival_pct\endcsname{76.3}
\expandafter\gdef\csname EERadj@landmark_trajectories.school3.0.7.survival_pct\endcsname{72.7}
\expandafter\gdef\csname EERadj@landmark_trajectories.school3.0.8.survival_pct\endcsname{69.1}
\expandafter\gdef\csname EERadj@landmark_trajectories.school3.0.9.survival_pct\endcsname{65.8}
\expandafter\gdef\csname EERadj@landmark_trajectories.school3.1.1.survival_pct\endcsname{94.9}
\expandafter\gdef\csname EERadj@landmark_trajectories.school3.1.2.survival_pct\endcsname{88.7}
\expandafter\gdef\csname EERadj@landmark_trajectories.school3.1.3.survival_pct\endcsname{81.1}
\expandafter\gdef\csname EERadj@landmark_trajectories.school3.1.4.survival_pct\endcsname{77.3}
\expandafter\gdef\csname EERadj@landmark_trajectories.school3.1.5.survival_pct\endcsname{72.5}
\expandafter\gdef\csname EERadj@landmark_trajectories.school3.1.6.survival_pct\endcsname{68.6}
\expandafter\gdef\csname EERadj@landmark_trajectories.school3.1.7.survival_pct\endcsname{63.0}
\expandafter\gdef\csname EERadj@landmark_trajectories.school3.1.8.survival_pct\endcsname{59.8}
\expandafter\gdef\csname EERadj@landmark_trajectories.school3.1.9.survival_pct\endcsname{56.1}
\expandafter\gdef\csname EERadj@landmark_trajectories.calendar4.0.1.survival_pct\endcsname{95.8}
\expandafter\gdef\csname EERadj@landmark_trajectories.calendar4.0.2.survival_pct\endcsname{91.9}
\expandafter\gdef\csname EERadj@landmark_trajectories.calendar4.0.3.survival_pct\endcsname{88.0}
\expandafter\gdef\csname EERadj@landmark_trajectories.calendar4.0.4.survival_pct\endcsname{84.0}
\expandafter\gdef\csname EERadj@landmark_trajectories.calendar4.0.5.survival_pct\endcsname{79.7}
\expandafter\gdef\csname EERadj@landmark_trajectories.calendar4.0.6.survival_pct\endcsname{76.5}
\expandafter\gdef\csname EERadj@landmark_trajectories.calendar4.0.7.survival_pct\endcsname{73.0}
\expandafter\gdef\csname EERadj@landmark_trajectories.calendar4.0.8.survival_pct\endcsname{69.3}
\expandafter\gdef\csname EERadj@landmark_trajectories.calendar4.0.9.survival_pct\endcsname{66.0}
\expandafter\gdef\csname EERadj@landmark_trajectories.calendar4.1.1.survival_pct\endcsname{95.2}
\expandafter\gdef\csname EERadj@landmark_trajectories.calendar4.1.2.survival_pct\endcsname{89.4}
\expandafter\gdef\csname EERadj@landmark_trajectories.calendar4.1.3.survival_pct\endcsname{81.7}
\expandafter\gdef\csname EERadj@landmark_trajectories.calendar4.1.4.survival_pct\endcsname{77.6}
\expandafter\gdef\csname EERadj@landmark_trajectories.calendar4.1.5.survival_pct\endcsname{72.9}
\expandafter\gdef\csname EERadj@landmark_trajectories.calendar4.1.6.survival_pct\endcsname{68.9}
\expandafter\gdef\csname EERadj@landmark_trajectories.calendar4.1.7.survival_pct\endcsname{63.2}
\expandafter\gdef\csname EERadj@landmark_trajectories.calendar4.1.8.survival_pct\endcsname{59.9}
\expandafter\gdef\csname EERadj@landmark_trajectories.calendar4.1.9.survival_pct\endcsname{55.9}
\expandafter\gdef\csname EERadj@landmark_trajectories.school3.0.1.cif_interruption\endcsname{0.0090}
\expandafter\gdef\csname EERadj@landmark_trajectories.school3.0.2.cif_interruption\endcsname{0.0165}
\expandafter\gdef\csname EERadj@landmark_trajectories.school3.0.3.cif_interruption\endcsname{0.0265}
\expandafter\gdef\csname EERadj@landmark_trajectories.school3.0.4.cif_interruption\endcsname{0.0332}
\expandafter\gdef\csname EERadj@landmark_trajectories.school3.0.5.cif_interruption\endcsname{0.0420}
\expandafter\gdef\csname EERadj@landmark_trajectories.school3.0.6.cif_interruption\endcsname{0.0495}
\expandafter\gdef\csname EERadj@landmark_trajectories.school3.0.7.cif_interruption\endcsname{0.0572}
\expandafter\gdef\csname EERadj@landmark_trajectories.school3.0.8.cif_interruption\endcsname{0.0653}
\expandafter\gdef\csname EERadj@landmark_trajectories.school3.0.9.cif_interruption\endcsname{0.0720}
\expandafter\gdef\csname EERadj@landmark_trajectories.school3.1.1.cif_interruption\endcsname{0.0244}
\expandafter\gdef\csname EERadj@landmark_trajectories.school3.1.2.cif_interruption\endcsname{0.0489}
\expandafter\gdef\csname EERadj@landmark_trajectories.school3.1.3.cif_interruption\endcsname{0.1015}
\expandafter\gdef\csname EERadj@landmark_trajectories.school3.1.4.cif_interruption\endcsname{0.1109}
\expandafter\gdef\csname EERadj@landmark_trajectories.school3.1.5.cif_interruption\endcsname{0.1352}
\expandafter\gdef\csname EERadj@landmark_trajectories.school3.1.6.cif_interruption\endcsname{0.1547}
\expandafter\gdef\csname EERadj@landmark_trajectories.school3.1.7.cif_interruption\endcsname{0.1889}
\expandafter\gdef\csname EERadj@landmark_trajectories.school3.1.8.cif_interruption\endcsname{0.2061}
\expandafter\gdef\csname EERadj@landmark_trajectories.school3.1.9.cif_interruption\endcsname{0.2308}
\expandafter\gdef\csname EERadj@landmark_trajectories.calendar4.0.1.cif_interruption\endcsname{0.0103}
\expandafter\gdef\csname EERadj@landmark_trajectories.calendar4.0.2.cif_interruption\endcsname{0.0183}
\expandafter\gdef\csname EERadj@landmark_trajectories.calendar4.0.3.cif_interruption\endcsname{0.0276}
\expandafter\gdef\csname EERadj@landmark_trajectories.calendar4.0.4.cif_interruption\endcsname{0.0338}
\expandafter\gdef\csname EERadj@landmark_trajectories.calendar4.0.5.cif_interruption\endcsname{0.0429}
\expandafter\gdef\csname EERadj@landmark_trajectories.calendar4.0.6.cif_interruption\endcsname{0.0510}
\expandafter\gdef\csname EERadj@landmark_trajectories.calendar4.0.7.cif_interruption\endcsname{0.0590}
\expandafter\gdef\csname EERadj@landmark_trajectories.calendar4.0.8.cif_interruption\endcsname{0.0674}
\expandafter\gdef\csname EERadj@landmark_trajectories.calendar4.0.9.cif_interruption\endcsname{0.0742}
\expandafter\gdef\csname EERadj@landmark_trajectories.calendar4.1.1.cif_interruption\endcsname{0.0252}
\expandafter\gdef\csname EERadj@landmark_trajectories.calendar4.1.2.cif_interruption\endcsname{0.0482}
\expandafter\gdef\csname EERadj@landmark_trajectories.calendar4.1.3.cif_interruption\endcsname{0.1022}
\expandafter\gdef\csname EERadj@landmark_trajectories.calendar4.1.4.cif_interruption\endcsname{0.1118}
\expandafter\gdef\csname EERadj@landmark_trajectories.calendar4.1.5.cif_interruption\endcsname{0.1341}
\expandafter\gdef\csname EERadj@landmark_trajectories.calendar4.1.6.cif_interruption\endcsname{0.1565}
\expandafter\gdef\csname EERadj@landmark_trajectories.calendar4.1.7.cif_interruption\endcsname{0.1913}
\expandafter\gdef\csname EERadj@landmark_trajectories.calendar4.1.8.cif_interruption\endcsname{0.2088}
\expandafter\gdef\csname EERadj@landmark_trajectories.calendar4.1.9.cif_interruption\endcsname{0.2364}
\expandafter\gdef\csname EERadj@landmark_trajectories.school3.0.1.cif_interruption_pct\endcsname{0.9}
\expandafter\gdef\csname EERadj@landmark_trajectories.school3.0.2.cif_interruption_pct\endcsname{1.6}
\expandafter\gdef\csname EERadj@landmark_trajectories.school3.0.3.cif_interruption_pct\endcsname{2.7}
\expandafter\gdef\csname EERadj@landmark_trajectories.school3.0.4.cif_interruption_pct\endcsname{3.3}
\expandafter\gdef\csname EERadj@landmark_trajectories.school3.0.5.cif_interruption_pct\endcsname{4.2}
\expandafter\gdef\csname EERadj@landmark_trajectories.school3.0.6.cif_interruption_pct\endcsname{4.9}
\expandafter\gdef\csname EERadj@landmark_trajectories.school3.0.7.cif_interruption_pct\endcsname{5.7}
\expandafter\gdef\csname EERadj@landmark_trajectories.school3.0.8.cif_interruption_pct\endcsname{6.5}
\expandafter\gdef\csname EERadj@landmark_trajectories.school3.0.9.cif_interruption_pct\endcsname{7.2}
\expandafter\gdef\csname EERadj@landmark_trajectories.school3.1.1.cif_interruption_pct\endcsname{2.4}
\expandafter\gdef\csname EERadj@landmark_trajectories.school3.1.2.cif_interruption_pct\endcsname{4.9}
\expandafter\gdef\csname EERadj@landmark_trajectories.school3.1.3.cif_interruption_pct\endcsname{10.1}
\expandafter\gdef\csname EERadj@landmark_trajectories.school3.1.4.cif_interruption_pct\endcsname{11.1}
\expandafter\gdef\csname EERadj@landmark_trajectories.school3.1.5.cif_interruption_pct\endcsname{13.5}
\expandafter\gdef\csname EERadj@landmark_trajectories.school3.1.6.cif_interruption_pct\endcsname{15.5}
\expandafter\gdef\csname EERadj@landmark_trajectories.school3.1.7.cif_interruption_pct\endcsname{18.9}
\expandafter\gdef\csname EERadj@landmark_trajectories.school3.1.8.cif_interruption_pct\endcsname{20.6}
\expandafter\gdef\csname EERadj@landmark_trajectories.school3.1.9.cif_interruption_pct\endcsname{23.1}
\expandafter\gdef\csname EERadj@landmark_trajectories.calendar4.0.1.cif_interruption_pct\endcsname{1.0}
\expandafter\gdef\csname EERadj@landmark_trajectories.calendar4.0.2.cif_interruption_pct\endcsname{1.8}
\expandafter\gdef\csname EERadj@landmark_trajectories.calendar4.0.3.cif_interruption_pct\endcsname{2.8}
\expandafter\gdef\csname EERadj@landmark_trajectories.calendar4.0.4.cif_interruption_pct\endcsname{3.4}
\expandafter\gdef\csname EERadj@landmark_trajectories.calendar4.0.5.cif_interruption_pct\endcsname{4.3}
\expandafter\gdef\csname EERadj@landmark_trajectories.calendar4.0.6.cif_interruption_pct\endcsname{5.1}
\expandafter\gdef\csname EERadj@landmark_trajectories.calendar4.0.7.cif_interruption_pct\endcsname{5.9}
\expandafter\gdef\csname EERadj@landmark_trajectories.calendar4.0.8.cif_interruption_pct\endcsname{6.7}
\expandafter\gdef\csname EERadj@landmark_trajectories.calendar4.0.9.cif_interruption_pct\endcsname{7.4}
\expandafter\gdef\csname EERadj@landmark_trajectories.calendar4.1.1.cif_interruption_pct\endcsname{2.5}
\expandafter\gdef\csname EERadj@landmark_trajectories.calendar4.1.2.cif_interruption_pct\endcsname{4.8}
\expandafter\gdef\csname EERadj@landmark_trajectories.calendar4.1.3.cif_interruption_pct\endcsname{10.2}
\expandafter\gdef\csname EERadj@landmark_trajectories.calendar4.1.4.cif_interruption_pct\endcsname{11.2}
\expandafter\gdef\csname EERadj@landmark_trajectories.calendar4.1.5.cif_interruption_pct\endcsname{13.4}
\expandafter\gdef\csname EERadj@landmark_trajectories.calendar4.1.6.cif_interruption_pct\endcsname{15.7}
\expandafter\gdef\csname EERadj@landmark_trajectories.calendar4.1.7.cif_interruption_pct\endcsname{19.1}
\expandafter\gdef\csname EERadj@landmark_trajectories.calendar4.1.8.cif_interruption_pct\endcsname{20.9}
\expandafter\gdef\csname EERadj@landmark_trajectories.calendar4.1.9.cif_interruption_pct\endcsname{23.6}
\expandafter\gdef\csname EERadj@landmark_trajectories.school3.0.1.cif_credential\endcsname{0.0343}
\expandafter\gdef\csname EERadj@landmark_trajectories.school3.0.2.cif_credential\endcsname{0.0684}
\expandafter\gdef\csname EERadj@landmark_trajectories.school3.0.3.cif_credential\endcsname{0.0968}
\expandafter\gdef\csname EERadj@landmark_trajectories.school3.0.4.cif_credential\endcsname{0.1317}
\expandafter\gdef\csname EERadj@landmark_trajectories.school3.0.5.cif_credential\endcsname{0.1643}
\expandafter\gdef\csname EERadj@landmark_trajectories.school3.0.6.cif_credential\endcsname{0.1875}
\expandafter\gdef\csname EERadj@landmark_trajectories.school3.0.7.cif_credential\endcsname{0.2155}
\expandafter\gdef\csname EERadj@landmark_trajectories.school3.0.8.cif_credential\endcsname{0.2433}
\expandafter\gdef\csname EERadj@landmark_trajectories.school3.0.9.cif_credential\endcsname{0.2697}
\expandafter\gdef\csname EERadj@landmark_trajectories.school3.1.1.cif_credential\endcsname{0.0266}
\expandafter\gdef\csname EERadj@landmark_trajectories.school3.1.2.cif_credential\endcsname{0.0645}
\expandafter\gdef\csname EERadj@landmark_trajectories.school3.1.3.cif_credential\endcsname{0.0873}
\expandafter\gdef\csname EERadj@landmark_trajectories.school3.1.4.cif_credential\endcsname{0.1156}
\expandafter\gdef\csname EERadj@landmark_trajectories.school3.1.5.cif_credential\endcsname{0.1400}
\expandafter\gdef\csname EERadj@landmark_trajectories.school3.1.6.cif_credential\endcsname{0.1595}
\expandafter\gdef\csname EERadj@landmark_trajectories.school3.1.7.cif_credential\endcsname{0.1814}
\expandafter\gdef\csname EERadj@landmark_trajectories.school3.1.8.cif_credential\endcsname{0.1962}
\expandafter\gdef\csname EERadj@landmark_trajectories.school3.1.9.cif_credential\endcsname{0.2085}
\expandafter\gdef\csname EERadj@landmark_trajectories.calendar4.0.1.cif_credential\endcsname{0.0313}
\expandafter\gdef\csname EERadj@landmark_trajectories.calendar4.0.2.cif_credential\endcsname{0.0632}
\expandafter\gdef\csname EERadj@landmark_trajectories.calendar4.0.3.cif_credential\endcsname{0.0920}
\expandafter\gdef\csname EERadj@landmark_trajectories.calendar4.0.4.cif_credential\endcsname{0.1262}
\expandafter\gdef\csname EERadj@landmark_trajectories.calendar4.0.5.cif_credential\endcsname{0.1600}
\expandafter\gdef\csname EERadj@landmark_trajectories.calendar4.0.6.cif_credential\endcsname{0.1837}
\expandafter\gdef\csname EERadj@landmark_trajectories.calendar4.0.7.cif_credential\endcsname{0.2111}
\expandafter\gdef\csname EERadj@landmark_trajectories.calendar4.0.8.cif_credential\endcsname{0.2393}
\expandafter\gdef\csname EERadj@landmark_trajectories.calendar4.0.9.cif_credential\endcsname{0.2659}
\expandafter\gdef\csname EERadj@landmark_trajectories.calendar4.1.1.cif_credential\endcsname{0.0229}
\expandafter\gdef\csname EERadj@landmark_trajectories.calendar4.1.2.cif_credential\endcsname{0.0574}
\expandafter\gdef\csname EERadj@landmark_trajectories.calendar4.1.3.cif_credential\endcsname{0.0809}
\expandafter\gdef\csname EERadj@landmark_trajectories.calendar4.1.4.cif_credential\endcsname{0.1122}
\expandafter\gdef\csname EERadj@landmark_trajectories.calendar4.1.5.cif_credential\endcsname{0.1370}
\expandafter\gdef\csname EERadj@landmark_trajectories.calendar4.1.6.cif_credential\endcsname{0.1544}
\expandafter\gdef\csname EERadj@landmark_trajectories.calendar4.1.7.cif_credential\endcsname{0.1768}
\expandafter\gdef\csname EERadj@landmark_trajectories.calendar4.1.8.cif_credential\endcsname{0.1918}
\expandafter\gdef\csname EERadj@landmark_trajectories.calendar4.1.9.cif_credential\endcsname{0.2043}
\expandafter\gdef\csname EERadj@landmark_trajectories.school3.0.1.cif_credential_pct\endcsname{3.4}
\expandafter\gdef\csname EERadj@landmark_trajectories.school3.0.2.cif_credential_pct\endcsname{6.8}
\expandafter\gdef\csname EERadj@landmark_trajectories.school3.0.3.cif_credential_pct\endcsname{9.7}
\expandafter\gdef\csname EERadj@landmark_trajectories.school3.0.4.cif_credential_pct\endcsname{13.2}
\expandafter\gdef\csname EERadj@landmark_trajectories.school3.0.5.cif_credential_pct\endcsname{16.4}
\expandafter\gdef\csname EERadj@landmark_trajectories.school3.0.6.cif_credential_pct\endcsname{18.7}
\expandafter\gdef\csname EERadj@landmark_trajectories.school3.0.7.cif_credential_pct\endcsname{21.5}
\expandafter\gdef\csname EERadj@landmark_trajectories.school3.0.8.cif_credential_pct\endcsname{24.3}
\expandafter\gdef\csname EERadj@landmark_trajectories.school3.0.9.cif_credential_pct\endcsname{27.0}
\expandafter\gdef\csname EERadj@landmark_trajectories.school3.1.1.cif_credential_pct\endcsname{2.7}
\expandafter\gdef\csname EERadj@landmark_trajectories.school3.1.2.cif_credential_pct\endcsname{6.4}
\expandafter\gdef\csname EERadj@landmark_trajectories.school3.1.3.cif_credential_pct\endcsname{8.7}
\expandafter\gdef\csname EERadj@landmark_trajectories.school3.1.4.cif_credential_pct\endcsname{11.6}
\expandafter\gdef\csname EERadj@landmark_trajectories.school3.1.5.cif_credential_pct\endcsname{14.0}
\expandafter\gdef\csname EERadj@landmark_trajectories.school3.1.6.cif_credential_pct\endcsname{15.9}
\expandafter\gdef\csname EERadj@landmark_trajectories.school3.1.7.cif_credential_pct\endcsname{18.1}
\expandafter\gdef\csname EERadj@landmark_trajectories.school3.1.8.cif_credential_pct\endcsname{19.6}
\expandafter\gdef\csname EERadj@landmark_trajectories.school3.1.9.cif_credential_pct\endcsname{20.9}
\expandafter\gdef\csname EERadj@landmark_trajectories.calendar4.0.1.cif_credential_pct\endcsname{3.1}
\expandafter\gdef\csname EERadj@landmark_trajectories.calendar4.0.2.cif_credential_pct\endcsname{6.3}
\expandafter\gdef\csname EERadj@landmark_trajectories.calendar4.0.3.cif_credential_pct\endcsname{9.2}
\expandafter\gdef\csname EERadj@landmark_trajectories.calendar4.0.4.cif_credential_pct\endcsname{12.6}
\expandafter\gdef\csname EERadj@landmark_trajectories.calendar4.0.5.cif_credential_pct\endcsname{16.0}
\expandafter\gdef\csname EERadj@landmark_trajectories.calendar4.0.6.cif_credential_pct\endcsname{18.4}
\expandafter\gdef\csname EERadj@landmark_trajectories.calendar4.0.7.cif_credential_pct\endcsname{21.1}
\expandafter\gdef\csname EERadj@landmark_trajectories.calendar4.0.8.cif_credential_pct\endcsname{23.9}
\expandafter\gdef\csname EERadj@landmark_trajectories.calendar4.0.9.cif_credential_pct\endcsname{26.6}
\expandafter\gdef\csname EERadj@landmark_trajectories.calendar4.1.1.cif_credential_pct\endcsname{2.3}
\expandafter\gdef\csname EERadj@landmark_trajectories.calendar4.1.2.cif_credential_pct\endcsname{5.7}
\expandafter\gdef\csname EERadj@landmark_trajectories.calendar4.1.3.cif_credential_pct\endcsname{8.1}
\expandafter\gdef\csname EERadj@landmark_trajectories.calendar4.1.4.cif_credential_pct\endcsname{11.2}
\expandafter\gdef\csname EERadj@landmark_trajectories.calendar4.1.5.cif_credential_pct\endcsname{13.7}
\expandafter\gdef\csname EERadj@landmark_trajectories.calendar4.1.6.cif_credential_pct\endcsname{15.4}
\expandafter\gdef\csname EERadj@landmark_trajectories.calendar4.1.7.cif_credential_pct\endcsname{17.7}
\expandafter\gdef\csname EERadj@landmark_trajectories.calendar4.1.8.cif_credential_pct\endcsname{19.2}
\expandafter\gdef\csname EERadj@landmark_trajectories.calendar4.1.9.cif_credential_pct\endcsname{20.4}
\expandafter\gdef\csname EERadj@landmark_trajectories.school3.0.1.raw_at_risk\endcsname{42,909}
\expandafter\gdef\csname EERadj@landmark_trajectories.school3.0.2.raw_at_risk\endcsname{41,410}
\expandafter\gdef\csname EERadj@landmark_trajectories.school3.0.3.raw_at_risk\endcsname{39,727}
\expandafter\gdef\csname EERadj@landmark_trajectories.school3.0.4.raw_at_risk\endcsname{38,138}
\expandafter\gdef\csname EERadj@landmark_trajectories.school3.0.5.raw_at_risk\endcsname{36,482}
\expandafter\gdef\csname EERadj@landmark_trajectories.school3.0.6.raw_at_risk\endcsname{34,738}
\expandafter\gdef\csname EERadj@landmark_trajectories.school3.0.7.raw_at_risk\endcsname{33,105}
\expandafter\gdef\csname EERadj@landmark_trajectories.school3.0.8.raw_at_risk\endcsname{31,429}
\expandafter\gdef\csname EERadj@landmark_trajectories.school3.0.9.raw_at_risk\endcsname{29,842}
\expandafter\gdef\csname EERadj@landmark_trajectories.school3.1.1.raw_at_risk\endcsname{451}
\expandafter\gdef\csname EERadj@landmark_trajectories.school3.1.2.raw_at_risk\endcsname{426}
\expandafter\gdef\csname EERadj@landmark_trajectories.school3.1.3.raw_at_risk\endcsname{388}
\expandafter\gdef\csname EERadj@landmark_trajectories.school3.1.4.raw_at_risk\endcsname{344}
\expandafter\gdef\csname EERadj@landmark_trajectories.school3.1.5.raw_at_risk\endcsname{318}
\expandafter\gdef\csname EERadj@landmark_trajectories.school3.1.6.raw_at_risk\endcsname{297}
\expandafter\gdef\csname EERadj@landmark_trajectories.school3.1.7.raw_at_risk\endcsname{281}
\expandafter\gdef\csname EERadj@landmark_trajectories.school3.1.8.raw_at_risk\endcsname{256}
\expandafter\gdef\csname EERadj@landmark_trajectories.school3.1.9.raw_at_risk\endcsname{242}
\expandafter\gdef\csname EERadj@landmark_trajectories.calendar4.0.1.raw_at_risk\endcsname{41,951}
\expandafter\gdef\csname EERadj@landmark_trajectories.calendar4.0.2.raw_at_risk\endcsname{40,472}
\expandafter\gdef\csname EERadj@landmark_trajectories.calendar4.0.3.raw_at_risk\endcsname{38,794}
\expandafter\gdef\csname EERadj@landmark_trajectories.calendar4.0.4.raw_at_risk\endcsname{37,212}
\expandafter\gdef\csname EERadj@landmark_trajectories.calendar4.0.5.raw_at_risk\endcsname{35,567}
\expandafter\gdef\csname EERadj@landmark_trajectories.calendar4.0.6.raw_at_risk\endcsname{33,823}
\expandafter\gdef\csname EERadj@landmark_trajectories.calendar4.0.7.raw_at_risk\endcsname{32,200}
\expandafter\gdef\csname EERadj@landmark_trajectories.calendar4.0.8.raw_at_risk\endcsname{30,543}
\expandafter\gdef\csname EERadj@landmark_trajectories.calendar4.0.9.raw_at_risk\endcsname{28,968}
\expandafter\gdef\csname EERadj@landmark_trajectories.calendar4.1.1.raw_at_risk\endcsname{437}
\expandafter\gdef\csname EERadj@landmark_trajectories.calendar4.1.2.raw_at_risk\endcsname{414}
\expandafter\gdef\csname EERadj@landmark_trajectories.calendar4.1.3.raw_at_risk\endcsname{381}
\expandafter\gdef\csname EERadj@landmark_trajectories.calendar4.1.4.raw_at_risk\endcsname{339}
\expandafter\gdef\csname EERadj@landmark_trajectories.calendar4.1.5.raw_at_risk\endcsname{313}
\expandafter\gdef\csname EERadj@landmark_trajectories.calendar4.1.6.raw_at_risk\endcsname{293}
\expandafter\gdef\csname EERadj@landmark_trajectories.calendar4.1.7.raw_at_risk\endcsname{277}
\expandafter\gdef\csname EERadj@landmark_trajectories.calendar4.1.8.raw_at_risk\endcsname{253}
\expandafter\gdef\csname EERadj@landmark_trajectories.calendar4.1.9.raw_at_risk\endcsname{239}
\expandafter\gdef\csname EERadj@landmark_trajectories.school3.0.1.raw_interruption\endcsname{316}
\expandafter\gdef\csname EERadj@landmark_trajectories.school3.0.2.raw_interruption\endcsname{327}
\expandafter\gdef\csname EERadj@landmark_trajectories.school3.0.3.raw_interruption\endcsname{339}
\expandafter\gdef\csname EERadj@landmark_trajectories.school3.0.4.raw_interruption\endcsname{306}
\expandafter\gdef\csname EERadj@landmark_trajectories.school3.0.5.raw_interruption\endcsname{352}
\expandafter\gdef\csname EERadj@landmark_trajectories.school3.0.6.raw_interruption\endcsname{346}
\expandafter\gdef\csname EERadj@landmark_trajectories.school3.0.7.raw_interruption\endcsname{337}
\expandafter\gdef\csname EERadj@landmark_trajectories.school3.0.8.raw_interruption\endcsname{333}
\expandafter\gdef\csname EERadj@landmark_trajectories.school3.0.9.raw_interruption\endcsname{285}
\expandafter\gdef\csname EERadj@landmark_trajectories.school3.1.1.raw_interruption\endcsname{11}
\expandafter\gdef\csname EERadj@landmark_trajectories.school3.1.2.raw_interruption\endcsname{11}
\expandafter\gdef\csname EERadj@landmark_trajectories.school3.1.3.raw_interruption\endcsname{23}
\expandafter\gdef\csname EERadj@landmark_trajectories.school3.1.4.raw_interruption\endcsname{4}
\expandafter\gdef\csname EERadj@landmark_trajectories.school3.1.5.raw_interruption\endcsname{10}
\expandafter\gdef\csname EERadj@landmark_trajectories.school3.1.6.raw_interruption\endcsname{8}
\expandafter\gdef\csname EERadj@landmark_trajectories.school3.1.7.raw_interruption\endcsname{14}
\expandafter\gdef\csname EERadj@landmark_trajectories.school3.1.8.raw_interruption\endcsname{7}
\expandafter\gdef\csname EERadj@landmark_trajectories.school3.1.9.raw_interruption\endcsname{10}
\expandafter\gdef\csname EERadj@landmark_trajectories.calendar4.0.1.raw_interruption\endcsname{324}
\expandafter\gdef\csname EERadj@landmark_trajectories.calendar4.0.2.raw_interruption\endcsname{338}
\expandafter\gdef\csname EERadj@landmark_trajectories.calendar4.0.3.raw_interruption\endcsname{347}
\expandafter\gdef\csname EERadj@landmark_trajectories.calendar4.0.4.raw_interruption\endcsname{305}
\expandafter\gdef\csname EERadj@landmark_trajectories.calendar4.0.5.raw_interruption\endcsname{354}
\expandafter\gdef\csname EERadj@landmark_trajectories.calendar4.0.6.raw_interruption\endcsname{356}
\expandafter\gdef\csname EERadj@landmark_trajectories.calendar4.0.7.raw_interruption\endcsname{344}
\expandafter\gdef\csname EERadj@landmark_trajectories.calendar4.0.8.raw_interruption\endcsname{343}
\expandafter\gdef\csname EERadj@landmark_trajectories.calendar4.0.9.raw_interruption\endcsname{289}
\expandafter\gdef\csname EERadj@landmark_trajectories.calendar4.1.1.raw_interruption\endcsname{11}
\expandafter\gdef\csname EERadj@landmark_trajectories.calendar4.1.2.raw_interruption\endcsname{10}
\expandafter\gdef\csname EERadj@landmark_trajectories.calendar4.1.3.raw_interruption\endcsname{23}
\expandafter\gdef\csname EERadj@landmark_trajectories.calendar4.1.4.raw_interruption\endcsname{4}
\expandafter\gdef\csname EERadj@landmark_trajectories.calendar4.1.5.raw_interruption\endcsname{9}
\expandafter\gdef\csname EERadj@landmark_trajectories.calendar4.1.6.raw_interruption\endcsname{9}
\expandafter\gdef\csname EERadj@landmark_trajectories.calendar4.1.7.raw_interruption\endcsname{14}
\expandafter\gdef\csname EERadj@landmark_trajectories.calendar4.1.8.raw_interruption\endcsname{7}
\expandafter\gdef\csname EERadj@landmark_trajectories.calendar4.1.9.raw_interruption\endcsname{11}
\expandafter\gdef\csname EERadj@landmark_trajectories.school3.0.1.raw_credential\endcsname{1,125}
\expandafter\gdef\csname EERadj@landmark_trajectories.school3.0.2.raw_credential\endcsname{1,303}
\expandafter\gdef\csname EERadj@landmark_trajectories.school3.0.3.raw_credential\endcsname{1,201}
\expandafter\gdef\csname EERadj@landmark_trajectories.school3.0.4.raw_credential\endcsname{1,286}
\expandafter\gdef\csname EERadj@landmark_trajectories.school3.0.5.raw_credential\endcsname{1,328}
\expandafter\gdef\csname EERadj@landmark_trajectories.school3.0.6.raw_credential\endcsname{1,216}
\expandafter\gdef\csname EERadj@landmark_trajectories.school3.0.7.raw_credential\endcsname{1,275}
\expandafter\gdef\csname EERadj@landmark_trajectories.school3.0.8.raw_credential\endcsname{1,193}
\expandafter\gdef\csname EERadj@landmark_trajectories.school3.0.9.raw_credential\endcsname{1,291}
\expandafter\gdef\csname EERadj@landmark_trajectories.school3.1.1.raw_credential\endcsname{12}
\expandafter\gdef\csname EERadj@landmark_trajectories.school3.1.2.raw_credential\endcsname{17}
\expandafter\gdef\csname EERadj@landmark_trajectories.school3.1.3.raw_credential\endcsname{10}
\expandafter\gdef\csname EERadj@landmark_trajectories.school3.1.4.raw_credential\endcsname{12}
\expandafter\gdef\csname EERadj@landmark_trajectories.school3.1.5.raw_credential\endcsname{10}
\expandafter\gdef\csname EERadj@landmark_trajectories.school3.1.6.raw_credential\endcsname{8}
\expandafter\gdef\csname EERadj@landmark_trajectories.school3.1.7.raw_credential\endcsname{9}
\expandafter\gdef\csname EERadj@landmark_trajectories.school3.1.8.raw_credential\endcsname{6}
\expandafter\gdef\csname EERadj@landmark_trajectories.school3.1.9.raw_credential\endcsname{5}
\expandafter\gdef\csname EERadj@landmark_trajectories.calendar4.0.1.raw_credential\endcsname{1,098}
\expandafter\gdef\csname EERadj@landmark_trajectories.calendar4.0.2.raw_credential\endcsname{1,287}
\expandafter\gdef\csname EERadj@landmark_trajectories.calendar4.0.3.raw_credential\endcsname{1,189}
\expandafter\gdef\csname EERadj@landmark_trajectories.calendar4.0.4.raw_credential\endcsname{1,279}
\expandafter\gdef\csname EERadj@landmark_trajectories.calendar4.0.5.raw_credential\endcsname{1,331}
\expandafter\gdef\csname EERadj@landmark_trajectories.calendar4.0.6.raw_credential\endcsname{1,202}
\expandafter\gdef\csname EERadj@landmark_trajectories.calendar4.0.7.raw_credential\endcsname{1,254}
\expandafter\gdef\csname EERadj@landmark_trajectories.calendar4.0.8.raw_credential\endcsname{1,178}
\expandafter\gdef\csname EERadj@landmark_trajectories.calendar4.0.9.raw_credential\endcsname{1,170}
\expandafter\gdef\csname EERadj@landmark_trajectories.calendar4.1.1.raw_credential\endcsname{10}
\expandafter\gdef\csname EERadj@landmark_trajectories.calendar4.1.2.raw_credential\endcsname{15}
\expandafter\gdef\csname EERadj@landmark_trajectories.calendar4.1.3.raw_credential\endcsname{10}
\expandafter\gdef\csname EERadj@landmark_trajectories.calendar4.1.4.raw_credential\endcsname{13}
\expandafter\gdef\csname EERadj@landmark_trajectories.calendar4.1.5.raw_credential\endcsname{10}
\expandafter\gdef\csname EERadj@landmark_trajectories.calendar4.1.6.raw_credential\endcsname{7}
\expandafter\gdef\csname EERadj@landmark_trajectories.calendar4.1.7.raw_credential\endcsname{9}
\expandafter\gdef\csname EERadj@landmark_trajectories.calendar4.1.8.raw_credential\endcsname{6}
\expandafter\gdef\csname EERadj@landmark_trajectories.calendar4.1.9.raw_credential\endcsname{5}
\expandafter\gdef\csname EERadj@landmark_trajectories.school3.0.1.raw_censor\endcsname{58}
\expandafter\gdef\csname EERadj@landmark_trajectories.school3.0.2.raw_censor\endcsname{53}
\expandafter\gdef\csname EERadj@landmark_trajectories.school3.0.3.raw_censor\endcsname{49}
\expandafter\gdef\csname EERadj@landmark_trajectories.school3.0.4.raw_censor\endcsname{64}
\expandafter\gdef\csname EERadj@landmark_trajectories.school3.0.5.raw_censor\endcsname{64}
\expandafter\gdef\csname EERadj@landmark_trajectories.school3.0.6.raw_censor\endcsname{71}
\expandafter\gdef\csname EERadj@landmark_trajectories.school3.0.7.raw_censor\endcsname{64}
\expandafter\gdef\csname EERadj@landmark_trajectories.school3.0.8.raw_censor\endcsname{61}
\expandafter\gdef\csname EERadj@landmark_trajectories.school3.0.9.raw_censor\endcsname{0}
\expandafter\gdef\csname EERadj@landmark_trajectories.school3.1.1.raw_censor\endcsname{2}
\expandafter\gdef\csname EERadj@landmark_trajectories.school3.1.2.raw_censor\endcsname{10}
\expandafter\gdef\csname EERadj@landmark_trajectories.school3.1.3.raw_censor\endcsname{11}
\expandafter\gdef\csname EERadj@landmark_trajectories.school3.1.4.raw_censor\endcsname{10}
\expandafter\gdef\csname EERadj@landmark_trajectories.school3.1.5.raw_censor\endcsname{1}
\expandafter\gdef\csname EERadj@landmark_trajectories.school3.1.6.raw_censor\endcsname{0}
\expandafter\gdef\csname EERadj@landmark_trajectories.school3.1.7.raw_censor\endcsname{2}
\expandafter\gdef\csname EERadj@landmark_trajectories.school3.1.8.raw_censor\endcsname{1}
\expandafter\gdef\csname EERadj@landmark_trajectories.school3.1.9.raw_censor\endcsname{0}
\expandafter\gdef\csname EERadj@landmark_trajectories.calendar4.0.1.raw_censor\endcsname{57}
\expandafter\gdef\csname EERadj@landmark_trajectories.calendar4.0.2.raw_censor\endcsname{53}
\expandafter\gdef\csname EERadj@landmark_trajectories.calendar4.0.3.raw_censor\endcsname{46}
\expandafter\gdef\csname EERadj@landmark_trajectories.calendar4.0.4.raw_censor\endcsname{61}
\expandafter\gdef\csname EERadj@landmark_trajectories.calendar4.0.5.raw_censor\endcsname{59}
\expandafter\gdef\csname EERadj@landmark_trajectories.calendar4.0.6.raw_censor\endcsname{65}
\expandafter\gdef\csname EERadj@landmark_trajectories.calendar4.0.7.raw_censor\endcsname{59}
\expandafter\gdef\csname EERadj@landmark_trajectories.calendar4.0.8.raw_censor\endcsname{54}
\expandafter\gdef\csname EERadj@landmark_trajectories.calendar4.0.9.raw_censor\endcsname{0}
\expandafter\gdef\csname EERadj@landmark_trajectories.calendar4.1.1.raw_censor\endcsname{2}
\expandafter\gdef\csname EERadj@landmark_trajectories.calendar4.1.2.raw_censor\endcsname{8}
\expandafter\gdef\csname EERadj@landmark_trajectories.calendar4.1.3.raw_censor\endcsname{9}
\expandafter\gdef\csname EERadj@landmark_trajectories.calendar4.1.4.raw_censor\endcsname{9}
\expandafter\gdef\csname EERadj@landmark_trajectories.calendar4.1.5.raw_censor\endcsname{1}
\expandafter\gdef\csname EERadj@landmark_trajectories.calendar4.1.6.raw_censor\endcsname{0}
\expandafter\gdef\csname EERadj@landmark_trajectories.calendar4.1.7.raw_censor\endcsname{1}
\expandafter\gdef\csname EERadj@landmark_trajectories.calendar4.1.8.raw_censor\endcsname{1}
\expandafter\gdef\csname EERadj@landmark_trajectories.calendar4.1.9.raw_censor\endcsname{0}
\expandafter\gdef\csname EERadj@landmark_counts.school3.calendar_only_people_excluded.value\endcsname{25}
\expandafter\gdef\csname EERadj@landmark_counts.school3.first_exposure_rows_before_retained_onset.value\endcsname{459}
\expandafter\gdef\csname EERadj@landmark_counts.school3.first_exposure_rows_without_retained_onset.value\endcsname{7}
\expandafter\gdef\csname EERadj@landmark_counts.school3.entrants_before_common_support.value\endcsname{452}
\expandafter\gdef\csname EERadj@landmark_counts.school3.entrants_excluded_no_age_calendar_controls.value\endcsname{1}
\expandafter\gdef\csname EERadj@landmark_counts.school3.control_index_rows_outside_entrant_blocks.value\endcsname{13,318}
\expandafter\gdef\csname EERadj@landmark_counts.school3.supported_age_calendar_blocks.value\endcsname{213}
\expandafter\gdef\csname EERadj@landmark_counts.school3.control_index_paths_with_later_conception.value\endcsname{2,089}
\expandafter\gdef\csname EERadj@landmark_counts.school3.entrant_index_rows_in_summer.value\endcsname{98}
\expandafter\gdef\csname EERadj@landmark_counts.calendar4.calendar_only_people_excluded.value\endcsname{24}
\expandafter\gdef\csname EERadj@landmark_counts.calendar4.first_exposure_rows_before_retained_onset.value\endcsname{445}
\expandafter\gdef\csname EERadj@landmark_counts.calendar4.first_exposure_rows_without_retained_onset.value\endcsname{7}
\expandafter\gdef\csname EERadj@landmark_counts.calendar4.entrants_before_common_support.value\endcsname{438}
\expandafter\gdef\csname EERadj@landmark_counts.calendar4.entrants_excluded_no_age_calendar_controls.value\endcsname{1}
\expandafter\gdef\csname EERadj@landmark_counts.calendar4.control_index_rows_outside_entrant_blocks.value\endcsname{13,384}
\expandafter\gdef\csname EERadj@landmark_counts.calendar4.supported_age_calendar_blocks.value\endcsname{210}
\expandafter\gdef\csname EERadj@landmark_counts.calendar4.control_index_paths_with_later_conception.value\endcsname{1,989}
\expandafter\gdef\csname EERadj@landmark_counts.calendar4.entrant_index_rows_in_summer.value\endcsname{93}

\providecommand{\EERjoint}[1]{\csname EERjoint@#1\endcsname}
\expandafter\def\csname EERjoint@results.full_linear.k1.rc\endcsname{0.00}
\expandafter\def\csname EERjoint@results.full_linear.k1.converged\endcsname{1.00}
\expandafter\def\csname EERjoint@results.full_linear.k1.b\endcsname{1.36}
\expandafter\def\csname EERjoint@results.full_linear.k1.se\endcsname{0.09}
\expandafter\def\csname EERjoint@results.full_linear.k1.se_cluster\endcsname{0.09}
\expandafter\def\csname EERjoint@results.full_linear.k1.hr\endcsname{3.91}
\expandafter\def\csname EERjoint@results.full_linear.k1.ci_lo\endcsname{3.30}
\expandafter\def\csname EERjoint@results.full_linear.k1.ci_hi\endcsname{4.64}
\expandafter\def\csname EERjoint@results.full_linear.k1.b_absent\endcsname{1.38}
\expandafter\def\csname EERjoint@results.full_linear.k1.se_absent\endcsname{0.14}
\expandafter\def\csname EERjoint@results.full_linear.k1.ll\endcsname{-6299.05}
\expandafter\def\csname EERjoint@results.full_linear.k1.kparams\endcsname{8.00}
\expandafter\def\csname EERjoint@results.full_linear.k1.women\endcsname{2,828}
\expandafter\def\csname EERjoint@results.full_linear.k1.person_months\endcsname{64,598}
\expandafter\def\csname EERjoint@results.full_linear.k1.dropout_events\endcsname{711}
\expandafter\def\csname EERjoint@results.full_linear.k1.pregnancy_events\endcsname{478}
\expandafter\def\csname EERjoint@results.full_linear.k1.exposed_months\endcsname{7,893}
\expandafter\def\csname EERjoint@results.full_linear.k1.rel_grad\endcsname{.}
\expandafter\def\csname EERjoint@results.full_linear.k1.v_pd\endcsname{.}
\expandafter\def\csname EERjoint@results.full_linear.k1.v_scaffold\endcsname{.}
\expandafter\def\csname EERjoint@results.full_linear.k1.clip_hits\endcsname{.}
\expandafter\def\csname EERjoint@results.full_linear.k1.n_starts\endcsname{.}
\expandafter\def\csname EERjoint@results.full_linear.k1.n_finite\endcsname{.}
\expandafter\def\csname EERjoint@results.full_linear.k1.n_bfgs_conv\endcsname{.}
\expandafter\def\csname EERjoint@results.full_linear.k1.n_aborted\endcsname{.}
\expandafter\def\csname EERjoint@results.full_linear.k1.best_start\endcsname{.}
\expandafter\def\csname EERjoint@results.full_linear.k1.lambda_p\endcsname{.}
\expandafter\def\csname EERjoint@results.full_linear.k1.lambda_d\endcsname{.}
\expandafter\def\csname EERjoint@results.full_linear.k1.min_share\endcsname{.}
\expandafter\def\csname EERjoint@results.full_linear.k1.reproduction_difference\endcsname{.}
\expandafter\def\csname EERjoint@results.full_linear.k1.iteration_ceiling\endcsname{.}
\expandafter\def\csname EERjoint@results.full_linear.k1.aic\endcsname{12614.10}
\expandafter\def\csname EERjoint@results.full_linear.k1.bic_persons\endcsname{12661.68}
\expandafter\def\csname EERjoint@results.full_linear.k1.cluster_ci_lo\endcsname{3.29}
\expandafter\def\csname EERjoint@results.full_linear.k1.cluster_ci_hi\endcsname{4.65}
\expandafter\def\csname EERjoint@results.full_linear.k2.rc\endcsname{0.00}
\expandafter\def\csname EERjoint@results.full_linear.k2.converged\endcsname{1.00}
\expandafter\def\csname EERjoint@results.full_linear.k2.b\endcsname{1.32}
\expandafter\def\csname EERjoint@results.full_linear.k2.se\endcsname{0.11}
\expandafter\def\csname EERjoint@results.full_linear.k2.se_cluster\endcsname{.}
\expandafter\def\csname EERjoint@results.full_linear.k2.hr\endcsname{3.76}
\expandafter\def\csname EERjoint@results.full_linear.k2.ci_lo\endcsname{3.00}
\expandafter\def\csname EERjoint@results.full_linear.k2.ci_hi\endcsname{4.71}
\expandafter\def\csname EERjoint@results.full_linear.k2.b_absent\endcsname{1.37}
\expandafter\def\csname EERjoint@results.full_linear.k2.se_absent\endcsname{0.14}
\expandafter\def\csname EERjoint@results.full_linear.k2.ll\endcsname{-6281.43}
\expandafter\def\csname EERjoint@results.full_linear.k2.kparams\endcsname{11.00}
\expandafter\def\csname EERjoint@results.full_linear.k2.women\endcsname{2,828}
\expandafter\def\csname EERjoint@results.full_linear.k2.person_months\endcsname{64,598}
\expandafter\def\csname EERjoint@results.full_linear.k2.dropout_events\endcsname{711}
\expandafter\def\csname EERjoint@results.full_linear.k2.pregnancy_events\endcsname{478}
\expandafter\def\csname EERjoint@results.full_linear.k2.exposed_months\endcsname{7,893}
\expandafter\def\csname EERjoint@results.full_linear.k2.rel_grad\endcsname{7.32e-08}
\expandafter\def\csname EERjoint@results.full_linear.k2.v_pd\endcsname{1.00}
\expandafter\def\csname EERjoint@results.full_linear.k2.v_scaffold\endcsname{0.00}
\expandafter\def\csname EERjoint@results.full_linear.k2.clip_hits\endcsname{0.00}
\expandafter\def\csname EERjoint@results.full_linear.k2.n_starts\endcsname{7}
\expandafter\def\csname EERjoint@results.full_linear.k2.n_finite\endcsname{7}
\expandafter\def\csname EERjoint@results.full_linear.k2.n_bfgs_conv\endcsname{7}
\expandafter\def\csname EERjoint@results.full_linear.k2.n_aborted\endcsname{0.00}
\expandafter\def\csname EERjoint@results.full_linear.k2.best_start\endcsname{1}
\expandafter\def\csname EERjoint@results.full_linear.k2.lambda_p\endcsname{-4.11}
\expandafter\def\csname EERjoint@results.full_linear.k2.lambda_d\endcsname{-0.16}
\expandafter\def\csname EERjoint@results.full_linear.k2.min_share\endcsname{0.034}
\expandafter\def\csname EERjoint@results.full_linear.k2.reproduction_difference\endcsname{.}
\expandafter\def\csname EERjoint@results.full_linear.k2.iteration_ceiling\endcsname{100.00}
\expandafter\def\csname EERjoint@results.full_linear.k2.aic\endcsname{12584.85}
\expandafter\def\csname EERjoint@results.full_linear.k2.bic_persons\endcsname{12650.27}
\expandafter\def\csname EERjoint@results.full_linear.k2.cluster_ci_lo\endcsname{.}
\expandafter\def\csname EERjoint@results.full_linear.k2.cluster_ci_hi\endcsname{.}
\expandafter\def\csname EERjoint@results.restricted_linear.k1.rc\endcsname{0.00}
\expandafter\def\csname EERjoint@results.restricted_linear.k1.converged\endcsname{1.00}
\expandafter\def\csname EERjoint@results.restricted_linear.k1.b\endcsname{1.37}
\expandafter\def\csname EERjoint@results.restricted_linear.k1.se\endcsname{0.09}
\expandafter\def\csname EERjoint@results.restricted_linear.k1.se_cluster\endcsname{0.09}
\expandafter\def\csname EERjoint@results.restricted_linear.k1.hr\endcsname{3.92}
\expandafter\def\csname EERjoint@results.restricted_linear.k1.ci_lo\endcsname{3.31}
\expandafter\def\csname EERjoint@results.restricted_linear.k1.ci_hi\endcsname{4.65}
\expandafter\def\csname EERjoint@results.restricted_linear.k1.b_absent\endcsname{.}
\expandafter\def\csname EERjoint@results.restricted_linear.k1.se_absent\endcsname{.}
\expandafter\def\csname EERjoint@results.restricted_linear.k1.ll\endcsname{-6039.67}
\expandafter\def\csname EERjoint@results.restricted_linear.k1.kparams\endcsname{7.00}
\expandafter\def\csname EERjoint@results.restricted_linear.k1.women\endcsname{2,708}
\expandafter\def\csname EERjoint@results.restricted_linear.k1.person_months\endcsname{62,478}
\expandafter\def\csname EERjoint@results.restricted_linear.k1.dropout_events\endcsname{648}
\expandafter\def\csname EERjoint@results.restricted_linear.k1.pregnancy_events\endcsname{478}
\expandafter\def\csname EERjoint@results.restricted_linear.k1.exposed_months\endcsname{5,773}
\expandafter\def\csname EERjoint@results.restricted_linear.k1.rel_grad\endcsname{.}
\expandafter\def\csname EERjoint@results.restricted_linear.k1.v_pd\endcsname{.}
\expandafter\def\csname EERjoint@results.restricted_linear.k1.v_scaffold\endcsname{.}
\expandafter\def\csname EERjoint@results.restricted_linear.k1.clip_hits\endcsname{.}
\expandafter\def\csname EERjoint@results.restricted_linear.k1.n_starts\endcsname{.}
\expandafter\def\csname EERjoint@results.restricted_linear.k1.n_finite\endcsname{.}
\expandafter\def\csname EERjoint@results.restricted_linear.k1.n_bfgs_conv\endcsname{.}
\expandafter\def\csname EERjoint@results.restricted_linear.k1.n_aborted\endcsname{.}
\expandafter\def\csname EERjoint@results.restricted_linear.k1.best_start\endcsname{.}
\expandafter\def\csname EERjoint@results.restricted_linear.k1.lambda_p\endcsname{.}
\expandafter\def\csname EERjoint@results.restricted_linear.k1.lambda_d\endcsname{.}
\expandafter\def\csname EERjoint@results.restricted_linear.k1.min_share\endcsname{.}
\expandafter\def\csname EERjoint@results.restricted_linear.k1.reproduction_difference\endcsname{.}
\expandafter\def\csname EERjoint@results.restricted_linear.k1.iteration_ceiling\endcsname{.}
\expandafter\def\csname EERjoint@results.restricted_linear.k1.aic\endcsname{12093.34}
\expandafter\def\csname EERjoint@results.restricted_linear.k1.bic_persons\endcsname{12134.67}
\expandafter\def\csname EERjoint@results.restricted_linear.k1.cluster_ci_lo\endcsname{3.30}
\expandafter\def\csname EERjoint@results.restricted_linear.k1.cluster_ci_hi\endcsname{4.66}
\expandafter\def\csname EERjoint@results.restricted_linear.k2.rc\endcsname{0.00}
\expandafter\def\csname EERjoint@results.restricted_linear.k2.converged\endcsname{1.00}
\expandafter\def\csname EERjoint@results.restricted_linear.k2.b\endcsname{1.33}
\expandafter\def\csname EERjoint@results.restricted_linear.k2.se\endcsname{0.11}
\expandafter\def\csname EERjoint@results.restricted_linear.k2.se_cluster\endcsname{.}
\expandafter\def\csname EERjoint@results.restricted_linear.k2.hr\endcsname{3.78}
\expandafter\def\csname EERjoint@results.restricted_linear.k2.ci_lo\endcsname{3.02}
\expandafter\def\csname EERjoint@results.restricted_linear.k2.ci_hi\endcsname{4.74}
\expandafter\def\csname EERjoint@results.restricted_linear.k2.b_absent\endcsname{.}
\expandafter\def\csname EERjoint@results.restricted_linear.k2.se_absent\endcsname{.}
\expandafter\def\csname EERjoint@results.restricted_linear.k2.ll\endcsname{-6022.07}
\expandafter\def\csname EERjoint@results.restricted_linear.k2.kparams\endcsname{10.00}
\expandafter\def\csname EERjoint@results.restricted_linear.k2.women\endcsname{2,708}
\expandafter\def\csname EERjoint@results.restricted_linear.k2.person_months\endcsname{62,478}
\expandafter\def\csname EERjoint@results.restricted_linear.k2.dropout_events\endcsname{648}
\expandafter\def\csname EERjoint@results.restricted_linear.k2.pregnancy_events\endcsname{478}
\expandafter\def\csname EERjoint@results.restricted_linear.k2.exposed_months\endcsname{5,773}
\expandafter\def\csname EERjoint@results.restricted_linear.k2.rel_grad\endcsname{6.04e-08}
\expandafter\def\csname EERjoint@results.restricted_linear.k2.v_pd\endcsname{1.00}
\expandafter\def\csname EERjoint@results.restricted_linear.k2.v_scaffold\endcsname{0.00}
\expandafter\def\csname EERjoint@results.restricted_linear.k2.clip_hits\endcsname{0.00}
\expandafter\def\csname EERjoint@results.restricted_linear.k2.n_starts\endcsname{7}
\expandafter\def\csname EERjoint@results.restricted_linear.k2.n_finite\endcsname{7}
\expandafter\def\csname EERjoint@results.restricted_linear.k2.n_bfgs_conv\endcsname{6}
\expandafter\def\csname EERjoint@results.restricted_linear.k2.n_aborted\endcsname{0.00}
\expandafter\def\csname EERjoint@results.restricted_linear.k2.best_start\endcsname{1}
\expandafter\def\csname EERjoint@results.restricted_linear.k2.lambda_p\endcsname{4.11}
\expandafter\def\csname EERjoint@results.restricted_linear.k2.lambda_d\endcsname{0.15}
\expandafter\def\csname EERjoint@results.restricted_linear.k2.min_share\endcsname{0.034}
\expandafter\def\csname EERjoint@results.restricted_linear.k2.reproduction_difference\endcsname{.}
\expandafter\def\csname EERjoint@results.restricted_linear.k2.iteration_ceiling\endcsname{100.00}
\expandafter\def\csname EERjoint@results.restricted_linear.k2.aic\endcsname{12064.14}
\expandafter\def\csname EERjoint@results.restricted_linear.k2.bic_persons\endcsname{12123.18}
\expandafter\def\csname EERjoint@results.restricted_linear.k2.cluster_ci_lo\endcsname{.}
\expandafter\def\csname EERjoint@results.restricted_linear.k2.cluster_ci_hi\endcsname{.}
\expandafter\def\csname EERjoint@results.debut_linear.k1.rc\endcsname{0.00}
\expandafter\def\csname EERjoint@results.debut_linear.k1.converged\endcsname{1.00}
\expandafter\def\csname EERjoint@results.debut_linear.k1.b\endcsname{1.37}
\expandafter\def\csname EERjoint@results.debut_linear.k1.se\endcsname{0.10}
\expandafter\def\csname EERjoint@results.debut_linear.k1.se_cluster\endcsname{0.10}
\expandafter\def\csname EERjoint@results.debut_linear.k1.hr\endcsname{3.94}
\expandafter\def\csname EERjoint@results.debut_linear.k1.ci_lo\endcsname{3.27}
\expandafter\def\csname EERjoint@results.debut_linear.k1.ci_hi\endcsname{4.75}
\expandafter\def\csname EERjoint@results.debut_linear.k1.b_absent\endcsname{.}
\expandafter\def\csname EERjoint@results.debut_linear.k1.se_absent\endcsname{.}
\expandafter\def\csname EERjoint@results.debut_linear.k1.ll\endcsname{-5437.28}
\expandafter\def\csname EERjoint@results.debut_linear.k1.kparams\endcsname{7.00}
\expandafter\def\csname EERjoint@results.debut_linear.k1.women\endcsname{2,623}
\expandafter\def\csname EERjoint@results.debut_linear.k1.person_months\endcsname{60,974}
\expandafter\def\csname EERjoint@results.debut_linear.k1.dropout_events\endcsname{602}
\expandafter\def\csname EERjoint@results.debut_linear.k1.pregnancy_events\endcsname{396}
\expandafter\def\csname EERjoint@results.debut_linear.k1.exposed_months\endcsname{4,434}
\expandafter\def\csname EERjoint@results.debut_linear.k1.rel_grad\endcsname{.}
\expandafter\def\csname EERjoint@results.debut_linear.k1.v_pd\endcsname{.}
\expandafter\def\csname EERjoint@results.debut_linear.k1.v_scaffold\endcsname{.}
\expandafter\def\csname EERjoint@results.debut_linear.k1.clip_hits\endcsname{.}
\expandafter\def\csname EERjoint@results.debut_linear.k1.n_starts\endcsname{.}
\expandafter\def\csname EERjoint@results.debut_linear.k1.n_finite\endcsname{.}
\expandafter\def\csname EERjoint@results.debut_linear.k1.n_bfgs_conv\endcsname{.}
\expandafter\def\csname EERjoint@results.debut_linear.k1.n_aborted\endcsname{.}
\expandafter\def\csname EERjoint@results.debut_linear.k1.best_start\endcsname{.}
\expandafter\def\csname EERjoint@results.debut_linear.k1.lambda_p\endcsname{.}
\expandafter\def\csname EERjoint@results.debut_linear.k1.lambda_d\endcsname{.}
\expandafter\def\csname EERjoint@results.debut_linear.k1.min_share\endcsname{.}
\expandafter\def\csname EERjoint@results.debut_linear.k1.reproduction_difference\endcsname{.}
\expandafter\def\csname EERjoint@results.debut_linear.k1.iteration_ceiling\endcsname{.}
\expandafter\def\csname EERjoint@results.debut_linear.k1.aic\endcsname{10888.57}
\expandafter\def\csname EERjoint@results.debut_linear.k1.bic_persons\endcsname{10929.67}
\expandafter\def\csname EERjoint@results.debut_linear.k1.cluster_ci_lo\endcsname{3.25}
\expandafter\def\csname EERjoint@results.debut_linear.k1.cluster_ci_hi\endcsname{4.78}
\expandafter\def\csname EERjoint@results.debut_linear.k2.rc\endcsname{0.00}
\expandafter\def\csname EERjoint@results.debut_linear.k2.converged\endcsname{1.00}
\expandafter\def\csname EERjoint@results.debut_linear.k2.b\endcsname{1.47}
\expandafter\def\csname EERjoint@results.debut_linear.k2.se\endcsname{0.24}
\expandafter\def\csname EERjoint@results.debut_linear.k2.se_cluster\endcsname{.}
\expandafter\def\csname EERjoint@results.debut_linear.k2.hr\endcsname{4.35}
\expandafter\def\csname EERjoint@results.debut_linear.k2.ci_lo\endcsname{2.73}
\expandafter\def\csname EERjoint@results.debut_linear.k2.ci_hi\endcsname{6.93}
\expandafter\def\csname EERjoint@results.debut_linear.k2.b_absent\endcsname{.}
\expandafter\def\csname EERjoint@results.debut_linear.k2.se_absent\endcsname{.}
\expandafter\def\csname EERjoint@results.debut_linear.k2.ll\endcsname{-5436.96}
\expandafter\def\csname EERjoint@results.debut_linear.k2.kparams\endcsname{10.00}
\expandafter\def\csname EERjoint@results.debut_linear.k2.women\endcsname{2,623}
\expandafter\def\csname EERjoint@results.debut_linear.k2.person_months\endcsname{60,974}
\expandafter\def\csname EERjoint@results.debut_linear.k2.dropout_events\endcsname{602}
\expandafter\def\csname EERjoint@results.debut_linear.k2.pregnancy_events\endcsname{396}
\expandafter\def\csname EERjoint@results.debut_linear.k2.exposed_months\endcsname{4,434}
\expandafter\def\csname EERjoint@results.debut_linear.k2.rel_grad\endcsname{5.85e-06}
\expandafter\def\csname EERjoint@results.debut_linear.k2.v_pd\endcsname{1.00}
\expandafter\def\csname EERjoint@results.debut_linear.k2.v_scaffold\endcsname{0.00}
\expandafter\def\csname EERjoint@results.debut_linear.k2.clip_hits\endcsname{0.00}
\expandafter\def\csname EERjoint@results.debut_linear.k2.n_starts\endcsname{7}
\expandafter\def\csname EERjoint@results.debut_linear.k2.n_finite\endcsname{7}
\expandafter\def\csname EERjoint@results.debut_linear.k2.n_bfgs_conv\endcsname{7}
\expandafter\def\csname EERjoint@results.debut_linear.k2.n_aborted\endcsname{0.00}
\expandafter\def\csname EERjoint@results.debut_linear.k2.best_start\endcsname{7}
\expandafter\def\csname EERjoint@results.debut_linear.k2.lambda_p\endcsname{-0.34}
\expandafter\def\csname EERjoint@results.debut_linear.k2.lambda_d\endcsname{1.78}
\expandafter\def\csname EERjoint@results.debut_linear.k2.min_share\endcsname{0.118}
\expandafter\def\csname EERjoint@results.debut_linear.k2.reproduction_difference\endcsname{.}
\expandafter\def\csname EERjoint@results.debut_linear.k2.iteration_ceiling\endcsname{100.00}
\expandafter\def\csname EERjoint@results.debut_linear.k2.aic\endcsname{10893.93}
\expandafter\def\csname EERjoint@results.debut_linear.k2.bic_persons\endcsname{10952.65}
\expandafter\def\csname EERjoint@results.debut_linear.k2.cluster_ci_lo\endcsname{.}
\expandafter\def\csname EERjoint@results.debut_linear.k2.cluster_ci_hi\endcsname{.}
\expandafter\def\csname EERjoint@counts.full.g0.women\endcsname{2,256}
\expandafter\def\csname EERjoint@counts.full.g0.calendar_months\endcsname{49,772}
\expandafter\def\csname EERjoint@counts.full.g0.school_months\endcsname{37,900}
\expandafter\def\csname EERjoint@counts.full.g0.interruption_events\endcsname{450}
\expandafter\def\csname EERjoint@counts.full.g0.conception_events\endcsname{26}
\expandafter\def\csname EERjoint@counts.full.g0.summer_conceptions\endcsname{7}
\expandafter\def\csname EERjoint@counts.full.g0.pregnancy_risk_months\endcsname{49,772}
\expandafter\def\csname EERjoint@counts.full.g0.exposed_school_months\endcsname{0}
\expandafter\def\csname EERjoint@counts.full.g0.onset_without_exposure_women\endcsname{26}
\expandafter\def\csname EERjoint@counts.full.g1.women\endcsname{452}
\expandafter\def\csname EERjoint@counts.full.g1.calendar_months\endcsname{12,706}
\expandafter\def\csname EERjoint@counts.full.g1.school_months\endcsname{9,508}
\expandafter\def\csname EERjoint@counts.full.g1.interruption_events\endcsname{198}
\expandafter\def\csname EERjoint@counts.full.g1.conception_events\endcsname{452}
\expandafter\def\csname EERjoint@counts.full.g1.summer_conceptions\endcsname{107}
\expandafter\def\csname EERjoint@counts.full.g1.pregnancy_risk_months\endcsname{6,933}
\expandafter\def\csname EERjoint@counts.full.g1.exposed_school_months\endcsname{4,394}
\expandafter\def\csname EERjoint@counts.full.g1.onset_without_exposure_women\endcsname{0}
\expandafter\def\csname EERjoint@counts.full.g2.women\endcsname{120}
\expandafter\def\csname EERjoint@counts.full.g2.calendar_months\endcsname{2,120}
\expandafter\def\csname EERjoint@counts.full.g2.school_months\endcsname{1,545}
\expandafter\def\csname EERjoint@counts.full.g2.interruption_events\endcsname{63}
\expandafter\def\csname EERjoint@counts.full.g2.conception_events\endcsname{0}
\expandafter\def\csname EERjoint@counts.full.g2.summer_conceptions\endcsname{0}
\expandafter\def\csname EERjoint@counts.full.g2.pregnancy_risk_months\endcsname{0}
\expandafter\def\csname EERjoint@counts.full.g2.exposed_school_months\endcsname{1,545}
\expandafter\def\csname EERjoint@counts.full.g2.onset_without_exposure_women\endcsname{0}
\expandafter\def\csname EERjoint@counts.full.g3.women\endcsname{0}
\expandafter\def\csname EERjoint@counts.full.g3.calendar_months\endcsname{0}
\expandafter\def\csname EERjoint@counts.full.g3.school_months\endcsname{0}
\expandafter\def\csname EERjoint@counts.full.g3.interruption_events\endcsname{0}
\expandafter\def\csname EERjoint@counts.full.g3.conception_events\endcsname{0}
\expandafter\def\csname EERjoint@counts.full.g3.summer_conceptions\endcsname{0}
\expandafter\def\csname EERjoint@counts.full.g3.pregnancy_risk_months\endcsname{0}
\expandafter\def\csname EERjoint@counts.full.g3.exposed_school_months\endcsname{0}
\expandafter\def\csname EERjoint@counts.full.g3.onset_without_exposure_women\endcsname{0}
\expandafter\def\csname EERjoint@counts.full.g4.women\endcsname{2,828}
\expandafter\def\csname EERjoint@counts.full.g4.calendar_months\endcsname{64,598}
\expandafter\def\csname EERjoint@counts.full.g4.school_months\endcsname{48,953}
\expandafter\def\csname EERjoint@counts.full.g4.interruption_events\endcsname{711}
\expandafter\def\csname EERjoint@counts.full.g4.conception_events\endcsname{478}
\expandafter\def\csname EERjoint@counts.full.g4.summer_conceptions\endcsname{114}
\expandafter\def\csname EERjoint@counts.full.g4.pregnancy_risk_months\endcsname{56,705}
\expandafter\def\csname EERjoint@counts.full.g4.exposed_school_months\endcsname{5,939}
\expandafter\def\csname EERjoint@counts.full.g4.onset_without_exposure_women\endcsname{26}
\expandafter\def\csname EERjoint@counts.restricted.g0.women\endcsname{2,256}
\expandafter\def\csname EERjoint@counts.restricted.g0.calendar_months\endcsname{49,772}
\expandafter\def\csname EERjoint@counts.restricted.g0.school_months\endcsname{37,900}
\expandafter\def\csname EERjoint@counts.restricted.g0.interruption_events\endcsname{450}
\expandafter\def\csname EERjoint@counts.restricted.g0.conception_events\endcsname{26}
\expandafter\def\csname EERjoint@counts.restricted.g0.summer_conceptions\endcsname{7}
\expandafter\def\csname EERjoint@counts.restricted.g0.pregnancy_risk_months\endcsname{49,772}
\expandafter\def\csname EERjoint@counts.restricted.g0.exposed_school_months\endcsname{0}
\expandafter\def\csname EERjoint@counts.restricted.g0.onset_without_exposure_women\endcsname{26}
\expandafter\def\csname EERjoint@counts.restricted.g1.women\endcsname{452}
\expandafter\def\csname EERjoint@counts.restricted.g1.calendar_months\endcsname{12,706}
\expandafter\def\csname EERjoint@counts.restricted.g1.school_months\endcsname{9,508}
\expandafter\def\csname EERjoint@counts.restricted.g1.interruption_events\endcsname{198}
\expandafter\def\csname EERjoint@counts.restricted.g1.conception_events\endcsname{452}
\expandafter\def\csname EERjoint@counts.restricted.g1.summer_conceptions\endcsname{107}
\expandafter\def\csname EERjoint@counts.restricted.g1.pregnancy_risk_months\endcsname{6,933}
\expandafter\def\csname EERjoint@counts.restricted.g1.exposed_school_months\endcsname{4,394}
\expandafter\def\csname EERjoint@counts.restricted.g1.onset_without_exposure_women\endcsname{0}
\expandafter\def\csname EERjoint@counts.restricted.g2.women\endcsname{0}
\expandafter\def\csname EERjoint@counts.restricted.g2.calendar_months\endcsname{0}
\expandafter\def\csname EERjoint@counts.restricted.g2.school_months\endcsname{0}
\expandafter\def\csname EERjoint@counts.restricted.g2.interruption_events\endcsname{0}
\expandafter\def\csname EERjoint@counts.restricted.g2.conception_events\endcsname{0}
\expandafter\def\csname EERjoint@counts.restricted.g2.summer_conceptions\endcsname{0}
\expandafter\def\csname EERjoint@counts.restricted.g2.pregnancy_risk_months\endcsname{0}
\expandafter\def\csname EERjoint@counts.restricted.g2.exposed_school_months\endcsname{0}
\expandafter\def\csname EERjoint@counts.restricted.g2.onset_without_exposure_women\endcsname{0}
\expandafter\def\csname EERjoint@counts.restricted.g3.women\endcsname{0}
\expandafter\def\csname EERjoint@counts.restricted.g3.calendar_months\endcsname{0}
\expandafter\def\csname EERjoint@counts.restricted.g3.school_months\endcsname{0}
\expandafter\def\csname EERjoint@counts.restricted.g3.interruption_events\endcsname{0}
\expandafter\def\csname EERjoint@counts.restricted.g3.conception_events\endcsname{0}
\expandafter\def\csname EERjoint@counts.restricted.g3.summer_conceptions\endcsname{0}
\expandafter\def\csname EERjoint@counts.restricted.g3.pregnancy_risk_months\endcsname{0}
\expandafter\def\csname EERjoint@counts.restricted.g3.exposed_school_months\endcsname{0}
\expandafter\def\csname EERjoint@counts.restricted.g3.onset_without_exposure_women\endcsname{0}
\expandafter\def\csname EERjoint@counts.restricted.g4.women\endcsname{2,708}
\expandafter\def\csname EERjoint@counts.restricted.g4.calendar_months\endcsname{62,478}
\expandafter\def\csname EERjoint@counts.restricted.g4.school_months\endcsname{47,408}
\expandafter\def\csname EERjoint@counts.restricted.g4.interruption_events\endcsname{648}
\expandafter\def\csname EERjoint@counts.restricted.g4.conception_events\endcsname{478}
\expandafter\def\csname EERjoint@counts.restricted.g4.summer_conceptions\endcsname{114}
\expandafter\def\csname EERjoint@counts.restricted.g4.pregnancy_risk_months\endcsname{56,705}
\expandafter\def\csname EERjoint@counts.restricted.g4.exposed_school_months\endcsname{4,394}
\expandafter\def\csname EERjoint@counts.restricted.g4.onset_without_exposure_women\endcsname{26}
\expandafter\def\csname EERjoint@counts.debut.g0.women\endcsname{2,250}
\expandafter\def\csname EERjoint@counts.debut.g0.calendar_months\endcsname{49,763}
\expandafter\def\csname EERjoint@counts.debut.g0.school_months\endcsname{37,892}
\expandafter\def\csname EERjoint@counts.debut.g0.interruption_events\endcsname{445}
\expandafter\def\csname EERjoint@counts.debut.g0.conception_events\endcsname{23}
\expandafter\def\csname EERjoint@counts.debut.g0.summer_conceptions\endcsname{6}
\expandafter\def\csname EERjoint@counts.debut.g0.pregnancy_risk_months\endcsname{49,763}
\expandafter\def\csname EERjoint@counts.debut.g0.exposed_school_months\endcsname{0}
\expandafter\def\csname EERjoint@counts.debut.g0.onset_without_exposure_women\endcsname{23}
\expandafter\def\csname EERjoint@counts.debut.g1.women\endcsname{373}
\expandafter\def\csname EERjoint@counts.debut.g1.calendar_months\endcsname{11,211}
\expandafter\def\csname EERjoint@counts.debut.g1.school_months\endcsname{8,372}
\expandafter\def\csname EERjoint@counts.debut.g1.interruption_events\endcsname{157}
\expandafter\def\csname EERjoint@counts.debut.g1.conception_events\endcsname{373}
\expandafter\def\csname EERjoint@counts.debut.g1.summer_conceptions\endcsname{87}
\expandafter\def\csname EERjoint@counts.debut.g1.pregnancy_risk_months\endcsname{6,777}
\expandafter\def\csname EERjoint@counts.debut.g1.exposed_school_months\endcsname{3,375}
\expandafter\def\csname EERjoint@counts.debut.g1.onset_without_exposure_women\endcsname{0}
\expandafter\def\csname EERjoint@counts.debut.g2.women\endcsname{0}
\expandafter\def\csname EERjoint@counts.debut.g2.calendar_months\endcsname{0}
\expandafter\def\csname EERjoint@counts.debut.g2.school_months\endcsname{0}
\expandafter\def\csname EERjoint@counts.debut.g2.interruption_events\endcsname{0}
\expandafter\def\csname EERjoint@counts.debut.g2.conception_events\endcsname{0}
\expandafter\def\csname EERjoint@counts.debut.g2.summer_conceptions\endcsname{0}
\expandafter\def\csname EERjoint@counts.debut.g2.pregnancy_risk_months\endcsname{0}
\expandafter\def\csname EERjoint@counts.debut.g2.exposed_school_months\endcsname{0}
\expandafter\def\csname EERjoint@counts.debut.g2.onset_without_exposure_women\endcsname{0}
\expandafter\def\csname EERjoint@counts.debut.g3.women\endcsname{0}
\expandafter\def\csname EERjoint@counts.debut.g3.calendar_months\endcsname{0}
\expandafter\def\csname EERjoint@counts.debut.g3.school_months\endcsname{0}
\expandafter\def\csname EERjoint@counts.debut.g3.interruption_events\endcsname{0}
\expandafter\def\csname EERjoint@counts.debut.g3.conception_events\endcsname{0}
\expandafter\def\csname EERjoint@counts.debut.g3.summer_conceptions\endcsname{0}
\expandafter\def\csname EERjoint@counts.debut.g3.pregnancy_risk_months\endcsname{0}
\expandafter\def\csname EERjoint@counts.debut.g3.exposed_school_months\endcsname{0}
\expandafter\def\csname EERjoint@counts.debut.g3.onset_without_exposure_women\endcsname{0}
\expandafter\def\csname EERjoint@counts.debut.g4.women\endcsname{2,623}
\expandafter\def\csname EERjoint@counts.debut.g4.calendar_months\endcsname{60,974}
\expandafter\def\csname EERjoint@counts.debut.g4.school_months\endcsname{46,264}
\expandafter\def\csname EERjoint@counts.debut.g4.interruption_events\endcsname{602}
\expandafter\def\csname EERjoint@counts.debut.g4.conception_events\endcsname{396}
\expandafter\def\csname EERjoint@counts.debut.g4.summer_conceptions\endcsname{93}
\expandafter\def\csname EERjoint@counts.debut.g4.pregnancy_risk_months\endcsname{56,540}
\expandafter\def\csname EERjoint@counts.debut.g4.exposed_school_months\endcsname{3,375}
\expandafter\def\csname EERjoint@counts.debut.g4.onset_without_exposure_women\endcsname{23}
\expandafter\def\csname EERjoint@diagnostics.full_linear.converged\endcsname{1.00}
\expandafter\def\csname EERjoint@diagnostics.full_linear.bfgs\endcsname{1.00}
\expandafter\def\csname EERjoint@diagnostics.full_linear.rel_grad\endcsname{7.32e-08}
\expandafter\def\csname EERjoint@diagnostics.full_linear.v_pd\endcsname{1.00}
\expandafter\def\csname EERjoint@diagnostics.full_linear.v_scaffold\endcsname{0.00}
\expandafter\def\csname EERjoint@diagnostics.full_linear.clip_hits\endcsname{0.00}
\expandafter\def\csname EERjoint@diagnostics.full_linear.min_eigen\endcsname{0.00}
\expandafter\def\csname EERjoint@diagnostics.full_linear.max_eigen\endcsname{0.19}
\expandafter\def\csname EERjoint@diagnostics.full_linear.eigen_ratio\endcsname{6.39e-04}
\expandafter\def\csname EERjoint@diagnostics.full_linear.saved_ll\endcsname{-6281.43}
\expandafter\def\csname EERjoint@diagnostics.full_linear.evaluated_ll\endcsname{-6281.43}
\expandafter\def\csname EERjoint@diagnostics.full_linear.relabeled_ll\endcsname{-6281.43}
\expandafter\def\csname EERjoint@diagnostics.full_linear.ll_difference\endcsname{0.00e+00}
\expandafter\def\csname EERjoint@diagnostics.full_linear.label_difference\endcsname{0.00e+00}
\expandafter\def\csname EERjoint@diagnostics.full_linear.lambda_p\endcsname{-4.11}
\expandafter\def\csname EERjoint@diagnostics.full_linear.lambda_y\endcsname{-0.16}
\expandafter\def\csname EERjoint@diagnostics.full_linear.pi1\endcsname{0.034}
\expandafter\def\csname EERjoint@diagnostics.full_linear.pi2\endcsname{0.966}
\expandafter\def\csname EERjoint@diagnostics.full_linear.oriented_lambda_p\endcsname{4.11}
\expandafter\def\csname EERjoint@diagnostics.full_linear.oriented_lambda_y\endcsname{0.16}
\expandafter\def\csname EERjoint@diagnostics.full_linear.higher_conception_share_pct\endcsname{3.42}
\expandafter\def\csname EERjoint@diagnostics.full_linear.conception_type_ratio\endcsname{61.13}
\expandafter\def\csname EERjoint@diagnostics.full_linear.outcome_type_ratio\endcsname{1.18}
\expandafter\def\csname EERjoint@diagnostics.restricted_linear.converged\endcsname{1.00}
\expandafter\def\csname EERjoint@diagnostics.restricted_linear.bfgs\endcsname{1.00}
\expandafter\def\csname EERjoint@diagnostics.restricted_linear.rel_grad\endcsname{6.04e-08}
\expandafter\def\csname EERjoint@diagnostics.restricted_linear.v_pd\endcsname{1.00}
\expandafter\def\csname EERjoint@diagnostics.restricted_linear.v_scaffold\endcsname{0.00}
\expandafter\def\csname EERjoint@diagnostics.restricted_linear.clip_hits\endcsname{0.00}
\expandafter\def\csname EERjoint@diagnostics.restricted_linear.min_eigen\endcsname{0.00}
\expandafter\def\csname EERjoint@diagnostics.restricted_linear.max_eigen\endcsname{0.10}
\expandafter\def\csname EERjoint@diagnostics.restricted_linear.eigen_ratio\endcsname{1.35e-03}
\expandafter\def\csname EERjoint@diagnostics.restricted_linear.saved_ll\endcsname{-6022.07}
\expandafter\def\csname EERjoint@diagnostics.restricted_linear.evaluated_ll\endcsname{-6022.07}
\expandafter\def\csname EERjoint@diagnostics.restricted_linear.relabeled_ll\endcsname{-6022.07}
\expandafter\def\csname EERjoint@diagnostics.restricted_linear.ll_difference\endcsname{0.00e+00}
\expandafter\def\csname EERjoint@diagnostics.restricted_linear.label_difference\endcsname{0.00e+00}
\expandafter\def\csname EERjoint@diagnostics.restricted_linear.lambda_p\endcsname{4.11}
\expandafter\def\csname EERjoint@diagnostics.restricted_linear.lambda_y\endcsname{0.15}
\expandafter\def\csname EERjoint@diagnostics.restricted_linear.pi1\endcsname{0.966}
\expandafter\def\csname EERjoint@diagnostics.restricted_linear.pi2\endcsname{0.034}
\expandafter\def\csname EERjoint@diagnostics.restricted_linear.oriented_lambda_p\endcsname{4.11}
\expandafter\def\csname EERjoint@diagnostics.restricted_linear.oriented_lambda_y\endcsname{0.15}
\expandafter\def\csname EERjoint@diagnostics.restricted_linear.higher_conception_share_pct\endcsname{3.41}
\expandafter\def\csname EERjoint@diagnostics.restricted_linear.conception_type_ratio\endcsname{61.19}
\expandafter\def\csname EERjoint@diagnostics.restricted_linear.outcome_type_ratio\endcsname{1.16}
\expandafter\def\csname EERjoint@diagnostics.debut_linear.converged\endcsname{1.00}
\expandafter\def\csname EERjoint@diagnostics.debut_linear.bfgs\endcsname{1.00}
\expandafter\def\csname EERjoint@diagnostics.debut_linear.rel_grad\endcsname{5.85e-06}
\expandafter\def\csname EERjoint@diagnostics.debut_linear.v_pd\endcsname{1.00}
\expandafter\def\csname EERjoint@diagnostics.debut_linear.v_scaffold\endcsname{0.00}
\expandafter\def\csname EERjoint@diagnostics.debut_linear.clip_hits\endcsname{0.00}
\expandafter\def\csname EERjoint@diagnostics.debut_linear.min_eigen\endcsname{0.00}
\expandafter\def\csname EERjoint@diagnostics.debut_linear.max_eigen\endcsname{106.16}
\expandafter\def\csname EERjoint@diagnostics.debut_linear.eigen_ratio\endcsname{1.36e-06}
\expandafter\def\csname EERjoint@diagnostics.debut_linear.saved_ll\endcsname{-5436.96}
\expandafter\def\csname EERjoint@diagnostics.debut_linear.evaluated_ll\endcsname{-5436.96}
\expandafter\def\csname EERjoint@diagnostics.debut_linear.relabeled_ll\endcsname{-5436.96}
\expandafter\def\csname EERjoint@diagnostics.debut_linear.ll_difference\endcsname{0.00e+00}
\expandafter\def\csname EERjoint@diagnostics.debut_linear.label_difference\endcsname{0.00e+00}
\expandafter\def\csname EERjoint@diagnostics.debut_linear.lambda_p\endcsname{-0.34}
\expandafter\def\csname EERjoint@diagnostics.debut_linear.lambda_y\endcsname{1.78}
\expandafter\def\csname EERjoint@diagnostics.debut_linear.pi1\endcsname{0.118}
\expandafter\def\csname EERjoint@diagnostics.debut_linear.pi2\endcsname{0.882}
\expandafter\def\csname EERjoint@diagnostics.debut_linear.oriented_lambda_p\endcsname{0.34}
\expandafter\def\csname EERjoint@diagnostics.debut_linear.oriented_lambda_y\endcsname{-1.78}
\expandafter\def\csname EERjoint@diagnostics.debut_linear.higher_conception_share_pct\endcsname{11.76}
\expandafter\def\csname EERjoint@diagnostics.debut_linear.conception_type_ratio\endcsname{1.40}
\expandafter\def\csname EERjoint@diagnostics.debut_linear.outcome_type_ratio\endcsname{0.17}
\expandafter\def\csname EERjoint@timing.all.exposed_events\endcsname{261}
\expandafter\def\csname EERjoint@timing.all.summer_start\endcsname{236}
\expandafter\def\csname EERjoint@timing.all.summer_before_p\endcsname{2}
\expandafter\def\csname EERjoint@timing.all.summer_same_p\endcsname{4}
\expandafter\def\csname EERjoint@timing.all.before_p_any\endcsname{2}
\expandafter\def\csname EERjoint@timing.all.events\endcsname{711}
\expandafter\def\csname EERjoint@confirmation.all.p_during_confirmation\endcsname{26}
\expandafter\def\csname EERjoint@confirmation.all.exposure_during_confirmation\endcsname{31}
\expandafter\def\csname EERjoint@confirmation.all.p_same_onset\endcsname{15}
\expandafter\def\csname EERjoint@confirmation.all.events\endcsname{711}

\providecommand{\EERjointflex}[1]{\csname EERjointflex@#1\endcsname}
\expandafter\def\csname EERjointflex@results.full_flexible.k1.rc\endcsname{0.00}
\expandafter\def\csname EERjointflex@results.full_flexible.k1.converged\endcsname{1.00}
\expandafter\def\csname EERjointflex@results.full_flexible.k1.b\endcsname{1.34}
\expandafter\def\csname EERjointflex@results.full_flexible.k1.se\endcsname{0.09}
\expandafter\def\csname EERjointflex@results.full_flexible.k1.se_cluster\endcsname{0.09}
\expandafter\def\csname EERjointflex@results.full_flexible.k1.hr\endcsname{3.81}
\expandafter\def\csname EERjointflex@results.full_flexible.k1.ci_lo\endcsname{3.21}
\expandafter\def\csname EERjointflex@results.full_flexible.k1.ci_hi\endcsname{4.53}
\expandafter\def\csname EERjointflex@results.full_flexible.k1.b_absent\endcsname{1.38}
\expandafter\def\csname EERjointflex@results.full_flexible.k1.se_absent\endcsname{0.14}
\expandafter\def\csname EERjointflex@results.full_flexible.k1.ll\endcsname{-6243.16}
\expandafter\def\csname EERjointflex@results.full_flexible.k1.kparams\endcsname{23.00}
\expandafter\def\csname EERjointflex@results.full_flexible.k1.women\endcsname{2,828}
\expandafter\def\csname EERjointflex@results.full_flexible.k1.person_months\endcsname{64,598}
\expandafter\def\csname EERjointflex@results.full_flexible.k1.dropout_events\endcsname{711}
\expandafter\def\csname EERjointflex@results.full_flexible.k1.pregnancy_events\endcsname{478}
\expandafter\def\csname EERjointflex@results.full_flexible.k1.exposed_months\endcsname{7,893}
\expandafter\def\csname EERjointflex@results.full_flexible.k1.rel_grad\endcsname{.}
\expandafter\def\csname EERjointflex@results.full_flexible.k1.v_pd\endcsname{.}
\expandafter\def\csname EERjointflex@results.full_flexible.k1.v_scaffold\endcsname{.}
\expandafter\def\csname EERjointflex@results.full_flexible.k1.clip_hits\endcsname{.}
\expandafter\def\csname EERjointflex@results.full_flexible.k1.n_starts\endcsname{.}
\expandafter\def\csname EERjointflex@results.full_flexible.k1.n_finite\endcsname{.}
\expandafter\def\csname EERjointflex@results.full_flexible.k1.n_bfgs_conv\endcsname{.}
\expandafter\def\csname EERjointflex@results.full_flexible.k1.n_aborted\endcsname{.}
\expandafter\def\csname EERjointflex@results.full_flexible.k1.best_start\endcsname{.}
\expandafter\def\csname EERjointflex@results.full_flexible.k1.lambda_p\endcsname{.}
\expandafter\def\csname EERjointflex@results.full_flexible.k1.lambda_d\endcsname{.}
\expandafter\def\csname EERjointflex@results.full_flexible.k1.min_share\endcsname{.}
\expandafter\def\csname EERjointflex@results.full_flexible.k1.reproduction_difference\endcsname{.}
\expandafter\def\csname EERjointflex@results.full_flexible.k1.iteration_ceiling\endcsname{.}
\expandafter\def\csname EERjointflex@results.full_flexible.k1.aic\endcsname{12532.31}
\expandafter\def\csname EERjointflex@results.full_flexible.k1.bic_persons\endcsname{12669.10}
\expandafter\def\csname EERjointflex@results.full_flexible.k1.cluster_ci_lo\endcsname{3.20}
\expandafter\def\csname EERjointflex@results.full_flexible.k1.cluster_ci_hi\endcsname{4.54}
\expandafter\def\csname EERjointflex@results.full_flexible.k2.rc\endcsname{0.00}
\expandafter\def\csname EERjointflex@results.full_flexible.k2.converged\endcsname{1.00}
\expandafter\def\csname EERjointflex@results.full_flexible.k2.b\endcsname{1.27}
\expandafter\def\csname EERjointflex@results.full_flexible.k2.se\endcsname{0.12}
\expandafter\def\csname EERjointflex@results.full_flexible.k2.se_cluster\endcsname{.}
\expandafter\def\csname EERjointflex@results.full_flexible.k2.hr\endcsname{3.56}
\expandafter\def\csname EERjointflex@results.full_flexible.k2.ci_lo\endcsname{2.80}
\expandafter\def\csname EERjointflex@results.full_flexible.k2.ci_hi\endcsname{4.52}
\expandafter\def\csname EERjointflex@results.full_flexible.k2.b_absent\endcsname{1.37}
\expandafter\def\csname EERjointflex@results.full_flexible.k2.se_absent\endcsname{0.14}
\expandafter\def\csname EERjointflex@results.full_flexible.k2.ll\endcsname{-6220.71}
\expandafter\def\csname EERjointflex@results.full_flexible.k2.kparams\endcsname{26.00}
\expandafter\def\csname EERjointflex@results.full_flexible.k2.women\endcsname{2,828}
\expandafter\def\csname EERjointflex@results.full_flexible.k2.person_months\endcsname{64,598}
\expandafter\def\csname EERjointflex@results.full_flexible.k2.dropout_events\endcsname{711}
\expandafter\def\csname EERjointflex@results.full_flexible.k2.pregnancy_events\endcsname{478}
\expandafter\def\csname EERjointflex@results.full_flexible.k2.exposed_months\endcsname{7,893}
\expandafter\def\csname EERjointflex@results.full_flexible.k2.rel_grad\endcsname{2.14e-07}
\expandafter\def\csname EERjointflex@results.full_flexible.k2.v_pd\endcsname{1.00}
\expandafter\def\csname EERjointflex@results.full_flexible.k2.v_scaffold\endcsname{0.00}
\expandafter\def\csname EERjointflex@results.full_flexible.k2.clip_hits\endcsname{0.00}
\expandafter\def\csname EERjointflex@results.full_flexible.k2.n_starts\endcsname{7}
\expandafter\def\csname EERjointflex@results.full_flexible.k2.n_finite\endcsname{7}
\expandafter\def\csname EERjointflex@results.full_flexible.k2.n_bfgs_conv\endcsname{7}
\expandafter\def\csname EERjointflex@results.full_flexible.k2.n_aborted\endcsname{0.00}
\expandafter\def\csname EERjointflex@results.full_flexible.k2.best_start\endcsname{4}
\expandafter\def\csname EERjointflex@results.full_flexible.k2.lambda_p\endcsname{4.13}
\expandafter\def\csname EERjointflex@results.full_flexible.k2.lambda_d\endcsname{0.22}
\expandafter\def\csname EERjointflex@results.full_flexible.k2.min_share\endcsname{0.046}
\expandafter\def\csname EERjointflex@results.full_flexible.k2.reproduction_difference\endcsname{.}
\expandafter\def\csname EERjointflex@results.full_flexible.k2.iteration_ceiling\endcsname{100.00}
\expandafter\def\csname EERjointflex@results.full_flexible.k2.aic\endcsname{12493.42}
\expandafter\def\csname EERjointflex@results.full_flexible.k2.bic_persons\endcsname{12648.05}
\expandafter\def\csname EERjointflex@results.full_flexible.k2.cluster_ci_lo\endcsname{.}
\expandafter\def\csname EERjointflex@results.full_flexible.k2.cluster_ci_hi\endcsname{.}
\expandafter\def\csname EERjointflex@results.restricted_flexible.k1.rc\endcsname{0.00}
\expandafter\def\csname EERjointflex@results.restricted_flexible.k1.converged\endcsname{1.00}
\expandafter\def\csname EERjointflex@results.restricted_flexible.k1.b\endcsname{1.34}
\expandafter\def\csname EERjointflex@results.restricted_flexible.k1.se\endcsname{0.09}
\expandafter\def\csname EERjointflex@results.restricted_flexible.k1.se_cluster\endcsname{0.09}
\expandafter\def\csname EERjointflex@results.restricted_flexible.k1.hr\endcsname{3.82}
\expandafter\def\csname EERjointflex@results.restricted_flexible.k1.ci_lo\endcsname{3.21}
\expandafter\def\csname EERjointflex@results.restricted_flexible.k1.ci_hi\endcsname{4.53}
\expandafter\def\csname EERjointflex@results.restricted_flexible.k1.b_absent\endcsname{.}
\expandafter\def\csname EERjointflex@results.restricted_flexible.k1.se_absent\endcsname{.}
\expandafter\def\csname EERjointflex@results.restricted_flexible.k1.ll\endcsname{-5995.36}
\expandafter\def\csname EERjointflex@results.restricted_flexible.k1.kparams\endcsname{22.00}
\expandafter\def\csname EERjointflex@results.restricted_flexible.k1.women\endcsname{2,708}
\expandafter\def\csname EERjointflex@results.restricted_flexible.k1.person_months\endcsname{62,478}
\expandafter\def\csname EERjointflex@results.restricted_flexible.k1.dropout_events\endcsname{648}
\expandafter\def\csname EERjointflex@results.restricted_flexible.k1.pregnancy_events\endcsname{478}
\expandafter\def\csname EERjointflex@results.restricted_flexible.k1.exposed_months\endcsname{5,773}
\expandafter\def\csname EERjointflex@results.restricted_flexible.k1.rel_grad\endcsname{.}
\expandafter\def\csname EERjointflex@results.restricted_flexible.k1.v_pd\endcsname{.}
\expandafter\def\csname EERjointflex@results.restricted_flexible.k1.v_scaffold\endcsname{.}
\expandafter\def\csname EERjointflex@results.restricted_flexible.k1.clip_hits\endcsname{.}
\expandafter\def\csname EERjointflex@results.restricted_flexible.k1.n_starts\endcsname{.}
\expandafter\def\csname EERjointflex@results.restricted_flexible.k1.n_finite\endcsname{.}
\expandafter\def\csname EERjointflex@results.restricted_flexible.k1.n_bfgs_conv\endcsname{.}
\expandafter\def\csname EERjointflex@results.restricted_flexible.k1.n_aborted\endcsname{.}
\expandafter\def\csname EERjointflex@results.restricted_flexible.k1.best_start\endcsname{.}
\expandafter\def\csname EERjointflex@results.restricted_flexible.k1.lambda_p\endcsname{.}
\expandafter\def\csname EERjointflex@results.restricted_flexible.k1.lambda_d\endcsname{.}
\expandafter\def\csname EERjointflex@results.restricted_flexible.k1.min_share\endcsname{.}
\expandafter\def\csname EERjointflex@results.restricted_flexible.k1.reproduction_difference\endcsname{.}
\expandafter\def\csname EERjointflex@results.restricted_flexible.k1.iteration_ceiling\endcsname{.}
\expandafter\def\csname EERjointflex@results.restricted_flexible.k1.aic\endcsname{12034.73}
\expandafter\def\csname EERjointflex@results.restricted_flexible.k1.bic_persons\endcsname{12164.61}
\expandafter\def\csname EERjointflex@results.restricted_flexible.k1.cluster_ci_lo\endcsname{3.20}
\expandafter\def\csname EERjointflex@results.restricted_flexible.k1.cluster_ci_hi\endcsname{4.55}
\expandafter\def\csname EERjointflex@results.restricted_flexible.k2.rc\endcsname{0.00}
\expandafter\def\csname EERjointflex@results.restricted_flexible.k2.converged\endcsname{1.00}
\expandafter\def\csname EERjointflex@results.restricted_flexible.k2.b\endcsname{1.27}
\expandafter\def\csname EERjointflex@results.restricted_flexible.k2.se\endcsname{0.12}
\expandafter\def\csname EERjointflex@results.restricted_flexible.k2.se_cluster\endcsname{.}
\expandafter\def\csname EERjointflex@results.restricted_flexible.k2.hr\endcsname{3.56}
\expandafter\def\csname EERjointflex@results.restricted_flexible.k2.ci_lo\endcsname{2.79}
\expandafter\def\csname EERjointflex@results.restricted_flexible.k2.ci_hi\endcsname{4.53}
\expandafter\def\csname EERjointflex@results.restricted_flexible.k2.b_absent\endcsname{.}
\expandafter\def\csname EERjointflex@results.restricted_flexible.k2.se_absent\endcsname{.}
\expandafter\def\csname EERjointflex@results.restricted_flexible.k2.ll\endcsname{-5972.92}
\expandafter\def\csname EERjointflex@results.restricted_flexible.k2.kparams\endcsname{25.00}
\expandafter\def\csname EERjointflex@results.restricted_flexible.k2.women\endcsname{2,708}
\expandafter\def\csname EERjointflex@results.restricted_flexible.k2.person_months\endcsname{62,478}
\expandafter\def\csname EERjointflex@results.restricted_flexible.k2.dropout_events\endcsname{648}
\expandafter\def\csname EERjointflex@results.restricted_flexible.k2.pregnancy_events\endcsname{478}
\expandafter\def\csname EERjointflex@results.restricted_flexible.k2.exposed_months\endcsname{5,773}
\expandafter\def\csname EERjointflex@results.restricted_flexible.k2.rel_grad\endcsname{3.31e-07}
\expandafter\def\csname EERjointflex@results.restricted_flexible.k2.v_pd\endcsname{1.00}
\expandafter\def\csname EERjointflex@results.restricted_flexible.k2.v_scaffold\endcsname{0.00}
\expandafter\def\csname EERjointflex@results.restricted_flexible.k2.clip_hits\endcsname{0.00}
\expandafter\def\csname EERjointflex@results.restricted_flexible.k2.n_starts\endcsname{7}
\expandafter\def\csname EERjointflex@results.restricted_flexible.k2.n_finite\endcsname{7}
\expandafter\def\csname EERjointflex@results.restricted_flexible.k2.n_bfgs_conv\endcsname{7}
\expandafter\def\csname EERjointflex@results.restricted_flexible.k2.n_aborted\endcsname{0.00}
\expandafter\def\csname EERjointflex@results.restricted_flexible.k2.best_start\endcsname{7}
\expandafter\def\csname EERjointflex@results.restricted_flexible.k2.lambda_p\endcsname{-4.13}
\expandafter\def\csname EERjointflex@results.restricted_flexible.k2.lambda_d\endcsname{-0.22}
\expandafter\def\csname EERjointflex@results.restricted_flexible.k2.min_share\endcsname{0.046}
\expandafter\def\csname EERjointflex@results.restricted_flexible.k2.reproduction_difference\endcsname{.}
\expandafter\def\csname EERjointflex@results.restricted_flexible.k2.iteration_ceiling\endcsname{100.00}
\expandafter\def\csname EERjointflex@results.restricted_flexible.k2.aic\endcsname{11995.83}
\expandafter\def\csname EERjointflex@results.restricted_flexible.k2.bic_persons\endcsname{12143.43}
\expandafter\def\csname EERjointflex@results.restricted_flexible.k2.cluster_ci_lo\endcsname{.}
\expandafter\def\csname EERjointflex@results.restricted_flexible.k2.cluster_ci_hi\endcsname{.}
\expandafter\def\csname EERjointflex@results.debut_flexible.k1.rc\endcsname{0.00}
\expandafter\def\csname EERjointflex@results.debut_flexible.k1.converged\endcsname{1.00}
\expandafter\def\csname EERjointflex@results.debut_flexible.k1.b\endcsname{1.35}
\expandafter\def\csname EERjointflex@results.debut_flexible.k1.se\endcsname{0.10}
\expandafter\def\csname EERjointflex@results.debut_flexible.k1.se_cluster\endcsname{0.10}
\expandafter\def\csname EERjointflex@results.debut_flexible.k1.hr\endcsname{3.84}
\expandafter\def\csname EERjointflex@results.debut_flexible.k1.ci_lo\endcsname{3.18}
\expandafter\def\csname EERjointflex@results.debut_flexible.k1.ci_hi\endcsname{4.64}
\expandafter\def\csname EERjointflex@results.debut_flexible.k1.b_absent\endcsname{.}
\expandafter\def\csname EERjointflex@results.debut_flexible.k1.se_absent\endcsname{.}
\expandafter\def\csname EERjointflex@results.debut_flexible.k1.ll\endcsname{-5399.22}
\expandafter\def\csname EERjointflex@results.debut_flexible.k1.kparams\endcsname{22.00}
\expandafter\def\csname EERjointflex@results.debut_flexible.k1.women\endcsname{2,623}
\expandafter\def\csname EERjointflex@results.debut_flexible.k1.person_months\endcsname{60,974}
\expandafter\def\csname EERjointflex@results.debut_flexible.k1.dropout_events\endcsname{602}
\expandafter\def\csname EERjointflex@results.debut_flexible.k1.pregnancy_events\endcsname{396}
\expandafter\def\csname EERjointflex@results.debut_flexible.k1.exposed_months\endcsname{4,434}
\expandafter\def\csname EERjointflex@results.debut_flexible.k1.rel_grad\endcsname{.}
\expandafter\def\csname EERjointflex@results.debut_flexible.k1.v_pd\endcsname{.}
\expandafter\def\csname EERjointflex@results.debut_flexible.k1.v_scaffold\endcsname{.}
\expandafter\def\csname EERjointflex@results.debut_flexible.k1.clip_hits\endcsname{.}
\expandafter\def\csname EERjointflex@results.debut_flexible.k1.n_starts\endcsname{.}
\expandafter\def\csname EERjointflex@results.debut_flexible.k1.n_finite\endcsname{.}
\expandafter\def\csname EERjointflex@results.debut_flexible.k1.n_bfgs_conv\endcsname{.}
\expandafter\def\csname EERjointflex@results.debut_flexible.k1.n_aborted\endcsname{.}
\expandafter\def\csname EERjointflex@results.debut_flexible.k1.best_start\endcsname{.}
\expandafter\def\csname EERjointflex@results.debut_flexible.k1.lambda_p\endcsname{.}
\expandafter\def\csname EERjointflex@results.debut_flexible.k1.lambda_d\endcsname{.}
\expandafter\def\csname EERjointflex@results.debut_flexible.k1.min_share\endcsname{.}
\expandafter\def\csname EERjointflex@results.debut_flexible.k1.reproduction_difference\endcsname{.}
\expandafter\def\csname EERjointflex@results.debut_flexible.k1.iteration_ceiling\endcsname{.}
\expandafter\def\csname EERjointflex@results.debut_flexible.k1.aic\endcsname{10842.45}
\expandafter\def\csname EERjointflex@results.debut_flexible.k1.bic_persons\endcsname{10971.63}
\expandafter\def\csname EERjointflex@results.debut_flexible.k1.cluster_ci_lo\endcsname{3.16}
\expandafter\def\csname EERjointflex@results.debut_flexible.k1.cluster_ci_hi\endcsname{4.67}
\expandafter\def\csname EERjointflex@results.debut_flexible.k2.rc\endcsname{0.00}
\expandafter\def\csname EERjointflex@results.debut_flexible.k2.converged\endcsname{1.00}
\expandafter\def\csname EERjointflex@results.debut_flexible.k2.b\endcsname{1.45}
\expandafter\def\csname EERjointflex@results.debut_flexible.k2.se\endcsname{0.23}
\expandafter\def\csname EERjointflex@results.debut_flexible.k2.se_cluster\endcsname{.}
\expandafter\def\csname EERjointflex@results.debut_flexible.k2.hr\endcsname{4.28}
\expandafter\def\csname EERjointflex@results.debut_flexible.k2.ci_lo\endcsname{2.71}
\expandafter\def\csname EERjointflex@results.debut_flexible.k2.ci_hi\endcsname{6.74}
\expandafter\def\csname EERjointflex@results.debut_flexible.k2.b_absent\endcsname{.}
\expandafter\def\csname EERjointflex@results.debut_flexible.k2.se_absent\endcsname{.}
\expandafter\def\csname EERjointflex@results.debut_flexible.k2.ll\endcsname{-5398.85}
\expandafter\def\csname EERjointflex@results.debut_flexible.k2.kparams\endcsname{25.00}
\expandafter\def\csname EERjointflex@results.debut_flexible.k2.women\endcsname{2,623}
\expandafter\def\csname EERjointflex@results.debut_flexible.k2.person_months\endcsname{60,974}
\expandafter\def\csname EERjointflex@results.debut_flexible.k2.dropout_events\endcsname{602}
\expandafter\def\csname EERjointflex@results.debut_flexible.k2.pregnancy_events\endcsname{396}
\expandafter\def\csname EERjointflex@results.debut_flexible.k2.exposed_months\endcsname{4,434}
\expandafter\def\csname EERjointflex@results.debut_flexible.k2.rel_grad\endcsname{1.20e-06}
\expandafter\def\csname EERjointflex@results.debut_flexible.k2.v_pd\endcsname{1.00}
\expandafter\def\csname EERjointflex@results.debut_flexible.k2.v_scaffold\endcsname{0.00}
\expandafter\def\csname EERjointflex@results.debut_flexible.k2.clip_hits\endcsname{0.00}
\expandafter\def\csname EERjointflex@results.debut_flexible.k2.n_starts\endcsname{7}
\expandafter\def\csname EERjointflex@results.debut_flexible.k2.n_finite\endcsname{7}
\expandafter\def\csname EERjointflex@results.debut_flexible.k2.n_bfgs_conv\endcsname{6}
\expandafter\def\csname EERjointflex@results.debut_flexible.k2.n_aborted\endcsname{0.00}
\expandafter\def\csname EERjointflex@results.debut_flexible.k2.best_start\endcsname{2}
\expandafter\def\csname EERjointflex@results.debut_flexible.k2.lambda_p\endcsname{-0.30}
\expandafter\def\csname EERjointflex@results.debut_flexible.k2.lambda_d\endcsname{1.80}
\expandafter\def\csname EERjointflex@results.debut_flexible.k2.min_share\endcsname{0.135}
\expandafter\def\csname EERjointflex@results.debut_flexible.k2.reproduction_difference\endcsname{.}
\expandafter\def\csname EERjointflex@results.debut_flexible.k2.iteration_ceiling\endcsname{100.00}
\expandafter\def\csname EERjointflex@results.debut_flexible.k2.aic\endcsname{10847.71}
\expandafter\def\csname EERjointflex@results.debut_flexible.k2.bic_persons\endcsname{10994.51}
\expandafter\def\csname EERjointflex@results.debut_flexible.k2.cluster_ci_lo\endcsname{.}
\expandafter\def\csname EERjointflex@results.debut_flexible.k2.cluster_ci_hi\endcsname{.}
\expandafter\def\csname EERjointflex@counts.full.g0.women\endcsname{2,256}
\expandafter\def\csname EERjointflex@counts.full.g0.calendar_months\endcsname{49,772}
\expandafter\def\csname EERjointflex@counts.full.g0.school_months\endcsname{37,900}
\expandafter\def\csname EERjointflex@counts.full.g0.interruption_events\endcsname{450}
\expandafter\def\csname EERjointflex@counts.full.g0.conception_events\endcsname{26}
\expandafter\def\csname EERjointflex@counts.full.g0.summer_conceptions\endcsname{7}
\expandafter\def\csname EERjointflex@counts.full.g0.pregnancy_risk_months\endcsname{49,772}
\expandafter\def\csname EERjointflex@counts.full.g0.exposed_school_months\endcsname{0}
\expandafter\def\csname EERjointflex@counts.full.g0.onset_without_exposure_women\endcsname{26}
\expandafter\def\csname EERjointflex@counts.full.g1.women\endcsname{452}
\expandafter\def\csname EERjointflex@counts.full.g1.calendar_months\endcsname{12,706}
\expandafter\def\csname EERjointflex@counts.full.g1.school_months\endcsname{9,508}
\expandafter\def\csname EERjointflex@counts.full.g1.interruption_events\endcsname{198}
\expandafter\def\csname EERjointflex@counts.full.g1.conception_events\endcsname{452}
\expandafter\def\csname EERjointflex@counts.full.g1.summer_conceptions\endcsname{107}
\expandafter\def\csname EERjointflex@counts.full.g1.pregnancy_risk_months\endcsname{6,933}
\expandafter\def\csname EERjointflex@counts.full.g1.exposed_school_months\endcsname{4,394}
\expandafter\def\csname EERjointflex@counts.full.g1.onset_without_exposure_women\endcsname{0}
\expandafter\def\csname EERjointflex@counts.full.g2.women\endcsname{120}
\expandafter\def\csname EERjointflex@counts.full.g2.calendar_months\endcsname{2,120}
\expandafter\def\csname EERjointflex@counts.full.g2.school_months\endcsname{1,545}
\expandafter\def\csname EERjointflex@counts.full.g2.interruption_events\endcsname{63}
\expandafter\def\csname EERjointflex@counts.full.g2.conception_events\endcsname{0}
\expandafter\def\csname EERjointflex@counts.full.g2.summer_conceptions\endcsname{0}
\expandafter\def\csname EERjointflex@counts.full.g2.pregnancy_risk_months\endcsname{0}
\expandafter\def\csname EERjointflex@counts.full.g2.exposed_school_months\endcsname{1,545}
\expandafter\def\csname EERjointflex@counts.full.g2.onset_without_exposure_women\endcsname{0}
\expandafter\def\csname EERjointflex@counts.full.g3.women\endcsname{0}
\expandafter\def\csname EERjointflex@counts.full.g3.calendar_months\endcsname{0}
\expandafter\def\csname EERjointflex@counts.full.g3.school_months\endcsname{0}
\expandafter\def\csname EERjointflex@counts.full.g3.interruption_events\endcsname{0}
\expandafter\def\csname EERjointflex@counts.full.g3.conception_events\endcsname{0}
\expandafter\def\csname EERjointflex@counts.full.g3.summer_conceptions\endcsname{0}
\expandafter\def\csname EERjointflex@counts.full.g3.pregnancy_risk_months\endcsname{0}
\expandafter\def\csname EERjointflex@counts.full.g3.exposed_school_months\endcsname{0}
\expandafter\def\csname EERjointflex@counts.full.g3.onset_without_exposure_women\endcsname{0}
\expandafter\def\csname EERjointflex@counts.full.g4.women\endcsname{2,828}
\expandafter\def\csname EERjointflex@counts.full.g4.calendar_months\endcsname{64,598}
\expandafter\def\csname EERjointflex@counts.full.g4.school_months\endcsname{48,953}
\expandafter\def\csname EERjointflex@counts.full.g4.interruption_events\endcsname{711}
\expandafter\def\csname EERjointflex@counts.full.g4.conception_events\endcsname{478}
\expandafter\def\csname EERjointflex@counts.full.g4.summer_conceptions\endcsname{114}
\expandafter\def\csname EERjointflex@counts.full.g4.pregnancy_risk_months\endcsname{56,705}
\expandafter\def\csname EERjointflex@counts.full.g4.exposed_school_months\endcsname{5,939}
\expandafter\def\csname EERjointflex@counts.full.g4.onset_without_exposure_women\endcsname{26}
\expandafter\def\csname EERjointflex@counts.restricted.g0.women\endcsname{2,256}
\expandafter\def\csname EERjointflex@counts.restricted.g0.calendar_months\endcsname{49,772}
\expandafter\def\csname EERjointflex@counts.restricted.g0.school_months\endcsname{37,900}
\expandafter\def\csname EERjointflex@counts.restricted.g0.interruption_events\endcsname{450}
\expandafter\def\csname EERjointflex@counts.restricted.g0.conception_events\endcsname{26}
\expandafter\def\csname EERjointflex@counts.restricted.g0.summer_conceptions\endcsname{7}
\expandafter\def\csname EERjointflex@counts.restricted.g0.pregnancy_risk_months\endcsname{49,772}
\expandafter\def\csname EERjointflex@counts.restricted.g0.exposed_school_months\endcsname{0}
\expandafter\def\csname EERjointflex@counts.restricted.g0.onset_without_exposure_women\endcsname{26}
\expandafter\def\csname EERjointflex@counts.restricted.g1.women\endcsname{452}
\expandafter\def\csname EERjointflex@counts.restricted.g1.calendar_months\endcsname{12,706}
\expandafter\def\csname EERjointflex@counts.restricted.g1.school_months\endcsname{9,508}
\expandafter\def\csname EERjointflex@counts.restricted.g1.interruption_events\endcsname{198}
\expandafter\def\csname EERjointflex@counts.restricted.g1.conception_events\endcsname{452}
\expandafter\def\csname EERjointflex@counts.restricted.g1.summer_conceptions\endcsname{107}
\expandafter\def\csname EERjointflex@counts.restricted.g1.pregnancy_risk_months\endcsname{6,933}
\expandafter\def\csname EERjointflex@counts.restricted.g1.exposed_school_months\endcsname{4,394}
\expandafter\def\csname EERjointflex@counts.restricted.g1.onset_without_exposure_women\endcsname{0}
\expandafter\def\csname EERjointflex@counts.restricted.g2.women\endcsname{0}
\expandafter\def\csname EERjointflex@counts.restricted.g2.calendar_months\endcsname{0}
\expandafter\def\csname EERjointflex@counts.restricted.g2.school_months\endcsname{0}
\expandafter\def\csname EERjointflex@counts.restricted.g2.interruption_events\endcsname{0}
\expandafter\def\csname EERjointflex@counts.restricted.g2.conception_events\endcsname{0}
\expandafter\def\csname EERjointflex@counts.restricted.g2.summer_conceptions\endcsname{0}
\expandafter\def\csname EERjointflex@counts.restricted.g2.pregnancy_risk_months\endcsname{0}
\expandafter\def\csname EERjointflex@counts.restricted.g2.exposed_school_months\endcsname{0}
\expandafter\def\csname EERjointflex@counts.restricted.g2.onset_without_exposure_women\endcsname{0}
\expandafter\def\csname EERjointflex@counts.restricted.g3.women\endcsname{0}
\expandafter\def\csname EERjointflex@counts.restricted.g3.calendar_months\endcsname{0}
\expandafter\def\csname EERjointflex@counts.restricted.g3.school_months\endcsname{0}
\expandafter\def\csname EERjointflex@counts.restricted.g3.interruption_events\endcsname{0}
\expandafter\def\csname EERjointflex@counts.restricted.g3.conception_events\endcsname{0}
\expandafter\def\csname EERjointflex@counts.restricted.g3.summer_conceptions\endcsname{0}
\expandafter\def\csname EERjointflex@counts.restricted.g3.pregnancy_risk_months\endcsname{0}
\expandafter\def\csname EERjointflex@counts.restricted.g3.exposed_school_months\endcsname{0}
\expandafter\def\csname EERjointflex@counts.restricted.g3.onset_without_exposure_women\endcsname{0}
\expandafter\def\csname EERjointflex@counts.restricted.g4.women\endcsname{2,708}
\expandafter\def\csname EERjointflex@counts.restricted.g4.calendar_months\endcsname{62,478}
\expandafter\def\csname EERjointflex@counts.restricted.g4.school_months\endcsname{47,408}
\expandafter\def\csname EERjointflex@counts.restricted.g4.interruption_events\endcsname{648}
\expandafter\def\csname EERjointflex@counts.restricted.g4.conception_events\endcsname{478}
\expandafter\def\csname EERjointflex@counts.restricted.g4.summer_conceptions\endcsname{114}
\expandafter\def\csname EERjointflex@counts.restricted.g4.pregnancy_risk_months\endcsname{56,705}
\expandafter\def\csname EERjointflex@counts.restricted.g4.exposed_school_months\endcsname{4,394}
\expandafter\def\csname EERjointflex@counts.restricted.g4.onset_without_exposure_women\endcsname{26}
\expandafter\def\csname EERjointflex@counts.debut.g0.women\endcsname{2,250}
\expandafter\def\csname EERjointflex@counts.debut.g0.calendar_months\endcsname{49,763}
\expandafter\def\csname EERjointflex@counts.debut.g0.school_months\endcsname{37,892}
\expandafter\def\csname EERjointflex@counts.debut.g0.interruption_events\endcsname{445}
\expandafter\def\csname EERjointflex@counts.debut.g0.conception_events\endcsname{23}
\expandafter\def\csname EERjointflex@counts.debut.g0.summer_conceptions\endcsname{6}
\expandafter\def\csname EERjointflex@counts.debut.g0.pregnancy_risk_months\endcsname{49,763}
\expandafter\def\csname EERjointflex@counts.debut.g0.exposed_school_months\endcsname{0}
\expandafter\def\csname EERjointflex@counts.debut.g0.onset_without_exposure_women\endcsname{23}
\expandafter\def\csname EERjointflex@counts.debut.g1.women\endcsname{373}
\expandafter\def\csname EERjointflex@counts.debut.g1.calendar_months\endcsname{11,211}
\expandafter\def\csname EERjointflex@counts.debut.g1.school_months\endcsname{8,372}
\expandafter\def\csname EERjointflex@counts.debut.g1.interruption_events\endcsname{157}
\expandafter\def\csname EERjointflex@counts.debut.g1.conception_events\endcsname{373}
\expandafter\def\csname EERjointflex@counts.debut.g1.summer_conceptions\endcsname{87}
\expandafter\def\csname EERjointflex@counts.debut.g1.pregnancy_risk_months\endcsname{6,777}
\expandafter\def\csname EERjointflex@counts.debut.g1.exposed_school_months\endcsname{3,375}
\expandafter\def\csname EERjointflex@counts.debut.g1.onset_without_exposure_women\endcsname{0}
\expandafter\def\csname EERjointflex@counts.debut.g2.women\endcsname{0}
\expandafter\def\csname EERjointflex@counts.debut.g2.calendar_months\endcsname{0}
\expandafter\def\csname EERjointflex@counts.debut.g2.school_months\endcsname{0}
\expandafter\def\csname EERjointflex@counts.debut.g2.interruption_events\endcsname{0}
\expandafter\def\csname EERjointflex@counts.debut.g2.conception_events\endcsname{0}
\expandafter\def\csname EERjointflex@counts.debut.g2.summer_conceptions\endcsname{0}
\expandafter\def\csname EERjointflex@counts.debut.g2.pregnancy_risk_months\endcsname{0}
\expandafter\def\csname EERjointflex@counts.debut.g2.exposed_school_months\endcsname{0}
\expandafter\def\csname EERjointflex@counts.debut.g2.onset_without_exposure_women\endcsname{0}
\expandafter\def\csname EERjointflex@counts.debut.g3.women\endcsname{0}
\expandafter\def\csname EERjointflex@counts.debut.g3.calendar_months\endcsname{0}
\expandafter\def\csname EERjointflex@counts.debut.g3.school_months\endcsname{0}
\expandafter\def\csname EERjointflex@counts.debut.g3.interruption_events\endcsname{0}
\expandafter\def\csname EERjointflex@counts.debut.g3.conception_events\endcsname{0}
\expandafter\def\csname EERjointflex@counts.debut.g3.summer_conceptions\endcsname{0}
\expandafter\def\csname EERjointflex@counts.debut.g3.pregnancy_risk_months\endcsname{0}
\expandafter\def\csname EERjointflex@counts.debut.g3.exposed_school_months\endcsname{0}
\expandafter\def\csname EERjointflex@counts.debut.g3.onset_without_exposure_women\endcsname{0}
\expandafter\def\csname EERjointflex@counts.debut.g4.women\endcsname{2,623}
\expandafter\def\csname EERjointflex@counts.debut.g4.calendar_months\endcsname{60,974}
\expandafter\def\csname EERjointflex@counts.debut.g4.school_months\endcsname{46,264}
\expandafter\def\csname EERjointflex@counts.debut.g4.interruption_events\endcsname{602}
\expandafter\def\csname EERjointflex@counts.debut.g4.conception_events\endcsname{396}
\expandafter\def\csname EERjointflex@counts.debut.g4.summer_conceptions\endcsname{93}
\expandafter\def\csname EERjointflex@counts.debut.g4.pregnancy_risk_months\endcsname{56,540}
\expandafter\def\csname EERjointflex@counts.debut.g4.exposed_school_months\endcsname{3,375}
\expandafter\def\csname EERjointflex@counts.debut.g4.onset_without_exposure_women\endcsname{23}
\expandafter\def\csname EERjointflex@diagnostics.full_flexible.converged\endcsname{1.00}
\expandafter\def\csname EERjointflex@diagnostics.full_flexible.bfgs\endcsname{1.00}
\expandafter\def\csname EERjointflex@diagnostics.full_flexible.rel_grad\endcsname{2.14e-07}
\expandafter\def\csname EERjointflex@diagnostics.full_flexible.v_pd\endcsname{1.00}
\expandafter\def\csname EERjointflex@diagnostics.full_flexible.v_scaffold\endcsname{0.00}
\expandafter\def\csname EERjointflex@diagnostics.full_flexible.clip_hits\endcsname{0.00}
\expandafter\def\csname EERjointflex@diagnostics.full_flexible.min_eigen\endcsname{0.00}
\expandafter\def\csname EERjointflex@diagnostics.full_flexible.max_eigen\endcsname{0.48}
\expandafter\def\csname EERjointflex@diagnostics.full_flexible.eigen_ratio\endcsname{1.04e-04}
\expandafter\def\csname EERjointflex@diagnostics.full_flexible.saved_ll\endcsname{-6220.71}
\expandafter\def\csname EERjointflex@diagnostics.full_flexible.evaluated_ll\endcsname{-6220.71}
\expandafter\def\csname EERjointflex@diagnostics.full_flexible.relabeled_ll\endcsname{-6220.71}
\expandafter\def\csname EERjointflex@diagnostics.full_flexible.ll_difference\endcsname{0.00e+00}
\expandafter\def\csname EERjointflex@diagnostics.full_flexible.label_difference\endcsname{0.00e+00}
\expandafter\def\csname EERjointflex@diagnostics.full_flexible.lambda_p\endcsname{4.13}
\expandafter\def\csname EERjointflex@diagnostics.full_flexible.lambda_y\endcsname{0.22}
\expandafter\def\csname EERjointflex@diagnostics.full_flexible.pi1\endcsname{0.954}
\expandafter\def\csname EERjointflex@diagnostics.full_flexible.pi2\endcsname{0.046}
\expandafter\def\csname EERjointflex@diagnostics.full_flexible.oriented_lambda_p\endcsname{4.13}
\expandafter\def\csname EERjointflex@diagnostics.full_flexible.oriented_lambda_y\endcsname{0.22}
\expandafter\def\csname EERjointflex@diagnostics.full_flexible.higher_conception_share_pct\endcsname{4.61}
\expandafter\def\csname EERjointflex@diagnostics.full_flexible.conception_type_ratio\endcsname{62.42}
\expandafter\def\csname EERjointflex@diagnostics.full_flexible.outcome_type_ratio\endcsname{1.24}
\expandafter\def\csname EERjointflex@diagnostics.restricted_flexible.converged\endcsname{1.00}
\expandafter\def\csname EERjointflex@diagnostics.restricted_flexible.bfgs\endcsname{1.00}
\expandafter\def\csname EERjointflex@diagnostics.restricted_flexible.rel_grad\endcsname{3.31e-07}
\expandafter\def\csname EERjointflex@diagnostics.restricted_flexible.v_pd\endcsname{1.00}
\expandafter\def\csname EERjointflex@diagnostics.restricted_flexible.v_scaffold\endcsname{0.00}
\expandafter\def\csname EERjointflex@diagnostics.restricted_flexible.clip_hits\endcsname{0.00}
\expandafter\def\csname EERjointflex@diagnostics.restricted_flexible.min_eigen\endcsname{0.00}
\expandafter\def\csname EERjointflex@diagnostics.restricted_flexible.max_eigen\endcsname{0.44}
\expandafter\def\csname EERjointflex@diagnostics.restricted_flexible.eigen_ratio\endcsname{1.23e-04}
\expandafter\def\csname EERjointflex@diagnostics.restricted_flexible.saved_ll\endcsname{-5972.92}
\expandafter\def\csname EERjointflex@diagnostics.restricted_flexible.evaluated_ll\endcsname{-5972.92}
\expandafter\def\csname EERjointflex@diagnostics.restricted_flexible.relabeled_ll\endcsname{-5972.92}
\expandafter\def\csname EERjointflex@diagnostics.restricted_flexible.ll_difference\endcsname{0.00e+00}
\expandafter\def\csname EERjointflex@diagnostics.restricted_flexible.label_difference\endcsname{9.09e-13}
\expandafter\def\csname EERjointflex@diagnostics.restricted_flexible.lambda_p\endcsname{-4.13}
\expandafter\def\csname EERjointflex@diagnostics.restricted_flexible.lambda_y\endcsname{-0.22}
\expandafter\def\csname EERjointflex@diagnostics.restricted_flexible.pi1\endcsname{0.046}
\expandafter\def\csname EERjointflex@diagnostics.restricted_flexible.pi2\endcsname{0.954}
\expandafter\def\csname EERjointflex@diagnostics.restricted_flexible.oriented_lambda_p\endcsname{4.13}
\expandafter\def\csname EERjointflex@diagnostics.restricted_flexible.oriented_lambda_y\endcsname{0.22}
\expandafter\def\csname EERjointflex@diagnostics.restricted_flexible.higher_conception_share_pct\endcsname{4.61}
\expandafter\def\csname EERjointflex@diagnostics.restricted_flexible.conception_type_ratio\endcsname{62.42}
\expandafter\def\csname EERjointflex@diagnostics.restricted_flexible.outcome_type_ratio\endcsname{1.24}
\expandafter\def\csname EERjointflex@diagnostics.debut_flexible.converged\endcsname{1.00}
\expandafter\def\csname EERjointflex@diagnostics.debut_flexible.bfgs\endcsname{1.00}
\expandafter\def\csname EERjointflex@diagnostics.debut_flexible.rel_grad\endcsname{1.20e-06}
\expandafter\def\csname EERjointflex@diagnostics.debut_flexible.v_pd\endcsname{1.00}
\expandafter\def\csname EERjointflex@diagnostics.debut_flexible.v_scaffold\endcsname{0.00}
\expandafter\def\csname EERjointflex@diagnostics.debut_flexible.clip_hits\endcsname{0.00}
\expandafter\def\csname EERjointflex@diagnostics.debut_flexible.min_eigen\endcsname{0.00}
\expandafter\def\csname EERjointflex@diagnostics.debut_flexible.max_eigen\endcsname{51.15}
\expandafter\def\csname EERjointflex@diagnostics.debut_flexible.eigen_ratio\endcsname{1.34e-06}
\expandafter\def\csname EERjointflex@diagnostics.debut_flexible.saved_ll\endcsname{-5398.85}
\expandafter\def\csname EERjointflex@diagnostics.debut_flexible.evaluated_ll\endcsname{-5398.85}
\expandafter\def\csname EERjointflex@diagnostics.debut_flexible.relabeled_ll\endcsname{-5398.85}
\expandafter\def\csname EERjointflex@diagnostics.debut_flexible.ll_difference\endcsname{0.00e+00}
\expandafter\def\csname EERjointflex@diagnostics.debut_flexible.label_difference\endcsname{0.00e+00}
\expandafter\def\csname EERjointflex@diagnostics.debut_flexible.lambda_p\endcsname{-0.30}
\expandafter\def\csname EERjointflex@diagnostics.debut_flexible.lambda_y\endcsname{1.80}
\expandafter\def\csname EERjointflex@diagnostics.debut_flexible.pi1\endcsname{0.135}
\expandafter\def\csname EERjointflex@diagnostics.debut_flexible.pi2\endcsname{0.865}
\expandafter\def\csname EERjointflex@diagnostics.debut_flexible.oriented_lambda_p\endcsname{0.30}
\expandafter\def\csname EERjointflex@diagnostics.debut_flexible.oriented_lambda_y\endcsname{-1.80}
\expandafter\def\csname EERjointflex@diagnostics.debut_flexible.higher_conception_share_pct\endcsname{13.50}
\expandafter\def\csname EERjointflex@diagnostics.debut_flexible.conception_type_ratio\endcsname{1.35}
\expandafter\def\csname EERjointflex@diagnostics.debut_flexible.outcome_type_ratio\endcsname{0.16}

\IfFileExists{evidence/refinement/joint/debut_entry/numbers_interruption_joint_entry.tex}{
\providecommand{\EERjointentry}[1]{\csname EERjointentry@#1\endcsname}
\expandafter\def\csname EERjointentry@results.entry_linear.k1.rc\endcsname{0.00}
\expandafter\def\csname EERjointentry@results.entry_linear.k1.converged\endcsname{1.00}
\expandafter\def\csname EERjointentry@results.entry_linear.k1.b\endcsname{1.36}
\expandafter\def\csname EERjointentry@results.entry_linear.k1.se\endcsname{0.09}
\expandafter\def\csname EERjointentry@results.entry_linear.k1.se_cluster\endcsname{0.10}
\expandafter\def\csname EERjointentry@results.entry_linear.k1.hr\endcsname{3.90}
\expandafter\def\csname EERjointentry@results.entry_linear.k1.ci_lo\endcsname{3.24}
\expandafter\def\csname EERjointentry@results.entry_linear.k1.ci_hi\endcsname{4.70}
\expandafter\def\csname EERjointentry@results.entry_linear.k1.b_absent\endcsname{1.38}
\expandafter\def\csname EERjointentry@results.entry_linear.k1.se_absent\endcsname{0.11}
\expandafter\def\csname EERjointentry@results.entry_linear.k1.ll\endcsname{-5891.54}
\expandafter\def\csname EERjointentry@results.entry_linear.k1.kparams\endcsname{8.00}
\expandafter\def\csname EERjointentry@results.entry_linear.k1.women\endcsname{2,828}
\expandafter\def\csname EERjointentry@results.entry_linear.k1.person_months\endcsname{64,598}
\expandafter\def\csname EERjointentry@results.entry_linear.k1.dropout_events\endcsname{711}
\expandafter\def\csname EERjointentry@results.entry_linear.k1.pregnancy_events\endcsname{396}
\expandafter\def\csname EERjointentry@results.entry_linear.k1.exposed_months\endcsname{7,893}
\expandafter\def\csname EERjointentry@results.entry_linear.k1.rel_grad\endcsname{.}
\expandafter\def\csname EERjointentry@results.entry_linear.k1.v_pd\endcsname{.}
\expandafter\def\csname EERjointentry@results.entry_linear.k1.v_scaffold\endcsname{.}
\expandafter\def\csname EERjointentry@results.entry_linear.k1.clip_hits\endcsname{.}
\expandafter\def\csname EERjointentry@results.entry_linear.k1.n_starts\endcsname{.}
\expandafter\def\csname EERjointentry@results.entry_linear.k1.n_finite\endcsname{.}
\expandafter\def\csname EERjointentry@results.entry_linear.k1.n_bfgs_conv\endcsname{.}
\expandafter\def\csname EERjointentry@results.entry_linear.k1.n_aborted\endcsname{.}
\expandafter\def\csname EERjointentry@results.entry_linear.k1.best_start\endcsname{.}
\expandafter\def\csname EERjointentry@results.entry_linear.k1.lambda_p\endcsname{.}
\expandafter\def\csname EERjointentry@results.entry_linear.k1.lambda_d\endcsname{.}
\expandafter\def\csname EERjointentry@results.entry_linear.k1.min_share\endcsname{.}
\expandafter\def\csname EERjointentry@results.entry_linear.k1.aic\endcsname{11799.07}
\expandafter\def\csname EERjointentry@results.entry_linear.k1.bic_persons\endcsname{11846.65}
\expandafter\def\csname EERjointentry@results.entry_linear.k1.cluster_ci_lo\endcsname{3.22}
\expandafter\def\csname EERjointentry@results.entry_linear.k1.cluster_ci_hi\endcsname{4.73}
\expandafter\def\csname EERjointentry@results.entry_linear.k1.hr_absent\endcsname{3.96}
\expandafter\def\csname EERjointentry@results.entry_linear.k1.conception_type_ratio\endcsname{.}
\expandafter\def\csname EERjointentry@results.entry_linear.k1.outcome_type_ratio\endcsname{.}
\expandafter\def\csname EERjointentry@results.entry_linear.k1.higher_conception_share_pct\endcsname{.}
\expandafter\def\csname EERjointentry@results.entry_linear.k2.rc\endcsname{0.00}
\expandafter\def\csname EERjointentry@results.entry_linear.k2.converged\endcsname{1.00}
\expandafter\def\csname EERjointentry@results.entry_linear.k2.b\endcsname{1.31}
\expandafter\def\csname EERjointentry@results.entry_linear.k2.se\endcsname{0.12}
\expandafter\def\csname EERjointentry@results.entry_linear.k2.se_cluster\endcsname{.}
\expandafter\def\csname EERjointentry@results.entry_linear.k2.hr\endcsname{3.72}
\expandafter\def\csname EERjointentry@results.entry_linear.k2.ci_lo\endcsname{2.94}
\expandafter\def\csname EERjointentry@results.entry_linear.k2.ci_hi\endcsname{4.70}
\expandafter\def\csname EERjointentry@results.entry_linear.k2.b_absent\endcsname{1.40}
\expandafter\def\csname EERjointentry@results.entry_linear.k2.se_absent\endcsname{0.12}
\expandafter\def\csname EERjointentry@results.entry_linear.k2.ll\endcsname{-5891.18}
\expandafter\def\csname EERjointentry@results.entry_linear.k2.kparams\endcsname{11.00}
\expandafter\def\csname EERjointentry@results.entry_linear.k2.women\endcsname{2,828}
\expandafter\def\csname EERjointentry@results.entry_linear.k2.person_months\endcsname{64,598}
\expandafter\def\csname EERjointentry@results.entry_linear.k2.dropout_events\endcsname{711}
\expandafter\def\csname EERjointentry@results.entry_linear.k2.pregnancy_events\endcsname{396}
\expandafter\def\csname EERjointentry@results.entry_linear.k2.exposed_months\endcsname{7,893}
\expandafter\def\csname EERjointentry@results.entry_linear.k2.rel_grad\endcsname{5.76e-07}
\expandafter\def\csname EERjointentry@results.entry_linear.k2.v_pd\endcsname{1.00}
\expandafter\def\csname EERjointentry@results.entry_linear.k2.v_scaffold\endcsname{0.00}
\expandafter\def\csname EERjointentry@results.entry_linear.k2.clip_hits\endcsname{0.00}
\expandafter\def\csname EERjointentry@results.entry_linear.k2.n_starts\endcsname{7}
\expandafter\def\csname EERjointentry@results.entry_linear.k2.n_finite\endcsname{7}
\expandafter\def\csname EERjointentry@results.entry_linear.k2.n_bfgs_conv\endcsname{7}
\expandafter\def\csname EERjointentry@results.entry_linear.k2.n_aborted\endcsname{0}
\expandafter\def\csname EERjointentry@results.entry_linear.k2.best_start\endcsname{6}
\expandafter\def\csname EERjointentry@results.entry_linear.k2.lambda_p\endcsname{1.67}
\expandafter\def\csname EERjointentry@results.entry_linear.k2.lambda_d\endcsname{1.69}
\expandafter\def\csname EERjointentry@results.entry_linear.k2.min_share\endcsname{0.018}
\expandafter\def\csname EERjointentry@results.entry_linear.k2.aic\endcsname{11804.36}
\expandafter\def\csname EERjointentry@results.entry_linear.k2.bic_persons\endcsname{11869.78}
\expandafter\def\csname EERjointentry@results.entry_linear.k2.cluster_ci_lo\endcsname{.}
\expandafter\def\csname EERjointentry@results.entry_linear.k2.cluster_ci_hi\endcsname{.}
\expandafter\def\csname EERjointentry@results.entry_linear.k2.hr_absent\endcsname{4.04}
\expandafter\def\csname EERjointentry@results.entry_linear.k2.conception_type_ratio\endcsname{5.31}
\expandafter\def\csname EERjointentry@results.entry_linear.k2.outcome_type_ratio\endcsname{5.40}
\expandafter\def\csname EERjointentry@results.entry_linear.k2.higher_conception_share_pct\endcsname{1.78}
\expandafter\def\csname EERjointentry@results.latercc_baseline_entry.k1.rc\endcsname{0.00}
\expandafter\def\csname EERjointentry@results.latercc_baseline_entry.k1.converged\endcsname{1.00}
\expandafter\def\csname EERjointentry@results.latercc_baseline_entry.k1.b\endcsname{1.35}
\expandafter\def\csname EERjointentry@results.latercc_baseline_entry.k1.se\endcsname{0.14}
\expandafter\def\csname EERjointentry@results.latercc_baseline_entry.k1.se_cluster\endcsname{0.15}
\expandafter\def\csname EERjointentry@results.latercc_baseline_entry.k1.hr\endcsname{3.85}
\expandafter\def\csname EERjointentry@results.latercc_baseline_entry.k1.ci_lo\endcsname{2.92}
\expandafter\def\csname EERjointentry@results.latercc_baseline_entry.k1.ci_hi\endcsname{5.09}
\expandafter\def\csname EERjointentry@results.latercc_baseline_entry.k1.b_absent\endcsname{1.23}
\expandafter\def\csname EERjointentry@results.latercc_baseline_entry.k1.se_absent\endcsname{0.26}
\expandafter\def\csname EERjointentry@results.latercc_baseline_entry.k1.ll\endcsname{-2981.13}
\expandafter\def\csname EERjointentry@results.latercc_baseline_entry.k1.kparams\endcsname{8.00}
\expandafter\def\csname EERjointentry@results.latercc_baseline_entry.k1.women\endcsname{1,733}
\expandafter\def\csname EERjointentry@results.latercc_baseline_entry.k1.person_months\endcsname{35,063}
\expandafter\def\csname EERjointentry@results.latercc_baseline_entry.k1.dropout_events\endcsname{336}
\expandafter\def\csname EERjointentry@results.latercc_baseline_entry.k1.pregnancy_events\endcsname{205}
\expandafter\def\csname EERjointentry@results.latercc_baseline_entry.k1.exposed_months\endcsname{2,661}
\expandafter\def\csname EERjointentry@results.latercc_baseline_entry.k1.rel_grad\endcsname{.}
\expandafter\def\csname EERjointentry@results.latercc_baseline_entry.k1.v_pd\endcsname{.}
\expandafter\def\csname EERjointentry@results.latercc_baseline_entry.k1.v_scaffold\endcsname{.}
\expandafter\def\csname EERjointentry@results.latercc_baseline_entry.k1.clip_hits\endcsname{.}
\expandafter\def\csname EERjointentry@results.latercc_baseline_entry.k1.n_starts\endcsname{.}
\expandafter\def\csname EERjointentry@results.latercc_baseline_entry.k1.n_finite\endcsname{.}
\expandafter\def\csname EERjointentry@results.latercc_baseline_entry.k1.n_bfgs_conv\endcsname{.}
\expandafter\def\csname EERjointentry@results.latercc_baseline_entry.k1.n_aborted\endcsname{.}
\expandafter\def\csname EERjointentry@results.latercc_baseline_entry.k1.best_start\endcsname{.}
\expandafter\def\csname EERjointentry@results.latercc_baseline_entry.k1.lambda_p\endcsname{.}
\expandafter\def\csname EERjointentry@results.latercc_baseline_entry.k1.lambda_d\endcsname{.}
\expandafter\def\csname EERjointentry@results.latercc_baseline_entry.k1.min_share\endcsname{.}
\expandafter\def\csname EERjointentry@results.latercc_baseline_entry.k1.aic\endcsname{5978.27}
\expandafter\def\csname EERjointentry@results.latercc_baseline_entry.k1.bic_persons\endcsname{6021.93}
\expandafter\def\csname EERjointentry@results.latercc_baseline_entry.k1.cluster_ci_lo\endcsname{2.89}
\expandafter\def\csname EERjointentry@results.latercc_baseline_entry.k1.cluster_ci_hi\endcsname{5.13}
\expandafter\def\csname EERjointentry@results.latercc_baseline_entry.k1.hr_absent\endcsname{3.41}
\expandafter\def\csname EERjointentry@results.latercc_baseline_entry.k1.conception_type_ratio\endcsname{.}
\expandafter\def\csname EERjointentry@results.latercc_baseline_entry.k1.outcome_type_ratio\endcsname{.}
\expandafter\def\csname EERjointentry@results.latercc_baseline_entry.k1.higher_conception_share_pct\endcsname{.}
\expandafter\def\csname EERjointentry@results.latercc_baseline_entry.k2.rc\endcsname{0.00}
\expandafter\def\csname EERjointentry@results.latercc_baseline_entry.k2.converged\endcsname{1.00}
\expandafter\def\csname EERjointentry@results.latercc_baseline_entry.k2.b\endcsname{1.29}
\expandafter\def\csname EERjointentry@results.latercc_baseline_entry.k2.se\endcsname{0.18}
\expandafter\def\csname EERjointentry@results.latercc_baseline_entry.k2.se_cluster\endcsname{.}
\expandafter\def\csname EERjointentry@results.latercc_baseline_entry.k2.hr\endcsname{3.64}
\expandafter\def\csname EERjointentry@results.latercc_baseline_entry.k2.ci_lo\endcsname{2.58}
\expandafter\def\csname EERjointentry@results.latercc_baseline_entry.k2.ci_hi\endcsname{5.13}
\expandafter\def\csname EERjointentry@results.latercc_baseline_entry.k2.b_absent\endcsname{1.26}
\expandafter\def\csname EERjointentry@results.latercc_baseline_entry.k2.se_absent\endcsname{0.27}
\expandafter\def\csname EERjointentry@results.latercc_baseline_entry.k2.ll\endcsname{-2980.69}
\expandafter\def\csname EERjointentry@results.latercc_baseline_entry.k2.kparams\endcsname{11.00}
\expandafter\def\csname EERjointentry@results.latercc_baseline_entry.k2.women\endcsname{1,733}
\expandafter\def\csname EERjointentry@results.latercc_baseline_entry.k2.person_months\endcsname{35,063}
\expandafter\def\csname EERjointentry@results.latercc_baseline_entry.k2.dropout_events\endcsname{336}
\expandafter\def\csname EERjointentry@results.latercc_baseline_entry.k2.pregnancy_events\endcsname{205}
\expandafter\def\csname EERjointentry@results.latercc_baseline_entry.k2.exposed_months\endcsname{2,661}
\expandafter\def\csname EERjointentry@results.latercc_baseline_entry.k2.rel_grad\endcsname{8.65e-07}
\expandafter\def\csname EERjointentry@results.latercc_baseline_entry.k2.v_pd\endcsname{1.00}
\expandafter\def\csname EERjointentry@results.latercc_baseline_entry.k2.v_scaffold\endcsname{0.00}
\expandafter\def\csname EERjointentry@results.latercc_baseline_entry.k2.clip_hits\endcsname{0.00}
\expandafter\def\csname EERjointentry@results.latercc_baseline_entry.k2.n_starts\endcsname{7}
\expandafter\def\csname EERjointentry@results.latercc_baseline_entry.k2.n_finite\endcsname{7}
\expandafter\def\csname EERjointentry@results.latercc_baseline_entry.k2.n_bfgs_conv\endcsname{1}
\expandafter\def\csname EERjointentry@results.latercc_baseline_entry.k2.n_aborted\endcsname{0}
\expandafter\def\csname EERjointentry@results.latercc_baseline_entry.k2.best_start\endcsname{5}
\expandafter\def\csname EERjointentry@results.latercc_baseline_entry.k2.lambda_p\endcsname{1.89}
\expandafter\def\csname EERjointentry@results.latercc_baseline_entry.k2.lambda_d\endcsname{2.24}
\expandafter\def\csname EERjointentry@results.latercc_baseline_entry.k2.min_share\endcsname{0.016}
\expandafter\def\csname EERjointentry@results.latercc_baseline_entry.k2.aic\endcsname{5983.38}
\expandafter\def\csname EERjointentry@results.latercc_baseline_entry.k2.bic_persons\endcsname{6043.42}
\expandafter\def\csname EERjointentry@results.latercc_baseline_entry.k2.cluster_ci_lo\endcsname{.}
\expandafter\def\csname EERjointentry@results.latercc_baseline_entry.k2.cluster_ci_hi\endcsname{.}
\expandafter\def\csname EERjointentry@results.latercc_baseline_entry.k2.hr_absent\endcsname{3.52}
\expandafter\def\csname EERjointentry@results.latercc_baseline_entry.k2.conception_type_ratio\endcsname{6.65}
\expandafter\def\csname EERjointentry@results.latercc_baseline_entry.k2.outcome_type_ratio\endcsname{9.42}
\expandafter\def\csname EERjointentry@results.latercc_baseline_entry.k2.higher_conception_share_pct\endcsname{1.55}
\expandafter\def\csname EERjointentry@results.latercc_rich_entry.k1.rc\endcsname{0.00}
\expandafter\def\csname EERjointentry@results.latercc_rich_entry.k1.converged\endcsname{1.00}
\expandafter\def\csname EERjointentry@results.latercc_rich_entry.k1.b\endcsname{1.18}
\expandafter\def\csname EERjointentry@results.latercc_rich_entry.k1.se\endcsname{0.15}
\expandafter\def\csname EERjointentry@results.latercc_rich_entry.k1.se_cluster\endcsname{0.16}
\expandafter\def\csname EERjointentry@results.latercc_rich_entry.k1.hr\endcsname{3.24}
\expandafter\def\csname EERjointentry@results.latercc_rich_entry.k1.ci_lo\endcsname{2.42}
\expandafter\def\csname EERjointentry@results.latercc_rich_entry.k1.ci_hi\endcsname{4.33}
\expandafter\def\csname EERjointentry@results.latercc_rich_entry.k1.b_absent\endcsname{0.79}
\expandafter\def\csname EERjointentry@results.latercc_rich_entry.k1.se_absent\endcsname{0.27}
\expandafter\def\csname EERjointentry@results.latercc_rich_entry.k1.ll\endcsname{-2880.00}
\expandafter\def\csname EERjointentry@results.latercc_rich_entry.k1.kparams\endcsname{32.00}
\expandafter\def\csname EERjointentry@results.latercc_rich_entry.k1.women\endcsname{1,733}
\expandafter\def\csname EERjointentry@results.latercc_rich_entry.k1.person_months\endcsname{35,063}
\expandafter\def\csname EERjointentry@results.latercc_rich_entry.k1.dropout_events\endcsname{336}
\expandafter\def\csname EERjointentry@results.latercc_rich_entry.k1.pregnancy_events\endcsname{205}
\expandafter\def\csname EERjointentry@results.latercc_rich_entry.k1.exposed_months\endcsname{2,661}
\expandafter\def\csname EERjointentry@results.latercc_rich_entry.k1.rel_grad\endcsname{.}
\expandafter\def\csname EERjointentry@results.latercc_rich_entry.k1.v_pd\endcsname{.}
\expandafter\def\csname EERjointentry@results.latercc_rich_entry.k1.v_scaffold\endcsname{.}
\expandafter\def\csname EERjointentry@results.latercc_rich_entry.k1.clip_hits\endcsname{.}
\expandafter\def\csname EERjointentry@results.latercc_rich_entry.k1.n_starts\endcsname{.}
\expandafter\def\csname EERjointentry@results.latercc_rich_entry.k1.n_finite\endcsname{.}
\expandafter\def\csname EERjointentry@results.latercc_rich_entry.k1.n_bfgs_conv\endcsname{.}
\expandafter\def\csname EERjointentry@results.latercc_rich_entry.k1.n_aborted\endcsname{.}
\expandafter\def\csname EERjointentry@results.latercc_rich_entry.k1.best_start\endcsname{.}
\expandafter\def\csname EERjointentry@results.latercc_rich_entry.k1.lambda_p\endcsname{.}
\expandafter\def\csname EERjointentry@results.latercc_rich_entry.k1.lambda_d\endcsname{.}
\expandafter\def\csname EERjointentry@results.latercc_rich_entry.k1.min_share\endcsname{.}
\expandafter\def\csname EERjointentry@results.latercc_rich_entry.k1.aic\endcsname{5823.99}
\expandafter\def\csname EERjointentry@results.latercc_rich_entry.k1.bic_persons\endcsname{5998.64}
\expandafter\def\csname EERjointentry@results.latercc_rich_entry.k1.cluster_ci_lo\endcsname{2.36}
\expandafter\def\csname EERjointentry@results.latercc_rich_entry.k1.cluster_ci_hi\endcsname{4.44}
\expandafter\def\csname EERjointentry@results.latercc_rich_entry.k1.hr_absent\endcsname{2.21}
\expandafter\def\csname EERjointentry@results.latercc_rich_entry.k1.conception_type_ratio\endcsname{.}
\expandafter\def\csname EERjointentry@results.latercc_rich_entry.k1.outcome_type_ratio\endcsname{.}
\expandafter\def\csname EERjointentry@results.latercc_rich_entry.k1.higher_conception_share_pct\endcsname{.}
\expandafter\def\csname EERjointentry@results.latercc_rich_entry.k2.rc\endcsname{0.00}
\expandafter\def\csname EERjointentry@results.latercc_rich_entry.k2.converged\endcsname{0.00}
\expandafter\def\csname EERjointentry@results.latercc_rich_entry.k2.b\endcsname{}
\expandafter\def\csname EERjointentry@results.latercc_rich_entry.k2.se\endcsname{}
\expandafter\def\csname EERjointentry@results.latercc_rich_entry.k2.se_cluster\endcsname{.}
\expandafter\def\csname EERjointentry@results.latercc_rich_entry.k2.hr\endcsname{}
\expandafter\def\csname EERjointentry@results.latercc_rich_entry.k2.ci_lo\endcsname{}
\expandafter\def\csname EERjointentry@results.latercc_rich_entry.k2.ci_hi\endcsname{}
\expandafter\def\csname EERjointentry@results.latercc_rich_entry.k2.b_absent\endcsname{}
\expandafter\def\csname EERjointentry@results.latercc_rich_entry.k2.se_absent\endcsname{}
\expandafter\def\csname EERjointentry@results.latercc_rich_entry.k2.ll\endcsname{-2875.49}
\expandafter\def\csname EERjointentry@results.latercc_rich_entry.k2.kparams\endcsname{35.00}
\expandafter\def\csname EERjointentry@results.latercc_rich_entry.k2.women\endcsname{1,733}
\expandafter\def\csname EERjointentry@results.latercc_rich_entry.k2.person_months\endcsname{35,063}
\expandafter\def\csname EERjointentry@results.latercc_rich_entry.k2.dropout_events\endcsname{336}
\expandafter\def\csname EERjointentry@results.latercc_rich_entry.k2.pregnancy_events\endcsname{205}
\expandafter\def\csname EERjointentry@results.latercc_rich_entry.k2.exposed_months\endcsname{2,661}
\expandafter\def\csname EERjointentry@results.latercc_rich_entry.k2.rel_grad\endcsname{5.73e-07}
\expandafter\def\csname EERjointentry@results.latercc_rich_entry.k2.v_pd\endcsname{0.00}
\expandafter\def\csname EERjointentry@results.latercc_rich_entry.k2.v_scaffold\endcsname{0.00}
\expandafter\def\csname EERjointentry@results.latercc_rich_entry.k2.clip_hits\endcsname{20106.00}
\expandafter\def\csname EERjointentry@results.latercc_rich_entry.k2.n_starts\endcsname{7}
\expandafter\def\csname EERjointentry@results.latercc_rich_entry.k2.n_finite\endcsname{7}
\expandafter\def\csname EERjointentry@results.latercc_rich_entry.k2.n_bfgs_conv\endcsname{0}
\expandafter\def\csname EERjointentry@results.latercc_rich_entry.k2.n_aborted\endcsname{0}
\expandafter\def\csname EERjointentry@results.latercc_rich_entry.k2.best_start\endcsname{2}
\expandafter\def\csname EERjointentry@results.latercc_rich_entry.k2.lambda_p\endcsname{}
\expandafter\def\csname EERjointentry@results.latercc_rich_entry.k2.lambda_d\endcsname{}
\expandafter\def\csname EERjointentry@results.latercc_rich_entry.k2.min_share\endcsname{}
\expandafter\def\csname EERjointentry@results.latercc_rich_entry.k2.aic\endcsname{5820.98}
\expandafter\def\csname EERjointentry@results.latercc_rich_entry.k2.bic_persons\endcsname{6012.00}
\expandafter\def\csname EERjointentry@results.latercc_rich_entry.k2.cluster_ci_lo\endcsname{.}
\expandafter\def\csname EERjointentry@results.latercc_rich_entry.k2.cluster_ci_hi\endcsname{.}
\expandafter\def\csname EERjointentry@results.latercc_rich_entry.k2.hr_absent\endcsname{}
\expandafter\def\csname EERjointentry@results.latercc_rich_entry.k2.conception_type_ratio\endcsname{}
\expandafter\def\csname EERjointentry@results.latercc_rich_entry.k2.outcome_type_ratio\endcsname{}
\expandafter\def\csname EERjointentry@results.latercc_rich_entry.k2.higher_conception_share_pct\endcsname{}
\expandafter\def\csname EERjointentry@posterior.entry_linear.month.month_1.women\endcsname{4}
\expandafter\def\csname EERjointentry@posterior.entry_linear.month.month_1.mass_post_high\endcsname{0.2}
\expandafter\def\csname EERjointentry@posterior.entry_linear.month.month_1.women_above_half\endcsname{0}
\expandafter\def\csname EERjointentry@posterior.entry_linear.month.month_1.total_mass\endcsname{50.3}
\expandafter\def\csname EERjointentry@posterior.entry_linear.month.month_1.mean_post_high\endcsname{0.039}
\expandafter\def\csname EERjointentry@posterior.entry_linear.month.month_1.share_of_high_mass\endcsname{0.003}
\expandafter\def\csname EERjointentry@posterior.entry_linear.month.month_2.women\endcsname{21}
\expandafter\def\csname EERjointentry@posterior.entry_linear.month.month_2.mass_post_high\endcsname{1.9}
\expandafter\def\csname EERjointentry@posterior.entry_linear.month.month_2.women_above_half\endcsname{0}
\expandafter\def\csname EERjointentry@posterior.entry_linear.month.month_2.total_mass\endcsname{50.3}
\expandafter\def\csname EERjointentry@posterior.entry_linear.month.month_2.mean_post_high\endcsname{0.092}
\expandafter\def\csname EERjointentry@posterior.entry_linear.month.month_2.share_of_high_mass\endcsname{0.038}
\expandafter\def\csname EERjointentry@posterior.entry_linear.month.month_3_6.women\endcsname{60}
\expandafter\def\csname EERjointentry@posterior.entry_linear.month.month_3_6.mass_post_high\endcsname{4.3}
\expandafter\def\csname EERjointentry@posterior.entry_linear.month.month_3_6.women_above_half\endcsname{0}
\expandafter\def\csname EERjointentry@posterior.entry_linear.month.month_3_6.total_mass\endcsname{50.3}
\expandafter\def\csname EERjointentry@posterior.entry_linear.month.month_3_6.mean_post_high\endcsname{0.072}
\expandafter\def\csname EERjointentry@posterior.entry_linear.month.month_3_6.share_of_high_mass\endcsname{0.086}
\expandafter\def\csname EERjointentry@posterior.entry_linear.month.month_7_plus.women\endcsname{311}
\expandafter\def\csname EERjointentry@posterior.entry_linear.month.month_7_plus.mass_post_high\endcsname{12.9}
\expandafter\def\csname EERjointentry@posterior.entry_linear.month.month_7_plus.women_above_half\endcsname{0}
\expandafter\def\csname EERjointentry@posterior.entry_linear.month.month_7_plus.total_mass\endcsname{50.3}
\expandafter\def\csname EERjointentry@posterior.entry_linear.month.month_7_plus.mean_post_high\endcsname{0.041}
\expandafter\def\csname EERjointentry@posterior.entry_linear.month.month_7_plus.share_of_high_mass\endcsname{0.256}
\expandafter\def\csname EERjointentry@posterior.full_linear.month.month_1.women\endcsname{6}
\expandafter\def\csname EERjointentry@posterior.full_linear.month.month_1.mass_post_high\endcsname{3.9}
\expandafter\def\csname EERjointentry@posterior.full_linear.month.month_1.women_above_half\endcsname{6}
\expandafter\def\csname EERjointentry@posterior.full_linear.month.month_1.total_mass\endcsname{96.7}
\expandafter\def\csname EERjointentry@posterior.full_linear.month.month_1.mean_post_high\endcsname{0.642}
\expandafter\def\csname EERjointentry@posterior.full_linear.month.month_1.share_of_high_mass\endcsname{0.040}
\expandafter\def\csname EERjointentry@posterior.full_linear.month.month_2.women\endcsname{101}
\expandafter\def\csname EERjointentry@posterior.full_linear.month.month_2.mass_post_high\endcsname{56.6}
\expandafter\def\csname EERjointentry@posterior.full_linear.month.month_2.women_above_half\endcsname{94}
\expandafter\def\csname EERjointentry@posterior.full_linear.month.month_2.total_mass\endcsname{96.7}
\expandafter\def\csname EERjointentry@posterior.full_linear.month.month_2.mean_post_high\endcsname{0.560}
\expandafter\def\csname EERjointentry@posterior.full_linear.month.month_2.share_of_high_mass\endcsname{0.585}
\expandafter\def\csname EERjointentry@posterior.full_linear.month.month_3_6.women\endcsname{60}
\expandafter\def\csname EERjointentry@posterior.full_linear.month.month_3_6.mass_post_high\endcsname{19.8}
\expandafter\def\csname EERjointentry@posterior.full_linear.month.month_3_6.women_above_half\endcsname{5}
\expandafter\def\csname EERjointentry@posterior.full_linear.month.month_3_6.total_mass\endcsname{96.7}
\expandafter\def\csname EERjointentry@posterior.full_linear.month.month_3_6.mean_post_high\endcsname{0.330}
\expandafter\def\csname EERjointentry@posterior.full_linear.month.month_3_6.share_of_high_mass\endcsname{0.205}
\expandafter\def\csname EERjointentry@posterior.full_linear.month.month_7_plus.women\endcsname{311}
\expandafter\def\csname EERjointentry@posterior.full_linear.month.month_7_plus.mass_post_high\endcsname{9.5}
\expandafter\def\csname EERjointentry@posterior.full_linear.month.month_7_plus.women_above_half\endcsname{0}
\expandafter\def\csname EERjointentry@posterior.full_linear.month.month_7_plus.total_mass\endcsname{96.7}
\expandafter\def\csname EERjointentry@posterior.full_linear.month.month_7_plus.mean_post_high\endcsname{0.030}
\expandafter\def\csname EERjointentry@posterior.full_linear.month.month_7_plus.share_of_high_mass\endcsname{0.098}
\expandafter\def\csname EERjointentry@posterior.restricted_l.month.month_1.women\endcsname{6}
\expandafter\def\csname EERjointentry@posterior.restricted_l.month.month_1.mass_post_high\endcsname{3.9}
\expandafter\def\csname EERjointentry@posterior.restricted_l.month.month_1.women_above_half\endcsname{6}
\expandafter\def\csname EERjointentry@posterior.restricted_l.month.month_1.total_mass\endcsname{92.5}
\expandafter\def\csname EERjointentry@posterior.restricted_l.month.month_1.mean_post_high\endcsname{0.643}
\expandafter\def\csname EERjointentry@posterior.restricted_l.month.month_1.share_of_high_mass\endcsname{0.042}
\expandafter\def\csname EERjointentry@posterior.restricted_l.month.month_2.women\endcsname{101}
\expandafter\def\csname EERjointentry@posterior.restricted_l.month.month_2.mass_post_high\endcsname{56.5}
\expandafter\def\csname EERjointentry@posterior.restricted_l.month.month_2.women_above_half\endcsname{96}
\expandafter\def\csname EERjointentry@posterior.restricted_l.month.month_2.total_mass\endcsname{92.5}
\expandafter\def\csname EERjointentry@posterior.restricted_l.month.month_2.mean_post_high\endcsname{0.560}
\expandafter\def\csname EERjointentry@posterior.restricted_l.month.month_2.share_of_high_mass\endcsname{0.611}
\expandafter\def\csname EERjointentry@posterior.restricted_l.month.month_3_6.women\endcsname{60}
\expandafter\def\csname EERjointentry@posterior.restricted_l.month.month_3_6.mass_post_high\endcsname{19.8}
\expandafter\def\csname EERjointentry@posterior.restricted_l.month.month_3_6.women_above_half\endcsname{5}
\expandafter\def\csname EERjointentry@posterior.restricted_l.month.month_3_6.total_mass\endcsname{92.5}
\expandafter\def\csname EERjointentry@posterior.restricted_l.month.month_3_6.mean_post_high\endcsname{0.330}
\expandafter\def\csname EERjointentry@posterior.restricted_l.month.month_3_6.share_of_high_mass\endcsname{0.214}
\expandafter\def\csname EERjointentry@posterior.restricted_l.month.month_7_plus.women\endcsname{311}
\expandafter\def\csname EERjointentry@posterior.restricted_l.month.month_7_plus.mass_post_high\endcsname{9.4}
\expandafter\def\csname EERjointentry@posterior.restricted_l.month.month_7_plus.women_above_half\endcsname{0}
\expandafter\def\csname EERjointentry@posterior.restricted_l.month.month_7_plus.total_mass\endcsname{92.5}
\expandafter\def\csname EERjointentry@posterior.restricted_l.month.month_7_plus.mean_post_high\endcsname{0.030}
\expandafter\def\csname EERjointentry@posterior.restricted_l.month.month_7_plus.share_of_high_mass\endcsname{0.102}
\expandafter\def\csname EERjointentry@posterior.entry_linear.exposed_at_entry.women\endcsname{199}
\expandafter\def\csname EERjointentry@posterior.entry_linear.exposed_at_entry.mass_post_high\endcsname{3.4}
\expandafter\def\csname EERjointentry@posterior.entry_linear.exposed_at_entry.women_above_half\endcsname{0}
\expandafter\def\csname EERjointentry@posterior.entry_linear.exposed_at_entry.total_mass\endcsname{50.3}
\expandafter\def\csname EERjointentry@posterior.entry_linear.exposed_at_entry.mean_post_high\endcsname{0.017}
\expandafter\def\csname EERjointentry@posterior.entry_linear.exposed_at_entry.share_of_high_mass\endcsname{0.069}
\expandafter\def\csname EERjointentry@posterior.entry_linear.never_exposed.women\endcsname{2,256}
\expandafter\def\csname EERjointentry@posterior.entry_linear.never_exposed.mass_post_high\endcsname{29.9}
\expandafter\def\csname EERjointentry@posterior.entry_linear.never_exposed.women_above_half\endcsname{0}
\expandafter\def\csname EERjointentry@posterior.entry_linear.never_exposed.total_mass\endcsname{50.3}
\expandafter\def\csname EERjointentry@posterior.entry_linear.never_exposed.mean_post_high\endcsname{0.013}
\expandafter\def\csname EERjointentry@posterior.entry_linear.never_exposed.share_of_high_mass\endcsname{0.595}
\expandafter\def\csname EERjointentry@posterior.entry_linear.onset_unmodified_debut.women\endcsname{373}
\expandafter\def\csname EERjointentry@posterior.entry_linear.onset_unmodified_debut.mass_post_high\endcsname{16.9}
\expandafter\def\csname EERjointentry@posterior.entry_linear.onset_unmodified_debut.women_above_half\endcsname{0}
\expandafter\def\csname EERjointentry@posterior.entry_linear.onset_unmodified_debut.total_mass\endcsname{50.3}
\expandafter\def\csname EERjointentry@posterior.entry_linear.onset_unmodified_debut.mean_post_high\endcsname{0.045}
\expandafter\def\csname EERjointentry@posterior.entry_linear.onset_unmodified_debut.share_of_high_mass\endcsname{0.336}
\expandafter\def\csname EERjointentry@posterior.full_linear.exposed_at_entry.women\endcsname{120}
\expandafter\def\csname EERjointentry@posterior.full_linear.exposed_at_entry.mass_post_high\endcsname{4.1}
\expandafter\def\csname EERjointentry@posterior.full_linear.exposed_at_entry.women_above_half\endcsname{0}
\expandafter\def\csname EERjointentry@posterior.full_linear.exposed_at_entry.total_mass\endcsname{96.7}
\expandafter\def\csname EERjointentry@posterior.full_linear.exposed_at_entry.mean_post_high\endcsname{0.034}
\expandafter\def\csname EERjointentry@posterior.full_linear.exposed_at_entry.share_of_high_mass\endcsname{0.042}
\expandafter\def\csname EERjointentry@posterior.full_linear.never_exposed.women\endcsname{2,256}
\expandafter\def\csname EERjointentry@posterior.full_linear.never_exposed.mass_post_high\endcsname{5.9}
\expandafter\def\csname EERjointentry@posterior.full_linear.never_exposed.women_above_half\endcsname{4}
\expandafter\def\csname EERjointentry@posterior.full_linear.never_exposed.total_mass\endcsname{96.7}
\expandafter\def\csname EERjointentry@posterior.full_linear.never_exposed.mean_post_high\endcsname{0.003}
\expandafter\def\csname EERjointentry@posterior.full_linear.never_exposed.share_of_high_mass\endcsname{0.061}
\expandafter\def\csname EERjointentry@posterior.full_linear.onset_modified_debut.women\endcsname{79}
\expandafter\def\csname EERjointentry@posterior.full_linear.onset_modified_debut.mass_post_high\endcsname{44.2}
\expandafter\def\csname EERjointentry@posterior.full_linear.onset_modified_debut.women_above_half\endcsname{72}
\expandafter\def\csname EERjointentry@posterior.full_linear.onset_modified_debut.total_mass\endcsname{96.7}
\expandafter\def\csname EERjointentry@posterior.full_linear.onset_modified_debut.mean_post_high\endcsname{0.559}
\expandafter\def\csname EERjointentry@posterior.full_linear.onset_modified_debut.share_of_high_mass\endcsname{0.457}
\expandafter\def\csname EERjointentry@posterior.full_linear.onset_unmodified_debut.women\endcsname{373}
\expandafter\def\csname EERjointentry@posterior.full_linear.onset_unmodified_debut.mass_post_high\endcsname{42.5}
\expandafter\def\csname EERjointentry@posterior.full_linear.onset_unmodified_debut.women_above_half\endcsname{29}
\expandafter\def\csname EERjointentry@posterior.full_linear.onset_unmodified_debut.total_mass\endcsname{96.7}
\expandafter\def\csname EERjointentry@posterior.full_linear.onset_unmodified_debut.mean_post_high\endcsname{0.114}
\expandafter\def\csname EERjointentry@posterior.full_linear.onset_unmodified_debut.share_of_high_mass\endcsname{0.440}
\expandafter\def\csname EERjointentry@posterior.latercc_baseline_entry.exposed_at_entry_or_reclassified.women\endcsname{35}
\expandafter\def\csname EERjointentry@posterior.latercc_baseline_entry.exposed_at_entry_or_reclassified.mass_post_high\endcsname{0.4}
\expandafter\def\csname EERjointentry@posterior.latercc_baseline_entry.exposed_at_entry_or_reclassified.women_above_half\endcsname{0}
\expandafter\def\csname EERjointentry@posterior.latercc_baseline_entry.exposed_at_entry_or_reclassified.total_mass\endcsname{26.9}
\expandafter\def\csname EERjointentry@posterior.latercc_baseline_entry.exposed_at_entry_or_reclassified.mean_post_high\endcsname{0.010}
\expandafter\def\csname EERjointentry@posterior.latercc_baseline_entry.exposed_at_entry_or_reclassified.share_of_high_mass\endcsname{0.014}
\expandafter\def\csname EERjointentry@posterior.latercc_baseline_entry.never_exposed.women\endcsname{1,507}
\expandafter\def\csname EERjointentry@posterior.latercc_baseline_entry.never_exposed.mass_post_high\endcsname{19.9}
\expandafter\def\csname EERjointentry@posterior.latercc_baseline_entry.never_exposed.women_above_half\endcsname{0}
\expandafter\def\csname EERjointentry@posterior.latercc_baseline_entry.never_exposed.total_mass\endcsname{26.9}
\expandafter\def\csname EERjointentry@posterior.latercc_baseline_entry.never_exposed.mean_post_high\endcsname{0.013}
\expandafter\def\csname EERjointentry@posterior.latercc_baseline_entry.never_exposed.share_of_high_mass\endcsname{0.740}
\expandafter\def\csname EERjointentry@posterior.latercc_baseline_entry.onset_retained.women\endcsname{191}
\expandafter\def\csname EERjointentry@posterior.latercc_baseline_entry.onset_retained.mass_post_high\endcsname{6.6}
\expandafter\def\csname EERjointentry@posterior.latercc_baseline_entry.onset_retained.women_above_half\endcsname{0}
\expandafter\def\csname EERjointentry@posterior.latercc_baseline_entry.onset_retained.total_mass\endcsname{26.9}
\expandafter\def\csname EERjointentry@posterior.latercc_baseline_entry.onset_retained.mean_post_high\endcsname{0.035}
\expandafter\def\csname EERjointentry@posterior.latercc_baseline_entry.onset_retained.share_of_high_mass\endcsname{0.246}
\expandafter\def\csname EERjointentry@posterior.latercc_rich_entry.exposed_at_entry_or_reclassified.women\endcsname{35}
\expandafter\def\csname EERjointentry@posterior.latercc_rich_entry.exposed_at_entry_or_reclassified.mass_post_high\endcsname{31.5}
\expandafter\def\csname EERjointentry@posterior.latercc_rich_entry.exposed_at_entry_or_reclassified.women_above_half\endcsname{34}
\expandafter\def\csname EERjointentry@posterior.latercc_rich_entry.exposed_at_entry_or_reclassified.total_mass\endcsname{1570.5}
\expandafter\def\csname EERjointentry@posterior.latercc_rich_entry.exposed_at_entry_or_reclassified.mean_post_high\endcsname{0.899}
\expandafter\def\csname EERjointentry@posterior.latercc_rich_entry.exposed_at_entry_or_reclassified.share_of_high_mass\endcsname{0.020}
\expandafter\def\csname EERjointentry@posterior.latercc_rich_entry.never_exposed.women\endcsname{1,507}
\expandafter\def\csname EERjointentry@posterior.latercc_rich_entry.never_exposed.mass_post_high\endcsname{1369.4}
\expandafter\def\csname EERjointentry@posterior.latercc_rich_entry.never_exposed.women_above_half\endcsname{1,507}
\expandafter\def\csname EERjointentry@posterior.latercc_rich_entry.never_exposed.total_mass\endcsname{1570.5}
\expandafter\def\csname EERjointentry@posterior.latercc_rich_entry.never_exposed.mean_post_high\endcsname{0.909}
\expandafter\def\csname EERjointentry@posterior.latercc_rich_entry.never_exposed.share_of_high_mass\endcsname{0.872}
\expandafter\def\csname EERjointentry@posterior.latercc_rich_entry.onset_retained.women\endcsname{191}
\expandafter\def\csname EERjointentry@posterior.latercc_rich_entry.onset_retained.mass_post_high\endcsname{169.7}
\expandafter\def\csname EERjointentry@posterior.latercc_rich_entry.onset_retained.women_above_half\endcsname{187}
\expandafter\def\csname EERjointentry@posterior.latercc_rich_entry.onset_retained.total_mass\endcsname{1570.5}
\expandafter\def\csname EERjointentry@posterior.latercc_rich_entry.onset_retained.mean_post_high\endcsname{0.888}
\expandafter\def\csname EERjointentry@posterior.latercc_rich_entry.onset_retained.share_of_high_mass\endcsname{0.108}
\expandafter\def\csname EERjointentry@posterior.restricted_l.never_exposed.women\endcsname{2,256}
\expandafter\def\csname EERjointentry@posterior.restricted_l.never_exposed.mass_post_high\endcsname{5.9}
\expandafter\def\csname EERjointentry@posterior.restricted_l.never_exposed.women_above_half\endcsname{4}
\expandafter\def\csname EERjointentry@posterior.restricted_l.never_exposed.total_mass\endcsname{92.5}
\expandafter\def\csname EERjointentry@posterior.restricted_l.never_exposed.mean_post_high\endcsname{0.003}
\expandafter\def\csname EERjointentry@posterior.restricted_l.never_exposed.share_of_high_mass\endcsname{0.063}
\expandafter\def\csname EERjointentry@posterior.restricted_l.onset_modified_debut.women\endcsname{79}
\expandafter\def\csname EERjointentry@posterior.restricted_l.onset_modified_debut.mass_post_high\endcsname{44.2}
\expandafter\def\csname EERjointentry@posterior.restricted_l.onset_modified_debut.women_above_half\endcsname{74}
\expandafter\def\csname EERjointentry@posterior.restricted_l.onset_modified_debut.total_mass\endcsname{92.5}
\expandafter\def\csname EERjointentry@posterior.restricted_l.onset_modified_debut.mean_post_high\endcsname{0.559}
\expandafter\def\csname EERjointentry@posterior.restricted_l.onset_modified_debut.share_of_high_mass\endcsname{0.478}
\expandafter\def\csname EERjointentry@posterior.restricted_l.onset_unmodified_debut.women\endcsname{373}
\expandafter\def\csname EERjointentry@posterior.restricted_l.onset_unmodified_debut.mass_post_high\endcsname{42.4}
\expandafter\def\csname EERjointentry@posterior.restricted_l.onset_unmodified_debut.women_above_half\endcsname{29}
\expandafter\def\csname EERjointentry@posterior.restricted_l.onset_unmodified_debut.total_mass\endcsname{92.5}
\expandafter\def\csname EERjointentry@posterior.restricted_l.onset_unmodified_debut.mean_post_high\endcsname{0.114}
\expandafter\def\csname EERjointentry@posterior.restricted_l.onset_unmodified_debut.share_of_high_mass\endcsname{0.459}
}{}
\IfFileExists{evidence/refinement/extensions/numbers_eer_sample.tex}{
\newcommand{\EERsample}[1]{\ifcsname EERsample.#1\endcsname\csname EERsample.#1\endcsname\else\PackageError{EER sample numbers}{Missing sample count #1}{Rerun export_sample_numbers.do.}\fi}
\expandafter\def\csname EERsample.broad.women\endcsname{4,235}
\expandafter\def\csname EERsample.broad.months\endcsname{147,956}
\expandafter\def\csname EERsample.broad.exits\endcsname{680}
\expandafter\def\csname EERsample.broad.completions\endcsname{3,232}
\expandafter\def\csname EERsample.broad.exposed_months\endcsname{8,593}
\expandafter\def\csname EERsample.person.women\endcsname{4,385}
\expandafter\def\csname EERsample.person.debut_before10\endcsname{38}
\expandafter\def\csname EERsample.person.recorded_exit\endcsname{4,278}
\expandafter\def\csname EERsample.person.dropout\endcsname{756}
\expandafter\def\csname EERsample.person.completion\endcsname{3,522}
\expandafter\def\csname EERsample.person.panel_with_exit\endcsname{4,174}
\expandafter\def\csname EERsample.person.panel_without_exit\endcsname{61}
\expandafter\def\csname EERsample.person.nopanel_with_exit\endcsname{104}
\expandafter\def\csname EERsample.person.nopanel_without_exit\endcsname{46}
\expandafter\def\csname EERsample.exposure.defining_women\endcsname{657}
\expandafter\def\csname EERsample.exposure.debut_corrected\endcsname{117}
\expandafter\def\csname EERsample.exposure.debut_imputed\endcsname{14}
\expandafter\def\csname EERsample.exposure.modified_exposed_months\endcsname{2,183}
\expandafter\def\csname EERsample.exposure.modified_exposed_share_pct\endcsname{25.4}
\expandafter\def\csname EERsample.postdebut.women\endcsname{2,951}
\expandafter\def\csname EERsample.postdebut.months\endcsname{61,649}
\expandafter\def\csname EERsample.postdebut.exits\endcsname{594}
\expandafter\def\csname EERsample.postdebut.completions\endcsname{2,139}
\expandafter\def\csname EERsample.postdebut.exposed_months\endcsname{8,593}
\expandafter\def\csname EERsample.onset.never_exposed.women\endcsname{2,326}
\expandafter\def\csname EERsample.onset.retained_onset.women\endcsname{423}
\expandafter\def\csname EERsample.onset.exposed_at_entry.women\endcsname{135}
\expandafter\def\csname EERsample.onset.missing_onset.women\endcsname{67}
\expandafter\def\csname EERsample.onset.exposed_women\endcsname{625}
\expandafter\def\csname EERsample.onset.without_retained_onset_women\endcsname{202}
\expandafter\def\csname EERsample.onset.missing_onset.pregnancy_risk_months\endcsname{765}
\expandafter\def\csname EERsample.restricted.women\endcsname{2,749}
\expandafter\def\csname EERsample.restricted.months\endcsname{57,991}
\expandafter\def\csname EERsample.restricted.exits\endcsname{501}
\expandafter\def\csname EERsample.debut.women\endcsname{2,673}
\expandafter\def\csname EERsample.debut.person_months\endcsname{56,471}
\expandafter\def\csname EERsample.debut.dropout_events\endcsname{467}
\expandafter\def\csname EERsample.debut.exposed_months\endcsname{4,327}
\expandafter\def\csname EERsample.exclusion.women\endcsname{76}
\expandafter\def\csname EERsample.exclusion.person_months\endcsname{1,520}
\expandafter\def\csname EERsample.exclusion.exposed_months\endcsname{1,373}
\expandafter\def\csname EERsample.exclusion.dropout_events\endcsname{34}
\expandafter\def\csname EERsample.exclusion.conception_events\endcsname{75}
\expandafter\def\csname EERsample.exclusion.adjusted_women\endcsname{68}
\expandafter\def\csname EERsample.exclusion.imputed_women\endcsname{8}
\expandafter\def\csname EERsample.exclusion.exposure_pct\endcsname{24.1}
\expandafter\def\csname EERsample.interruption.school3.main_baseline.women\endcsname{2,828}
\expandafter\def\csname EERsample.interruption.school3.main_baseline.months\endcsname{48,953}
\expandafter\def\csname EERsample.interruption.school3.main_baseline.events\endcsname{711}
\expandafter\def\csname EERsample.interruption.school3.main_baseline.exposed_months\endcsname{5,939}
\expandafter\def\csname EERsample.interruption.school3.main_baseline.exposed_events\endcsname{261}
\expandafter\def\csname EERsample.interruption.school3.later_all_core.women\endcsname{2,313}
\expandafter\def\csname EERsample.interruption.school3.later_all_core.months\endcsname{35,595}
\expandafter\def\csname EERsample.interruption.school3.later_all_core.events\endcsname{521}
\expandafter\def\csname EERsample.interruption.school3.later_all_core.exposed_months\endcsname{3,002}
\expandafter\def\csname EERsample.interruption.school3.later_all_core.exposed_events\endcsname{136}
\expandafter\def\csname EERsample.interruption.school3.later_cc_rich.women\endcsname{1,733}
\expandafter\def\csname EERsample.interruption.school3.later_cc_rich.months\endcsname{26,777}
\expandafter\def\csname EERsample.interruption.school3.later_cc_rich.events\endcsname{336}
\expandafter\def\csname EERsample.interruption.school3.later_cc_rich.exposed_months\endcsname{2,055}
\expandafter\def\csname EERsample.interruption.school3.later_cc_rich.exposed_events\endcsname{84}
\expandafter\def\csname EERsample.interruption.calendar4.main_baseline.women\endcsname{2,809}
\expandafter\def\csname EERsample.interruption.calendar4.main_baseline.months\endcsname{48,120}
\expandafter\def\csname EERsample.interruption.calendar4.main_baseline.events\endcsname{726}
\expandafter\def\csname EERsample.interruption.calendar4.main_baseline.exposed_months\endcsname{5,804}
\expandafter\def\csname EERsample.interruption.calendar4.main_baseline.exposed_events\endcsname{258}
\expandafter\def\csname EERsample.interruption.calendar4.later_all_core.women\endcsname{2,291}
\expandafter\def\csname EERsample.interruption.calendar4.later_all_core.months\endcsname{34,828}
\expandafter\def\csname EERsample.interruption.calendar4.later_all_core.events\endcsname{530}
\expandafter\def\csname EERsample.interruption.calendar4.later_all_core.exposed_months\endcsname{2,918}
\expandafter\def\csname EERsample.interruption.calendar4.later_all_core.exposed_events\endcsname{130}
\expandafter\def\csname EERsample.interruption.calendar4.later_cc_rich.women\endcsname{1,715}
\expandafter\def\csname EERsample.interruption.calendar4.later_cc_rich.months\endcsname{26,147}
\expandafter\def\csname EERsample.interruption.calendar4.later_cc_rich.events\endcsname{346}
\expandafter\def\csname EERsample.interruption.calendar4.later_cc_rich.exposed_months\endcsname{2,010}
\expandafter\def\csname EERsample.interruption.calendar4.later_cc_rich.exposed_events\endcsname{79}
\expandafter\def\csname EERsample.interruption.school3.treated0.women\endcsname{2,690}
\expandafter\def\csname EERsample.interruption.school3.treated0.months\endcsname{43,014}
\expandafter\def\csname EERsample.interruption.school3.treated0.events\endcsname{450}
\expandafter\def\csname EERsample.interruption.school3.treated1.women\endcsname{562}
\expandafter\def\csname EERsample.interruption.school3.treated1.months\endcsname{5,939}
\expandafter\def\csname EERsample.interruption.school3.treated1.events\endcsname{261}
\expandafter\def\csname EERsample.interruption.school3.main_women\endcsname{2,828}
\expandafter\def\csname EERsample.interruption.school3.history_supported_women\endcsname{2,571}
\expandafter\def\csname EERsample.interruption.school3.later_eligible_women\endcsname{2,313}
\expandafter\def\csname EERsample.interruption.school3.later_missing_mother_grade97\endcsname{176}
\expandafter\def\csname EERsample.interruption.school3.later_missing_household_group\endcsname{9}
\expandafter\def\csname EERsample.interruption.school3.later_missing_asvab10\endcsname{456}
\expandafter\def\csname EERsample.interruption.calendar4.main_women\endcsname{2,809}
\expandafter\def\csname EERsample.interruption.calendar4.history_supported_women\endcsname{2,553}
\expandafter\def\csname EERsample.interruption.calendar4.later_eligible_women\endcsname{2,291}
\expandafter\def\csname EERsample.interruption.calendar4.later_missing_mother_grade97\endcsname{173}
\expandafter\def\csname EERsample.interruption.calendar4.later_missing_household_group\endcsname{9}
\expandafter\def\csname EERsample.interruption.calendar4.later_missing_asvab10\endcsname{454}
\expandafter\def\csname EERsample.joint.full.women\endcsname{2,828}
\expandafter\def\csname EERsample.joint.full.calendar_months\endcsname{64,598}
\expandafter\def\csname EERsample.joint.full.school_months\endcsname{48,953}
\expandafter\def\csname EERsample.joint.full.interruption_events\endcsname{711}
\expandafter\def\csname EERsample.joint.full.conception_events\endcsname{478}
\expandafter\def\csname EERsample.joint.full.summer_conceptions\endcsname{114}
\expandafter\def\csname EERsample.joint.full.pregnancy_risk_months\endcsname{56,705}
\expandafter\def\csname EERsample.joint.full.exposed_school_months\endcsname{5,939}
\expandafter\def\csname EERsample.joint.full.onset_without_exposure_women\endcsname{26}
\expandafter\def\csname EERsample.joint.full.summer_conception_share_pct\endcsname{23.8}
\expandafter\def\csname EERsample.joint.timing.exposed_events\endcsname{261}
\expandafter\def\csname EERsample.joint.timing.summer_start\endcsname{236}
\expandafter\def\csname EERsample.joint.timing.summer_before_p\endcsname{2}
\expandafter\def\csname EERsample.joint.timing.summer_same_p\endcsname{4}
\expandafter\def\csname EERsample.joint.timing.before_p_any\endcsname{2}
\expandafter\def\csname EERsample.joint.timing.events\endcsname{711}
\expandafter\def\csname EERsample.joint.confirmation.p_during_confirmation\endcsname{26}
\expandafter\def\csname EERsample.joint.confirmation.exposure_during_confirmation\endcsname{31}
\expandafter\def\csname EERsample.joint.confirmation.p_same_onset\endcsname{15}
\expandafter\def\csname EERsample.joint.confirmation.events\endcsname{711}
\expandafter\def\csname EERsample.latent.full_linear.high_share_pct\endcsname{3.0}
\expandafter\def\csname EERsample.latent.full_linear.high_share_pct_rounded\endcsname{3}
\expandafter\def\csname EERsample.latent.full_linear.conception_ratio\endcsname{62.32}
\expandafter\def\csname EERsample.latent.full_linear.conception_ratio_rounded\endcsname{62}
\expandafter\def\csname EERsample.latent.full_linear.exit_ratio\endcsname{1.17}
\expandafter\def\csname EERsample.latent.restricted_linear.high_share_pct\endcsname{3.0}
\expandafter\def\csname EERsample.latent.restricted_linear.high_share_pct_rounded\endcsname{3}
\expandafter\def\csname EERsample.latent.restricted_linear.conception_ratio\endcsname{61.72}
\expandafter\def\csname EERsample.latent.restricted_linear.conception_ratio_rounded\endcsname{62}
\expandafter\def\csname EERsample.latent.restricted_linear.exit_ratio\endcsname{1.26}
\expandafter\def\csname EERsample.latent.debut_linear.high_share_pct\endcsname{56.4}
\expandafter\def\csname EERsample.latent.debut_linear.high_share_pct_rounded\endcsname{56}
\expandafter\def\csname EERsample.latent.debut_linear.conception_ratio\endcsname{2.66}
\expandafter\def\csname EERsample.latent.debut_linear.conception_ratio_rounded\endcsname{3}
\expandafter\def\csname EERsample.latent.debut_linear.exit_ratio\endcsname{0.08}
}{}
\IfFileExists{evidence/refinement/extensions/numbers_interruption_summary.tex}{
\newcommand{\EERsum}[1]{\ifcsname EERsum.#1\endcsname\csname EERsum.#1\endcsname\else\PackageError{EER summary numbers}{Missing summary value #1}{Rerun interruption_summary_stats.do.}\fi}
\expandafter\def\csname EERsum.main.all.race1.value\endcsname{29.9}
\expandafter\def\csname EERsum.main.all.race1.n\endcsname{2,828}
\expandafter\def\csname EERsum.main.all.race2.value\endcsname{19.7}
\expandafter\def\csname EERsum.main.all.race2.n\endcsname{2,828}
\expandafter\def\csname EERsum.main.all.race3.value\endcsname{1.1}
\expandafter\def\csname EERsum.main.all.race3.n\endcsname{2,828}
\expandafter\def\csname EERsum.main.all.race4.value\endcsname{49.3}
\expandafter\def\csname EERsum.main.all.race4.n\endcsname{2,828}
\expandafter\def\csname EERsum.main.all.household1.value\endcsname{42.0}
\expandafter\def\csname EERsum.main.all.household1.n\endcsname{2,818}
\expandafter\def\csname EERsum.main.all.household2.value\endcsname{14.7}
\expandafter\def\csname EERsum.main.all.household2.n\endcsname{2,818}
\expandafter\def\csname EERsum.main.all.household3.value\endcsname{43.3}
\expandafter\def\csname EERsum.main.all.household3.n\endcsname{2,818}
\expandafter\def\csname EERsum.main.all.interrupted.value\endcsname{25.1}
\expandafter\def\csname EERsum.main.all.interrupted.n\endcsname{2,828}
\expandafter\def\csname EERsum.main.all.credential.value\endcsname{69.8}
\expandafter\def\csname EERsum.main.all.credential.n\endcsname{2,828}
\expandafter\def\csname EERsum.main.all.mother_grade97.value\endcsname{12.3}
\expandafter\def\csname EERsum.main.all.mother_grade97.n\endcsname{2,600}
\expandafter\def\csname EERsum.main.all.mother_grade97.sd\endcsname{2.8}
\expandafter\def\csname EERsum.main.all.asvab_pct.value\endcsname{43.0}
\expandafter\def\csname EERsum.main.all.asvab_pct.n\endcsname{2,268}
\expandafter\def\csname EERsum.main.all.asvab_pct.sd\endcsname{27.7}
\expandafter\def\csname EERsum.main.all.age_entry.value\endcsname{16.3}
\expandafter\def\csname EERsum.main.all.age_entry.n\endcsname{2,828}
\expandafter\def\csname EERsum.main.all.age_entry.sd\endcsname{1.2}
\expandafter\def\csname EERsum.main.all.year_entry.value\endcsname{1998.3}
\expandafter\def\csname EERsum.main.all.year_entry.n\endcsname{2,828}
\expandafter\def\csname EERsum.main.all.year_entry.sd\endcsname{1.5}
\expandafter\def\csname EERsum.main.all.months.value\endcsname{17.3}
\expandafter\def\csname EERsum.main.all.months.n\endcsname{2,828}
\expandafter\def\csname EERsum.main.all.months.sd\endcsname{10.5}
\expandafter\def\csname EERsum.main.exposed.race1.value\endcsname{46.8}
\expandafter\def\csname EERsum.main.exposed.race1.n\endcsname{562}
\expandafter\def\csname EERsum.main.exposed.race2.value\endcsname{23.0}
\expandafter\def\csname EERsum.main.exposed.race2.n\endcsname{562}
\expandafter\def\csname EERsum.main.exposed.race3.value\endcsname{1.6}
\expandafter\def\csname EERsum.main.exposed.race3.n\endcsname{562}
\expandafter\def\csname EERsum.main.exposed.race4.value\endcsname{28.6}
\expandafter\def\csname EERsum.main.exposed.race4.n\endcsname{562}
\expandafter\def\csname EERsum.main.exposed.household1.value\endcsname{25.5}
\expandafter\def\csname EERsum.main.exposed.household1.n\endcsname{560}
\expandafter\def\csname EERsum.main.exposed.household2.value\endcsname{13.9}
\expandafter\def\csname EERsum.main.exposed.household2.n\endcsname{560}
\expandafter\def\csname EERsum.main.exposed.household3.value\endcsname{60.5}
\expandafter\def\csname EERsum.main.exposed.household3.n\endcsname{560}
\expandafter\def\csname EERsum.main.exposed.interrupted.value\endcsname{46.4}
\expandafter\def\csname EERsum.main.exposed.interrupted.n\endcsname{562}
\expandafter\def\csname EERsum.main.exposed.credential.value\endcsname{43.6}
\expandafter\def\csname EERsum.main.exposed.credential.n\endcsname{562}
\expandafter\def\csname EERsum.main.exposed.mother_grade97.value\endcsname{11.3}
\expandafter\def\csname EERsum.main.exposed.mother_grade97.n\endcsname{496}
\expandafter\def\csname EERsum.main.exposed.mother_grade97.sd\endcsname{2.6}
\expandafter\def\csname EERsum.main.exposed.asvab_pct.value\endcsname{26.6}
\expandafter\def\csname EERsum.main.exposed.asvab_pct.n\endcsname{420}
\expandafter\def\csname EERsum.main.exposed.asvab_pct.sd\endcsname{21.8}
\expandafter\def\csname EERsum.main.exposed.age_entry.value\endcsname{15.9}
\expandafter\def\csname EERsum.main.exposed.age_entry.n\endcsname{562}
\expandafter\def\csname EERsum.main.exposed.age_entry.sd\endcsname{1.2}
\expandafter\def\csname EERsum.main.exposed.year_entry.value\endcsname{1997.9}
\expandafter\def\csname EERsum.main.exposed.year_entry.n\endcsname{562}
\expandafter\def\csname EERsum.main.exposed.year_entry.sd\endcsname{1.2}
\expandafter\def\csname EERsum.main.exposed.months.value\endcsname{19.4}
\expandafter\def\csname EERsum.main.exposed.months.n\endcsname{562}
\expandafter\def\csname EERsum.main.exposed.months.sd\endcsname{11.3}
\expandafter\def\csname EERsum.main.exposed.age_conception.value\endcsname{16.6}
\expandafter\def\csname EERsum.main.exposed.age_conception.n\endcsname{562}
\expandafter\def\csname EERsum.main.exposed.age_conception.sd\endcsname{1.4}
\expandafter\def\csname EERsum.main.unexposed.race1.value\endcsname{25.7}
\expandafter\def\csname EERsum.main.unexposed.race1.n\endcsname{2,266}
\expandafter\def\csname EERsum.main.unexposed.race2.value\endcsname{18.9}
\expandafter\def\csname EERsum.main.unexposed.race2.n\endcsname{2,266}
\expandafter\def\csname EERsum.main.unexposed.race3.value\endcsname{1.0}
\expandafter\def\csname EERsum.main.unexposed.race3.n\endcsname{2,266}
\expandafter\def\csname EERsum.main.unexposed.race4.value\endcsname{54.4}
\expandafter\def\csname EERsum.main.unexposed.race4.n\endcsname{2,266}
\expandafter\def\csname EERsum.main.unexposed.household1.value\endcsname{46.1}
\expandafter\def\csname EERsum.main.unexposed.household1.n\endcsname{2,258}
\expandafter\def\csname EERsum.main.unexposed.household2.value\endcsname{14.9}
\expandafter\def\csname EERsum.main.unexposed.household2.n\endcsname{2,258}
\expandafter\def\csname EERsum.main.unexposed.household3.value\endcsname{39.0}
\expandafter\def\csname EERsum.main.unexposed.household3.n\endcsname{2,258}
\expandafter\def\csname EERsum.main.unexposed.interrupted.value\endcsname{19.9}
\expandafter\def\csname EERsum.main.unexposed.interrupted.n\endcsname{2,266}
\expandafter\def\csname EERsum.main.unexposed.credential.value\endcsname{76.3}
\expandafter\def\csname EERsum.main.unexposed.credential.n\endcsname{2,266}
\expandafter\def\csname EERsum.main.unexposed.mother_grade97.value\endcsname{12.6}
\expandafter\def\csname EERsum.main.unexposed.mother_grade97.n\endcsname{2,104}
\expandafter\def\csname EERsum.main.unexposed.mother_grade97.sd\endcsname{2.7}
\expandafter\def\csname EERsum.main.unexposed.asvab_pct.value\endcsname{46.7}
\expandafter\def\csname EERsum.main.unexposed.asvab_pct.n\endcsname{1,848}
\expandafter\def\csname EERsum.main.unexposed.asvab_pct.sd\endcsname{27.6}
\expandafter\def\csname EERsum.main.unexposed.age_entry.value\endcsname{16.4}
\expandafter\def\csname EERsum.main.unexposed.age_entry.n\endcsname{2,266}
\expandafter\def\csname EERsum.main.unexposed.age_entry.sd\endcsname{1.2}
\expandafter\def\csname EERsum.main.unexposed.year_entry.value\endcsname{1998.5}
\expandafter\def\csname EERsum.main.unexposed.year_entry.n\endcsname{2,266}
\expandafter\def\csname EERsum.main.unexposed.year_entry.sd\endcsname{1.5}
\expandafter\def\csname EERsum.main.unexposed.months.value\endcsname{16.8}
\expandafter\def\csname EERsum.main.unexposed.months.n\endcsname{2,266}
\expandafter\def\csname EERsum.main.unexposed.months.sd\endcsname{10.3}
\expandafter\def\csname EERsum.later_cc.all.race1.value\endcsname{28.0}
\expandafter\def\csname EERsum.later_cc.all.race1.n\endcsname{1,733}
\expandafter\def\csname EERsum.later_cc.all.race2.value\endcsname{18.2}
\expandafter\def\csname EERsum.later_cc.all.race2.n\endcsname{1,733}
\expandafter\def\csname EERsum.later_cc.all.race3.value\endcsname{0.9}
\expandafter\def\csname EERsum.later_cc.all.race3.n\endcsname{1,733}
\expandafter\def\csname EERsum.later_cc.all.race4.value\endcsname{52.8}
\expandafter\def\csname EERsum.later_cc.all.race4.n\endcsname{1,733}
\expandafter\def\csname EERsum.later_cc.all.household1.value\endcsname{47.7}
\expandafter\def\csname EERsum.later_cc.all.household1.n\endcsname{1,733}
\expandafter\def\csname EERsum.later_cc.all.household2.value\endcsname{13.8}
\expandafter\def\csname EERsum.later_cc.all.household2.n\endcsname{1,733}
\expandafter\def\csname EERsum.later_cc.all.household3.value\endcsname{38.4}
\expandafter\def\csname EERsum.later_cc.all.household3.n\endcsname{1,733}
\expandafter\def\csname EERsum.later_cc.all.interrupted.value\endcsname{19.4}
\expandafter\def\csname EERsum.later_cc.all.interrupted.n\endcsname{1,733}
\expandafter\def\csname EERsum.later_cc.all.credential.value\endcsname{77.0}
\expandafter\def\csname EERsum.later_cc.all.credential.n\endcsname{1,733}
\expandafter\def\csname EERsum.later_cc.all.mother_grade97.value\endcsname{12.6}
\expandafter\def\csname EERsum.later_cc.all.mother_grade97.n\endcsname{1,733}
\expandafter\def\csname EERsum.later_cc.all.mother_grade97.sd\endcsname{2.8}
\expandafter\def\csname EERsum.later_cc.all.asvab_pct.value\endcsname{44.9}
\expandafter\def\csname EERsum.later_cc.all.asvab_pct.n\endcsname{1,733}
\expandafter\def\csname EERsum.later_cc.all.asvab_pct.sd\endcsname{27.6}
\expandafter\def\csname EERsum.later_cc.all.age_entry.value\endcsname{16.7}
\expandafter\def\csname EERsum.later_cc.all.age_entry.n\endcsname{1,733}
\expandafter\def\csname EERsum.later_cc.all.age_entry.sd\endcsname{1.1}
\expandafter\def\csname EERsum.later_cc.all.year_entry.value\endcsname{1998.9}
\expandafter\def\csname EERsum.later_cc.all.year_entry.n\endcsname{1,733}
\expandafter\def\csname EERsum.later_cc.all.year_entry.sd\endcsname{1.2}
\expandafter\def\csname EERsum.later_cc.all.months.value\endcsname{15.5}
\expandafter\def\csname EERsum.later_cc.all.months.n\endcsname{1,733}
\expandafter\def\csname EERsum.later_cc.all.months.sd\endcsname{9.6}
\expandafter\def\csname EERsum.later_cc.exposed.race1.value\endcsname{47.3}
\expandafter\def\csname EERsum.later_cc.exposed.race1.n\endcsname{222}
\expandafter\def\csname EERsum.later_cc.exposed.race2.value\endcsname{21.2}
\expandafter\def\csname EERsum.later_cc.exposed.race2.n\endcsname{222}
\expandafter\def\csname EERsum.later_cc.exposed.race3.value\endcsname{0.9}
\expandafter\def\csname EERsum.later_cc.exposed.race3.n\endcsname{222}
\expandafter\def\csname EERsum.later_cc.exposed.race4.value\endcsname{30.6}
\expandafter\def\csname EERsum.later_cc.exposed.race4.n\endcsname{222}
\expandafter\def\csname EERsum.later_cc.exposed.household1.value\endcsname{29.3}
\expandafter\def\csname EERsum.later_cc.exposed.household1.n\endcsname{222}
\expandafter\def\csname EERsum.later_cc.exposed.household2.value\endcsname{14.0}
\expandafter\def\csname EERsum.later_cc.exposed.household2.n\endcsname{222}
\expandafter\def\csname EERsum.later_cc.exposed.household3.value\endcsname{56.8}
\expandafter\def\csname EERsum.later_cc.exposed.household3.n\endcsname{222}
\expandafter\def\csname EERsum.later_cc.exposed.interrupted.value\endcsname{37.8}
\expandafter\def\csname EERsum.later_cc.exposed.interrupted.n\endcsname{222}
\expandafter\def\csname EERsum.later_cc.exposed.credential.value\endcsname{50.5}
\expandafter\def\csname EERsum.later_cc.exposed.credential.n\endcsname{222}
\expandafter\def\csname EERsum.later_cc.exposed.mother_grade97.value\endcsname{11.5}
\expandafter\def\csname EERsum.later_cc.exposed.mother_grade97.n\endcsname{222}
\expandafter\def\csname EERsum.later_cc.exposed.mother_grade97.sd\endcsname{2.8}
\expandafter\def\csname EERsum.later_cc.exposed.asvab_pct.value\endcsname{27.6}
\expandafter\def\csname EERsum.later_cc.exposed.asvab_pct.n\endcsname{222}
\expandafter\def\csname EERsum.later_cc.exposed.asvab_pct.sd\endcsname{21.2}
\expandafter\def\csname EERsum.later_cc.exposed.age_entry.value\endcsname{16.2}
\expandafter\def\csname EERsum.later_cc.exposed.age_entry.n\endcsname{222}
\expandafter\def\csname EERsum.later_cc.exposed.age_entry.sd\endcsname{1.1}
\expandafter\def\csname EERsum.later_cc.exposed.year_entry.value\endcsname{1998.7}
\expandafter\def\csname EERsum.later_cc.exposed.year_entry.n\endcsname{222}
\expandafter\def\csname EERsum.later_cc.exposed.year_entry.sd\endcsname{1.0}
\expandafter\def\csname EERsum.later_cc.exposed.months.value\endcsname{20.0}
\expandafter\def\csname EERsum.later_cc.exposed.months.n\endcsname{222}
\expandafter\def\csname EERsum.later_cc.exposed.months.sd\endcsname{10.3}
\expandafter\def\csname EERsum.later_cc.exposed.age_conception.value\endcsname{17.3}
\expandafter\def\csname EERsum.later_cc.exposed.age_conception.n\endcsname{222}
\expandafter\def\csname EERsum.later_cc.exposed.age_conception.sd\endcsname{1.2}
\expandafter\def\csname EERsum.later_cc.unexposed.race1.value\endcsname{25.2}
\expandafter\def\csname EERsum.later_cc.unexposed.race1.n\endcsname{1,511}
\expandafter\def\csname EERsum.later_cc.unexposed.race2.value\endcsname{17.8}
\expandafter\def\csname EERsum.later_cc.unexposed.race2.n\endcsname{1,511}
\expandafter\def\csname EERsum.later_cc.unexposed.race3.value\endcsname{0.9}
\expandafter\def\csname EERsum.later_cc.unexposed.race3.n\endcsname{1,511}
\expandafter\def\csname EERsum.later_cc.unexposed.race4.value\endcsname{56.1}
\expandafter\def\csname EERsum.later_cc.unexposed.race4.n\endcsname{1,511}
\expandafter\def\csname EERsum.later_cc.unexposed.household1.value\endcsname{50.4}
\expandafter\def\csname EERsum.later_cc.unexposed.household1.n\endcsname{1,511}
\expandafter\def\csname EERsum.later_cc.unexposed.household2.value\endcsname{13.8}
\expandafter\def\csname EERsum.later_cc.unexposed.household2.n\endcsname{1,511}
\expandafter\def\csname EERsum.later_cc.unexposed.household3.value\endcsname{35.7}
\expandafter\def\csname EERsum.later_cc.unexposed.household3.n\endcsname{1,511}
\expandafter\def\csname EERsum.later_cc.unexposed.interrupted.value\endcsname{16.7}
\expandafter\def\csname EERsum.later_cc.unexposed.interrupted.n\endcsname{1,511}
\expandafter\def\csname EERsum.later_cc.unexposed.credential.value\endcsname{80.9}
\expandafter\def\csname EERsum.later_cc.unexposed.credential.n\endcsname{1,511}
\expandafter\def\csname EERsum.later_cc.unexposed.mother_grade97.value\endcsname{12.7}
\expandafter\def\csname EERsum.later_cc.unexposed.mother_grade97.n\endcsname{1,511}
\expandafter\def\csname EERsum.later_cc.unexposed.mother_grade97.sd\endcsname{2.7}
\expandafter\def\csname EERsum.later_cc.unexposed.asvab_pct.value\endcsname{47.5}
\expandafter\def\csname EERsum.later_cc.unexposed.asvab_pct.n\endcsname{1,511}
\expandafter\def\csname EERsum.later_cc.unexposed.asvab_pct.sd\endcsname{27.5}
\expandafter\def\csname EERsum.later_cc.unexposed.age_entry.value\endcsname{16.7}
\expandafter\def\csname EERsum.later_cc.unexposed.age_entry.n\endcsname{1,511}
\expandafter\def\csname EERsum.later_cc.unexposed.age_entry.sd\endcsname{1.1}
\expandafter\def\csname EERsum.later_cc.unexposed.year_entry.value\endcsname{1999.0}
\expandafter\def\csname EERsum.later_cc.unexposed.year_entry.n\endcsname{1,511}
\expandafter\def\csname EERsum.later_cc.unexposed.year_entry.sd\endcsname{1.2}
\expandafter\def\csname EERsum.later_cc.unexposed.months.value\endcsname{14.8}
\expandafter\def\csname EERsum.later_cc.unexposed.months.n\endcsname{1,511}
\expandafter\def\csname EERsum.later_cc.unexposed.months.sd\endcsname{9.3}
\expandafter\def\csname EERsum.count.main.all.women\endcsname{2,828}
\expandafter\def\csname EERsum.count.main.exposed.women\endcsname{562}
\expandafter\def\csname EERsum.count.main.unexposed.women\endcsname{2,266}
\expandafter\def\csname EERsum.count.main.unexposed.women_with_later_conception\endcsname{1,358}
\expandafter\def\csname EERsum.count.main.all.interrupted_women\endcsname{711}
\expandafter\def\csname EERsum.count.main.all.missing_mother_grade\endcsname{228}
\expandafter\def\csname EERsum.count.main.all.missing_household\endcsname{10}
\expandafter\def\csname EERsum.count.main.all.missing_asvab\endcsname{560}
\expandafter\def\csname EERsum.count.main.all.school_months\endcsname{48,953}
\expandafter\def\csname EERsum.count.main.all.exposed_school_months\endcsname{5,939}
\expandafter\def\csname EERsum.count.later_cc.all.women\endcsname{1,733}
\expandafter\def\csname EERsum.count.later_cc.exposed.women\endcsname{222}
\expandafter\def\csname EERsum.count.later_cc.unexposed.women\endcsname{1,511}
\expandafter\def\csname EERsum.count.later_cc.unexposed.women_with_later_conception\endcsname{872}
\expandafter\def\csname EERsum.count.later_cc.all.interrupted_women\endcsname{336}
\expandafter\def\csname EERsum.count.later_cc.all.missing_mother_grade\endcsname{0}
\expandafter\def\csname EERsum.count.later_cc.all.missing_household\endcsname{0}
\expandafter\def\csname EERsum.count.later_cc.all.missing_asvab\endcsname{0}
\expandafter\def\csname EERsum.count.later_cc.all.school_months\endcsname{26,777}
\expandafter\def\csname EERsum.count.later_cc.all.exposed_school_months\endcsname{2,055}
}{}
\IfFileExists{evidence/refinement/extensions/numbers_interruption_heterogeneity.tex}{
\newcommand{\EERhet}[1]{\ifcsname EERhet.#1\endcsname\csname EERhet.#1\endcsname\else\PackageError{EER heterogeneity numbers}{Missing heterogeneity value #1}{Rerun interruption_heterogeneity.do.}\fi}
\expandafter\def\csname EERhet.tercile_rich.t1.women\endcsname{578}
\expandafter\def\csname EERhet.tercile_rich.t1.months\endcsname{9,458}
\expandafter\def\csname EERhet.tercile_rich.t1.fitted_months\endcsname{9,458}
\expandafter\def\csname EERhet.tercile_rich.t1.events\endcsname{184}
\expandafter\def\csname EERhet.tercile_rich.t1.exposed_months\endcsname{1,318}
\expandafter\def\csname EERhet.tercile_rich.t1.exposed_events\endcsname{57}
\expandafter\def\csname EERhet.tercile_rich.t1.unexposed_months\endcsname{8,140}
\expandafter\def\csname EERhet.tercile_rich.t1.unexposed_events\endcsname{127}
\expandafter\def\csname EERhet.tercile_rich.t1.rate_unexposed_per_1000\endcsname{15.6}
\expandafter\def\csname EERhet.tercile_rich.t1.rate_exposed_per_1000\endcsname{43.2}
\expandafter\def\csname EERhet.tercile_rich.t1.hr\endcsname{2.51}
\expandafter\def\csname EERhet.tercile_rich.t1.ci_lo\endcsname{1.76}
\expandafter\def\csname EERhet.tercile_rich.t1.ci_hi\endcsname{3.60}
\expandafter\def\csname EERhet.tercile_rich.t1.ci\endcsname{[1.76, 3.60]}
\expandafter\def\csname EERhet.tercile_rich.t1.b\endcsname{0.922}
\expandafter\def\csname EERhet.tercile_rich.t1.se\endcsname{0.183}
\expandafter\def\csname EERhet.tercile_rich.t1.status\endcsname{converged}
\expandafter\def\csname EERhet.tercile_rich.t2.women\endcsname{578}
\expandafter\def\csname EERhet.tercile_rich.t2.months\endcsname{8,990}
\expandafter\def\csname EERhet.tercile_rich.t2.fitted_months\endcsname{8,897}
\expandafter\def\csname EERhet.tercile_rich.t2.events\endcsname{106}
\expandafter\def\csname EERhet.tercile_rich.t2.exposed_months\endcsname{549}
\expandafter\def\csname EERhet.tercile_rich.t2.exposed_events\endcsname{25}
\expandafter\def\csname EERhet.tercile_rich.t2.unexposed_months\endcsname{8,441}
\expandafter\def\csname EERhet.tercile_rich.t2.unexposed_events\endcsname{81}
\expandafter\def\csname EERhet.tercile_rich.t2.rate_unexposed_per_1000\endcsname{9.6}
\expandafter\def\csname EERhet.tercile_rich.t2.rate_exposed_per_1000\endcsname{45.5}
\expandafter\def\csname EERhet.tercile_rich.t2.hr\endcsname{4.87}
\expandafter\def\csname EERhet.tercile_rich.t2.ci_lo\endcsname{3.10}
\expandafter\def\csname EERhet.tercile_rich.t2.ci_hi\endcsname{7.64}
\expandafter\def\csname EERhet.tercile_rich.t2.ci\endcsname{[3.10, 7.64]}
\expandafter\def\csname EERhet.tercile_rich.t2.b\endcsname{1.582}
\expandafter\def\csname EERhet.tercile_rich.t2.se\endcsname{0.230}
\expandafter\def\csname EERhet.tercile_rich.t2.status\endcsname{converged_reduced}
\expandafter\def\csname EERhet.tercile_rich.t3.women\endcsname{577}
\expandafter\def\csname EERhet.tercile_rich.t3.months\endcsname{8,329}
\expandafter\def\csname EERhet.tercile_rich.t3.fitted_months\endcsname{8,299}
\expandafter\def\csname EERhet.tercile_rich.t3.events\endcsname{46}
\expandafter\def\csname EERhet.tercile_rich.t3.exposed_months\endcsname{188}
\expandafter\def\csname EERhet.tercile_rich.t3.exposed_events\endcsname{2}
\expandafter\def\csname EERhet.tercile_rich.t3.unexposed_months\endcsname{8,141}
\expandafter\def\csname EERhet.tercile_rich.t3.unexposed_events\endcsname{44}
\expandafter\def\csname EERhet.tercile_rich.t3.rate_unexposed_per_1000\endcsname{5.4}
\expandafter\def\csname EERhet.tercile_rich.t3.rate_exposed_per_1000\endcsname{10.6}
\expandafter\def\csname EERhet.tercile_rich.t3.hr\endcsname{1.28}
\expandafter\def\csname EERhet.tercile_rich.t3.ci_lo\endcsname{0.35}
\expandafter\def\csname EERhet.tercile_rich.t3.ci_hi\endcsname{4.66}
\expandafter\def\csname EERhet.tercile_rich.t3.ci\endcsname{[0.35, 4.66]}
\expandafter\def\csname EERhet.tercile_rich.t3.b\endcsname{0.249}
\expandafter\def\csname EERhet.tercile_rich.t3.se\endcsname{0.658}
\expandafter\def\csname EERhet.tercile_rich.t3.status\endcsname{converged_reduced}
\expandafter\def\csname EERhet.pooled_interaction.t1.women\endcsname{1,733}
\expandafter\def\csname EERhet.pooled_interaction.t1.months\endcsname{26,777}
\expandafter\def\csname EERhet.pooled_interaction.t1.fitted_months\endcsname{26,777}
\expandafter\def\csname EERhet.pooled_interaction.t1.events\endcsname{336}
\expandafter\def\csname EERhet.pooled_interaction.t1.exposed_months\endcsname{2,055}
\expandafter\def\csname EERhet.pooled_interaction.t1.exposed_events\endcsname{84}
\expandafter\def\csname EERhet.pooled_interaction.t1.unexposed_months\endcsname{24,722}
\expandafter\def\csname EERhet.pooled_interaction.t1.unexposed_events\endcsname{252}
\expandafter\def\csname EERhet.pooled_interaction.t1.hr\endcsname{2.48}
\expandafter\def\csname EERhet.pooled_interaction.t1.ci_lo\endcsname{1.71}
\expandafter\def\csname EERhet.pooled_interaction.t1.ci_hi\endcsname{3.61}
\expandafter\def\csname EERhet.pooled_interaction.t1.ci\endcsname{[1.71, 3.61]}
\expandafter\def\csname EERhet.pooled_interaction.t1.b\endcsname{0.910}
\expandafter\def\csname EERhet.pooled_interaction.t1.se\endcsname{0.191}
\expandafter\def\csname EERhet.pooled_interaction.t1.status\endcsname{converged}
\expandafter\def\csname EERhet.pooled_interaction.t2.women\endcsname{1,733}
\expandafter\def\csname EERhet.pooled_interaction.t2.months\endcsname{26,777}
\expandafter\def\csname EERhet.pooled_interaction.t2.fitted_months\endcsname{26,777}
\expandafter\def\csname EERhet.pooled_interaction.t2.events\endcsname{336}
\expandafter\def\csname EERhet.pooled_interaction.t2.exposed_months\endcsname{2,055}
\expandafter\def\csname EERhet.pooled_interaction.t2.exposed_events\endcsname{84}
\expandafter\def\csname EERhet.pooled_interaction.t2.unexposed_months\endcsname{24,722}
\expandafter\def\csname EERhet.pooled_interaction.t2.unexposed_events\endcsname{252}
\expandafter\def\csname EERhet.pooled_interaction.t2.hr\endcsname{4.44}
\expandafter\def\csname EERhet.pooled_interaction.t2.ci_lo\endcsname{2.83}
\expandafter\def\csname EERhet.pooled_interaction.t2.ci_hi\endcsname{6.97}
\expandafter\def\csname EERhet.pooled_interaction.t2.ci\endcsname{[2.83, 6.97]}
\expandafter\def\csname EERhet.pooled_interaction.t2.b\endcsname{1.491}
\expandafter\def\csname EERhet.pooled_interaction.t2.se\endcsname{0.230}
\expandafter\def\csname EERhet.pooled_interaction.t2.status\endcsname{converged}
\expandafter\def\csname EERhet.pooled_interaction.t3.women\endcsname{1,733}
\expandafter\def\csname EERhet.pooled_interaction.t3.months\endcsname{26,777}
\expandafter\def\csname EERhet.pooled_interaction.t3.fitted_months\endcsname{26,777}
\expandafter\def\csname EERhet.pooled_interaction.t3.events\endcsname{336}
\expandafter\def\csname EERhet.pooled_interaction.t3.exposed_months\endcsname{2,055}
\expandafter\def\csname EERhet.pooled_interaction.t3.exposed_events\endcsname{84}
\expandafter\def\csname EERhet.pooled_interaction.t3.unexposed_months\endcsname{24,722}
\expandafter\def\csname EERhet.pooled_interaction.t3.unexposed_events\endcsname{252}
\expandafter\def\csname EERhet.pooled_interaction.t3.hr\endcsname{1.52}
\expandafter\def\csname EERhet.pooled_interaction.t3.ci_lo\endcsname{0.42}
\expandafter\def\csname EERhet.pooled_interaction.t3.ci_hi\endcsname{5.56}
\expandafter\def\csname EERhet.pooled_interaction.t3.ci\endcsname{[0.42, 5.56]}
\expandafter\def\csname EERhet.pooled_interaction.t3.b\endcsname{0.419}
\expandafter\def\csname EERhet.pooled_interaction.t3.se\endcsname{0.661}
\expandafter\def\csname EERhet.pooled_interaction.t3.status\endcsname{converged}
\expandafter\def\csname EERhet.test.separate_fits_equality.chi2\endcsname{6.93}
\expandafter\def\csname EERhet.test.separate_fits_equality.df\endcsname{2}
\expandafter\def\csname EERhet.test.separate_fits_equality.p\endcsname{0.031}
\expandafter\def\csname EERhet.test.pooled_interaction.chi2\endcsname{5.06}
\expandafter\def\csname EERhet.test.pooled_interaction.df\endcsname{2}
\expandafter\def\csname EERhet.test.pooled_interaction.p\endcsname{0.080}
\expandafter\def\csname EERhet.test.pooled_interaction_t2_vs.chi2\endcsname{3.88}
\expandafter\def\csname EERhet.test.pooled_interaction_t2_vs.df\endcsname{1}
\expandafter\def\csname EERhet.test.pooled_interaction_t2_vs.p\endcsname{0.049}
\expandafter\def\csname EERhet.test.pooled_interaction_t3_vs.chi2\endcsname{0.51}
\expandafter\def\csname EERhet.test.pooled_interaction_t3_vs.df\endcsname{1}
\expandafter\def\csname EERhet.test.pooled_interaction_t3_vs.p\endcsname{0.477}
\expandafter\def\csname EERhet.count.tercile_upper_cutoff_percentile.t1\endcsname{28.5}
\expandafter\def\csname EERhet.count.tercile_upper_cutoff_percentile.t2\endcsname{59.2}
\expandafter\def\csname EERhet.count.tercile_upper_cutoff_percentile.t3\endcsname{100.0}
\expandafter\def\csname EERhet.count.women.t1\endcsname{578}
\expandafter\def\csname EERhet.count.asvab_percentile_mean.t1\endcsname{14.1}
\expandafter\def\csname EERhet.count.asvab_percentile_min.t1\endcsname{0.0}
\expandafter\def\csname EERhet.count.asvab_percentile_max.t1\endcsname{28.5}
\expandafter\def\csname EERhet.count.women.t2\endcsname{578}
\expandafter\def\csname EERhet.count.asvab_percentile_mean.t2\endcsname{43.3}
\expandafter\def\csname EERhet.count.asvab_percentile_min.t2\endcsname{28.6}
\expandafter\def\csname EERhet.count.asvab_percentile_max.t2\endcsname{59.2}
\expandafter\def\csname EERhet.count.women.t3\endcsname{577}
\expandafter\def\csname EERhet.count.asvab_percentile_mean.t3\endcsname{77.4}
\expandafter\def\csname EERhet.count.asvab_percentile_min.t3\endcsname{59.3}
\expandafter\def\csname EERhet.count.asvab_percentile_max.t3\endcsname{100.0}
\expandafter\def\csname EERhet.count.landmark_first_exposure_rows.t0\endcsname{225}
\expandafter\def\csname EERhet.count.landmark_first_exposure_rows_without_retained_onset.t0\endcsname{0}
\expandafter\def\csname EERhet.count.landmark_entrants_before_common_support.t1\endcsname{135}
\expandafter\def\csname EERhet.count.landmark_entrants_excluded_no_controls.t1\endcsname{2}
\expandafter\def\csname EERhet.count.landmark_control_index_rows_outside_entrant_blocks.t1\endcsname{6,431}
\expandafter\def\csname EERhet.count.landmark_entrants_before_common_support.t2\endcsname{70}
\expandafter\def\csname EERhet.count.landmark_entrants_excluded_no_controls.t2\endcsname{0}
\expandafter\def\csname EERhet.count.landmark_control_index_rows_outside_entrant_blocks.t2\endcsname{7,395}
\expandafter\def\csname EERhet.count.landmark_entrants_before_common_support.t3\endcsname{20}
\expandafter\def\csname EERhet.count.landmark_entrants_excluded_no_controls.t3\endcsname{0}
\expandafter\def\csname EERhet.count.landmark_control_index_rows_outside_entrant_blocks.t3\endcsname{9,158}
\expandafter\def\csname EERhet.count.landmark_supported_blocks.t0\endcsname{161}
\expandafter\def\csname EERhet.count.landmark_control_index_paths_with_later_conception.t1\endcsname{284}
\expandafter\def\csname EERhet.count.landmark_entrant_index_rows_in_summer.t1\endcsname{24}
\expandafter\def\csname EERhet.count.landmark_control_index_paths_with_later_conception.t2\endcsname{152}
\expandafter\def\csname EERhet.count.landmark_entrant_index_rows_in_summer.t2\endcsname{16}
\expandafter\def\csname EERhet.count.landmark_control_index_paths_with_later_conception.t3\endcsname{13}
\expandafter\def\csname EERhet.count.landmark_entrant_index_rows_in_summer.t3\endcsname{2}
\expandafter\def\csname EERhet.landmark.t1.g0.cif_interruption_pct\endcsname{9.2}
\expandafter\def\csname EERhet.landmark.t1.g0.cif_credential_pct\endcsname{26.5}
\expandafter\def\csname EERhet.landmark.t1.g0.survival_pct\endcsname{64.4}
\expandafter\def\csname EERhet.landmark.t1.g0.raw_interruption_fraction_pct\endcsname{8.8}
\expandafter\def\csname EERhet.landmark.t1.g0.weighted_observed_fraction_pct\endcsname{9.0}
\expandafter\def\csname EERhet.landmark.t1.g0.censor_bound_lo_pct\endcsname{9.0}
\expandafter\def\csname EERhet.landmark.t1.g0.censor_bound_hi_pct\endcsname{11.4}
\expandafter\def\csname EERhet.landmark.t1.g0.index_rows\endcsname{4,239}
\expandafter\def\csname EERhet.landmark.t1.g0.unique_women\endcsname{532}
\expandafter\def\csname EERhet.landmark.t1.g0.raw_interruptions\endcsname{371}
\expandafter\def\csname EERhet.landmark.t1.g0.raw_credentials\endcsname{997}
\expandafter\def\csname EERhet.landmark.t1.g0.raw_early_censors\endcsname{48}
\expandafter\def\csname EERhet.landmark.t1.g0.raw_full_observed\endcsname{4,191}
\expandafter\def\csname EERhet.landmark.t1.g0.months_with_risk\endcsname{9}
\expandafter\def\csname EERhet.landmark.t1.g1.cif_interruption_pct\endcsname{24.2}
\expandafter\def\csname EERhet.landmark.t1.g1.cif_credential_pct\endcsname{20.5}
\expandafter\def\csname EERhet.landmark.t1.g1.survival_pct\endcsname{55.4}
\expandafter\def\csname EERhet.landmark.t1.g1.raw_interruption_fraction_pct\endcsname{21.8}
\expandafter\def\csname EERhet.landmark.t1.g1.weighted_observed_fraction_pct\endcsname{21.8}
\expandafter\def\csname EERhet.landmark.t1.g1.censor_bound_lo_pct\endcsname{21.8}
\expandafter\def\csname EERhet.landmark.t1.g1.censor_bound_hi_pct\endcsname{33.8}
\expandafter\def\csname EERhet.landmark.t1.g1.index_rows\endcsname{133}
\expandafter\def\csname EERhet.landmark.t1.g1.unique_women\endcsname{133}
\expandafter\def\csname EERhet.landmark.t1.g1.raw_interruptions\endcsname{29}
\expandafter\def\csname EERhet.landmark.t1.g1.raw_credentials\endcsname{25}
\expandafter\def\csname EERhet.landmark.t1.g1.raw_early_censors\endcsname{16}
\expandafter\def\csname EERhet.landmark.t1.g1.raw_full_observed\endcsname{117}
\expandafter\def\csname EERhet.landmark.t1.g1.months_with_risk\endcsname{9}
\expandafter\def\csname EERhet.landmark.t2.g0.cif_interruption_pct\endcsname{6.6}
\expandafter\def\csname EERhet.landmark.t2.g0.cif_credential_pct\endcsname{34.9}
\expandafter\def\csname EERhet.landmark.t2.g0.survival_pct\endcsname{58.5}
\expandafter\def\csname EERhet.landmark.t2.g0.raw_interruption_fraction_pct\endcsname{6.4}
\expandafter\def\csname EERhet.landmark.t2.g0.weighted_observed_fraction_pct\endcsname{6.6}
\expandafter\def\csname EERhet.landmark.t2.g0.censor_bound_lo_pct\endcsname{6.6}
\expandafter\def\csname EERhet.landmark.t2.g0.censor_bound_hi_pct\endcsname{7.2}
\expandafter\def\csname EERhet.landmark.t2.g0.index_rows\endcsname{3,550}
\expandafter\def\csname EERhet.landmark.t2.g0.unique_women\endcsname{548}
\expandafter\def\csname EERhet.landmark.t2.g0.raw_interruptions\endcsname{228}
\expandafter\def\csname EERhet.landmark.t2.g0.raw_credentials\endcsname{1,123}
\expandafter\def\csname EERhet.landmark.t2.g0.raw_early_censors\endcsname{28}
\expandafter\def\csname EERhet.landmark.t2.g0.raw_full_observed\endcsname{3,522}
\expandafter\def\csname EERhet.landmark.t2.g0.months_with_risk\endcsname{9}
\expandafter\def\csname EERhet.landmark.t2.g1.cif_interruption_pct\endcsname{25.0}
\expandafter\def\csname EERhet.landmark.t2.g1.cif_credential_pct\endcsname{31.1}
\expandafter\def\csname EERhet.landmark.t2.g1.survival_pct\endcsname{43.9}
\expandafter\def\csname EERhet.landmark.t2.g1.raw_interruption_fraction_pct\endcsname{24.3}
\expandafter\def\csname EERhet.landmark.t2.g1.weighted_observed_fraction_pct\endcsname{24.3}
\expandafter\def\csname EERhet.landmark.t2.g1.censor_bound_lo_pct\endcsname{24.3}
\expandafter\def\csname EERhet.landmark.t2.g1.censor_bound_hi_pct\endcsname{30.0}
\expandafter\def\csname EERhet.landmark.t2.g1.index_rows\endcsname{70}
\expandafter\def\csname EERhet.landmark.t2.g1.unique_women\endcsname{70}
\expandafter\def\csname EERhet.landmark.t2.g1.raw_interruptions\endcsname{17}
\expandafter\def\csname EERhet.landmark.t2.g1.raw_credentials\endcsname{21}
\expandafter\def\csname EERhet.landmark.t2.g1.raw_early_censors\endcsname{4}
\expandafter\def\csname EERhet.landmark.t2.g1.raw_full_observed\endcsname{66}
\expandafter\def\csname EERhet.landmark.t2.g1.months_with_risk\endcsname{9}
\expandafter\def\csname EERhet.landmark.t3.g0.cif_interruption_pct\endcsname{3.1}
\expandafter\def\csname EERhet.landmark.t3.g0.cif_credential_pct\endcsname{54.7}
\expandafter\def\csname EERhet.landmark.t3.g0.survival_pct\endcsname{42.2}
\expandafter\def\csname EERhet.landmark.t3.g0.raw_interruption_fraction_pct\endcsname{2.9}
\expandafter\def\csname EERhet.landmark.t3.g0.weighted_observed_fraction_pct\endcsname{3.1}
\expandafter\def\csname EERhet.landmark.t3.g0.censor_bound_lo_pct\endcsname{3.1}
\expandafter\def\csname EERhet.landmark.t3.g0.censor_bound_hi_pct\endcsname{3.5}
\expandafter\def\csname EERhet.landmark.t3.g0.index_rows\endcsname{1,389}
\expandafter\def\csname EERhet.landmark.t3.g0.unique_women\endcsname{509}
\expandafter\def\csname EERhet.landmark.t3.g0.raw_interruptions\endcsname{40}
\expandafter\def\csname EERhet.landmark.t3.g0.raw_credentials\endcsname{712}
\expandafter\def\csname EERhet.landmark.t3.g0.raw_early_censors\endcsname{8}
\expandafter\def\csname EERhet.landmark.t3.g0.raw_full_observed\endcsname{1,381}
\expandafter\def\csname EERhet.landmark.t3.g0.months_with_risk\endcsname{9}
\expandafter\def\csname EERhet.landmark.t3.g1.cif_interruption_pct\endcsname{5.0}
\expandafter\def\csname EERhet.landmark.t3.g1.cif_credential_pct\endcsname{45.0}
\expandafter\def\csname EERhet.landmark.t3.g1.survival_pct\endcsname{50.0}
\expandafter\def\csname EERhet.landmark.t3.g1.raw_interruption_fraction_pct\endcsname{5.0}
\expandafter\def\csname EERhet.landmark.t3.g1.weighted_observed_fraction_pct\endcsname{5.0}
\expandafter\def\csname EERhet.landmark.t3.g1.censor_bound_lo_pct\endcsname{5.0}
\expandafter\def\csname EERhet.landmark.t3.g1.censor_bound_hi_pct\endcsname{5.0}
\expandafter\def\csname EERhet.landmark.t3.g1.index_rows\endcsname{20}
\expandafter\def\csname EERhet.landmark.t3.g1.unique_women\endcsname{20}
\expandafter\def\csname EERhet.landmark.t3.g1.raw_interruptions\endcsname{1}
\expandafter\def\csname EERhet.landmark.t3.g1.raw_credentials\endcsname{9}
\expandafter\def\csname EERhet.landmark.t3.g1.raw_early_censors\endcsname{0}
\expandafter\def\csname EERhet.landmark.t3.g1.raw_full_observed\endcsname{20}
\expandafter\def\csname EERhet.landmark.t3.g1.months_with_risk\endcsname{9}
}{}
\IfFileExists{evidence/refinement/extensions/numbers_interruption_expectations.tex}{
\newcommand{\EERexp}[1]{\ifcsname EERexp.#1\endcsname\csname EERexp.#1\endcsname\else\PackageError{EER expectation numbers}{Missing expectation value #1}{Rerun interruption_expectations.do.}\fi}
\expandafter\def\csname EERexp.later_all_core.low.women\endcsname{600}
\expandafter\def\csname EERexp.later_all_core.low.months\endcsname{6,033}
\expandafter\def\csname EERexp.later_all_core.low.fitted_months\endcsname{5,952}
\expandafter\def\csname EERexp.later_all_core.low.events\endcsname{88}
\expandafter\def\csname EERexp.later_all_core.low.exposed_months\endcsname{340}
\expandafter\def\csname EERexp.later_all_core.low.exposed_events\endcsname{19}
\expandafter\def\csname EERexp.later_all_core.low.hr\endcsname{4.21}
\expandafter\def\csname EERexp.later_all_core.low.ci_lo\endcsname{2.45}
\expandafter\def\csname EERexp.later_all_core.low.ci_hi\endcsname{7.23}
\expandafter\def\csname EERexp.later_all_core.low.ci\endcsname{[2.45, 7.23]}
\expandafter\def\csname EERexp.later_all_core.low.b\endcsname{1.437}
\expandafter\def\csname EERexp.later_all_core.low.se\endcsname{0.276}
\expandafter\def\csname EERexp.later_all_core.low.status\endcsname{converged_reduced}
\expandafter\def\csname EERexp.later_all_core.high.women\endcsname{78}
\expandafter\def\csname EERexp.later_all_core.high.months\endcsname{866}
\expandafter\def\csname EERexp.later_all_core.high.fitted_months\endcsname{853}
\expandafter\def\csname EERexp.later_all_core.high.events\endcsname{20}
\expandafter\def\csname EERexp.later_all_core.high.exposed_months\endcsname{120}
\expandafter\def\csname EERexp.later_all_core.high.exposed_events\endcsname{2}
\expandafter\def\csname EERexp.later_all_core.high.status\endcsname{not_valid}
\expandafter\def\csname EERexp.later_all_core_pooled.low.women\endcsname{678}
\expandafter\def\csname EERexp.later_all_core_pooled.low.months\endcsname{6,899}
\expandafter\def\csname EERexp.later_all_core_pooled.low.fitted_months\endcsname{6,883}
\expandafter\def\csname EERexp.later_all_core_pooled.low.events\endcsname{108}
\expandafter\def\csname EERexp.later_all_core_pooled.low.exposed_months\endcsname{460}
\expandafter\def\csname EERexp.later_all_core_pooled.low.exposed_events\endcsname{21}
\expandafter\def\csname EERexp.later_all_core_pooled.low.hr\endcsname{4.14}
\expandafter\def\csname EERexp.later_all_core_pooled.low.ci_lo\endcsname{2.42}
\expandafter\def\csname EERexp.later_all_core_pooled.low.ci_hi\endcsname{7.09}
\expandafter\def\csname EERexp.later_all_core_pooled.low.ci\endcsname{[2.42, 7.09]}
\expandafter\def\csname EERexp.later_all_core_pooled.low.b\endcsname{1.421}
\expandafter\def\csname EERexp.later_all_core_pooled.low.se\endcsname{0.275}
\expandafter\def\csname EERexp.later_all_core_pooled.low.status\endcsname{converged_reduced}
\expandafter\def\csname EERexp.later_all_core_pooled.high.women\endcsname{678}
\expandafter\def\csname EERexp.later_all_core_pooled.high.months\endcsname{6,899}
\expandafter\def\csname EERexp.later_all_core_pooled.high.fitted_months\endcsname{6,883}
\expandafter\def\csname EERexp.later_all_core_pooled.high.events\endcsname{108}
\expandafter\def\csname EERexp.later_all_core_pooled.high.exposed_months\endcsname{460}
\expandafter\def\csname EERexp.later_all_core_pooled.high.exposed_events\endcsname{21}
\expandafter\def\csname EERexp.later_all_core_pooled.high.hr\endcsname{0.64}
\expandafter\def\csname EERexp.later_all_core_pooled.high.ci_lo\endcsname{0.20}
\expandafter\def\csname EERexp.later_all_core_pooled.high.ci_hi\endcsname{2.03}
\expandafter\def\csname EERexp.later_all_core_pooled.high.ci\endcsname{[0.20, 2.03]}
\expandafter\def\csname EERexp.later_all_core_pooled.high.b\endcsname{-0.447}
\expandafter\def\csname EERexp.later_all_core_pooled.high.se\endcsname{0.590}
\expandafter\def\csname EERexp.later_all_core_pooled.high.status\endcsname{converged_reduced}
\expandafter\def\csname EERexp.later_all_core_reported.both.women\endcsname{678}
\expandafter\def\csname EERexp.later_all_core_reported.both.months\endcsname{6,899}
\expandafter\def\csname EERexp.later_all_core_reported.both.fitted_months\endcsname{6,883}
\expandafter\def\csname EERexp.later_all_core_reported.both.events\endcsname{108}
\expandafter\def\csname EERexp.later_all_core_reported.both.exposed_months\endcsname{460}
\expandafter\def\csname EERexp.later_all_core_reported.both.exposed_events\endcsname{21}
\expandafter\def\csname EERexp.later_all_core_reported.both.hr\endcsname{3.00}
\expandafter\def\csname EERexp.later_all_core_reported.both.ci_lo\endcsname{1.83}
\expandafter\def\csname EERexp.later_all_core_reported.both.ci_hi\endcsname{4.93}
\expandafter\def\csname EERexp.later_all_core_reported.both.ci\endcsname{[1.83, 4.93]}
\expandafter\def\csname EERexp.later_all_core_reported.both.b\endcsname{1.100}
\expandafter\def\csname EERexp.later_all_core_reported.both.se\endcsname{0.253}
\expandafter\def\csname EERexp.later_all_core_reported.both.status\endcsname{converged_reduced}
\expandafter\def\csname EERexp.later_cc_rich.low.women\endcsname{452}
\expandafter\def\csname EERexp.later_cc_rich.low.months\endcsname{4,511}
\expandafter\def\csname EERexp.later_cc_rich.low.fitted_months\endcsname{4,369}
\expandafter\def\csname EERexp.later_cc_rich.low.events\endcsname{50}
\expandafter\def\csname EERexp.later_cc_rich.low.exposed_months\endcsname{208}
\expandafter\def\csname EERexp.later_cc_rich.low.exposed_events\endcsname{11}
\expandafter\def\csname EERexp.later_cc_rich.low.hr\endcsname{4.00}
\expandafter\def\csname EERexp.later_cc_rich.low.ci_lo\endcsname{1.98}
\expandafter\def\csname EERexp.later_cc_rich.low.ci_hi\endcsname{8.06}
\expandafter\def\csname EERexp.later_cc_rich.low.ci\endcsname{[1.98, 8.06]}
\expandafter\def\csname EERexp.later_cc_rich.low.b\endcsname{1.386}
\expandafter\def\csname EERexp.later_cc_rich.low.se\endcsname{0.358}
\expandafter\def\csname EERexp.later_cc_rich.low.status\endcsname{converged_reduced}
\expandafter\def\csname EERexp.later_cc_rich.high.women\endcsname{60}
\expandafter\def\csname EERexp.later_cc_rich.high.months\endcsname{696}
\expandafter\def\csname EERexp.later_cc_rich.high.fitted_months\endcsname{680}
\expandafter\def\csname EERexp.later_cc_rich.high.events\endcsname{12}
\expandafter\def\csname EERexp.later_cc_rich.high.exposed_months\endcsname{102}
\expandafter\def\csname EERexp.later_cc_rich.high.exposed_events\endcsname{2}
\expandafter\def\csname EERexp.later_cc_rich.high.status\endcsname{not_valid}
\expandafter\def\csname EERexp.later_cc_rich_pooled.low.women\endcsname{512}
\expandafter\def\csname EERexp.later_cc_rich_pooled.low.months\endcsname{5,207}
\expandafter\def\csname EERexp.later_cc_rich_pooled.low.fitted_months\endcsname{5,128}
\expandafter\def\csname EERexp.later_cc_rich_pooled.low.events\endcsname{62}
\expandafter\def\csname EERexp.later_cc_rich_pooled.low.exposed_months\endcsname{310}
\expandafter\def\csname EERexp.later_cc_rich_pooled.low.exposed_events\endcsname{13}
\expandafter\def\csname EERexp.later_cc_rich_pooled.low.hr\endcsname{4.18}
\expandafter\def\csname EERexp.later_cc_rich_pooled.low.ci_lo\endcsname{2.09}
\expandafter\def\csname EERexp.later_cc_rich_pooled.low.ci_hi\endcsname{8.36}
\expandafter\def\csname EERexp.later_cc_rich_pooled.low.ci\endcsname{[2.09, 8.36]}
\expandafter\def\csname EERexp.later_cc_rich_pooled.low.b\endcsname{1.431}
\expandafter\def\csname EERexp.later_cc_rich_pooled.low.se\endcsname{0.353}
\expandafter\def\csname EERexp.later_cc_rich_pooled.low.status\endcsname{converged_reduced}
\expandafter\def\csname EERexp.later_cc_rich_pooled.high.women\endcsname{512}
\expandafter\def\csname EERexp.later_cc_rich_pooled.high.months\endcsname{5,207}
\expandafter\def\csname EERexp.later_cc_rich_pooled.high.fitted_months\endcsname{5,128}
\expandafter\def\csname EERexp.later_cc_rich_pooled.high.events\endcsname{62}
\expandafter\def\csname EERexp.later_cc_rich_pooled.high.exposed_months\endcsname{310}
\expandafter\def\csname EERexp.later_cc_rich_pooled.high.exposed_events\endcsname{13}
\expandafter\def\csname EERexp.later_cc_rich_pooled.high.hr\endcsname{1.30}
\expandafter\def\csname EERexp.later_cc_rich_pooled.high.ci_lo\endcsname{0.45}
\expandafter\def\csname EERexp.later_cc_rich_pooled.high.ci_hi\endcsname{3.81}
\expandafter\def\csname EERexp.later_cc_rich_pooled.high.ci\endcsname{[0.45, 3.81]}
\expandafter\def\csname EERexp.later_cc_rich_pooled.high.b\endcsname{0.266}
\expandafter\def\csname EERexp.later_cc_rich_pooled.high.se\endcsname{0.547}
\expandafter\def\csname EERexp.later_cc_rich_pooled.high.status\endcsname{converged_reduced}
\expandafter\def\csname EERexp.later_cc_rich_reported.both.women\endcsname{512}
\expandafter\def\csname EERexp.later_cc_rich_reported.both.months\endcsname{5,207}
\expandafter\def\csname EERexp.later_cc_rich_reported.both.fitted_months\endcsname{5,128}
\expandafter\def\csname EERexp.later_cc_rich_reported.both.events\endcsname{62}
\expandafter\def\csname EERexp.later_cc_rich_reported.both.exposed_months\endcsname{310}
\expandafter\def\csname EERexp.later_cc_rich_reported.both.exposed_events\endcsname{13}
\expandafter\def\csname EERexp.later_cc_rich_reported.both.hr\endcsname{3.25}
\expandafter\def\csname EERexp.later_cc_rich_reported.both.ci_lo\endcsname{1.83}
\expandafter\def\csname EERexp.later_cc_rich_reported.both.ci_hi\endcsname{5.75}
\expandafter\def\csname EERexp.later_cc_rich_reported.both.ci\endcsname{[1.83, 5.75]}
\expandafter\def\csname EERexp.later_cc_rich_reported.both.b\endcsname{1.178}
\expandafter\def\csname EERexp.later_cc_rich_reported.both.se\endcsname{0.292}
\expandafter\def\csname EERexp.later_cc_rich_reported.both.status\endcsname{converged_reduced}
\expandafter\def\csname EERexp.test.later_all_core.chi2\endcsname{8.36}
\expandafter\def\csname EERexp.test.later_all_core.df\endcsname{1}
\expandafter\def\csname EERexp.test.later_all_core.p\endcsname{0.004}
\expandafter\def\csname EERexp.test.later_cc_rich.chi2\endcsname{3.02}
\expandafter\def\csname EERexp.test.later_cc_rich.df\endcsname{1}
\expandafter\def\csname EERexp.test.later_cc_rich.p\endcsname{0.082}
\expandafter\def\csname EERexp.count.raw_file.all.respondents\endcsname{8,984}
\expandafter\def\csname EERexp.count.raw_file.all.valid_reports\endcsname{1,742}
\expandafter\def\csname EERexp.count.later_all_core.missing.women\endcsname{1,635}
\expandafter\def\csname EERexp.count.later_all_core.missing.months\endcsname{28,696}
\expandafter\def\csname EERexp.count.later_all_core.missing.events\endcsname{413}
\expandafter\def\csname EERexp.count.later_all_core.missing.exposed_months\endcsname{2,542}
\expandafter\def\csname EERexp.count.later_all_core.missing.exposed_events\endcsname{115}
\expandafter\def\csname EERexp.count.later_all_core.missing.women_with_retained_conception\endcsname{1,029}
\expandafter\def\csname EERexp.count.later_all_core.low.women\endcsname{600}
\expandafter\def\csname EERexp.count.later_all_core.low.months\endcsname{6,033}
\expandafter\def\csname EERexp.count.later_all_core.low.events\endcsname{88}
\expandafter\def\csname EERexp.count.later_all_core.low.exposed_months\endcsname{340}
\expandafter\def\csname EERexp.count.later_all_core.low.exposed_events\endcsname{19}
\expandafter\def\csname EERexp.count.later_all_core.low.women_with_retained_conception\endcsname{406}
\expandafter\def\csname EERexp.count.later_all_core.low.expect_pct_mean\endcsname{1.3}
\expandafter\def\csname EERexp.count.later_all_core.high.women\endcsname{78}
\expandafter\def\csname EERexp.count.later_all_core.high.months\endcsname{866}
\expandafter\def\csname EERexp.count.later_all_core.high.events\endcsname{20}
\expandafter\def\csname EERexp.count.later_all_core.high.exposed_months\endcsname{120}
\expandafter\def\csname EERexp.count.later_all_core.high.exposed_events\endcsname{2}
\expandafter\def\csname EERexp.count.later_all_core.high.women_with_retained_conception\endcsname{57}
\expandafter\def\csname EERexp.count.later_all_core.high.expect_pct_mean\endcsname{49.5}
\expandafter\def\csname EERexp.count.later_cc_rich.missing.women\endcsname{1,221}
\expandafter\def\csname EERexp.count.later_cc_rich.missing.months\endcsname{21,570}
\expandafter\def\csname EERexp.count.later_cc_rich.missing.events\endcsname{274}
\expandafter\def\csname EERexp.count.later_cc_rich.missing.exposed_months\endcsname{1,745}
\expandafter\def\csname EERexp.count.later_cc_rich.missing.exposed_events\endcsname{71}
\expandafter\def\csname EERexp.count.later_cc_rich.missing.women_with_retained_conception\endcsname{741}
\expandafter\def\csname EERexp.count.later_cc_rich.low.women\endcsname{452}
\expandafter\def\csname EERexp.count.later_cc_rich.low.months\endcsname{4,511}
\expandafter\def\csname EERexp.count.later_cc_rich.low.events\endcsname{50}
\expandafter\def\csname EERexp.count.later_cc_rich.low.exposed_months\endcsname{208}
\expandafter\def\csname EERexp.count.later_cc_rich.low.exposed_events\endcsname{11}
\expandafter\def\csname EERexp.count.later_cc_rich.low.women_with_retained_conception\endcsname{310}
\expandafter\def\csname EERexp.count.later_cc_rich.low.expect_pct_mean\endcsname{1.4}
\expandafter\def\csname EERexp.count.later_cc_rich.high.women\endcsname{60}
\expandafter\def\csname EERexp.count.later_cc_rich.high.months\endcsname{696}
\expandafter\def\csname EERexp.count.later_cc_rich.high.events\endcsname{12}
\expandafter\def\csname EERexp.count.later_cc_rich.high.exposed_months\endcsname{102}
\expandafter\def\csname EERexp.count.later_cc_rich.high.exposed_events\endcsname{2}
\expandafter\def\csname EERexp.count.later_cc_rich.high.women_with_retained_conception\endcsname{43}
\expandafter\def\csname EERexp.count.later_cc_rich.high.expect_pct_mean\endcsname{49.1}
}{}
\IfFileExists{evidence/numbers_conception_screen.tex}{
\providecommand{\ConceptionScreenN}{9}
\providecommand{\ConceptionScreenRecovered}{8}
}{}
\IfFileExists{evidence/numbers_exit_gap.tex}{
\providecommand{\ExitGapExposed}{246}

\providecommand{\ExitGapConceive}{14}
\providecommand{\ExitGapShare}{5.7}
}{}
\IfFileExists{evidence/numbers_sample.tex}{

\providecommand{\RecExitRateFullUnexp}{0.31}
\providecommand{\RecExitRatePostUnexp}{0.66}
\providecommand{\RecExitRateExposed}{2.86}
\providecommand{\RecExitRatePreDebut}{0.10}
}{}
\IfFileExists{evidence/numbers_lag_sensitivity.tex}{
\providecommand{\RecExitHRFull}{5.67}
\providecommand{\RecExitHRPostDebut}{3.75}
}{}
\IfFileExists{evidence/numbers_unobserved_onset.tex}{

\newcommand{\SplitCloglogCleanHR}{3.34}

\newcommand{\SplitCloglogUnobsHR}{4.66}

}{}
\IfFileExists{evidence/refinement/timing/numbers_interruption_timing.tex}{
\providecommand{\EERtiming}[1]{\csname EERtiming@#1\endcsname}
\expandafter\def\csname EERtiming@single.base.hr\endcsname{3.85}
\expandafter\def\csname EERtiming@single.base.ci_lo\endcsname{3.29}
\expandafter\def\csname EERtiming@single.base.ci_hi\endcsname{4.51}
\expandafter\def\csname EERtiming@single.base.women\endcsname{2,828}
\expandafter\def\csname EERtiming@single.base.months\endcsname{48,953}
\expandafter\def\csname EERtiming@single.base.events\endcsname{711}
\expandafter\def\csname EERtiming@single.base.exposed_events\endcsname{261}
\expandafter\def\csname EERtiming@single.base.term_events\endcsname{261}
\expandafter\def\csname EERtiming@single.base.status\endcsname{converged}
\expandafter\def\csname EERtiming@single.lag0.hr\endcsname{3.96}
\expandafter\def\csname EERtiming@single.lag0.ci_lo\endcsname{3.39}
\expandafter\def\csname EERtiming@single.lag0.ci_hi\endcsname{4.63}
\expandafter\def\csname EERtiming@single.lag0.women\endcsname{2,828}
\expandafter\def\csname EERtiming@single.lag0.months\endcsname{48,953}
\expandafter\def\csname EERtiming@single.lag0.events\endcsname{711}
\expandafter\def\csname EERtiming@single.lag0.exposed_events\endcsname{276}
\expandafter\def\csname EERtiming@single.lag0.term_events\endcsname{276}
\expandafter\def\csname EERtiming@single.lag0.status\endcsname{converged}
\expandafter\def\csname EERtiming@single.lag2.hr\endcsname{3.82}
\expandafter\def\csname EERtiming@single.lag2.ci_lo\endcsname{3.25}
\expandafter\def\csname EERtiming@single.lag2.ci_hi\endcsname{4.48}
\expandafter\def\csname EERtiming@single.lag2.women\endcsname{2,828}
\expandafter\def\csname EERtiming@single.lag2.months\endcsname{48,953}
\expandafter\def\csname EERtiming@single.lag2.events\endcsname{711}
\expandafter\def\csname EERtiming@single.lag2.exposed_events\endcsname{249}
\expandafter\def\csname EERtiming@single.lag2.term_events\endcsname{249}
\expandafter\def\csname EERtiming@single.lag2.status\endcsname{converged}
\expandafter\def\csname EERtiming@single.lag3.hr\endcsname{3.79}
\expandafter\def\csname EERtiming@single.lag3.ci_lo\endcsname{3.22}
\expandafter\def\csname EERtiming@single.lag3.ci_hi\endcsname{4.46}
\expandafter\def\csname EERtiming@single.lag3.women\endcsname{2,828}
\expandafter\def\csname EERtiming@single.lag3.months\endcsname{48,953}
\expandafter\def\csname EERtiming@single.lag3.events\endcsname{711}
\expandafter\def\csname EERtiming@single.lag3.exposed_events\endcsname{238}
\expandafter\def\csname EERtiming@single.lag3.term_events\endcsname{238}
\expandafter\def\csname EERtiming@single.lag3.status\endcsname{converged}
\expandafter\def\csname EERtiming@single.omit0.hr\endcsname{3.95}
\expandafter\def\csname EERtiming@single.omit0.ci_lo\endcsname{3.37}
\expandafter\def\csname EERtiming@single.omit0.ci_hi\endcsname{4.64}
\expandafter\def\csname EERtiming@single.omit0.women\endcsname{2,828}
\expandafter\def\csname EERtiming@single.omit0.months\endcsname{48,589}
\expandafter\def\csname EERtiming@single.omit0.events\endcsname{696}
\expandafter\def\csname EERtiming@single.omit0.exposed_events\endcsname{261}
\expandafter\def\csname EERtiming@single.omit0.term_events\endcsname{261}
\expandafter\def\csname EERtiming@single.omit0.status\endcsname{converged}
\expandafter\def\csname EERtiming@single.omit0to1.hr\endcsname{4.00}
\expandafter\def\csname EERtiming@single.omit0to1.ci_lo\endcsname{3.41}
\expandafter\def\csname EERtiming@single.omit0to1.ci_hi\endcsname{4.71}
\expandafter\def\csname EERtiming@single.omit0to1.women\endcsname{2,828}
\expandafter\def\csname EERtiming@single.omit0to1.months\endcsname{48,230}
\expandafter\def\csname EERtiming@single.omit0to1.events\endcsname{684}
\expandafter\def\csname EERtiming@single.omit0to1.exposed_events\endcsname{249}
\expandafter\def\csname EERtiming@single.omit0to1.term_events\endcsname{249}
\expandafter\def\csname EERtiming@single.omit0to1.status\endcsname{converged}
\expandafter\def\csname EERtiming@single.omitm1to1.hr\endcsname{4.09}
\expandafter\def\csname EERtiming@single.omitm1to1.ci_lo\endcsname{3.48}
\expandafter\def\csname EERtiming@single.omitm1to1.ci_hi\endcsname{4.81}
\expandafter\def\csname EERtiming@single.omitm1to1.women\endcsname{2,816}
\expandafter\def\csname EERtiming@single.omitm1to1.months\endcsname{47,842}
\expandafter\def\csname EERtiming@single.omitm1to1.events\endcsname{671}
\expandafter\def\csname EERtiming@single.omitm1to1.exposed_events\endcsname{249}
\expandafter\def\csname EERtiming@single.omitm1to1.term_events\endcsname{249}
\expandafter\def\csname EERtiming@single.omitm1to1.status\endcsname{converged}
\expandafter\def\csname EERtiming@single.omitm1to2.hr\endcsname{4.13}
\expandafter\def\csname EERtiming@single.omitm1to2.ci_lo\endcsname{3.50}
\expandafter\def\csname EERtiming@single.omitm1to2.ci_hi\endcsname{4.88}
\expandafter\def\csname EERtiming@single.omitm1to2.women\endcsname{2,812}
\expandafter\def\csname EERtiming@single.omitm1to2.months\endcsname{47,518}
\expandafter\def\csname EERtiming@single.omitm1to2.events\endcsname{660}
\expandafter\def\csname EERtiming@single.omitm1to2.exposed_events\endcsname{238}
\expandafter\def\csname EERtiming@single.omitm1to2.term_events\endcsname{238}
\expandafter\def\csname EERtiming@single.omitm1to2.status\endcsname{converged}
\expandafter\def\csname EERtiming@single.omitm2to2.hr\endcsname{4.20}
\expandafter\def\csname EERtiming@single.omitm2to2.ci_lo\endcsname{3.55}
\expandafter\def\csname EERtiming@single.omitm2to2.ci_hi\endcsname{4.96}
\expandafter\def\csname EERtiming@single.omitm2to2.women\endcsname{2,809}
\expandafter\def\csname EERtiming@single.omitm2to2.months\endcsname{47,202}
\expandafter\def\csname EERtiming@single.omitm2to2.events\endcsname{650}
\expandafter\def\csname EERtiming@single.omitm2to2.exposed_events\endcsname{238}
\expandafter\def\csname EERtiming@single.omitm2to2.term_events\endcsname{238}
\expandafter\def\csname EERtiming@single.omitm2to2.status\endcsname{converged}
\expandafter\def\csname EERtiming@single.bin1to3.hr\endcsname{4.45}
\expandafter\def\csname EERtiming@single.bin1to3.ci_lo\endcsname{3.30}
\expandafter\def\csname EERtiming@single.bin1to3.ci_hi\endcsname{6.02}
\expandafter\def\csname EERtiming@single.bin1to3.women\endcsname{2,828}
\expandafter\def\csname EERtiming@single.bin1to3.months\endcsname{48,953}
\expandafter\def\csname EERtiming@single.bin1to3.events\endcsname{711}
\expandafter\def\csname EERtiming@single.bin1to3.exposed_events\endcsname{261}
\expandafter\def\csname EERtiming@single.bin1to3.term_events\endcsname{46}
\expandafter\def\csname EERtiming@single.bin1to3.status\endcsname{converged}
\expandafter\def\csname EERtiming@single.bin4to6.hr\endcsname{2.88}
\expandafter\def\csname EERtiming@single.bin4to6.ci_lo\endcsname{1.90}
\expandafter\def\csname EERtiming@single.bin4to6.ci_hi\endcsname{4.36}
\expandafter\def\csname EERtiming@single.bin4to6.women\endcsname{2,828}
\expandafter\def\csname EERtiming@single.bin4to6.months\endcsname{48,953}
\expandafter\def\csname EERtiming@single.bin4to6.events\endcsname{711}
\expandafter\def\csname EERtiming@single.bin4to6.exposed_events\endcsname{261}
\expandafter\def\csname EERtiming@single.bin4to6.term_events\endcsname{24}
\expandafter\def\csname EERtiming@single.bin4to6.status\endcsname{converged}
\expandafter\def\csname EERtiming@single.bin7to12.hr\endcsname{4.74}
\expandafter\def\csname EERtiming@single.bin7to12.ci_lo\endcsname{3.62}
\expandafter\def\csname EERtiming@single.bin7to12.ci_hi\endcsname{6.20}
\expandafter\def\csname EERtiming@single.bin7to12.women\endcsname{2,828}
\expandafter\def\csname EERtiming@single.bin7to12.months\endcsname{48,953}
\expandafter\def\csname EERtiming@single.bin7to12.events\endcsname{711}
\expandafter\def\csname EERtiming@single.bin7to12.exposed_events\endcsname{261}
\expandafter\def\csname EERtiming@single.bin7to12.term_events\endcsname{67}
\expandafter\def\csname EERtiming@single.bin7to12.status\endcsname{converged}
\expandafter\def\csname EERtiming@single.bin13p.hr\endcsname{3.53}
\expandafter\def\csname EERtiming@single.bin13p.ci_lo\endcsname{2.89}
\expandafter\def\csname EERtiming@single.bin13p.ci_hi\endcsname{4.31}
\expandafter\def\csname EERtiming@single.bin13p.women\endcsname{2,828}
\expandafter\def\csname EERtiming@single.bin13p.months\endcsname{48,953}
\expandafter\def\csname EERtiming@single.bin13p.events\endcsname{711}
\expandafter\def\csname EERtiming@single.bin13p.exposed_events\endcsname{261}
\expandafter\def\csname EERtiming@single.bin13p.term_events\endcsname{124}
\expandafter\def\csname EERtiming@single.bin13p.status\endcsname{converged}
\expandafter\def\csname EERtiming@single.dropconf.hr\endcsname{4.11}
\expandafter\def\csname EERtiming@single.dropconf.ci_lo\endcsname{3.50}
\expandafter\def\csname EERtiming@single.dropconf.ci_hi\endcsname{4.82}
\expandafter\def\csname EERtiming@single.dropconf.women\endcsname{2,797}
\expandafter\def\csname EERtiming@single.dropconf.months\endcsname{48,559}
\expandafter\def\csname EERtiming@single.dropconf.events\endcsname{680}
\expandafter\def\csname EERtiming@single.dropconf.exposed_events\endcsname{261}
\expandafter\def\csname EERtiming@single.dropconf.term_events\endcsname{261}
\expandafter\def\csname EERtiming@single.dropconf.status\endcsname{converged}
\expandafter\def\csname EERtiming@single.redate.hr\endcsname{3.90}
\expandafter\def\csname EERtiming@single.redate.ci_lo\endcsname{3.33}
\expandafter\def\csname EERtiming@single.redate.ci_hi\endcsname{4.57}
\expandafter\def\csname EERtiming@single.redate.women\endcsname{2,821}
\expandafter\def\csname EERtiming@single.redate.months\endcsname{64,043}
\expandafter\def\csname EERtiming@single.redate.events\endcsname{704}
\expandafter\def\csname EERtiming@single.redate.exposed_events\endcsname{255}
\expandafter\def\csname EERtiming@single.redate.term_events\endcsname{255}
\expandafter\def\csname EERtiming@single.redate.status\endcsname{converged}
\expandafter\def\csname EERtiming@single.dropsummer.hr\endcsname{3.76}
\expandafter\def\csname EERtiming@single.dropsummer.ci_lo\endcsname{3.21}
\expandafter\def\csname EERtiming@single.dropsummer.ci_hi\endcsname{4.41}
\expandafter\def\csname EERtiming@single.dropsummer.women\endcsname{2,822}
\expandafter\def\csname EERtiming@single.dropsummer.months\endcsname{48,895}
\expandafter\def\csname EERtiming@single.dropsummer.events\endcsname{705}
\expandafter\def\csname EERtiming@single.dropsummer.exposed_events\endcsname{255}
\expandafter\def\csname EERtiming@single.dropsummer.term_events\endcsname{255}
\expandafter\def\csname EERtiming@single.dropsummer.status\endcsname{converged}
\expandafter\def\csname EERtiming@single.birthm1.hr\endcsname{3.92}
\expandafter\def\csname EERtiming@single.birthm1.ci_lo\endcsname{3.35}
\expandafter\def\csname EERtiming@single.birthm1.ci_hi\endcsname{4.58}
\expandafter\def\csname EERtiming@single.birthm1.women\endcsname{2,828}
\expandafter\def\csname EERtiming@single.birthm1.months\endcsname{48,953}
\expandafter\def\csname EERtiming@single.birthm1.events\endcsname{711}
\expandafter\def\csname EERtiming@single.birthm1.exposed_events\endcsname{271}
\expandafter\def\csname EERtiming@single.birthm1.term_events\endcsname{271}
\expandafter\def\csname EERtiming@single.birthm1.status\endcsname{converged}
\expandafter\def\csname EERtiming@single.birthp1.hr\endcsname{3.81}
\expandafter\def\csname EERtiming@single.birthp1.ci_lo\endcsname{3.25}
\expandafter\def\csname EERtiming@single.birthp1.ci_hi\endcsname{4.47}
\expandafter\def\csname EERtiming@single.birthp1.women\endcsname{2,828}
\expandafter\def\csname EERtiming@single.birthp1.months\endcsname{48,953}
\expandafter\def\csname EERtiming@single.birthp1.events\endcsname{711}
\expandafter\def\csname EERtiming@single.birthp1.exposed_events\endcsname{252}
\expandafter\def\csname EERtiming@single.birthp1.term_events\endcsname{252}
\expandafter\def\csname EERtiming@single.birthp1.status\endcsname{converged}
\expandafter\def\csname EERtiming@single.intm1.hr\endcsname{3.89}
\expandafter\def\csname EERtiming@single.intm1.ci_lo\endcsname{3.32}
\expandafter\def\csname EERtiming@single.intm1.ci_hi\endcsname{4.55}
\expandafter\def\csname EERtiming@single.intm1.women\endcsname{2,828}
\expandafter\def\csname EERtiming@single.intm1.months\endcsname{48,953}
\expandafter\def\csname EERtiming@single.intm1.events\endcsname{711}
\expandafter\def\csname EERtiming@single.intm1.exposed_events\endcsname{266}
\expandafter\def\csname EERtiming@single.intm1.term_events\endcsname{266}
\expandafter\def\csname EERtiming@single.intm1.status\endcsname{converged}
\expandafter\def\csname EERtiming@single.intp1.hr\endcsname{3.85}
\expandafter\def\csname EERtiming@single.intp1.ci_lo\endcsname{3.29}
\expandafter\def\csname EERtiming@single.intp1.ci_hi\endcsname{4.52}
\expandafter\def\csname EERtiming@single.intp1.women\endcsname{2,828}
\expandafter\def\csname EERtiming@single.intp1.months\endcsname{48,953}
\expandafter\def\csname EERtiming@single.intp1.events\endcsname{711}
\expandafter\def\csname EERtiming@single.intp1.exposed_events\endcsname{258}
\expandafter\def\csname EERtiming@single.intp1.term_events\endcsname{258}
\expandafter\def\csname EERtiming@single.intp1.status\endcsname{converged}
\expandafter\def\csname EERtiming@joint.base.full.k1.hr\endcsname{3.91}
\expandafter\def\csname EERtiming@joint.base.full.k1.ci_lo\endcsname{3.30}
\expandafter\def\csname EERtiming@joint.base.full.k1.ci_hi\endcsname{4.64}
\expandafter\def\csname EERtiming@joint.base.full.k1.ll\endcsname{-6299.05}
\expandafter\def\csname EERtiming@joint.base.full.k1.status\endcsname{converged}
\expandafter\def\csname EERtiming@joint.base.full.k2.hr\endcsname{3.76}
\expandafter\def\csname EERtiming@joint.base.full.k2.ci_lo\endcsname{3.00}
\expandafter\def\csname EERtiming@joint.base.full.k2.ci_hi\endcsname{4.71}
\expandafter\def\csname EERtiming@joint.base.full.k2.ll\endcsname{-6281.43}
\expandafter\def\csname EERtiming@joint.base.full.k2.status\endcsname{converged}
\expandafter\def\csname EERtiming@joint.base.full.women\endcsname{2,828}
\expandafter\def\csname EERtiming@joint.base.full.months\endcsname{64,598}
\expandafter\def\csname EERtiming@joint.base.full.outrisk_months\endcsname{48,953}
\expandafter\def\csname EERtiming@joint.base.full.events\endcsname{711}
\expandafter\def\csname EERtiming@joint.base.full.exposed_onset_events\endcsname{198}
\expandafter\def\csname EERtiming@joint.base.full.exposed_absent_events\endcsname{63}
\expandafter\def\csname EERtiming@joint.base.full.pregnancy_events\endcsname{478}
\expandafter\def\csname EERtiming@joint.base.full.k2.rel_grad\endcsname{7.32e-08}
\expandafter\def\csname EERtiming@joint.base.full.k2.n_bfgs_conv\endcsname{7}
\expandafter\def\csname EERtiming@joint.base.full.k2.n_starts\endcsname{7}
\expandafter\def\csname EERtiming@joint.base.full.k2.clip_hits\endcsname{0}
\expandafter\def\csname EERtiming@joint.base.full.k2.min_share\endcsname{0.034}
\expandafter\def\csname EERtiming@joint.base.restricted.k1.hr\endcsname{3.92}
\expandafter\def\csname EERtiming@joint.base.restricted.k1.ci_lo\endcsname{3.31}
\expandafter\def\csname EERtiming@joint.base.restricted.k1.ci_hi\endcsname{4.65}
\expandafter\def\csname EERtiming@joint.base.restricted.k1.ll\endcsname{-6039.67}
\expandafter\def\csname EERtiming@joint.base.restricted.k1.status\endcsname{converged}
\expandafter\def\csname EERtiming@joint.base.restricted.k2.hr\endcsname{3.78}
\expandafter\def\csname EERtiming@joint.base.restricted.k2.ci_lo\endcsname{3.02}
\expandafter\def\csname EERtiming@joint.base.restricted.k2.ci_hi\endcsname{4.74}
\expandafter\def\csname EERtiming@joint.base.restricted.k2.ll\endcsname{-6022.07}
\expandafter\def\csname EERtiming@joint.base.restricted.k2.status\endcsname{converged}
\expandafter\def\csname EERtiming@joint.base.restricted.women\endcsname{2,708}
\expandafter\def\csname EERtiming@joint.base.restricted.months\endcsname{62,478}
\expandafter\def\csname EERtiming@joint.base.restricted.outrisk_months\endcsname{47,408}
\expandafter\def\csname EERtiming@joint.base.restricted.events\endcsname{648}
\expandafter\def\csname EERtiming@joint.base.restricted.exposed_onset_events\endcsname{198}
\expandafter\def\csname EERtiming@joint.base.restricted.exposed_absent_events\endcsname{0}
\expandafter\def\csname EERtiming@joint.base.restricted.pregnancy_events\endcsname{478}
\expandafter\def\csname EERtiming@joint.base.restricted.k2.rel_grad\endcsname{6.04e-08}
\expandafter\def\csname EERtiming@joint.base.restricted.k2.n_bfgs_conv\endcsname{6}
\expandafter\def\csname EERtiming@joint.base.restricted.k2.n_starts\endcsname{7}
\expandafter\def\csname EERtiming@joint.base.restricted.k2.clip_hits\endcsname{0}
\expandafter\def\csname EERtiming@joint.base.restricted.k2.min_share\endcsname{0.034}
\expandafter\def\csname EERtiming@joint.lag0.full.k1.hr\endcsname{4.01}
\expandafter\def\csname EERtiming@joint.lag0.full.k1.ci_lo\endcsname{3.40}
\expandafter\def\csname EERtiming@joint.lag0.full.k1.ci_hi\endcsname{4.74}
\expandafter\def\csname EERtiming@joint.lag0.full.k1.ll\endcsname{-6289.64}
\expandafter\def\csname EERtiming@joint.lag0.full.k1.status\endcsname{converged}
\expandafter\def\csname EERtiming@joint.lag0.full.k2.hr\endcsname{4.00}
\expandafter\def\csname EERtiming@joint.lag0.full.k2.ci_lo\endcsname{3.24}
\expandafter\def\csname EERtiming@joint.lag0.full.k2.ci_hi\endcsname{4.94}
\expandafter\def\csname EERtiming@joint.lag0.full.k2.ll\endcsname{-6272.16}
\expandafter\def\csname EERtiming@joint.lag0.full.k2.status\endcsname{converged}
\expandafter\def\csname EERtiming@joint.lag0.full.women\endcsname{2,828}
\expandafter\def\csname EERtiming@joint.lag0.full.months\endcsname{64,598}
\expandafter\def\csname EERtiming@joint.lag0.full.outrisk_months\endcsname{48,953}
\expandafter\def\csname EERtiming@joint.lag0.full.events\endcsname{711}
\expandafter\def\csname EERtiming@joint.lag0.full.exposed_onset_events\endcsname{213}
\expandafter\def\csname EERtiming@joint.lag0.full.exposed_absent_events\endcsname{63}
\expandafter\def\csname EERtiming@joint.lag0.full.pregnancy_events\endcsname{478}
\expandafter\def\csname EERtiming@joint.lag0.full.k2.rel_grad\endcsname{4.76e-08}
\expandafter\def\csname EERtiming@joint.lag0.full.k2.n_bfgs_conv\endcsname{7}
\expandafter\def\csname EERtiming@joint.lag0.full.k2.n_starts\endcsname{7}
\expandafter\def\csname EERtiming@joint.lag0.full.k2.clip_hits\endcsname{0}
\expandafter\def\csname EERtiming@joint.lag0.full.k2.min_share\endcsname{0.034}
\expandafter\def\csname EERtiming@joint.lag0.restricted.k1.hr\endcsname{4.02}
\expandafter\def\csname EERtiming@joint.lag0.restricted.k1.ci_lo\endcsname{3.41}
\expandafter\def\csname EERtiming@joint.lag0.restricted.k1.ci_hi\endcsname{4.76}
\expandafter\def\csname EERtiming@joint.lag0.restricted.k1.ll\endcsname{-6030.24}
\expandafter\def\csname EERtiming@joint.lag0.restricted.k1.status\endcsname{converged}
\expandafter\def\csname EERtiming@joint.lag0.restricted.k2.hr\endcsname{4.02}
\expandafter\def\csname EERtiming@joint.lag0.restricted.k2.ci_lo\endcsname{3.26}
\expandafter\def\csname EERtiming@joint.lag0.restricted.k2.ci_hi\endcsname{4.97}
\expandafter\def\csname EERtiming@joint.lag0.restricted.k2.ll\endcsname{-6012.76}
\expandafter\def\csname EERtiming@joint.lag0.restricted.k2.status\endcsname{converged}
\expandafter\def\csname EERtiming@joint.lag0.restricted.women\endcsname{2,708}
\expandafter\def\csname EERtiming@joint.lag0.restricted.months\endcsname{62,478}
\expandafter\def\csname EERtiming@joint.lag0.restricted.outrisk_months\endcsname{47,408}
\expandafter\def\csname EERtiming@joint.lag0.restricted.events\endcsname{648}
\expandafter\def\csname EERtiming@joint.lag0.restricted.exposed_onset_events\endcsname{213}
\expandafter\def\csname EERtiming@joint.lag0.restricted.exposed_absent_events\endcsname{0}
\expandafter\def\csname EERtiming@joint.lag0.restricted.pregnancy_events\endcsname{478}
\expandafter\def\csname EERtiming@joint.lag0.restricted.k2.rel_grad\endcsname{3.06e-07}
\expandafter\def\csname EERtiming@joint.lag0.restricted.k2.n_bfgs_conv\endcsname{6}
\expandafter\def\csname EERtiming@joint.lag0.restricted.k2.n_starts\endcsname{7}
\expandafter\def\csname EERtiming@joint.lag0.restricted.k2.clip_hits\endcsname{0}
\expandafter\def\csname EERtiming@joint.lag0.restricted.k2.min_share\endcsname{0.034}
\expandafter\def\csname EERtiming@joint.lag2.full.k1.hr\endcsname{3.90}
\expandafter\def\csname EERtiming@joint.lag2.full.k1.ci_lo\endcsname{3.28}
\expandafter\def\csname EERtiming@joint.lag2.full.k1.ci_hi\endcsname{4.64}
\expandafter\def\csname EERtiming@joint.lag2.full.k1.ll\endcsname{-6304.27}
\expandafter\def\csname EERtiming@joint.lag2.full.k1.status\endcsname{converged}
\expandafter\def\csname EERtiming@joint.lag2.full.k2.hr\endcsname{3.63}
\expandafter\def\csname EERtiming@joint.lag2.full.k2.ci_lo\endcsname{2.86}
\expandafter\def\csname EERtiming@joint.lag2.full.k2.ci_hi\endcsname{4.62}
\expandafter\def\csname EERtiming@joint.lag2.full.k2.ll\endcsname{-6286.41}
\expandafter\def\csname EERtiming@joint.lag2.full.k2.status\endcsname{converged}
\expandafter\def\csname EERtiming@joint.lag2.full.women\endcsname{2,828}
\expandafter\def\csname EERtiming@joint.lag2.full.months\endcsname{64,598}
\expandafter\def\csname EERtiming@joint.lag2.full.outrisk_months\endcsname{48,953}
\expandafter\def\csname EERtiming@joint.lag2.full.events\endcsname{711}
\expandafter\def\csname EERtiming@joint.lag2.full.exposed_onset_events\endcsname{186}
\expandafter\def\csname EERtiming@joint.lag2.full.exposed_absent_events\endcsname{63}
\expandafter\def\csname EERtiming@joint.lag2.full.pregnancy_events\endcsname{478}
\expandafter\def\csname EERtiming@joint.lag2.full.k2.rel_grad\endcsname{4.28e-08}
\expandafter\def\csname EERtiming@joint.lag2.full.k2.n_bfgs_conv\endcsname{7}
\expandafter\def\csname EERtiming@joint.lag2.full.k2.n_starts\endcsname{7}
\expandafter\def\csname EERtiming@joint.lag2.full.k2.clip_hits\endcsname{0}
\expandafter\def\csname EERtiming@joint.lag2.full.k2.min_share\endcsname{0.035}
\expandafter\def\csname EERtiming@joint.lag2.restricted.k1.hr\endcsname{3.91}
\expandafter\def\csname EERtiming@joint.lag2.restricted.k1.ci_lo\endcsname{3.29}
\expandafter\def\csname EERtiming@joint.lag2.restricted.k1.ci_hi\endcsname{4.65}
\expandafter\def\csname EERtiming@joint.lag2.restricted.k1.ll\endcsname{-6045.02}
\expandafter\def\csname EERtiming@joint.lag2.restricted.k1.status\endcsname{converged}
\expandafter\def\csname EERtiming@joint.lag2.restricted.k2.hr\endcsname{3.66}
\expandafter\def\csname EERtiming@joint.lag2.restricted.k2.ci_lo\endcsname{2.88}
\expandafter\def\csname EERtiming@joint.lag2.restricted.k2.ci_hi\endcsname{4.65}
\expandafter\def\csname EERtiming@joint.lag2.restricted.k2.ll\endcsname{-6027.20}
\expandafter\def\csname EERtiming@joint.lag2.restricted.k2.status\endcsname{converged}
\expandafter\def\csname EERtiming@joint.lag2.restricted.women\endcsname{2,708}
\expandafter\def\csname EERtiming@joint.lag2.restricted.months\endcsname{62,478}
\expandafter\def\csname EERtiming@joint.lag2.restricted.outrisk_months\endcsname{47,408}
\expandafter\def\csname EERtiming@joint.lag2.restricted.events\endcsname{648}
\expandafter\def\csname EERtiming@joint.lag2.restricted.exposed_onset_events\endcsname{186}
\expandafter\def\csname EERtiming@joint.lag2.restricted.exposed_absent_events\endcsname{0}
\expandafter\def\csname EERtiming@joint.lag2.restricted.pregnancy_events\endcsname{478}
\expandafter\def\csname EERtiming@joint.lag2.restricted.k2.rel_grad\endcsname{9.25e-08}
\expandafter\def\csname EERtiming@joint.lag2.restricted.k2.n_bfgs_conv\endcsname{7}
\expandafter\def\csname EERtiming@joint.lag2.restricted.k2.n_starts\endcsname{7}
\expandafter\def\csname EERtiming@joint.lag2.restricted.k2.clip_hits\endcsname{0}
\expandafter\def\csname EERtiming@joint.lag2.restricted.k2.min_share\endcsname{0.035}
\expandafter\def\csname EERtiming@joint.lag3.full.k1.hr\endcsname{3.89}
\expandafter\def\csname EERtiming@joint.lag3.full.k1.ci_lo\endcsname{3.26}
\expandafter\def\csname EERtiming@joint.lag3.full.k1.ci_hi\endcsname{4.64}
\expandafter\def\csname EERtiming@joint.lag3.full.k1.ll\endcsname{-6309.16}
\expandafter\def\csname EERtiming@joint.lag3.full.k1.status\endcsname{converged}
\expandafter\def\csname EERtiming@joint.lag3.full.k2.hr\endcsname{3.46}
\expandafter\def\csname EERtiming@joint.lag3.full.k2.ci_lo\endcsname{2.66}
\expandafter\def\csname EERtiming@joint.lag3.full.k2.ci_hi\endcsname{4.50}
\expandafter\def\csname EERtiming@joint.lag3.full.k2.ll\endcsname{-6290.86}
\expandafter\def\csname EERtiming@joint.lag3.full.k2.status\endcsname{converged}
\expandafter\def\csname EERtiming@joint.lag3.full.women\endcsname{2,828}
\expandafter\def\csname EERtiming@joint.lag3.full.months\endcsname{64,598}
\expandafter\def\csname EERtiming@joint.lag3.full.outrisk_months\endcsname{48,953}
\expandafter\def\csname EERtiming@joint.lag3.full.events\endcsname{711}
\expandafter\def\csname EERtiming@joint.lag3.full.exposed_onset_events\endcsname{175}
\expandafter\def\csname EERtiming@joint.lag3.full.exposed_absent_events\endcsname{63}
\expandafter\def\csname EERtiming@joint.lag3.full.pregnancy_events\endcsname{478}
\expandafter\def\csname EERtiming@joint.lag3.full.k2.rel_grad\endcsname{1.32e-07}
\expandafter\def\csname EERtiming@joint.lag3.full.k2.n_bfgs_conv\endcsname{7}
\expandafter\def\csname EERtiming@joint.lag3.full.k2.n_starts\endcsname{7}
\expandafter\def\csname EERtiming@joint.lag3.full.k2.clip_hits\endcsname{0}
\expandafter\def\csname EERtiming@joint.lag3.full.k2.min_share\endcsname{0.036}
\expandafter\def\csname EERtiming@joint.lag3.restricted.k1.hr\endcsname{3.90}
\expandafter\def\csname EERtiming@joint.lag3.restricted.k1.ci_lo\endcsname{3.26}
\expandafter\def\csname EERtiming@joint.lag3.restricted.k1.ci_hi\endcsname{4.65}
\expandafter\def\csname EERtiming@joint.lag3.restricted.k1.ll\endcsname{-6050.18}
\expandafter\def\csname EERtiming@joint.lag3.restricted.k1.status\endcsname{converged}
\expandafter\def\csname EERtiming@joint.lag3.restricted.k2.hr\endcsname{3.48}
\expandafter\def\csname EERtiming@joint.lag3.restricted.k2.ci_lo\endcsname{2.68}
\expandafter\def\csname EERtiming@joint.lag3.restricted.k2.ci_hi\endcsname{4.52}
\expandafter\def\csname EERtiming@joint.lag3.restricted.k2.ll\endcsname{-6031.94}
\expandafter\def\csname EERtiming@joint.lag3.restricted.k2.status\endcsname{converged}
\expandafter\def\csname EERtiming@joint.lag3.restricted.women\endcsname{2,708}
\expandafter\def\csname EERtiming@joint.lag3.restricted.months\endcsname{62,478}
\expandafter\def\csname EERtiming@joint.lag3.restricted.outrisk_months\endcsname{47,408}
\expandafter\def\csname EERtiming@joint.lag3.restricted.events\endcsname{648}
\expandafter\def\csname EERtiming@joint.lag3.restricted.exposed_onset_events\endcsname{175}
\expandafter\def\csname EERtiming@joint.lag3.restricted.exposed_absent_events\endcsname{0}
\expandafter\def\csname EERtiming@joint.lag3.restricted.pregnancy_events\endcsname{478}
\expandafter\def\csname EERtiming@joint.lag3.restricted.k2.rel_grad\endcsname{6.84e-08}
\expandafter\def\csname EERtiming@joint.lag3.restricted.k2.n_bfgs_conv\endcsname{7}
\expandafter\def\csname EERtiming@joint.lag3.restricted.k2.n_starts\endcsname{7}
\expandafter\def\csname EERtiming@joint.lag3.restricted.k2.clip_hits\endcsname{0}
\expandafter\def\csname EERtiming@joint.lag3.restricted.k2.min_share\endcsname{0.036}
\expandafter\def\csname EERtiming@joint.dropconf.full.k1.hr\endcsname{4.17}
\expandafter\def\csname EERtiming@joint.dropconf.full.k1.ci_lo\endcsname{3.51}
\expandafter\def\csname EERtiming@joint.dropconf.full.k1.ci_hi\endcsname{4.95}
\expandafter\def\csname EERtiming@joint.dropconf.full.k1.ll\endcsname{-6079.03}
\expandafter\def\csname EERtiming@joint.dropconf.full.k1.status\endcsname{converged}
\expandafter\def\csname EERtiming@joint.dropconf.full.k2.hr\endcsname{4.16}
\expandafter\def\csname EERtiming@joint.dropconf.full.k2.ci_lo\endcsname{3.33}
\expandafter\def\csname EERtiming@joint.dropconf.full.k2.ci_hi\endcsname{5.18}
\expandafter\def\csname EERtiming@joint.dropconf.full.k2.ll\endcsname{-6060.83}
\expandafter\def\csname EERtiming@joint.dropconf.full.k2.status\endcsname{converged}
\expandafter\def\csname EERtiming@joint.dropconf.full.women\endcsname{2,797}
\expandafter\def\csname EERtiming@joint.dropconf.full.months\endcsname{64,068}
\expandafter\def\csname EERtiming@joint.dropconf.full.outrisk_months\endcsname{48,559}
\expandafter\def\csname EERtiming@joint.dropconf.full.events\endcsname{680}
\expandafter\def\csname EERtiming@joint.dropconf.full.exposed_onset_events\endcsname{198}
\expandafter\def\csname EERtiming@joint.dropconf.full.exposed_absent_events\endcsname{63}
\expandafter\def\csname EERtiming@joint.dropconf.full.pregnancy_events\endcsname{463}
\expandafter\def\csname EERtiming@joint.dropconf.full.k2.rel_grad\endcsname{1.82e-07}
\expandafter\def\csname EERtiming@joint.dropconf.full.k2.n_bfgs_conv\endcsname{7}
\expandafter\def\csname EERtiming@joint.dropconf.full.k2.n_starts\endcsname{7}
\expandafter\def\csname EERtiming@joint.dropconf.full.k2.clip_hits\endcsname{0}
\expandafter\def\csname EERtiming@joint.dropconf.full.k2.min_share\endcsname{0.035}
\expandafter\def\csname EERtiming@joint.dropconf.restricted.k1.hr\endcsname{4.18}
\expandafter\def\csname EERtiming@joint.dropconf.restricted.k1.ci_lo\endcsname{3.52}
\expandafter\def\csname EERtiming@joint.dropconf.restricted.k1.ci_hi\endcsname{4.97}
\expandafter\def\csname EERtiming@joint.dropconf.restricted.k1.ll\endcsname{-5819.60}
\expandafter\def\csname EERtiming@joint.dropconf.restricted.k1.status\endcsname{converged}
\expandafter\def\csname EERtiming@joint.dropconf.restricted.k2.hr\endcsname{4.18}
\expandafter\def\csname EERtiming@joint.dropconf.restricted.k2.ci_lo\endcsname{3.35}
\expandafter\def\csname EERtiming@joint.dropconf.restricted.k2.ci_hi\endcsname{5.21}
\expandafter\def\csname EERtiming@joint.dropconf.restricted.k2.ll\endcsname{-5801.40}
\expandafter\def\csname EERtiming@joint.dropconf.restricted.k2.status\endcsname{converged}
\expandafter\def\csname EERtiming@joint.dropconf.restricted.women\endcsname{2,677}
\expandafter\def\csname EERtiming@joint.dropconf.restricted.months\endcsname{61,948}
\expandafter\def\csname EERtiming@joint.dropconf.restricted.outrisk_months\endcsname{47,014}
\expandafter\def\csname EERtiming@joint.dropconf.restricted.events\endcsname{617}
\expandafter\def\csname EERtiming@joint.dropconf.restricted.exposed_onset_events\endcsname{198}
\expandafter\def\csname EERtiming@joint.dropconf.restricted.exposed_absent_events\endcsname{0}
\expandafter\def\csname EERtiming@joint.dropconf.restricted.pregnancy_events\endcsname{463}
\expandafter\def\csname EERtiming@joint.dropconf.restricted.k2.rel_grad\endcsname{2.57e-08}
\expandafter\def\csname EERtiming@joint.dropconf.restricted.k2.n_bfgs_conv\endcsname{6}
\expandafter\def\csname EERtiming@joint.dropconf.restricted.k2.n_starts\endcsname{7}
\expandafter\def\csname EERtiming@joint.dropconf.restricted.k2.clip_hits\endcsname{0}
\expandafter\def\csname EERtiming@joint.dropconf.restricted.k2.min_share\endcsname{0.035}
\expandafter\def\csname EERtiming@joint.redate.full.k1.hr\endcsname{3.86}
\expandafter\def\csname EERtiming@joint.redate.full.k1.ci_lo\endcsname{3.25}
\expandafter\def\csname EERtiming@joint.redate.full.k1.ci_hi\endcsname{4.59}
\expandafter\def\csname EERtiming@joint.redate.full.k1.ll\endcsname{-6416.79}
\expandafter\def\csname EERtiming@joint.redate.full.k1.status\endcsname{converged}
\expandafter\def\csname EERtiming@joint.redate.full.k2.hr\endcsname{3.69}
\expandafter\def\csname EERtiming@joint.redate.full.k2.ci_lo\endcsname{2.94}
\expandafter\def\csname EERtiming@joint.redate.full.k2.ci_hi\endcsname{4.65}
\expandafter\def\csname EERtiming@joint.redate.full.k2.ll\endcsname{-6398.86}
\expandafter\def\csname EERtiming@joint.redate.full.k2.status\endcsname{converged}
\expandafter\def\csname EERtiming@joint.redate.full.women\endcsname{2,821}
\expandafter\def\csname EERtiming@joint.redate.full.months\endcsname{64,043}
\expandafter\def\csname EERtiming@joint.redate.full.outrisk_months\endcsname{64,043}
\expandafter\def\csname EERtiming@joint.redate.full.events\endcsname{704}
\expandafter\def\csname EERtiming@joint.redate.full.exposed_onset_events\endcsname{192}
\expandafter\def\csname EERtiming@joint.redate.full.exposed_absent_events\endcsname{63}
\expandafter\def\csname EERtiming@joint.redate.full.pregnancy_events\endcsname{471}
\expandafter\def\csname EERtiming@joint.redate.full.k2.rel_grad\endcsname{2.87e-07}
\expandafter\def\csname EERtiming@joint.redate.full.k2.n_bfgs_conv\endcsname{7}
\expandafter\def\csname EERtiming@joint.redate.full.k2.n_starts\endcsname{7}
\expandafter\def\csname EERtiming@joint.redate.full.k2.clip_hits\endcsname{0}
\expandafter\def\csname EERtiming@joint.redate.full.k2.min_share\endcsname{0.034}
\expandafter\def\csname EERtiming@joint.redate.restricted.k1.hr\endcsname{3.87}
\expandafter\def\csname EERtiming@joint.redate.restricted.k1.ci_lo\endcsname{3.26}
\expandafter\def\csname EERtiming@joint.redate.restricted.k1.ci_hi\endcsname{4.60}
\expandafter\def\csname EERtiming@joint.redate.restricted.k1.ll\endcsname{-6144.69}
\expandafter\def\csname EERtiming@joint.redate.restricted.k1.status\endcsname{converged}
\expandafter\def\csname EERtiming@joint.redate.restricted.k2.hr\endcsname{3.71}
\expandafter\def\csname EERtiming@joint.redate.restricted.k2.ci_lo\endcsname{2.95}
\expandafter\def\csname EERtiming@joint.redate.restricted.k2.ci_hi\endcsname{4.67}
\expandafter\def\csname EERtiming@joint.redate.restricted.k2.ll\endcsname{-6126.79}
\expandafter\def\csname EERtiming@joint.redate.restricted.k2.status\endcsname{converged}
\expandafter\def\csname EERtiming@joint.redate.restricted.women\endcsname{2,701}
\expandafter\def\csname EERtiming@joint.redate.restricted.months\endcsname{61,999}
\expandafter\def\csname EERtiming@joint.redate.restricted.outrisk_months\endcsname{61,999}
\expandafter\def\csname EERtiming@joint.redate.restricted.events\endcsname{641}
\expandafter\def\csname EERtiming@joint.redate.restricted.exposed_onset_events\endcsname{192}
\expandafter\def\csname EERtiming@joint.redate.restricted.exposed_absent_events\endcsname{0}
\expandafter\def\csname EERtiming@joint.redate.restricted.pregnancy_events\endcsname{471}
\expandafter\def\csname EERtiming@joint.redate.restricted.k2.rel_grad\endcsname{9.02e-08}
\expandafter\def\csname EERtiming@joint.redate.restricted.k2.n_bfgs_conv\endcsname{7}
\expandafter\def\csname EERtiming@joint.redate.restricted.k2.n_starts\endcsname{7}
\expandafter\def\csname EERtiming@joint.redate.restricted.k2.clip_hits\endcsname{0}
\expandafter\def\csname EERtiming@joint.redate.restricted.k2.min_share\endcsname{0.034}
\expandafter\def\csname EERtiming@single.range.min\endcsname{3.76}
\expandafter\def\csname EERtiming@single.range.minkey\endcsname{dropsummer}
\expandafter\def\csname EERtiming@single.range.max\endcsname{4.20}
\expandafter\def\csname EERtiming@single.range.maxkey\endcsname{omitm2to2}
\expandafter\def\csname EERtiming@single.range.rows\endcsname{16}
\expandafter\def\csname EERtiming@single.bins.wald.chi2\endcsname{6.66}
\expandafter\def\csname EERtiming@single.bins.wald.df\endcsname{3}
\expandafter\def\csname EERtiming@single.bins.wald.p\endcsname{0.084}
\expandafter\def\csname EERtiming@joint.k2range.min\endcsname{3.46}
\expandafter\def\csname EERtiming@joint.k2range.max\endcsname{4.18}
\expandafter\def\csname EERtiming@joint.k2range.rows\endcsname{12}
}{}
\IfFileExists{evidence/refinement/debut_rule/numbers_debut_rule.tex}{
\providecommand{\EERdebutrule}[1]{\ifcsname EERdebutrule@#1\endcsname\csname EERdebutrule@#1\endcsname\else\PackageError{EER debut rule}{Missing value #1}{Rerun debut_rule_sensitivity.do.}\fi}
\expandafter\def\csname EERdebutrule@changed.min\endcsname{36}
\expandafter\def\csname EERdebutrule@changed.max\endcsname{339}
\expandafter\def\csname EERdebutrule@left.min\endcsname{15}
\expandafter\def\csname EERdebutrule@left.max\endcsname{54}
\expandafter\def\csname EERdebutrule@base.women\endcsname{2,828}
\expandafter\def\csname EERdebutrule@alt.hr_min\endcsname{3.80}
\expandafter\def\csname EERdebutrule@alt.hr_max\endcsname{3.86}
\expandafter\def\csname EERdebutrule@base.hr\endcsname{3.85}
}{}
\IfFileExists{evidence/refinement/timing_split/numbers_timing_split.tex}{
\providecommand{\EERsplit}[1]{\csname EERsplit@#1\endcsname}
\expandafter\def\csname EERsplit@joint.gest.full.k1.m13.hr\endcsname{4.67}
\expandafter\def\csname EERsplit@joint.gest.full.k1.m13.lo\endcsname{3.44}
\expandafter\def\csname EERsplit@joint.gest.full.k1.m13.hi\endcsname{6.33}
\expandafter\def\csname EERsplit@joint.gest.full.k1.m13.ci\endcsname{[3.44, 6.33]}
\expandafter\def\csname EERsplit@joint.gest.full.k1.m13.events\endcsname{46}
\expandafter\def\csname EERsplit@joint.gest.full.k1.m49.hr\endcsname{3.78}
\expandafter\def\csname EERsplit@joint.gest.full.k1.m49.lo\endcsname{2.84}
\expandafter\def\csname EERsplit@joint.gest.full.k1.m49.hi\endcsname{5.03}
\expandafter\def\csname EERsplit@joint.gest.full.k1.m49.ci\endcsname{[2.84, 5.03]}
\expandafter\def\csname EERsplit@joint.gest.full.k1.m49.events\endcsname{53}
\expandafter\def\csname EERsplit@joint.gest.full.k1.m49.diffp\endcsname{0.29}
\expandafter\def\csname EERsplit@joint.gest.full.k1.m10p.hr\endcsname{3.88}
\expandafter\def\csname EERsplit@joint.gest.full.k1.m10p.lo\endcsname{3.10}
\expandafter\def\csname EERsplit@joint.gest.full.k1.m10p.hi\endcsname{4.86}
\expandafter\def\csname EERsplit@joint.gest.full.k1.m10p.ci\endcsname{[3.10, 4.86]}
\expandafter\def\csname EERsplit@joint.gest.full.k1.m10p.events\endcsname{99}
\expandafter\def\csname EERsplit@joint.gest.full.k1.m10p.diffp\endcsname{0.31}
\expandafter\def\csname EERsplit@joint.gest.full.k1.k0.hr\endcsname{4.01}
\expandafter\def\csname EERsplit@joint.gest.full.k1.k0.lo\endcsname{2.39}
\expandafter\def\csname EERsplit@joint.gest.full.k1.k0.hi\endcsname{6.70}
\expandafter\def\csname EERsplit@joint.gest.full.k1.k0.ci\endcsname{[2.39, 6.70]}
\expandafter\def\csname EERsplit@joint.gest.full.k1.k0.events\endcsname{15}
\expandafter\def\csname EERsplit@joint.gest.full.k1.expabsent.hr\endcsname{4.09}
\expandafter\def\csname EERsplit@joint.gest.full.k1.expabsent.lo\endcsname{3.09}
\expandafter\def\csname EERsplit@joint.gest.full.k1.expabsent.hi\endcsname{5.40}
\expandafter\def\csname EERsplit@joint.gest.full.k1.expabsent.ci\endcsname{[3.09, 5.40]}
\expandafter\def\csname EERsplit@joint.gest.full.k1.expabsent.events\endcsname{63}
\expandafter\def\csname EERsplit@joint.gest.full.k2.m13.hr\endcsname{4.62}
\expandafter\def\csname EERsplit@joint.gest.full.k2.m13.lo\endcsname{3.36}
\expandafter\def\csname EERsplit@joint.gest.full.k2.m13.hi\endcsname{6.36}
\expandafter\def\csname EERsplit@joint.gest.full.k2.m13.ci\endcsname{[3.36, 6.36]}
\expandafter\def\csname EERsplit@joint.gest.full.k2.m13.events\endcsname{46}
\expandafter\def\csname EERsplit@joint.gest.full.k2.m49.hr\endcsname{3.73}
\expandafter\def\csname EERsplit@joint.gest.full.k2.m49.lo\endcsname{2.74}
\expandafter\def\csname EERsplit@joint.gest.full.k2.m49.hi\endcsname{5.08}
\expandafter\def\csname EERsplit@joint.gest.full.k2.m49.ci\endcsname{[2.74, 5.08]}
\expandafter\def\csname EERsplit@joint.gest.full.k2.m49.events\endcsname{53}
\expandafter\def\csname EERsplit@joint.gest.full.k2.m49.diffp\endcsname{0.29}
\expandafter\def\csname EERsplit@joint.gest.full.k2.m10p.hr\endcsname{3.81}
\expandafter\def\csname EERsplit@joint.gest.full.k2.m10p.lo\endcsname{2.88}
\expandafter\def\csname EERsplit@joint.gest.full.k2.m10p.hi\endcsname{5.04}
\expandafter\def\csname EERsplit@joint.gest.full.k2.m10p.ci\endcsname{[2.88, 5.04]}
\expandafter\def\csname EERsplit@joint.gest.full.k2.m10p.events\endcsname{99}
\expandafter\def\csname EERsplit@joint.gest.full.k2.m10p.diffp\endcsname{0.30}
\expandafter\def\csname EERsplit@joint.gest.full.k2.k0.hr\endcsname{3.97}
\expandafter\def\csname EERsplit@joint.gest.full.k2.k0.lo\endcsname{2.36}
\expandafter\def\csname EERsplit@joint.gest.full.k2.k0.hi\endcsname{6.69}
\expandafter\def\csname EERsplit@joint.gest.full.k2.k0.ci\endcsname{[2.36, 6.69]}
\expandafter\def\csname EERsplit@joint.gest.full.k2.k0.events\endcsname{15}
\expandafter\def\csname EERsplit@joint.gest.full.k2.expabsent.hr\endcsname{4.08}
\expandafter\def\csname EERsplit@joint.gest.full.k2.expabsent.lo\endcsname{3.08}
\expandafter\def\csname EERsplit@joint.gest.full.k2.expabsent.hi\endcsname{5.40}
\expandafter\def\csname EERsplit@joint.gest.full.k2.expabsent.ci\endcsname{[3.08, 5.40]}
\expandafter\def\csname EERsplit@joint.gest.full.k2.expabsent.events\endcsname{63}
\expandafter\def\csname EERsplit@joint.gest.restricted.k1.m13.hr\endcsname{4.67}
\expandafter\def\csname EERsplit@joint.gest.restricted.k1.m13.lo\endcsname{3.45}
\expandafter\def\csname EERsplit@joint.gest.restricted.k1.m13.hi\endcsname{6.33}
\expandafter\def\csname EERsplit@joint.gest.restricted.k1.m13.ci\endcsname{[3.45, 6.33]}
\expandafter\def\csname EERsplit@joint.gest.restricted.k1.m13.events\endcsname{46}
\expandafter\def\csname EERsplit@joint.gest.restricted.k1.m49.hr\endcsname{3.79}
\expandafter\def\csname EERsplit@joint.gest.restricted.k1.m49.lo\endcsname{2.85}
\expandafter\def\csname EERsplit@joint.gest.restricted.k1.m49.hi\endcsname{5.04}
\expandafter\def\csname EERsplit@joint.gest.restricted.k1.m49.ci\endcsname{[2.85, 5.04]}
\expandafter\def\csname EERsplit@joint.gest.restricted.k1.m49.events\endcsname{53}
\expandafter\def\csname EERsplit@joint.gest.restricted.k1.m49.diffp\endcsname{0.30}
\expandafter\def\csname EERsplit@joint.gest.restricted.k1.m10p.hr\endcsname{3.89}
\expandafter\def\csname EERsplit@joint.gest.restricted.k1.m10p.lo\endcsname{3.11}
\expandafter\def\csname EERsplit@joint.gest.restricted.k1.m10p.hi\endcsname{4.88}
\expandafter\def\csname EERsplit@joint.gest.restricted.k1.m10p.ci\endcsname{[3.11, 4.88]}
\expandafter\def\csname EERsplit@joint.gest.restricted.k1.m10p.events\endcsname{99}
\expandafter\def\csname EERsplit@joint.gest.restricted.k1.m10p.diffp\endcsname{0.31}
\expandafter\def\csname EERsplit@joint.gest.restricted.k1.k0.hr\endcsname{4.01}
\expandafter\def\csname EERsplit@joint.gest.restricted.k1.k0.lo\endcsname{2.40}
\expandafter\def\csname EERsplit@joint.gest.restricted.k1.k0.hi\endcsname{6.71}
\expandafter\def\csname EERsplit@joint.gest.restricted.k1.k0.ci\endcsname{[2.40, 6.71]}
\expandafter\def\csname EERsplit@joint.gest.restricted.k1.k0.events\endcsname{15}
\expandafter\def\csname EERsplit@joint.gest.restricted.k2.m13.hr\endcsname{4.64}
\expandafter\def\csname EERsplit@joint.gest.restricted.k2.m13.lo\endcsname{3.37}
\expandafter\def\csname EERsplit@joint.gest.restricted.k2.m13.hi\endcsname{6.38}
\expandafter\def\csname EERsplit@joint.gest.restricted.k2.m13.ci\endcsname{[3.37, 6.38]}
\expandafter\def\csname EERsplit@joint.gest.restricted.k2.m13.events\endcsname{46}
\expandafter\def\csname EERsplit@joint.gest.restricted.k2.m49.hr\endcsname{3.75}
\expandafter\def\csname EERsplit@joint.gest.restricted.k2.m49.lo\endcsname{2.76}
\expandafter\def\csname EERsplit@joint.gest.restricted.k2.m49.hi\endcsname{5.10}
\expandafter\def\csname EERsplit@joint.gest.restricted.k2.m49.ci\endcsname{[2.76, 5.10]}
\expandafter\def\csname EERsplit@joint.gest.restricted.k2.m49.events\endcsname{53}
\expandafter\def\csname EERsplit@joint.gest.restricted.k2.m49.diffp\endcsname{0.29}
\expandafter\def\csname EERsplit@joint.gest.restricted.k2.m10p.hr\endcsname{3.84}
\expandafter\def\csname EERsplit@joint.gest.restricted.k2.m10p.lo\endcsname{2.90}
\expandafter\def\csname EERsplit@joint.gest.restricted.k2.m10p.hi\endcsname{5.09}
\expandafter\def\csname EERsplit@joint.gest.restricted.k2.m10p.ci\endcsname{[2.90, 5.09]}
\expandafter\def\csname EERsplit@joint.gest.restricted.k2.m10p.events\endcsname{99}
\expandafter\def\csname EERsplit@joint.gest.restricted.k2.m10p.diffp\endcsname{0.31}
\expandafter\def\csname EERsplit@joint.gest.restricted.k2.k0.hr\endcsname{3.98}
\expandafter\def\csname EERsplit@joint.gest.restricted.k2.k0.lo\endcsname{2.36}
\expandafter\def\csname EERsplit@joint.gest.restricted.k2.k0.hi\endcsname{6.71}
\expandafter\def\csname EERsplit@joint.gest.restricted.k2.k0.ci\endcsname{[2.36, 6.71]}
\expandafter\def\csname EERsplit@joint.gest.restricted.k2.k0.events\endcsname{15}
\expandafter\def\csname EERsplit@joint.profile.full.k1.m1.hr\endcsname{3.28}
\expandafter\def\csname EERsplit@joint.profile.full.k1.m1.lo\endcsname{1.85}
\expandafter\def\csname EERsplit@joint.profile.full.k1.m1.hi\endcsname{5.82}
\expandafter\def\csname EERsplit@joint.profile.full.k1.m1.ci\endcsname{[1.85, 5.82]}
\expandafter\def\csname EERsplit@joint.profile.full.k1.m1.events\endcsname{12}
\expandafter\def\csname EERsplit@joint.profile.full.k1.m23.hr\endcsname{5.49}
\expandafter\def\csname EERsplit@joint.profile.full.k1.m23.lo\endcsname{3.87}
\expandafter\def\csname EERsplit@joint.profile.full.k1.m23.hi\endcsname{7.78}
\expandafter\def\csname EERsplit@joint.profile.full.k1.m23.ci\endcsname{[3.87, 7.78]}
\expandafter\def\csname EERsplit@joint.profile.full.k1.m23.events\endcsname{34}
\expandafter\def\csname EERsplit@joint.profile.full.k1.m23.diffp\endcsname{0.12}
\expandafter\def\csname EERsplit@joint.profile.full.k1.m46.hr\endcsname{2.91}
\expandafter\def\csname EERsplit@joint.profile.full.k1.m46.lo\endcsname{1.90}
\expandafter\def\csname EERsplit@joint.profile.full.k1.m46.hi\endcsname{4.47}
\expandafter\def\csname EERsplit@joint.profile.full.k1.m46.ci\endcsname{[1.90, 4.47]}
\expandafter\def\csname EERsplit@joint.profile.full.k1.m46.events\endcsname{22}
\expandafter\def\csname EERsplit@joint.profile.full.k1.m46.diffp\endcsname{0.74}
\expandafter\def\csname EERsplit@joint.profile.full.k1.m79.hr\endcsname{4.80}
\expandafter\def\csname EERsplit@joint.profile.full.k1.m79.lo\endcsname{3.33}
\expandafter\def\csname EERsplit@joint.profile.full.k1.m79.hi\endcsname{6.92}
\expandafter\def\csname EERsplit@joint.profile.full.k1.m79.ci\endcsname{[3.33, 6.92]}
\expandafter\def\csname EERsplit@joint.profile.full.k1.m79.events\endcsname{31}
\expandafter\def\csname EERsplit@joint.profile.full.k1.m79.diffp\endcsname{0.26}
\expandafter\def\csname EERsplit@joint.profile.full.k1.m1015.hr\endcsname{3.62}
\expandafter\def\csname EERsplit@joint.profile.full.k1.m1015.lo\endcsname{2.57}
\expandafter\def\csname EERsplit@joint.profile.full.k1.m1015.hi\endcsname{5.09}
\expandafter\def\csname EERsplit@joint.profile.full.k1.m1015.ci\endcsname{[2.57, 5.09]}
\expandafter\def\csname EERsplit@joint.profile.full.k1.m1015.events\endcsname{36}
\expandafter\def\csname EERsplit@joint.profile.full.k1.m1015.diffp\endcsname{0.77}
\expandafter\def\csname EERsplit@joint.profile.full.k1.m16p.hr\endcsname{4.06}
\expandafter\def\csname EERsplit@joint.profile.full.k1.m16p.lo\endcsname{3.09}
\expandafter\def\csname EERsplit@joint.profile.full.k1.m16p.hi\endcsname{5.33}
\expandafter\def\csname EERsplit@joint.profile.full.k1.m16p.ci\endcsname{[3.09, 5.33]}
\expandafter\def\csname EERsplit@joint.profile.full.k1.m16p.events\endcsname{63}
\expandafter\def\csname EERsplit@joint.profile.full.k1.m16p.diffp\endcsname{0.50}
\expandafter\def\csname EERsplit@joint.profile.full.k1.k0.hr\endcsname{4.01}
\expandafter\def\csname EERsplit@joint.profile.full.k1.k0.lo\endcsname{2.39}
\expandafter\def\csname EERsplit@joint.profile.full.k1.k0.hi\endcsname{6.70}
\expandafter\def\csname EERsplit@joint.profile.full.k1.k0.ci\endcsname{[2.39, 6.70]}
\expandafter\def\csname EERsplit@joint.profile.full.k1.k0.events\endcsname{15}
\expandafter\def\csname EERsplit@joint.profile.full.k1.expabsent.hr\endcsname{4.08}
\expandafter\def\csname EERsplit@joint.profile.full.k1.expabsent.lo\endcsname{3.08}
\expandafter\def\csname EERsplit@joint.profile.full.k1.expabsent.hi\endcsname{5.40}
\expandafter\def\csname EERsplit@joint.profile.full.k1.expabsent.ci\endcsname{[3.08, 5.40]}
\expandafter\def\csname EERsplit@joint.profile.full.k1.expabsent.events\endcsname{63}
\expandafter\def\csname EERsplit@joint.k0only.full.k1.all.hr\endcsname{4.02}
\expandafter\def\csname EERsplit@joint.k0only.full.k1.all.lo\endcsname{3.38}
\expandafter\def\csname EERsplit@joint.k0only.full.k1.all.hi\endcsname{4.77}
\expandafter\def\csname EERsplit@joint.k0only.full.k1.all.ci\endcsname{[3.38, 4.77]}
\expandafter\def\csname EERsplit@joint.k0only.full.k1.all.events\endcsname{198}
\expandafter\def\csname EERsplit@joint.k0only.full.k1.k0.hr\endcsname{4.01}
\expandafter\def\csname EERsplit@joint.k0only.full.k1.k0.lo\endcsname{2.39}
\expandafter\def\csname EERsplit@joint.k0only.full.k1.k0.hi\endcsname{6.70}
\expandafter\def\csname EERsplit@joint.k0only.full.k1.k0.ci\endcsname{[2.39, 6.70]}
\expandafter\def\csname EERsplit@joint.k0only.full.k1.k0.events\endcsname{15}
\expandafter\def\csname EERsplit@joint.k0only.full.k1.expabsent.hr\endcsname{4.08}
\expandafter\def\csname EERsplit@joint.k0only.full.k1.expabsent.lo\endcsname{3.08}
\expandafter\def\csname EERsplit@joint.k0only.full.k1.expabsent.hi\endcsname{5.39}
\expandafter\def\csname EERsplit@joint.k0only.full.k1.expabsent.ci\endcsname{[3.08, 5.39]}
\expandafter\def\csname EERsplit@joint.k0only.full.k1.expabsent.events\endcsname{63}
\expandafter\def\csname EERsplit@joint.k0only.full.k2.all.hr\endcsname{4.00}
\expandafter\def\csname EERsplit@joint.k0only.full.k2.all.lo\endcsname{3.22}
\expandafter\def\csname EERsplit@joint.k0only.full.k2.all.hi\endcsname{4.98}
\expandafter\def\csname EERsplit@joint.k0only.full.k2.all.ci\endcsname{[3.22, 4.98]}
\expandafter\def\csname EERsplit@joint.k0only.full.k2.all.events\endcsname{198}
\expandafter\def\csname EERsplit@joint.k0only.full.k2.k0.hr\endcsname{4.00}
\expandafter\def\csname EERsplit@joint.k0only.full.k2.k0.lo\endcsname{2.37}
\expandafter\def\csname EERsplit@joint.k0only.full.k2.k0.hi\endcsname{6.73}
\expandafter\def\csname EERsplit@joint.k0only.full.k2.k0.ci\endcsname{[2.37, 6.73]}
\expandafter\def\csname EERsplit@joint.k0only.full.k2.k0.events\endcsname{15}
\expandafter\def\csname EERsplit@joint.k0only.full.k2.expabsent.hr\endcsname{4.08}
\expandafter\def\csname EERsplit@joint.k0only.full.k2.expabsent.lo\endcsname{3.08}
\expandafter\def\csname EERsplit@joint.k0only.full.k2.expabsent.hi\endcsname{5.39}
\expandafter\def\csname EERsplit@joint.k0only.full.k2.expabsent.ci\endcsname{[3.08, 5.39]}
\expandafter\def\csname EERsplit@joint.k0only.full.k2.expabsent.events\endcsname{63}
\expandafter\def\csname EERsplit@joint.k0only.restricted.k1.all.hr\endcsname{4.03}
\expandafter\def\csname EERsplit@joint.k0only.restricted.k1.all.lo\endcsname{3.39}
\expandafter\def\csname EERsplit@joint.k0only.restricted.k1.all.hi\endcsname{4.78}
\expandafter\def\csname EERsplit@joint.k0only.restricted.k1.all.ci\endcsname{[3.39, 4.78]}
\expandafter\def\csname EERsplit@joint.k0only.restricted.k1.all.events\endcsname{198}
\expandafter\def\csname EERsplit@joint.k0only.restricted.k1.k0.hr\endcsname{4.01}
\expandafter\def\csname EERsplit@joint.k0only.restricted.k1.k0.lo\endcsname{2.40}
\expandafter\def\csname EERsplit@joint.k0only.restricted.k1.k0.hi\endcsname{6.72}
\expandafter\def\csname EERsplit@joint.k0only.restricted.k1.k0.ci\endcsname{[2.40, 6.72]}
\expandafter\def\csname EERsplit@joint.k0only.restricted.k1.k0.events\endcsname{15}
\expandafter\def\csname EERsplit@joint.k0only.restricted.k2.all.hr\endcsname{4.02}
\expandafter\def\csname EERsplit@joint.k0only.restricted.k2.all.lo\endcsname{3.24}
\expandafter\def\csname EERsplit@joint.k0only.restricted.k2.all.hi\endcsname{5.01}
\expandafter\def\csname EERsplit@joint.k0only.restricted.k2.all.ci\endcsname{[3.24, 5.01]}
\expandafter\def\csname EERsplit@joint.k0only.restricted.k2.all.events\endcsname{198}
\expandafter\def\csname EERsplit@joint.k0only.restricted.k2.k0.hr\endcsname{4.01}
\expandafter\def\csname EERsplit@joint.k0only.restricted.k2.k0.lo\endcsname{2.38}
\expandafter\def\csname EERsplit@joint.k0only.restricted.k2.k0.hi\endcsname{6.76}
\expandafter\def\csname EERsplit@joint.k0only.restricted.k2.k0.ci\endcsname{[2.38, 6.76]}
\expandafter\def\csname EERsplit@joint.k0only.restricted.k2.k0.events\endcsname{15}
\expandafter\def\csname EERsplit@joint.gest.full.k1.status\endcsname{converged}
\expandafter\def\csname EERsplit@joint.gest.full.k1.ll\endcsname{-6288.99}
\expandafter\def\csname EERsplit@joint.gest.full.k1.lr\endcsname{20.1}
\expandafter\def\csname EERsplit@joint.gest.full.k1.lrdf\endcsname{3}
\expandafter\def\csname EERsplit@joint.gest.full.k1.lrp\endcsname{0.0002}
\expandafter\def\csname EERsplit@joint.gest.full.k1.waldchi2\endcsname{1.34}
\expandafter\def\csname EERsplit@joint.gest.full.k1.walddf\endcsname{2}
\expandafter\def\csname EERsplit@joint.gest.full.k1.waldp\endcsname{0.51}
\expandafter\def\csname EERsplit@joint.gest.full.k1.women\endcsname{2,828}
\expandafter\def\csname EERsplit@joint.gest.full.k1.months\endcsname{64,598}
\expandafter\def\csname EERsplit@joint.gest.full.k2.status\endcsname{converged}
\expandafter\def\csname EERsplit@joint.gest.full.k2.ll\endcsname{-6271.49}
\expandafter\def\csname EERsplit@joint.gest.full.k2.lr\endcsname{19.9}
\expandafter\def\csname EERsplit@joint.gest.full.k2.lrdf\endcsname{3}
\expandafter\def\csname EERsplit@joint.gest.full.k2.lrp\endcsname{0.0002}
\expandafter\def\csname EERsplit@joint.gest.full.k2.waldchi2\endcsname{1.38}
\expandafter\def\csname EERsplit@joint.gest.full.k2.walddf\endcsname{2}
\expandafter\def\csname EERsplit@joint.gest.full.k2.waldp\endcsname{0.50}
\expandafter\def\csname EERsplit@joint.gest.full.k2.women\endcsname{2,828}
\expandafter\def\csname EERsplit@joint.gest.full.k2.months\endcsname{64,598}
\expandafter\def\csname EERsplit@joint.gest.restricted.k1.status\endcsname{converged}
\expandafter\def\csname EERsplit@joint.gest.restricted.k1.ll\endcsname{-6029.60}
\expandafter\def\csname EERsplit@joint.gest.restricted.k1.lr\endcsname{20.1}
\expandafter\def\csname EERsplit@joint.gest.restricted.k1.lrdf\endcsname{3}
\expandafter\def\csname EERsplit@joint.gest.restricted.k1.lrp\endcsname{0.0002}
\expandafter\def\csname EERsplit@joint.gest.restricted.k1.waldchi2\endcsname{1.32}
\expandafter\def\csname EERsplit@joint.gest.restricted.k1.walddf\endcsname{2}
\expandafter\def\csname EERsplit@joint.gest.restricted.k1.waldp\endcsname{0.52}
\expandafter\def\csname EERsplit@joint.gest.restricted.k1.women\endcsname{2,708}
\expandafter\def\csname EERsplit@joint.gest.restricted.k1.months\endcsname{62,478}
\expandafter\def\csname EERsplit@joint.gest.restricted.k2.status\endcsname{converged}
\expandafter\def\csname EERsplit@joint.gest.restricted.k2.ll\endcsname{-6012.10}
\expandafter\def\csname EERsplit@joint.gest.restricted.k2.lr\endcsname{19.9}
\expandafter\def\csname EERsplit@joint.gest.restricted.k2.lrdf\endcsname{3}
\expandafter\def\csname EERsplit@joint.gest.restricted.k2.lrp\endcsname{0.0002}
\expandafter\def\csname EERsplit@joint.gest.restricted.k2.waldchi2\endcsname{1.35}
\expandafter\def\csname EERsplit@joint.gest.restricted.k2.walddf\endcsname{2}
\expandafter\def\csname EERsplit@joint.gest.restricted.k2.waldp\endcsname{0.51}
\expandafter\def\csname EERsplit@joint.gest.restricted.k2.women\endcsname{2,708}
\expandafter\def\csname EERsplit@joint.gest.restricted.k2.months\endcsname{62,478}
\expandafter\def\csname EERsplit@joint.k0only.full.k1.status\endcsname{converged}
\expandafter\def\csname EERsplit@joint.k0only.full.k1.ll\endcsname{-6289.64}
\expandafter\def\csname EERsplit@joint.k0only.full.k1.lr\endcsname{18.8}
\expandafter\def\csname EERsplit@joint.k0only.full.k1.lrdf\endcsname{1}
\expandafter\def\csname EERsplit@joint.k0only.full.k1.lrp\endcsname{0.0000}
\expandafter\def\csname EERsplit@joint.k0only.full.k1.women\endcsname{2,828}
\expandafter\def\csname EERsplit@joint.k0only.full.k1.months\endcsname{64,598}
\expandafter\def\csname EERsplit@joint.k0only.full.k2.status\endcsname{converged}
\expandafter\def\csname EERsplit@joint.k0only.full.k2.ll\endcsname{-6272.16}
\expandafter\def\csname EERsplit@joint.k0only.full.k2.lr\endcsname{18.5}
\expandafter\def\csname EERsplit@joint.k0only.full.k2.lrdf\endcsname{1}
\expandafter\def\csname EERsplit@joint.k0only.full.k2.lrp\endcsname{0.0000}
\expandafter\def\csname EERsplit@joint.k0only.full.k2.women\endcsname{2,828}
\expandafter\def\csname EERsplit@joint.k0only.full.k2.months\endcsname{64,598}
\expandafter\def\csname EERsplit@joint.k0only.restricted.k1.status\endcsname{converged}
\expandafter\def\csname EERsplit@joint.k0only.restricted.k1.ll\endcsname{-6030.24}
\expandafter\def\csname EERsplit@joint.k0only.restricted.k1.lr\endcsname{18.9}
\expandafter\def\csname EERsplit@joint.k0only.restricted.k1.lrdf\endcsname{1}
\expandafter\def\csname EERsplit@joint.k0only.restricted.k1.lrp\endcsname{0.0000}
\expandafter\def\csname EERsplit@joint.k0only.restricted.k1.women\endcsname{2,708}
\expandafter\def\csname EERsplit@joint.k0only.restricted.k1.months\endcsname{62,478}
\expandafter\def\csname EERsplit@joint.k0only.restricted.k2.status\endcsname{converged}
\expandafter\def\csname EERsplit@joint.k0only.restricted.k2.ll\endcsname{-6012.76}
\expandafter\def\csname EERsplit@joint.k0only.restricted.k2.lr\endcsname{18.6}
\expandafter\def\csname EERsplit@joint.k0only.restricted.k2.lrdf\endcsname{1}
\expandafter\def\csname EERsplit@joint.k0only.restricted.k2.lrp\endcsname{0.0000}
\expandafter\def\csname EERsplit@joint.k0only.restricted.k2.women\endcsname{2,708}
\expandafter\def\csname EERsplit@joint.k0only.restricted.k2.months\endcsname{62,478}
\expandafter\def\csname EERsplit@joint.profile.full.k1.status\endcsname{converged}
\expandafter\def\csname EERsplit@joint.profile.full.k1.ll\endcsname{-6285.93}
\expandafter\def\csname EERsplit@joint.profile.full.k1.lr\endcsname{26.3}
\expandafter\def\csname EERsplit@joint.profile.full.k1.lrdf\endcsname{6}
\expandafter\def\csname EERsplit@joint.profile.full.k1.lrp\endcsname{0.0002}
\expandafter\def\csname EERsplit@joint.profile.full.k1.waldchi2\endcsname{7.41}
\expandafter\def\csname EERsplit@joint.profile.full.k1.walddf\endcsname{5}
\expandafter\def\csname EERsplit@joint.profile.full.k1.waldp\endcsname{0.19}
\expandafter\def\csname EERsplit@joint.profile.full.k1.women\endcsname{2,828}
\expandafter\def\csname EERsplit@joint.profile.full.k1.months\endcsname{64,598}
\expandafter\def\csname EERsplit@absence.ref.women\endcsname{2,676}
\expandafter\def\csname EERsplit@absence.ref.months\endcsname{42,650}
\expandafter\def\csname EERsplit@absence.ref.starts\endcsname{584}
\expandafter\def\csname EERsplit@absence.ref.interruptions\endcsname{435}
\expandafter\def\csname EERsplit@absence.ref.startpct\endcsname{1.4}
\expandafter\def\csname EERsplit@absence.ref.sharepct\endcsname{74}
\expandafter\def\csname EERsplit@absence.ref.intpct\endcsname{1.0}
\expandafter\def\csname EERsplit@absence.ref.summerstarts\endcsname{187}
\expandafter\def\csname EERsplit@absence.ref.credential\endcsname{28}
\expandafter\def\csname EERsplit@absence.k0.women\endcsname{364}
\expandafter\def\csname EERsplit@absence.k0.months\endcsname{364}
\expandafter\def\csname EERsplit@absence.k0.starts\endcsname{17}
\expandafter\def\csname EERsplit@absence.k0.interruptions\endcsname{15}
\expandafter\def\csname EERsplit@absence.k0.startpct\endcsname{4.7}
\expandafter\def\csname EERsplit@absence.k0.sharepct\endcsname{88}
\expandafter\def\csname EERsplit@absence.k0.intpct\endcsname{4.1}
\expandafter\def\csname EERsplit@absence.k0.startratio\endcsname{3.4}
\expandafter\def\csname EERsplit@absence.k0.startlo\endcsname{2.1}
\expandafter\def\csname EERsplit@absence.k0.starthi\endcsname{5.5}
\expandafter\def\csname EERsplit@absence.k0.startci\endcsname{[2.1, 5.5]}
\expandafter\def\csname EERsplit@absence.k0.shareratio\endcsname{1.18}
\expandafter\def\csname EERsplit@absence.k0.sharelo\endcsname{0.99}
\expandafter\def\csname EERsplit@absence.k0.sharehi\endcsname{1.42}
\expandafter\def\csname EERsplit@absence.k0.shareci\endcsname{[0.99, 1.42]}
\expandafter\def\csname EERsplit@absence.k0.intratio\endcsname{4.0}
\expandafter\def\csname EERsplit@absence.k0.summerstarts\endcsname{6}
\expandafter\def\csname EERsplit@absence.k0.credential\endcsname{0}
\expandafter\def\csname EERsplit@absence.k0.startratiops\endcsname{3.2}
\expandafter\def\csname EERsplit@absence.m13.women\endcsname{426}
\expandafter\def\csname EERsplit@absence.m13.months\endcsname{962}
\expandafter\def\csname EERsplit@absence.m13.starts\endcsname{54}
\expandafter\def\csname EERsplit@absence.m13.interruptions\endcsname{46}
\expandafter\def\csname EERsplit@absence.m13.startpct\endcsname{5.6}
\expandafter\def\csname EERsplit@absence.m13.sharepct\endcsname{85}
\expandafter\def\csname EERsplit@absence.m13.intpct\endcsname{4.8}
\expandafter\def\csname EERsplit@absence.m13.startratio\endcsname{4.1}
\expandafter\def\csname EERsplit@absence.m13.startlo\endcsname{3.1}
\expandafter\def\csname EERsplit@absence.m13.starthi\endcsname{5.4}
\expandafter\def\csname EERsplit@absence.m13.startci\endcsname{[3.1, 5.4]}
\expandafter\def\csname EERsplit@absence.m13.shareratio\endcsname{1.14}
\expandafter\def\csname EERsplit@absence.m13.sharelo\endcsname{1.01}
\expandafter\def\csname EERsplit@absence.m13.sharehi\endcsname{1.29}
\expandafter\def\csname EERsplit@absence.m13.shareci\endcsname{[1.01, 1.29]}
\expandafter\def\csname EERsplit@absence.m13.intratio\endcsname{4.7}
\expandafter\def\csname EERsplit@absence.m13.summerstarts\endcsname{11}
\expandafter\def\csname EERsplit@absence.m13.credential\endcsname{3}
\expandafter\def\csname EERsplit@absence.m13.startratiops\endcsname{4.8}
\expandafter\def\csname EERsplit@absence.m49.women\endcsname{335}
\expandafter\def\csname EERsplit@absence.m49.months\endcsname{1,315}
\expandafter\def\csname EERsplit@absence.m49.starts\endcsname{65}
\expandafter\def\csname EERsplit@absence.m49.interruptions\endcsname{53}
\expandafter\def\csname EERsplit@absence.m49.startpct\endcsname{4.9}
\expandafter\def\csname EERsplit@absence.m49.sharepct\endcsname{82}
\expandafter\def\csname EERsplit@absence.m49.intpct\endcsname{4.0}
\expandafter\def\csname EERsplit@absence.m49.startratio\endcsname{3.6}
\expandafter\def\csname EERsplit@absence.m49.startlo\endcsname{2.8}
\expandafter\def\csname EERsplit@absence.m49.starthi\endcsname{4.6}
\expandafter\def\csname EERsplit@absence.m49.startci\endcsname{[2.8, 4.6]}
\expandafter\def\csname EERsplit@absence.m49.shareratio\endcsname{1.09}
\expandafter\def\csname EERsplit@absence.m49.sharelo\endcsname{0.97}
\expandafter\def\csname EERsplit@absence.m49.sharehi\endcsname{1.24}
\expandafter\def\csname EERsplit@absence.m49.shareci\endcsname{[0.97, 1.24]}
\expandafter\def\csname EERsplit@absence.m49.intratio\endcsname{4.0}
\expandafter\def\csname EERsplit@absence.m49.summerstarts\endcsname{20}
\expandafter\def\csname EERsplit@absence.m49.credential\endcsname{1}
\expandafter\def\csname EERsplit@absence.m49.startratiops\endcsname{3.7}
\expandafter\def\csname EERsplit@absence.m10p.women\endcsname{220}
\expandafter\def\csname EERsplit@absence.m10p.months\endcsname{2,117}
\expandafter\def\csname EERsplit@absence.m10p.starts\endcsname{120}
\expandafter\def\csname EERsplit@absence.m10p.interruptions\endcsname{99}
\expandafter\def\csname EERsplit@absence.m10p.startpct\endcsname{5.7}
\expandafter\def\csname EERsplit@absence.m10p.sharepct\endcsname{82}
\expandafter\def\csname EERsplit@absence.m10p.intpct\endcsname{4.7}
\expandafter\def\csname EERsplit@absence.m10p.startratio\endcsname{4.1}
\expandafter\def\csname EERsplit@absence.m10p.startlo\endcsname{3.4}
\expandafter\def\csname EERsplit@absence.m10p.starthi\endcsname{5.0}
\expandafter\def\csname EERsplit@absence.m10p.startci\endcsname{[3.4, 5.0]}
\expandafter\def\csname EERsplit@absence.m10p.shareratio\endcsname{1.11}
\expandafter\def\csname EERsplit@absence.m10p.sharelo\endcsname{1.01}
\expandafter\def\csname EERsplit@absence.m10p.sharehi\endcsname{1.22}
\expandafter\def\csname EERsplit@absence.m10p.shareci\endcsname{[1.01, 1.22]}
\expandafter\def\csname EERsplit@absence.m10p.intratio\endcsname{4.6}
\expandafter\def\csname EERsplit@absence.m10p.summerstarts\endcsname{43}
\expandafter\def\csname EERsplit@absence.m10p.credential\endcsname{3}
\expandafter\def\csname EERsplit@absence.m10p.startratiops\endcsname{3.9}
\expandafter\def\csname EERsplit@absence.expabsent.women\endcsname{120}
\expandafter\def\csname EERsplit@absence.expabsent.months\endcsname{1,545}
\expandafter\def\csname EERsplit@absence.expabsent.starts\endcsname{76}
\expandafter\def\csname EERsplit@absence.expabsent.interruptions\endcsname{63}
\expandafter\def\csname EERsplit@absence.expabsent.startpct\endcsname{4.9}
\expandafter\def\csname EERsplit@absence.expabsent.sharepct\endcsname{83}
\expandafter\def\csname EERsplit@absence.expabsent.intpct\endcsname{4.1}
\expandafter\def\csname EERsplit@absence.expabsent.startratio\endcsname{3.6}
\expandafter\def\csname EERsplit@absence.expabsent.startlo\endcsname{2.8}
\expandafter\def\csname EERsplit@absence.expabsent.starthi\endcsname{4.5}
\expandafter\def\csname EERsplit@absence.expabsent.startci\endcsname{[2.8, 4.5]}
\expandafter\def\csname EERsplit@absence.expabsent.shareratio\endcsname{1.11}
\expandafter\def\csname EERsplit@absence.expabsent.sharelo\endcsname{0.99}
\expandafter\def\csname EERsplit@absence.expabsent.sharehi\endcsname{1.25}
\expandafter\def\csname EERsplit@absence.expabsent.shareci\endcsname{[0.99, 1.25]}
\expandafter\def\csname EERsplit@absence.expabsent.intratio\endcsname{4.0}
\expandafter\def\csname EERsplit@absence.expabsent.summerstarts\endcsname{35}
\expandafter\def\csname EERsplit@absence.expabsent.credential\endcsname{2}
\expandafter\def\csname EERsplit@absence.expabsent.startratiops\endcsname{2.9}
\expandafter\def\csname EERsplit@absmodel.start.k0.ratio\endcsname{3.45}
\expandafter\def\csname EERsplit@absmodel.start.k0.lo\endcsname{2.13}
\expandafter\def\csname EERsplit@absmodel.start.k0.hi\endcsname{5.60}
\expandafter\def\csname EERsplit@absmodel.start.k0.ci\endcsname{[2.13, 5.60]}
\expandafter\def\csname EERsplit@absmodel.start.k0.p\endcsname{0.000}
\expandafter\def\csname EERsplit@absmodel.start.k0.events\endcsname{17}
\expandafter\def\csname EERsplit@absmodel.start.m13.ratio\endcsname{4.11}
\expandafter\def\csname EERsplit@absmodel.start.m13.lo\endcsname{3.13}
\expandafter\def\csname EERsplit@absmodel.start.m13.hi\endcsname{5.41}
\expandafter\def\csname EERsplit@absmodel.start.m13.ci\endcsname{[3.13, 5.41]}
\expandafter\def\csname EERsplit@absmodel.start.m13.p\endcsname{0.000}
\expandafter\def\csname EERsplit@absmodel.start.m13.events\endcsname{54}
\expandafter\def\csname EERsplit@absmodel.start.m49.ratio\endcsname{3.38}
\expandafter\def\csname EERsplit@absmodel.start.m49.lo\endcsname{2.62}
\expandafter\def\csname EERsplit@absmodel.start.m49.hi\endcsname{4.36}
\expandafter\def\csname EERsplit@absmodel.start.m49.ci\endcsname{[2.62, 4.36]}
\expandafter\def\csname EERsplit@absmodel.start.m49.p\endcsname{0.000}
\expandafter\def\csname EERsplit@absmodel.start.m49.events\endcsname{65}
\expandafter\def\csname EERsplit@absmodel.start.m10p.ratio\endcsname{3.49}
\expandafter\def\csname EERsplit@absmodel.start.m10p.lo\endcsname{2.83}
\expandafter\def\csname EERsplit@absmodel.start.m10p.hi\endcsname{4.31}
\expandafter\def\csname EERsplit@absmodel.start.m10p.ci\endcsname{[2.83, 4.31]}
\expandafter\def\csname EERsplit@absmodel.start.m10p.p\endcsname{0.000}
\expandafter\def\csname EERsplit@absmodel.start.m10p.events\endcsname{120}
\expandafter\def\csname EERsplit@absmodel.start.expabsent.ratio\endcsname{3.82}
\expandafter\def\csname EERsplit@absmodel.start.expabsent.lo\endcsname{2.94}
\expandafter\def\csname EERsplit@absmodel.start.expabsent.hi\endcsname{4.97}
\expandafter\def\csname EERsplit@absmodel.start.expabsent.ci\endcsname{[2.94, 4.97]}
\expandafter\def\csname EERsplit@absmodel.start.expabsent.p\endcsname{0.000}
\expandafter\def\csname EERsplit@absmodel.start.expabsent.events\endcsname{76}
\expandafter\def\csname EERsplit@absmodel.startns.k0.ratio\endcsname{3.29}
\expandafter\def\csname EERsplit@absmodel.startns.k0.lo\endcsname{1.81}
\expandafter\def\csname EERsplit@absmodel.startns.k0.hi\endcsname{6.01}
\expandafter\def\csname EERsplit@absmodel.startns.k0.ci\endcsname{[1.81, 6.01]}
\expandafter\def\csname EERsplit@absmodel.startns.k0.p\endcsname{0.000}
\expandafter\def\csname EERsplit@absmodel.startns.k0.events\endcsname{11}
\expandafter\def\csname EERsplit@absmodel.startns.m13.ratio\endcsname{4.93}
\expandafter\def\csname EERsplit@absmodel.startns.m13.lo\endcsname{3.62}
\expandafter\def\csname EERsplit@absmodel.startns.m13.hi\endcsname{6.71}
\expandafter\def\csname EERsplit@absmodel.startns.m13.ci\endcsname{[3.62, 6.71]}
\expandafter\def\csname EERsplit@absmodel.startns.m13.p\endcsname{0.000}
\expandafter\def\csname EERsplit@absmodel.startns.m13.events\endcsname{43}
\expandafter\def\csname EERsplit@absmodel.startns.m49.ratio\endcsname{3.61}
\expandafter\def\csname EERsplit@absmodel.startns.m49.lo\endcsname{2.67}
\expandafter\def\csname EERsplit@absmodel.startns.m49.hi\endcsname{4.88}
\expandafter\def\csname EERsplit@absmodel.startns.m49.ci\endcsname{[2.67, 4.88]}
\expandafter\def\csname EERsplit@absmodel.startns.m49.p\endcsname{0.000}
\expandafter\def\csname EERsplit@absmodel.startns.m49.events\endcsname{45}
\expandafter\def\csname EERsplit@absmodel.startns.m10p.ratio\endcsname{3.45}
\expandafter\def\csname EERsplit@absmodel.startns.m10p.lo\endcsname{2.66}
\expandafter\def\csname EERsplit@absmodel.startns.m10p.hi\endcsname{4.47}
\expandafter\def\csname EERsplit@absmodel.startns.m10p.ci\endcsname{[2.66, 4.47]}
\expandafter\def\csname EERsplit@absmodel.startns.m10p.p\endcsname{0.000}
\expandafter\def\csname EERsplit@absmodel.startns.m10p.events\endcsname{77}
\expandafter\def\csname EERsplit@absmodel.startns.expabsent.ratio\endcsname{3.27}
\expandafter\def\csname EERsplit@absmodel.startns.expabsent.lo\endcsname{2.32}
\expandafter\def\csname EERsplit@absmodel.startns.expabsent.hi\endcsname{4.62}
\expandafter\def\csname EERsplit@absmodel.startns.expabsent.ci\endcsname{[2.32, 4.62]}
\expandafter\def\csname EERsplit@absmodel.startns.expabsent.p\endcsname{0.000}
\expandafter\def\csname EERsplit@absmodel.startns.expabsent.events\endcsname{41}
\expandafter\def\csname EERsplit@absmodel.startsep.k0.ratio\endcsname{3.34}
\expandafter\def\csname EERsplit@absmodel.startsep.k0.lo\endcsname{2.07}
\expandafter\def\csname EERsplit@absmodel.startsep.k0.hi\endcsname{5.42}
\expandafter\def\csname EERsplit@absmodel.startsep.k0.ci\endcsname{[2.07, 5.42]}
\expandafter\def\csname EERsplit@absmodel.startsep.k0.p\endcsname{0.000}
\expandafter\def\csname EERsplit@absmodel.startsep.k0.events\endcsname{17}
\expandafter\def\csname EERsplit@absmodel.startsep.m13.ratio\endcsname{4.06}
\expandafter\def\csname EERsplit@absmodel.startsep.m13.lo\endcsname{3.08}
\expandafter\def\csname EERsplit@absmodel.startsep.m13.hi\endcsname{5.34}
\expandafter\def\csname EERsplit@absmodel.startsep.m13.ci\endcsname{[3.08, 5.34]}
\expandafter\def\csname EERsplit@absmodel.startsep.m13.p\endcsname{0.000}
\expandafter\def\csname EERsplit@absmodel.startsep.m13.events\endcsname{54}
\expandafter\def\csname EERsplit@absmodel.startsep.m49.ratio\endcsname{3.36}
\expandafter\def\csname EERsplit@absmodel.startsep.m49.lo\endcsname{2.61}
\expandafter\def\csname EERsplit@absmodel.startsep.m49.hi\endcsname{4.32}
\expandafter\def\csname EERsplit@absmodel.startsep.m49.ci\endcsname{[2.61, 4.32]}
\expandafter\def\csname EERsplit@absmodel.startsep.m49.p\endcsname{0.000}
\expandafter\def\csname EERsplit@absmodel.startsep.m49.events\endcsname{65}
\expandafter\def\csname EERsplit@absmodel.startsep.m10p.ratio\endcsname{3.44}
\expandafter\def\csname EERsplit@absmodel.startsep.m10p.lo\endcsname{2.79}
\expandafter\def\csname EERsplit@absmodel.startsep.m10p.hi\endcsname{4.24}
\expandafter\def\csname EERsplit@absmodel.startsep.m10p.ci\endcsname{[2.79, 4.24]}
\expandafter\def\csname EERsplit@absmodel.startsep.m10p.p\endcsname{0.000}
\expandafter\def\csname EERsplit@absmodel.startsep.m10p.events\endcsname{120}
\expandafter\def\csname EERsplit@absmodel.startsep.expabsent.ratio\endcsname{3.79}
\expandafter\def\csname EERsplit@absmodel.startsep.expabsent.lo\endcsname{2.92}
\expandafter\def\csname EERsplit@absmodel.startsep.expabsent.hi\endcsname{4.92}
\expandafter\def\csname EERsplit@absmodel.startsep.expabsent.ci\endcsname{[2.92, 4.92]}
\expandafter\def\csname EERsplit@absmodel.startsep.expabsent.p\endcsname{0.000}
\expandafter\def\csname EERsplit@absmodel.startsep.expabsent.events\endcsname{76}
\expandafter\def\csname EERsplit@absmodel.qualify.k0.ratio\endcsname{1.19}
\expandafter\def\csname EERsplit@absmodel.qualify.k0.lo\endcsname{0.99}
\expandafter\def\csname EERsplit@absmodel.qualify.k0.hi\endcsname{1.42}
\expandafter\def\csname EERsplit@absmodel.qualify.k0.ci\endcsname{[0.99, 1.42]}
\expandafter\def\csname EERsplit@absmodel.qualify.k0.p\endcsname{0.063}
\expandafter\def\csname EERsplit@absmodel.qualify.k0.events\endcsname{15}
\expandafter\def\csname EERsplit@absmodel.qualify.m13.ratio\endcsname{1.12}
\expandafter\def\csname EERsplit@absmodel.qualify.m13.lo\endcsname{0.99}
\expandafter\def\csname EERsplit@absmodel.qualify.m13.hi\endcsname{1.27}
\expandafter\def\csname EERsplit@absmodel.qualify.m13.ci\endcsname{[0.99, 1.27]}
\expandafter\def\csname EERsplit@absmodel.qualify.m13.p\endcsname{0.068}
\expandafter\def\csname EERsplit@absmodel.qualify.m13.events\endcsname{46}
\expandafter\def\csname EERsplit@absmodel.qualify.m49.ratio\endcsname{1.10}
\expandafter\def\csname EERsplit@absmodel.qualify.m49.lo\endcsname{0.97}
\expandafter\def\csname EERsplit@absmodel.qualify.m49.hi\endcsname{1.24}
\expandafter\def\csname EERsplit@absmodel.qualify.m49.ci\endcsname{[0.97, 1.24]}
\expandafter\def\csname EERsplit@absmodel.qualify.m49.p\endcsname{0.143}
\expandafter\def\csname EERsplit@absmodel.qualify.m49.events\endcsname{53}
\expandafter\def\csname EERsplit@absmodel.qualify.m10p.ratio\endcsname{1.09}
\expandafter\def\csname EERsplit@absmodel.qualify.m10p.lo\endcsname{1.00}
\expandafter\def\csname EERsplit@absmodel.qualify.m10p.hi\endcsname{1.20}
\expandafter\def\csname EERsplit@absmodel.qualify.m10p.ci\endcsname{[1.00, 1.20]}
\expandafter\def\csname EERsplit@absmodel.qualify.m10p.p\endcsname{0.060}
\expandafter\def\csname EERsplit@absmodel.qualify.m10p.events\endcsname{99}
\expandafter\def\csname EERsplit@absmodel.qualify.expabsent.ratio\endcsname{1.08}
\expandafter\def\csname EERsplit@absmodel.qualify.expabsent.lo\endcsname{0.97}
\expandafter\def\csname EERsplit@absmodel.qualify.expabsent.hi\endcsname{1.20}
\expandafter\def\csname EERsplit@absmodel.qualify.expabsent.ci\endcsname{[0.97, 1.20]}
\expandafter\def\csname EERsplit@absmodel.qualify.expabsent.p\endcsname{0.183}
\expandafter\def\csname EERsplit@absmodel.qualify.expabsent.events\endcsname{63}
\expandafter\def\csname EERsplit@dating.all.k0.women\endcsname{364}
\expandafter\def\csname EERsplit@dating.all.k0.months\endcsname{364}
\expandafter\def\csname EERsplit@dating.all.k0.starts\endcsname{17}
\expandafter\def\csname EERsplit@dating.all.k0.interruptions\endcsname{15}
\expandafter\def\csname EERsplit@dating.all.k0.intpct\endcsname{4.1}
\expandafter\def\csname EERsplit@dating.all.k0.intlo\endcsname{2.3}
\expandafter\def\csname EERsplit@dating.all.k0.inthi\endcsname{6.7}
\expandafter\def\csname EERsplit@dating.all.k0.intratio\endcsname{4.04}
\expandafter\def\csname EERsplit@dating.all.k0.intratiolo\endcsname{2.44}
\expandafter\def\csname EERsplit@dating.all.k0.intratiohi\endcsname{6.69}
\expandafter\def\csname EERsplit@dating.all.k0.intratioci\endcsname{[2.44, 6.69]}
\expandafter\def\csname EERsplit@dating.roster.k0.women\endcsname{253}
\expandafter\def\csname EERsplit@dating.roster.k0.months\endcsname{253}
\expandafter\def\csname EERsplit@dating.roster.k0.starts\endcsname{12}
\expandafter\def\csname EERsplit@dating.roster.k0.interruptions\endcsname{10}
\expandafter\def\csname EERsplit@dating.roster.k0.intpct\endcsname{4.0}
\expandafter\def\csname EERsplit@dating.roster.k0.intlo\endcsname{1.9}
\expandafter\def\csname EERsplit@dating.roster.k0.inthi\endcsname{7.1}
\expandafter\def\csname EERsplit@dating.roster.k0.intratio\endcsname{3.88}
\expandafter\def\csname EERsplit@dating.roster.k0.intratiolo\endcsname{2.10}
\expandafter\def\csname EERsplit@dating.roster.k0.intratiohi\endcsname{7.16}
\expandafter\def\csname EERsplit@dating.roster.k0.intratioci\endcsname{[2.10, 7.16]}
\expandafter\def\csname EERsplit@dating.int9.k0.women\endcsname{38}
\expandafter\def\csname EERsplit@dating.int9.k0.months\endcsname{38}
\expandafter\def\csname EERsplit@dating.int9.k0.starts\endcsname{1}
\expandafter\def\csname EERsplit@dating.int9.k0.interruptions\endcsname{1}
\expandafter\def\csname EERsplit@dating.int9.k0.intpct\endcsname{2.6}
\expandafter\def\csname EERsplit@dating.int9.k0.intlo\endcsname{0.1}
\expandafter\def\csname EERsplit@dating.int9.k0.inthi\endcsname{13.8}
\expandafter\def\csname EERsplit@dating.int9.k0.intratio\endcsname{2.58}
\expandafter\def\csname EERsplit@dating.int9.k0.intratiolo\endcsname{0.37}
\expandafter\def\csname EERsplit@dating.int9.k0.intratiohi\endcsname{17.89}
\expandafter\def\csname EERsplit@dating.int9.k0.intratioci\endcsname{[0.37, 17.89]}
\expandafter\def\csname EERsplit@dating.int10p.k0.women\endcsname{69}
\expandafter\def\csname EERsplit@dating.int10p.k0.months\endcsname{69}
\expandafter\def\csname EERsplit@dating.int10p.k0.starts\endcsname{4}
\expandafter\def\csname EERsplit@dating.int10p.k0.interruptions\endcsname{4}
\expandafter\def\csname EERsplit@dating.int10p.k0.intpct\endcsname{5.8}
\expandafter\def\csname EERsplit@dating.int10p.k0.intlo\endcsname{1.6}
\expandafter\def\csname EERsplit@dating.int10p.k0.inthi\endcsname{14.2}
\expandafter\def\csname EERsplit@dating.int10p.k0.intratio\endcsname{5.68}
\expandafter\def\csname EERsplit@dating.int10p.k0.intratiolo\endcsname{2.19}
\expandafter\def\csname EERsplit@dating.int10p.k0.intratiohi\endcsname{14.78}
\expandafter\def\csname EERsplit@dating.int10p.k0.intratioci\endcsname{[2.19, 14.78]}
\expandafter\def\csname EERsplit@dating.intnobirth.k0.women\endcsname{4}
\expandafter\def\csname EERsplit@dating.intnobirth.k0.months\endcsname{4}
\expandafter\def\csname EERsplit@dating.intnobirth.k0.starts\endcsname{0}
\expandafter\def\csname EERsplit@dating.intnobirth.k0.interruptions\endcsname{0}
\expandafter\def\csname EERsplit@dating.intnobirth.k0.intpct\endcsname{0.0}
\expandafter\def\csname EERsplit@dating.intnobirth.k0.intlo\endcsname{0.0}
\expandafter\def\csname EERsplit@dating.intnobirth.k0.inthi\endcsname{60.2}
\expandafter\def\csname EERsplit@dating.all.k13.women\endcsname{426}
\expandafter\def\csname EERsplit@dating.all.k13.months\endcsname{962}
\expandafter\def\csname EERsplit@dating.all.k13.starts\endcsname{54}
\expandafter\def\csname EERsplit@dating.all.k13.interruptions\endcsname{46}
\expandafter\def\csname EERsplit@dating.all.k13.intpct\endcsname{4.8}
\expandafter\def\csname EERsplit@dating.all.k13.intlo\endcsname{3.5}
\expandafter\def\csname EERsplit@dating.all.k13.inthi\endcsname{6.3}
\expandafter\def\csname EERsplit@dating.all.k13.intratio\endcsname{4.69}
\expandafter\def\csname EERsplit@dating.all.k13.intratiolo\endcsname{3.48}
\expandafter\def\csname EERsplit@dating.all.k13.intratiohi\endcsname{6.31}
\expandafter\def\csname EERsplit@dating.all.k13.intratioci\endcsname{[3.48, 6.31]}
\expandafter\def\csname EERsplit@dating.roster.k13.women\endcsname{304}
\expandafter\def\csname EERsplit@dating.roster.k13.months\endcsname{701}
\expandafter\def\csname EERsplit@dating.roster.k13.starts\endcsname{39}
\expandafter\def\csname EERsplit@dating.roster.k13.interruptions\endcsname{34}
\expandafter\def\csname EERsplit@dating.roster.k13.intpct\endcsname{4.9}
\expandafter\def\csname EERsplit@dating.roster.k13.intlo\endcsname{3.4}
\expandafter\def\csname EERsplit@dating.roster.k13.inthi\endcsname{6.7}
\expandafter\def\csname EERsplit@dating.roster.k13.intratio\endcsname{4.76}
\expandafter\def\csname EERsplit@dating.roster.k13.intratiolo\endcsname{3.38}
\expandafter\def\csname EERsplit@dating.roster.k13.intratiohi\endcsname{6.69}
\expandafter\def\csname EERsplit@dating.roster.k13.intratioci\endcsname{[3.38, 6.69]}
\expandafter\def\csname EERsplit@dating.int9.k13.women\endcsname{44}
\expandafter\def\csname EERsplit@dating.int9.k13.months\endcsname{98}
\expandafter\def\csname EERsplit@dating.int9.k13.starts\endcsname{7}
\expandafter\def\csname EERsplit@dating.int9.k13.interruptions\endcsname{7}
\expandafter\def\csname EERsplit@dating.int9.k13.intpct\endcsname{7.1}
\expandafter\def\csname EERsplit@dating.int9.k13.intlo\endcsname{2.9}
\expandafter\def\csname EERsplit@dating.int9.k13.inthi\endcsname{14.2}
\expandafter\def\csname EERsplit@dating.int9.k13.intratio\endcsname{7.00}
\expandafter\def\csname EERsplit@dating.int9.k13.intratiolo\endcsname{3.41}
\expandafter\def\csname EERsplit@dating.int9.k13.intratiohi\endcsname{14.39}
\expandafter\def\csname EERsplit@dating.int9.k13.intratioci\endcsname{[3.41, 14.39]}
\expandafter\def\csname EERsplit@dating.int10p.k13.women\endcsname{74}
\expandafter\def\csname EERsplit@dating.int10p.k13.months\endcsname{151}
\expandafter\def\csname EERsplit@dating.int10p.k13.starts\endcsname{6}
\expandafter\def\csname EERsplit@dating.int10p.k13.interruptions\endcsname{4}
\expandafter\def\csname EERsplit@dating.int10p.k13.intpct\endcsname{2.6}
\expandafter\def\csname EERsplit@dating.int10p.k13.intlo\endcsname{0.7}
\expandafter\def\csname EERsplit@dating.int10p.k13.inthi\endcsname{6.6}
\expandafter\def\csname EERsplit@dating.int10p.k13.intratio\endcsname{2.60}
\expandafter\def\csname EERsplit@dating.int10p.k13.intratiolo\endcsname{0.98}
\expandafter\def\csname EERsplit@dating.int10p.k13.intratiohi\endcsname{6.86}
\expandafter\def\csname EERsplit@dating.int10p.k13.intratioci\endcsname{[0.98, 6.86]}
\expandafter\def\csname EERsplit@dating.intnobirth.k13.women\endcsname{4}
\expandafter\def\csname EERsplit@dating.intnobirth.k13.months\endcsname{12}
\expandafter\def\csname EERsplit@dating.intnobirth.k13.starts\endcsname{2}
\expandafter\def\csname EERsplit@dating.intnobirth.k13.interruptions\endcsname{1}
\expandafter\def\csname EERsplit@dating.intnobirth.k13.intpct\endcsname{8.3}
\expandafter\def\csname EERsplit@dating.intnobirth.k13.intlo\endcsname{0.2}
\expandafter\def\csname EERsplit@dating.intnobirth.k13.inthi\endcsname{38.5}
\expandafter\def\csname EERsplit@dating.intnobirth.k13.intratio\endcsname{8.17}
\expandafter\def\csname EERsplit@dating.intnobirth.k13.intratiolo\endcsname{1.25}
\expandafter\def\csname EERsplit@dating.intnobirth.k13.intratiohi\endcsname{53.48}
\expandafter\def\csname EERsplit@dating.intnobirth.k13.intratioci\endcsname{[1.25, 53.48]}
\expandafter\def\csname EERsplit@dating.all.k4p.women\endcsname{335}
\expandafter\def\csname EERsplit@dating.all.k4p.months\endcsname{3,432}
\expandafter\def\csname EERsplit@dating.all.k4p.starts\endcsname{185}
\expandafter\def\csname EERsplit@dating.all.k4p.interruptions\endcsname{152}
\expandafter\def\csname EERsplit@dating.all.k4p.intpct\endcsname{4.4}
\expandafter\def\csname EERsplit@dating.all.k4p.intlo\endcsname{3.8}
\expandafter\def\csname EERsplit@dating.all.k4p.inthi\endcsname{5.2}
\expandafter\def\csname EERsplit@dating.all.k4p.intratio\endcsname{4.34}
\expandafter\def\csname EERsplit@dating.all.k4p.intratiolo\endcsname{3.62}
\expandafter\def\csname EERsplit@dating.all.k4p.intratiohi\endcsname{5.21}
\expandafter\def\csname EERsplit@dating.all.k4p.intratioci\endcsname{[3.62, 5.21]}
\expandafter\def\csname EERsplit@dating.roster.k4p.women\endcsname{235}
\expandafter\def\csname EERsplit@dating.roster.k4p.months\endcsname{2,632}
\expandafter\def\csname EERsplit@dating.roster.k4p.starts\endcsname{134}
\expandafter\def\csname EERsplit@dating.roster.k4p.interruptions\endcsname{107}
\expandafter\def\csname EERsplit@dating.roster.k4p.intpct\endcsname{4.1}
\expandafter\def\csname EERsplit@dating.roster.k4p.intlo\endcsname{3.3}
\expandafter\def\csname EERsplit@dating.roster.k4p.inthi\endcsname{4.9}
\expandafter\def\csname EERsplit@dating.roster.k4p.intratio\endcsname{3.99}
\expandafter\def\csname EERsplit@dating.roster.k4p.intratiolo\endcsname{3.24}
\expandafter\def\csname EERsplit@dating.roster.k4p.intratiohi\endcsname{4.91}
\expandafter\def\csname EERsplit@dating.roster.k4p.intratioci\endcsname{[3.24, 4.91]}
\expandafter\def\csname EERsplit@dating.int9.k4p.women\endcsname{33}
\expandafter\def\csname EERsplit@dating.int9.k4p.months\endcsname{245}
\expandafter\def\csname EERsplit@dating.int9.k4p.starts\endcsname{12}
\expandafter\def\csname EERsplit@dating.int9.k4p.interruptions\endcsname{11}
\expandafter\def\csname EERsplit@dating.int9.k4p.intpct\endcsname{4.5}
\expandafter\def\csname EERsplit@dating.int9.k4p.intlo\endcsname{2.3}
\expandafter\def\csname EERsplit@dating.int9.k4p.inthi\endcsname{7.9}
\expandafter\def\csname EERsplit@dating.int9.k4p.intratio\endcsname{4.40}
\expandafter\def\csname EERsplit@dating.int9.k4p.intratiolo\endcsname{2.45}
\expandafter\def\csname EERsplit@dating.int9.k4p.intratiohi\endcsname{7.90}
\expandafter\def\csname EERsplit@dating.int9.k4p.intratioci\endcsname{[2.45, 7.90]}
\expandafter\def\csname EERsplit@dating.int10p.k4p.women\endcsname{64}
\expandafter\def\csname EERsplit@dating.int10p.k4p.months\endcsname{535}
\expandafter\def\csname EERsplit@dating.int10p.k4p.starts\endcsname{38}
\expandafter\def\csname EERsplit@dating.int10p.k4p.interruptions\endcsname{33}
\expandafter\def\csname EERsplit@dating.int10p.k4p.intpct\endcsname{6.2}
\expandafter\def\csname EERsplit@dating.int10p.k4p.intlo\endcsname{4.3}
\expandafter\def\csname EERsplit@dating.int10p.k4p.inthi\endcsname{8.6}
\expandafter\def\csname EERsplit@dating.int10p.k4p.intratio\endcsname{6.05}
\expandafter\def\csname EERsplit@dating.int10p.k4p.intratiolo\endcsname{4.29}
\expandafter\def\csname EERsplit@dating.int10p.k4p.intratiohi\endcsname{8.53}
\expandafter\def\csname EERsplit@dating.int10p.k4p.intratioci\endcsname{[4.29, 8.53]}
\expandafter\def\csname EERsplit@dating.intnobirth.k4p.women\endcsname{3}
\expandafter\def\csname EERsplit@dating.intnobirth.k4p.months\endcsname{20}
\expandafter\def\csname EERsplit@dating.intnobirth.k4p.starts\endcsname{1}
\expandafter\def\csname EERsplit@dating.intnobirth.k4p.interruptions\endcsname{1}
\expandafter\def\csname EERsplit@dating.intnobirth.k4p.intpct\endcsname{5.0}
\expandafter\def\csname EERsplit@dating.intnobirth.k4p.intlo\endcsname{0.1}
\expandafter\def\csname EERsplit@dating.intnobirth.k4p.inthi\endcsname{24.9}
\expandafter\def\csname EERsplit@dating.intnobirth.k4p.intratio\endcsname{4.90}
\expandafter\def\csname EERsplit@dating.intnobirth.k4p.intratiolo\endcsname{0.72}
\expandafter\def\csname EERsplit@dating.intnobirth.k4p.intratiohi\endcsname{33.19}
\expandafter\def\csname EERsplit@dating.intnobirth.k4p.intratioci\endcsname{[0.72, 33.19]}
\expandafter\def\csname EERsplit@dating.ref.ref.months\endcsname{42,650}
\expandafter\def\csname EERsplit@dating.ref.ref.starts\endcsname{584}
\expandafter\def\csname EERsplit@dating.ref.ref.interruptions\endcsname{435}
\expandafter\def\csname EERsplit@dating.ref.ref.intpct\endcsname{1.0}
}{}
\newcommand{\cellstack}[2][r]{\begin{tabular}[t]{@{}#1@{}}#2\end{tabular}}
\newcommand{\notetext}[1]{\begin{tablenotes}[flushleft]\footnotesize\item #1\end{tablenotes}}

\usepackage{float}
\title{Pregnancy and Interrupted Schooling\\Evidence from Monthly Histories\thanks{We thank seminar participants at Auburn University for helpful comments. Park acknowledges support from the Department of Economics. All errors are our own.}}
\author{Jonghoon Park\thanks{Department of Economics, Auburn University. \texttt{jzp0150@auburn.edu}}\hspace*{0.75em}and R. Alan Seals\thanks{Department of Economics, Auburn University. \texttt{alan.seals@auburn.edu}}}
\date{October 2026}
\begin{document}
\hypersetup{pageanchor=false}
\pagenumbering{gobble}
\maketitle
\begin{abstract}
\singlespacing
\noindent Pregnancy can disrupt schooling before infant care begins. We use monthly histories from the National Longitudinal Survey of Youth 1997 to study school interruptions during pregnancy and after childbirth and whether students later return to school or earn a diploma or GED. We jointly model conception and sustained interruptions in enrollment, allowing persistent unobserved differences to affect both processes. Under the timing-of-events identifying assumptions, pregnancy raises the monthly interruption rate to nearly four times the rate without pregnancy. Interruption rates are elevated before delivery and remain elevated afterward. After an interruption, students already exposed to pregnancy return to school and earn a diploma or GED less often within two years. These findings show that schooling disruption is already present during pregnancy, before the additional demands of infant care begin.
\end{abstract}

\clearpage
\pagenumbering{arabic}
\hypersetup{pageanchor=true}
\section{Introduction}

In 2023, women aged 15 to 19 gave birth to about 140,000 babies in the United States.\footnote{The National Vital Statistics System reports 140,977 births to mothers aged 15--19 in 2023. See Table 3 of its final 2023 births report (\url{https://www.cdc.gov/nchs/data/nvsr/nvsr74/nvsr74-1.pdf}).} For a student who conceives during the school year, health demands and preparations for parenthood can disrupt schooling months before an infant needs care. Yet some students may already be disengaging from school before they conceive, and for them a departure after conception may continue an earlier withdrawal.

\citet{kearney_why_2012} conclude that a low economic trajectory leads many teenage girls to have children and that the birth itself adds little to later disadvantage. Among Black and Latina students in New York City public schools, \citet{heiland_school_2022} find that future teenage mothers miss more school than other students years before they conceive. By the time they conceive, 50 percent of them have at least four absences a month, up from about 10 percent at age 12. The authors find that these absences can account for as much as half of the high school completion gap associated with teenage childbearing.\footnote{\citet{klepinger_adolescent_1995,klepinger_how_1999} estimate larger schooling losses when they use instrumental variables to address selection than when they use ordinary least squares.} However, these completion comparisons do not show when students leave school relative to conception and childbirth.

We use monthly histories from the National Longitudinal Survey of Youth 1997 (NLSY97) to follow women's schooling before, during, and after pregnancy. Our main comparisons use months from sexual debut onward among students who enter observation enrolled and without a high school diploma or General Educational Development (GED) credential. We observe their schooling until a sustained interruption, receipt of either credential, or the end of eligible follow-up. For students who interrupt, we track return to school and diploma or GED receipt over the next two years.

Women with weaker attachment to school may be more likely both to become pregnant and to interrupt their education. We jointly model conception and sustained interruption, allowing persistent unobserved differences among women to affect both processes. A causal interpretation rests on the timing-of-events assumptions \citep{abbring_nonparametric_2003}. Sexually active students may expect to become pregnant without knowing the month in which conception will occur. Under the model's assumptions, the timing of conception allows us to identify a change in interruption rates after pregnancy begins separately from persistent differences that affect schooling both before and after conception. However, students must not change their schooling in anticipation of becoming pregnant. Conception need not be unplanned, but a student must not begin withdrawing from school because she intends to become pregnant. Identification also requires that no unobserved change, such as declining commitment to school, jointly affects conception timing and school interruption after accounting for the differences represented in the model.

Under these assumptions, the model estimates that pregnancy raises the monthly interruption rate to nearly four times the rate without pregnancy, for women who conceive during the estimation period. The estimate also remains near four when we exclude women already exposed to pregnancy at the start of observation. The model accounts for age, time since entry, and persistent unobserved differences. We find that the pregnancy--interruption association changes little when we require women who conceive to have been enrolled in each of the three preceding school months. This reduces concern that the result reflects students who were already missing school in the months before conception.

Family background and academic preparation may affect both pregnancy and schooling in ways that the joint model does not fully capture. We examine this concern by adding observed student characteristics to regressions of monthly interruption risk. Controlling for race and calendar period leaves the estimated interruption rate after conception at roughly four times the rate without pregnancy. We next examine whether maternal education, household structure, and academic preparation account for part of this difference. We restrict this comparison to women who had not yet conceived when the NLSY97 measured these characteristics. For these women, the estimated monthly rate of school interruption after conception is \EERadj{adjusted_results.school3.later_cc_core.hr} times the rate without pregnancy. Adding controls for family background and academic preparation reduces this ratio to \EERadj{adjusted_results.school3.later_cc_rich.hr}. Family and academic differences thus explain part of the association, but students with similar observed backgrounds still face a substantially higher interruption rate after conception. Our conception dates can also miss earlier pregnancies that ended in miscarriage, abortion, or stillbirth. In regressions that do not model conception, adding these pregnancies lowers the main-sample ratio from \PregHistoryBaselineHR{} to \PregHistoryHR{}.

Over the nine months after conception, an estimated \EERadj{landmark_results.school3.1.cif_interruption_pct} percent of pregnant students interrupt school. The estimate is \EERadj{landmark_results.school3.0.cif_interruption_pct} percent for students of the same age and calendar month who have not yet conceived. We use recorded birth dates to examine whether schooling disruptions begin before the demands of infant care. Compared with months without pregnancy exposure, the estimated monthly interruption rate is about four times as high during pregnancy and remains elevated after delivery. The postbirth comparison, however, includes only students who have reached delivery without already interrupting or completing school. Students who interrupted during pregnancy no longer contribute to the postbirth rate, so similar rates before and after delivery do not imply that caring for a newborn adds no further difficulty.

An interruption need not end a student's education. Among students who interrupt at least one month after conception, \EERReturnExposed{} percent return within two years. The share is \EERReturnUnexposed{} percent among students who interrupt in the conception month or earlier, or have no observed pregnancy. Diploma or GED receipt is also lower in the first group. The groups may differ in their circumstances and reasons for leaving, so these gaps do not identify pregnancy's effect on eventual attainment. In either group, most students have no recorded diploma or GED two years after an interruption.

\subsection*{\normalfont\itshape Contribution to the Literature}

\citet{upchurch_timing_1990} use monthly NLSY79 histories to distinguish leaving school from completing school after a departure. They conclude that motherhood lowers completion mainly because mothers who have already left school are less likely to return and graduate. \citet{olsen_endogenous_1989} address selection into pregnancy using instrumental variables in a duration model of school leaving among low-income Black adolescent women. Their estimated pregnancy effect is small and imprecise. We revisit these questions in the NLSY97 cohort, jointly modeling conception and school interruption. Our interruption measure requires absences containing at least three school months, compared with the one-month absences outside summer studied by Upchurch and McCarthy. In this later cohort, interruption risk remains elevated after conception even when we allow persistent unobserved differences to affect both pregnancy and school leaving.

\citet{berthelon_early_2025} likewise document persistent enrollment declines around conception in Chile. Their annual observations, taken each March, can place departures during pregnancy and after delivery within the same interval. Using monthly histories, we distinguish these departures and find that interruption risk is already elevated during pregnancy relative to months without pregnancy exposure.\footnote{\citet{heiland_school_2022} find a related pattern in attendance among Black and Latina students in New York City public schools. The gap in absences between future teenage mothers and other students continues to widen during pregnancy.} This prenatal disruption also matters for interpreting \citet{schulkind_timing_2019}. They find lower completion among teenage mothers who give birth in the six months before expected high school graduation than among those who give birth in the six months afterward. Both groups, however, become pregnant before expected graduation. The earlier-birth group also conceives earlier and must remain in school longer after conception to reach expected graduation. Because interruption risk is already elevated during pregnancy, earlier conception means more months of exposure to that risk before graduation.

\citet{geronimus_socioeconomic_1992} compare sisters in three U.S.\ surveys and find that teenage mothers are less likely to finish high school in two of them. \citet{heiland_estimating_2019}, however, find that sibling comparisons in Add Health reduce the estimated loss in completed schooling to nearly zero. \citet{holmlund_estimating_2005} and \citet{carvajal_early_2026} find fewer completed years among teenage mothers in Sweden and Chile, respectively, even in sibling comparisons that account for school performance before motherhood. In studies that compare teenage births with teenage pregnancies that ended in miscarriage, conclusions also depend on which educational outcome is measured. Comparing women who became pregnant as teenagers, \citet{hotz_teenage_2005} estimate lower diploma receipt but higher GED receipt, yielding a small and imprecise estimate for receipt of either credential. \citet{ashcraft_consequences_2013} estimate small, imprecise reductions in completed schooling and GED receipt. \citet{fletcher_education_2009} find lower diploma receipt, while their estimates for completed years are close to zero. These findings make the distinction between interruption and eventual attainment consequential. A student may leave school yet later complete a credential, and the credential she earns may differ from the one she originally pursued. We follow this sequence from ongoing enrollment through interruption and then observe return and credential receipt for two years.

\section{Data}

\subsection{Schooling histories and outcomes}\label{sec:histories}

The NLSY97 began in 1997 with a nationally representative sample of U.S.\ youths born between 1980 and 1984. We combine women's monthly school enrollment histories for 1997--2007 with reports of first sexual intercourse, pregnancies, births, and high school credentials. We use two measures of school leaving. Our main measure comes from the monthly enrollment calendar, which shows for each month whether a student was enrolled in school.\footnote{The calendar covers kindergarten through twelfth grade. We classify each month as enrolled, not enrolled, or unknown. We count attendance and reported vacation as enrollment, and no enrollment or expulsion as nonenrollment. We treat the survey's ``other'' category and missing responses as unknown.} From it we identify sustained interruptions, long runs of months out of school. The second measure is the high school leaving date in the NLSY97. For a student who leaves without a diploma, this date often marks when she completed her last grade, not when she stopped going to school. We call it a recorded exit when she leaves without a diploma or GED.\footnote{The NLSY97 constructs the high school leaving date from reports across several interviews. For graduates, it is the graduation date. For other students, it is the date she completed her last grade, or the date she stopped attending if that date is missing. A leaving date counts as a recorded exit when her last completed grade is below 12 and no diploma or GED is dated on or before the month after it. A diploma or GED dated later does not change this classification.}

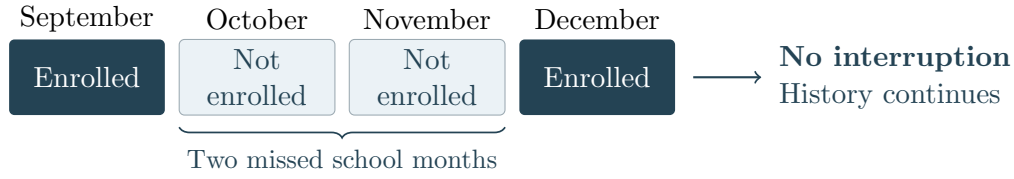
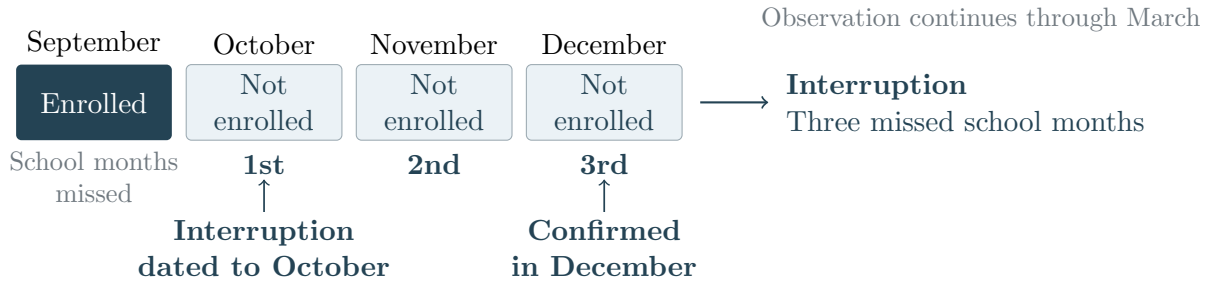
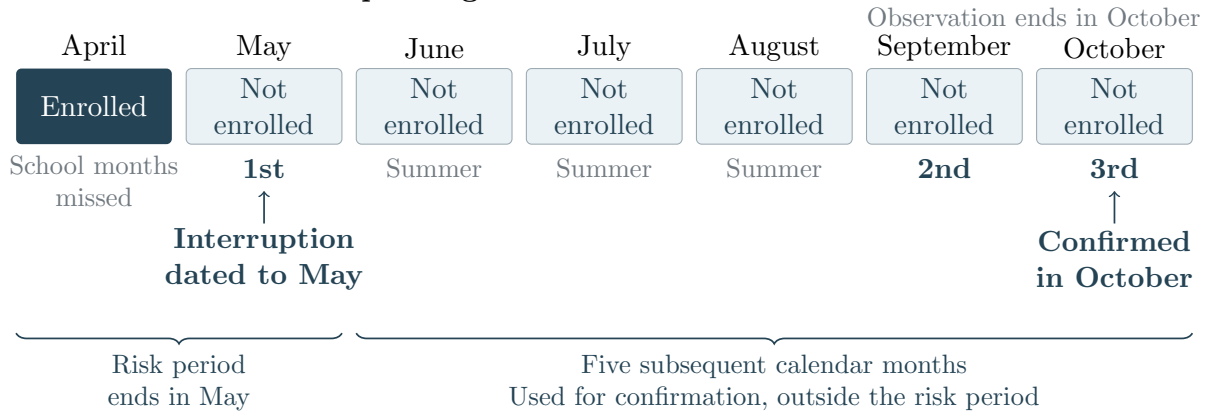
\begin{figure}[t]
\centering
\singlespacing
\caption{Dating and Confirming Sustained School Interruptions}
\label{fig:interruptiontimeline}
\definecolor{timelineink}{RGB}{36,67,84}
\definecolor{timelinepale}{RGB}{235,242,246}
\definecolor{timelinegray}{RGB}{112,121,128}
\usetikzlibrary{decorations.pathreplacing}
\begin{tikzpicture}[
  x=1cm,y=1cm,
  font=\small,
  month/.style={font=\small,inner sep=0pt},
  enrolled/.style={rectangle,rounded corners=2pt,fill=timelineink,
    text=white,minimum width=2.06cm,minimum height=1.00cm,
    align=center,inner sep=2pt,font=\small},
  absent/.style={rectangle,rounded corners=2pt,fill=timelinepale,
    draw=timelineink!45,line width=0.4pt,text=timelineink,
    minimum width=2.06cm,minimum height=1.00cm,
    align=center,inner sep=2pt,font=\small},
  detail/.style={align=center,font=\footnotesize,inner sep=0pt},
  brace/.style={decorate,decoration={brace,mirror,amplitude=4pt},
    draw=timelineink,line width=0.65pt}
]
\path[use as bounding box] (0,1.3) rectangle (16,16.4);

\node[anchor=west,font=\small\bfseries,inner sep=0pt] at (0.15,16.0)
  {A\quad Brief absence followed by return};
\foreach \x/\m in {1.2/September,3.45/October,5.7/November,7.95/December}
  \node[month] at (\x,15.37) {\m};
\node[enrolled] at (1.2,14.60) {Enrolled};
\node[absent] at (3.45,14.60) {Not\\enrolled};
\node[absent] at (5.7,14.60) {Not\\enrolled};
\node[enrolled] at (7.95,14.60) {Enrolled};
\draw[brace] (2.42,13.94) -- (6.73,13.94);
\node[detail,text=timelineink] at (4.575,13.52) {Two missed school months};
\draw[->,draw=timelineink,line width=0.8pt] (9.22,14.60) -- (10.12,14.60);
\node[anchor=west,align=left,inner sep=0pt,text=timelineink]
  at (10.35,14.60) {\textbf{No interruption}\\History continues};

\draw[black!18,line width=0.4pt] (0.15,13.02) -- (15.85,13.02);
\node[anchor=west,font=\small\bfseries,inner sep=0pt] at (0.15,12.55)
  {B\quad Three consecutive missed school months};
\node[anchor=east,font=\footnotesize,inner sep=0pt,text=timelinegray]
  at (15.85,12.10) {Observation continues through March};
\foreach \x/\m in {1.2/September,3.45/October,5.7/November,7.95/December}
  \node[month] at (\x,11.77) {\m};
\node[enrolled] at (1.2,11.00) {Enrolled};
\foreach \x in {3.45,5.7,7.95}
  \node[absent] at (\x,11.00) {Not\\enrolled};
\node[detail,text=timelinegray] at (1.2,10.02) {School months\\missed};
\foreach \x/\ordinal in {3.45/1st,5.7/2nd,7.95/3rd}
  \node[detail,text=timelineink,font=\small\bfseries] at (\x,10.18) {\ordinal};
\foreach \x in {3.45,7.95}
  \draw[->,draw=timelineink,line width=0.65pt] (\x,9.55) -- (\x,9.98);
\node[detail,text=timelineink,font=\small\bfseries] at (3.45,9.08)
  {Interruption\\dated to October};
\node[detail,text=timelineink,font=\small\bfseries] at (7.95,9.08)
  {Confirmed\\in December};
\draw[->,draw=timelineink,line width=0.8pt] (9.22,11.00) -- (10.12,11.00);
\node[anchor=west,align=left,inner sep=0pt,text=timelineink]
  at (10.35,11.00) {\textbf{Interruption}\\Three missed school months};

\draw[black!18,line width=0.4pt] (0.15,8.02) -- (15.85,8.02);
\node[anchor=west,font=\small\bfseries,inner sep=0pt] at (0.15,7.55)
  {C\quad Sustained absence spanning summer};
\node[anchor=east,font=\footnotesize,inner sep=0pt,text=timelinegray]
  at (15.85,7.18) {Observation ends in October};
\foreach \x/\m in {1.2/April,3.45/May,5.7/June,7.95/July,10.2/August,12.45/September,14.7/October}
  \node[month] at (\x,6.77) {\m};
\node[enrolled] at (1.2,6.00) {Enrolled};
\foreach \x in {3.45,5.7,7.95,10.2,12.45,14.7}
  \node[absent] at (\x,6.00) {Not\\enrolled};
\node[detail,text=timelinegray] at (1.2,5.02) {School months\\missed};
\node[detail,text=timelineink,font=\small\bfseries] at (3.45,5.18) {1st};
\foreach \x in {5.7,7.95,10.2}
  \node[detail,text=timelinegray] at (\x,5.18) {Summer};
\node[detail,text=timelineink,font=\small\bfseries] at (12.45,5.18) {2nd};
\node[detail,text=timelineink,font=\small\bfseries] at (14.7,5.18) {3rd};
\foreach \x in {3.45,14.7}
  \draw[->,draw=timelineink,line width=0.65pt] (\x,4.45) -- (\x,4.88);
\node[detail,text=timelineink,font=\small\bfseries] at (3.45,3.98)
  {Interruption\\dated to May};
\node[detail,text=timelineink,font=\small\bfseries] at (14.7,3.98)
  {Confirmed\\in October};

\draw[brace] (0.17,3.00) -- (4.48,3.00);
\node[detail,text=timelineink] at (2.325,2.35) {Risk period\\ends in May};
\draw[brace] (4.67,3.00) -- (15.73,3.00);
\node[detail,text=timelineink] at (10.2,2.35)
  {Five subsequent calendar months\\Used for confirmation, outside the risk period};
\end{tikzpicture}

\vspace{0.3em}
\begin{minipage}{\textwidth}
\footnotesize
\textit{Notes.} Illustrative histories under the main definition, which requires a continuous absence containing at least three September--May months. All three students meet the entry and sexual-debut requirements, with no credential receipt or unknown enrollment. In Panel A, at least five additional months are observed after December. In Panel B, October, November, and December are the three missed school months. The interruption is dated to October and confirmed in December. Observation continues through March, providing five calendar months after October. In Panel C, nonenrollment is recorded throughout summer, and October is the earliest observation limit. May is therefore the last possible interruption month. The five-month allowance applies to every history, including those without an interruption. The online appendix specifies the observation limits.
\end{minipage}
\end{figure}

We define a sustained school interruption as a continuous absence that includes at least three school months, September through May (Figure~\ref{fig:interruptiontimeline}). An absence from May through October qualifies, but an absence during summer alone does not. The student must be enrolled in the month before the absence, and we date the interruption to its first missed school month. If a diploma or GED is dated on or before the month the absence qualifies, we treat the student as having finished school and do not count the absence. Our alternative definition requires at least four consecutive months of absence, including at least two school months, under the same rules.\footnote{Each woman's history starts in her first eligible enrollment, the first month she is enrolled at age 13 or older and before any dated diploma or GED. This month can fall before ninth grade. The history ends in the month she turns 21 or in the month before her enrollment status first becomes unknown, if either comes before the end of 2007. Confirming an absence can take five more months, because an absence that begins in May reaches its third school month only in October. We therefore use the last five months of each history only to confirm absences. A woman's months at risk of a first interruption end at her first interruption, at a credential, or five months before her history ends, whichever comes first.}

For recorded exits, we keep only the months in which a woman is enrolled, from age 13 through the month she turns 21, until her recorded exit or credential. We skip months out of school or with unknown status, so her months need not be consecutive. A recorded exit can fall in any month, including summer, and needs no later months to confirm it.

\subsection{Pregnancy and sexual-debut dates}\label{sec:dates}

We estimate conception months from reports of current pregnancies and dates of live births. For pregnancies reported at an interview, we convert the reported weeks of pregnancy to months and subtract this duration from the interview month. For live births, we subtract nine months from the recorded birth month, assuming a nine-month pregnancy for every birth. We exclude any estimated conception month before the month a woman turns 13. For each woman, we select the earliest remaining month across both sources. We call this month measured conception.\footnote{The age restriction excludes at least one candidate conception month for \ConceptionScreenN{} women. Of these women, \ConceptionScreenRecovered{} have a later candidate month that meets the age requirement. Excluding a date under this age restriction does not establish that it is a reporting error.} These sources can miss an earlier pregnancy that ended in miscarriage, abortion, or stillbirth, so measured conception need not mark a woman's first pregnancy.

We classify months as exposed to pregnancy starting in the month after measured conception. We do not observe when a woman recognizes her pregnancy, and the one-month delay allows time for her to do so. A school interruption or recorded exit in the conception month therefore falls in an unexposed month. Once exposure starts, all later observed months remain exposed, including months after the pregnancy ends. A woman whose first observed month already falls after conception is exposed at entry.

We use sexual debut, the month of first sexual intercourse, to mark when pregnancy risk begins. The NLSY97 asks respondents how old they were at first intercourse and, from 2000 on, also the month and year. We turn each complete answer into a candidate month. For a month-and-year answer, we use the stated month. From an age answer, we know only the year of age, so we use the month six months after the birthday at that age. For example, we date an answer of 15 to six months after her fifteenth birthday. We set debut to the first of these candidate months across all of a woman's answers in the 1997--2007 rounds, so it does not depend on which months she contributes to an analysis. If her first age-based month comes before her first month-and-year month, we therefore date debut from an age answer. We impose no lower age limit. For \EERsample{person.debut_before10} of the \EERsample{person.women} women in the survey, debut falls before age ten. Because each history begins at age 13 or later, all of these women's observed months fall after debut.

Pregnancy risk must begin before the measured pregnancy, so we adjust debut dates that conflict with measured conception or are missing. If debut falls in or after the conception month, we move it to the month before conception. If debut is missing and conception is known, we assign debut to the month before conception. We call these corrected and imputed debut dates. We do not assign a debut date from conception when a woman has no measured conception date.

In the recorded-exit analysis, a conception defines exposure only if it precedes any dated recorded exit or credential, or if neither is dated. In the interruption analysis, any measured conception defines exposure. Both schooling analyses use the same adjusted debut dates.

\subsection{Main samples}

Our main interruption analysis keeps only months from each woman's measured sexual-debut month onward. Months before debut would add comparison months in which we assume no pregnancy risk. Women with no debut date contribute no months, but none of them has a measured conception, so no exposed months are lost. Because a sustained interruption cannot start in summer, we also keep only September--May months. We call each retained observation a school-risk month. It represents one woman in one month when a first interruption can occur. The main interruption sample contains \EERadj{adjusted_results.school3.main_baseline.women} women, \EERadj{adjusted_results.school3.main_baseline.months} school-risk months, and \EERadj{adjusted_results.school3.main_baseline.events} interruptions.

The recorded-exit analysis starts with \EERsample{broad.women} women, each with at least one enrolled month that meets its age and stopping rules. These women contribute \EERsample{broad.months} person-months, including months before measured sexual debut. The interruption analysis uses only women from this group, and its own calendar and confirmation rules leave some of them with no eligible month. We apply the same debut restriction to the recorded-exit analysis. Keeping only enrolled months from measured debut onward leaves \EERsample{postdebut.women} women, \EERsample{postdebut.months} person-months, and \EERsample{postdebut.exits} recorded exits. These counts differ from the interruption counts because the two analyses keep different months.

Including months before debut, \EERsample{exposure.defining_women} of the recorded-exit women have a conception that defines exposure. We correct \EERsample{exposure.debut_corrected} of their debut dates and impute \EERsample{exposure.debut_imputed}. Women with either change contribute \EERsample{exposure.modified_exposed_share_pct} percent of exposed months in the recorded-exit analysis.

\subsection{Background characteristics}\label{sec:background}

We measure family background at the first interview, using the mother's highest grade completed and household composition. Household composition distinguishes both biological parents, other two-parent households, and other arrangements. We measure academic preparation with the combined math and verbal percentile from the Armed Services Vocational Aptitude Battery (ASVAB). This percentile reflects preparation at the time of testing rather than a fixed ability. We also record race and ethnicity.

These measures must precede conception, because pregnancy could change a student's circumstances or test performance. First interviews took place in 1997--1998, and ASVAB testing ran from summer 1997 through spring 1998. For each woman, we therefore start a later observation period at the later of June 1998 and her first interview month. We keep only later months and exclude women whose measured conception falls on or before that start. The first possible school month is therefore September 1998. Of the \EERadj{adjusted_counts.school3.later_eligible_women.value} women who remain, \EERadj{adjusted_results.school3.later_cc_rich.women} have every required background measure. We call them the later complete-case sample (online appendix, Section D).

Women ever exposed to pregnancy during their school-risk months differ in background from women never exposed (Table~\ref{tab:summarystats}). They are more often Black and less likely to have lived with both biological parents at the first interview. Their mothers completed fewer grades, and their average ASVAB percentile is \EERsum{main.exposed.asvab_pct.value}, compared with \EERsum{main.unexposed.asvab_pct.value} among never-exposed women. During observation, they interrupt school more often and receive a diploma or GED less often. These differences describe the groups and do not establish effects of pregnancy. We define the groups by school-risk months only, so \EERsum{count.main.unexposed.women_with_later_conception} never-exposed women have a measured conception, with pregnancy exposure beginning only after those months end.

\section{Pregnancy and Interrupted Schooling}

\subsection{Monthly rates and nine-month shares}\label{sec:interruptions}

We compare monthly interruption rates with and without pregnancy exposure in a single-equation model, which does not model conception. The model uses each woman's school-risk months. For woman $i$ in calendar month $c$, $X_{ic}$ contains the controls and $E_{ic}$ marks months with pregnancy exposure, as defined in Section~\ref{sec:dates}. With intercept $a$ and coefficients $\beta$ and $\theta$, we specify
\begin{equation}
\lambda_{ic}=\exp\{a+X_{ic}'\beta+\theta E_{ic}\}.
\label{eq:singleinterruption}
\end{equation}
The intensity $\lambda_{ic}$ is the monthly rate underlying $h_{ic}=1-\exp\{-\lambda_{ic}\}$, the probability that she starts a sustained interruption in month $c$. We estimate equation~(\ref{eq:singleinterruption}) by complementary log-log regression with duration categories, school-term indicators, and quadratic age. School term separates the fall and spring terms. The duration categories group the months since first eligible enrollment, a count that continues through summer and does not restart at sexual debut. Single-equation models use standard errors clustered by woman.\footnote{We estimate all regressions without survey weights. Regression inference does not account for the survey's strata or primary sampling units.} The hazard ratio $\exp(\theta)$ divides the intensity with pregnancy exposure by the intensity without it at the same age, school term, and duration. The model imposes the same ratio in every month with pregnancy exposure, at every age, duration, and school term.

The ratio compares students who are still at risk of a first sustained interruption. The months without pregnancy exposure include earlier months of women later exposed and all months of women who remain unexposed. A woman can therefore contribute months both with and without pregnancy exposure. The ratio would measure pregnancy's effect if students who become pregnant would otherwise interrupt at the same rate as students with the same controls. Students who are less attached to school may also be more likely both to conceive and to leave. The ratio can include those prior differences, so we treat it as an association.

After we control for age, school term, and months since first eligible enrollment, the hazard ratio is \EERInterruptionHR{} (Table~\ref{tab:interruption}).\footnote{Debut errors never change which months are exposed, because exposure starts after measured conception and debut always precedes measured conception. A debut error changes only which months without pregnancy exposure enter the comparison, and it can move the hazard ratio in either direction. A debut dated too early adds months in which interruptions are rare, which tends to lower the rate without pregnancy exposure and raise the ratio. In the main interruption model, dropping the debut restriction adds every otherwise eligible school-risk month before debut and every such month of women with no debut date. With this change, the ratio rises from \EERnum{alternative.school3.postdebut.hr} to \EERnum{alternative.school3.full.hr} (online appendix, Section B). A corrected or imputed debut is set to the month before conception, but the true debut may be earlier. If it is earlier and the assigned date falls after the woman's history begins, her eligible months from the true debut, or from the start of her history if that is later, until the month before the assigned date are missing from the comparison.} The alternative definition, which requires four calendar months of absence, gives a ratio of \EERnum{alternative.calendar4.postdebut.hr}. The hazard ratio compares intensities rather than monthly probabilities, but for rare events the two are nearly the same. An intensity is the rate at which interruptions start, and the probability is the chance that one starts in a given month. When that chance is small, the two numbers almost coincide. Students start an interruption in about 1 percent of school-risk months without pregnancy exposure and about 4 percent with it (online appendix, Section D). A hazard ratio near four therefore means that a student with pregnancy exposure is about four times as likely to start an interruption in a given month as a comparable student without it. These monthly risks accumulate, so over the nine months after conception an estimated \EERadj{landmark_results.school3.1.cif_interruption_pct} percent of pregnant students interrupt school, compared with \EERadj{landmark_results.school3.0.cif_interruption_pct} percent of comparison students who have not conceived (Table~\ref{tab:landmarkscale}).

To compute these nine-month shares, we compare follow-up periods for two groups of students from the same sample. A pregnant student's period starts in the month after her measured conception, the index month, so each pregnant student has one period. She contributes it only if her conception month falls within her observed history, which excludes students who are already pregnant when we first observe them. A comparison period starts in the same calendar month as a pregnant student's period, for any student who has not yet conceived by then. The same student can therefore serve as a comparison in many different months, including months before she herself conceives. The \EERadj{landmark_results.school3.1.unique_women} pregnant students contribute \EERadj{landmark_results.school3.1.index_rows} periods, and the \EERadj{landmark_results.school3.0.unique_women} comparison students contribute \EERadj{landmark_results.school3.0.index_rows}. A period can start only if the student has not yet had a first interruption or received a diploma or GED. Each period covers the starting month and the next eight calendar months, including summer.\footnote{We weight comparison periods to the pregnant periods' distribution of completed age in years and starting calendar month. A comparison period stays in that group even if the student conceives during it. A diploma or GED ends interruption risk, so we treat it as a competing event.}

Some periods end before the nine months are over, because the student's history ends first, for example when she turns 21 or her enrollment status becomes unknown. This happens for \EERadj{landmark_results.school3.1.raw_early_censors} pregnant periods and \EERadj{landmark_results.school3.0.raw_early_censors} comparison periods. We do not drop periods that end early. A period we observe for five months counts in those five months, and for the remaining four we assume the student would have interrupted at the same rate as students in her group whose periods we still observe. To check whether this assumption drives the gap, we also compute the extreme cases, in which none or all of the periods that end early contain an interruption. Even if none of the pregnant periods that end early contain an interruption and all of the comparison periods that end early do, \EERadj{landmark_results.school3.1.censor_bound_lo_pct} percent of pregnant periods contain an interruption, against a weighted \EERadj{landmark_results.school3.0.censor_bound_hi_pct} percent of comparison periods, so incomplete follow-up cannot explain the difference. Pregnant students can still differ from comparison students in ways other than pregnancy, so the comparison describes schooling around conception rather than an effect of pregnancy.

\subsection{Joint conception and interruption model}\label{sec:joint}

Unobserved differences between women can affect both pregnancy and schooling. A woman who expects little benefit from school may be more likely to conceive early and leave without a credential. Comparing months with and without pregnancy exposure could then attribute part of those prior differences to pregnancy. We model conception and schooling together to allow for differences that persist through a woman's history.\footnote{A related NLSY97 study jointly models delinquency, arrest, and school leaving \citep{ward_delinquency_2021}.}

The model allows two unobserved groups of women, which we call latent types \citep{heckman_method_1984}. Each group has its own conception rate and schooling rate without pregnancy exposure. A group more likely to conceive can be either more or less likely to leave school without pregnancy exposure. A woman's type stays fixed, but we do not observe it. The population share of each type is the same for all women, whatever their controls. With $k$ denoting latent type, and with $c$ now running over every calendar month of a woman's history, the joint model extends equation~(\ref{eq:singleinterruption}) as
\begin{align}
\lambda^P_{ic}(k) &= \exp\{a_P+X^P_{ic}{}'\beta_P+u_{Pk}\},\label{eq:conception}\\
\lambda^Y_{ic}(k) &= \exp\{a_Y+X^Y_{ic}{}'\beta_Y+u_{Yk}
                         +\gamma E^O_{ic}+\delta E^U_{ic}\}.
\label{eq:intensities}
\end{align}
The letter $P$ denotes conception, and $Y$ denotes the schooling event, either recorded exit or the start of a sustained interruption. In equations~(\ref{eq:conception}) and~(\ref{eq:intensities}), $X^P_{ic}$ and $X^Y_{ic}$ contain the controls, with coefficients $\beta_P$ and $\beta_Y$ and intercepts $a_P$ and $a_Y$. The main joint models use linear terms for age and for months since first eligible enrollment, capped at 48 months, which we call linear controls. Flexible controls instead use the duration categories, school-term indicators, and quadratic age of equation~(\ref{eq:singleinterruption}). The terms $u_{Pk}$ and $u_{Yk}$ allow the two types to have different rates at the same controls and without pregnancy exposure. We call these type-specific terms loadings. Compared with equation~(\ref{eq:singleinterruption}), equation~(\ref{eq:intensities}) adds the type term $u_{Yk}$ and splits the pregnancy exposure term $\theta E_{ic}$ into two terms, described next.

The joint model uses the month in which each woman conceives, but for some exposed women we do not observe that month. Some are already pregnant when we first observe them, and in the recorded-exit analysis a gap in observation can skip the conception month. We therefore split pregnancy exposure into two indicators. The indicator $E^O_{ic}$ marks pregnancy exposure after a conception month we observe, which we call a retained onset. The indicator $E^U_{ic}$ marks pregnancy exposure when we do not observe the conception month. The coefficient $\gamma$ applies after a retained onset. The separate coefficient $\delta$ applies to exposed months without an observed conception, so the interruption rate in those months does not enter $\gamma$ directly, although those months still contribute to the estimated controls and latent types that $\gamma$ shares. Both coefficients compare months with pregnancy exposure with months without it. As in the single-equation model, the months without pregnancy exposure include the earlier months of women who later conceive and every month of women who do not conceive during observation. As a check, we also estimate the model on a retained-onset sample, which drops the exposed women whose conception month we do not observe. Women who are never exposed remain. This sample needs no $\delta$, so it shows whether keeping the exposed women without an observed conception month, with their own coefficient, changes the estimate of $\gamma$.

For event $j\in\{P,Y\}$, the probability that the event occurs in month $c$ is $h^j_{ic}(k)=1-\exp\{-\lambda^j_{ic}(k)\}$. In each month we observe her, a woman contributes $h^j_{ic}(k)$ if the event occurs and $1-h^j_{ic}(k)$ if it does not. Multiplying these terms over all her observed months gives the probability of her whole history if she is of type $k$. Because we do not observe her type, we average this probability over the two types, weighting each by its population share. We observe conception until it occurs or her observation ends, and schooling until her first schooling event, a credential, or the end of observation. Summer months enter the conception model, because conception can occur in summer. They do not enter the interruption model, because an interruption cannot start in summer.\footnote{We estimate the model by maximum likelihood with \texttt{hsmixture}, a Stata package we wrote (version 2.4.0, available at \url{https://github.com/alanseals/hsmixture}). For the interruption models, we add one option to its joint estimator so that a month can enter the conception model but not the interruption model, as summer months do. The replication package contains both versions.}

A causal interpretation rests on two timing-of-events conditions \citep{abbring_nonparametric_2003}. The first is no anticipation. Given the controls and latent type, future conception must not change the current schooling rate, so pregnancy plans must not change schooling before conception. Uncertainty about conception timing provides a reason this condition might hold.\footnote{\citet{frimmel_grandmothers_2022} use this reasoning in a timing-of-events model of grandmothers' exit from work. They find no abrupt change in exits at conception and no statistically detectable coordination effect in the preceding months. These grandmother results do not show that the condition holds for students.} Students, however, choose their own contraception and attendance, and a student planning pregnancy could reduce attendance before conception.\footnote{Our measure of interruption adds a further requirement. We date an interruption to its first missed school month, but we count it only if the absence lasts long enough, which we learn from later months. Suppose a student misses a school month and conceives the following month. If pregnancy then keeps her out of school, her absence can grow long enough to qualify, and we date the interruption to the month before she conceived. A later conception would then help create an interruption that we record as starting without pregnancy exposure. Such cases add interruptions to months without pregnancy exposure and push the hazard ratio down. Dropping the women whose pregnancy exposure begins while an absence is being confirmed raises the single-equation hazard ratio from \EERtiming{single.base.hr} to \EERtiming{single.dropconf.hr} (online appendix, Section B).}

The second condition concerns persistent differences. Women can differ persistently in contraception and school attachment. We assume that the controls and latent type account for those persistent differences. Conditional on them, conception timing must have no remaining unobserved determinants in common with schooling in the absence of pregnancy. The model must also account correctly for selection into histories that start late or end early.

The second condition fails if a short-lived event raises both conception and the chance of leaving school. \citet{weitzman_exposure_2023} show that conception can respond to such an event within a week. Following young women in Flint, Michigan, week by week, they find that a homicide within a quarter mile of a woman's home raises her chance of conceiving that week by 2 percentage points. When she has sex in that week, her chance of using any contraception is 23 percentage points lower, and recent pill users are less likely to take the pill, while her reported desire for pregnancy shows no statistically detectable change. Most of these effects fade within two weeks. They do not study schooling, but the same kind of event, or a relationship change or a sudden loss of family income, could also keep a student out of school. A woman's latent type is fixed, so it cannot absorb a shock that lasts only weeks, and the model would then attribute part of the shock's effect on schooling to pregnancy. Controlling for observed characteristics cannot rule out such shocks.

Under these conditions, $\exp(\gamma)$ is the factor by which pregnancy changes a student's monthly interruption intensity. It compares her intensity in the months after a conception we observe with the intensity she would have had in those months without pregnancy, at the same controls and latent type. The model separates this change from persistent differences because the two act at different times. A woman's latent type raises or lowers her interruption intensity in every month, before and after she conceives, while $\gamma$ applies only after conception. A student who is less attached to school therefore has a higher interruption intensity even before she conceives, and the model attributes that difference to her type rather than to pregnancy. Her conception history also carries information about her type, because the types differ in how quickly they conceive.

With one type, the model has no latent differences, so the conception and schooling equations can be estimated separately. With two types, the two histories together reveal the differences they share. Comparing one-type and two-type estimates on the same women, months, and controls shows how much allowing for these differences changes the pregnancy estimate.

The model imposes the same ratio, $\exp(\gamma)$, in every month after a retained onset, for both types, and at every age and duration. This ratio therefore summarizes the change across all months after conception, during pregnancy and after birth. It does not show whether that change differs between pregnancy and the months after delivery, or across students. The ratio applies only to students still at risk of a first sustained interruption. It therefore does not give the share of pregnant students who interrupt, and it does not measure effects on return to school or on credentials.

We read only $\exp(\gamma)$ causally, and it has two further limits. It relies on measured conception to date pregnancy exposure, so an earlier pregnancy missing from our sources, such as one that ended in miscarriage, abortion, or stillbirth (Section~\ref{sec:dates}), leaves months that we treat as unexposed. It also measures how much pregnancy raises the interruption rate, not why.

Comparing the linear and flexible specifications checks whether the estimated pregnancy effect depends on how we model age, duration, and the school calendar. Both hazards use the same controls.\footnote{We calculate intervals for one-type and two-type models from the model's covariance matrix. We fit each two-type model from seven starting values and keep the solution whose likelihood is at least that of every other start. We report it only if it passes numerical checks on convergence, the gradient, and the covariance matrix (online appendix, Section B).} Neither this comparison nor the comparison across types tests anticipation or excludes common shocks.

Under the timing-of-events conditions, pregnancy raises a student's monthly interruption rate to nearly four times the rate she would have in the same month without pregnancy (Table~\ref{tab:matchedmodels}). This comparison holds the controls and latent type fixed. The two-type estimate is \EERjoint{results.full_linear.k2.hr}, and the one-type estimate is \EERjoint{results.full_linear.k1.hr}, so allowing a second type changes the estimate little in the main samples. The one-type estimate need not equal the single-equation hazard ratio, because the controls and the coefficients for pregnancy exposure differ. In the retained-onset sample, which for interruption excludes only women exposed at entry, the two-type estimate is \EERjoint{results.restricted_linear.k2.hr}. With flexible controls, the two-type ratio is \EERjointflex{results.full_flexible.k2.hr} in the full sample and also \EERjointflex{results.restricted_flexible.k2.hr} in the retained-onset sample (online appendix, Section B).

The Akaike and Bayesian information criteria (AIC and BIC), which penalize additional parameters, both favor two types over one in the full and retained-onset samples, with either set of controls (online appendix, Section B). The two types differ mainly in how quickly they conceive. One type, which we call the higher-conception type, makes up \EERjoint{diagnostics.full_linear.higher_conception_share_pct} percent of women, and its conception intensity is \EERjoint{diagnostics.full_linear.conception_type_ratio} times the other type's. Its interruption intensity without pregnancy exposure is \EERjoint{diagnostics.full_linear.outcome_type_ratio} times the other type's, a difference statistically indistinguishable from zero. The second type therefore separates a small group of women who conceive quickly but adds little about differences in schooling, which is why allowing it changes the pregnancy estimate little.

The joint estimate stays close to four when we drop the conception histories of women whose conception month we observe and whose debut dates we corrected or imputed, whether or not they later have months with pregnancy exposure.\footnote{In the joint model, debut also marks when conception risk begins, so a debut error also changes which months enter the conception equation.} In the main two-type fit, the \EERjointentry{posterior.full_linear.onset_modified_debut.women} of these women who later have months with pregnancy exposure account for about half of the estimated membership of the higher-conception type. Women who conceive in their first two observed months account for more than three fifths of it (online appendix, Section B). We remove the affected women's conception events and conception-risk months but keep every woman and every schooling month, and we call this the modified-onset grouping. The affected women's months with pregnancy exposure receive the coefficient $\delta$ in equation~(\ref{eq:intensities}). The ratio for pregnancy exposure after the conceptions that remain is \EERjointentry{results.entry_linear.k1.hr} with one type and \EERjointentry{results.entry_linear.k2.hr} with two types (Table~\ref{tab:matchedmodels}). Under this grouping, the higher-conception type also has a higher interruption intensity without pregnancy exposure, and AIC and BIC both favor one type. Removing these conception observations therefore changes the estimated differences between types but moves the pregnancy estimate little.

\subsection{Selection, observed background, and measurement checks}\label{sec:checks}

The single-equation hazard ratio stays near four when we add race and calendar-period controls or require enrollment or observation before conception. It also stays near four when we change the timing of pregnancy exposure. It falls when we add family background and academic preparation in the later complete-case sample, and when we add pregnancies that ended without a live birth.

Adding race and calendar-period controls does little to change the estimated hazard ratio (Table~\ref{tab:adjustedschooling}). We keep the main sample's women and months fixed. The hazard ratio changes from \EERadj{adjusted_results.school3.main_baseline.hr} to \EERadj{adjusted_results.school3.main_core.hr}.

Adding family background and academic preparation lowers the estimated hazard ratio in the later complete-case sample (Section~\ref{sec:background}). Adding race and calendar-period controls changes the ratio from \EERadj{adjusted_results.school3.later_cc_baseline.hr} to \EERadj{adjusted_results.school3.later_cc_core.hr}. Adding the mother's highest grade completed, household composition, and the ASVAB percentile then lowers it to \EERadj{adjusted_results.school3.later_cc_rich.hr}. All three models use these same women and months, so differences across them reflect the added controls. The four-calendar-month definition gives a similar reduction. Even with all these controls, the estimated interruption rate is more than twice as high with pregnancy exposure.\footnote{Adding the same background measures to both equations of the joint model gives a similar result. We fit the joint model to the same \EERadj{adjusted_results.school3.later_cc_rich.women} women as the single-equation comparison, who have every background measure and had not yet conceived when those measures were collected. As in the last check of Section~\ref{sec:joint}, we drop the conception events of women whose debut dates we corrected or imputed. With one latent type, the estimate falls from \EERjointentry{results.latercc_baseline_entry.k1.hr} to \EERjointentry{results.latercc_rich_entry.k1.hr} when we add the background measures (online appendix, Section B). We cannot add a second type to this model, because the two-type fit with background measures fails the numerical checks. This comparison therefore accounts for observed background but not for latent differences between women.}

Students who were already drifting away from school before they conceived could drive the estimate. We check this in two ways. Each check restricts only women who conceive or become exposed to pregnancy during observation, and all other women remain in the comparison. The first requires enrollment in each of the three school months before conception, which also removes women already pregnant when we first observe them. The hazard ratio is then \EERInterruptionPriorHR{}, close to the main estimate of \EERInterruptionHR{} (see Table~\ref{tab:adjustedschooling}). The second requires that we observe the conception month and at least three earlier school-risk months, which may include short absences. We call histories that meet this requirement supported histories, and the hazard ratio among them is \EERadj{adjusted_results.school3.history_baseline.hr}, against \EERadj{adjusted_results.school3.main_baseline.hr} for the full sample (see Table~\ref{tab:adjustedschooling}). However, students who pass either check could still have changed their plans for school before conceiving, so neither check shows that conception timing is unrelated to those plans.

Measured conception can miss an earlier pregnancy that ended in miscarriage, abortion, or stillbirth (Section~\ref{sec:dates}). Separate pregnancy-history reports date some of these pregnancies. When we use them to date a woman's conception earlier, or to date it for a woman with no measured conception, the single-equation hazard ratio falls from \PregHistoryBaselineHR{} to \PregHistoryHR{} (95 percent confidence interval \PregHistoryCILower{}--\PregHistoryCIUpper{}). We keep the same women, schooling months, and controls. Alternative treatments of conflicting reports and uncertain dates give hazard ratios between \PregHistoryAltMin{} and \PregHistoryAltMax{} (online appendix, Section B). All other estimates in the paper use the main conception dates.

We also vary how we date pregnancy exposure, interruptions, and conception. We start pregnancy exposure in the conception month or one to three months after it, drop up to two months on each side of conception, and drop women whose pregnancy exposure begins while an absence is being confirmed. We also date each interruption to its first nonenrolled calendar month, drop women whose exposed interruption continues an absence that began in the summer at or before conception, and shift birth-based or interview-based conception dates by one month. Across these checks, the single-equation hazard ratio ranges from \EERtiming{single.range.min} to \EERtiming{single.range.max}.\footnote{We also estimate the main single-equation interruption model under three other rules for combining a woman's answers. The rules use only the answers from her first round with an answer, only those from her last such round, or her month-and-year answers when she gives them and her age answers otherwise. Under each rule, we again take the first candidate month and apply the adjustment described in Section~\ref{sec:dates} for debut dates that conflict with measured conception or are missing. Each rule changes debut for between \EERdebutrule{changed.min} and \EERdebutrule{changed.max} of the \EERdebutrule{base.women} women in the main interruption sample, and between \EERdebutrule{left.min} and \EERdebutrule{left.max} of them leave the sample because the new debut leaves them no school-risk months. The hazard ratio ranges from \EERdebutrule{alt.hr_min} to \EERdebutrule{alt.hr_max}, against \EERdebutrule{base.hr} under our rule. Because debut always precedes measured conception, no rule changes which months are exposed.} The joint models repeat the exposure-lag, confirmation, and interruption-dating checks, and their two-type estimates range from \EERtiming{joint.k2range.min} to \EERtiming{joint.k2range.max} (online appendix, Section B).\footnote{\citet{upchurch_timing_1990} date motherhood from birth in their main models and find no statistically significant effect of a first birth on dropping out. The result is unchanged when they date motherhood from estimated conception. We instead find a higher rate of starting a sustained interruption with pregnancy exposure. The studies use different schooling outcomes, cohorts, and models, so we cannot attribute the different findings to conception dating alone.}

\subsection{Interruptions during pregnancy and after delivery}\label{sec:birth}

To separate months of pregnancy from months after delivery, we need a birth date for each pregnancy we study. We link each woman's measured conception to her first recorded birth and keep the link only if the birth falls eight to ten months after conception, within one month of conception plus nine months. If a woman's pregnancy falls in her observed history but no birth falls in that window, we exclude her entire schooling history from this analysis. Women with neither a measured conception month nor pregnancy exposure in their observed risk history remain as comparisons. The online appendix repeats the analysis with a three-month tolerance and with exact nine-month agreement (Section B).\footnote{For recorded exit, the pregnancy must begin before her recorded exit or credential. For sustained interruption, conception or pregnancy exposure must fall within her observed history. We apply these rules before keeping only months after sexual debut, and we use her first birth whose date implies conception at age 13 or later. If conception is estimated by subtracting nine months from the same birth month, the dates meet the matching rule automatically, so the match gives no independent confirmation of when conception occurred. Agreement between dates also does not verify that the records refer to the same pregnancy.}

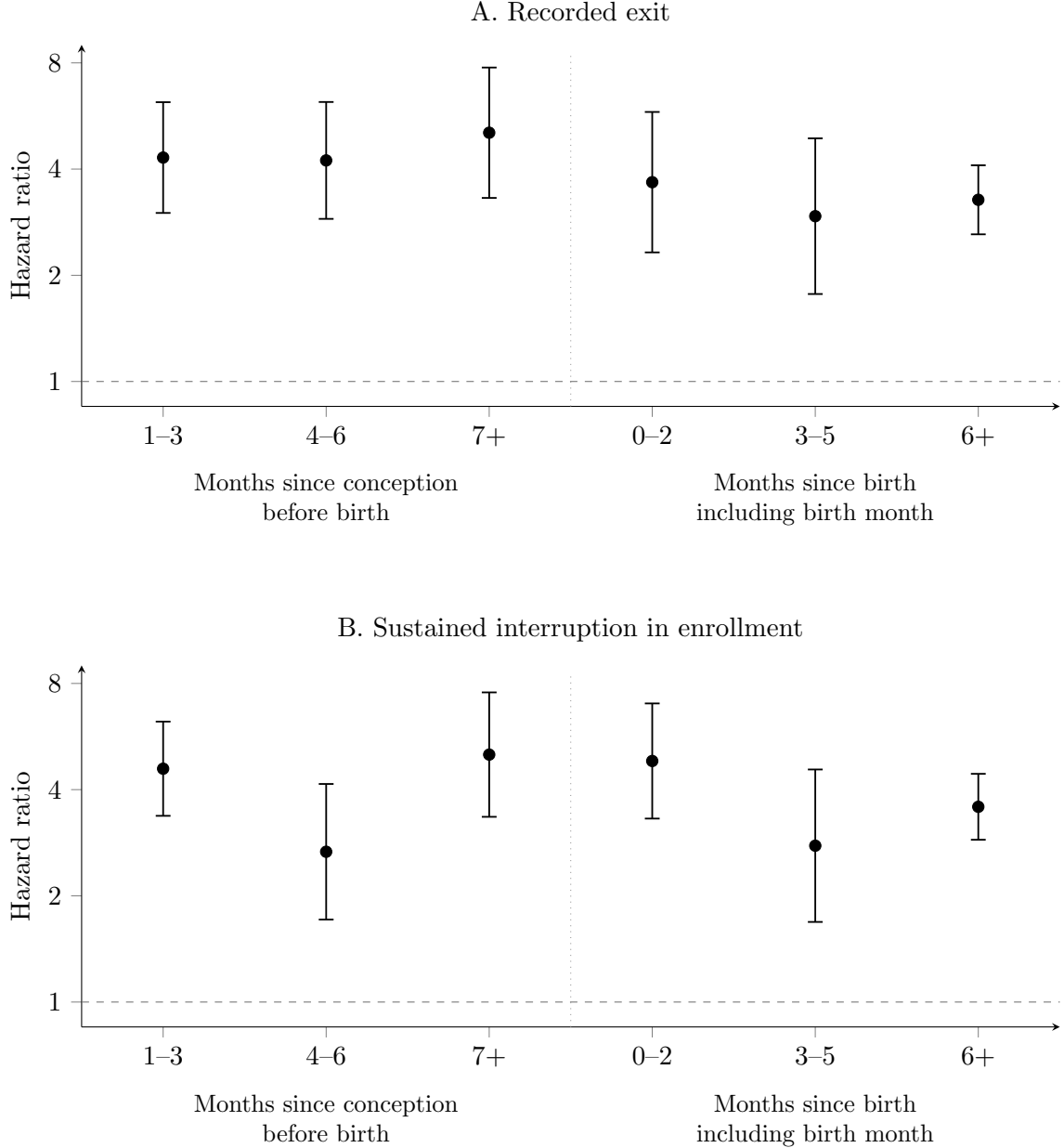
\begin{figure}[p]\centering\singlespacing\small
\caption{Schooling Disruptions during Pregnancy and after Birth}\label{fig:birthprofile}
\begin{tikzpicture}
\begin{axis}[
width=0.94\textwidth,height=2.65in,ymode=log,log basis y=2,ymin=0.85,ymax=9,
xmin=0.5,xmax=6.5,xtick={1,2,3,4,5,6},xticklabels={1--3,4--6,7+,0--2,3--5,6+},
ytick={1,2,4,8},yticklabels={1,2,4,8},ylabel={Hazard ratio},
title={A. Recorded exit},title style={font=\small},
axis lines=left,tick align=outside,clip=false,grid=none,
label style={font=\small},tick label style={font=\small}]
\addplot[black!55,dashed,domain=0.5:6.5,samples=2] {1};
\draw[black!40,dotted] (axis cs:3.5,0.85) -- (axis cs:3.5,8.5);
\draw[black, line width=0.7pt] (axis cs:1,3.004269939766506) -- (axis cs:1,6.190713897650735);
\draw[black, line width=0.7pt] ([xshift=-3pt]axis cs:1,3.004269939766506) -- ([xshift=3pt]axis cs:1,3.004269939766506);
\draw[black, line width=0.7pt] ([xshift=-3pt]axis cs:1,6.190713897650735) -- ([xshift=3pt]axis cs:1,6.190713897650735);
\draw[black, line width=0.7pt] (axis cs:2,2.890297861292847) -- (axis cs:2,6.198622900989819);
\draw[black, line width=0.7pt] ([xshift=-3pt]axis cs:2,2.890297861292847) -- ([xshift=3pt]axis cs:2,2.890297861292847);
\draw[black, line width=0.7pt] ([xshift=-3pt]axis cs:2,6.198622900989819) -- ([xshift=3pt]axis cs:2,6.198622900989819);
\draw[black, line width=0.7pt] (axis cs:3,3.313745178554679) -- (axis cs:3,7.762315798083264);
\draw[black, line width=0.7pt] ([xshift=-3pt]axis cs:3,3.313745178554679) -- ([xshift=3pt]axis cs:3,3.313745178554679);
\draw[black, line width=0.7pt] ([xshift=-3pt]axis cs:3,7.762315798083264) -- ([xshift=3pt]axis cs:3,7.762315798083264);
\draw[black, line width=0.7pt] (axis cs:4,2.322281731581135) -- (axis cs:4,5.807716321899279);
\draw[black, line width=0.7pt] ([xshift=-3pt]axis cs:4,2.322281731581135) -- ([xshift=3pt]axis cs:4,2.322281731581135);
\draw[black, line width=0.7pt] ([xshift=-3pt]axis cs:4,5.807716321899279) -- ([xshift=3pt]axis cs:4,5.807716321899279);
\draw[black, line width=0.7pt] (axis cs:5,1.769992904453576) -- (axis cs:5,4.889300441729643);
\draw[black, line width=0.7pt] ([xshift=-3pt]axis cs:5,1.769992904453576) -- ([xshift=3pt]axis cs:5,1.769992904453576);
\draw[black, line width=0.7pt] ([xshift=-3pt]axis cs:5,4.889300441729643) -- ([xshift=3pt]axis cs:5,4.889300441729643);
\draw[black, line width=0.7pt] (axis cs:6,2.61309768277166) -- (axis cs:6,4.10074471279839);
\draw[black, line width=0.7pt] ([xshift=-3pt]axis cs:6,2.61309768277166) -- ([xshift=3pt]axis cs:6,2.61309768277166);
\draw[black, line width=0.7pt] ([xshift=-3pt]axis cs:6,4.10074471279839) -- ([xshift=3pt]axis cs:6,4.10074471279839);
\addplot[only marks,mark=*,mark size=2.3pt,black] coordinates {(1,4.312606597918113) (2,4.232713847366927) (3,5.071719289384738) (4,3.672486013118672) (5,2.941772780077211) (6,3.273476211407623)};
\node[anchor=north,font=\footnotesize,align=center] at (axis description cs:0.25,-0.16) {Months since conception\\before birth};
\node[anchor=north,font=\footnotesize,align=center] at (axis description cs:0.75,-0.16) {Months since birth\\including birth month};
\end{axis}\end{tikzpicture}

\vspace{0.4in}
\begin{tikzpicture}
\begin{axis}[
width=0.94\textwidth,height=2.65in,ymode=log,log basis y=2,ymin=0.85,ymax=9,
xmin=0.5,xmax=6.5,xtick={1,2,3,4,5,6},xticklabels={1--3,4--6,7+,0--2,3--5,6+},
ytick={1,2,4,8},yticklabels={1,2,4,8},ylabel={Hazard ratio},
title={B. Sustained interruption in enrollment},title style={font=\small},
axis lines=left,tick align=outside,clip=false,grid=none,
label style={font=\small},tick label style={font=\small}]
\addplot[black!55,dashed,domain=0.5:6.5,samples=2] {1};
\draw[black!40,dotted] (axis cs:3.5,0.85) -- (axis cs:3.5,8.5);
\draw[black, line width=0.7pt] (axis cs:1,3.371793353137336) -- (axis cs:1,6.236712106669039);
\draw[black, line width=0.7pt] ([xshift=-3pt]axis cs:1,3.371793353137336) -- ([xshift=3pt]axis cs:1,3.371793353137336);
\draw[black, line width=0.7pt] ([xshift=-3pt]axis cs:1,6.236712106669039) -- ([xshift=3pt]axis cs:1,6.236712106669039);
\draw[black, line width=0.7pt] (axis cs:2,1.713194799138486) -- (axis cs:2,4.15119895266558);
\draw[black, line width=0.7pt] ([xshift=-3pt]axis cs:2,1.713194799138486) -- ([xshift=3pt]axis cs:2,1.713194799138486);
\draw[black, line width=0.7pt] ([xshift=-3pt]axis cs:2,4.15119895266558) -- ([xshift=3pt]axis cs:2,4.15119895266558);
\draw[black, line width=0.7pt] (axis cs:3,3.349948796892333) -- (axis cs:3,7.546975020806133);
\draw[black, line width=0.7pt] ([xshift=-3pt]axis cs:3,3.349948796892333) -- ([xshift=3pt]axis cs:3,3.349948796892333);
\draw[black, line width=0.7pt] ([xshift=-3pt]axis cs:3,7.546975020806133) -- ([xshift=3pt]axis cs:3,7.546975020806133);
\draw[black, line width=0.7pt] (axis cs:4,3.31203989772999) -- (axis cs:4,7.02163941294402);
\draw[black, line width=0.7pt] ([xshift=-3pt]axis cs:4,3.31203989772999) -- ([xshift=3pt]axis cs:4,3.31203989772999);
\draw[black, line width=0.7pt] ([xshift=-3pt]axis cs:4,7.02163941294402) -- ([xshift=3pt]axis cs:4,7.02163941294402);
\draw[black, line width=0.7pt] (axis cs:5,1.686965719424324) -- (axis cs:5,4.562877143265161);
\draw[black, line width=0.7pt] ([xshift=-3pt]axis cs:5,1.686965719424324) -- ([xshift=3pt]axis cs:5,1.686965719424324);
\draw[black, line width=0.7pt] ([xshift=-3pt]axis cs:5,4.562877143265161) -- ([xshift=3pt]axis cs:5,4.562877143265161);
\draw[black, line width=0.7pt] (axis cs:6,2.884197118526554) -- (axis cs:6,4.434828588702787);
\draw[black, line width=0.7pt] ([xshift=-3pt]axis cs:6,2.884197118526554) -- ([xshift=3pt]axis cs:6,2.884197118526554);
\draw[black, line width=0.7pt] ([xshift=-3pt]axis cs:6,4.434828588702787) -- ([xshift=3pt]axis cs:6,4.434828588702787);
\addplot[only marks,mark=*,mark size=2.3pt,black] coordinates {(1,4.58572834200826) (2,2.666798165571553) (3,5.028118921736636) (4,4.822442315170186) (5,2.774421979914576) (6,3.576439547468371)};
\node[anchor=north,font=\footnotesize,align=center] at (axis description cs:0.25,-0.16) {Months since conception\\before birth};
\node[anchor=north,font=\footnotesize,align=center] at (axis description cs:0.75,-0.16) {Months since birth\\including birth month};
\end{axis}\end{tikzpicture}

\vspace{0.3in}
\begin{minipage}{\textwidth}\footnotesize
Points are hazard ratios and bars are 95\% confidence intervals on a logarithmic scale. The dashed line marks the unexposed reference. The dotted divider separates prebirth months from the birth month and later months. All six exposure periods enter one complementary log-log model in each panel. A woman with a pregnancy in her history is kept only if her first recorded birth falls eight to ten months after conception. Controls are duration categories, school term, and quadratic age, with confidence intervals clustered by woman. Recorded exit uses 2,914 women, 60,719 retained enrolled months, and 577 events. Sustained interruption in enrollment uses 2,796 women, 48,299 school months, and 694 events. The first three periods measure time since conception while still before recorded birth. The final period is open ended. The panels use different schooling outcomes and risk periods. Neither panel identifies separate causal effects of pregnancy and infant care.
\end{minipage}\end{figure}

The single-equation models estimate higher interruption and school-exit rates before birth than in months without pregnancy exposure (Figure~\ref{fig:birthprofile}). Panel A measures recorded exit, and Panel B measures sustained interruption. Each model extends equation~(\ref{eq:singleinterruption}) with six coefficients for pregnancy exposure. Three cover periods before birth, measured in months since conception, and three cover periods from the birth month onward, measured in months since delivery. Months without pregnancy exposure are the reference, and the controls are those of equation~(\ref{eq:singleinterruption}). A second model for each outcome combines these periods into one period before birth and one from the birth month onward. The hazard ratio for interruption is \EERPrebirthInterruptionHR{} before birth and \EERPostbirthInterruptionHR{} from the birth month onward (online appendix, Section B). For recorded exit, the two ratios are \EERPrebirthHR{} and \EERPostbirthHR{}. The higher rates are already present before newborn care begins.

Compared with months without pregnancy exposure, sustained interruption rises about as much after birth as during pregnancy. The two ratios are statistically indistinguishable, whether we require the birth to fall exactly nine months after conception, within one month of that date, or within three months of it. For recorded exit, the ratio after birth is lower, and the difference is significant at the five-percent level unless the birth must fall exactly nine months after conception (online appendix, Section B).

In the first three months after conception, the interruption rate is already \EERnum{birth.school3.1.phase1.hr} times the rate without pregnancy exposure, long before a newborn needs care. In every later period, during pregnancy and after delivery, the rate is between \EERnum{birth.school3.1.later.min} and \EERnum{birth.school3.1.later.max} times as high. The joint model of Section~\ref{sec:joint} shows the same timing without linking pregnancies to births. When we let its pregnancy ratio differ by months since conception, the interruption rate is \EERsplit{joint.gest.full.k2.m13.hr} times the rate without pregnancy in months 1 to 3, \EERsplit{joint.gest.full.k2.m49.hr} in months 4 to 9, and \EERsplit{joint.gest.full.k2.m10p.hr} from month 10 onward, and the three ratios are not statistically distinguishable (online appendix, Section B). Schooling disruption is therefore already present early in pregnancy and continues after birth. The rate is about as high in the conception month itself, \EERsplit{joint.gest.full.k2.k0.hr} times the rate without pregnancy, although we count that month as a month without pregnancy exposure (Section~\ref{sec:dates}).\footnote{The ratios by months since conception give the conception month its own term, which removes it from the months without pregnancy exposure. That change alone raises the single ratio of Section~\ref{sec:joint} from \EERjoint{results.full_linear.k2.hr} to \EERsplit{joint.k0only.full.k2.all.hr}.} These data therefore cannot show that the disruption begins only after conception. A student who leaves school may conceive sooner, or a short-lived event of the kind discussed in Section~\ref{sec:joint} may affect both. The students still enrolled at delivery are those who did not interrupt during pregnancy, and they may be more attached to school, so similar ratios before and after delivery do not show that caring for a newborn adds no difficulty. The profiles also cannot tell us why students interrupt during pregnancy. Health demands, responses by families or schools, preparation for parenting, and changing personal circumstances could each lead a pregnant student to leave school.

\subsection{Recorded exit as a secondary outcome}

Measured with the leaving date students report, instead of the enrollment calendar, the monthly rate of leaving school with pregnancy exposure is again about four times the rate without it. As in Section~\ref{sec:interruptions}, we compare months from sexual debut onward. Students with pregnancy exposure leave school at \RecExitHRPostDebut{} times the rate of students without it, close to the ratio of \EERInterruptionHR{} for sustained interruption (Table~\ref{tab:riskcomparison}). The ratio depends far more on which months form the comparison. Before sexual debut a student cannot be pregnant, and students leave school in only \RecExitRatePreDebut{} percent of those months. Counting them as months without pregnancy exposure lowers the comparison rate from \RecExitRatePostUnexp{} to \RecExitRateFullUnexp{} percent, while the rate with pregnancy exposure stays at \RecExitRateExposed{} percent. With these months included, students with pregnancy exposure appear to leave school at \RecExitHRFull{} times the rate of students without it, instead of \RecExitHRPostDebut{}. Adding the same months to the interruption comparison raises that ratio from \EERInterruptionHR{} to \EERnum{alternative.school3.full.hr}. Comparing pregnant students with all students, instead of with students who have begun having sex, therefore makes the ratio about 40 to 50 percent larger.

Each recorded exit takes its month from the high school leaving date that students report in their NLSY97 interviews, and that date can place an exit during pregnancy even when the student was out of school before she conceived. Of the \ExitGapExposed{} recorded exits in months with pregnancy exposure, \ExitGapConceive{} (\ExitGapShare{} percent) belong to students who conceived during an absence from school of at least four months that began before their recorded exit. Because their recorded exits fall after conception, they count as exits during pregnancy and raise the exit rate with pregnancy exposure. The enrollment calendar shows whether a student attended in each month, so it dates the start of the absence, before pregnancy exposure begins. For this reason the calendar is our main measure.

Modeling conception and recorded exit together, as in Section~\ref{sec:joint}, gives a ratio of about three (Table~\ref{tab:riskcomparison}). With one type, the exit rate after an observed conception month is \EERnum{model.full.linear.k1.hr} times the rate without pregnancy exposure, and with two types it is \EERMainHR{}. In the retained-onset sample, which drops the exposed women whose conception month we do not observe, the two-type ratio is \EERRestrictedHR{}.\footnote{Of the \EERsample{onset.exposed_women} exposed women in the recorded-exit sample after sexual debut, \EERsample{onset.retained_onset.women} have an observed conception month. Another \EERsample{onset.exposed_at_entry.women} are already exposed when we first observe them, and \EERsample{onset.missing_onset.women} become exposed across a gap in observation (online appendix, Section A). Flexible controls give two-type ratios of \EERnum{model.full.flexible.k2.hr} in the full sample and \EERnum{model.restricted.flexible.k2.hr} in the retained-onset sample.} As for interruption, allowing two types changes the ratio little because the types differ in conception and hardly at all in school leaving. The higher-conception type makes up \EERsample{latent.full_linear.high_share_pct_rounded} percent of women, and its conception intensity is \EERsample{latent.full_linear.conception_ratio_rounded} times the other type's, but its exit intensity without pregnancy exposure is only \EERsample{latent.full_linear.exit_ratio} times the other type's, a difference statistically indistinguishable from zero.

Most of the drop from the single-equation ratio of \RecExitHRPostDebut{} comes from separating the \EERsample{onset.without_retained_onset_women} exposed women whose conception month we do not observe. These women leave school at a higher rate. A single-equation model that gives them their own coefficient yields \SplitCloglogCleanHR{} after an observed conception month and \SplitCloglogUnobsHR{} for these women (online appendix, Section C). The joint estimates use the same recorded exits, so the dating problem described above affects them too.

\subsection{Enrollment and credentials after interruption}

The monthly rates in the earlier sections show how often students leave school, but a student who leaves can still return or earn a diploma or GED. In the main interruption sample, \EERadj{adjusted_results.school3.main_baseline.events} of the \EERadj{adjusted_results.school3.main_baseline.women} women have a sustained interruption, and we follow each of them for 24 months after it begins. A woman returns to school if the enrollment calendar shows her enrolled in any later school month, even if she leaves again. We also record whether she receives a diploma or GED during those 24 months. We leave out the few women whose outcome we cannot observe, \EERnum{followup.original_postdebut.school3.24.all.observed_resumption.missing_all} for return and \EERnum{followup.original_postdebut.school3.24.all.any_credential.missing_all} for credentials, instead of counting them as not returning or not earning a credential.\footnote{The online appendix (Section B) gives the coverage rules, outcomes at 12 months, and follow-up after recorded exit and under the alternative interruption definition.} Of the \EERadj{adjusted_results.school3.main_baseline.events} women, \EERnum{followup.original_postdebut.school3.24.1.observed_resumption.total} already had pregnancy exposure when they interrupted and \EERnum{followup.original_postdebut.school3.24.0.observed_resumption.total} did not, and a woman who conceives later stays in the second group.

Women who already had pregnancy exposure when they interrupted return to school less often (Table~\ref{tab:followup}). Within 24 months, \EERReturnExposed{} percent of them return, against \EERReturnUnexposed{} percent of women without pregnancy exposure at interruption.

Women who already had pregnancy exposure when they interrupted also earn a diploma or GED less often. Within 24 months, \EERCredentialExposed{} percent of them receive one, against \EERCredentialUnexposed{} percent of women without pregnancy exposure at interruption. Almost all of this difference is in diplomas, which \EERnum{followup.original_postdebut.school3.24.1.diploma.pct} percent of the first group receive against \EERnum{followup.original_postdebut.school3.24.0.diploma.pct} percent of the second. GED receipt is nearly the same, \EERnum{followup.original_postdebut.school3.24.1.ged.pct} against \EERnum{followup.original_postdebut.school3.24.0.ged.pct} percent.\footnote{Studies that compare teenage births with teenage pregnancies that ended in miscarriage also find the loss in diplomas. \citet{fletcher_education_2009} estimate that a teenage birth lowers diploma receipt by 8 to 9 percentage points among women in their early twenties and find no effect on GED receipt. \citet{hotz_teenage_2005} estimate a diploma loss of 5 to 12 percentage points by age 28, but a GED gain of 11 to 13 points offsets it, so their estimates for receipt of either credential are close to zero. Our follow-up ends two years after the interruption, so it cannot show whether a later rise in GED receipt closes the gap in credentials.} A woman can earn a GED without returning to school, so in both groups more women earn a credential than return. Even so, about three quarters of the women with pregnancy exposure at interruption and two thirds of the others have no credential two years after they interrupt.

\section{Conclusion}

Students who expect little benefit from school may be more likely both to conceive early and to leave school without a credential. Students who become pregnant would then leave school more often even if pregnancy had no effect, and a comparison of months with and without pregnancy would credit pregnancy with part of that difference. Following the timing-of-events approach of \citet{abbring_nonparametric_2003}, we use the timing of conception to separate the effect of pregnancy from differences between students. A persistent difference between students raises or lowers a student's interruption rate in every month, before and after she conceives, while pregnancy can change the rate only after conception.

We model conception and sustained interruption together and allow two latent types of women, each with its own conception rate and interruption rate. The model attributes a difference that is present before conception to a woman's type rather than to pregnancy. Because the types can conceive at different rates, it also uses when she conceives to infer her type. This separation requires two conditions. First, students must not change their schooling in anticipation of a pregnancy they plan. Second, apart from the controls and each woman's persistent type, nothing may affect both when she conceives and whether she would leave school without pregnancy.

Under these conditions, pregnancy raises a student's monthly interruption rate to \EERjoint{results.full_linear.k2.hr} times the rate she would have in the same month without pregnancy. In the main estimates, the two types differ in how quickly they conceive but hardly at all in how often they interrupt school, so allowing for them changes the estimate little. Two comparisons in single-equation models, which do not model conception, lower the ratio toward three. For women whose family background and test scores were measured before they conceived, adding these measures lowers it from \EERadj{adjusted_results.school3.later_cc_core.hr} to \EERadj{adjusted_results.school3.later_cc_rich.hr}. Counting pregnancies that ended without a live birth, which our conception dates can miss, lowers the main-sample ratio from \PregHistoryBaselineHR{} to \PregHistoryHR{}. The estimates therefore place the monthly interruption rate with pregnancy between about three and four times the rate without it.

The disruption is already present early in pregnancy and continues after birth. When we let the joint model's ratio differ by months since conception, it is \EERsplit{joint.gest.full.k2.m13.hr} in the first three months, long before a newborn needs care, \EERsplit{joint.gest.full.k2.m49.hr} in months 4 to 9, and \EERsplit{joint.gest.full.k2.m10p.hr} from month 10 onward. These monthly risks accumulate. In a descriptive comparison over the nine months after conception, roughly the length of a pregnancy, an estimated \EERadj{landmark_results.school3.1.cif_interruption_pct} percent of pregnant students interrupt school, compared with \EERadj{landmark_results.school3.0.cif_interruption_pct} percent of students of the same age and calendar month who have not yet conceived.

Students who interrupted at least one month after conception also return less often. Within two years, \EERReturnExposed{} percent of them return to school, against \EERReturnUnexposed{} percent of students who interrupted without pregnancy exposure. They also earn diplomas less often, \EERnum{followup.original_postdebut.school3.24.1.diploma.pct} against \EERnum{followup.original_postdebut.school3.24.0.diploma.pct} percent, while GED receipt is similar. Most students in both groups have no diploma or GED two years after they interrupt.

Because the risk is already elevated in the first months of pregnancy, support that begins only at childbirth would come after months of elevated risk. In the conception month, which we count as a month without pregnancy exposure, the interruption rate is \EERsplit{joint.gest.full.k2.k0.hr} times the rate without pregnancy, about as high as in the months after conception. A student who leaves school may conceive sooner, or a short-lived event such as a relationship change or a sudden loss of family income may affect both, so we cannot place the start of the disruption after conception. We also do not observe whether health demands, responses by families or schools, preparation for parenting, or changing personal circumstances lead pregnant students to interrupt.
\clearpage
\begingroup\singlespacing
\bibliographystyle{aer}
\bibliography{references}
\endgroup

\clearpage
\begin{table}[tbp]
\centering\singlespacing\small
\caption{Sample Characteristics by Pregnancy Exposure}\label{tab:summarystats}
\begin{threeparttable}
\begin{tabular*}{\textwidth}{@{\extracolsep{\fill}}@{}>{\raggedright\arraybackslash}p{3.2in}rrr@{}}
\toprule
Characteristic & \shortstack{Never exposed\\N = 2,266} & \shortstack{Ever exposed\\N = 562} & \shortstack{All women\\N = 2,828} \\
\midrule
Black (\%) & 25.7 & 46.8 & 29.9 \\
Hispanic (\%) & 18.9 & 23.0 & 19.7 \\
Mixed race, non-Hispanic (\%) & 1.0 & 1.6 & 1.1 \\
Non-Black, non-Hispanic (\%) & 54.4 & 28.6 & 49.3 \\
\addlinespace[5pt]
Mother's highest grade completed & 12.6 (2.7) & 11.3 (2.6) & 12.3 (2.8) \\
\addlinespace[5pt]
Both biological parents (\%) & 46.1 & 25.5 & 42.0 \\
Other two-parent household (\%) & 14.9 & 13.9 & 14.7 \\
Other household arrangement (\%) & 39.0 & 60.5 & 43.3 \\
\addlinespace[5pt]
ASVAB math--verbal percentile & 46.7 (27.6) & 26.6 (21.8) & 43.0 (27.7) \\
\addlinespace[5pt]
Age at first school-risk month & 16.4 (1.2) & 15.9 (1.2) & 16.3 (1.2) \\
School-risk months observed & 16.8 (10.3) & 19.4 (11.3) & 17.3 (10.5) \\
\addlinespace[5pt]
Interrupted schooling during observation (\%) & 19.9 & 46.4 & 25.1 \\
Diploma or GED by end of observation (\%) & 76.3 & 43.6 & 69.8 \\
\addlinespace[5pt]
Age at conception, exposed women & -- & 16.6 (1.4) & -- \\
\bottomrule
\end{tabular*}
\notetext{The sample contains the women in the post-debut school-risk months of the main three-school-month interruption model. A woman is ever exposed when at least one of her school-risk months in this sample falls after her measured conception, and never exposed otherwise. Entries are person-level percentages of the women in the column or means with standard deviations in parentheses. Race, household group, and mother's grade are measured at the first interview. ASVAB is the combined math and verbal percentile. Age at conception is reported for exposed women only. Background information is not imputed, and a row with a missing measure uses the women who have it. In this sample 228 women lack mother's grade, 10 lack the household group, and 560 lack the ASVAB score. Interruption and credential shares refer to each woman's retained observation window, so a credential earned after that window is not counted. Never exposed does not mean never pregnant, since 1,358 of the never-exposed women have a measured conception date. The differences describe the two exposure groups and are not effects of pregnancy.}
\end{threeparttable}
\end{table}

\clearpage
\begin{table}[tbp]
\centering\singlespacing\small
\caption{Pregnancy and Sustained Interruptions in Enrollment}\label{tab:interruption}
\begin{threeparttable}
\begin{tabular*}{\textwidth}{@{\extracolsep{\fill}}@{}>{\raggedright\arraybackslash}p{1.0in}>{\raggedright\arraybackslash}p{1.5in}rrrr@{}}
\toprule
Definition & Sample & \shortstack{HR\\{[95\% CI]}} & Women & \shortstack{School\\months} & Events \\
\midrule
Three school months & From sexual debut & \cellstack[r]{3.85\\{[3.29, 4.51]}} & 2,828 & 48,953 & 711 \\
\addlinespace[5pt]
Four calendar months & From sexual debut & \cellstack[r]{3.72\\{[3.18, 4.35]}} & 2,809 & 48,120 & 726 \\
\bottomrule
\end{tabular*}
\notetext{HR is the hazard ratio and CI its 95\% confidence interval. The first definition requires a continuous run of observed nonenrollment containing three September--May months. The second requires four calendar months including at least two September--May months. Events occur at the first school month in the qualifying run, after prior observed enrollment and without a credential by confirmation. The sample is drawn from the women who contribute eligible months to the recorded-exit sample. School months are September--May observations, while duration advances in calendar time from entry. All models use seven duration categories, school term, and quadratic age, with confidence intervals clustered by woman. Exposure uses the measured conception date with a one-month lag.}
\end{threeparttable}
\end{table}

\clearpage
\begin{table}[tbp]
\centering\singlespacing\small
\caption{Schooling over the Nine Calendar Months after Conception}\label{tab:landmarkscale}
\begin{threeparttable}
\begin{tabular*}{\textwidth}{@{\extracolsep{\fill}}@{}>{\raggedright\arraybackslash}p{3.35in}rr@{}}
\toprule
Measure & Pregnant students & Comparison students \\
\midrule
Interruption incidence (\%) & 23.1 & 7.2 \\
Extreme cases for periods that end early (\%) & 21.7--29.9 & 7.1--9.0 \\
\addlinespace[5pt]
Follow-up periods & 451 & 42,909 \\
Students & 451 & 2,658 \\
Periods with an observed interruption & 98 & 2,941 \\
Periods with an observed credential & 89 & 11,218 \\
Periods that end before nine months & 37 & 484 \\
\bottomrule
\end{tabular*}
\notetext{An interruption requires a continuous run of nonenrollment containing three September--May months. A pregnant student's period begins in the month after her measured conception and covers that month and the next eight calendar months, including summer. Only students whose conception month falls within their observed history contribute a pregnant period. A comparison period begins in the same calendar month for a student who has not yet conceived, and comparison periods are weighted to the pregnant periods' completed age in years and starting calendar month. Periods in an age and starting-month group without a period from the other group are excluded. A period starts only if the student has had no interruption or credential before its first month. A student can contribute many comparison periods and can later conceive, so one interruption or credential can appear in several comparison periods. A comparison period stays in its group if the student conceives during it. A diploma or GED is a competing event. A period ends early when the student's history ends first. The incidence assumes that students in periods that end early interrupt at the rate of students in the same group whose periods continue. The extreme cases assume that none or all of the periods that end early contain an interruption. Counts are unweighted. The comparison is descriptive, and the appendix gives the full accounting and construction rules.}
\end{threeparttable}
\end{table}

\clearpage
\begin{table}[tbp]
\centering\singlespacing\small
\caption{Pregnancy and Sustained Interruption in Matched Joint Models}\label{tab:matchedmodels}
\begin{threeparttable}
\begin{tabular*}{\textwidth}{@{\extracolsep{\fill}}@{}>{\raggedright\arraybackslash}p{2.25in}rrr@{}}
\toprule
Sample & Women & \cellstack[r]{{One type}\\{HR [95\% CI]}} & \cellstack[r]{{Two types}\\{HR [95\% CI]}} \\
\midrule
Full sample & 2,828 & \cellstack[r]{{3.91}\\{[3.30, 4.64]}} & \cellstack[r]{{3.76}\\{[3.00, 4.71]}} \\
Retained onset & 2,708 & \cellstack[r]{{3.92}\\{[3.31, 4.65]}} & \cellstack[r]{{3.78}\\{[3.02, 4.74]}} \\
\addlinespace[5pt]
Modified-onset grouping & 2,828 & \cellstack[r]{{3.90}\\{[3.24, 4.70]}} & \cellstack[r]{{3.72}\\{[2.94, 4.70]}} \\
\bottomrule
\end{tabular*}
\notetext{HR is the hazard ratio for pregnancy exposure after a retained onset, $\exp(\gamma)$, with a model-based 95\% interval. The outcome is the start of a continuous run of nonenrollment containing three September--May months, and exposure begins one month after measured conception. Both hazards use linear terms for age and for months since first eligible enrollment, capped at 48, and each pair of one-type and two-type models uses the same observations. In the full sample, exposed months without an observed conception month have their own coefficient, $\delta$. The retained-onset sample drops the women already exposed when we first observe them. Summer months enter only the conception hazard. The final row keeps the full sample but removes the conception event and conception-risk months of women whose conception month is retained and whose debut date was set from that conception, so their exposed months receive $\delta$. The appendix reports sample counts, flexible controls, and type diagnostics.}
\end{threeparttable}
\end{table}

\clearpage
\begin{table}[tbp]
\centering\singlespacing\small
\caption{Pregnancy and Interruption after Accounting for Observed Background}\label{tab:adjustedschooling}
\begin{threeparttable}
\begin{tabular*}{\textwidth}{@{\extracolsep{\fill}}@{}>{\raggedright\arraybackslash}p{2.0in}rrrr@{}}
\toprule
Sample and controls & \shortstack{Three school months\\HR [95\% CI]} & Women & \shortstack{Four calendar months\\HR [95\% CI]} & Women \\
\midrule
Post-debut sample, baseline & \cellstack[r]{3.85\\{[3.29, 4.51]}} & 2,828 & \cellstack[r]{3.72\\{[3.18, 4.35]}} & 2,809 \\
\quad Race and calendar controls & \cellstack[r]{3.97\\{[3.36, 4.69]}} & 2,828 & \cellstack[r]{3.79\\{[3.21, 4.47]}} & 2,809 \\
\addlinespace[5pt]
Recent enrollment, baseline & \cellstack[r]{3.79\\{[3.16, 4.56]}} & 2,672 & \cellstack[r]{3.61\\{[3.01, 4.35]}} & 2,655 \\
\addlinespace[5pt]
Supported histories, baseline & \cellstack[r]{3.63\\{[2.91, 4.53]}} & 2,571 & \cellstack[r]{3.46\\{[2.78, 4.32]}} & 2,553 \\
\quad Race and calendar controls & \cellstack[r]{3.79\\{[3.02, 4.75]}} & 2,571 & \cellstack[r]{3.56\\{[2.84, 4.46]}} & 2,553 \\
\addlinespace[5pt]
Later sample, race and calendar controls & \cellstack[r]{3.52\\{[2.83, 4.38]}} & 2,313 & \cellstack[r]{3.31\\{[2.66, 4.12]}} & 2,291 \\
\addlinespace[5pt]
Later complete cases, baseline & \cellstack[r]{3.53\\{[2.71, 4.61]}} & 1,733 & \cellstack[r]{3.18\\{[2.42, 4.16]}} & 1,715 \\
\quad Race and calendar controls & \cellstack[r]{3.70\\{[2.79, 4.91]}} & 1,733 & \cellstack[r]{3.28\\{[2.46, 4.36]}} & 1,715 \\
\quad Family background and ASVAB added & \cellstack[r]{2.85\\{[2.12, 3.82]}} & 1,733 & \cellstack[r]{2.47\\{[1.83, 3.31]}} & 1,715 \\
\bottomrule
\end{tabular*}
\notetext{Hazard ratios with woman-clustered 95\% intervals. Baseline controls are seven duration categories, school term, and quadratic age. Race and calendar controls add four race and ethnicity groups and calendar-year indicators with 2003--2005 pooled. Recent enrollment requires each woman who conceives or is exposed to pregnancy during observation to be enrolled in all three school months before conception. Women already pregnant when we first observe them are therefore excluded. Supported histories require retained onset and three earlier school-risk months among women with observed conception or exposure. Women without a pregnancy remain as comparisons. Later risk months begin in September 1998, after first-round interviews and ASVAB testing, and the later sample excludes women whose measured conception falls on or before the later of June 1998 and the first interview. The last three rows use identical complete cases. Family controls are the mother's highest grade completed, household composition, and the ASVAB percentile. Missing covariates are not imputed. The first outcome requires a continuous run of nonenrollment containing three September--May months, and the second requires four calendar months including at least two September--May months. Exposure begins one month after measured conception. The appendix reports construction details and full coefficients for every row except recent enrollment.}
\end{threeparttable}
\end{table}

\clearpage
\begin{table}[tbp]
\centering\singlespacing\small
\caption{Pregnancy and Recorded Exit}\label{tab:riskcomparison}
\begin{threeparttable}
\begin{tabular*}{\textwidth}{@{\extracolsep{\fill}}@{}>{\raggedright\arraybackslash}p{3.3in}rr@{}}
\toprule
Measure & All eligible months & From sexual debut \\
\midrule
Women & 4,235 & 2,951 \\
Person-months & 147,956 & 61,649 \\
Recorded exits & 680 & 594 \\
Credential completions & 3,232 & 2,139 \\
\addlinespace[5pt]
Exposed person-months & 8,593 & 8,593 \\
Exits in exposed months & 246 & 246 \\
Monthly exit rate, exposed & 2.86\% & 2.86\% \\
\addlinespace[5pt]
Unexposed person-months & 139,363 & 53,056 \\
Exits in unexposed months & 434 & 348 \\
Monthly exit rate, unexposed & 0.31\% & 0.66\% \\
\addlinespace[5pt]
Adjusted hazard ratio & 5.67 & 3.75 \\
95\% confidence interval & [4.73, 6.80] & [3.15, 4.47] \\
\addlinespace[5pt]
Joint model, one type & -- & \cellstack[r]{3.35\\{[2.75, 4.08]}} \\
Joint model, two types & -- & \cellstack[r]{3.23\\{[2.48, 4.21]}} \\
Joint model, retained onset, one type & -- & \cellstack[r]{3.26\\{[2.67, 3.97]}} \\
Joint model, retained onset, two types & -- & \cellstack[r]{3.08\\{[2.36, 4.04]}} \\
\bottomrule
\end{tabular*}
\notetext{Each column retains eligible enrolled months through recorded exit or censoring. The column from sexual debut includes the measured month of sexual debut and later months. Exposure begins one month after measured conception. The monthly exit rate is recorded exits divided by person-months in the indicated exposure category. The adjusted hazard ratios come from separate complementary log-log models with duration categories, school term, and quadratic age. Their confidence intervals cluster by woman. Debut dates are corrected or imputed when debut falls in or after measured conception or is missing while conception is known, as described in the text. The last four rows report hazard ratios for pregnancy exposure after a retained onset, with model-based 95\% intervals, from joint models of conception and recorded exit estimated on the months from sexual debut onward. A dash marks a model not estimated on all eligible months. Both equations control for age and for months since first eligible enrollment, capped at 48. In the full post-debut joint-model sample, months with pregnancy exposure but no observed conception month have their own coefficient. The retained-onset models use 2,749 women and exclude exposed women whose conception month we do not observe.}
\end{threeparttable}
\end{table}

\clearpage
\begin{table}[tbp]
\centering\singlespacing\small
\caption{Schooling within Two Years of a Sustained Interruption}\label{tab:followup}
\begin{threeparttable}
\begin{tabular*}{\textwidth}{@{\extracolsep{\fill}}@{}>{\raggedright\arraybackslash}p{2.25in}rrrr@{}}
\toprule
Outcome & \shortstack{Exposed\\Percent} & Yes/observed & \shortstack{Unexposed\\Percent} & Yes/observed \\
\midrule
Returned to enrollment & 20.4 & 52/255 & 27.6 & 123/445 \\
Diploma or GED & 26.2 & 68/260 & 35.0 & 157/448 \\
\addlinespace[5pt]
\quad Diploma & 11.5 & 30/260 & 19.6 & 88/448 \\
\quad GED & 14.6 & 38/260 & 15.4 & 69/448 \\
\bottomrule
\end{tabular*}
\notetext{The sample contains 261 exposed and 450 unexposed women with a three-school-month interruption at or after measured sexual debut. Exposure is fixed at interruption using a one-month lag after conception. Return requires September--May nonenrollment at or after the event followed by enrollment in a later September--May month. Enrollment denominators require complete September--May status and interview and calendar coverage through 24 months. Credential denominators require interview coverage and use recorded diploma and GED dates. Each percentage uses the displayed numerator and observed denominator. These comparisons describe women who interrupted schooling and do not identify pregnancy effects on eventual attainment. The online appendix's post-debut follow-up table reports both horizons, missing counts, and full conventions.}
\end{threeparttable}
\end{table}

\clearpage

\clearpage

\clearpage

\clearpage
\setcounter{section}{0}\setcounter{table}{0}\setcounter{figure}{0}
\setcounter{equation}{0}\setcounter{footnote}{0}
\renewcommand{\thesection}{\Alph{section}}
\renewcommand{\thetable}{A\arabic{table}}
\renewcommand{\theequation}{A\arabic{equation}}
\renewcommand{\theHsection}{appendix.\Alph{section}}
\renewcommand{\theHtable}{appendix.\arabic{table}}
\renewcommand{\theHequation}{appendix.\arabic{equation}}
\let\appendixtable\table
\let\endappendixtable\endtable
\renewenvironment{table}[1][]{\appendixtable[H]}{\endappendixtable}
\begin{center}
{\Large Online Appendix}\\[0.4em]
{\large Pregnancy and Interrupted Schooling}
\end{center}
\section{Recorded-Exit Dates and Sample Construction}

The person-level file from the National Longitudinal Survey of Youth 1997 (NLSY97) contains records for \EERsample{person.women} female respondents. We classify \EERsample{person.recorded_exit} women as having a dropout or credential endpoint, comprising \EERsample{person.dropout} dropouts and \EERsample{person.completion} completions. Estimation also requires eligible age and enrollment observations, including the event month if an event is to be counted.

The broad enrollment panel contains \EERsample{broad.women} women, \EERsample{broad.months} eligible person-months, \EERsample{broad.exits} recorded exits, and \EERsample{broad.completions} completions. Of these women, \EERsample{person.panel_with_exit} have a classified dropout or credential endpoint and \EERsample{person.panel_without_exit} do not. Another \EERsample{person.nopanel_with_exit} women with such an endpoint and \EERsample{person.nopanel_without_exit} without one contribute no eligible months. Having a classified endpoint and contributing to the enrollment panel are therefore separate conditions.

We retain recorded school enrollment from the thirteenth birthday month through the earliest of the twenty-first birthday month, observed dropout, or credential receipt. Missing enrollment does not count as attendance. An event contributes to estimation only if its month is eligible. Keeping months from sexual debut onward gives \EERsample{postdebut.women} women and \EERsample{postdebut.months} person-months. This post-debut sample contains \EERsample{postdebut.exits} recorded exits, \EERsample{postdebut.completions} completions, and \EERsample{postdebut.exposed_months} exposed months.

The NLSY97 constructs the school-leaving date $L_i$ from reports across survey rounds and supplies it as \texttt{CVC\_HS\_LEFT\_DATE} (Z90848.00). For graduates, this variable uses the graduation date. For nongraduates, it uses the date the last grade was completed or, when that date is unavailable, the date attendance stopped. The survey also supplies the highest grade completed at leaving, $H_i$, in \texttt{CVC\_HS\_LEFT\_HGC} (Z90849.00).\footnote{The NLSY97 codebook entry for Z90848.00 documents the three-step date construction, and the entry for Z90849.00 defines grade completed at leaving. See also the NLSY97 \href{https://nlsinfo.org/content/cohorts/nlsy97/other-documentation/codebook-supplement/appendix-1-education-variable}{Appendix 1, Education Variable Creation}, under ``Date left high school and highest high school grade completed/attended (collapsed).''}

We construct the credential date $G_i$ as the earlier valid date from \texttt{CVC\_HS\_DIPLOMA} (Z90842.00) and \texttt{CVC\_GED} (Z90841.00), which report receipt of a high school diploma and a General Educational Development (GED) credential. These survey variables express dates as continuous month numbers. We convert negative missing-value codes to missing and retain only positive dates. If neither credential has a valid date, $G_i$ is missing. The NLSY97 can record credential receipt without a valid receipt date, so a missing $G_i$ does not establish that no credential was reported.\footnote{The ``Date received diploma or degree (collapsed)'' section of NLSY97 Appendix 1 explains that a reported credential without a receipt date can have a date coded $-3$. Our date-based construction does not assign a completion month from an undated receipt report.}

We use these survey variables to define the dropout date. For an observed leaving date and a valid grade report, we set
\begin{equation}
D_i=L_i\quad\text{if }H_i<12\text{ and }(G_i\text{ is missing or }G_i>L_i+1).
\end{equation}
The grade cutoff and one-month allowance are our classification rules. A diploma or GED dated on or before the month following departure prevents that departure from counting as a dropout. A credential dated later does not remove the earlier dropout. Missing, invalid, or ungraded grade reports do not satisfy the definition.

Conception can define exposure if it precedes the earlier observed $D_i$ or $G_i$. If both dates are missing, conception can still define exposure during eligible observed months.

The recorded leaving date and monthly enrollment calendar can identify different transitions. An absence before recorded leaving may be followed by return. A leaving date based on grade completion can also precede the end of attendance. Our separate interruption measure uses the monthly calendar, with rules for summer, short absences, and confirmation.

Conception dates come from current-pregnancy reports and recorded live-birth dates less nine months. We exclude candidate dates before age 13 and choose the earliest remaining date. Exposure begins one month later in retained observations, provided conception meets the cutoff above. The measured date may omit an earlier pregnancy ending without a live birth or refer to a different pregnancy from a later recorded birth. A woman's eventual live-birth status therefore does not establish the outcome of her measured pregnancy.

We classify exposed women by whether conception appears in their retained observations (Table~\ref{tab:sample}). Women with a retained onset have a conception-month observation followed by at least one exposed month. Other women are already exposed at their first observation or become exposed across a gap that omits the conception month. Observation can begin late or start again after a gap, so exposure at entry need not imply conception before the calendar begins. Some women conceive in their last retained month and have no later exposed month under the one-month lag. They belong to the never-exposed group, and their conceptions enter the conception likelihood.

\begin{table}[tbp]
\centering\singlespacing\small
\caption{Recorded-Exit Sample by Observation of Pregnancy Onset}\label{tab:sample}
\begin{threeparttable}
\begin{tabular*}{\textwidth}{@{\extracolsep{\fill}}@{}>{\raggedright\arraybackslash}p{1.9in}rrrr@{}}
\toprule
Group & Women & Person-months & Exposed months & Exits \\
\midrule
Conception month retained & 423 & 11,528 & 5,700 & 153 \\
Exposed at entry & 135 & 2,130 & 2,130 & 58 \\
Onset between retained months & 67 & 1,528 & 763 & 35 \\
Never exposed & 2,326 & 46,463 & 0 & 348 \\
Total & 2,951 & 61,649 & 8,593 & 594 \\
\bottomrule
\end{tabular*}
\notetext{The sample retains eligible enrolled months from sexual debut onward. Exposure begins one month after measured conception. The onset groups refer to whether the conception-month row survives the sample restrictions. Exposed-at-entry women need not have conceived before the enrollment calendar begins. Never exposed includes women who conceive in their last retained month and have no exposed month under the lag. Exits are recorded school leaving without a diploma or GED by the following month.}
\end{threeparttable}
\end{table}

\section{Sustained Interruption and Subsequent Schooling}

\subsection{Direct measurement from the enrollment calendar}

The monthly school enrollment calendar codes attendance and vacation separately. We treat codes 2 and 4 as enrollment, codes 1 and 5 as observed nonenrollment, and code 6 or missing responses as unknown. Entry is the first enrolled month from the thirteenth through the twenty-first birthday month, before a credential. We measure elapsed duration from this entry month. An absence that ends before meeting either interruption definition does not end the history or restart the duration measure. We stop observation at the month before enrollment status first becomes unknown and do not resume the history after that gap.

A sustained interruption requires enrollment immediately before a continuous absence. The main definition requires at least three September--May months within the absence. The alternative requires at least four consecutive calendar months, including at least two September--May months. Both date the interruption to the first September--May month of the absence. A diploma or GED received on or before confirmation prevents the absence from qualifying. A credential received afterward does not erase the interruption.

We allow an interruption to begin no later than five months before the earliest observation limit. That limit is the calendar endpoint, last interview, twenty-first birthday month, or month before status first becomes unknown. An absence beginning in May reaches its third missed school month in October if nonenrollment continues through summer. We therefore reserve the final five calendar months for confirming interruptions and exclude those months from the regression's risk period. The same allowance applies to every history, including those without an interruption. The definition dates a sustained absence retrospectively. It does not establish at departure that a student has left permanently.

We follow consecutive calendar months from entry through interruption, a credential, or the observation limit with its confirmation allowance. Measured conception defines exposure one month later, without the recorded-exit analysis's conception cutoff. Single-equation regressions use September--May observations because interruptions cannot begin in summer under either definition. Elapsed duration still advances through summer. Restricting the main sample to months from sexual debut onward does not reset duration at debut. Controls are seven duration categories, quadratic age, and a September--December versus January--May term indicator.

Both definitions are estimated on each of three samples (Table~\ref{tab:interruptionall}). The sample of all eligible months includes months before debut. The post-debut sample begins at the constructed debut date. The recent-enrollment sample requires recorded enrollment in the three school months before conception for women with a conception or pregnancy exposure during the observed risk period. This requirement also excludes histories already exposed at entry. Women without a pregnancy in that period remain in the comparison group. This restriction selects histories with recent enrollment but does not eliminate anticipation or changing circumstances shared by conception and schooling.

\begin{table}[tbp]
\centering\singlespacing\small
\caption{Complete Comparisons for the Two Enrollment-Interruption Measures}\label{tab:interruptionall}
\begin{threeparttable}
\begin{tabular*}{\textwidth}{@{\extracolsep{\fill}}@{}>{\raggedright\arraybackslash}p{1.0in}>{\raggedright\arraybackslash}p{1.5in}rrrr@{}}
\toprule
Definition & Sample & \shortstack{HR\\{[95\% CI]}} & Women & \shortstack{School\\months} & Events \\
\midrule
Three school months & All eligible months & \cellstack[r]{5.52\\{[4.71, 6.46]}} & 4,205 & 120,582 & 910 \\
 & From sexual debut & \cellstack[r]{3.85\\{[3.29, 4.51]}} & 2,828 & 48,953 & 711 \\
 & Recent enrollment & \cellstack[r]{3.79\\{[3.16, 4.56]}} & 2,672 & 46,811 & 628 \\
\addlinespace[5pt]
Four calendar months & All eligible months & \cellstack[r]{5.37\\{[4.59, 6.28]}} & 4,205 & 119,261 & 950 \\
 & From sexual debut & \cellstack[r]{3.72\\{[3.18, 4.35]}} & 2,809 & 48,120 & 726 \\
 & Recent enrollment & \cellstack[r]{3.61\\{[3.01, 4.35]}} & 2,655 & 46,059 & 641 \\
\bottomrule
\end{tabular*}
\notetext{HR is the hazard ratio and CI its 95\% confidence interval. The first definition requires a continuous run of observed nonenrollment containing three September--May months. The second requires four calendar months including at least two September--May months. Events occur at the first school month in the qualifying run, after prior observed enrollment and without a credential by confirmation. The sample is drawn from the women who contribute eligible months to the recorded-exit sample. School months are September--May observations, while duration advances in calendar time from entry. All models use seven duration categories, school term, and quadratic age, with confidence intervals clustered by woman. Exposure uses the measured conception date with a one-month lag. Recent enrollment requires enrollment in the three preceding school months for women with pregnancy during their observed risk period. This last restriction selects histories and is a sensitivity comparison, not an exogeneity test.}
\end{threeparttable}
\end{table}

\subsection{Joint estimation with conception observed through summer}\label{sec:jointestimation}

The joint model uses the same women and school-risk months as the main post-debut interruption model, adding summer observations for conception (Table~\ref{tab:interruptionjointcounts}). In this full post-debut sample, \EERjoint{counts.full.g4.summer_conceptions} of \EERjoint{counts.full.g4.conception_events} observed conceptions occur during summer. We keep these events and all observed conception-risk months. Summer months contribute nothing to the schooling log likelihood because an interruption cannot begin then.

\begin{table}[tbp]
\centering\singlespacing\small
\caption{Calendar and Event Counts in the Joint Interruption Samples}\label{tab:interruptionjointcounts}
\begin{threeparttable}
\begin{tabular*}{\textwidth}{@{\extracolsep{\fill}}@{}>{\raggedright\arraybackslash}p{1.35in}rrrrrr@{}}
\toprule
Sample & Women & \cellstack[r]{{Calendar}\\{months}} & \cellstack[r]{{School}\\{months}} & Interruptions & Conceptions & \cellstack[r]{{Summer}\\{conceptions}} \\
\midrule
Full sample & 2,828 & 64,598 & 48,953 & 711 & 478 & 114 \\
Retained onset & 2,708 & 62,478 & 47,408 & 648 & 478 & 114 \\
\bottomrule
\end{tabular*}
\notetext{The full sample is the main post-debut interruption sample. Calendar months retain summer for conception. School months are September--May and supply the interruption likelihood. Retained onset removes women already exposed at entry. The histories are continuous, so no conception is skipped across a retained-panel gap. Some women conceive in their last observed month and supply a conception event without a later exposed month.}
\end{threeparttable}
\end{table}

The full post-debut joint sample contains \EERjoint{counts.full.g1.women} women with a retained onset and later exposure, \EERjoint{counts.full.g2.women} already exposed at entry, and \EERjoint{counts.full.g0.women} without an exposed month. The last group includes \EERjoint{counts.full.g0.onset_without_exposure_women} women who conceive at the end of observation. They contribute a conception event but no later exposed month. Exposure here counts every calendar month. For \the\numexpr\EERjoint{counts.full.g1.women}+\EERjoint{counts.full.g2.women}-\EERsum{count.main.exposed.women}\relax{} women with a retained onset, every exposed month falls in summer. The main text's table of sample characteristics counts exposure only in school-risk months, so it lists \EERsum{count.main.exposed.women} ever-exposed women rather than \the\numexpr\EERjoint{counts.full.g1.women}+\EERjoint{counts.full.g2.women}\relax. The retained-onset sample excludes women exposed at entry. Because these histories are continuous, conception cannot be skipped across a gap as it can in the recorded-exit sample.

We calculate the probability of each woman's observed history under each type. We then combine those probabilities using the estimated population share of each type. This calculation forms her likelihood contribution.

An isolated copy of the estimator uses an indicator to determine which months contribute to the schooling likelihood. Setting this indicator to one in every month exactly reproduces the saved recorded-exit likelihoods. With heterogeneity loadings set to zero, the joint log likelihood equals the sum of the separate conception and school-month interruption log likelihoods. An independent calculation across women and types also agrees when loadings are nonzero. Separate checks reproduce the selected saved likelihoods and verify that consistently relabeling the types leaves them unchanged.

We fit each two-type model from seven fixed starting values and keep the solution whose likelihood is at least as large as that of every other start. That solution passes the convergence checks only if the optimizer reports convergence and the norm of the gradient divided by one plus the absolute log likelihood does not exceed $10^{-5}$. Its covariance matrix must also be estimated from the model rather than supplied by a fallback, and it must be positive definite, with its smallest eigenvalue above $10^{-12}$ times its largest. When the selected solution fails these checks, we withhold its coefficients and intervals and retain every candidate in the replication files. The recorded-exit models use the same checks.

Table~\ref{tab:interruptionjoint} reports the one-type and two-type pregnancy estimates for each sample and control set. Table~\ref{tab:interruptionjointfit} gives the likelihoods and the Akaike and Bayesian information criteria (AIC and BIC), and Table~\ref{tab:interruptionjointtypes} the fitted type distributions. Tables~\ref{tab:interruptionjointcoefficientslinearall}--\ref{tab:interruptionjointcoefficientsflexibleschooling} report the complete two-type coefficient vectors.

\begin{table}[tbp]
\centering\singlespacing\small
\caption{Matched Joint Models of Conception and Sustained Interruption}\label{tab:interruptionjoint}
\begin{threeparttable}
\begin{tabular*}{\textwidth}{@{\extracolsep{\fill}}@{}>{\raggedright\arraybackslash}p{2.25in}rrr@{}}
\toprule
Sample and controls & Women & One type & Two types \\
\midrule
Full sample, linear & 2,828 & \cellstack[r]{{3.91}\\{[3.30, 4.64]}} & \cellstack[r]{{3.76}\\{[3.00, 4.71]}} \\
Retained onset, linear & 2,708 & \cellstack[r]{{3.92}\\{[3.31, 4.65]}} & \cellstack[r]{{3.78}\\{[3.02, 4.74]}} \\
\addlinespace[5pt]
Full sample, flexible & 2,828 & \cellstack[r]{{3.81}\\{[3.21, 4.53]}} & \cellstack[r]{{3.56}\\{[2.80, 4.52]}} \\
Retained onset, flexible & 2,708 & \cellstack[r]{{3.82}\\{[3.21, 4.53]}} & \cellstack[r]{{3.56}\\{[2.79, 4.53]}} \\
\addlinespace[5pt]
Full sample, modified-onset grouping & 2,828 & \cellstack[r]{{3.90}\\{[3.24, 4.70]}} & \cellstack[r]{{3.72}\\{[2.94, 4.70]}} \\
Later complete cases, linear & 1,733 & \cellstack[r]{{3.85}\\{[2.92, 5.09]}} & \cellstack[r]{{3.64}\\{[2.58, 5.13]}} \\
Later complete cases, background controls & 1,733 & \cellstack[r]{{3.24}\\{[2.42, 4.33]}} & -- \\
\bottomrule
\end{tabular*}
\notetext{Cells report hazard ratios for pregnancy exposure after a retained onset, with model-based 95\% intervals. The schooling outcome is the first missed school month of a run containing three nonenrolled school months. Summer observations remain in the conception likelihood and have structurally zero schooling-event risk. A dash indicates that the selected two-type fit failed the strict convergence checks, and its coefficient and interval are withheld. Linear controls use age and capped elapsed duration. Flexible controls use duration categories, quadratic age, and school term in both hazards, with a summer term only in conception. The last three rows use the modified-onset grouping, in which women with a retained conception month whose debut date we set from that conception contribute no conception event, and their exposed school months, if any, receive the absent-onset coefficient. Background controls add race, calendar period, mother's highest grade, household structure, and the Armed Services Vocational Aptitude Battery (ASVAB) percentile to both hazards on the later complete cases.}
\end{threeparttable}
\end{table}

\begin{table}[tbp]
\centering\singlespacing\small
\caption{Likelihood and Information Criteria for the Joint Interruption Models}\label{tab:interruptionjointfit}
\begin{threeparttable}
\begin{tabular*}{\textwidth}{@{\extracolsep{\fill}}@{}>{\raggedright\arraybackslash}p{2.0in}rrrr@{}}
\toprule
Sample and controls & Types & Log likelihood & AIC & BIC \\
\midrule
Full sample, linear & 1 & -6299.05 & 12614.10 & 12661.68 \\
 & 2 & -6281.43 & 12584.85 & 12650.27 \\
\addlinespace[5pt]
Retained onset, linear & 1 & -6039.67 & 12093.34 & 12134.67 \\
 & 2 & -6022.07 & 12064.14 & 12123.18 \\
\addlinespace[5pt]
Full sample, flexible & 1 & -6243.16 & 12532.31 & 12669.10 \\
 & 2 & -6220.71 & 12493.42 & 12648.05 \\
\addlinespace[5pt]
Retained onset, flexible & 1 & -5995.36 & 12034.73 & 12164.61 \\
 & 2 & -5972.92 & 11995.83 & 12143.43 \\
\addlinespace[5pt]
Full sample, modified-onset grouping & 1 & -5891.54 & 11799.07 & 11846.65 \\
 & 2 & -5891.18 & 11804.36 & 11869.78 \\
\addlinespace[5pt]
Later complete cases, linear & 1 & -2981.13 & 5978.27 & 6021.93 \\
 & 2 & -2980.69 & 5983.38 & 6043.42 \\
\addlinespace[5pt]
Later complete cases, background controls & 1 & -2880.00 & 5823.99 & 5998.64 \\
 & 2 & -- & -- & -- \\
\bottomrule
\end{tabular*}
\notetext{AIC and BIC are reported only for numerically validated fits. BIC uses the number of women. Lower values favor a model within a fixed sample and control set. A dash denotes a failed strict convergence check, not a zero effect or evidence against two types. The replication retains the candidate likelihood and all starting-value attempts for failed fits. Neither criterion evaluates the causal identifying assumptions.}
\end{threeparttable}
\end{table}

\begin{table}[tbp]
\centering\singlespacing\small
\caption{Persistent Heterogeneity in the Joint Interruption Models}\label{tab:interruptionjointtypes}
\begin{threeparttable}
\begin{tabular*}{\textwidth}{@{\extracolsep{\fill}}@{}>{\raggedright\arraybackslash}p{2.1in}rrr@{}}
\toprule
Sample and controls & \cellstack[r]{{Higher-conception}\\{type share (\%)}} & \cellstack[r]{{Conception}\\{intensity ratio}} & \cellstack[r]{{Interruption}\\{intensity ratio}} \\
\midrule
Full sample, linear & 3.42 & 61.13 & 1.18 \\
Retained onset, linear & 3.41 & 61.19 & 1.16 \\
\addlinespace[5pt]
Full sample, flexible & 4.61 & 62.42 & 1.24 \\
Retained onset, flexible & 4.61 & 62.42 & 1.24 \\
\addlinespace[5pt]
Full sample, modified-onset grouping & 1.78 & 5.31 & 5.40 \\
Later complete cases, linear & 1.55 & 6.65 & 9.42 \\
Later complete cases, background controls & -- & -- & -- \\
\bottomrule
\end{tabular*}
\notetext{Each row describes the fitted two-type distribution. Types are oriented by conception intensity. The share belongs to the higher-conception type, and both ratios compare that type with the other type at the same observed covariates and exposure status. These are baseline type differences, not pregnancy effects. A dash withholds the latent distribution when the selected fit fails strict convergence. The coefficient tables report the underlying loadings and model-based standard errors under the original estimation labels. The last three rows use the modified-onset grouping, as in the matched comparisons.}
\end{threeparttable}
\end{table}

\begin{table}[tbp]
\centering\singlespacing\small
\caption{Joint Interruption Coefficients with Linear Controls}\label{tab:interruptionjointcoefficientslinearall}
\begin{threeparttable}
\begin{tabular*}{\textwidth}{@{\extracolsep{\fill}}@{}>{\raggedright\arraybackslash}p{2.65in}rr@{}}
\toprule
Parameter & Full sample & Retained onset \\
\midrule
\textit{Conception equation} &  &  \\
Age centered at sixteen & \cellstack[r]{{0.167}\\{(0.057)}} & \cellstack[r]{{0.167}\\{(0.057)}} \\
Elapsed duration & \cellstack[r]{{0.031}\\{(0.053)}} & \cellstack[r]{{0.031}\\{(0.053)}} \\
Constant & \cellstack[r]{{-1.113}\\{(0.280)}} & \cellstack[r]{{-5.225}\\{(0.160)}} \\
\textit{Interruption equation} &  &  \\
Age centered at sixteen & \cellstack[r]{{0.238}\\{(0.039)}} & \cellstack[r]{{0.226}\\{(0.041)}} \\
Elapsed duration & \cellstack[r]{{0.032}\\{(0.036)}} & \cellstack[r]{{0.038}\\{(0.037)}} \\
Exposure without retained onset & \cellstack[r]{{1.375}\\{(0.143)}} & -- \\
Constant & \cellstack[r]{{-4.742}\\{(0.296)}} & \cellstack[r]{{-4.904}\\{(0.096)}} \\
\textit{Pregnancy and latent distribution} &  &  \\
Pregnancy exposure, retained onset & \cellstack[r]{{1.325}\\{(0.115)}} & \cellstack[r]{{1.331}\\{(0.115)}} \\
Conception loading & \cellstack[r]{{-4.113}\\{(0.226)}} & \cellstack[r]{{4.114}\\{(0.226)}} \\
Interruption loading & \cellstack[r]{{-0.162}\\{(0.298)}} & \cellstack[r]{{0.147}\\{(0.297)}} \\
Mixing logit & \cellstack[r]{{3.341}\\{(0.234)}} & \cellstack[r]{{-3.342}\\{(0.234)}} \\
\bottomrule
\end{tabular*}
\notetext{Two-type log-intensity coefficients and model-based standard errors. Both equations and the latent distribution belong to one likelihood. Type one is normalized to zero and type two to one. The mixing logit gives the log odds of type two. An absent-onset exposure term appears only in the full sample. A dash denotes an absent term or a model withheld for failed strict convergence. Type labels follow each fit's estimation order, so the constants, the signs of the loadings, and the mixing logit can differ between columns. Table~\ref{tab:interruptionjointtypes} orients the types by conception intensity. All one-type coefficient vectors and failed-fit candidates are retained in the replication exports.}
\end{threeparttable}
\end{table}

\begin{table}[tbp]
\centering\singlespacing\small
\caption{Joint Interruption Coefficients with Flexible Controls, Conception}\label{tab:interruptionjointcoefficientsflexibleconception}
\begin{threeparttable}
\begin{tabular*}{\textwidth}{@{\extracolsep{\fill}}@{}>{\raggedright\arraybackslash}p{2.65in}rr@{}}
\toprule
Parameter & Full sample & Retained onset \\
\midrule
\textit{Conception equation} &  &  \\
Duration, months 7--12 & \cellstack[r]{{0.826}\\{(0.229)}} & \cellstack[r]{{0.826}\\{(0.229)}} \\
Duration, months 13--18 & \cellstack[r]{{1.003}\\{(0.275)}} & \cellstack[r]{{1.004}\\{(0.275)}} \\
Duration, months 19--24 & \cellstack[r]{{0.828}\\{(0.302)}} & \cellstack[r]{{0.829}\\{(0.302)}} \\
Duration, months 25--36 & \cellstack[r]{{0.877}\\{(0.290)}} & \cellstack[r]{{0.878}\\{(0.290)}} \\
Duration, months 37--48 & \cellstack[r]{{0.930}\\{(0.300)}} & \cellstack[r]{{0.931}\\{(0.300)}} \\
Duration, over 48 months & \cellstack[r]{{0.896}\\{(0.312)}} & \cellstack[r]{{0.897}\\{(0.312)}} \\
January--May & \cellstack[r]{{0.058}\\{(0.112)}} & \cellstack[r]{{0.058}\\{(0.112)}} \\
June--August & \cellstack[r]{{0.050}\\{(0.130)}} & \cellstack[r]{{0.050}\\{(0.130)}} \\
Age centered at sixteen & \cellstack[r]{{0.002}\\{(0.073)}} & \cellstack[r]{{0.002}\\{(0.073)}} \\
Centered age squared & \cellstack[r]{{0.083}\\{(0.025)}} & \cellstack[r]{{0.083}\\{(0.025)}} \\
Constant & \cellstack[r]{{-6.141}\\{(0.298)}} & \cellstack[r]{{-2.008}\\{(0.276)}} \\
\bottomrule
\end{tabular*}
\notetext{Two-type log-intensity coefficients and model-based standard errors. Both equations and the latent distribution belong to one likelihood. Type one is normalized to zero and type two to one. The associated schooling table reports the exposure and latent-distribution parameters. Type labels follow each fit's estimation order, so the constants, the signs of the loadings, and the mixing logit can differ between columns. Table~\ref{tab:interruptionjointtypes} orients the types by conception intensity. All one-type coefficient vectors and failed-fit candidates are retained in the replication exports.}
\end{threeparttable}
\end{table}

\begin{table}[tbp]
\centering\singlespacing\small
\caption{Joint Interruption Coefficients with Flexible Controls, Schooling}\label{tab:interruptionjointcoefficientsflexibleschooling}
\begin{threeparttable}
\begin{tabular*}{\textwidth}{@{\extracolsep{\fill}}@{}>{\raggedright\arraybackslash}p{2.65in}rr@{}}
\toprule
Parameter & Full sample & Retained onset \\
\midrule
\textit{Interruption equation} &  &  \\
Duration, months 7--12 & \cellstack[r]{{-0.522}\\{(0.194)}} & \cellstack[r]{{-0.336}\\{(0.226)}} \\
Duration, months 13--18 & \cellstack[r]{{0.149}\\{(0.174)}} & \cellstack[r]{{0.182}\\{(0.207)}} \\
Duration, months 19--24 & \cellstack[r]{{-0.077}\\{(0.173)}} & \cellstack[r]{{0.059}\\{(0.200)}} \\
Duration, months 25--36 & \cellstack[r]{{-0.030}\\{(0.161)}} & \cellstack[r]{{0.052}\\{(0.188)}} \\
Duration, months 37--48 & \cellstack[r]{{-0.027}\\{(0.170)}} & \cellstack[r]{{0.093}\\{(0.194)}} \\
Duration, over 48 months & \cellstack[r]{{0.011}\\{(0.179)}} & \cellstack[r]{{0.113}\\{(0.202)}} \\
January--May & \cellstack[r]{{-0.644}\\{(0.077)}} & \cellstack[r]{{-0.620}\\{(0.081)}} \\
Age centered at sixteen & \cellstack[r]{{0.206}\\{(0.061)}} & \cellstack[r]{{0.192}\\{(0.063)}} \\
Centered age squared & \cellstack[r]{{0.021}\\{(0.019)}} & \cellstack[r]{{0.022}\\{(0.020)}} \\
Exposure without retained onset & \cellstack[r]{{1.373}\\{(0.143)}} & -- \\
Constant & \cellstack[r]{{-4.463}\\{(0.139)}} & \cellstack[r]{{-4.337}\\{(0.290)}} \\
\textit{Pregnancy and latent distribution} &  &  \\
Pregnancy exposure, retained onset & \cellstack[r]{{1.269}\\{(0.123)}} & \cellstack[r]{{1.269}\\{(0.123)}} \\
Conception loading & \cellstack[r]{{4.134}\\{(0.198)}} & \cellstack[r]{{-4.134}\\{(0.198)}} \\
Interruption loading & \cellstack[r]{{0.218}\\{(0.260)}} & \cellstack[r]{{-0.218}\\{(0.259)}} \\
Mixing logit & \cellstack[r]{{-3.029}\\{(0.191)}} & \cellstack[r]{{3.029}\\{(0.191)}} \\
\bottomrule
\end{tabular*}
\notetext{Two-type log-intensity coefficients and model-based standard errors. Both equations and the latent distribution belong to one likelihood. Type one is normalized to zero and type two to one. The mixing logit gives the log odds of type two. An absent-onset exposure term appears only in the full sample. A dash denotes an absent term or a model withheld for failed strict convergence. Type labels follow each fit's estimation order, so the constants, the signs of the loadings, and the mixing logit can differ between columns. Table~\ref{tab:interruptionjointtypes} orients the types by conception intensity. All one-type coefficient vectors and failed-fit candidates are retained in the replication exports.}
\end{threeparttable}
\end{table}

Under the modified-onset grouping, we drop the conception event and conception-risk months of a woman whose conception month is retained and whose debut date we set from that conception. Her exposed school months, if any, receive the absent-onset coefficient. A woman with no subsequent exposed month remains unexposed in the schooling likelihood. Every other woman's observations and the exposure variable are unchanged. We fit this grouping to the full sample and to the later complete cases with and without background controls, which enter both hazards (Tables~\ref{tab:interruptionjoint}, \ref{tab:interruptionjointfit}, and~\ref{tab:interruptionjointtypes}). Table~\ref{tab:interruptionjointcoefficientsentry} reports the two-type coefficient vectors for the full sample and for the later complete cases with linear controls.

\begin{table}[tbp]
\centering\singlespacing\small
\caption{Joint Interruption Coefficients under the Modified-Onset Grouping}\label{tab:interruptionjointcoefficientsentry}
\begin{threeparttable}
\begin{tabular*}{\textwidth}{@{\extracolsep{\fill}}@{}>{\raggedright\arraybackslash}p{2.45in}rr@{}}
\toprule
Parameter & \cellstack[r]{{Full sample,}\\{modified-onset grouping}} & \cellstack[r]{{Later complete cases,}\\{linear}} \\
\midrule
\textit{Conception equation} &  &  \\
Age centered at sixteen & \cellstack[r]{{0.216}\\{(0.054)}} & \cellstack[r]{{0.326}\\{(0.077)}} \\
Elapsed duration & \cellstack[r]{{-0.016}\\{(0.046)}} & \cellstack[r]{{0.020}\\{(0.083)}} \\
Constant & \cellstack[r]{{-5.195}\\{(0.141)}} & \cellstack[r]{{-5.594}\\{(0.265)}} \\
\textit{Interruption equation} &  &  \\
Age centered at sixteen & \cellstack[r]{{0.251}\\{(0.045)}} & \cellstack[r]{{0.214}\\{(0.068)}} \\
Elapsed duration & \cellstack[r]{{0.040}\\{(0.038)}} & \cellstack[r]{{0.025}\\{(0.076)}} \\
Exposure without retained onset & \cellstack[r]{{1.395}\\{(0.118)}} & \cellstack[r]{{1.258}\\{(0.269)}} \\
Constant & \cellstack[r]{{-4.980}\\{(0.155)}} & \cellstack[r]{{-4.993}\\{(0.288)}} \\
\textit{Pregnancy and latent distribution} &  &  \\
Pregnancy exposure, retained onset & \cellstack[r]{{1.313}\\{(0.120)}} & \cellstack[r]{{1.291}\\{(0.175)}} \\
Conception loading & \cellstack[r]{{1.670}\\{(1.082)}} & \cellstack[r]{{1.894}\\{(0.982)}} \\
Interruption loading & \cellstack[r]{{1.686}\\{(0.692)}} & \cellstack[r]{{2.243}\\{(0.988)}} \\
Mixing logit & \cellstack[r]{{-4.012}\\{(1.787)}} & \cellstack[r]{{-4.149}\\{(1.564)}} \\
\bottomrule
\end{tabular*}
\notetext{Two-type log-intensity coefficients and model-based standard errors. Both equations and the latent distribution belong to one likelihood. Type one is normalized to zero and type two to one. The mixing logit gives the log odds of type two. The two-type fits for the later complete cases with background controls fail the convergence checks, so their coefficients are withheld. Type labels follow each fit's estimation order, so the constants, the signs of the loadings, and the mixing logit can differ between columns. Table~\ref{tab:interruptionjointtypes} orients the types by conception intensity. All one-type coefficient vectors are retained in the replication exports.}
\end{threeparttable}
\end{table}

Adding background controls lowers the rate ratio in the later complete-case sample with the modified-onset grouping. The one-type estimate falls from \EERjointentry{results.latercc_baseline_entry.k1.hr} with linear controls to \EERjointentry{results.latercc_rich_entry.k1.hr} with background controls in both processes. With linear controls, the two-type fit passes the numerical checks and gives \EERjointentry{results.latercc_baseline_entry.k2.hr}, but AIC and BIC both favor one type.\footnote{Only one of the seven starting values converges for this fit.} With background controls, the two-type fit fails the convergence checks, so we withhold it. Because only the one-type fit is available, this comparison accounts for observed background but not latent differences between women.\footnote{On the same women, the single-equation comparison falls from \EERadj{adjusted_results.school3.later_cc_baseline.hr} to \EERadj{adjusted_results.school3.later_cc_rich.hr}. Those models use one coefficient for all months with pregnancy exposure and different baseline controls. The modified-onset models instead distinguish the two exposure groups and report the retained-onset coefficient.}

Table~\ref{tab:interruptionjointmembership} reports, by group, the mean posterior probability of the higher-conception type under the main fit and under the modified-onset fit. The posterior applies Bayes' rule to each woman's type-specific likelihood contributions at the fitted parameters. Averaging the posteriors across women reproduces the fitted type share. Under the main fit, the \EERjointentry{posterior.full_linear.onset_modified_debut.women} exposed women with a retained onset and a debut set from conception supply about half of the higher-conception type's posterior mass. Women who conceive in their first two observed months supply more than three fifths. Assigned debut precedes conception by one month, but the retained enrollment history can begin later. Among women exposed after a retained onset, those with debut set from conception have a higher average membership probability than women with reported dates. These are estimated probabilities, rather than observed group assignments. Under the main fit, the types' baseline interruption intensities are not statistically distinguishable. Under the modified-onset fit, no woman has a posterior above one half.

\begin{table}[tbp]
\centering\singlespacing\small
\caption{Posterior Type Membership by Debut Modification and Conception Timing}\label{tab:interruptionjointmembership}
\begin{threeparttable}
\begin{tabular*}{\textwidth}{@{\extracolsep{\fill}}@{}>{\raggedright\arraybackslash}p{1.7in}rrrrrr@{}}
\toprule
Group & \cellstack[r]{{Women,}\\{main}} & \cellstack[r]{{Mean}\\{posterior}} & \cellstack[r]{{Share of}\\{mass}} & \cellstack[r]{{Women,}\\{modified}} & \cellstack[r]{{Mean}\\{posterior}} & \cellstack[r]{{Share of}\\{mass}} \\
\midrule
Exposed after onset, debut set from conception & 79 & 0.559 & 0.457 & -- & -- & -- \\
Exposed after onset, reported debut & 373 & 0.114 & 0.440 & 373 & 0.045 & 0.336 \\
Exposure without retained onset & 120 & 0.034 & 0.042 & 199 & 0.017 & 0.069 \\
Never exposed & 2,256 & 0.003 & 0.061 & 2,256 & 0.013 & 0.595 \\
\addlinespace[5pt]
\textit{By retained month of conception} &  &  &  &  &  &  \\
Conception in first retained month & 6 & 0.642 & 0.040 & 4 & 0.039 & 0.003 \\
Conception in second retained month & 101 & 0.560 & 0.585 & 21 & 0.092 & 0.038 \\
Conception in months three to six & 60 & 0.330 & 0.205 & 60 & 0.072 & 0.086 \\
Conception in month seven or later & 311 & 0.030 & 0.098 & 311 & 0.041 & 0.256 \\
\bottomrule
\end{tabular*}
\notetext{The first three numeric columns describe the main two-type interruption fit with linear controls and the last three the modified-onset fit. Women columns give the group size under each fit. Each woman's posterior probability of the higher-conception type is computed from her own history at the fitted parameters. Mean posterior averages that probability within the group, and share of type mass divides the group's summed posterior by the sample total, which equals the fitted type share times the number of women. Conception-month rows include only women whose conception event is in that fit's likelihood. Women whose conception occurs in their last retained month have no exposed month and appear in the never-exposed group, including those whose debut was set from conception. A dash marks a group that does not exist under that fit.}
\end{threeparttable}
\end{table}

Conception can occur after an interruption's dated start but before the absence is confirmed. This occurs in \EERjoint{confirmation.all.p_during_confirmation} of the main \EERjoint{confirmation.all.events} interruptions, from the month after the start through the confirming month. First exposure falls within that interval for \EERjoint{confirmation.all.exposure_during_confirmation} events. That count adds the \EERjoint{confirmation.all.p_same_onset} conceptions in the start month itself, whose exposure begins the next month, and leaves out the \the\numexpr\EERjoint{confirmation.all.p_during_confirmation}+\EERjoint{confirmation.all.p_same_onset}-\EERjoint{confirmation.all.exposure_during_confirmation}\relax{} conceptions in the confirming month, whose exposure begins after it. All count as unexposed at the interruption's start. If pregnancy causes an earlier absence to continue, it could determine whether that absence qualifies as sustained. The causal interpretation requires this not to occur. Recorded exit provides a complementary outcome with a different dating rule.

Some qualifying absences begin in summer before their first missed school month. Among exposed interruptions, \EERjoint{timing.all.summer_before_p} have observed nonenrollment beginning in summer before measured conception, and \EERjoint{timing.all.summer_same_p} begin in the conception month. We still date the interruption to the first missed September--May month. The event is therefore the start of a sustained school-year absence. The first calendar month of nonenrollment need not follow conception.

\FloatBarrier
\subsection{Timing checks}

Table~\ref{tab:timingsingle} changes one timing choice at a time in the single-equation interruption model. Starting exposure in the conception month, or two or three months after it, gives ratios of \EERtiming{single.lag0.hr}, \EERtiming{single.lag2.hr}, and \EERtiming{single.lag3.hr}, compared with \EERtiming{single.base.hr} at baseline. Dropping months around measured conception raises the ratio slightly, to \EERtiming{single.omitm2to2.hr} when months $-2$ through 2 are removed. Dropping the women whose exposure begins during confirmation gives \EERtiming{single.dropconf.hr}. Dating each interruption at its first nonenrolled calendar month gives \EERtiming{single.redate.hr}. Dropping the women whose exposed interruption continues an absence that began in summer at or before measured conception gives \EERtiming{single.dropsummer.hr}. We also shift birth-based or interview-based conception dates by one month in either direction. Across the \EERtiming{single.range.rows} rows outside the time-since-conception model, the ratio ranges from \EERtiming{single.range.min} to \EERtiming{single.range.max}.

The ratio does not rise steadily with time since conception. The ratios for months 1--3, 4--6, 7--12, and 13 onward are \EERtiming{single.bin1to3.hr}, \EERtiming{single.bin4to6.hr}, \EERtiming{single.bin7to12.hr}, and \EERtiming{single.bin13p.hr}, and they are not statistically distinguishable at the five-percent level.

\begin{table}[tbp]
\centering\singlespacing\footnotesize
\caption{Timing Checks for Sustained Interruption, Single-Equation Models}\label{tab:timingsingle}
\begin{threeparttable}
\begin{tabular*}{\textwidth}{@{\extracolsep{\fill}}@{}>{\raggedright\arraybackslash}p{2.75in}rrrrrr@{}}
\toprule
Check & HR & 95\% CI & Women & \shortstack{Risk\\months} & Events & \shortstack{Exposed\\events} \\
\midrule
\textit{Exposure start} &  &  &  &  &  &  \\
\quad One month after conception (baseline) & 3.85 & [3.29, 4.51] & 2,828 & 48,953 & 711 & 261 \\
\quad Conception month & 3.96 & [3.39, 4.63] & 2,828 & 48,953 & 711 & 276 \\
\quad Two months after conception & 3.82 & [3.25, 4.48] & 2,828 & 48,953 & 711 & 249 \\
\quad Three months after conception & 3.79 & [3.22, 4.46] & 2,828 & 48,953 & 711 & 238 \\
\textit{Months omitted around measured conception} &  &  &  &  &  &  \\
\quad Month 0 & 3.95 & [3.37, 4.64] & 2,828 & 48,589 & 696 & 261 \\
\quad Months 0 to 1 & 4.00 & [3.41, 4.71] & 2,828 & 48,230 & 684 & 249 \\
\quad Months $-1$ to 1 & 4.09 & [3.48, 4.81] & 2,816 & 47,842 & 671 & 249 \\
\quad Months $-1$ to 2 & 4.13 & [3.50, 4.88] & 2,812 & 47,518 & 660 & 238 \\
\quad Months $-2$ to 2 & 4.20 & [3.55, 4.96] & 2,809 & 47,202 & 650 & 238 \\
\textit{Time since conception, one model} &  &  &  &  &  &  \\
\quad Months 1--3 & 4.45 & [3.30, 6.02] &  &  &  &  \\
\quad Months 4--6 & 2.88 & [1.90, 4.36] &  &  &  &  \\
\quad Months 7--12 & 4.74 & [3.62, 6.20] &  &  &  &  \\
\quad Month 13 onward & 3.53 & [2.89, 4.31] &  &  &  &  \\
\textit{Interruption dating} &  &  &  &  &  &  \\
\quad Drop women exposed during confirmation & 4.11 & [3.50, 4.82] & 2,797 & 48,559 & 680 & 261 \\
\quad Date at first nonenrolled calendar month & 3.90 & [3.33, 4.57] & 2,821 & 64,043 & 704 & 255 \\
\quad Drop summer starts by conception & 3.76 & [3.21, 4.41] & 2,822 & 48,895 & 705 & 255 \\
\textit{Conception dating} &  &  &  &  &  &  \\
\quad Birth-based dates one month earlier & 3.92 & [3.35, 4.58] & 2,828 & 48,953 & 711 & 271 \\
\quad Birth-based dates one month later & 3.81 & [3.25, 4.47] & 2,828 & 48,953 & 711 & 252 \\
\quad Interview-based dates one month earlier & 3.89 & [3.32, 4.55] & 2,828 & 48,953 & 711 & 266 \\
\quad Interview-based dates one month later & 3.85 & [3.29, 4.52] & 2,828 & 48,953 & 711 & 258 \\
\bottomrule
\end{tabular*}
\notetext{Complementary log-log models for the three-school-month interruption in the post-debut sample, using September--May risk months except where noted, with seven duration categories, school term, quadratic age, and woman-clustered 95\% intervals. Exposure-start rows move the month in which exposure begins without changing the sample. Omission rows drop the listed calendar months relative to measured conception, including any event in them. The time-since-conception rows come from one model on the baseline sample, so their women, months, and events are those of the baseline row. The model replaces the exposure indicator with four periods, which contain \EERtiming{single.bin1to3.term_events}, \EERtiming{single.bin4to6.term_events}, \EERtiming{single.bin7to12.term_events}, and \EERtiming{single.bin13p.term_events} exposed interruptions, and the Wald test of equal ratios across the periods gives p = \EERtiming{single.bins.wald.p}. Women exposed during confirmation are those whose first exposed month falls after the dated start of the interruption but by its confirming month. Dating at the first nonenrolled calendar month keeps summer months in the risk period and adds a summer term, so its month count includes summer. The summer-start row drops women whose exposed interruption continues a June--August absence that began at or before measured conception. Conception-dating rows move the selected date by one month for women whose date comes from the named source. None of these checks tests whether anticipated conception changes schooling.}
\end{threeparttable}
\end{table}

Table~\ref{tab:timingjoint} repeats the exposure-lag, confirmation, and dating checks in the joint model. The two-type estimate ranges from \EERtiming{joint.k2range.min} to \EERtiming{joint.k2range.max} across these fits and falls to \EERtiming{joint.lag3.full.k2.hr} in the full sample when exposure starts three months after conception. These checks vary how conception, exposure, and interruption are dated. They do not show whether an anticipated pregnancy changes schooling before conception, and omitting months around conception also removes the months that show which event came first.

\begin{table}[tbp]
\centering\singlespacing\footnotesize
\caption{Timing Checks for Sustained Interruption, Joint Models}\label{tab:timingjoint}
\begin{threeparttable}
\begin{tabular*}{\textwidth}{@{\extracolsep{\fill}}@{}>{\raggedright\arraybackslash}p{1.05in}lrrrrrr@{}}
\toprule
Check & Sample & Women & Events & \shortstack{One type\\HR} & 95\% CI & \shortstack{Two types\\HR} & 95\% CI \\
\midrule
Baseline, lag 1 & Full & 2,828 & 711 & 3.91 & [3.30, 4.64] & 3.76 & [3.00, 4.71] \\
 & Retained onset & 2,708 & 648 & 3.92 & [3.31, 4.65] & 3.78 & [3.02, 4.74] \\
\addlinespace[5pt]
Exposure lag 0 & Full & 2,828 & 711 & 4.01 & [3.40, 4.74] & 4.00 & [3.24, 4.94] \\
 & Retained onset & 2,708 & 648 & 4.02 & [3.41, 4.76] & 4.02 & [3.26, 4.97] \\
\addlinespace[5pt]
Exposure lag 2 & Full & 2,828 & 711 & 3.90 & [3.28, 4.64] & 3.63 & [2.86, 4.62] \\
 & Retained onset & 2,708 & 648 & 3.91 & [3.29, 4.65] & 3.66 & [2.88, 4.65] \\
\addlinespace[5pt]
Exposure lag 3 & Full & 2,828 & 711 & 3.89 & [3.26, 4.64] & 3.46 & [2.66, 4.50] \\
 & Retained onset & 2,708 & 648 & 3.90 & [3.26, 4.65] & 3.48 & [2.68, 4.52] \\
\addlinespace[5pt]
Drop women exposed during confirmation & Full & 2,797 & 680 & 4.17 & [3.51, 4.95] & 4.16 & [3.33, 5.18] \\
 & Retained onset & 2,677 & 617 & 4.18 & [3.52, 4.97] & 4.18 & [3.35, 5.21] \\
\addlinespace[5pt]
Date at first nonenrolled calendar month & Full & 2,821 & 704 & 3.86 & [3.25, 4.59] & 3.69 & [2.94, 4.65] \\
 & Retained onset & 2,701 & 641 & 3.87 & [3.26, 4.60] & 3.71 & [2.95, 4.67] \\
\bottomrule
\end{tabular*}
\notetext{HR is the hazard ratio for pregnancy exposure after a retained onset. Joint conception and interruption models with linear age and capped elapsed duration in both hazards and model-based 95\% intervals, as in the main joint table. The retained-onset sample excludes women already exposed at entry. Every two-type fit passes the strict convergence checks, and each selected fit has a likelihood at least as large as each of the other six fixed starts. Exposure-lag rows move the month in which exposure begins and leave conception events unchanged. The row dated at the first nonenrolled calendar month keeps summer months in the interruption risk period and adds a summer term to both hazards. The joint models do not repeat the omission windows, the summer-start exclusion, or the conception-dating shifts.}
\end{threeparttable}
\end{table}

Table~\ref{tab:timingsplit} lets the joint model's pregnancy ratio differ with months since an observed conception. The interruption equation has one ratio for months 1 to 3 and added terms for months 4 to 9 and for month 10 onward. The conception month, which the one-month lag leaves without pregnancy exposure, receives its own term. Samples and controls follow the main joint models. One of the seven starting points of each two-type fit is the optimum of the same model without the split. In the full sample with two types, a joint test that the three post-conception ratios are equal gives p = \EERsplit{joint.gest.full.k2.waldp}. Adding only the conception-month term to the model without the split raises its two-type ratio from \EERjoint{results.full_linear.k2.hr} to \EERsplit{joint.k0only.full.k2.all.hr}, close to the \EERtiming{joint.lag0.full.k2.hr} obtained when exposure starts in the conception month (Table~\ref{tab:timingjoint}).

\begin{table}[tbp]
\centering\singlespacing\small
\caption{Joint Models by Months Since Conception}\label{tab:timingsplit}
\begin{threeparttable}
\begin{tabular*}{\textwidth}{@{\extracolsep{\fill}}@{}>{\raggedright\arraybackslash}p{1.9in}rrrr@{}}
\toprule
Months since conception & \shortstack{Full\\one type} & \shortstack{Full\\two types} & \shortstack{Retained onset\\one type} & \shortstack{Retained onset\\two types} \\
\midrule
Conception month & \cellstack[r]{4.01\\{[2.39, 6.70]}} & \cellstack[r]{3.97\\{[2.36, 6.69]}} & \cellstack[r]{4.01\\{[2.40, 6.71]}} & \cellstack[r]{3.98\\{[2.36, 6.71]}} \\
Months 1--3 & \cellstack[r]{4.67\\{[3.44, 6.33]}} & \cellstack[r]{4.62\\{[3.36, 6.36]}} & \cellstack[r]{4.67\\{[3.45, 6.33]}} & \cellstack[r]{4.64\\{[3.37, 6.38]}} \\
Months 4--9 & \cellstack[r]{3.78\\{[2.84, 5.03]}} & \cellstack[r]{3.73\\{[2.74, 5.08]}} & \cellstack[r]{3.79\\{[2.85, 5.04]}} & \cellstack[r]{3.75\\{[2.76, 5.10]}} \\
Month 10 onward & \cellstack[r]{3.88\\{[3.10, 4.86]}} & \cellstack[r]{3.81\\{[2.88, 5.04]}} & \cellstack[r]{3.89\\{[3.11, 4.88]}} & \cellstack[r]{3.84\\{[2.90, 5.09]}} \\
Exposure without an observed conception month & \cellstack[r]{4.09\\{[3.09, 5.40]}} & \cellstack[r]{4.08\\{[3.08, 5.40]}} & -- & -- \\
\bottomrule
\end{tabular*}
\notetext{Joint conception and interruption models with linear age and capped elapsed duration in both hazards and model-based 95\% intervals, as in the main joint table. Each cell gives the hazard ratio relative to months without pregnancy exposure other than the conception month, with its interval below. The interruption equation has one ratio for months 1--3 after an observed conception and added terms for later months. The conception month, which the one-month exposure lag leaves without pregnancy exposure, has its own term. The full sample has \EERsplit{joint.gest.full.k2.women} women and \EERsplit{joint.gest.full.k2.months} calendar months. The retained-onset sample excludes women exposed at entry and has \EERsplit{joint.gest.restricted.k2.women} women. One of the seven starting points of each two-type fit is the optimum of the same model without the split, and every two-type fit passes the strict convergence checks. In the full sample with two types, the test that the three post-conception ratios are equal gives p = \EERsplit{joint.gest.full.k2.waldp}, and the likelihood-ratio statistic against the model without the split is \EERsplit{joint.gest.full.k2.lr} on \EERsplit{joint.gest.full.k2.lrdf} degrees of freedom.}
\end{threeparttable}
\end{table}

Table~\ref{tab:absencesplit} shows how the rates rise. An interruption begins with an absence after an enrolled school month, and it counts only if the absence lasts at least three school months. Absences begin \EERsplit{absmodel.start.m13.ratio} times as often in months 1 to 3 after conception as in months without pregnancy exposure, and a similar multiple holds in later months. The share of absences that last is only \EERsplit{absmodel.qualify.m13.ratio} times as high. The conception month shows the same pattern, so its interruptions do not mainly come from pregnancy keeping a student out after an ordinary absence. Dropping absences that began in summer, or adding a September indicator, leaves every ratio for absences beginning above three. These single-equation comparisons do not allow for persistent differences between women.

\begin{table}[tbp]
\centering\singlespacing\small
\caption{Absences Beginning and Lasting by Months Since Conception}\label{tab:absencesplit}
\begin{threeparttable}
\begin{tabular*}{\textwidth}{@{\extracolsep{\fill}}@{}>{\raggedright\arraybackslash}p{1.9in}rrrr@{}}
\toprule
Months since conception & \shortstack{Absences\\begin} & \shortstack{Share\\that last} & \shortstack{Begun\\absences} & Interruptions \\
\midrule
Conception month & \cellstack[r]{3.45\\{[2.13, 5.60]}} & \cellstack[r]{1.19\\{[0.99, 1.42]}} & 17 & 15 \\
Months 1--3 & \cellstack[r]{4.11\\{[3.13, 5.41]}} & \cellstack[r]{1.12\\{[0.99, 1.27]}} & 54 & 46 \\
Months 4--9 & \cellstack[r]{3.38\\{[2.62, 4.36]}} & \cellstack[r]{1.10\\{[0.97, 1.24]}} & 65 & 53 \\
Month 10 onward & \cellstack[r]{3.49\\{[2.83, 4.31]}} & \cellstack[r]{1.09\\{[1.00, 1.20]}} & 120 & 99 \\
Exposure without an observed conception month & \cellstack[r]{3.82\\{[2.94, 4.97]}} & \cellstack[r]{1.08\\{[0.97, 1.20]}} & 76 & 63 \\
\bottomrule
\end{tabular*}
\notetext{Single-equation models on the post-debut interruption sample with seven duration categories, school term, quadratic age, and woman-clustered 95\% intervals. An absence begins in a school month in which a student is not enrolled after an enrolled month and is dated to its first school month. It becomes an interruption if it lasts at least three school months with no diploma or GED dated first. The first column is a complementary log-log hazard ratio for absences beginning. The second is a risk ratio, from a Poisson model with a log link, for the share of begun absences that become interruptions. Both compare with months without pregnancy exposure other than the conception month, the same comparison months as in Table~\ref{tab:timingsplit}. In these months, \EERsplit{absence.ref.starts} absences begin and \EERsplit{absence.ref.sharepct} percent of them last. The comparisons do not allow for persistent differences between women.}
\end{threeparttable}
\end{table}

Late conception dates explain little of the elevated conception month. A pregnancy lasts a little under nine months from conception, so a conception dated nine months before a live birth usually falls in the true month or the month before, and only a long pregnancy places it a month late. For these birth-dated conceptions, the conception-month interruption rate is \EERsplit{dating.roster.k0.intpct} percent, against \EERsplit{dating.all.k0.intpct} percent for all conceptions and \EERsplit{dating.ref.ref.intpct} percent in months without pregnancy exposure.

\FloatBarrier
\subsection{Agreement between conception and birth dates}

The main date-agreement rule links measured conception to the earliest roster birth kept after the conception-age restriction. The dates must be nine months apart, allowing a one-month discrepancy in either direction. Alternative samples require exact agreement or allow a three-month discrepancy (Table~\ref{tab:birthagreement}). We exclude the whole history when an eligible pregnancy fails the rule and retain women without an eligible pregnancy.

Pregnancy eligibility follows the schooling outcome being estimated. For recorded exit, conception must meet that analysis's cutoff based on dropout or completion. For interruption, conception or an exposed month must fall within the observed calendar risk period. We apply this condition before restricting to post-debut and school-month observations.

The main text's models of schooling around birth assign every exposed month to one of six periods. Before birth, the periods are months 1--3, 4--6, and 7 onward since conception. After delivery, they are the birth month through month 2, months 3--5, and month 6 onward. Unexposed months are the reference group. A separate model directly estimates one coefficient before birth and one from birth onward. We test their equality using woman-clustered inference. These pooled coefficients are estimated directly, not averaged from the six period ratios.

Many conception dates equal the linked birth date less nine months by construction. Agreement therefore does not independently verify pregnancy identity or clinical gestation length. Students who leave or complete earlier also cease to contribute to later periods. The prebirth and postbirth comparison can thus reflect changes in who remains at risk as well as timing relative to delivery.

\begin{table}[tbp]
\centering\singlespacing\small
\caption{Pregnancy and Birth Timing under Alternative Date-Agreement Rules}\label{tab:birthagreement}
\begin{threeparttable}
\begin{tabular*}{\textwidth}{@{\extracolsep{\fill}}@{}>{\raggedright\arraybackslash}p{1.1in}rrrrrr@{}}
\toprule
Outcome & \shortstack{Tolerance\\(months)} & \shortstack{Before birth\\HR [95\% CI]} & \shortstack{Birth or later\\HR [95\% CI]} & $p$ & Women & Months \\
\midrule
Recorded exit & 0 & \cellstack[r]{4.27\\{[3.29, 5.55]}} & \cellstack[r]{3.19\\{[2.58, 3.94]}} & 0.054 & 2,839 & 58,746 \\
 & 1 & \cellstack[r]{4.46\\{[3.51, 5.69]}} & \cellstack[r]{3.28\\{[2.67, 4.03]}} & 0.027 & 2,914 & 60,719 \\
 & 3 & \cellstack[r]{4.46\\{[3.51, 5.67]}} & \cellstack[r]{3.34\\{[2.73, 4.10]}} & 0.036 & 2,927 & 61,035 \\
\addlinespace[5pt]
Three school months & 0 & \cellstack[r]{3.71\\{[2.88, 4.79]}} & \cellstack[r]{3.52\\{[2.90, 4.27]}} & 0.715 & 2,718 & 46,670 \\
 & 1 & \cellstack[r]{4.02\\{[3.20, 5.06]}} & \cellstack[r]{3.66\\{[3.03, 4.42]}} & 0.486 & 2,796 & 48,299 \\
 & 3 & \cellstack[r]{4.03\\{[3.21, 5.05]}} & \cellstack[r]{3.73\\{[3.10, 4.50]}} & 0.571 & 2,809 & 48,564 \\
\addlinespace[5pt]
Four calendar months & 0 & \cellstack[r]{3.54\\{[2.75, 4.56]}} & \cellstack[r]{3.53\\{[2.91, 4.29]}} & 0.982 & 2,703 & 45,954 \\
 & 1 & \cellstack[r]{3.81\\{[3.03, 4.80]}} & \cellstack[r]{3.58\\{[2.96, 4.33]}} & 0.654 & 2,777 & 47,466 \\
 & 3 & \cellstack[r]{3.82\\{[3.04, 4.79]}} & \cellstack[r]{3.65\\{[3.03, 4.41]}} & 0.746 & 2,790 & 47,731 \\
\bottomrule
\end{tabular*}
\notetext{The sample retains post-debut women without an eligible pregnancy and those whose measured conception date agrees with the earliest retained roster birth minus nine months within the stated tolerance. Agreement is partly mechanical because birth minus nine months is a conception-date source. It does not establish clinical gestation length or verify a pregnancy identifier. The two coefficients are estimated directly in one complementary log-log model and compare each period with unexposed months. The birth month belongs to birth or later. The reported p-value tests equality of the two coefficients with woman-clustered inference. All models use duration categories, school term, and quadratic age. Recorded exit uses eligible enrolled months. The interruption measures use September--May calendar-risk months. The groups remaining at risk can differ before and after birth, so the equality test does not identify a change in the causal effect.}
\end{threeparttable}
\end{table}

\FloatBarrier
\subsection{Observed schooling after each endpoint}

Follow-up begins at recorded exit or the first qualifying interruption under each definition. Exposure is fixed at that date using the one-month conception lag. Table~\ref{tab:followupschool3postdebut} reports both horizons for the main post-debut interruptions. Tables~\ref{tab:followuprecorded}--\ref{tab:followupcalendar4} include events before debut. All four tables use women who contribute eligible months to the recorded-exit sample. The replication also retains follow-up for women with a schooling event outside that sample.

Enrollment follow-up requires known status in every September--May month and interview and calendar coverage through the horizon. Return requires observed nonenrollment in a school month at or after the schooling event, followed by enrollment in a later school month. Attendance after recorded exit alone can reflect continued enrollment, so it does not establish return. Return does not require enrollment at the end of follow-up.

Credential follow-up requires interview coverage through the horizon and uses recorded diploma and GED dates. With adequate interview coverage, we record no credential by the horizon when neither a diploma nor a GED is dated on or before it. We do not infer receipt from the highest grade completed.

Each percentage uses women with observed follow-up for that outcome and horizon. Women with unknown follow-up enter the missing count but not the percentage. The observed sample can differ at 12 and 24 months, so a change between horizons need not describe progress by the same women. Follow-up can extend beyond age 21 when the required records remain available, although both schooling analyses stop at that age. Diploma and GED receipt need not be mutually exclusive. The comparisons condition on a recorded exit or an interruption. They describe subsequent schooling and do not identify pregnancy effects on recovery or final attainment.

\begin{table}[tbp]
\centering\singlespacing\small
\caption{Schooling after Interruption in the Post-Debut Sample}\label{tab:followupschool3postdebut}
\begin{threeparttable}
\begin{tabular*}{\textwidth}{@{\extracolsep{\fill}}@{}rp{1.7in}rrrr@{}}
\toprule
Months & Outcome & \shortstack{Exposed\\Percent} & Yes/observed & \shortstack{Unexposed\\Percent} & Yes/observed \\
\midrule
12 & Returned to enrollment & 20.5 & 53/259 & 27.0 & 121/448 \\
 & Diploma or GED & 15.7 & 41/261 & 22.0 & 99/450 \\
 & \quad Diploma & 7.7 & 20/261 & 14.0 & 63/450 \\
 & \quad GED & 8.0 & 21/261 & 8.0 & 36/450 \\
\addlinespace[5pt]
24 & Returned to enrollment & 20.4 & 52/255 & 27.6 & 123/445 \\
 & Diploma or GED & 26.2 & 68/260 & 35.0 & 157/448 \\
 & \quad Diploma & 11.5 & 30/260 & 19.6 & 88/448 \\
 & \quad GED & 14.6 & 38/260 & 15.4 & 69/448 \\
\bottomrule
\end{tabular*}
\notetext{Exposure is fixed at the event date using a one-month lag after conception. The sample consists of the women who contribute eligible months to the recorded-exit sample and who experienced the indicated schooling event. Each percentage uses the displayed numerator and observed denominator. Enrollment follow-up requires observed status in every September--May month through the horizon and interview and calendar coverage. Return requires nonenrollment in a September--May month at or after the event followed by enrollment in a later September--May month. Credential follow-up requires interview coverage through the horizon and uses recorded diploma and GED dates. A woman with adequate follow-up and no recorded credential date is counted as having no recorded credential. The comparisons describe subsequent schooling among women with the observed event. They do not identify pregnancy effects on eventual attainment. Diploma and GED rows report receipt of each credential by the horizon. The observed denominator can change between horizons. The two horizons are not a common-sample cumulative curve. At 12 months, enrollment follow-up is missing for 2 exposed and 2 unexposed women. Credential follow-up is missing for 0 and 0, respectively. At 24 months, enrollment follow-up is missing for 6 exposed and 5 unexposed women. Credential follow-up is missing for 1 and 2, respectively. The interruption requires three September--May months. Events occur at or after the constructed debut date, matching the post-debut interruption sample.}
\end{threeparttable}
\end{table}

\begin{table}[tbp]
\centering\singlespacing\small
\caption{Schooling after Recorded Exit, No Debut Restriction}\label{tab:followuprecorded}
\begin{threeparttable}
\begin{tabular*}{\textwidth}{@{\extracolsep{\fill}}@{}rp{1.7in}rrrr@{}}
\toprule
Months & Outcome & \shortstack{Exposed\\Percent} & Yes/observed & \shortstack{Unexposed\\Percent} & Yes/observed \\
\midrule
12 & Returned to enrollment & 1.7 & 4/242 & 1.2 & 5/419 \\
 & Diploma or GED & 10.2 & 25/246 & 14.5 & 62/428 \\
 & \quad Diploma & 0.0 & 0/246 & 0.0 & 0/428 \\
 & \quad GED & 10.2 & 25/246 & 14.5 & 62/428 \\
\addlinespace[5pt]
24 & Returned to enrollment & 2.5 & 6/241 & 1.2 & 5/416 \\
 & Diploma or GED & 19.1 & 47/246 & 24.1 & 103/427 \\
 & \quad Diploma & 0.4 & 1/246 & 0.0 & 0/427 \\
 & \quad GED & 18.7 & 46/246 & 24.1 & 103/427 \\
\bottomrule
\end{tabular*}
\notetext{Exposure is fixed at the event date using a one-month lag after conception. The sample consists of the women who contribute eligible months to the recorded-exit sample and who experienced the indicated schooling event. Each percentage uses the displayed numerator and observed denominator. Enrollment follow-up requires observed status in every September--May month through the horizon and interview and calendar coverage. Return requires nonenrollment in a September--May month at or after the event followed by enrollment in a later September--May month. Credential follow-up requires interview coverage through the horizon and uses recorded diploma and GED dates. A woman with adequate follow-up and no recorded credential date is counted as having no recorded credential. The comparisons describe subsequent schooling among women with the observed event. They do not identify pregnancy effects on eventual attainment. Diploma and GED rows report receipt of each credential by the horizon. The observed denominator can change between horizons. The two horizons are not a common-sample cumulative curve. At 12 months, enrollment follow-up is missing for 4 exposed and 15 unexposed women. Credential follow-up is missing for 0 and 6, respectively. At 24 months, enrollment follow-up is missing for 5 exposed and 18 unexposed women. Credential follow-up is missing for 0 and 7, respectively. The sample includes all eligible events without the sexual-debut restriction, so it adds events before a known debut date and events of women with no debut date. Following the recorded exit date, observed return requires affirmative nonenrollment at or after that date before later enrollment. Continued attendance is not counted as return.}
\end{threeparttable}
\end{table}

\begin{table}[tbp]
\centering\singlespacing\small
\caption{Schooling after a Three-School-Month Interruption, No Debut Restriction}\label{tab:followupschool3}
\begin{threeparttable}
\begin{tabular*}{\textwidth}{@{\extracolsep{\fill}}@{}rp{1.7in}rrrr@{}}
\toprule
Months & Outcome & \shortstack{Exposed\\Percent} & Yes/observed & \shortstack{Unexposed\\Percent} & Yes/observed \\
\midrule
12 & Returned to enrollment & 20.5 & 53/259 & 32.4 & 209/646 \\
 & Diploma or GED & 15.7 & 41/261 & 21.3 & 138/649 \\
 & \quad Diploma & 7.7 & 20/261 & 14.8 & 96/649 \\
 & \quad GED & 8.0 & 21/261 & 6.5 & 42/649 \\
\addlinespace[5pt]
24 & Returned to enrollment & 20.4 & 52/255 & 34.1 & 218/640 \\
 & Diploma or GED & 26.2 & 68/260 & 35.2 & 227/645 \\
 & \quad Diploma & 11.5 & 30/260 & 22.2 & 143/645 \\
 & \quad GED & 14.6 & 38/260 & 13.0 & 84/645 \\
\bottomrule
\end{tabular*}
\notetext{Exposure is fixed at the event date using a one-month lag after conception. The sample consists of the women who contribute eligible months to the recorded-exit sample and who experienced the indicated schooling event. Each percentage uses the displayed numerator and observed denominator. Enrollment follow-up requires observed status in every September--May month through the horizon and interview and calendar coverage. Return requires nonenrollment in a September--May month at or after the event followed by enrollment in a later September--May month. Credential follow-up requires interview coverage through the horizon and uses recorded diploma and GED dates. A woman with adequate follow-up and no recorded credential date is counted as having no recorded credential. The comparisons describe subsequent schooling among women with the observed event. They do not identify pregnancy effects on eventual attainment. Diploma and GED rows report receipt of each credential by the horizon. The observed denominator can change between horizons. The two horizons are not a common-sample cumulative curve. At 12 months, enrollment follow-up is missing for 2 exposed and 3 unexposed women. Credential follow-up is missing for 0 and 0, respectively. At 24 months, enrollment follow-up is missing for 6 exposed and 9 unexposed women. Credential follow-up is missing for 1 and 4, respectively. The sample includes all eligible events without the sexual-debut restriction, so it adds events before a known debut date and events of women with no debut date.}
\end{threeparttable}
\end{table}

\begin{table}[tbp]
\centering\singlespacing\small
\caption{Schooling after a Four-Calendar-Month Interruption, No Debut Restriction}\label{tab:followupcalendar4}
\begin{threeparttable}
\begin{tabular*}{\textwidth}{@{\extracolsep{\fill}}@{}rp{1.7in}rrrr@{}}
\toprule
Months & Outcome & \shortstack{Exposed\\Percent} & Yes/observed & \shortstack{Unexposed\\Percent} & Yes/observed \\
\midrule
12 & Returned to enrollment & 22.0 & 56/255 & 38.4 & 265/690 \\
 & Diploma or GED & 14.3 & 37/258 & 20.2 & 140/692 \\
 & \quad Diploma & 7.0 & 18/258 & 13.7 & 95/692 \\
 & \quad GED & 7.4 & 19/258 & 6.5 & 45/692 \\
\addlinespace[5pt]
24 & Returned to enrollment & 22.1 & 56/253 & 39.5 & 269/681 \\
 & Diploma or GED & 24.9 & 64/257 & 34.3 & 236/688 \\
 & \quad Diploma & 11.3 & 29/257 & 21.9 & 151/688 \\
 & \quad GED & 13.6 & 35/257 & 12.4 & 85/688 \\
\bottomrule
\end{tabular*}
\notetext{Exposure is fixed at the event date using a one-month lag after conception. The sample consists of the women who contribute eligible months to the recorded-exit sample and who experienced the indicated schooling event. Each percentage uses the displayed numerator and observed denominator. Enrollment follow-up requires observed status in every September--May month through the horizon and interview and calendar coverage. Return requires nonenrollment in a September--May month at or after the event followed by enrollment in a later September--May month. Credential follow-up requires interview coverage through the horizon and uses recorded diploma and GED dates. A woman with adequate follow-up and no recorded credential date is counted as having no recorded credential. The comparisons describe subsequent schooling among women with the observed event. They do not identify pregnancy effects on eventual attainment. Diploma and GED rows report receipt of each credential by the horizon. The observed denominator can change between horizons. The two horizons are not a common-sample cumulative curve. At 12 months, enrollment follow-up is missing for 3 exposed and 2 unexposed women. Credential follow-up is missing for 0 and 0, respectively. At 24 months, enrollment follow-up is missing for 5 exposed and 11 unexposed women. Credential follow-up is missing for 1 and 4, respectively. The sample includes all eligible events without the sexual-debut restriction, so it adds events before a known debut date and events of women with no debut date.}
\end{threeparttable}
\end{table}

\subsection{Pregnancies that ended without a live birth}

Adding pregnancies that ended in miscarriage, abortion, or stillbirth lowers the hazard ratio from \PregHistoryBaselineHR{} to \PregHistoryHR{} (95 percent confidence interval \PregHistoryCILower{}--\PregHistoryCIUpper{}). The added reports come from survey waves 1997--2023. Separate survey reports give the month each such pregnancy ended and its length, so we date conception by subtracting the length from the ending month. We keep reports with valid dates and a length of one to ten months. The implied conception must fall at age 13 or older, and the pregnancy must end by the interview. For each woman, we use the earliest conception from these reports and the main sources, keeping the same women, months, and controls. Alternative treatments of conflicting reports and uncertain dates yield hazard ratios between \PregHistoryAltMin{} and \PregHistoryAltMax{}. All other estimates and sample counts use the main conception dates.

\section{Recorded-Exit Specifications and Diagnostics}

Each matched comparison uses the same controls in both hazards and in its one-type and two-type models (Table~\ref{tab:matchedall}). Linear controls are age centered at 16 and elapsed calendar months since first eligible enrollment, capped at 48 and divided by 12. We define entry before restricting to post-debut months. The full sample, which Table~\ref{tab:matchedall} labels ``All post-debut women,'' has separate exposure coefficients for retained and absent conception months. The retained-onset sample excludes exposed women without a conception-month observation.

Flexible controls use duration categories of 0--6, 7--12, 13--18, 19--24, 25--36, 37--48, and more than 48 months. They also include September--December, January--May, and June--August terms, with centered age and its square. The supplementary single-equation models use these controls. Matched comparisons report model-based intervals for both one-type and two-type models, while supplementary single-equation models cluster by woman. The replication also retains woman-clustered intervals for the one-type matched models. All two-type intervals condition on the specified heterogeneity distribution.

We fit each two-type model from seven starting values, initially allowing 100 iterations per start, and keep the start whose finite likelihood is at least as large as that of every other start. If that solution fails the convergence checks defined in Section~\ref{sec:jointestimation}, we continue it with a larger iteration budget and retain its history. Exports record the sample, rank, convergence, likelihood, available uncertainty, and checks for numerical clipping. The table reports the full and retained-onset specifications. We withhold intervals when numerical validation fails.

\begin{table}[tbp]
\centering\singlespacing\small
\caption{Latent Heterogeneity and Baseline Flexibility in the Recorded-Exit Joint Models}\label{tab:matchedall}
\begin{threeparttable}
\begin{tabular*}{\textwidth}{@{\extracolsep{\fill}}@{}>{\raggedright\arraybackslash}p{1.8in}lrrrl@{}}
\toprule
Sample & Controls & Types & HR & 95\% CI & Status \\
\midrule
All post-debut women & Linear & 1 & 3.35 & [2.75, 4.08] & Converged \\
 & Linear & 2 & 3.23 & [2.48, 4.21] & Converged \\
 & Flexible & 1 & 3.34 & [2.74, 4.07] & Converged \\
 & Flexible & 2 & 3.23 & [2.47, 4.21] & Converged \\
\addlinespace[5pt]
Retained-onset histories & Linear & 1 & 3.26 & [2.67, 3.97] & Converged \\
 & Linear & 2 & 3.08 & [2.36, 4.04] & Converged \\
 & Flexible & 1 & 3.25 & [2.66, 3.97] & Converged \\
 & Flexible & 2 & 3.10 & [2.37, 4.07] & Converged \\
\bottomrule
\end{tabular*}
\notetext{HR is the hazard ratio for pregnancy exposure after a retained onset in joint models of conception and recorded exit. All selected solutions pass the convergence and numerical checks. Intervals are model based. Linear controls are centered age and elapsed duration capped at 48 months. Flexible controls are seven duration categories, three school terms, and quadratic age. Each pair has identical controls in both hazards and the same observations. No-mixture woman-clustered intervals are retained in the replication exports. All post-debut women includes 2,951 women, 61,649 months, and 594 exits. Retained-onset histories includes 2,749 women, 57,991 months, and 501 exits.}
\end{threeparttable}
\end{table}

We label the types by conception propensity so their shares and loadings are comparable across models (Table~\ref{tab:latenttypes}). The higher type always has the higher conception intensity. Its share and both loadings refer to that same type. Exponentiating each loading gives an intensity ratio at common covariates. These are not ratios of monthly event probabilities, which also depend on the baseline hazard.

In the full recorded-exit sample, the two types differ much more in conception than in their school-exit rate without pregnancy exposure. With linear controls, the model estimates that \EERsample{latent.full_linear.high_share_pct} percent of women belong to the higher-conception type. Its conception intensity is roughly \EERsample{latent.full_linear.conception_ratio_rounded} times that of the other type at the same controls. Its estimated exit intensity without exposure is only slightly higher. The exit loading is statistically indistinguishable from zero, so we do not establish a difference in baseline exit intensity. In this sample, adding the second type changes the pregnancy estimate little.

\begin{table}[tbp]
\centering\singlespacing\small
\caption{Latent Types across Samples and Baseline Specifications}\label{tab:latenttypes}
\begin{threeparttable}
\begin{tabular*}{\textwidth}{@{\extracolsep{\fill}}@{}>{\raggedright\arraybackslash}p{1.55in}rrrrrr@{}}
\toprule
Sample and controls & \cellstack[r]{Lower type\\share (\%)} & \cellstack[r]{Higher type\\share (\%)} & \cellstack[r]{Conception\\loading} & \cellstack[r]{Exit\\loading} & \cellstack[r]{Conception\\ratio} & \cellstack[r]{Exit\\ratio} \\
\midrule
\cellstack[l]{Full sample\\Linear controls} & 97.0 & 3.0 & 4.132 & 0.158 & 62.32 & 1.17 \\
\cellstack[l]{Full sample\\Flexible controls} & 96.6 & 3.4 & 4.175 & 0.129 & 65.03 & 1.14 \\
\addlinespace[5pt]
\cellstack[l]{Retained onset\\Linear controls} & 97.0 & 3.0 & 4.123 & 0.230 & 61.72 & 1.26 \\
\cellstack[l]{Retained onset\\Flexible controls} & 96.5 & 3.5 & 4.161 & 0.172 & 64.11 & 1.19 \\
\bottomrule
\end{tabular*}
\notetext{Higher and lower types refer to conception propensity. The lower type is normalized to zero and the higher type to one, with its conception loading positive. When the original fitted conception loading is negative, both loadings are reversed and the shares exchanged for this table. The exit loading can be negative. Ratios are the exponentiated loadings and compare the higher-conception type with the lower-conception type at the same covariates and pregnancy exposure. They are intensity ratios, not ratios of monthly event probabilities. Shares are estimated mixing probabilities, not observed group frequencies or classifications of individual women. Full sample allows a separate coefficient for exposure without a retained onset. Retained onset excludes those exposed histories.}
\end{threeparttable}
\end{table}

AIC and BIC compare model fit while penalizing additional parameters within a fixed sample (Table~\ref{tab:modelcriteria}). BIC uses the number of women because each woman supplies one independent likelihood contribution with a fixed latent type. Both criteria favor two types in the full and retained-onset samples. Neither criterion tests anticipation or common shocks. The usual chi-square likelihood-ratio test is not valid for choosing the number of mixture types because its regularity conditions fail under the null.

\begin{table}[tbp]
\centering\singlespacing\small
\caption{Fit and Complexity in the Matched One-Type and Two-Type Models}\label{tab:modelcriteria}
\begin{threeparttable}
\begin{tabular*}{\textwidth}{@{\extracolsep{\fill}}@{}>{\raggedright\arraybackslash}p{1.65in}rrrrr@{}}
\toprule
Sample and controls & Types & Parameters & Log likelihood & AIC & BIC \\
\midrule
\cellstack[l]{Full sample\\Linear controls} & 1 & 8 & -5732.51 & 11481.02 & 11528.93 \\
 & 2 & 11 & -5718.69 & 11459.38 & 11525.26 \\
\addlinespace[5pt]
\cellstack[l]{Full sample\\Flexible controls} & 1 & 24 & -5714.83 & 11477.65 & 11621.41 \\
 & 2 & 27 & -5698.65 & 11451.30 & 11613.02 \\
\addlinespace[5pt]
\cellstack[l]{Retained onset\\Linear controls} & 1 & 7 & -5311.85 & 10637.69 & 10679.13 \\
 & 2 & 10 & -5297.92 & 10615.84 & 10675.03 \\
\addlinespace[5pt]
\cellstack[l]{Retained onset\\Flexible controls} & 1 & 23 & -5296.34 & 10638.68 & 10774.82 \\
 & 2 & 26 & -5280.13 & 10612.25 & 10766.15 \\
\bottomrule
\end{tabular*}
\notetext{Each adjacent pair uses identical observations and covariates in the conception and exit equations. The one-type joint likelihood is the sum of the separate conception and exit likelihoods. AIC equals minus twice the log likelihood plus twice the number of parameters. BIC replaces the last term with the number of parameters times the log number of women, the independent likelihood unit. Lower values indicate the preferred fit--complexity tradeoff within a comparison. AIC favors two types in every pair. BIC favors two types in every pair. Neither criterion tests the causal assumptions. An ordinary chi-square likelihood-ratio test across the number of types is not valid under the nonregular mixture null.}
\end{threeparttable}
\end{table}

Tables~\ref{tab:jointcoefflinear}--\ref{tab:jointcoeffflexexit} report every coefficient and its uncertainty, including the controls and latent distribution. They retain the original type labels, while the summary above orders types by conception propensity. Relabeling changes the intercepts, loading signs, and mixing logit together. It leaves the likelihood and pregnancy coefficient unchanged.

\begin{table}[tbp]
\centering\singlespacing\small
\caption{Complete Joint-Model Coefficients with Linear Controls}\label{tab:jointcoefflinear}
\begin{threeparttable}
\begin{tabular*}{\textwidth}{@{\extracolsep{\fill}}@{}>{\raggedright\arraybackslash}p{2.8in}rr@{}}
\toprule
Parameter & Full sample & Retained onset \\
\midrule
\textit{Conception equation} &  &  \\
Age centered at sixteen & \cellstack[r]{0.139\\(0.058)} & \cellstack[r]{0.130\\(0.058)} \\
Elapsed duration & \cellstack[r]{0.080\\(0.057)} & \cellstack[r]{0.080\\(0.057)} \\
Constant & \cellstack[r]{-1.194\\(0.293)} & \cellstack[r]{-5.302\\(0.175)} \\
\addlinespace[5pt]
\textit{Exit equation} &  &  \\
Age centered at sixteen & \cellstack[r]{0.292\\(0.041)} & \cellstack[r]{0.318\\(0.045)} \\
Elapsed duration & \cellstack[r]{0.035\\(0.039)} & \cellstack[r]{0.037\\(0.043)} \\
Exposure without retained onset & \cellstack[r]{1.533\\(0.121)} & -- \\
Constant & \cellstack[r]{-5.303\\(0.367)} & \cellstack[r]{-5.491\\(0.115)} \\
\addlinespace[5pt]
\textit{Pregnancy and latent distribution} &  &  \\
Pregnancy exposure, retained onset & \cellstack[r]{1.173\\(0.135)} & \cellstack[r]{1.127\\(0.137)} \\
Conception loading & \cellstack[r]{-4.132\\(0.238)} & \cellstack[r]{4.123\\(0.237)} \\
Exit loading & \cellstack[r]{-0.158\\(0.372)} & \cellstack[r]{0.230\\(0.380)} \\
Mixing logit & \cellstack[r]{3.476\\(0.260)} & \cellstack[r]{-3.462\\(0.259)} \\
\bottomrule
\end{tabular*}
\notetext{Cells report coefficients with model-based standard errors in parentheses. These are coefficients of the log intensity, not exponentiated hazard ratios. All equations belong to the same joint likelihood. The retained-onset sample excludes exposure histories without a conception-month observation. In the exit equation, an absent-onset exposure coefficient is included only in the full sample. Elapsed duration is calendar months since first eligible enrollment, capped at forty-eight months and divided by twelve. Exposure starts one month after conception. The original fitted normalization fixes type one at zero and type two at one. The mixing logit is the log odds of type two relative to type one. The tables reporting loadings and the logit retain their original labels. Table~\ref{tab:latenttypes} orients all models by conception propensity. }
\end{threeparttable}
\end{table}

\begin{table}[tbp]
\centering\singlespacing\small
\caption{Complete Conception-Equation Coefficients with Flexible Controls}\label{tab:jointcoeffflexconception}
\begin{threeparttable}
\begin{tabular*}{\textwidth}{@{\extracolsep{\fill}}@{}>{\raggedright\arraybackslash}p{2.8in}rr@{}}
\toprule
Parameter & Full sample & Retained onset \\
\midrule
Duration, months 7--12 & \cellstack[r]{0.862\\(0.248)} & \cellstack[r]{0.861\\(0.248)} \\
Duration, months 13--18 & \cellstack[r]{0.746\\(0.296)} & \cellstack[r]{0.744\\(0.297)} \\
Duration, months 19--24 & \cellstack[r]{0.748\\(0.305)} & \cellstack[r]{0.748\\(0.305)} \\
Duration, months 25--36 & \cellstack[r]{0.743\\(0.294)} & \cellstack[r]{0.745\\(0.294)} \\
Duration, months 37--48 & \cellstack[r]{0.935\\(0.306)} & \cellstack[r]{0.939\\(0.306)} \\
Duration, over 48 months & \cellstack[r]{0.944\\(0.315)} & \cellstack[r]{0.942\\(0.315)} \\
\addlinespace[5pt]
January--May & \cellstack[r]{0.099\\(0.111)} & \cellstack[r]{0.102\\(0.111)} \\
June--August & \cellstack[r]{0.116\\(0.152)} & \cellstack[r]{0.117\\(0.152)} \\
\addlinespace[5pt]
Age centered at sixteen & \cellstack[r]{-0.024\\(0.075)} & \cellstack[r]{-0.039\\(0.075)} \\
Centered age squared & \cellstack[r]{0.070\\(0.024)} & \cellstack[r]{0.073\\(0.024)} \\
Constant & \cellstack[r]{-6.025\\(0.297)} & \cellstack[r]{-6.002\\(0.298)} \\
\bottomrule
\end{tabular*}
\notetext{Cells report coefficients with model-based standard errors in parentheses. These are coefficients of the log intensity, not exponentiated hazard ratios. All equations belong to the same joint likelihood. The retained-onset sample excludes exposure histories without a conception-month observation. In the exit equation, an absent-onset exposure coefficient is included only in the full sample. Reference categories are elapsed duration of zero through six months and September--December. Duration is calendar time since first eligible enrollment. The clock is defined before the post-debut restriction. Pregnancy exposure starts one month after conception. The associated exit equation and latent-distribution parameters appear in Table~\ref{tab:jointcoeffflexexit}. The original fitted normalization fixes type one at zero and type two at one. The mixing logit is the log odds of type two relative to type one. The tables reporting loadings and the logit retain their original labels. Table~\ref{tab:latenttypes} orients all models by conception propensity. }
\end{threeparttable}
\end{table}

\begin{table}[tbp]
\centering\singlespacing\footnotesize
\caption{Complete Exit and Latent-Distribution Coefficients with Flexible Controls}\label{tab:jointcoeffflexexit}
\begin{threeparttable}
\begin{tabular*}{\textwidth}{@{\extracolsep{\fill}}@{}>{\raggedright\arraybackslash}p{2.8in}rr@{}}
\toprule
Parameter & Full sample & Retained onset \\
\midrule
Duration, months 7--12 & \cellstack[r]{-0.524\\(0.224)} & \cellstack[r]{-0.314\\(0.271)} \\
Duration, months 13--18 & \cellstack[r]{-0.102\\(0.201)} & \cellstack[r]{-0.042\\(0.249)} \\
Duration, months 19--24 & \cellstack[r]{-0.140\\(0.197)} & \cellstack[r]{-0.079\\(0.242)} \\
Duration, months 25--36 & \cellstack[r]{0.063\\(0.175)} & \cellstack[r]{0.092\\(0.217)} \\
Duration, months 37--48 & \cellstack[r]{-0.108\\(0.186)} & \cellstack[r]{-0.021\\(0.226)} \\
Duration, over 48 months & \cellstack[r]{-0.011\\(0.190)} & \cellstack[r]{0.040\\(0.231)} \\
\addlinespace[5pt]
January--May & \cellstack[r]{-0.110\\(0.093)} & \cellstack[r]{-0.175\\(0.102)} \\
June--August & \cellstack[r]{0.180\\(0.116)} & \cellstack[r]{0.192\\(0.126)} \\
\addlinespace[5pt]
Age centered at sixteen & \cellstack[r]{0.296\\(0.072)} & \cellstack[r]{0.336\\(0.081)} \\
Centered age squared & \cellstack[r]{-0.002\\(0.019)} & \cellstack[r]{-0.005\\(0.021)} \\
Exposure without retained onset & \cellstack[r]{1.536\\(0.122)} & -- \\
Constant & \cellstack[r]{-5.276\\(0.160)} & \cellstack[r]{-5.345\\(0.199)} \\
\addlinespace[5pt]
\textit{Pregnancy and latent distribution} &  &  \\
Pregnancy exposure, retained onset & \cellstack[r]{1.171\\(0.136)} & \cellstack[r]{1.133\\(0.139)} \\
Conception loading & \cellstack[r]{4.175\\(0.223)} & \cellstack[r]{4.161\\(0.221)} \\
Exit loading & \cellstack[r]{0.129\\(0.335)} & \cellstack[r]{0.172\\(0.340)} \\
Mixing logit & \cellstack[r]{-3.337\\(0.228)} & \cellstack[r]{-3.321\\(0.227)} \\
\bottomrule
\end{tabular*}
\notetext{Cells report coefficients with model-based standard errors in parentheses. These are coefficients of the log intensity, not exponentiated hazard ratios. All equations belong to the same joint likelihood. The retained-onset sample excludes exposure histories without a conception-month observation. In the exit equation, an absent-onset exposure coefficient is included only in the full sample. Reference categories are elapsed duration of zero through six months and September--December. Duration is calendar time since first eligible enrollment. The clock is defined before the post-debut restriction. Pregnancy exposure starts one month after conception. The original fitted normalization fixes type one at zero and type two at one. The mixing logit is the log odds of type two relative to type one. The tables reporting loadings and the logit retain their original labels. Table~\ref{tab:latenttypes} orients all models by conception propensity. }
\end{threeparttable}
\end{table}

The pooled joint model assigns one coefficient to all exposed months (Table~\ref{tab:onsetcomparison}). The split model estimates a separate coefficient for women without a retained onset. Both use the entire post-debut recorded-exit sample. The coefficient is larger when onset is absent. This difference could reflect selection or timing measurement, and the comparison cannot distinguish them.

\begin{table}[tbp]
\centering\singlespacing\small
\caption{Pooled and Separate Exposure Coefficients}\label{tab:onsetcomparison}
\begin{threeparttable}
\begin{tabular*}{\textwidth}{@{\extracolsep{\fill}}@{}lrr@{}}
\toprule
Specification and exposure term & HR & 95\% CI \\
\midrule
Joint model, pooled exposure & 3.91 & [3.22, 4.75] \\
Joint model, retained onset & 3.23 & [2.48, 4.21] \\
Joint model, onset absent & 4.63 & [3.65, 5.88] \\
Single equation, retained onset & 3.34 & [2.71, 4.10] \\
Single equation, onset absent & 4.66 & [3.64, 5.97] \\
\bottomrule
\end{tabular*}
\notetext{All models use 2,951 women, 61,649 person-months, and 594 exits. The pooled two-type joint model estimates one exposure coefficient. The next two rows come from one two-type joint model with separate exposure coefficients. The final two rows come from one single-equation model with the same split. Onset absent refers only to exposed women without a retained onset. Joint models use linear age and capped elapsed duration. Single-equation models use duration categories, school term, and quadratic age. Joint intervals use model-based standard errors. Single-equation intervals cluster by woman.}
\end{threeparttable}
\end{table}

Separate exposure coefficients do not restore missing conception events. The \EERsample{onset.exposed_at_entry.women} women exposed at entry contribute no conception-risk observations. The \EERsample{onset.missing_onset.women} who become exposed across a gap contribute \EERsample{onset.missing_onset.pregnancy_risk_months} earlier risk months but no conception event. Women with retained onset and later exposure contribute both earlier risk months and the event. Some never-exposed women conceive in their last retained month, including women who leave school or receive a credential in that month. The one-month lag leaves that month unexposed, and the conception still enters the conception likelihood.

The retained-onset estimates exclude all observations for the \EERsample{onset.without_retained_onset_women} exposed women without a conception-month observation. They keep \EERsample{onset.retained_onset.women} exposed women with retained onset and \EERsample{onset.never_exposed.women} never-exposed women. The resulting sample contains \EERsample{restricted.women} women, \EERsample{restricted.months} person-months, and \EERsample{restricted.exits} recorded exits. Reported sample sizes include the never-exposed women as well as the women supplying exposed months.

We also estimate single-equation models with four periods since conception (Table~\ref{tab:dynamics}). The reference group includes all unexposed months, including months before conception for women later exposed. Every exposed month enters a period, and the final period begins at month 13 with no upper limit. These estimates use the single-equation controls and do not condition on pregnancy outcome or impose the birth-date agreement rule.

\begin{table}[tbp]
\centering\singlespacing\small
\caption{Recorded Exit by Time Since Measured Conception}\label{tab:dynamics}
\begin{threeparttable}
\begin{tabular*}{\textwidth}{@{\extracolsep{\fill}}@{}lrrr@{}}
\toprule
Months since conception & HR & 95\% CI & Exits \\
\midrule
Months 1--3 & 4.40 & [3.11, 6.22] & 36 \\
Months 4--6 & 4.43 & [3.08, 6.38] & 32 \\
Months 7--12 & 4.40 & [3.28, 5.91] & 54 \\
Month 13 onward & 3.22 & [2.60, 3.99] & 124 \\
\bottomrule
\end{tabular*}
\notetext{Complementary log-log model on 61,649 post-debut person-months with 594 exits. The reference category is unexposed person-months. All exposed months enter a period, including months of women whose conception-month row is absent. The model includes duration categories, school term, and quadratic age, with standard errors clustered by woman. Periods measure time since conception, not gestation or time since birth. The final period is open ended. These are single-equation estimates pooling the onset groups, rather than dynamic joint-model estimates.}
\end{threeparttable}
\end{table}

The lag comparison starts exposure zero, one, two, or three months after measured conception (Table~\ref{tab:lag}). Each model uses the pooled single-equation specification and the same post-debut observations. Only conceptions before recorded exit or a credential define exposure, as in the main sample, so a conception in the exit or credential month stays unexposed at every lag. Shifting the threshold changes which exits count as exposed but leaves their dates unchanged. The exercise assesses exposure coding within fixed event histories. Reported standard errors apply to log hazard ratios.

\begin{table}[tbp]
\centering\singlespacing\small
\caption{Sensitivity to the Exposure Lag}\label{tab:lag}
\begin{threeparttable}
\begin{tabular*}{\textwidth}{@{\extracolsep{\fill}}@{}lrrrrr@{}}
\toprule
Lag & HR & 95\% CI & SE (log HR) & Person-months & Exits \\
\midrule
0 months & 3.55 & [2.99, 4.23] & (0.089) & 61,649 & 594 \\
1 month (baseline) & 3.75 & [3.15, 4.47] & (0.090) & 61,649 & 594 \\
2 months & 3.57 & [2.99, 4.27] & (0.091) & 61,649 & 594 \\
3 months & 3.32 & [2.77, 3.97] & (0.092) & 61,649 & 594 \\
\bottomrule
\end{tabular*}
\notetext{Each row uses the same post-debut risk set in a pooled single-equation complementary log-log model. Exposure begins the stated number of months after measured conception. Only conceptions before recorded exit or a credential define exposure, so a conception in the exit or credential month stays unexposed at every lag. All models include duration categories, school term, and quadratic age. Standard errors cluster by woman. SE is the standard error of the log hazard ratio. This comparison does not re-estimate the joint retained-onset model.}
\end{threeparttable}
\end{table}

We separately omit observations near measured conception in the broad sample, including months before sexual debut (Table~\ref{tab:omission}). Each row removes the indicated months and estimates a pooled exposure coefficient with seven duration categories, three school terms, and quadratic age. All models retain \EERsample{broad.women} women and cluster standard errors by woman. Windows are counted from conceptions before recorded exit or a credential, so a conception in the exit or credential month defines no window. The four-month window removes the month before conception through the second month afterward, and Table~\ref{tab:omission} reports the resulting person-months and recorded exits.

\begin{table}[tbp]
\centering\singlespacing\small
\caption{Omitting Months Near Measured Conception}\label{tab:omission}
\begin{threeparttable}
\begin{tabular*}{\textwidth}{@{\extracolsep{\fill}}@{}lrrrr@{}}
\toprule
Months omitted & Person-months & Exits & HR & 95\% CI \\
\midrule
None & 147,956 & 680 & 5.67 & [4.73, 6.80] \\
$0$ & 147,531 & 680 & 5.64 & [4.71, 6.77] \\
$0,1$ & 147,098 & 666 & 5.56 & [4.61, 6.69] \\
$-1,0,1$ & 146,674 & 666 & 5.53 & [4.59, 6.66] \\
$-1,0,1,2$ & 146,255 & 649 & 5.29 & [4.37, 6.40] \\
\bottomrule
\end{tabular*}
\notetext{Month zero is measured conception. Negative values precede conception and positive values follow it. Windows are counted from conceptions before recorded exit or a credential, so a conception in the exit or credential month defines no window. Each row fits a pooled complementary log-log model to the broad sample, including months before sexual debut. Exposure begins one month after conception. All specifications include duration categories, school term, and quadratic age, with standard errors clustered by woman. The omitted windows change the estimation sample. This exercise assesses sensitivity to observations near measured onset and does not re-estimate the joint model or remove persistent omitted shocks.}
\end{threeparttable}
\end{table}

Omitting months near conception assesses whether the recorded-exit association depends on observations close to the exposure threshold. A shared shock could affect schooling beyond the omitted months, so this comparison does not rule out coincident changes in pregnancy and exit risk.

The cause-specific models estimate exit without a credential and completion separately in the broad sample (Table~\ref{tab:completion}). Controls are linear age, six-month duration categories capped at the eighth category, and a September--May indicator. Standard errors cluster by woman. These models include neither a joint conception equation nor the main model's latent types. The completion coefficient is therefore a conditional association among students still enrolled.

\begin{table}[tbp]
\centering\singlespacing\small
\caption{Pregnancy Exposure and Two Ways of Leaving the Risk Set}\label{tab:completion}
\begin{threeparttable}
\begin{tabular*}{\textwidth}{@{\extracolsep{\fill}}@{}>{\raggedright\arraybackslash}p{3.0in}rrr@{}}
\toprule
Outcome & HR & 95\% CI & Events \\
\midrule
Recorded exit without a credential & 5.55 & [4.61, 6.67] & 680 \\
Diploma or GED completion & 0.44 & [0.35, 0.56] & 3,232 \\
\bottomrule
\end{tabular*}
\notetext{Separate cause-specific complementary log-log models use 4,235 women and 147,956 person-months in the broad sample. Each model treats the other outcome as a competing exit. Exposure begins one month after measured conception. Controls are centered linear age, eight duration categories with six-month intervals and a final open category, and a September--May indicator. Confidence intervals use standard errors clustered by woman. These pooled models include months before sexual debut and do not jointly model conception. The estimates describe exit intensities among women still enrolled, rather than effects on eventual credential attainment.}
\end{threeparttable}
\end{table}

\section{Background Adjustment and Schooling Risk}

\subsection{Covariate measurement and sample selection}

The full-sample adjustment adds four combined race and ethnicity categories and calendar-period indicators. We pool the sparse final three years and use the same calendar categories for both interruption definitions. Supported histories require a retained onset and three earlier school-risk months for women with conception or exposure during observation. Women without either remain in the comparison group. This requirement establishes time at risk before conception. The separate recent-enrollment check instead requires recorded enrollment in the three preceding school months.

Household composition is measured at the first interview. We combine the original responses into both biological parents, other two-parent households, and other arrangements. Maternal education uses the mother's highest grade completed, including information on nonresident biological mothers. Out-of-range values and ungraded schooling are missing. The combined ASVAB math--verbal percentile has three implied decimal places in the raw data. Its regression coefficient is scaled to ten percentile points.

The richer comparison starts after both the first interview and June 1998, a conservative endpoint for ASVAB testing. We exclude women whose measured conception falls on or before the later of these dates. After removing summer observations, school-risk months begin in September 1998. Background is therefore measured before retained risk months and conception, although prior schooling may still affect the test score.

Among the \EERadj{adjusted_counts.school3.later_eligible_women.value} eligible women in the three-school-month sample, maternal education is missing for \EERadj{adjusted_counts.school3.later_missing_mother_grade97.value}, household composition for \EERadj{adjusted_counts.school3.later_missing_household_group.value}, and ASVAB for \EERadj{adjusted_counts.school3.later_missing_asvab10.value}. These groups overlap. The complete-case sample contains \EERadj{adjusted_results.school3.later_cc_rich.women} women. A separate demographic model retains all later-eligible women, allowing us to assess selection into complete cases.

Each baseline and adjusted comparison uses the same observations, as do all three complete-case specifications. All sixteen models converge and retain the expected estimation observations. Tables~\ref{tab:adjcoeffschool3main}--\ref{tab:adjcoeffcalendar4later} report every coefficient and reference category, with standard errors clustered by woman.

\begin{table}[tbp]
\centering\singlespacing\footnotesize
\caption{Interruption Coefficients for Three School Months}\label{tab:adjcoeffschool3main}
\begin{threeparttable}
\begin{tabular*}{\textwidth}{@{\extracolsep{\fill}}@{}>{\raggedright\arraybackslash}p{1.95in}rrrr@{}}
\toprule
Covariate & \shortstack{Post-debut\\Baseline} & \shortstack{Post-debut\\Race and\\calendar} & \shortstack{Supported history\\Baseline} & \shortstack{Supported history\\Race and\\calendar} \\
\midrule
Pregnancy exposure & 1.348 (0.081) & 1.378 (0.085) & 1.290 (0.112) & 1.331 (0.116) \\
Duration, months 0--6 & Reference & Reference & Reference & Reference \\
Duration, months 7--12 & -0.526 (0.192) & -0.761 (0.229) & -0.427 (0.239) & -0.934 (0.290) \\
Duration, months 13--18 & 0.141 (0.172) & -0.491 (0.252) & 0.140 (0.221) & -0.946 (0.318) \\
Duration, months 19--24 & -0.090 (0.168) & -0.941 (0.262) & -0.026 (0.214) & -1.410 (0.335) \\
Duration, months 25--36 & -0.048 (0.154) & -1.292 (0.290) & 0.042 (0.199) & -1.905 (0.371) \\
Duration, months 37--48 & -0.052 (0.164) & -1.577 (0.336) & 0.066 (0.208) & -2.377 (0.420) \\
Duration, over 48 months & -0.020 (0.170) & -1.787 (0.396) & 0.071 (0.216) & -2.809 (0.488) \\
September--December & Reference & Reference & Reference & Reference \\
January--May & -0.644 (0.078) & -0.904 (0.096) & -0.590 (0.088) & -0.981 (0.105) \\
Age centered at sixteen & 0.207 (0.060) & 0.206 (0.062) & 0.188 (0.067) & 0.204 (0.067) \\
Centered age squared & 0.019 (0.019) & 0.009 (0.020) & 0.019 (0.022) & 0.006 (0.023) \\
Black & -- & -0.217 (0.092) & -- & -0.198 (0.102) \\
Hispanic & -- & 0.157 (0.099) & -- & 0.158 (0.110) \\
Mixed race, non-Hispanic & -- & 0.029 (0.293) & -- & -0.220 (0.350) \\
Non-Black, non-Hispanic & -- & Reference & -- & Reference \\
Calendar 1997 & -- & Reference & -- & Reference \\
Calendar 1998 & -- & 0.731 (0.236) & -- & 1.149 (0.293) \\
Calendar 1999 & -- & 1.334 (0.286) & -- & 1.957 (0.359) \\
Calendar 2000 & -- & 1.682 (0.331) & -- & 2.510 (0.403) \\
Calendar 2001 & -- & 1.873 (0.388) & -- & 2.927 (0.458) \\
Calendar 2002 & -- & 2.093 (0.435) & -- & 3.236 (0.518) \\
Calendar 2003--2005 & -- & 2.290 (0.496) & -- & 3.452 (0.576) \\
Constant & -4.441 (0.131) & -4.505 (0.149) & -4.513 (0.169) & -4.562 (0.186) \\
\addlinespace[5pt]
Women & 2,828 & 2,828 & 2,571 & 2,571 \\
School months & 48,953 & 48,953 & 45,471 & 45,471 \\
Interruptions & 711 & 711 & 570 & 570 \\
\bottomrule
\end{tabular*}
\notetext{Entries are log-intensity coefficients with woman-clustered standard errors in parentheses. Reference denotes the omitted category. A dash denotes a control absent from that specification. Duration advances in calendar months from first eligible enrollment. The main text's observed-background table defines the sample restrictions and controls. These coefficients correspond to the hazard ratios reported there.}
\end{threeparttable}
\end{table}

\begin{table}[tbp]
\centering\singlespacing\footnotesize
\caption{Interruption Coefficients for Three School Months in the Later Sample}\label{tab:adjcoeffschool3later}
\begin{threeparttable}
\begin{tabular*}{\textwidth}{@{\extracolsep{\fill}}@{}>{\raggedright\arraybackslash}p{1.95in}rrrr@{}}
\toprule
Covariate & \shortstack{All later women\\Race and\\calendar} & \shortstack{Complete cases\\Baseline} & \shortstack{Complete cases\\Race and\\calendar} & \shortstack{Complete cases\\Background\\added} \\
\midrule
Pregnancy exposure & 1.259 (0.111) & 1.262 (0.136) & 1.308 (0.145) & 1.046 (0.150) \\
Duration, months 0--6 & Reference & Reference & Reference & Reference \\
Duration, months 7--12 & -0.555 (0.711) & -1.052 (0.922) & -0.851 (0.904) & -0.533 (0.853) \\
Duration, months 13--18 & -0.252 (0.544) & -1.295 (0.674) & -1.042 (0.674) & -0.685 (0.600) \\
Duration, months 19--24 & -0.690 (0.536) & -1.371 (0.673) & -1.321 (0.710) & -0.984 (0.637) \\
Duration, months 25--36 & -0.958 (0.519) & -1.258 (0.664) & -1.732 (0.760) & -1.341 (0.698) \\
Duration, months 37--48 & -1.245 (0.512) & -1.200 (0.668) & -2.219 (0.833) & -1.747 (0.785) \\
Duration, over 48 months & -1.422 (0.522) & -1.186 (0.674) & -2.629 (0.925) & -2.146 (0.895) \\
September--December & Reference & Reference & Reference & Reference \\
January--May & -0.821 (0.118) & -0.579 (0.118) & -0.938 (0.157) & -0.863 (0.157) \\
Age centered at sixteen & 0.131 (0.084) & -0.014 (0.094) & 0.002 (0.097) & 0.066 (0.100) \\
Centered age squared & 0.033 (0.026) & 0.081 (0.028) & 0.077 (0.030) & 0.037 (0.032) \\
Black & -0.210 (0.108) & -- & -0.339 (0.139) & -0.799 (0.146) \\
Hispanic & 0.134 (0.115) & -- & 0.124 (0.145) & -0.235 (0.159) \\
Mixed race, non-Hispanic & -0.145 (0.424) & -- & -0.625 (0.571) & -0.695 (0.615) \\
Non-Black, non-Hispanic & Reference & -- & Reference & Reference \\
Calendar 1998 & Reference & -- & Reference & Reference \\
Calendar 1999 & 0.523 (0.221) & -- & 0.737 (0.296) & 0.700 (0.302) \\
Calendar 2000 & 0.917 (0.288) & -- & 1.243 (0.394) & 1.162 (0.402) \\
Calendar 2001 & 1.115 (0.361) & -- & 1.830 (0.521) & 1.720 (0.533) \\
Calendar 2002 & 1.284 (0.426) & -- & 1.941 (0.611) & 1.817 (0.632) \\
Calendar 2003--2005 & 1.448 (0.493) & -- & 2.086 (0.694) & 1.899 (0.715) \\
Mother's highest grade completed & -- & -- & -- & -0.044 (0.027) \\
Both biological parents & -- & -- & -- & Reference \\
Other two-parent household & -- & -- & -- & 0.435 (0.175) \\
Other household arrangement & -- & -- & -- & 0.468 (0.134) \\
ASVAB, ten percentile points & -- & -- & -- & -0.167 (0.026) \\
Constant & -4.032 (0.533) & -3.257 (0.645) & -3.515 (0.648) & -2.756 (0.633) \\
\addlinespace[5pt]
Women & 2,313 & 1,733 & 1,733 & 1,733 \\
School months & 35,595 & 26,777 & 26,777 & 26,777 \\
Interruptions & 521 & 336 & 336 & 336 \\
\bottomrule
\end{tabular*}
\notetext{Entries are log-intensity coefficients with woman-clustered standard errors in parentheses. Reference denotes the omitted category. A dash denotes a control absent from that specification. Duration advances in calendar months from first eligible enrollment. The main text's observed-background table defines the sample restrictions and controls. The later complete-case sample uses information measured before its retained risk months and conception. These coefficients correspond to the hazard ratios reported there.}
\end{threeparttable}
\end{table}

\begin{table}[tbp]
\centering\singlespacing\footnotesize
\caption{Interruption Coefficients for Four Calendar Months}\label{tab:adjcoeffcalendar4main}
\begin{threeparttable}
\begin{tabular*}{\textwidth}{@{\extracolsep{\fill}}@{}>{\raggedright\arraybackslash}p{1.95in}rrrr@{}}
\toprule
Covariate & \shortstack{Post-debut\\Baseline} & \shortstack{Post-debut\\Race and\\calendar} & \shortstack{Supported history\\Baseline} & \shortstack{Supported history\\Race and\\calendar} \\
\midrule
Pregnancy exposure & 1.313 (0.080) & 1.332 (0.084) & 1.242 (0.113) & 1.271 (0.115) \\
Duration, months 0--6 & Reference & Reference & Reference & Reference \\
Duration, months 7--12 & -0.561 (0.191) & -0.858 (0.224) & -0.458 (0.236) & -0.995 (0.279) \\
Duration, months 13--18 & 0.227 (0.166) & -0.475 (0.238) & 0.281 (0.210) & -0.800 (0.300) \\
Duration, months 19--24 & -0.115 (0.167) & -0.969 (0.250) & -0.054 (0.211) & -1.350 (0.321) \\
Duration, months 25--36 & -0.048 (0.152) & -1.221 (0.280) & 0.052 (0.195) & -1.687 (0.364) \\
Duration, months 37--48 & -0.002 (0.161) & -1.436 (0.327) & 0.125 (0.204) & -2.028 (0.420) \\
Duration, over 48 months & -0.019 (0.168) & -1.688 (0.386) & 0.099 (0.212) & -2.457 (0.489) \\
September--December & Reference & Reference & Reference & Reference \\
January--May & -0.663 (0.078) & -0.896 (0.096) & -0.622 (0.087) & -0.953 (0.107) \\
Age centered at sixteen & 0.180 (0.057) & 0.177 (0.058) & 0.156 (0.063) & 0.168 (0.062) \\
Centered age squared & 0.024 (0.019) & 0.016 (0.020) & 0.026 (0.021) & 0.016 (0.022) \\
Black & -- & -0.172 (0.090) & -- & -0.145 (0.099) \\
Hispanic & -- & 0.170 (0.098) & -- & 0.142 (0.110) \\
Mixed race, non-Hispanic & -- & 0.174 (0.294) & -- & -0.041 (0.331) \\
Non-Black, non-Hispanic & -- & Reference & -- & Reference \\
Calendar 1997 & -- & Reference & -- & Reference \\
Calendar 1998 & -- & 0.832 (0.228) & -- & 1.185 (0.281) \\
Calendar 1999 & -- & 1.288 (0.280) & -- & 1.801 (0.351) \\
Calendar 2000 & -- & 1.582 (0.325) & -- & 2.231 (0.403) \\
Calendar 2001 & -- & 1.801 (0.378) & -- & 2.618 (0.461) \\
Calendar 2002 & -- & 2.004 (0.430) & -- & 2.933 (0.523) \\
Calendar 2003--2005 & -- & 2.098 (0.504) & -- & 3.006 (0.589) \\
Constant & -4.366 (0.128) & -4.466 (0.147) & -4.448 (0.165) & -4.533 (0.185) \\
\addlinespace[5pt]
Women & 2,809 & 2,809 & 2,553 & 2,553 \\
School months & 48,120 & 48,120 & 44,724 & 44,724 \\
Interruptions & 726 & 726 & 586 & 586 \\
\bottomrule
\end{tabular*}
\notetext{Entries are log-intensity coefficients with woman-clustered standard errors in parentheses. Reference denotes the omitted category. A dash denotes a control absent from that specification. Duration advances in calendar months from first eligible enrollment. The main text's observed-background table defines the sample restrictions and controls. These coefficients correspond to the hazard ratios reported there.}
\end{threeparttable}
\end{table}

\begin{table}[tbp]
\centering\singlespacing\footnotesize
\caption{Interruption Coefficients for Four Calendar Months in the Later Sample}\label{tab:adjcoeffcalendar4later}
\begin{threeparttable}
\begin{tabular*}{\textwidth}{@{\extracolsep{\fill}}@{}>{\raggedright\arraybackslash}p{1.95in}rrrr@{}}
\toprule
Covariate & \shortstack{All later women\\Race and\\calendar} & \shortstack{Complete cases\\Baseline} & \shortstack{Complete cases\\Race and\\calendar} & \shortstack{Complete cases\\Background\\added} \\
\midrule
Pregnancy exposure & 1.196 (0.111) & 1.155 (0.138) & 1.187 (0.145) & 0.903 (0.151) \\
Duration, months 0--6 & Reference & Reference & Reference & Reference \\
Duration, months 7--12 & -0.467 (0.640) & -0.688 (0.820) & -0.555 (0.793) & -0.196 (0.731) \\
Duration, months 13--18 & -0.255 (0.498) & -1.050 (0.656) & -0.882 (0.636) & -0.472 (0.554) \\
Duration, months 19--24 & -0.800 (0.496) & -1.437 (0.662) & -1.391 (0.676) & -1.002 (0.594) \\
Duration, months 25--36 & -0.873 (0.477) & -1.229 (0.653) & -1.536 (0.724) & -1.091 (0.654) \\
Duration, months 37--48 & -1.055 (0.474) & -1.143 (0.656) & -1.867 (0.793) & -1.326 (0.736) \\
Duration, over 48 months & -1.255 (0.490) & -1.134 (0.662) & -2.187 (0.881) & -1.625 (0.842) \\
September--December & Reference & Reference & Reference & Reference \\
January--May & -0.764 (0.121) & -0.541 (0.117) & -0.796 (0.161) & -0.715 (0.160) \\
Age centered at sixteen & 0.096 (0.079) & -0.034 (0.086) & -0.029 (0.088) & 0.037 (0.091) \\
Centered age squared & 0.043 (0.026) & 0.086 (0.028) & 0.087 (0.029) & 0.043 (0.031) \\
Black & -0.161 (0.105) & -- & -0.240 (0.134) & -0.778 (0.141) \\
Hispanic & 0.126 (0.115) & -- & 0.146 (0.144) & -0.255 (0.157) \\
Mixed race, non-Hispanic & 0.063 (0.391) & -- & -0.170 (0.470) & -0.194 (0.447) \\
Non-Black, non-Hispanic & Reference & -- & Reference & Reference \\
Calendar 1998 & Reference & -- & Reference & Reference \\
Calendar 1999 & 0.328 (0.214) & -- & 0.480 (0.288) & 0.432 (0.293) \\
Calendar 2000 & 0.623 (0.288) & -- & 0.831 (0.395) & 0.735 (0.399) \\
Calendar 2001 & 0.818 (0.361) & -- & 1.323 (0.527) & 1.197 (0.532) \\
Calendar 2002 & 0.956 (0.432) & -- & 1.411 (0.619) & 1.270 (0.631) \\
Calendar 2003--2005 & 1.002 (0.511) & -- & 1.365 (0.716) & 1.134 (0.725) \\
Mother's highest grade completed & -- & -- & -- & -0.039 (0.026) \\
Both biological parents & -- & -- & -- & Reference \\
Other two-parent household & -- & -- & -- & 0.326 (0.179) \\
Other household arrangement & -- & -- & -- & 0.534 (0.131) \\
ASVAB, ten percentile points & -- & -- & -- & -0.193 (0.026) \\
Constant & -3.847 (0.487) & -3.227 (0.633) & -3.402 (0.613) & -2.640 (0.588) \\
\addlinespace[5pt]
Women & 2,291 & 1,715 & 1,715 & 1,715 \\
School months & 34,828 & 26,147 & 26,147 & 26,147 \\
Interruptions & 530 & 346 & 346 & 346 \\
\bottomrule
\end{tabular*}
\notetext{Entries are log-intensity coefficients with woman-clustered standard errors in parentheses. Reference denotes the omitted category. A dash denotes a control absent from that specification. Duration advances in calendar months from first eligible enrollment. The main text's observed-background table defines the sample restrictions and controls. The later complete-case sample uses information measured before its retained risk months and conception. These coefficients correspond to the hazard ratios reported there.}
\end{threeparttable}
\end{table}

\FloatBarrier
\subsection{Monthly rates and nine-month follow-up}

Observed monthly rates use the same post-debut school months as the main interruption models (Table~\ref{tab:interruptionrates}). Women can contribute to both exposure states, so their counts overlap. Each risk month and event belongs to only one exposure state, and those counts sum to the fitted sample totals.

\begin{table}[tbp]
\centering\singlespacing\small
\caption{Observed Monthly Interruption Rates}\label{tab:interruptionrates}
\begin{threeparttable}
\begin{tabular*}{\textwidth}{@{\extracolsep{\fill}}@{}>{\raggedright\arraybackslash}p{1.75in}lrrrr@{}}
\toprule
Definition & Exposure & Women & School months & Events & \shortstack{Events per\\1,000 months} \\
\midrule
Three school months & Unexposed & 2,690 & 43,014 & 450 & 10.5 \\
 & Exposed & 562 & 5,939 & 261 & 43.9 \\
\addlinespace[5pt]
Four calendar months & Unexposed & 2,671 & 42,316 & 468 & 11.1 \\
 & Exposed & 549 & 5,804 & 258 & 44.5 \\
\bottomrule
\end{tabular*}
\notetext{The exact post-debut model samples supply these counts. Exposure begins one month after measured conception. Risk months are September--May. The underlying calendar clock continues through summer. Women can contribute to both exposure states, so the two women counts do not sum to the sample size. Rates are unadjusted events per 1,000 school months, not nine-month probabilities.}
\end{threeparttable}
\end{table}

For nine-month follow-up, each pregnant period starts one month after a retained onset. Comparison periods start in eligible months before conception or in months of women with no measured conception. We match exact calendar month and completed age in years. Within each matching group, every comparison period receives a weight equal to the number of pregnant periods divided by the number of comparison periods. We exclude groups lacking either. A woman can supply several comparison periods and later a pregnant period, so we report both periods and students. A school interruption can appear in more than one period for the same woman.

Both groups are followed from the first month of each period through eight calendar months afterward. Summer remains part of elapsed time even though interruptions cannot begin then. Later conception does not censor a comparison period. Interruption and credential receipt are competing events. Other censoring follows the original observation limits, including the confirmation allowance. We keep periods that end early rather than require complete follow-up at entry.

We update interruption incidence each month using the probability of remaining event-free through the previous month and the weighted interruption fraction among those still at risk. Credential incidence uses the same risk set. Event-free survival, interruption incidence, and credential incidence sum to one in every month. Censoring occurs after the last observed risk month. The nine-month risks require independent follow-up within the weighted groups. The extreme cases instead assign every period that ends early either no interruption or an interruption by the horizon.

The two interruption definitions give similar nine-month comparisons (Table~\ref{tab:landmarkscaleall}). Under both, the lower end of the pregnant group's range exceeds the upper end of the comparison group's range. Incomplete follow-up therefore cannot eliminate the descriptive difference under the recorded dates and endpoint rules. The extreme cases do not address selection into pregnancy, measurement error, or retrospective confirmation of an absence. We report no significance test for this standardized comparison with repeated periods.

\begin{table}[tbp]
\centering\singlespacing\footnotesize
\caption{Complete Accounting for the Nine-Month Schooling Comparison}\label{tab:landmarkscaleall}
\begin{threeparttable}
\begin{tabular*}{\textwidth}{@{\extracolsep{\fill}}@{}>{\raggedright\arraybackslash}p{2.0in}rrrr@{}}
\toprule
Measure & \shortstack{Three school\\months\\Pregnancy} & \shortstack{Three school\\months\\Comparison} & \shortstack{Four calendar\\months\\Pregnancy} & \shortstack{Four calendar\\months\\Comparison} \\
\midrule
Interruption incidence (\%) & 23.1 & 7.2 & 23.6 & 7.4 \\
Extreme cases for periods that end early (\%) & 21.7--29.9 & 7.1--9.0 & 22.4--29.5 & 7.4--8.9 \\
\addlinespace[5pt]
Follow-up periods & 451 & 42,909 & 437 & 41,951 \\
Students & 451 & 2,658 & 437 & 2,637 \\
Periods with an observed interruption & 98 & 2,941 & 98 & 3,000 \\
Periods with an observed credential & 89 & 11,218 & 85 & 10,988 \\
Periods that end before nine months & 37 & 484 & 31 & 454 \\
Periods with an endpoint or full follow-up & 414 & 42,425 & 406 & 41,497 \\
\addlinespace[5pt]
Observed interruption share, unweighted (\%) & 21.7 & 6.9 & 22.4 & 7.2 \\
Observed interruption share, weighted (\%) & 21.7 & 7.1 & 22.4 & 7.4 \\
Credential incidence (\%) & 20.9 & 27.0 & 20.4 & 26.6 \\
Event-free survival (\%) & 56.1 & 65.8 & 55.9 & 66.0 \\
\addlinespace[5pt]
Total initial weight & 451.000 & 451.000 & 437.000 & 437.000 \\
Weight of periods that end before nine months & 37.000 & 8.350 & 31.000 & 6.793 \\
\bottomrule
\end{tabular*}
\notetext{A pregnant student's period begins in the month after a retained onset. A comparison period begins in the same calendar month for a student who has not yet conceived. Comparison periods are weighted to reproduce the pregnant periods' completed age in years and starting calendar month, and each pregnant period has weight one. A student can contribute several comparison periods and can later conceive, so one interruption or credential can appear in several comparison periods. Each period covers the same nine calendar months, including summer. A comparison period stays in its group if the student conceives during it. A diploma or GED is a competing event. Periods end early under each outcome's observation rules. Cumulative incidence uses weighted discrete Aalen--Johansen accounting and assumes that periods that end early would follow the rate of periods in the same group that continue. The extreme cases assume that none or all of the periods that end early contain an interruption. Counts are unweighted unless stated. The comparisons are descriptive.}
\end{threeparttable}
\end{table}

\section[Summary Statistics, Heterogeneity, and Expectations]{Summary Statistics, Heterogeneity,\\ and Expectations}

The richer models and their ASVAB splits use women with all three background measures (Table~\ref{tab:summarystatslater}). Any measured conception follows both the first interview and the ASVAB field period. The richer expectation comparisons further require a reported pregnancy expectation.

\begin{table}[tbp]
\centering\singlespacing\small
\caption{Sample Characteristics by Pregnancy Exposure in the Later Complete Cases}\label{tab:summarystatslater}
\begin{threeparttable}
\begin{tabular*}{\textwidth}{@{\extracolsep{\fill}}@{}>{\raggedright\arraybackslash}p{3.2in}rrr@{}}
\toprule
Characteristic & \shortstack{Never exposed\\N = 1,511} & \shortstack{Ever exposed\\N = 222} & \shortstack{All women\\N = 1,733} \\
\midrule
Black (\%) & 25.2 & 47.3 & 28.0 \\
Hispanic (\%) & 17.8 & 21.2 & 18.2 \\
Mixed race, non-Hispanic (\%) & 0.9 & 0.9 & 0.9 \\
Non-Black, non-Hispanic (\%) & 56.1 & 30.6 & 52.8 \\
\addlinespace[5pt]
Mother's highest grade completed & 12.7 (2.7) & 11.5 (2.8) & 12.6 (2.8) \\
\addlinespace[5pt]
Both biological parents (\%) & 50.4 & 29.3 & 47.7 \\
Other two-parent household (\%) & 13.8 & 14.0 & 13.8 \\
Other household arrangement (\%) & 35.7 & 56.8 & 38.4 \\
\addlinespace[5pt]
ASVAB math--verbal percentile & 47.5 (27.5) & 27.6 (21.2) & 44.9 (27.6) \\
\addlinespace[5pt]
Age at first school-risk month & 16.7 (1.1) & 16.2 (1.1) & 16.7 (1.1) \\
School-risk months observed & 14.8 (9.3) & 20.0 (10.3) & 15.5 (9.6) \\
\addlinespace[5pt]
Interrupted schooling during observation (\%) & 16.7 & 37.8 & 19.4 \\
Diploma or GED by end of observation (\%) & 80.9 & 50.5 & 77.0 \\
\addlinespace[5pt]
Age at conception, exposed women & -- & 17.3 (1.2) & -- \\
\bottomrule
\end{tabular*}
\notetext{The sample contains the women in the later complete-case sample of the richer adjusted interruption model, whose school-risk months begin in September 1998 after the first interview and the ASVAB field period and who have no measured conception on or before the later of June 1998 and the first interview. A woman is ever exposed when at least one of her school-risk months in this sample falls after her measured conception, and never exposed otherwise. Entries are person-level percentages of the women in the column or means with standard deviations in parentheses. Race, household group, and mother's grade are measured at the first interview. ASVAB is the combined math and verbal percentile. Age at conception is reported for exposed women only. Every woman in this sample has all three background measures. Interruption and credential shares refer to each woman's retained observation window, so a credential earned after that window is not counted. Never exposed does not mean never pregnant, since 872 of the never-exposed women have a measured conception date. The differences describe the two exposure groups and are not effects of pregnancy.}
\end{threeparttable}
\end{table}

Women in the middle third of the ASVAB distribution have a larger estimated hazard ratio than women in the bottom third (Table~\ref{tab:heterogeneity}). We divide women in the later complete-case sample into three approximately equal groups by ASVAB percentile. We call these groups terciles and estimate the model with all background controls separately in each. The bottom tercile has a hazard ratio of \EERhet{tercile_rich.t1.hr} and an observed rate in months without pregnancy exposure of \EERhet{tercile_rich.t1.rate_unexposed_per_1000} interruptions per 1,000 school-risk months. The middle tercile has a ratio of \EERhet{tercile_rich.t2.hr} and a lower observed rate in months without pregnancy exposure of \EERhet{tercile_rich.t2.rate_unexposed_per_1000}. The top tercile has only \EERhet{tercile_rich.t3.exposed_events} interruptions during pregnancy exposure. We withhold its ratio because so few events do not support a reliable estimate.

Two tests disagree about whether the hazard ratios differ across the three groups. A separate-fit test, which compares the separate models, rejects equality at the five-percent level. A pooled-interaction test, which uses one model with a separate exposure coefficient for each tercile, does not. Both tests include the sparse top tercile, so neither tests only the middle-versus-bottom difference.\footnote{The separate-fit and pooled-interaction tests give p-values of \EERhet{test.separate_fits_equality.p} and \EERhet{test.pooled_interaction.p}, respectively. Both depend on the top tercile's sparse interruptions during exposure. The middle-tercile fit omits observations in covariate cells that perfectly predict no interruption. Table~\ref{tab:heterogeneity} reports eligible counts before these exclusions and explains the fitted sample.}

\begin{table}[tbp]
\centering\singlespacing\footnotesize
\caption{Pregnancy and Interruption by ASVAB Tercile}\label{tab:heterogeneity}
\begin{threeparttable}
\begin{tabular*}{\textwidth}{@{\extracolsep{\fill}}@{}>{\raggedright\arraybackslash}p{1.45in}rrrrrrr@{}}
\toprule
ASVAB tercile & HR & 95\% CI & Women & Events & \shortstack{Exposed\\events} & \shortstack{Unexposed\\per 1,000} & \shortstack{Exposed\\per 1,000} \\
\midrule
\textit{Separate fits by tercile} &  &  &  &  &  &  &  \\
Bottom tercile & 2.51 & [1.76, 3.60] & 578 & 184 & 57 & 15.6 & 43.2 \\
Middle tercile & 4.87$^{a}$ & [3.10, 7.64] & 578 & 106 & 25 & 9.6 & 45.5 \\
Top tercile & -- & -- & 577 & 46 & 2 & 5.4 & 10.6 \\
\addlinespace[5pt]
\textit{Pooled interaction model} &  &  &  &  &  &  &  \\
Bottom tercile (reference) & 2.48 & [1.71, 3.61] & 1,733 & 336 & 84 &  &  \\
Middle tercile & 4.44 & [2.83, 6.97] &  &  &  &  &  \\
Top tercile & -- & -- &  &  &  &  &  \\
\addlinespace[5pt]
Joint interaction test, p & 0.080 &  &  &  &  &  &  \\
Equal separate ratios, p & 0.031 &  &  &  &  &  &  \\
\bottomrule
\end{tabular*}
\notetext{Complementary log-log interruption models for three school months of nonenrollment on the later complete-case sample, with woman-clustered 95\% confidence intervals. Terciles are cut on women by the ASVAB math and verbal percentile measured before the retained risk months, with upper cutoffs of 28.5 and 59.2 percentile points, and they hold 578, 578, and 577 women. The separate fits use the full richer specification within each tercile. The pooled model fits the same controls on all 1,733 women with exposure interacted with tercile, and the bottom tercile is its reference. The joint test is the Wald test of the two interaction terms with the clustered covariance, on 2 degrees of freedom. The equality test compares the three separate estimates using their clustered variances, on 2 degrees of freedom. Rates are interruptions per 1,000 school-risk months in the unexposed and exposed months of each tercile. Women and event counts describe the eligible sample before perfect-prediction exclusions. These exclusions can remove entire women as well as months. A superscript a marks a fit in which Stata omitted cells that predict no interruption, fitted on 8,897 of 8,990 school months in the middle tercile. A dash withholds a ratio that rests on fewer than five exposed interruptions.}
\end{threeparttable}
\end{table}

Pregnant students in the bottom and middle ASVAB groups have similar nine-month interruption shares (Table~\ref{tab:heterogeneitylandmark}). Within each tercile of the later complete-case sample, we compare pregnant students with comparison students standardized to the same age and calendar month. In the bottom tercile, the estimated shares are \EERhet{landmark.t1.g1.cif_interruption_pct} percent for pregnant students and \EERhet{landmark.t1.g0.cif_interruption_pct} percent for comparison students. In the middle tercile, they are \EERhet{landmark.t2.g1.cif_interruption_pct} and \EERhet{landmark.t2.g0.cif_interruption_pct} percent. About one in four pregnant students interrupts in either group. The lower comparison share contributes to the larger relative difference over nine months in the middle tercile. These results remain descriptive because selection into pregnancy can explain part of the difference.

\begin{table}[tbp]
\centering\singlespacing\footnotesize
\caption{Nine-Month Interruption Incidence by ASVAB Tercile}\label{tab:heterogeneitylandmark}
\begin{threeparttable}
\begin{tabular*}{\textwidth}{@{\extracolsep{\fill}}@{}>{\raggedright\arraybackslash}p{1.4in}rrrrrr@{}}
\toprule
Measure & \shortstack{Bottom\\Pregnancy} & \shortstack{Bottom\\Comparison} & \shortstack{Middle\\Pregnancy} & \shortstack{Middle\\Comparison} & \shortstack{Top\\Pregnancy} & \shortstack{Top\\Comparison} \\
\midrule
Interruption incidence (\%) & 24.2 & 9.2 & 25.0 & 6.6 & 5.0 & 3.1 \\
Extreme cases (\%) & 21.8--33.8 & 9.0--11.4 & 24.3--30.0 & 6.6--7.2 & 5.0--5.0 & 3.1--3.5 \\
Credential incidence (\%) & 20.5 & 26.5 & 31.1 & 34.9 & 45.0 & 54.7 \\
\addlinespace[5pt]
Follow-up periods & 133 & 4,239 & 70 & 3,550 & 20 & 1,389 \\
Students & 133 & 532 & 70 & 548 & 20 & 509 \\
Periods with an observed interruption & 29 & 371 & 17 & 228 & 1 & 40 \\
Periods with an observed credential & 25 & 997 & 21 & 1,123 & 9 & 712 \\
Periods that end before nine months & 16 & 48 & 4 & 28 & 0 & 8 \\
\bottomrule
\end{tabular*}
\notetext{Each period covers nine calendar months in the later complete-case sample, by ASVAB tercile, for the three-school-month interruption. A pregnant student's period begins in the month after a retained onset, in a calendar month after the later of June 1998 and her first interview. Comparison periods begin in the same calendar months for women with no measured conception by then and are weighted within calendar month, completed age in years, and tercile to reproduce the pregnant periods' distribution. A student can contribute several comparison periods and can later conceive, so one interruption or credential can appear in several comparison periods. A comparison period stays in its group if the student conceives during it. A credential is a competing event. Incidence is the weighted discrete Aalen--Johansen cumulative incidence by month nine. The extreme cases assume that none or all of the periods that end before nine months contain an interruption. Counts are unweighted. Pregnant periods without an age and calendar match within their tercile are excluded, 2 in the bottom tercile, 0 in the middle, and 0 in the top. The comparison uses 161 matching blocks. These are descriptive comparisons without standard errors. The top tercile has 1 interruption among its 20 pregnant periods, so its incidence is not informative.}
\end{threeparttable}
\end{table}

The reported pregnancy expectations cannot reliably test anticipation in our sample (Table~\ref{tab:expectations}). In 1997, women aged 15 and older reported their percent chance of pregnancy within a year. These reports precede conception in the later observation period. They do not measure expectations immediately before each conception. Among all women eligible for that later period, \EERexp{count.later_all_core.missing.women} have no report. Only \EERexp{later_all_core.high.women} report a chance above 20 percent. They contribute only \EERexp{later_all_core.high.exposed_events} interruptions during pregnancy exposure. We divide women with reports by their stated chance, but the high-expectation group has too few such events to test the identifying condition reliably.

\begin{table}[tbp]
\centering\singlespacing\small
\caption{Pregnancy and Interruption by 1997 Pregnancy Expectation}\label{tab:expectations}
\begin{threeparttable}
\begin{tabular*}{\textwidth}{@{\extracolsep{\fill}}@{}>{\raggedright\arraybackslash}p{2.6in}rrrrr@{}}
\toprule
Sample and 1997 expectation & HR & 95\% CI & Women & Events & \shortstack{Exposed\\events} \\
\midrule
\textit{Later sample, race and calendar controls} &  &  &  &  &  \\
Low expectation, 0 to 20 percent & 4.21$^{a}$ & [2.45, 7.23] & 600 & 88 & 19 \\
High expectation, 21 to 100 percent & -- & -- & 78 & 20 & 2 \\
Both groups pooled & 3.00$^{a}$ & [1.83, 4.93] & 678 & 108 & 21 \\
Interaction of exposure and high expectation, p & 0.004 &  &  &  &  \\
\addlinespace[5pt]
\textit{Later complete cases, family background and ASVAB added} &  &  &  &  &  \\
Low expectation, 0 to 20 percent & 4.00$^{a}$ & [1.98, 8.06] & 452 & 50 & 11 \\
High expectation, 21 to 100 percent & -- & -- & 60 & 12 & 2 \\
Both groups pooled & 3.25$^{a}$ & [1.83, 5.75] & 512 & 62 & 13 \\
Interaction of exposure and high expectation, p & 0.082 &  &  &  &  \\
\bottomrule
\end{tabular*}
\notetext{Complementary log-log interruption models for three school months of nonenrollment on the later samples, with woman-clustered 95\% confidence intervals. The 1997 report is the respondent's stated percent chance of becoming pregnant within one year, asked of women aged 15 and older at that interview, and it precedes every retained onset in these samples. The first block uses all later-eligible women with baseline, race, and calendar controls. The second block uses the complete cases and adds mother's grade, household group, and ASVAB percentile. Each expectation row is a separate fit, and the pooled row fits all women with a report. The interaction test is the clustered Wald test of exposure interacted with the high group in a pooled model. Women without a report are absent from every row. There are 1,635 such women in the first sample, with 413 interruptions of which 115 are exposed, and 1,221 in the second, with 274 interruptions of which 71 are exposed. Women and event counts describe the eligible sample before perfect-prediction exclusions. In the reporting checks, the sparse-event screen requires at least five exposed interruptions, and the large-coefficient screen flags an absolute coefficient above 10 on any nonconstant term. The high-expectation estimate is withheld in the first block by the sparse-event screen (2 exposed interruptions), and in the second block by the sparse-event screen (2 exposed interruptions) and the large-coefficient screen. The interaction tests in the first and second blocks also depend on the high-expectation group's 2 exposed interruptions and should be interpreted cautiously. A superscript a marks a fit in which Stata omitted cells that predict no interruption, so the fitted sample can contain fewer women as well as fewer months. A dash marks an estimate withheld by the reporting checks for sparse exposed events or numerical reliability.}
\end{threeparttable}
\end{table}

\end{document}